\documentclass[12pt,a4paper,twoside,openright]{book}
\usepackage{amsmath}
\usepackage{amssymb}
\usepackage[dvips]{graphicx}
\usepackage{epsf}
\usepackage{hyperref}
\usepackage{slashed}
\usepackage{verbatim}
\usepackage{accents} 
\usepackage{simplewick}
\usepackage{cyrillic}
\usepackage{subfigure}
\usepackage{stmaryrd} 

\usepackage[czech,english]{babel}
\usepackage[cp1250]{inputenc}

\usepackage[intoc,noprefix]{nomencl}
\makenomenclature

\usepackage{wasysym} 

\newdimen\InnerMargin
\newdimen\OuterMargin

\oddsidemargin=\InnerMargin   
\evensidemargin=\OuterMargin  
\textwidth=\paperwidth
\advance\textwidth by-\InnerMargin
\advance\textwidth by-\OuterMargin

\newdimen\BottomMargin

\textheight=\paperheight
\advance\textheight by-\topmargin
\advance\textheight by-\headheight
\advance\textheight by-\headsep
\advance\textheight by-\topskip
\advance\textheight by-\footskip
\advance\textheight by-\BottomMargin

\newcommand{\bra}[1]{\langle#1\vert}
\newcommand{\ket}[1]{\vert#1\rangle}
\newcommand{\Tr}{\mathop{\rm Tr}\nolimits}
\newcommand{\adj}{\mathop{\rm adj}\nolimits}
\newcommand{\diag}{\mathop{\rm diag}\nolimits}
\newcommand{\I}{\ensuremath{\mathrm{i}}}
\newcommand{\e}{\ensuremath{\mathrm{e}}}
\renewcommand{\Re}{\mathop{\rm Re}\nolimits}
\renewcommand{\Im}{\mathop{\rm Im}\nolimits}
\newcommand{\eL}{\mathcal{L}}
\newcommand{\D}{\ensuremath{\mathrm{D}}}
\renewcommand{\d}{\ensuremath{\mathrm{d}}}
\newcommand{\hc}{\ensuremath{\mathrm{h.c.}}}
\newcommand{\cc}{\ensuremath{\mathrm{c.c.}}}
\newcommand{\threevector}[1]{\boldsymbol{#1}}
\newcommand{\qm}[1]{``#1''} 
\newcommand{\vecleftright}[1]{\accentset{\leftrightarrow}{#1}}

\newcommand{\exclamation}[1]{\stackrel{\text{!}}{#1}}
\newcommand{\cyrrm}{\fontencoding{OT2}\selectfont\textcyrup} 
\newcommand{\T}{\ensuremath{\mathrm{T}}}
\newcommand{\C}{\ensuremath{\mathrm{c}}}
\newcommand{\BR}{\mathop{\rm BR}\nolimits}
\newcommand{\MeV}{\ensuremath{\mathrm{\,MeV}}}
\newcommand{\GeV}{\ensuremath{\mathrm{\,GeV}}}

\newcommand{\bbl}{\ensuremath{\llbracket}}
\newcommand{\bbr}{\ensuremath{\rrbracket}}

\newcommand{\TransformsTo}{\ensuremath{\longrightarrow}}

\newcommand{\group}[1]{\mathbb{#1}}

\newcommand{\algebra}[1]{\mathfrak{#1}}

\DeclareFontFamily{OT1}{cmrx}{}
\DeclareFontShape{OT1}{cmrx}{m}{n}{<->cmr10}{}
\let\saveLongrightarrow\Longrightarrow
\makeatletter
\renewcommand*{\Longrightarrow}{%
    \mathrel{\rlap{\fontfamily{cmrx}\fontencoding{OT1}\selectfont=}%
    \hphantom{\saveLongrightarrow}%
    \llap{$\m@th\Rightarrow$}}}
\makeatother

\newcommand{\intro}[1]{\textsl{#1}}

\usepackage{bbm} 
\def\unitmatrix{\mathbbm{1}} 
\def\zeromatrix{\includegraphics[height=0.69em]{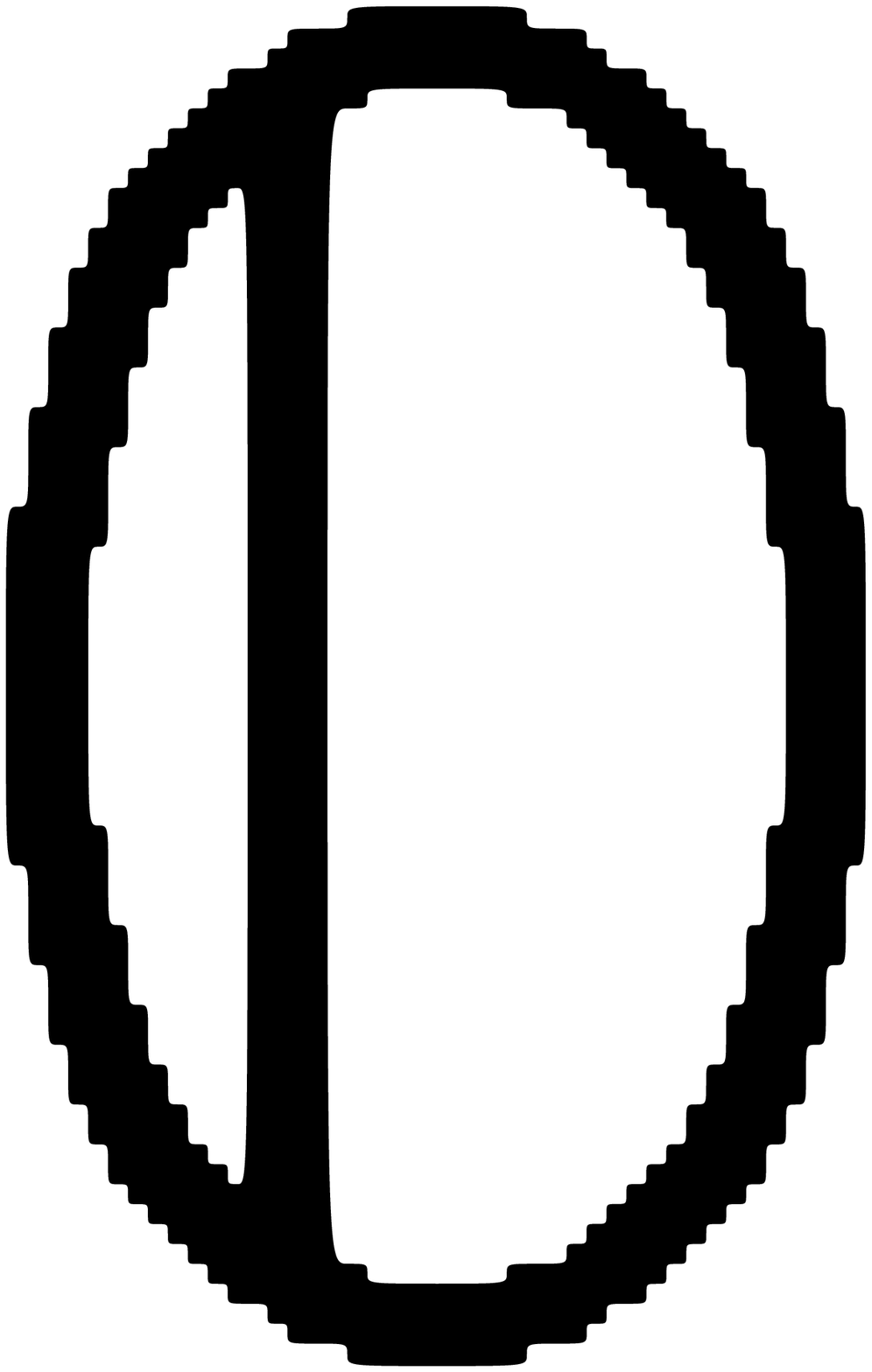}}

\newcommand{\chapterstar}[1]{
    \chapter*{#1}
    \addcontentsline{toc}{chapter}{#1}
    \markboth{\uppercase{#1}}{\uppercase{#1}}
    }

\begin{document}


\pagestyle{empty}

\begin{center}
{\Large \sc Charles University in~Prague\\
Faculty of Mathematics and Physics}

\vspace{0.1cm}

{\large and \\}

\vspace{0.1cm}

{\Large \sc Academy of Sciences of the Czech Republic\\
Nuclear Physics Institute}

\vspace{0.1cm}

\end{center}
\hrule

\vspace*{2cm}
\begin{figure}[h]
\centering \epsfxsize=0.55\hsize \mbox{\epsfbox{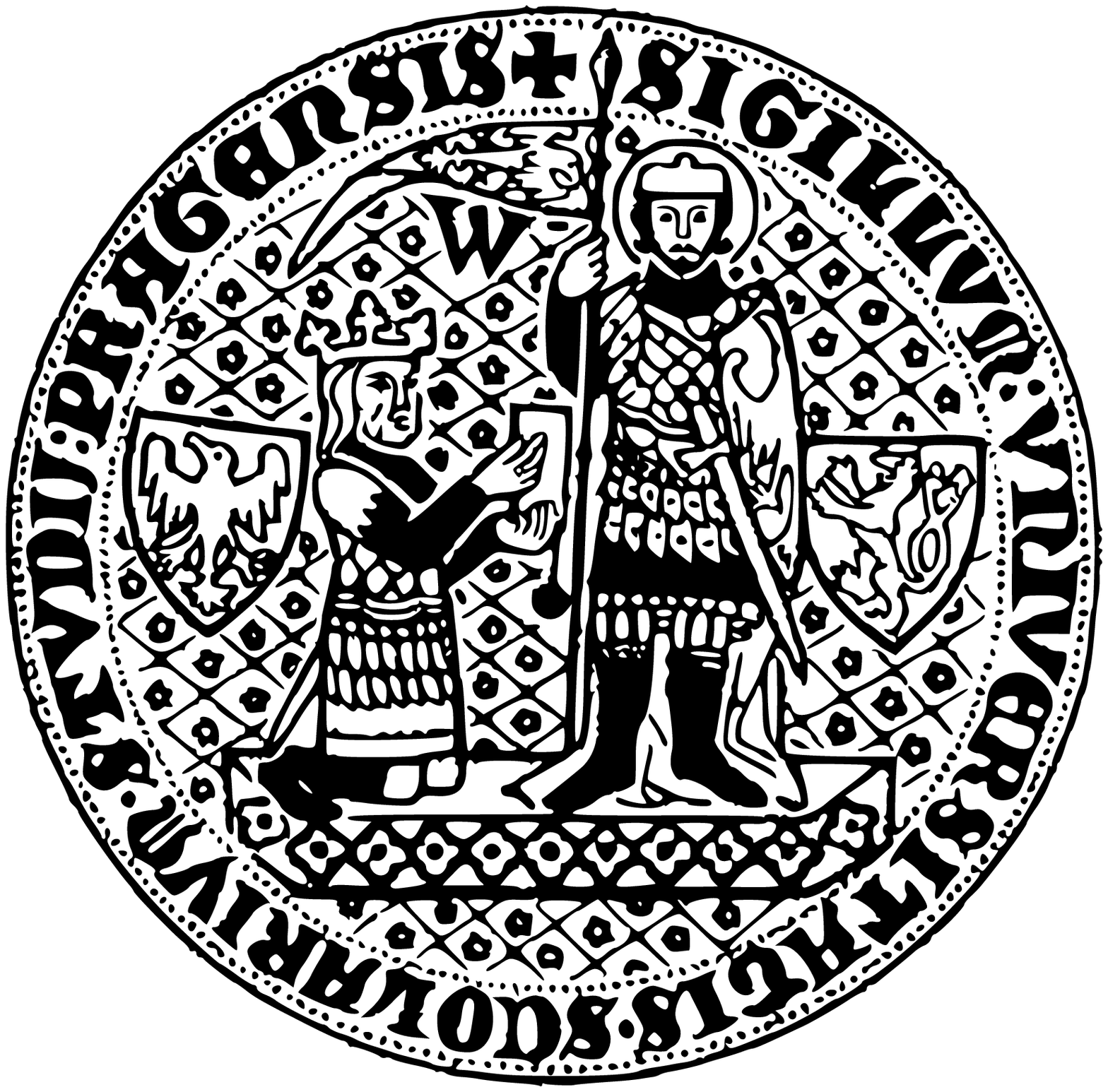}}
\end{figure}
\vspace{2cm}


\begin{center}

{\Large \sc Petr Beneš}\\
\vspace*{1cm}
{\LARGE \bf Dynamical symmetry breaking in models with strong Yukawa interactions}\\
\vspace*{0.5cm}
Thesis submitted for the degree of Philosophi{\ae} Doctor\\
\vspace*{1cm}
Thesis advisor:\quad\quad  \textbf{Ing.~Jiří Hošek,~DrSc.},\quad NPI AS CR\\


\vspace{1cm}

Prague, February 2012

\end{center}


\newpage
\pagestyle{headings}
\cleardoublepage


\section*{Acknowledgements}

First of all, I would like to thank Jiří Hošek to whom I am indebted for his patient supervision of this thesis and for many enlightening discussions. I am grateful to Jiří Hořejší for the support during the years and to Tomáš Brauner for tirelessly clarifying many QFT ideas to me. I wish to thank also Michal Malinský for suggestions leading to improvement of the quality of the thesis and Bruno Machet for useful discussions concerning the issue of quark flavor mixing.

I would like to express my thanks to my Řež colleagues, fellow Ph.D. students and friends Adam Smetana, Hynek Bíla and Filip Křížek for all those scientific, as well as not-so-scientific discussions we have had in Řež.

Finally, many personal thanks go to Borjanka for carefully reading the manuscript and for her moral support and encouragement over the past years.

I acknowledge the IPNP, Charles University in Prague, and the ECT* Trento for the support during the ECT* Doctoral Training Programme 2005. The presented work was supported in part by the Institutional Research Plan AV0Z10480505, by the GACR grant No.~202/06/0734 and by the Grant LA08015 of the Ministry of Education of the Czech Republic.

All Feynman diagrams were drawn using the JaxoDraw \cite{Binosi:2003yf,Binosi:2008ig}.

\vglue 0pt plus 1fill

\noindent
I declare that I carried out this doctoral thesis independently, and only with the cited sources, literature and other professional sources.

\medskip\noindent
I understand that my work relates to the rights and obligations under the Act No.~121/2000 Coll., the Copyright Act, as amended, in particular the fact that the Charles University in Prague has the right to conclude a license agreement on the use of this work as a school work pursuant to Section 60 paragraph 1 of the Copyright Act.

\vspace{10mm}

\hbox{\hbox to 0.5\hsize{%
In ........................ date ........................
\hss}\hbox to 0.5\hsize{%
signature
\hss}}

\vspace{20mm}
\newpage


\newpage
\pagestyle{headings}



\newdimen\LeftColumn
\newdimen\MiddleColumn
\newdimen\RightColumn
\LeftColumn=3.8cm
\MiddleColumn=0.5cm
\RightColumn=\textwidth
\advance\RightColumn by-\LeftColumn
\advance\RightColumn by-\MiddleColumn

\noindent
\begin{tabular*}{\textwidth}{@{}p{\LeftColumn}@{} @{}p{\MiddleColumn}@{} @{}p{\RightColumn}@{}}
\textbf{Název práce:}              && Dynamické narušení symetrie v modelech se silnými yukawovskými interakcemi
\\ &&\\
\textbf{Autor:}                    && Petr Beneš
\\ &&\\
\textbf{Katedra:}                  && Ústav teoretické fyziky MFF UK
\\ &&\\
\textbf{Vedoucí disertační práce:} && Ing.~Jiří Hošek,~DrSc., Oddělení teoretické fyziky, Ústav jaderné fyziky, AV ČR, v.~v.~i.
\\ &&\\
\textbf{Abstrakt:}                 && Hlavním cílem disertace je prozkoumat možnost spontánního narušení symetrie prostřednictvím silné yukawovské interakce. Technicky se předpokládá narušení symetrie skalárními a fermionovými propagátory spíše než skalárními vakuovými středními hodnotami. Myšlenka je nejdřív ukázána na příkladu jednoduchého modelu s abelovskou symetrií a posléze aplikována na realistický model elektroslabých interakcí. Disertace se navíc zabývá některými obecnějšími, modelově nezávislými otázkami, týkajícími se nejen diskutovaného modelu silné yukawovské dynamiky, ale rovněž širší třídy modelů s dynamickým generováním fermionových hmot. První z těchto otázek je problém mixingu fermionových \uv{flavorů} za přítomnosti fermionových self-energií s obecnou impulsovou závislostí. Konkrétně je diskutováno, jak v takových modelech definovat CKM matici, která, jak je ukázáno, může vyjít v principu neunitární. Další otázkou je problém počítání hmot kalibračních bosonů v případě, že je symetrie narušena fermionovými self-energiemi. Kromě odvození vzorce pro hmotovou matici kalibračních bosonů jsou také nalezeny korekce k související Pagelsově--Stokarově formuli.
\\ &&\\
\textbf{Klíčová slova:}            && Dynamické narušení elektroslabé symetrie
\end{tabular*}

\newpage

\noindent
\begin{tabular*}{\textwidth}{@{}p{\LeftColumn}@{} @{}p{\MiddleColumn}@{} @{}p{\RightColumn}@{}}
\textbf{Title:}                    && Dynamical symmetry breaking in models with strong Yukawa interactions
\\ &&\\
\textbf{Author:}                   && Petr Beneš
\\ &&\\
\textbf{Department:}               && Institute of Theoretical Physics, Faculty of Mathematics and Physics, Charles University
\\ &&\\
\textbf{Supervisor:}               && Ing.~Jiří Hošek,~DrSc., Department of Theoretical Physics, Nuclear Physics Institute, Academy of Sciences of the Czech Republic
\\ &&\\
\textbf{Abstract:}                 && The primary aim of the thesis is to explore the possibility of spontaneous symmetry breaking by strong Yukawa dynamics. Technically, the symmetry is assumed to be broken by formation of symmetry-breaking parts of both the scalar and the fermion propagators, rather than by the scalar vacuum expectation values. The idea is first introduced on an example of a toy model with the underlying symmetry being an Abelian one and later applied to a realistic model of electroweak interaction. In addition, the thesis also deals with some more general, model-independent issues, applicable not only to the discussed model of strong Yukawa dynamics, but to a wider class of models with dynamical mass generation. First of these issues is the problem of fermion flavor mixing in the presence of fermion self-energies with a general momentum dependence. It is in particular shown how to define the CKM matrix in such models and argued that it can come out in principle non-unitary. Second issue is the problem of calculating the gauge boson masses when the symmetry is broken by fermion self-energies. On top of deriving the formula for the gauge boson mass matrix we also find corrections to the related Pagels--Stokar formula.
\\ &&\\
\textbf{Keywords:}                 && Dynamical electroweak symmetry breaking
\end{tabular*}

%
%
%
%
%
%
%
%
%
%
%
%
%
%
%
%
%
%
%
%


\newpage

\tableofcontents

\cleardoublepage
\phantomsection
\addcontentsline{toc}{chapter}{List of figures}
\renewcommand\listfigurename{List of figures}
\listoffigures

\cleardoublepage
\phantomsection
\addcontentsline{toc}{chapter}{List of tables}
\renewcommand\listtablename{List of tables}
\listoftables



\cleardoublepage
\phantomsection
\markboth{\uppercase{List of acronyms}}{\uppercase{List of acronyms}}
\renewcommand{\nomname}{List of acronyms}
\printnomenclature[2cm]

\nomenclature{1PI}{one-particle irreducible}
\nomenclature{2PI}{two-particle irreducible}
\nomenclature{2HDM}{Two-Higgs-Doublet Model}
\nomenclature[cc]{$\cc$}{complex conjugate}
\nomenclature{CJT}{Cornwall--Jackiw--Tomboulis}
\nomenclature{CKM}{Cabibbo--Kobayashi--Maskawa}
\nomenclature[etc]{(E)TC}{(extended) technicolor}
\nomenclature{EW}{electroweak}
\nomenclature{EWSB}{electroweak symmetry breaking}
\nomenclature{FCNC}{flavor-changing neutral currents}
\nomenclature[hc]{$\hc$}{Hermitian conjugate}
\nomenclature{LSZ}{Lehmann--Symanzik--Zimmermann}
\nomenclature{MCS}{models with condensing scalars}
\nomenclature{NG}{Nambu--Goldstone}
\nomenclature{NJL}{Nambu--Jona-Lasinio}
\nomenclature{PS}{Pagels--Stokar}
\nomenclature{QCD}{quantum chromodynamics}
\nomenclature{QED}{quantum electrodynamics}
\nomenclature{QFT}{quantum field theory/theoretical}
\nomenclature{SD}{Schwinger--Dyson}
\nomenclature{SM}{Standard Model (of electroweak interactions)}
\nomenclature{SSB}{spontaneous symmetry breaking/breakdown}
\nomenclature{SUSY}{supersymmetry/supersymmetric}
\nomenclature{VEV(s)}{vacuum expectation value(s)}
\nomenclature{WT}{Ward--Takahashi}

\noindent
All these acronyms are common in the literature, except for the MCS idiosyncratic one.


\chapterstar{Conventions and notations}

For reader's convenience we list here some of the conventions and notations which are used throughout the text:
\begin{itemize}
\item We use the \qm{natural} units, i.e., we set $c=\hbar=1$.

\item A four-vector is denoted as $p=(p_0,p_1,p_2,p_3)$ and a three-vector as $\threevector{p}=(p_1,p_2,p_3)$.

\item For the Minkowski metric tensor we use the \qm{West Coast} convention, i.e.,
\begin{eqnarray}
\label{symbols:gmunu}
g_{\mu\nu} \ = \  g^{\mu\nu} &=&
\left(\begin{array}{rrrr}
1 & 0 & 0 & 0 \\
0 & -1 & 0 & 0 \\
0 & 0 & -1 & 0 \\
0 & 0 & 0 & -1
\end{array}\right)
\,.
\end{eqnarray}
Thus, for a dot-product of two four-vectors $p$ and $k$ we have
\begin{eqnarray}
p \cdot k &=& g_{\mu\nu}p^\mu k^\nu \ =\  p_0k_0-\threevector{p}\cdot\threevector{k} \,.
\end{eqnarray}
According to the sign of the quadrate $p^2=p \cdot p$, we call a four-vector $p$
\begin{eqnarray*}
\hbox{time-like} & \Leftrightarrow& p^2>0 \,, \\
\hbox{light-like (null)} & \Leftrightarrow& p^2=0 \,, \\
\hbox{space-like} & \Leftrightarrow& p^2<0 \,.
\end{eqnarray*}

\item The $\gamma_5$ matrix is defined as $\gamma_5=\I\gamma^0\gamma^1\gamma^2\gamma^3$.

\item We will frequently use the chiral projectors
\begin{equation}
P_L = \frac{1-\gamma_5}{2} \,, \quad\quad P_R = \frac{1+\gamma_5}{2}
\end{equation}
and correspondingly the \emph{left-handed} and \emph{right-handed} fermion fields $\psi_L=P_L\psi$ and $\psi_R=P_R\psi$.

\item For the totally antisymmetric Levi-Civita tensor (symbol) $\varepsilon_{\mu\nu\rho\sigma}$ we adopt the sign convention $\varepsilon_{0123}=+1$.

\item Charge conjugation $\psi^\C$ of a bispinor $\psi$ is defined as (for details see appendix~\ref{app:charge})
\begin{eqnarray}
\label{symbols:psiC}
  \psi^\C &\equiv& C \bar\psi^\T \,.
\end{eqnarray}

\item Analogously to \eqref{symbols:psiC}, we define also the \qm{charge transpose} $A^\C$ of a \emph{matrix} $A$ in the Dirac (bispinor) space as
\begin{eqnarray}
\label{symbols:AC}
  A^\C &\equiv& C A^\T C^{-1} \,.
\end{eqnarray}



\item The Pauli matrices are denoted by $\sigma$'s (rather than by $\tau$'s):
\begin{equation}
\sigma_1 = \left(\begin{array}{rr} 0 &   1  \\  1 &  0 \end{array}\right) \,,\quad
\sigma_2 = \left(\begin{array}{rr} 0 & -\I  \\ \I &  0 \end{array}\right) \,,\quad
\sigma_3 = \left(\begin{array}{rr} 1 &   0  \\  0 & -1 \end{array}\right) \,.\quad
\end{equation}

\item We define operator $\vecleftright{\partial}$ as $f\,\vecleftright{\partial}_\mu g\equiv f(\partial_\mu g) - (\partial_\mu f)g$.

\item The Feynman \qm{slash} notation for four-vectors ($\slashed{p}=p_\mu\gamma^\mu$) or partial derivatives ($\slashed{\partial}=\partial_\mu\gamma^\mu$) will be extensively used throughout the text.




\item In analogy with the standard Dirac conjugation for bispinors $\bar\psi=\psi^\dag\gamma_0$ we also define \qm{Dirac conjugation} for matrices:
\begin{eqnarray}
\label{symbols:barA}
  \bar A &\equiv& \gamma_0 A^\dag \gamma_0 \,.
\end{eqnarray}

\item Apart from the normal commutator $[A,B] = AB - BA$ of two matrices $A$ and $B$, we define also the \qm{generalized commutator}
\begin{eqnarray}
\label{symbols:comAB}
  \bbl A,B \bbr &\equiv& AB - \bar B \bar A \,.
\end{eqnarray}





\item The trace is denoted by $\Tr$ and is always taken over all indices. If some indices are not traced over, it is explicitly indicated.

\item The zero and unit matrices are most often denoted simply as $0$ and $1$, respectively. Occasionally, the symbols $\zeromatrix$ and $\unitmatrix$ are used as well, in order to emphasize their matrix character. If the matrix dimension is not clear from the context, we indicate it by a subscript.

\item Convention for representing the Green's functions in Feynman diagrams is the following:

\begin{tabular}{p{0.5cm}ll}
& Full Green's functions: a black blob  &
$\begin{array}{c}\includegraphics[scale=0.9]{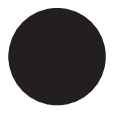}\end{array}$ \\
& 1PI Green's functions:  a shaded blob &
$\begin{array}{c}\includegraphics[scale=0.9]{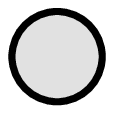}\end{array}$
\end{tabular}



\end{itemize}

\chapter{Introduction}

\section{Electroweak and chiral symmetry breaking}

One of today's experimental certainties is the following observed elementary particle spectrum. First of all, there are three generations of massive and electrically charged fermions\footnote{By \qm{fermions} we will always mean \qm{spin-$\frac{1}{2}$ fermions}.}, the quarks and the charged leptons. On top of these, there are three electrically neutral fermions, the neutrinos, at least some of which having non-zero, though as yet undetermined masses. And finally, these \qm{standard} fermions interact in a specific way with the vector (i.e., spin-$1$) bosons: with eight massless gluons and with four \emph{electroweak} (EW) vector bosons: with the massless photon, the massive $Z$ boson and two equally heavy $W^+$ and $W^-$ bosons.

So much for what experimentalists tell us, let us now focus on theorists' achievements. The way how to arrive at a consistent\footnote{By \qm{consistent} we mean merely \qm{renormalizable}. We neglect here the question of short distance behavior of the theory, as well as much deeper question whether an interacting QFT does exist at all \cite{Haag:1955ev}.} interacting quantum field theory (QFT) of fermions and vector bosons has been known already for a long time. It is the gauge principle, whose essence is, broadly speaking, the requirement of invariance of the Lagrangian under local (position-dependent) action of some Lie group. This requirement leads to necessity of introducing the appropriate affine connection -- the coveted vector gauge bosons. The gauge principle was originally formulated with Abelian $\group{U}(1)$ group and eventually generalized by Yang and Mills \cite{Yang:1954ek} to non-Abelian groups.

As formulated, the gauge principle can be directly applied only to interactions of fermions with the photon and with the gluons. The former case is the famous quantum electrodynamics (QED) with the underlying symmetry group being the Abelian electromagnetic $\group{U}(1)_{\mathrm{em}}$ one, whereas the latter case is the no less famous quantum chromodynamics (QCD) with the non-Abelian symmetry group $\group{SU}(3)_{\mathrm{c}}$.

However, the $Z$ and $W^\pm$ bosons cannot be incorporated into this scheme that straightforwardly. Generally speaking, the problem is in their massiveness: Directly applied gauge principle yields strictly massless vector bosons. In order to overcome this problem, another deep QFT result has to be invoked: the Nambu--Goldstone (NG) theorem \cite{Nambu:1960tm,Goldstone:1961eq,Goldstone:1962es}. It considers the situation of the \emph{spontaneous symmetry breaking} (SSB), i.e., the situation when the symmetry of the equations of motion (i.e., of the Lagrangian) is higher than the symmetry of their solutions (i.e., in particular of the vacuum state and of the Green's functions). The NG theorem states that if the symmetry in question is global (position-independent), then there emerge certain number of massless scalar (i.e., spin-$0$) states in the spectrum, the NG bosons.

The crucial non-trivial finding \cite{Higgs:1964pj,Englert:1964et,Guralnik:1964eu}, called the Englert--Brout--Higgs--Guralnik--Hagen--Kibble mechanism or shortly just the \emph{Higgs mechanism}, is that the NG theorem can be fruitfully combined with the gauge principle. That is to say, one can consider the situation of spontaneous breaking of a local rather than a global symmetry. In such a case no NG bosons appear, but instead some of the gauge bosons obtain mass. Number of such massive gauge boson is the same as the number of the NG bosons, present if the broken symmetry were global. In terms of the degrees of freedom it can be interpreted as that the \qm{would-be} NG bosons are transformed into the longitudinal polarization states of the gauge bosons. This is often paraphrased as that the gauge bosons become heavy by \qm{eating} the NG bosons.

The observed spectrum of the EW gauge bosons together with the pattern of their mutual interactions, as well as their interactions with fermions, can be accommodated by the assumption of \emph{electroweak symmetry breaking} (EWSB). That is to say, one first assumes existence of the EW gauge symmetry with the underlying group $\group{SU}(2)_{\mathrm{L}} \times \group{U}(1)_{\mathrm{Y}}$, corresponding to four massless gauge bosons. Subsequently, this gauge symmetry is assumed to be spontaneously broken down to its electromagnetic subgroup $\group{U}(1)_{\mathrm{em}}$, leaving only one gauge boson, the photon, massless, whereas the other three ones, $Z$ and $W^\pm$, obtain this way non-vanishing masses.

Another issue are the fermion masses. They are protected by any symmetry that treats independently the left-handed and the right-handed components of the fermion fields.\footnote{Strictly speaking, this is true only for the fermion masses of the Dirac type, not of the Majorana type.} A symmetry of such properties is called \emph{chiral}. The electroweak symmetry is chiral, hence not only the gauge boson, but also the fermions are massless at the level of Lagrangian.

However, mere EWSB does not necessarily imply that the fermion masses are no longer protected. The problem is that there may exist a larger chiral symmetry than the electroweak one. In fact, the \qm{minimal} Lagrangian consisting only of the standard fermions and their electroweakly symmetric interactions enjoys the rather large symmetry\footnote{For simplicity we assume at the moment that there are no right-handed neutrinos.} $\group{U}(3)_{q_L} \times \group{U}(3)_{u_R} \times \group{U}(3)_{d_R} \times \group{U}(3)_{\ell_L} \times \group{U}(3)_{e_R}$, which is obviously chiral. Of course, it contains the gauged symmetry $\group{U}(1)_{\mathrm{Y}}$ as a subgroup and also some of its subgroups are anomalously broken. Still, however, even though the electroweak symmetry gets broken, there remains enough chiral symmetry to protect the fermion masses. Of course, unless something breaks somehow (explicitly or spontaneously) this residual chiral symmetry.

Thus, (almost) all that remains to satisfy a theorist's mind in the quest of finding a consistent QFT framework describing the Nature is to invent a suitable mechanism(s) of the electroweak and chiral symmetry breaking. That is to say, to enhance the currently observed particle spectrum and its electroweakly symmetric interactions with some new dynamics (to be eventually experimentally observed), being at the level of Lagrangian electroweakly symmetric too and making the two required jobs: First, to spontaneously break the electroweak symmetry and second, to break (explicitly or spontaneously) the chiral symmetry. Needless to say that the latter implies the former, but the reverse is not true.

\section{Ways out}

A suitable mechanism of EWSB (and of chiral symmetry breaking) remains an open question, experimental as well as theoretical. To date there are no experimental clues. On the other hand, being one of the most urgent issues of the last decades' particle physics, there are naturally many theoretical proposals of such a mechanism, though none of them being completely satisfactory and widely accepted.

The most prominent example of models aspiring to account for the EWSB is no doubt the Standard Model (SM) \cite{Glashow:1961tr,Weinberg:1967tq,Salam:1968rm}. There are at least three reasons for its popularity: It is historically the first such model, it is in a way a minimal EWSB model and finally it is \qm{user-friendly} in the sense of allowing for perturbative calculations. The SM introduces an $\group{SU}(2)_{\mathrm{L}}$ scalar doublet, known as the \emph{Higgs doublet}. The key ingredient is the form of its potential, chosen in such a peculiar way that the electrically neutral real component of the Higgs doublet develops a non-vanishing vacuum expectation value (VEV), breaking the electroweak symmetry down to the electromagnetic one. The Higgs doublet also inevitably\footnote{According to the Gell-Mann's \emph{Totalitarian principle}: \qm{Everything not forbidden is compulsory.}} couples to fermions. The corresponding Yukawa interactions happen to break explicitly all chiral symmetries except for those being a subgroup of the EW symmetry. Thus, after EWSB no residual chiral symmetry remains to protect the fermion masses, which indeed emerge as products of the Yukawa coupling constants and the Higgs field VEV.

The SM can be generalized by assuming other scalar representations than one doublet. The common feature of such generalizations is more free parameters, allowing for better parameterization of observed phenomena such as neutrino masses, $\mathcal{CP}$ violation, etc. Most straightforwardly, one can consider two scalar doublets and arrive at the Two-Higgs-Doublet Model (2HDM) \cite{Lee:1973iz,Gunion:1989we}. While in SM there remains after EWSB only one real scalar degree of freedom (the \emph{Higgs boson}), whose mass is the only free parameter of the EWSB sector, in 2HDM the situation is considerably more complicated and allowing for richer phenomenology. Further, instead of adding just more doublets, one can also consider a scalar triplet \cite{Schechter:1980gr,Cheng:1980qt,Gelmini:1980re}, charged singlet \cite{Zee:1980ai,Cheung:1999az} and doubly charged singlet \cite{Zee:1985rj,Babu:1988ki}. All possibilities can be naturally combined.

One of the virtues of these \emph{models with condensing scalars} (MCS) is their \qm{transparency} in the sense that the particle spectrum is directly readable from the Lagrangian. This is connected with another positive feature, that they are weakly coupled\footnote{Perhaps with the exception of the top quark Yukawa coupling(s).} and thus practical, as one can use the perturbation theory. On the other hand, these models have also certain drawbacks. For instance, they always possess at least as many free parameters as there are distinct massive fermions, since their masses are proportional to the Yukawa coupling constants. In other words, the hierarchy of fermion masses is not \emph{explained}, but merely \emph{parameterized}.\footnote{A philosopher would assert that after all \emph{any} physical theory is merely a parameterization of experimental data and there is nothing such as an \qm{explanation}. Still, there are better parameterizations and worse parameterizations and one of the criteria how to distinguish one from another is the number of their tunable parameters.}



However, the most serious problem of the MCS is probably the one connected with the very assumption of the existence of elementary scalars. Unless there is some special symmetry, the scalar masses are not stable against quadratic radiative corrections. In other words, they tend to be of order of the theory's cut-off, which is presumably either the grand unification scale ($10^{15}-10^{16}\GeV$) or even the Planck scale ($10^{19}\GeV$). On the other hand, the scalar masses should be at the same time well below the theory's cut-off. This follows from the requirement that the Landau poles of the scalar self-couplings, which are proportional to the scalar masses, do not occur below the theory's cut-off. In order to keep scalar masses reasonably low one has to fine-tune their bare masses with an incredible accuracy, which is considered \emph{unnatural} \cite{'tHooft:1979bh}. This mismatch between the \qm{natural} and the \qm{required} values of the scalar masses is just the \emph{hierarchy problem} of the SM (and generally of all MCS).

One way out is to invent some kind of symmetry protecting the scalar masses, in much the same way as the chiral symmetry protects the fermion masses. Such symmetry has been really invented \cite{Ramond:1971gb,Neveu:1971rx,Gervais:1971ji,Golfand:1971iw,Volkov:1973ix,Wess:1974tw} and is known as the \emph{supersymmetry} (SUSY). Its characteristic feature is presence of both fermions and bosons in the same irreducible representations, the \emph{supermultiplets}. Thus, if SUSY is unbroken, the protection of fermion masses by the chiral symmetry implies protection of masses of the scalars from corresponding supermultiplet.

The first and most straightforward attempt to apply the general idea of SUSY on the SM is the Minimal Supersymmetric Standard Model (MSSM) \cite{Haber:1984rc,Nilles:1983ge,Chung:2003fi}. It invokes the $\mathcal{N}=1$ SUSY algebra and puts the standard fermions and the SM gauge bosons into the chiral and gauge supermultiplets, respectively. In order to avoid the gauge anomaly it postulates two Higgs doublets (thus, the MSSM includes as a part the 2HDM) and puts them into chiral supermultiplets.


If SUSY were exact, we would observe for each particle also its superpartner with the same mass and the spin differing by $1/2$. However, none of those superpartners has been observed. Thus, SUSY has to be broken. Moreover, it has to be broken \emph{softly}, i.e., in such a way that the very reason for using SUSY, i.e., stabilizing the Higgs mass, is not jeopardized. Actually, finding a reliable mechanism for such SUSY breaking appears to be probably the most serious theoretical problem of the MSSM (and of its various non-minimal extensions), although proposals of solutions do exist. Anyway, from phenomenological point of view the best one can do at the moment is to merely \emph{parameterize} this soft SUSY breaking. This is achieved by breaking SUSY explicitly by adding operators with \emph{positive} mass dimension into the Lagrangian. Such SUSY breaking has the desired property that the scalar masses are renormalized only logarithmically. On the other hand, it also introduces many new free parameters into the model, as is after all common when dealing with a phenomenological Lagrangian. In fact, these form the vast majority of those infamous $124$ free parameters \cite{Haber:1997if} of MSSM.

Another way of tackling the problem of EWSB is to realize that EWSB actually \emph{does happen} due to already known dynamics, namely due to the QCD dynamics of quarks and gluons. Broadly speaking, as the QCD dynamics becomes strong at the scale $\Lambda_{\mathrm{QCD}} \sim 200 \, \mathrm{MeV}$, the quarks form condensates that break their chiral symmetry. The point is that at the same time these condensates break also the EW symmetry, just according to the correct pattern, i.e., down to the $\group{U}(1)_{\mathrm{em}}$. Moreover, the ratio of the resulting masses of the EW gauge bosons is correct in the sense that $\rho = 1$.\footnote{For the precise definition of the $\rho$-parameter see Eq.~\eqref{ewd:rho} thereinafter.} However, there is a slight problem that the very magnitude of these masses is about $2\,600$ times smaller than the experimentally measured values. Another problem is the fermion mass spectrum: As the QCD dynamics breaks the quark chiral symmetry down to the vectorial subgroup $\group{SU}(N_{\mathrm{f}})$ (in the case of $N_{\mathrm{f}}$ quarks), whereas the chiral symmetry in the lepton sector remains unbroken, the net result of QCD is that the quarks come out all equally massive and the leptons remain all equally massless, both with flagrant contradiction with experiment.

Nevertheless, the inspiration is obvious. Most straightforwardly, one can assume \cite{Weinberg:1979bn,Susskind:1978ms} that on top of the color gauge symmetry $\group{SU}(3)_{\mathrm{c}}$ there exists also its \qm{scaled-up copy}. That is to say, there exists so called \emph{technicolor} (TC) gauge symmetry with the corresponding group $\group{G}_{\mathrm{TC}}$ being $\group{SU}(N_{\mathrm{TC}})$ and new fermions called the \emph{technifermions} (sometimes referred to as the \emph{techniquarks}, in order to emphasize the analogy with QCD), being, analogously to the ordinary quarks, charged under both TC and EW groups, so that their condensates contribute to the EW gauge boson masses. The scale $\Lambda_{\mathrm{TC}}$, at which TC dynamics becomes strong, must be about roughly $\Lambda_{\mathrm{TC}} \sim 500 \, \mathrm{GeV}$ in order to account for the measured magnitudes of EW gauge boson masses.


However, since the TC dynamics couples only to the technifermions, the mass spectrum of the standard fermions remains unaffected by it. In order to fix this problem the class of Extended TC (ETC) models was invented \cite{Dimopoulos:1979es,Eichten:1979ah}. The basic idea is to gauge the flavor (generation) symmetry of the standard fermions and include in the representations of the corresponding gauge group $\group{G}_{\mathrm{ETC}}$ also the technifermions. Obviously, by construction it is $\group{G}_{\mathrm{TC}} \subset \group{G}_{\mathrm{ETC}}$ and actually it is assumed that $\group{G}_{\mathrm{ETC}}$ is spontaneously broken down to $\group{G}_{\mathrm{TC}}$. As the standard fermions and the technifermions are coupled to each other, the consequent technifermion chiral symmetry breaking gives rise also to standard fermion masses.

The simple picture sketched above, with the TC dynamics being just a scaled-up version of QCD, turned out to be a bit too na\"{\i}ve from the phenomenological point of view. Thus, the idea of \emph{walking} was proposed \cite{Holdom:1981rm,Akiba:1985rr,Yamawaki:1985zg,Appelquist:1986an,Appelquist:1987fc}: The TC dynamics is such that the corresponding coupling constant does not run, like in QCD, but rather walks, i.e., stays almost constant over a large extent of scales.

The two paradigms described above, the SUSY extensions of SM and the ETC theories, are probably the most popular classes of models, describing the anticipated (and experimentally searched for) physics beyond SM. This by no means means that no other proposals exist. To mention at least some them: There are ideas like the Top quark condensate \cite{Miransky:1988xi,Bardeen:1989ds} and Topcolor \cite{Hill:1991at,Hill:1994hp} models, inspired by the surprisingly large mass of the top quark. There are attempts known as the Little Higgs models \cite{Weinberg:1972fn,Georgi:1974yw,Georgi:1975tz} trying to explain the lightness of the Higgs boson by assuming that it is a pseudo-NG boson of some spontaneously broken approximate global symmetry. There are models of EWSB based on the assumption of existence of extra dimensions \cite{Csaki:2005vy}. And finally, is has been also recently attempted to gauge the fermion flavor symmetries in an ETC manner, but without introducing the technifermions and the corresponding TC dynamics \cite{Hosek:2009ys,Benes:2011gi,Smetana:2011tj}.

\section{This thesis}

This thesis concerns with three main topics, being mutually thematically related, but possessing a different level of originality and generality.

\subsection{Strong Yukawa dynamics}

Most of the models of EWSB mentioned in the previous section contains either weakly coupled elementary scalars, or strongly coupled gauge bosons. There is also a third logical possibility -- the strongly coupled elementary scalars. One particular realization of this possibility was proposed in Refs.~\cite{Brauner:2004kg,Brauner:2005hw}.\footnote{Strongly coupled scalars have been employed, for instance, also in the context of SUSY in Ref.~\cite{Luty:2000fj}.} The key idea is that it is not solely the scalar dynamics (i.e., the self-couplings in the scalar \qm{potential}) which is responsible for the EWSB, but rather the Yukawa dynamics. More precisely, mutual interactions of scalars and fermions are assumed to form EWSB propagators of both the scalars and the fermions. In order to do so, the Yukawa dynamics must be presumably strong.

As can be inferred from the thesis' name, building of such a model of EWSB based on a strong Yukawa dynamics is its leading (but by no means the only) subject. However, instead of jumping directly into the realm of EW interaction, first in part~\ref{part:abel} we show the main ideas on a simple toy model, in which only an Abelian $\group{U}(1)$ symmetry is broken dynamically via the non-perturbative solutions to the equations of motion. Since the full model of EW interactions will be rather complicated (at least numerically), the Abelian toy model can serve as a laboratory for exploring some general features, which hold even in the more complicated EW $\group{SU}(2)_{\mathrm{L}} \times \group{U}(1)_{\mathrm{Y}}$ model.

The \qm{Abelian} introduction in part~\ref{part:abel} is done in two consecutive steps: First, in chapter~\ref{chp:inf} we give a more intuitive and diagrammatical introduction to the very idea with the emphasis on the differences from the MCS. Second, in chapter~\ref{chp:frm} we redo the analysis from the previous chapter in a more formal way, relying less on intuitive diagrammatical considerations and allowing better for eventual generalization in next part.


Only in part~\ref{part:ew} we apply the idea of strong Yukawa dynamics on a realistic model of EWSB. First, in chapter~\ref{chp:ew1} we define the model in terms of its Lagrangian and the particle content and present in this context also a convenient parameterization of the fields. In chapter~\ref{chp:ewa} we prepare the ground for the eventual demonstration that the EWSB by strong Yukawa dynamics is possible; technically, we construct the space of the propagators (the Ansatz) on which we will be looking for the EWSB solutions. Finally, in chapter~\ref{chp:ewdyn} we write down the relevant equations of motion whose solutions are expected to exhibit the coveted EWSB. We give also some numerical evidence that the proposed scenario is viable and the concept is not empty.

The parts~\ref{part:abel} and \ref{part:ew} are based on Refs.~\cite{Brauner:2005hw,Benes:2006ny} and \cite{Brauner:2004kg,Benes:2008ir}, respectively, but treat the subject in more detailed and technical way.

\subsection{Fermion flavor mixing in models with dynamical mass generation}

The model of EWSB with strong Yukawa dynamics, discussed in part~\ref{part:ew}, brings some more general, model-independent questions, which are common for a wider class of models with dynamical fermion mass generation, including in particular also the ETC models.

The first of such problems, discussed in this thesis, is the problem of fermion flavor mixing. Let us first briefly review how it is treated in MCS. Once the scalars develop their VEVs, the Yukawa coupling terms give rise to fermion bilinear terms -- the mass terms. However, as the Yukawa interactions tie together fermions from different generations, so do consequently also the resulting fermion mass terms. In other words, one ends up with fermion mass matrices which are in principle arbitrary complex $3 \times 3$ matrices. In particular, they are generally not diagonal. However, the mass spectrum is easily revealed by looking for their eigenvalues. It is also comfortable to have the Lagrangian expressed directly in terms of the fermion fields, creating and annihilating the fermions with definite masses. Such a basis of fermion fields is called the \emph{mass eigenstate basis} and obviously it is the basis in which the mass matrices are diagonal (and non-negative). It can be obtained by unitary rotations of the original basis.

The original basis is commonly referred to as the \emph{weak eigenstate basis}. The reason for that is that the interaction terms of the fermions and the EW gauge bosons are in this basis flavor-diagonal (i.e., they do not link together fermions from different families). However, in the course of mass-diagonalization of the Lagrangian this changes: Applying the above mentioned unitary transformations of the fermion fields leads to the emergence of non-diagonal flavor transitions in the charged current interaction Lagrangian, i.e., in the interaction Lagrangian of fermions with the $W^\pm$ bosons.\footnote{Interestingly enough, the interactions with photon and the $Z$ boson remain flavor-diagonal, so that there are no flavor changing neutral currents (FCNC) at the tree level.} Strength of such inter-flavor interactions is in the quark sector parameterized by the $3 \times 3$ Cabibbo--Kobayashi--Maskawa (CKM) matrix in the flavor space, which is by construction automatically unitary.

So much for the situation in the MCS. The main lesson is that in these models the treatment of the fermion flavor mixing relies on the presence of mass matrices in the Lagrangian. In models with dynamical fermion mass generation, however, the situation is different. Typically, instead of constant, momentum-independent mass matrices in the Lagrangian one obtains rather their momentum-dependent generalizations -- the fermion one-particle irreducible (1PI) parts of the propagators, the self-energies. Due to their momentum dependence they cannot be interpreted as Lagrangian quantities and hence it is not \emph{a priori} clear how to treat the fermion flavor mixing: How to define the mass eigenstate basis and how to (in the case of quarks) define and calculate the CKM matrix.

This question is discussed in part~\ref{part:mx} and a solution is proposed. It is shown that depending on details of momentum dependencies of quark self-energies the appropriately defined CKM matrix can be in general non-unitary. As this subject is discussed thoroughly already in Ref.~\cite{Benes:2009iz}, the part~\ref{part:mx} is relatively concise and consists only of chapter~\ref{chp:mx}.


\subsection{Gauge boson masses}

Another model-independent problem common in various models with dynamical fermion mass generation is the problem of the gauge boson masses. Typically the situation is as follows: There are some fermion fields, sitting in representations of some gauge group (not necessarily a simple one). Some dynamics (whose precise details are not essential for the present discussion) generate self-energies of these fermions, which in turn induce breaking of the gauge symmetry down to some of its subgroup (not necessarily the trivial one). Thus, as the SSB is \qm{proportional} to the fermion self-energies, so must be also the resulting non-vanishing masses of some of the gauge bosons, arising due to the Higgs mechanism.

The question how to calculate the gauge boson masses in terms of the fermion self-energies is discussed in detail in part~\ref{part:gauge}. Although the issue has already been discussed in the literature, we present more systematic and more general treatment and find some flaws in the way it has been treated in the literature so far. Namely, we point out the problem of symmetricity of the gauge boson mass matrix. Although we improve the situation at least to the extent that we can calculate the mass matrix of the EW gauge bosons as symmetric (assuming arbitrary number of fermion generations and the most general fermion mixing, as well as contribution from massive Majorana neutrinos), in more general theories (depending on the gauge group and the fermion representations) the problem resists. This is one of the reasons why the results obtained in part~\ref{part:gauge} have not been published yet.

Part~\ref{part:gauge} is organized as follows: First, in chapter~\ref{chp:gbp} we review, primarily for the sake of establishing the notation, some \qm{textbook} facts and state the key assumptions under which we in the subsequent chapter~\ref{chp:gbm} derive the master formula for the gauge boson mass matrix in terms of the fermion self-energies. The chapters~\ref{chp:ablgg} and \ref{ewM} are then dedicated to specific application of the general gauge boson mass matrix formula on the Abelian toy model and EWSB model from parts~\ref{part:abel} and \ref{part:ew}, respectively.

\subsection{Appendices}

In order to make the text reasonably self-contained, we also, after summarizing and concluding in chapter~\ref{chp:conclusions}, provide for the reader's convenience several appendices. In appendix~\ref{app:charge} we define the notion of fermion charge conjugation and state some of its properties. Appendix~\ref{app:quant} is devoted to reviewing the way how to quantize a general fermion field. We introduce for this purpose the method of Faddeev and Jackiw, which we later, in appendix~\ref{app:major}, apply also to the more constrained Majorana fermion field. In appendix~\ref{app:fermi propag} we discuss possible parameterizations of multicomponent fermion fields with the emphasis on the Nambu--Gorkov formalism, which is used extensively throughout the main text. Similar analysis, although in less detail, is done also for multicomponent scalar fields in appendix~\ref{app:sclr}.



























\part{Abelian toy model}
\label{part:abel}


\chapter{An informal introduction}
\label{chp:inf}

\intro{In this chapter we give a brief, less formal but more intuitive introduction to the very idea of breaking a symmetry by scalar two-point functions, rather than by a one-point function. For this purpose we employ a toy model with underlying Abelian symmetry. The idea will be, still on the example of an Abelian symmetry, rephrased more formally in the next chapter~\ref{chp:frm} and eventually, in the subsequent part~\ref{part:ew}, applied on a realistic model of spontaneous breaking of the electroweak symmetry.\intro}

\intro{This chapter, as well as the following one, is based on Refs.~\cite{Brauner:2005hw,Benes:2006ny}.}


\section{Motivation}

We consider a complex scalar field $\phi$ and a massless fermion field $\psi$. Their Lagrangian reads
\begin{eqnarray}
\label{chp:prelim:eq:eL}
\eL &=&
\bar\psi\I \slashed{\partial} \psi +
(\partial_{\mu}\phi)^\dag(\partial^{\mu}\phi)
- V(\phi)
+ \eL_{\mathrm{Yukawa}} \,,
\end{eqnarray}
with the scalar potential given by
\begin{eqnarray}
\label{chp:prelim:eq:V}
V(\phi) &=& M^2\phi^{\dagger}\phi + \frac{1}{2}\lambda(\phi^{\dagger}\phi)^2
\end{eqnarray}
and the Yukawa part assumed to be
\begin{eqnarray}
\label{chp:prelim:eq:Yukawa}
\eL_{\mathrm{Yukawa}} &=& y \bar\psi_{L}\psi_{R}\phi + y^* \bar\psi_{R}\psi_{L}\phi^{\dagger} \,.
\end{eqnarray}
The Yukawa coupling constant $y$ can be in fact considered real without loss of generality. Indeed, if we write $y=|y|\e^{\I\alpha}$, we can always eliminate the phase $\e^{\I\alpha}$ by redefining, e.g., $\phi \rightarrow \e^{\I\alpha} \phi$. We will deliberately keep $y$ complex, however, as it will help us to keep track of which of the two interaction terms in \eqref{chp:prelim:eq:Yukawa} will be actually used in the particular vertices of the loop diagrams later on.

Notice that the Yukawa interactions \eqref{chp:prelim:eq:Yukawa} are not the most general ones, they are postulated to have a rather special form. In particular, the terms
\begin{eqnarray}
\label{chp:prelim:eq:Yukawaprime}
\eL_{\mathrm{Yukawa}}^\prime &=& y^\prime \bar\psi_{L}\psi_{R}\phi^{\dagger} + y^{\prime*} \bar\psi_{R}\psi_{L}\phi
\end{eqnarray}
would have to be included in order to have the most general Yukawa interactions.

Let us investigate the symmetries of the Lagrangian. First, we observe that the Lagrangian remains invariant under the phase transformation of the fermion field, $\psi \rightarrow [\psi]^\prime = \e^{ \I \alpha } \, \psi$, i.e., under the vectorial symmetry $\group{U}(1)_{\mathrm{V}}$. This corresponds to the fermion number conservation. Apart from this (rather uninteresting) vectorial symmetry, the Lagrangian is invariant also under axial $\group{U}(1)_{\mathrm{A}}$ symmetry, which is going to play a more important r\^{o}le in our considerations. Unlike the vectorial symmetry, the axial symmetry acts not only on the fermion,
\begin{subequations}
\label{chp:prelim:eq:U1A}
\begin{eqnarray}
\group{U}(1)_{\mathrm{A}}\,:\qquad \psi & \TransformsTo & [\psi]^\prime \>=\> \e^{ \I Q \theta \gamma_5} \, \psi \,,
\end{eqnarray}
but also on the scalar:
\begin{eqnarray}
\group{U}(1)_{\mathrm{A}}\,:\qquad \phi & \TransformsTo & [\phi]^\prime \>=\> \e^{ -2\I Q \theta} \, \phi \,,
\end{eqnarray}
\end{subequations}
where $\theta$ is the parameter of the $\group{U}(1)_{\mathrm{A}}$ transformation and $Q$ is the axial charge. Notice that the Lagrangian \eqref{chp:prelim:eq:eL} is invariant under $\group{U}(1)_{\mathrm{A}}$, \eqref{chp:prelim:eq:U1A}, due to the absence of the Yukawa terms of the type \eqref{chp:prelim:eq:Yukawaprime}. One can view it also from the opposite perspective: The terms \eqref{chp:prelim:eq:Yukawaprime} are forbidden by the requirement of $\group{U}(1)_{\mathrm{A}}$ invariance.

Notice that the axial $\group{U}(1)_{\mathrm{A}}$ symmetry is in fact anomalously violated. We will ignore this problem in this chapter; in fact it can be (and will be, in the next chapter) easily fixed by introducing additional fermions with appropriately chosen axial charges.


The basic observation is that potential fermion mass terms
\begin{eqnarray}
\label{chp:prelim:eL:mass}
\eL_{\mathrm{mass}}  & = & {}- m \bar \psi_L \psi_R + \hc
\end{eqnarray}
are forbidden by the underlying axial symmetry $\group{U}(1)_{\mathrm{A}}$. Thus, in order to generate the fermion mass, the axial symmetry has to be broken somehow. This breaking may be either explicit (i.e., by suitable symmetry-breaking terms, added to the Lagrangian) or spontaneous (i.e., by symmetry-breaking solutions of the equations of motion). Here we are going to explore the latter possibility, because later on, in part~\ref{part:ew}, we will apply the ideas of the present Abelian toy model to the realistic model of electroweak interaction, where the spontaneous breaking of the symmetry is a must if one insists on a renormalizable theory of massive vector bosons.

We are now going to check what are the actual possibilities of breaking spontaneously the axial symmetry. Before doing that let us just remark that the spontaneous symmetry breaking (SSB) will be in any case a non-perturbative effect: If a Lagrangian (in particular its interaction part) possesses some symmetry, then the symmetry is preserved at any order of the perturbative expansion.

\section{SSB by a one-point function}
\label{chp:prelim:SSB1point}

First of all, let us see how the task of spontaneous breaking of the axial symmetry and the associated fermion mass generation is solved traditionally: We review here basically the $\group{O}(2) \sim \group{U}(1)$ linear $\sigma$-model \cite{GellMann:1960np,Goldstone:1961eq} (whose generalizations lie in the very heart of the MCS). The key assumption is that scalar \qm{mass} squared (or more precisely, the scalar mass parameter in the potential \eqref{chp:prelim:eq:V}) is negative: $M^2<0$ (but still with $\lambda>0$, in order to have the Hamiltonian bounded from below). In consequence the classical scalar field configuration which minimizes the Hamiltonian, the \qm{vacuum}, is not $\phi_0 = 0$, but rather\footnote{We deliberately choose $\phi_0$ to be real.}
\begin{eqnarray}
\phi_0 & = & \frac{v}{\sqrt{2}} \,,
\end{eqnarray}
with
\begin{eqnarray}
\label{chp:prelim:eq:v}
v & \equiv & \sqrt{\frac{-2M^2}{\lambda}} \,.
\end{eqnarray}

The quantization process basically consists of quantizing the field fluctuation around the vacuum -- the classical minimum $\phi_0$. Thus, in our case, the true dynamical variable to be quantized is not $\phi$, but rather its shifted value $\phi-\phi_0$. In the language of the quantum field theory we say that the scalar field $\phi$ develops the non-vanishing vacuum expectation value (VEV)
\begin{eqnarray}
\label{chp:prelim:eq:VEV}
\langle \phi \rangle \ \equiv \ \bra{0}\phi(x)\ket{0} &=& \frac{v}{\sqrt{2}} \,.
\end{eqnarray}
Now we can conveniently rewrite the original complex field $\phi$ as
\begin{eqnarray}
\phi &=& \frac{1}{\sqrt{2}}\big(v+\sigma+\I\pi\big) \,,
\end{eqnarray}
where $\sigma$ and $\pi$ are real fields, whose VEVs are by construction vanishing. Upon plugging this decomposition into the Lagrangian \eqref{chp:prelim:eq:eL} and using the definition \eqref{chp:prelim:eq:v} of $v$ we find that $\sigma$ (whose analogue in the SM is known as the \emph{Higgs boson}) has the non-vanishing mass
\begin{subequations}
\begin{eqnarray}
M_{\sigma} &=& \sqrt{-2M^2} \\ &=&  \sqrt{\lambda} v \,,
\end{eqnarray}
\end{subequations}
while $\pi$ is massless,
\begin{eqnarray}
M_{\pi} &=& 0 \,.
\end{eqnarray}

The Green's one-point function \eqref{chp:prelim:eq:VEV} is obviously non-invariant under the $\group{U}(1)_{\mathrm{A}}$. Thus, the axial symmetry is spontaneously broken (with the corresponding NG boson being just the massless pseudo-scalar field $\pi$) and the fermion's masslessness is no longer protected. Indeed, upon performing the shift $\phi \rightarrow \phi - \phi_0$ in the Yukawa Lagrangian \eqref{chp:prelim:eq:Yukawa} the fermion mass terms \eqref{chp:prelim:eL:mass} emerge, with the mass\footnote{What we inconsistently call here the fermion \qm{mass}, should be more appropriately called merely a fermion \qm{mass parameter}. The actual mass, i.e., the pole of the fermion propagator, is of course given by $|m|$.} $m$ given by
\begin{eqnarray}
m & = & - \frac{v}{\sqrt{2}} y \,.
\end{eqnarray}
Finally, note that the SSB of the axial symmetry is really a non-perturbative effect, as advertised above, since the expression \eqref{chp:prelim:eq:v} for $v$ is non-analytical at $\lambda=0$.

\section{SSB by a two-point function}
\label{chp:prelim:SSB2point}

In the previous section the axial symmetry was broken by formation of the scalar's one-point function $\langle\phi\rangle$, \eqref{chp:prelim:eq:VEV}. It is natural to ask whether it is possible to break the axial symmetry also by some other Green's function, non-consistent with the axial symmetry. Apart from the one-point function, the next-to-simplest possibility is a two-point function -- the propagator. The ordinary two-point function of the type $\langle \phi\phi^\dag \rangle$ (or, equivalently, $\langle \phi^\dag\phi \rangle$), however, does not serve well for this purpose, since it is invariant under the axial symmetry. However, there is another possibility: the function $\langle \phi\phi \rangle$ (or $\langle \phi^\dag \phi^\dag \rangle$), which clearly violates the axial symmetry.

A detailed and more formal discussion of the very mechanism of generating such \qm{anomalous} two-point functions is going to be the topic of the next chapter. Now we choose to discuss these issues at more intuitive and heuristic level, focusing mainly on the consequences for the particle spectrum.



Let us begin with the scalar itself; the following reasoning is adopted from \cite{Brauner:2005hw}. For the sake of present considerations, we will consider the one-particle irreducible (1PI) part of $\langle \phi\phi \rangle$ to be momentum-independent; later on when formalizing our considerations we will take into account a general momentum dependence. Thus, let us for the moment \emph{assume} that the 1PI parts of the symmetry-breaking propagators of the type $\langle \phi\phi \rangle$, $\langle \phi^\dag\phi^\dag \rangle$ are \emph{somehow} generated by means of the dynamics of the theory. Namely, we assume the 1PI scalar propagators (and the corresponding Feynman rules) to have form
\begin{subequations}
\label{chp:prelim:prpgscal:1PI}
\begin{eqnarray}
\langle \phi\phi \rangle_{\mathrm{1PI}}
\ = \
\begin{array}{c}
\scalebox{0.85}{\includegraphics[trim = 10bp 12bp 19bp 11bp,clip]{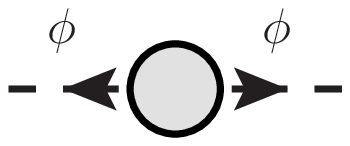}}
\end{array}
&=& -\I \mu^{2\hphantom{*}} \,,
\\
\langle \phi^\dag\phi^\dag \rangle_{\mathrm{1PI}}
\ = \
\begin{array}{c}
\scalebox{0.85}{\includegraphics[trim = 10bp 12bp 19bp 11bp,clip]{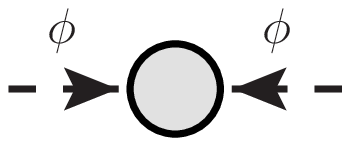}}
\end{array}
&=& -\I \mu^{2*} \,,
\end{eqnarray}
\end{subequations}
with $\mu$ being a complex constant with the dimension of mass. The corresponding full propagators are
\begin{subequations}
\label{chp:prelim:prpgscal:full}
\begin{eqnarray}
\langle \phi\phi \rangle
\ = \
\begin{array}{c}
\scalebox{0.85}{\includegraphics[trim = 10bp 12bp 19bp 11bp,clip]{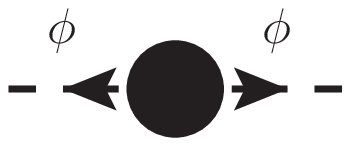}}
\end{array}
&=& \I \frac{\mu^2}{(p^2-M^2)^2-|\mu^2|^2} \,,
\label{chp:prelim:prpgscal:full:phiphi}
\\
\langle \phi^\dag\phi^\dag \rangle
\ = \
\begin{array}{c}
\scalebox{0.85}{\includegraphics[trim = 10bp 12bp 19bp 11bp,clip]{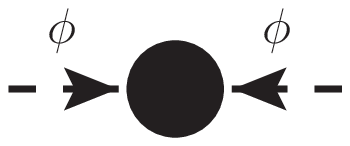}}
\end{array}
&=& \I \frac{\mu^{2*}}{(p^2-M^2)^2-|\mu^2|^2} \,.
\end{eqnarray}
\end{subequations}
In derivation of the full propagators \eqref{chp:prelim:prpgscal:full} we assumed that the 1PI corrections to the \qm{normal} propagators $\langle \phi\phi^\dag \rangle$, $\langle \phi^\dag\phi \rangle$ vanished:
\begin{subequations}
\begin{eqnarray}
\langle \phi\phi^\dag \rangle_{\mathrm{1PI}}
\ = \
\begin{array}{c}
\scalebox{0.85}{\includegraphics[trim = 10bp 12bp 19bp 11bp,clip]{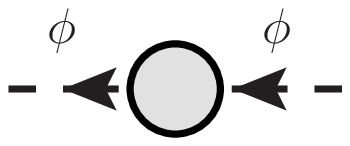}}
\end{array}
&=& 0 \,,
\\
\langle \phi^\dag\phi \rangle_{\mathrm{1PI}}
\ = \
\begin{array}{c}
\scalebox{0.85}{\includegraphics[trim = 10bp 12bp 19bp 11bp,clip]{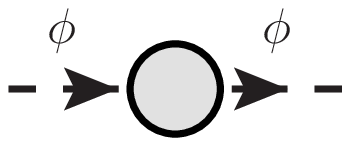}}
\end{array}
&=& 0 \,.
\end{eqnarray}
\end{subequations}
The assumption about the existence of the constant 1PI propagators $\langle \phi\phi \rangle_{\mathrm{1PI}}$, $\langle \phi^\dag\phi^\dag \rangle_{\mathrm{1PI}}$, \eqref{chp:prelim:prpgscal:1PI}, is equivalent to the assumption about the existence of the effective scalar quadratic terms of the type $\phi\phi$, $\phi^\dag\phi^\dag$ in the Lagrangian:
\begin{eqnarray}
\label{chp:prelim:eL:scl:free}
\eL_{\mathrm{scalar,free}} &=&
(\partial_{\mu}\phi)^\dag(\partial^{\mu}\phi)
-M^2\phi^\dag\phi
-\frac{1}{2}\mu^{*2}\phi\phi
-\frac{1}{2}\mu^{2}\phi^\dag\phi^\dag \,.
\end{eqnarray}
Decomposing now the complex field $\phi$ to its real and imaginary part as
\begin{eqnarray}
\phi & = & \frac{1}{\sqrt{2}} \big(\phi_1 + \I\phi_2\big)
\end{eqnarray}
and appropriately rotating the real fields $\phi_{1,2}$, one can diagonalize the free scalar Lagrangian \eqref{chp:prelim:eL:scl:free} and one finds the spectrum to be
\begin{eqnarray}
\label{chp:prelim:def:varphiM12}
M_{1,2}^2 & = & M^2 \pm |\mu|^2 \,.
\end{eqnarray}
The corresponding mass eigenstates $\varphi_{1,2}$ are real scalar fields and can be expressed as certain linear combinations of the original $\phi_{1,2}$ fields:
\begin{eqnarray}
\label{chp:prelim:def:varphi}
\left(\begin{array}{c}
\phi_1 \\
\phi_2 \\
\end{array}\right)
& = &
\left(\begin{array}{rr}
\cos \theta & -\sin \theta \\
\sin \theta &  \cos \theta \\
\end{array}\right)
\left(\begin{array}{c}
\varphi_1 \\
\varphi_2 \\
\end{array}\right) \,,
\end{eqnarray}
or, more compactly, as \cite{Brauner:2005hw}
\begin{eqnarray}
\label{chp:prelim:def:varphicmpct}
\phi & = & \frac{1}{\sqrt{2}} \e^{\I \theta} \big(\varphi_1 + \I\varphi_2\big) \,,
\end{eqnarray}
where the mixing angle $\theta$ is given by
\begin{eqnarray}
\label{chp:prelim:def:varphitheta}
\tan 2 \theta & = & \frac{\Im \mu^2}{\Re \mu^2} \,.
\end{eqnarray}
Thus, in a nutshell, we conclude that the assumption about the existence of non-vanishing scalar two-point functions of the type $\langle \phi \phi \rangle$, $\langle \phi^\dag \phi^\dag \rangle$ inevitably leads to splitting of the complex scalar $\phi$ with the mass $M^2$ into two real scalars $\varphi_1$, $\varphi_2$ with \emph{different} masses $M_1^2$ and $M_2^2$, respectively.

Let us now turn our attention to the fermion. Once the axial symmetry is broken by formation of the scalar propagators $\langle \phi \phi \rangle$, $\langle \phi^\dag \phi^\dag \rangle$, nothing protects the fermion from acquiring a mass. Recall that the potential fermion mass terms \eqref{chp:prelim:eL:mass} read
\begin{eqnarray}
\eL_{\mathrm{fermion,mass}} & = & {}- m \bar \psi_L \psi_R - m^* \bar \psi_R \psi_L \,.
\end{eqnarray}
Such effective mass terms are actually equivalent to the formation of 1PI parts of the fermion propagators, connecting the left-handed and right-handed chiral fields:
\begin{subequations}
\begin{eqnarray}
\langle \psi_L \bar\psi_R \rangle_{\mathrm{1PI}} & = & - \I m   P_R \,,
\\
\langle \psi_R \bar\psi_L \rangle_{\mathrm{1PI}} & = & - \I m^* P_L \,.
\end{eqnarray}
\end{subequations}
However, instead of seeking directly for the fermion mass $m$ itself, let us consider its generalization: The momentum-dependent complex self-energy $\Sigma(p^2)$. The fermion 1PI propagators are therefore assumed to have form
\begin{subequations}
\label{chp:prelim:psi:sgm}
\begin{eqnarray}
\langle \psi_L \bar \psi_R \rangle_{\mathrm{1PI}}
\ = \
\begin{array}{c}
\scalebox{0.85}{\includegraphics[trim = 10bp 12bp 19bp 11bp,clip]{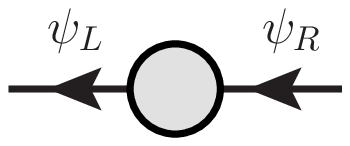}}
\end{array}
&=& - \I \, \Sigma(p^2)\,P_R \,,
\\
\langle \psi_R \bar \psi_L \rangle_{\mathrm{1PI}}
\ = \
\begin{array}{c}
\scalebox{0.85}{\includegraphics[trim = 10bp 12bp 19bp 11bp,clip]{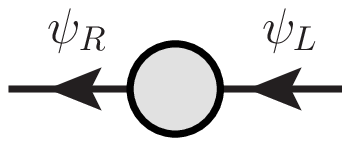}}
\end{array}
&=& - \I \, \Sigma^*\!(p^2)\,P_L
\end{eqnarray}
\end{subequations}
and the corresponding full propagators read
\begin{subequations}
\label{chp:prelim:psi:fll}
\begin{eqnarray}
\langle \psi_L \bar \psi_R \rangle
\ = \
\begin{array}{c}
\scalebox{0.85}{\includegraphics[trim = 10bp 12bp 19bp 11bp,clip]{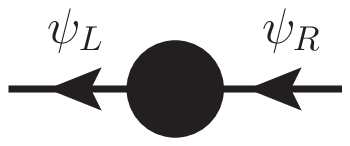}}
\end{array}
&=& \I \frac{\Sigma(p^2)}{p^2-|\Sigma(p^2)|^2} P_L \,,
\\
\langle \psi_R \bar \psi_L \rangle
\ = \
\begin{array}{c}
\scalebox{0.85}{\includegraphics[trim = 10bp 12bp 19bp 11bp,clip]{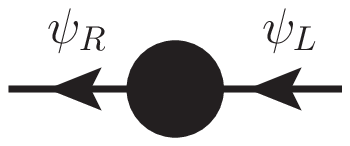}}
\end{array}
&=& \I \frac{\Sigma^*\!(p^2)}{p^2-|\Sigma(p^2)|^2} P_R \,.
\end{eqnarray}
\end{subequations}
Analogously to the scalar case, we again assumed that there are no 1PI corrections to the fermion propagators $\langle \psi_L \bar\psi_L \rangle$, $\langle \psi_R \bar\psi_R \rangle$ (i.e., proportional to $\slashed{p}$, see appendix~\ref{app:fermi propag}):
\begin{subequations}
\begin{eqnarray}
\langle \psi_L \bar \psi_L \rangle_{\mathrm{1PI}}
\ = \
\begin{array}{c}
\scalebox{0.85}{\includegraphics[trim = 10bp 12bp 19bp 11bp,clip]{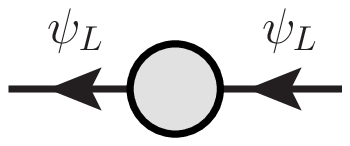}}
\end{array}
&=& 0 \,,
\\
\langle \psi_R \bar \psi_R \rangle_{\mathrm{1PI}}
\ = \
\begin{array}{c}
\scalebox{0.85}{\includegraphics[trim = 10bp 12bp 19bp 11bp,clip]{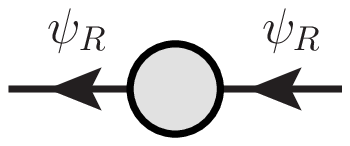}}
\end{array}
&=& 0 \,.
\end{eqnarray}
\end{subequations}
Finally, it is also useful to see how the 1PI and full propagator of the fermion field $\psi = \psi_L + \psi_R$ look like:

\begin{subequations}
\begin{eqnarray}
\langle \psi \bar \psi \rangle_{\mathrm{1PI}}
\ = \
\begin{array}{c}
\scalebox{0.85}{\includegraphics[trim = 10bp 12bp 19bp 11bp,clip]{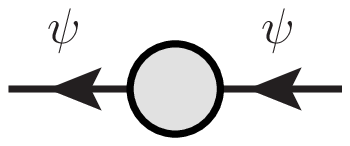}}
\end{array}
&=& - \I \big[ \Sigma^*\!(p^2)P_L + \Sigma(p^2)P_R \big] \,,
\\
\langle \psi \bar \psi \rangle
\ = \
\begin{array}{c}
\scalebox{0.85}{\includegraphics[trim = 10bp 12bp 19bp 11bp,clip]{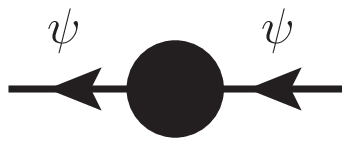}}
\end{array}
&=& \I \frac{\slashed{p} + \Sigma(p^2)P_L + \Sigma^*\!(p^2)P_R}{p^2-|\Sigma(p^2)|^2} \,.
\end{eqnarray}
\end{subequations}
Now we can see that with the self-energy $\Sigma(p^2)$ the fermion mass squared $m^2$ can be easily revealed as the pole of the propagator. That is, as the solution of the equation
\begin{eqnarray}
\label{chp:prelim:psi:m2}
m^2 &=& |\Sigma(m^2)|^2 \,.
\end{eqnarray}

\begin{figure}[t]
\begin{center}
\includegraphics[width=0.6\textwidth]{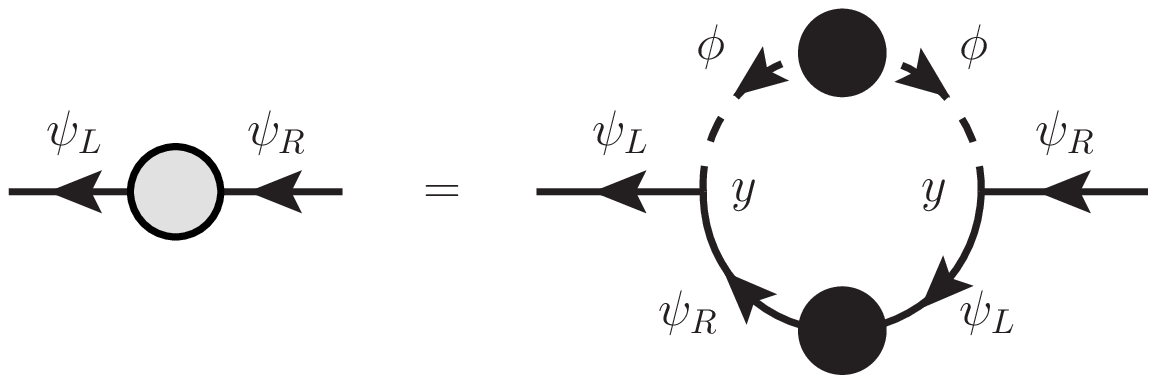}
\caption[SD equation for $\langle \psi_L \bar \psi_R \rangle_{\mathrm{1PI}} = -\I \Sigma \, P_R$.]{Diagrammatical representation of the equation \eqref{chp:prelim:SDpsi}. Cf.~the Feynman rules \eqref{chp:prelim:prpgscal:full}, \eqref{chp:prelim:psi:sgm}, \eqref{chp:prelim:psi:fll}.}
\label{chp:prelim:fig:SDpsi}
\end{center}
\end{figure}

Now the question is how the fermion self-energy $\Sigma(p^2)$ can be actually generated. The key r\^{o}le is here played by the symmetry-breaking scalar propagators $\langle\phi\phi\rangle$, $\langle\phi^\dag\phi^\dag\rangle$. For instance, the 1PI propagator $\langle \psi_L \bar\psi_R \rangle_{\mathrm{1PI}}$ can be calculated via the loop diagram containing the full scalar propagator $\langle\phi\phi\rangle$, as depicted in Fig.~\ref{chp:prelim:fig:SDpsi}. Using the explicit formul{\ae} \eqref{chp:prelim:prpgscal:full}, \eqref{chp:prelim:psi:sgm}, \eqref{chp:prelim:psi:fll} for the propagators, the diagram in Fig.~\ref{chp:prelim:fig:SDpsi} can be translated as
\begin{eqnarray}
\label{chp:prelim:SDpsi}
-\I \Sigma(p^2) & = & y^2 \int\!\frac{\d^4 k}{(2\pi)^4}\,
\frac{\Sigma^*\!(k^2)}{k^2-|\Sigma(k^2)|^2}
\frac{\mu^2}{[(k-p)^2-M^2]^2-|\mu^2|^2} \,.
\end{eqnarray}
A similar diagram as that in Fig.~\ref{chp:prelim:fig:SDpsi} can be drawn also for $\langle \psi_R \bar\psi_L \rangle_{\mathrm{1PI}}$, just instead of $\langle\phi\phi\rangle$ there would be rather $\langle\phi^\dag\phi^\dag\rangle$. Therefore only the substitution $\Sigma \leftrightarrow \Sigma^*$ and $\mu \rightarrow \mu^*$ would have to be done in Eq.~\eqref{chp:prelim:SDpsi}, together with $y \rightarrow y^*$, as this time the second term of the Yukawa interactions \eqref{chp:prelim:eq:Yukawa} would come into play.

The equation \eqref{chp:prelim:SDpsi} is an integral equation for the unknown complex function $\Sigma(p^2)$. The equation is non-linear and homogenous: It obviously possesses the trivial solution $\Sigma(p^2) \equiv 0$. We are of course seeking for a non-trivial solution.

Notice the convergence properties of the integral in \eqref{chp:prelim:SDpsi}: It converges even for a constant fermion self-energy, since the kernel (the scalar propagator $\langle \phi \phi \rangle$, \eqref{chp:prelim:prpgscal:full:phiphi}) asymptotically behaves like $1/k^4$ and the whole integrand thus as $1/k^6$. The physical reason for this is the following: The scalar propagator $\langle \phi \phi \rangle$ is in fact a difference of the propagators of the scalar mass eigenstates $\varphi_1$, $\varphi_2$, introduced in \eqref{chp:prelim:def:varphi}:
\begin{eqnarray}
\label{chp:prelim:prpg:diff}
\langle \phi \phi \rangle & = & \frac{1}{2}\e^{2\I\theta} \big( \langle \varphi_1\varphi_1 \rangle - \langle \varphi_2\varphi_2 \rangle \big) \,,
\end{eqnarray}
as can be shown using \eqref{chp:prelim:def:varphicmpct} (note also that, by assumption, $\langle \varphi_1\varphi_2 \rangle = \langle \varphi_2\varphi_1 \rangle = 0$). Indeed, taking into account the relations \eqref{chp:prelim:def:varphiM12} and \eqref{chp:prelim:def:varphitheta}, the explicit form \eqref{chp:prelim:prpgscal:full:phiphi} of $\langle \phi \phi \rangle$ in terms of $\mu$ can be rewritten as
\begin{eqnarray}
\frac{\mu^2}{(p^2-M^2)^2-|\mu^2|^2} & = & \frac{1}{2}\e^{2\I\theta}
\bigg( \frac{1}{p^2-M_1^2} - \frac{1}{p^2-M_2^2} \bigg) \,.
\end{eqnarray}
which is nothing else than \eqref{chp:prelim:prpg:diff}. These convergence properties of the integral suggest that the resulting non-trivial solution $\Sigma(p^2)$, if it exists, should be UV-finite. In fact, we can estimate from the behavior of the kernel in \eqref{chp:prelim:SDpsi} for large exterior momentum that $\Sigma(p^2)$ should behave like $1/p^4$ for large $p^2$.

\begin{figure}[t]
\begin{center}
\includegraphics[width=0.6\textwidth]{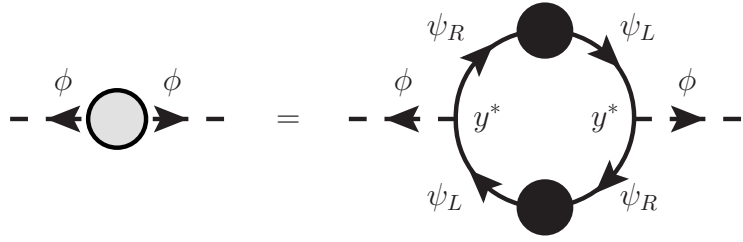}
\caption[SD equation for $\langle \phi \phi \rangle_{\mathrm{1PI}} = -\I \Pi$.]{Diagrammatical representation of the equation \eqref{chp:prelim:SDphi}.}
\label{chp:prelim:fig:SDphi}
\end{center}
\end{figure}

Now that we have generated the fermion propagators $\langle \psi_L \bar\psi_R \rangle$, $\langle \psi_R \bar\psi_L \rangle$, we can turn back to the question how the scalar propagators $\langle \phi \phi \rangle$, $\langle \phi^\dag \phi^\dag \rangle$, whose existence was only assumed so far, can be generated. One could ask whether it is possible to draw Feynman diagrams for the scalar 1PI functions $\langle \phi \phi \rangle_{\mathrm{1PI}}$, $\langle \phi^\dag \phi^\dag \rangle_{\mathrm{1PI}}$ in a similar way, as we have just done for the fermion propagators. Indeed, it turns out that with the fermion chirality-changing functions $\langle \psi_L \bar\psi_R \rangle$, $\langle \psi_R \bar\psi_L \rangle$ at disposal such diagrams can really be drawn. An example of such a diagram is depicted in Fig.~\ref{chp:prelim:fig:SDphi} for the Green's function $\langle \phi \phi \rangle_{\mathrm{1PI}}$.

Before proceeding further it should be noted that the very existence of diagrams such as that in Fig.~\ref{chp:prelim:fig:SDphi} forces us to revise our assumption that the 1PI functions $\langle \phi \phi \rangle_{\mathrm{1PI}}$, $\langle \phi^\dag \phi^\dag \rangle_{\mathrm{1PI}}$ are momentum-independent (i.e., proportional to complex \emph{constants} $\mu^{2}$, $\mu^{2*}$, respectively). Instead, let us generalize the Eqs.~\eqref{chp:prelim:prpgscal:1PI} in a similar manner as we did before for the fermion 1PI propagators: Let us promote the constants $\mu^{2}$, $\mu^{2*}$ to complex functions, i.e., to the momentum-dependent self-energies $\Pi(p^2)$, $\Pi^*(p^2)$. That is, we assume that Eqs.~\eqref{chp:prelim:prpgscal:1PI} now modify as
\begin{subequations}
\label{chp:prelim:prpgscal:1PIdep}
\begin{eqnarray}
\langle \phi\phi \rangle_{\mathrm{1PI}} & = & -\I \, \Pi(p^2) \,,
\\
\langle \phi^\dag\phi^\dag \rangle_{\mathrm{1PI}} & = & -\I \, \Pi^*(p^2)
\end{eqnarray}
\end{subequations}
and the corresponding full propagators are given by
\begin{subequations}
\begin{eqnarray}
\langle \phi\phi \rangle & = & \I \frac{\Pi(p^2)}{(p^2-M^2)^2-|\Pi(p^2)|^2} \,,
\\
\langle \phi^\dag\phi^\dag \rangle & = & \I \frac{\Pi^*(p^2)}{(p^2-M^2)^2-|\Pi(p^2)|^2} \,,
\end{eqnarray}
\end{subequations}
cf.~\eqref{chp:prelim:prpgscal:full}. (We do not present the Feynman rules again, as they are of course the same as those in \eqref{chp:prelim:prpgscal:1PI}, \eqref{chp:prelim:prpgscal:full}.) The scalar spectrum is now given by the equation
\begin{eqnarray}
\label{chp:prelim:phi:m2}
M_{1,2}^2 & = & M^2 \pm |\Pi(M_{1,2}^2)| \,,
\end{eqnarray}
which is just a generalization of \eqref{chp:prelim:def:varphiM12}.

Now we can write down the equation depicted diagrammatically in Fig.~\ref{chp:prelim:fig:SDphi}:
\begin{eqnarray}
\label{chp:prelim:SDphi}
-\I \, \Pi(p^2) & = & -2  y^{*2} \int\!\frac{\d^4 k}{(2\pi)^4}\,
\frac{\Sigma(k^2)}{k^2-|\Sigma(k^2)|^2}
\frac{\Sigma((k-p)^2)}{(k-p)^2-|\Sigma((k-p)^2)|^2} \,.
\end{eqnarray}
The minus sign on the right-hand side is for the fermion loop, while the factor of $2$ comes from $\Tr[P_L P_L] = 2$ (there is no combinatorial factor). Note that since the fermion self-energy $\Sigma(p^2)$ presumably behaves like $1/p^4$ at large $p^2$, the integral in \eqref{chp:prelim:SDphi} does converge. In fact, it does converge as long as the $\Sigma(p^2)$ decreases, no matter how slowly (the limit case is $\Sigma(p^2) = \mathrm{const.}$, in which case the integral \eqref{chp:prelim:SDphi} diverges logarithmically). Consequently, since $\Sigma(p^2)$ is a decreasing function, the equation \eqref{chp:prelim:SDphi} suggests that $\Pi(p^2)$ is a decreasing function too.

Having upgraded the constant scalar propagators $\langle \phi \phi \rangle_{\mathrm{1PI}}$, $\langle \phi^\dag \phi^\dag \rangle_{\mathrm{1PI}}$, \eqref{chp:prelim:prpgscal:1PI}, to momentum-dependent ones \eqref{chp:prelim:prpgscal:1PIdep}, we should also accordingly revise the Eq.~\eqref{chp:prelim:SDpsi} for the fermion self-energy. Not surprisingly, the result is
\begin{eqnarray}
\label{chp:prelim:SDpsidep}
-\I \, \Sigma(p^2) & = & y^2 \int\!\frac{\d^4 k}{(2\pi)^4}\,
\frac{\Sigma^*\!(k^2)}{k^2-|\Sigma(k^2)|^2}
\frac{\Pi((k-p)^2)}{[(k-p)^2-M^2]^2-|\Pi((k-p)^2)|^2} \,.
\end{eqnarray}
Notice that since $\Pi(p^2)$ is assumed to be a decreasing function, the convergence behavior of the integral \eqref{chp:prelim:SDpsi} has been actually improved.

\section{Summary}

Let us recapitulate the results of this chapter. While in the case $M^2<0$ the axial symmetry is broken down by the dynamics of the scalar itself, through its VEV (i.e., constant one-point function $\langle \phi \rangle$), we have shown that in the case $M^2>0$ the axial symmetry can be presumably broken by the common (i.e., Yukawa) dynamics of \emph{both} the scalar and the fermion. The SSB of the axial symmetry is driven by the two-point functions of the type $\langle \phi\phi \rangle$ and $\langle \psi_L\bar\psi_R \rangle$, or more precisely, by their 1PI parts $\Pi$ and $\Sigma$. They are functions of momentum squared and are tight together by the system of equations \eqref{chp:prelim:SDphi} and \eqref{chp:prelim:SDpsidep}, which we state here again for the reader's convenience:\footnote{Recall that $y=|y|\e^{\I\alpha}$ can be considered real, without loss of generality. The elimination of the $\e^{\I\alpha}$ by redefinition $\phi \rightarrow \e^{\I\alpha} \phi$, mentioned above, corresponds to redefinition $\Pi \rightarrow \e^{2\I\alpha} \Pi$ in equations \eqref{chp:prelim:SD}.}
\begin{subequations}
\label{chp:prelim:SD}
\begin{eqnarray}
-\I \,\Pi(p^2) & = & -2  y^{*2} \int\!\frac{\d^4 k}{(2\pi)^4}\,
\frac{\Sigma(k^2)}{k^2-|\Sigma(k^2)|^2}
\frac{\Sigma((k-p)^2)}{(k-p)^2-|\Sigma((k-p)^2)|^2} \,,
\\
-\I \,\Sigma(p^2) & = & y^2 \int\!\frac{\d^4 k}{(2\pi)^4}\,
\frac{\Sigma^*\!(k^2)}{k^2-|\Sigma(k^2)|^2}
\frac{\Pi((k-p)^2)}{[(k-p)^2-M^2]^2-|\Pi((k-p)^2)|^2} \,.
\end{eqnarray}
\end{subequations}
In order to have SSB, these equations must posses some non-trivial solution (apart from the obvious trivial solution $\Pi(p^2) = \Sigma(p^2) \equiv 0$, corresponding to no SSB).

The equations \eqref{chp:prelim:SD} are subset of the Schwinger--Dyson (SD) equations \cite{Dyson:1949ha,Schwinger:1951ex,Schwinger:1951hq}. They can be understood as a formal summation of all orders of the perturbation theory, therefore they themselves are non-perturbative. This is in accordance with our previous claim that any SSB must be a non-perturbative effect.

The SD equations constitute in principle infinite \qm{tower} of coupled integral equations for all Green's functions of the theory, not only the two-point functions. For practical calculation one usually has to truncate this \qm{tower} at some level. We truncated it at the level of three-point Green's functions, which we approximate by the bare ones. Although we will in the following chapters derive the SD equations in a more formal way, we will still use the same truncation scheme, i.e., we will always neglect the three- and more-point functions in non-perturbative calculations. (Nevertheless, there will be some \emph{perturbative} calculations of the three-point functions.)

Since the scalar mass $M$ is the only mass scale in the Lagrangian, the fermion mass $m$, as calculated from the Eq.~\eqref{chp:prelim:psi:m2}, will necessarily have the form
\begin{eqnarray}
m & = & M \, f(y) \,.
\end{eqnarray}
Here $f(y)$ must be a function only of the Yukawa coupling constant $y$, since it is the only dimensionless parameter of the SD equations \eqref{chp:prelim:SD}, whose solution $\Sigma$ is. This function is non-perturbative, i.e., non-analytic in $y$. Moreover, inspired by the situation in the Nambu--Jona-Lasinio (NJL) model \cite{Nambu:1961tp,Nambu:1961fr}, where schematically $f(y) \sim \exp(-1/y)$, one hopes that the a small change in $y$ (within the same order of magnitude) might produce a much larger (by several orders of magnitude) change in $f(y)$. Put another way, different Yukawa coupling constants, yet of the same order of magnitude, can potentially produce a large hierarchy in the fermion spectrum. This is to be compared with the situation in case of condensing scalar, Sec.~\ref{chp:prelim:SSB1point}, where the fermion mass $m$ depends linearly on $y$:
\begin{eqnarray}
m & = & -\sqrt{\frac{-M^2}{\lambda}} \, y \,.
\end{eqnarray}

\chapter{Formal developments}
\label{chp:frm}

\intro{While the previous chapter served as a rather intuitive and informal introduction, now we are going to treat the same subject more rigorously and in more detail. In particular, we discuss here in detail which parts of the scalar and fermion self-energies are actually needed for the sake of demonstration of spontaneous breaking of the axial symmetry and we show how to arrive more decently at the Schwinger--Dyson equations, derived in the previous chapter in a rather clumsy way. Finally, we also give some numeric evidence of viability of the presented scheme.}


\section{The model}

\subsection{Lagrangian}

We consider a complex scalar field $\phi$ and two species of massless fermions, $\psi_1$ and $\psi_2$, with the Lagrangian
\begin{eqnarray}
\label{chp:frm:eL}
\eL & = &
\bar\psi_1\I \slashed{\partial} \psi_1 +
\bar\psi_2\I \slashed{\partial} \psi_2 +
(\partial_{\mu}\phi)^\dag(\partial^{\mu}\phi)
-M^2\phi^{\dagger}\phi
+ \eL_{\mathrm{Yukawa}} \,.
\end{eqnarray}
The Yukawa interactions are again not the most general ones:
\begin{eqnarray}
\label{chp:frm:eL:Yukawa}
\eL_{\mathrm{Yukawa}} &=&  \hphantom{+\,}
y_1 \bar\psi_{1L}\psi_{1R}\phi + y_1^* \bar\psi_{1R}\psi_{1L}\phi^{\dagger}
\nonumber \\ && + \,
y_2 \bar\psi_{2R}\psi_{2L}\phi + y_2^* \bar\psi_{2L}\psi_{2R}\phi^{\dagger} \,.
\end{eqnarray}
In particular, the terms with interchanged $\phi \leftrightarrow \phi^\dag$,
\begin{eqnarray}
\label{chp:frm:eL:tildeYukawa}
\tilde\eL_{\mathrm{Yukawa}} &=&  \hphantom{+\,}
\tilde y_1 \bar\psi_{1L}\psi_{1R}\phi^{\dagger} + \tilde y_1^{*} \bar\psi_{1R}\psi_{1L}\phi
\nonumber \\ && + \,
\tilde y_2 \bar\psi_{2R}\psi_{2L}\phi^{\dagger} + \tilde y_2^{*} \bar\psi_{2L}\psi_{2R}\phi \,,
\end{eqnarray}
are absent. On top of it, also the Yukawa interaction terms mixing both fermion species (i.e., the terms proportional to, e.g., $\bar\psi_{1L}\psi_{2R}$) are missing.

The Yukawa coupling constants $y_1$, $y_2$ can be again without loss of generality considered real, since the phase can be eliminated by a redefinition of the corresponding fermion fields (e.g., by phase transformations of $\psi_{1R}$ and $\psi_{2R}$). Nevertheless, we keep them deliberately complex for similar reasons as in the previous chapter.

Notice that we do not consider in the Lagrangian \eqref{chp:frm:eL} the scalar self-interaction term
\begin{eqnarray}
\label{chp:frm:sclr:selfint}
\eL_{\mathrm{selfint.}} &=& -\frac{1}{2}\lambda(\phi^{\dagger}\phi)^2 \,.
\end{eqnarray}
This is due to the lesson learned in previous chapter that within the present scheme of breaking the symmetry solely by the Yukawa dynamics, through formation of appropriate fermion and scalar two-point functions, the pure scalar dynamics of the type \eqref{chp:frm:sclr:selfint} is dispensable, in contrast to the breaking of the symmetry by scalar VEV, discussed in Sec.~\ref{chp:prelim:SSB1point}. Thus, from now on we will systematically neglect the scalar self-interactions of the type \eqref{chp:frm:sclr:selfint} in the rest of this text. Of course, in more phenomenologically oriented treatment the scalar self-interactions would have to be included, as they are not protected by any symmetry and they would be generated anyway by means of radiative corrections.


\subsection{Symmetries}

Let us investigate the symmetries of the Lagrangian. First, we observe that the both fermion numbers are separately conserved, which corresponds to the vectorial symmetry $\group{U}(1)_{\mathrm{V}_{\!1}} \times \group{U}(1)_{\mathrm{V}_{\!2}}$. Just for the sake of later references let us write the transformation of $\psi_{i}$ under $\group{U}(1)_{\mathrm{V}_{\!j}}$ as\footnote{\label{footnote:nosum}No sum over the fermion specie index $j$ is assumed, here, as well as in the rest of the text.}
\begin{eqnarray}
\group{U}(1)_{\mathrm{V}_{\!j}}\,:\qquad \psi_{i}  &\TransformsTo&
{[\psi_{i}]}^\prime \>=\> \e^{ \I \theta_{\mathrm{V}_{\!j}} t_{i,\mathrm{V}_{\!j}}} \, \psi_{i} \,,
\end{eqnarray}
where $\theta_{\mathrm{V}_{\!j}}$ are the parameters of the transformation and the generators $t_{i,\mathrm{V}_{\!j}}$ are given by
\begin{eqnarray}
\label{chp:frm:U1Vipsigen}
t_{i,\mathrm{V}_{\!j}} &=&
\begin{cases}
\ Q_{\mathrm{V}_{\!i}} & \text{if $i=j$} \,,\\
\ 0 & \text{if $i \neq j$} \,,
\end{cases}
\end{eqnarray}
where $Q_{\mathrm{V}_{\!i}}$ are some non-vanishing real numbers. (Needless to say that $\phi$ transforms trivially under $\group{U}(1)_{\mathrm{V}_{\!i}}$, i.e., $t_{\phi,\mathrm{V}_{\!i}} = 0$.) The fermion numbers are conserved separately for both fermion species due to the specific form of the Yukawa interactions, namely due to the lack of the mixing terms; otherwise there would be only one $\group{U}(1)_{\mathrm{V}}$ symmetry, corresponding to the global fermion number conservation of both fermion species.

Apart from the vectorial symmetry, there is also axial symmetry, which is going to play a more important r\^{o}le in our considerations. In contrast to the case of vectorial symmetries, this time, instead of two independent axial symmetries $\group{U}(1)_{\mathrm{A}_{\!1}} \times \group{U}(1)_{\mathrm{A}_{\!2}}$ (which would be present in absence of the Yukawa interactions), there is rather a single axial symmetry $\group{U}(1)_{\mathrm{A}}$. It acts on the fermions as
\begin{subequations}
\label{chp:frm:U1A}
\begin{eqnarray}
\label{chp:frm:U1Apsi}
\group{U}(1)_{\mathrm{A}}\,:\qquad \psi_{i}  &\TransformsTo&
{[\psi_{i}]}^\prime \>=\> \e^{ \I \theta_{\mathrm{A}} t_{i,\mathrm{A}}} \, \psi_{i}
\end{eqnarray}
and on the scalar as
\begin{eqnarray}
\label{chp:frm:U1Aphi}
\group{U}(1)_{\mathrm{A}}\,:\qquad \phi      &\TransformsTo&
[\phi]^\prime \>=\> \e^{ \I \theta_{\mathrm{A}} t_{\phi,\mathrm{A}}} \, \phi \,.
\end{eqnarray}
\end{subequations}
The fermion $\group{U}(1)_{\mathrm{A}}$ generators $t_{i,\mathrm{A}}$ are given by
\begin{eqnarray}
\label{chp:frm:Apsigen}
t_{i,\mathrm{A}} &=& Q_{i,\mathrm{A}} \gamma_5 \,,
\end{eqnarray}
with the axial charges $Q_{1,\mathrm{A}}$ and $Q_{2,\mathrm{A}}$ (being of course non-vanishing real numbers) constrained by
\begin{eqnarray}
\label{chp:frm:anomalyfree}
Q_{1,\mathrm{A}} + Q_{2,\mathrm{A}} & = & 0 \,.
\end{eqnarray}
The scalar $\group{U}(1)_{\mathrm{A}}$ generator $t_{\phi,\mathrm{A}}$ then reads
\begin{subequations}
\label{chp:frm:U1Aphigen}
\begin{eqnarray}
t_{\phi,\mathrm{A}} &=& -2 Q_{1,\mathrm{A}} \\ &=& + 2 Q_{2,\mathrm{A}} \,.
\end{eqnarray}
\end{subequations}
Note that the axial symmetry $\group{U}(1)_{\mathrm{A}}$ with the action \eqref{chp:frm:U1A} in fact forbids the Yukawa interactions $\tilde\eL_{\mathrm{Yukawa}}$, \eqref{chp:frm:eL:tildeYukawa}.

Due to the existence of the axial symmetry, one might concern whether this symmetry is not anomalous. At the moment this question is actually not too urgent, as the axial symmetry is global. However, later on in chapter~\ref{chp:ablgg} (more precisely, in Sec.~\ref{chp:ablgg:sec:abl}) we will gauge it and consequently it will become obligatory to remove the axial anomaly in order to have a consistent gauge quantum field theory. Nevertheless, the theory is in fact anomaly free already at this moment. This is thanks to the introduction of the two fermion species $\psi_1$ and $\psi_2$ with opposite axial charges, see Eq.~\eqref{chp:frm:anomalyfree}.\footnote{In this respect the two fermion species $\psi_1$ and $\psi_2$ can be regarded as analogues of the leptons and quarks, respectively.} The condition \eqref{chp:frm:anomalyfree}, primarily necessary for the theory to be invariant under $\group{U}(1)_{\mathrm{A}}$, is precisely the condition for cancelation of the axial anomaly. This is ultimately the reason why we have introduced two fermion species instead of only one and why we have chosen the Yukawa interactions to have the special form \eqref{chp:frm:eL:Yukawa}.

\subsection{Nambu--Gorkov formalism}

As noted in the previous chapter, the axial symmetry has to be (spontaneously) broken in order to allow for the generation of the fermion masses. We assume that this symmetry breakdown will be driven by formation of the scalar propagators of the type $\langle \phi\phi\rangle$ and $\langle \phi^\dag\phi^\dag\rangle$. Thus, it turns out to be convenient to reparameterize the theory in terms of new degree of freedom: The Nambu--Gorkov doublet $\Phi$, defined as\footnote{This is special case of more general definition \eqref{app:sclr:NG}, discussed in detail in appendix~\ref{app:sclr}.}
\begin{eqnarray}
\label{chp:frm:NambuGorkov}
\Phi &\equiv& \left(\begin{array}{c} \phi \\ \phi^\dag \end{array}\right)
\end{eqnarray}
and introduced originally for fermions in Refs.~\cite{Nambu:1960tm,Gorkov} in the context of the theory of superconductivity. The point is that now the propagator $\langle \Phi\Phi^\dag\rangle$ contains the two symmetry-breaking propagators $\langle \phi\phi\rangle$, $\langle \phi^\dag\phi^\dag\rangle$, together with the two symmetry-conserving propagators $\langle \phi\phi^\dag\rangle$, $\langle \phi^\dag\phi\rangle$, and allows this way to treat them all on the same footing:
\begin{eqnarray}
\label{chp:frm:NambuGorkovprop}
\langle \Phi\Phi^\dag\rangle
&=&
\left(\begin{array}{cc}
\langle \phi     \phi^\dag \rangle  &  \langle \phi     \phi \rangle \\
\langle \phi^\dag\phi^\dag \rangle  &  \langle \phi^\dag\phi \rangle
\end{array}\right)
\,.
\end{eqnarray}
The free, full and 1PI scalar propagators of the form \eqref{chp:frm:NambuGorkovprop} will be discussed in more detail in Sec.~\ref{chp:frm:sbsec:sclrprop}. Now let us note the key property of the Nambu--Gorkov field $\Phi$: It is real in the sense that its charge conjugate (i.e., basically the Hermitian conjugate) is proportional to itself:
\begin{eqnarray}
\label{chp:frm:NGcond}
\Phi &=& \sigma_1 \, \Phi^{\dag\T} \,,
\end{eqnarray}
where the Pauli matrix $\sigma_1$ acts in the two-dimensional space of the Nambu--Gorkov doublet \eqref{chp:frm:NambuGorkov}.

Let us rewrite the action of $\group{U}(1)_{\mathrm{A}}$ on the scalar field from the basis $\phi$, \eqref{chp:frm:U1Aphi}, to the Nambu--Gorkov basis $\Phi$:
\begin{eqnarray}
\label{chp:frm:U1APhi}
\group{U}(1)_{\mathrm{A}}\,:\qquad \Phi      &\TransformsTo&
[\Phi]^\prime \>=\> \e^{ \I \theta_{\mathrm{A}} t_{\Phi,\mathrm{A}}} \, \Phi \,,
\end{eqnarray}
where the generator $t_{\Phi,\mathrm{A}}$ is expressed in terms of $t_{\phi,\mathrm{A}}$, \eqref{chp:frm:U1Aphigen}, as
\begin{eqnarray}
\label{chp:frm:U1APhigen}
t_{\Phi,\mathrm{A}} &=& \left(\begin{array}{cc} t_{\phi,\mathrm{A}} & 0 \\ 0 & -t_{\phi,\mathrm{A}} \end{array}\right) \,.
\end{eqnarray}

We now rewrite the theory in terms of the Nambu--Gorkov field $\Phi$. The free scalar part of the Lagrangian \eqref{chp:frm:eL},
\begin{subequations}
\label{chp:frm:eLsclr}
\begin{eqnarray}
\eL_{\mathrm{scalar}} &=&
(\partial_{\mu}\phi)^\dag(\partial^{\mu}\phi) - M^2\phi^{\dagger}\phi
\,,
\end{eqnarray}
is easily rewritten in terms of $\Phi$ as
\begin{eqnarray}
\eL_{\mathrm{scalar}} &=&
\frac{1}{2}(\partial_{\mu}\Phi)^\dag(\partial^{\mu}\Phi) - \frac{1}{2}M^2 \Phi^{\dag}\Phi
\,.
\end{eqnarray}
\end{subequations}
The Yukawa Lagrangian \eqref{chp:frm:eL:Yukawa} can be written compactly in terms of $\Phi$ as
\begin{subequations}
\begin{eqnarray}
\label{chp:frm:eL:Ya}
\eL_{\mathrm{Yukawa}} &=& \sum_{i=1,2} \bar\psi_i\,\bar Y_i\,\psi_i\,\Phi \,,
\end{eqnarray}
or equivalently as
\begin{eqnarray}
\label{chp:frm:eL:Yb}
\eL_{\mathrm{Yukawa}} &=& \sum_{i=1,2} \Phi^\dag\,\bar\psi_i\,Y_i\,\psi_i \,.
\end{eqnarray}
\end{subequations}
The equivalence of the two apparently different expressions \eqref{chp:frm:eL:Ya} and \eqref{chp:frm:eL:Yb} is just a consequence of the reality of the field $\Phi$. The coupling constants $Y_i$ are doublets operating in the space of the Nambu--Gorkov field $\Phi$ and are defined as
\begin{equation}
\label{chp:frm:Y1Y2def}
Y_1 \>\equiv\> \left(\begin{array}{c} y_1^* P_L \\ y_1 P_R \end{array}\right)\,,
\quad\quad
Y_2 \>\equiv\> \left(\begin{array}{c} y_2^* P_R \\ y_2 P_L \end{array}\right)\,.
\end{equation}
The conjugate coupling constants $\bar Y_i$ are defined in accordance with \eqref{symbols:barA}, i.e., as
\begin{eqnarray}
\bar Y_i & \equiv & \gamma_0 Y_i^\dag \gamma_0 \,,
\end{eqnarray}
so that we have explicitly
\begin{equation}
\bar Y_1 \>=\> \Big( y_1 P_R , y_1^* P_L \Big)\,,
\quad\quad
\bar Y_2 \>=\> \Big( y_2 P_L , y_2^* P_R \Big)\,.
\end{equation}

\section{Propagators}

In this section we will first introduce the notation for the scalar and fermion propagators, then we will state the form of the propagators that we will be looking for in the next section in the quest for demonstrating the SSB and finally we will say something about what kind of spectrum is to be expected.

\subsection{Scalar propagators}
\label{chp:frm:sbsec:sclrprop}

Let us begin with the scalar. We denote the full scalar propagator (in the Nambu--Gorkov basis $\Phi$) as\footnote{From now on we will usually not explicitly indicate the momentum arguments at propagators and self-energies, unless they will not be obvious from the context.}
\begin{eqnarray}
\I \, G_\Phi &=& \langle \Phi \Phi^\dag \rangle \ = \
\begin{array}{c}
\scalebox{0.85}{\includegraphics[trim = 10bp 12bp 19bp 11bp,clip]{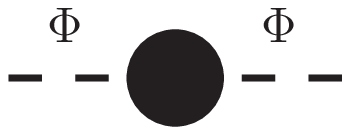}}
\end{array}
\,.
\end{eqnarray}
Notice that there are no arrows on the scalar line as a consequence of the reality of the field $\Phi$. The free propagator
\begin{eqnarray}
\I \, D_\Phi &=& \langle \Phi \Phi^\dag \rangle_0 \,,
\end{eqnarray}
determined by the free scalar Lagrangian \eqref{chp:frm:eLsclr}, is in the momentum representation given by
\begin{eqnarray}
\label{chp:frm:DPhi}
D_\Phi &=& \left(\begin{array}{cc} \dfrac{1}{p^2-M^2} & 0 \\ 0 & \dfrac{1}{p^2-M^2} \end{array}\right) \,.
\end{eqnarray}
The scalar self-energy $\boldsymbol{\Pi}$, defined as
\begin{eqnarray}
-\I \, \boldsymbol{\Pi} &=& \langle \Phi \Phi^\dag \rangle_{\mathrm{1PI}} \ = \
\begin{array}{c}
\scalebox{0.85}{\includegraphics[trim = 10bp 12bp 19bp 11bp,clip]{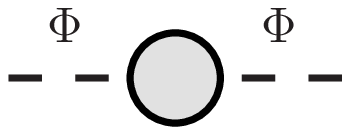}}
\end{array}
\,,
\end{eqnarray}
is now given in terms of the bare and full propagators by
\begin{eqnarray}
\label{chp:frm:PiDPhiGphi}
\boldsymbol{\Pi} &=& D_\Phi^{-1} - G_\Phi^{-1} \,.
\end{eqnarray}
The meaning of this expression is that it actually corresponds to the geometric series
\begin{subequations}
\begin{eqnarray}
G_\Phi &=& D_\Phi + D_\Phi\,\boldsymbol{\Pi}\,D_\Phi + D_\Phi\,\boldsymbol{\Pi}\,D_\Phi\,\boldsymbol{\Pi}\,D_\Phi + \ldots
\\ &=&
\big( D_\Phi^{-1} - \boldsymbol{\Pi} \big)^{-1} \,.
\end{eqnarray}
\end{subequations}
In other words, $\boldsymbol{\Pi}$ is indeed the 1PI part of the full propagator $G_\Phi$.

The reality condition \eqref{chp:frm:NGcond} of $\Phi$ has important impacts on the form of the propagators. It induces a non-trivial symmetry of the scalar propagator $G_\Phi$,
\begin{eqnarray}
\label{chp:frm:NGcondprop}
G_\Phi &=& \sigma_1 \, G_\Phi^\T \, \sigma_1 \,.
\end{eqnarray}
This condition must be satisfied also by the free propagator $D_\Phi$ (and, indeed, it is satisfied, see explicit form \eqref{chp:frm:DPhi} of $D_\Phi$), as it is just a special case of $G_\Phi$. Thus, the self-energy $\boldsymbol{\Pi}$ must satisfy the analogous condition too:
\begin{eqnarray}
\label{chp:frm:NGcondpropPi}
\boldsymbol{\Pi} &=& \sigma_1 \, \boldsymbol{\Pi}^\T \, \sigma_1 \,,
\end{eqnarray}
as can be inferred from the expression \eqref{chp:frm:PiDPhiGphi}.

\subsection{Fermion propagators}

Similarly in the fermion sector, the full propagators of the fermion fields are\footnote{We assume here implicitly that the fermion numbers $\group{U}(1)_{\mathrm{V}_{\!1}}$ and $\group{U}(1)_{\mathrm{V}_{\!2}}$ are \emph{separately} conserved even once the dynamics is taken into account. Otherwise we would have to consider propagator of the field $\psi \equiv \bigl(\begin{smallmatrix} \psi_1 \\ \psi_2 \end{smallmatrix} \bigr)$ with non-vanishing off-diagonal elements, breaking $\group{U}(1)_{\mathrm{V}_{\!1}} \times \group{U}(1)_{\mathrm{V}_{\!2}}$ spontaneously down to common fermion number symmetry $\group{U}(1)_{\mathrm{V}}$.}
\begin{eqnarray}
\I \, G_{\psi_i} &=& \langle \psi_i \bar\psi_i \rangle \ = \
\begin{array}{c}
\scalebox{0.85}{\includegraphics[trim = 10bp 12bp 19bp 11bp,clip]{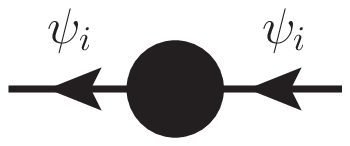}}
\end{array}
\,.
\end{eqnarray}
Note that since $\psi_i = \psi_{iL}+\psi_{iR}$, the propagators $\langle \psi_i \bar\psi_i \rangle$ contain all the particular propagators $\langle\psi_{iL}\bar\psi_{iL}\rangle$, $\langle\psi_{iR}\bar\psi_{iR}\rangle$, $\langle\psi_{iL}\bar\psi_{iR}\rangle$, $\langle\psi_{iR}\bar\psi_{iL}\rangle$, which have been treated in the previous chapter separately. The Lagrangian \eqref{chp:frm:eL} contains no fermion mass terms, consequently the free propagators
\begin{eqnarray}
\I \, S_{i} &=& \langle \psi_i \bar\psi_i \rangle_0
\end{eqnarray}
are in momentum space given simply by
\begin{eqnarray}
S_i^{-1} &=& \slashed{p} \,.
\end{eqnarray}
The fermion self-energies $\boldsymbol{\Sigma}_i$,
\begin{eqnarray}
\label{chp:frm:psi1PI}
- \I \, \boldsymbol{\Sigma}_i &=& \langle \psi_i \bar\psi_i \rangle_{\mathrm{1PI}} \ = \
\begin{array}{c}
\scalebox{0.85}{\includegraphics[trim = 10bp 12bp 19bp 11bp,clip]{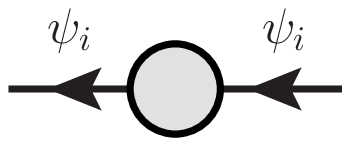}}
\end{array}
\,,
\end{eqnarray}
are now given by
\begin{eqnarray}
\boldsymbol{\Sigma}_i &=& S_i^{-1} - G_{\psi_i}^{-1} \,,
\end{eqnarray}
which again correspond to the geometric series
\begin{subequations}
\begin{eqnarray}
G_{\psi_i} &=&
S_i +  S_i\,\boldsymbol{\Sigma}_i\,S_i + S_i\,\boldsymbol{\Sigma}_i\,S_i\,\boldsymbol{\Sigma}_i\,S_i + \ldots
\\ &=& \big( S_{i}^{-1} - \boldsymbol{\Sigma}_i \big)^{-1} \,,
\end{eqnarray}
\end{subequations}
so that $\boldsymbol{\Sigma}_i$ are indeed nothing else than the 1PI parts of the full propagators, as indicated in \eqref{chp:frm:psi1PI}.

\subsection{Ans\"{a}tze for the self-energies}

The self-energies $\boldsymbol{\Sigma}_i$ and $\boldsymbol{\Pi}$, as the agents of the SSB of the axial symmetry, will be subject of our searching in the next section. In fact, our aim will be merely to demonstrate the possibility the SSB itself, we will not be interested in, e.g., the symmetry-preserving results. Therefore it is unnecessary to treat the self-energies in the full generality, it is sufficient to focus only on their symmetry-breaking parts. In other words, is useful to consider a suitable Ansatz for the self-energies.

The first step in constraining the otherwise in principle (almost) completely arbitrary self-energies is to impose the Hermiticity conditions
\begin{subequations}
\label{chp:frm:Herm}
\begin{eqnarray}
\boldsymbol{\Sigma}_i & = & \boldsymbol{\bar\Sigma}_i \,, \\
\boldsymbol{\Pi}      & = & \boldsymbol{\Pi}^\dag
\end{eqnarray}
\end{subequations}
(where $\boldsymbol{\bar\Sigma}_i \equiv \gamma_0\,\boldsymbol{\Sigma}_i^\dag\,\gamma_0$, cf.~\eqref{symbols:barA}). These conditions have no relation to the pattern of the SSB, in fact they are merely of a technical character. They will serve us for two purposes: First, they will reduce the number of unconstrained parts of the self-energies (i.e., reduce the number of independent SD equations) and second, they will eventually ensure that the resulting fermion and scalar spectrum (more precisely, the masses \emph{squared}) will be real. In order to understand better the Hermiticity conditions \eqref{chp:frm:Herm}, it is also useful to realize that the same conditions would hold if the self-energies were (momentum-independent) mass parameters in a \emph{Hermitian} Lagrangian.

The Hermiticity conditions \eqref{chp:frm:Herm}, together with the scalar symmetry condition \eqref{chp:frm:NGcondpropPi}, lead to the self-energies of the specific form
\begin{subequations}
\label{chp:frm:anz:intermed}
\begin{eqnarray}
\boldsymbol{\Sigma}_i &=& \slashed{p}\,\big(A_L\,P_L + A_R\,P_L\big) + \big(\Sigma^* P_L + \Sigma\,P_R\big) \,,
\\
\boldsymbol{\Pi} &=& \left(\begin{array}{cc} \Pi_N & \Pi \\ \Pi^* & \Pi_N \end{array}\right) \,,
\end{eqnarray}
\end{subequations}
where $A_L$, $A_R$, $\Pi_N$ and $\Sigma$, $\Pi$ are respectively real and complex, but otherwise arbitrary functions of $p^2$.

In order to further meaningfully constrain the self-energies, it is worth considering their relation to the axial symmetry $\group{U}(1)_{\mathrm{A}}$, which is assumed to be broken by them. In particular, it is important to know that the non-invariance of the self-energies under $\group{U}(1)_{\mathrm{A}}$ is measured by the quantities
\begin{subequations}
\label{chp:frm:noninv}
\begin{eqnarray}
\boldsymbol{\Sigma}_i\,t_{i,\mathrm{A}}-\bar t_{i,\mathrm{A}}\,\boldsymbol{\Sigma}_i &=& \bbl\boldsymbol{\Sigma}_i,t_{i,\mathrm{A}}\bbr \,, \\
\boldsymbol{\Pi}\,t_{\Phi,\mathrm{A}} - t_{\Phi,\mathrm{A}}\,\boldsymbol{\Pi} &=& [\boldsymbol{\Pi},t_{\Phi,\mathrm{A}}]
\end{eqnarray}
\end{subequations}
(cf.~definition \eqref{symbols:comAB}). This can be seen in two ways. First, more formally, one can directly study the transformation behavior of the self-energies under $\group{U}(1)_{\mathrm{A}}$, induced by the transformation rules \eqref{chp:frm:U1Apsi} and \eqref{chp:frm:U1APhi} of $\psi_i$ and $\Phi$, respectively. The corresponding self-energies $\boldsymbol{\Sigma}_{i}$ and $\boldsymbol{\Pi}$ then transform as
\begin{subequations}
\begin{eqnarray}
\group{U}(1)_{\mathrm{A}}\,:\qquad \boldsymbol{\Sigma}_{i}  \ \TransformsTo \  {[\boldsymbol{\Sigma}_{i}]}^\prime
&=& \e^{\I\theta_{\mathrm{A}}\bar t_{i,\mathrm{A}}}\,\boldsymbol{\Sigma}_{i}\,\e^{-\I\theta_{\mathrm{A}}t_{i,\mathrm{A}}}
\\ &=& \boldsymbol{\Sigma}_{i} - \I\theta_{\mathrm{A}}\big(\boldsymbol{\Sigma}_i\,t_{i,\mathrm{A}}-\bar t_{i,\mathrm{A}}\,\boldsymbol{\Sigma}_i\big) + \mathcal{O}(\theta_{\mathrm{A}}^2)
\end{eqnarray}
\end{subequations}
and
\begin{subequations}
\begin{eqnarray}
\group{U}(1)_{\mathrm{A}}\,:\qquad \boldsymbol{\Pi}      \ \TransformsTo \ [\boldsymbol{\Pi}]^\prime
&=& \e^{ \I \theta_{\mathrm{A}} t_{\phi,\mathrm{A}}}\,\boldsymbol{\Pi}\,\e^{- \I \theta_{\mathrm{A}} t_{\phi,\mathrm{A}}}
\\ &=& \boldsymbol{\Pi} - \I \theta_{\mathrm{A}} \big(\boldsymbol{\Pi}\,t_{\Phi,\mathrm{A}} - t_{\Phi,\mathrm{A}}\,\boldsymbol{\Pi}\big) + \mathcal{O}(\theta_{\mathrm{A}}^2) \,,
\end{eqnarray}
\end{subequations}
respectively. We see that the non-invariance of the self-energies under $\group{U}(1)_{\mathrm{A}}$ is indeed proportional to the corresponding quantities \eqref{chp:frm:noninv}. Another way of seeing it, less formal but perhaps more illuminating, is to imagine that the self-energies are momentum-independent (except for the $\slashed{p}$ in $\boldsymbol{\Sigma}_i$) and thus being interpretable as mass parameters of some effective Lagrangian:
\begin{eqnarray}
\label{chp:frm:eLeff}
\eL_{\mathrm{eff}} &\equiv&
- \sum_{i=1,2} \bar\psi_i\,\boldsymbol{\Sigma}_i\,\psi_i - \frac{1}{2}\Phi^\dag\,\boldsymbol{\Pi}\,\Phi \,.
\end{eqnarray}
(Accordingly, this effective Lagrangian is basically the mass Lagrangian for the fermions and scalar and it also contains corrections to the kinetic terms for the fermions, due to above mentioned $\slashed{p}$ being substituted by $-\I\slashed{\partial}$.) Upon performing the $\group{U}(1)_{\mathrm{A}}$ transformations \eqref{chp:frm:U1Apsi} and \eqref{chp:frm:U1APhi}, the Lagrangian \eqref{chp:frm:eLeff} transforms as
\begin{subequations}
\begin{eqnarray}
\group{U}(1)_{\mathrm{A}}\,:\qquad \eL_{\mathrm{eff}}  \ \TransformsTo \  {[\eL_{\mathrm{eff}}]}^\prime
&=&
- \sum_{i=1,2} \bar\psi_i\,\e^{-\I\theta_{\mathrm{A}}\bar t_{i,\mathrm{A}}}\,\boldsymbol{\Sigma}_{i}\,\e^{\I\theta_{\mathrm{A}}t_{i,\mathrm{A}}}\,\psi_i
- \frac{1}{2}\Phi^\dag\,\e^{-\I \theta_{\mathrm{A}} t_{\phi,\mathrm{A}}}\,\boldsymbol{\Pi}\,\e^{\I \theta_\mathrm{A} t_{\phi,\mathrm{A}}}\,\Phi
\nonumber \\  && \\
&=& \eL_{\mathrm{eff}} -
\I\theta_{\mathrm{A}} \sum_{i=1,2} \bar\psi_i \big(\boldsymbol{\Sigma}_i\,t_{i,\mathrm{A}} - \bar t_{i,\mathrm{A}}\,\boldsymbol{\Sigma}_i\big) \psi_i
\nonumber \\ && \phantom{\eL_{\mathrm{eff}}}
- \frac{1}{2}\I\theta_{\mathrm{A}}\Phi^\dag \big(\boldsymbol{\Pi}\,t_{\Phi,\mathrm{A}} - t_{\Phi,\mathrm{A}}\,\boldsymbol{\Pi}\big) \Phi
{}+\mathcal{O}(\theta_{\mathrm{A}}^2) \,.
\end{eqnarray}
\end{subequations}
Again, the change of the Lagrangian \eqref{chp:frm:eLeff}, i.e., the model's non-invariance under the axial symmetry $\group{U}(1)_{\mathrm{A}}$, driven by the self-energies $\boldsymbol{\Sigma}_i$, $\boldsymbol{\Pi}$, is again proportional to the quantities \eqref{chp:frm:noninv}.

We can now check explicitly how the self-energies of the specific form \eqref{chp:frm:anz:intermed} break the axial symmetry $\group{U}(1)_{\mathrm{A}}$. Short calculation reveals the symmetry-breaking quantities \eqref{chp:frm:noninv} to be
\begin{subequations}
\label{chp:frm:noninv:intermed}
\begin{eqnarray}
\bbl \boldsymbol{\Sigma}_i,t_{i,\mathrm{A}} \bbr &=& 2Q_{i,\mathrm{A}}\big(\Sigma\,P_R-\Sigma^* P_L\big) \,, \\
{[\boldsymbol{\Pi},t_{\Phi,\mathrm{A}}]}
&=&   -4Q_{1,\mathrm{A}} \left(\begin{array}{cc} 0 & -\Pi \\ \Pi^* & 0 \\ \end{array}\right)
\>=\> +4Q_{2,\mathrm{A}} \left(\begin{array}{cc} 0 & -\Pi \\ \Pi^* & 0 \\ \end{array}\right) \,.
\end{eqnarray}
\end{subequations}
We see that in equations \eqref{chp:frm:noninv:intermed} some form-factors from the self-energies \eqref{chp:frm:anz:intermed} are projected out. Namely, the form-factors $A_L$, $A_R$ and $\Pi_N$ are missing, which means that they do not break the symmetry. However, our aim here is to focus on the very mechanism of the SSB, or more precisely, to demonstrate that the SSB can happen. For this purpose the symmetry-preserving parts of the self-energies, while important for a more phenomenologically oriented analysis, are not essential. We will therefore systematically neglect them and rather consider only the symmetry-breaking part of self-energies \eqref{chp:frm:anz:intermed}, i.e., the parts $\Sigma$ and $\Pi$. The Ansatz for the self-energies will be therefore considered to be
\begin{subequations}
\label{chp:frm:Ansatz}
\begin{eqnarray}
\boldsymbol{\Sigma}_i &=& \Sigma^* P_L + \Sigma\,P_R \,,
\\
\boldsymbol{\Pi} &=& \left(\begin{array}{cc} 0 & \Pi \\ \Pi^* & 0 \end{array}\right) \,.
\end{eqnarray}
\end{subequations}
Notice that this Ansatz is in accordance with the Ansatz considered in the previous chapter. The corresponding full propagators follow immediately:
\begin{subequations}
\begin{eqnarray}
G_{\psi_i} &=& \frac{\slashed{p} + \boldsymbol{\Sigma}^\dag_{i}}{p^2-|\Sigma_{i}|^2} \,,
\label{chp:frm:Gpsii}
\\
G_\Phi &=&
\frac{1}{(p^2-M^2)^2-|\Pi|^2} \left(\begin{array}{cc} p^2-M^2 & \Pi \\ \Pi^* & p^2-M^2 \end{array}\right) \,.
\end{eqnarray}
\end{subequations}
This of course corresponds to the same equations for the spectrum as in the previous chapter:
\begin{subequations}
\label{chp:frm:eqsSpectrum}
\begin{eqnarray}
m_i^2 &=& |\Sigma_{i}(m_i^2)|^2 \,, \\
M_{1,2}^2 &=& M^2 \pm |\Pi(M_{1,2}^2)| \,,
\end{eqnarray}
\end{subequations}
cf.~Eqs.~\eqref{chp:prelim:psi:m2}, \eqref{chp:prelim:phi:m2}.

Let us finally remark that we could analyze in the same way also the vectorial symmetries $\group{U}(1)_{\mathrm{V}_{\!i}}$. It is evident from the form \eqref{chp:frm:U1Vipsigen} of the corresponding generators $t_{i,\mathrm{V}_{\!j}}$, which are just pure real numbers without any $\gamma_5$, that this time we would have
\begin{eqnarray}
\bbl \boldsymbol{\Sigma}_i,t_{i,\mathrm{V}_{\!j}} \bbr &=& t_{i,\mathrm{V}_{\!j}} (\boldsymbol{\Sigma}_i - \boldsymbol{\Sigma}_i) \ = \ 0
\end{eqnarray}
for any $\boldsymbol{\Sigma}_i$. Therefore the vectorial symmetries cannot be broken by the fermion self-energies $\langle \psi_i\bar\psi_i \rangle \sim \boldsymbol{\Sigma}_i$ (not to mention the self-energy of the scalar, which does not couple to the vectorial symmetries at all), which is after all expected. The only possibility to break the vectorial symmetries would be to consider the fermion self-energies also of the type $\langle \psi_i^\C\bar\psi_i \rangle$, $\langle \psi_i\bar\psi_i^\C \rangle$, where $\psi_i^\C$ denotes the charge conjugate\footnote{For more detail on charge conjugation of fermions see appendix~\ref{app:charge}.} of $\psi_i$. This would lead to the generation of the Majorana self-energies. We will actually explore this possibility later on in the context of neutrinos.

\section{Dynamics}
\label{sec:dynabel}

Our general strategy in demonstrating the spontaneous breakdown of the axial $\group{U}(1)_{\mathrm{A}}$ symmetry will be to search for the symmetry-breaking parts of the propagators, i.e., the parts $\Sigma_{i}$ and $\Pi$ of the self-energies $\boldsymbol{\Sigma}_{i}$ and $\boldsymbol{\Pi}$, respectively, as shown in the previous section. We have already mentioned the important observation that at any finite order of perturbative expansion the $\group{U}(1)_{\mathrm{A}}$ symmetry remains preserved and the self-energies $\Sigma_{i}$ and $\Pi$ vanish. The SSB is therefore necessarily a \emph{non-perturbative effect} and to treat it one has to employ some non-perturbative technique. The technique used here are the \emph{Schwinger--Dyson} (SD) equations, which represent a formal summation of all orders of perturbative expansion and as such they provide the desired non-perturbative treatment.

\begin{figure}[t]
\begin{center}
\includegraphics[width=0.35\textwidth]{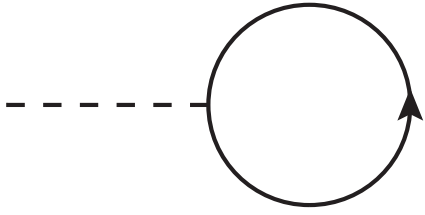}
\caption[A tadpole diagram, contributing to the scalar VEV.]{A tadpole diagram, contributing to scalar VEV.}
\label{chp:frm:fig:vev}
\end{center}
\end{figure}

At this point a remark concerning the scalar VEV and the associated issue of tadpole terms in the SD equations is in order. Recall that we neglected the scalar self-coupling \eqref{chp:frm:sclr:selfint}. This was done on the basis of argument that the scalar self-coupling is not essential for the proposed mechanism of SSB, as it is not driven by the scalar VEV, formed solely by scalar dynamics, but rather by the scalar and fermion propagators, formed by the Yukawa dynamics. However, the Yukawa dynamics can give rise to the scalar VEV too, as shown \emph{schematically} in Fig.~\ref{chp:frm:fig:vev}. Thus, to be be consistent with the motivation of neglecting the scalar self-coupling \eqref{chp:frm:sclr:selfint}, we will neglect the possible tadpole diagrams as well. More precisely, we will neglect them from the very beginning, i.e., we will derive the SD equations already \emph{under the assumption} of vanishing scalar VEV. Again, we stress that in principle the scalar VEV should be taken into account, as it is not protected by any symmetry.

\subsection{Cornwall--Jackiw--Tomboulis formalism}

There are various methods how to derive the SD equations. Here we are going use the method based on the Cornwall--Jackiw--Tomboulis (CJT) formalism \cite{Cornwall:1974vz}. We first define the appropriate effective potential and then we search for its stationary points with respect to the variations of the full propagators (or, equivalently, the self-energies, since the free propagators are fixed). This will lead to the integral SD equations, by solving which one can find the full propagators. We follow this program first for general self-energies $\boldsymbol{\Sigma}_i$, $\boldsymbol{\Pi}$ and only then we will take into account the specific Ansatz \eqref{chp:frm:Ansatz}.

The CJT effective potential is defined as
\begin{eqnarray}
V[G_\Phi,G_{\psi_1},G_{\psi_2}] &\equiv& V_\Phi[G_\Phi] + \sum_{i=1,2} V_{\psi_i}[G_{\psi_i}] + V_2[G_\Phi,G_{\psi_1},G_{\psi_2}] \,,
\end{eqnarray}
where
\begin{subequations}
\label{chp:frm:VpsiiVPhi}
\begin{eqnarray}
V_{\psi_i}[G_{\psi_i}] &\equiv& -\I \int\!\frac{\d^4 k}{(2\pi)^4}\,
\Tr \Big\{\ln(S_i^{-1}G_{\psi_i})-S_i^{-1}G_{\psi_i}+1 \Big\} \,,
\\
V_\Phi[G_\Phi] &\equiv& \frac{1}{2}\I \int\!\frac{\d^4 k}{(2\pi)^4}\,
\Tr \Big\{\ln(D^{-1}G_\Phi)-D^{-1}G_\Phi+1 \Big\} \,.
\end{eqnarray}
\end{subequations}
The factor of $1/2$ at $V_\Phi$ is due to the reality of the Nambu--Gorkov field $\Phi$, otherwise there would be the factor of $1$ in the case of complex $\Phi$. Similarly for the fermions, since they are complex fields, the factor at $V_{\psi_i}$ is $1$. For the real (i.e., Majorana) fermions there would be the factor of $1/2$ too (this will actually be the case later on when dealing with Majorana neutrinos). Finally, the minus sign at $V_{\psi_i}$ is due to the fermion nature of $\psi_i$.

\begin{figure}[t]
\begin{center}
\includegraphics[width=0.7\textwidth]{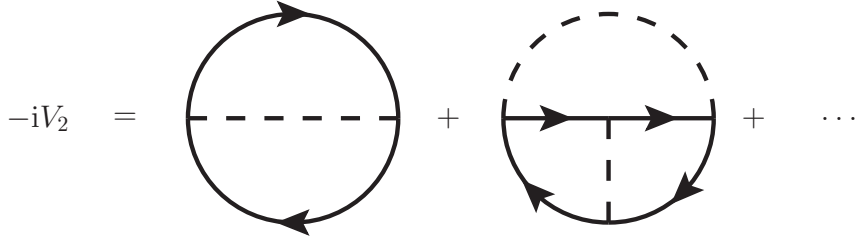}
\caption[The \qm{bubble} diagrams, contributing to the effective potential $V_2$.]{The two-particle irreducible (2PI) diagrams contributing to $V_2$.}
\label{chp:frm:fig:bubbles}
\end{center}
\end{figure}

The quantity $V_2$ is what actually defines the dynamics. It is the sum of all two-particle irreducible (2PI) vacuum diagrams (\qm{bubbles}), see Fig.~\ref{chp:frm:fig:bubbles}. More formally, it is given by the vacuum-to-vacuum amplitude
\begin{eqnarray}
\bra{0} S \ket{0} &=& 1-(2\pi)^4\,\delta^4(0)\,\I V_2 \,,
\end{eqnarray}
where $S$ is the $S$-matrix given by Yukawa interaction Lagrangian \eqref{chp:frm:eL:Yukawa}.

The effective potential $V[G]$ is a functional of the full propagators of all fields in the theory, denoted collectively as $G\equiv(G_\Phi,G_{\psi_1},G_{\psi_2})$. According to Ref.~\cite{Cornwall:1974vz}, the SD equations correspond to its stationary point, i.e., to the point in the space of the full propagators where the partial functional derivatives of the effective potential vanish. The SD equations are thus na{\"{\i}}vely given by
\begin{eqnarray}
\frac{\delta V[G]}{\delta G} &=& 0 \,.
\end{eqnarray}
However, one has to be careful. An attention has to be taken concerning the direction of allowed variation. It may happen that the fields in question have some symmetry, which induces also the symmetry of the corresponding propagator. Thus, while looking for the stationary point of the effective potential $V$, one has to make sure that the variations of the propagator respect this symmetry.

Indeed, in our case it is the scalar Nambu--Gorkov field $\Phi$, which possesses the non-trivial symmetry \eqref{chp:frm:NGcond}, inducing the symmetry \eqref{chp:frm:NGcondprop} of the propagator. Therefore we will search for the stationary point of $V$ not on the whole space of all propagators, but rather only on its subspace, defined by the constraint \eqref{chp:frm:NGcondprop}.

Technically, extremizing of a function $V(G)$ (generalization to the functional $V[G]$ is straightforward) over a multivariable $G$, constrained, e.g., by the condition $G = f(G)$, is achieved by means of the method of Lagrange multipliers: One first constructs a new function (the Lagrange function) $V(G,\lambda) \equiv V(G) - \lambda (G - f(G))$. Now, instead of minimizing $V(G)$ with respect to the constrained set of variables $G$, one minimizes $V(G,\lambda)$ with respect to the whole (unconstrained) set of both the variables $G$ and $\lambda$. Having obtained the result, one can eliminate the $\lambda$ from the result in favor of $G$ and get this way the final result satisfying the prescribed condition $G = f(G)$.

Let us now apply the method of Lagrange multipliers to the problem of extremizing the functional $V[G]$ on the subset constrained by \eqref{chp:frm:NGcondprop}. Thus, instead of extremizing just $V[G]$ with respect to variations of the propagators, we introduce the new functional $V_\lambda[G,\lambda]$, defined as
\begin{eqnarray}
V_\lambda[G,\lambda] &\equiv& V[G] + V_{\Phi,\lambda}[G_\Phi,\lambda] \,,
\end{eqnarray}
where
\begin{eqnarray}
V_{\Phi,\lambda}[G_\Phi,\lambda] &\equiv&
\int\!\frac{\d^4 k}{(2\pi)^4}\,
\Tr \Big\{ \lambda \Big( G_\Phi - \sigma_1 \, G_\Phi^\T \, \sigma_1 \Big) \Big\} \,,
\end{eqnarray}
and extremize it with respect to both the propagators $G$ and the Lagrange multiplier $\lambda$:
\begin{subequations}
\label{chp:frm:extremizing}
\begin{eqnarray}
\frac{\delta \, V_\lambda}{\delta \, G} &=& 0 \,,
\label{chp:frm:extrmz:G} \\
\frac{\delta \, V_\lambda}{\delta \, \lambda} &=& 0 \,.
\label{chp:frm:extrmz:lambda}
\end{eqnarray}
\end{subequations}
Recall that the Lagrange multiplier $\lambda$ is not a number, but rather a momentum-dependent $2 \times 2$ matrix, operating in the two-dimension Nambu--Gorkov space (in other words, it has the same structure as $G_\Phi$).

Before continuing, let us make a technical aside. It turns out to be more convenient to calculate not directly the functional derivatives \eqref{chp:frm:extremizing}, but rather their matrix transpose. Put another way, since $V_\lambda$ (as well as all $V$, $V_\Phi$, etc.) is a pure number (not a matrix), we make all differentiations with respect to $G^\T$ and $\lambda^\T$, rather than with respect to $G$ and $\lambda$. This is because of the matrix identities
\begin{subequations}
\begin{eqnarray}
\frac{\partial}{\partial A^\T}\det A &=& A^{-1}\det A \,,
\label{chp:matrix_identity1}
\\
\frac{\partial}{\partial A^\T}\Tr(AB) &=& B \,,
\end{eqnarray}
\end{subequations}
holding for any matrices\footnote{Provided $A^{-1}$ does exist, otherwise the right-hand side of \eqref{chp:matrix_identity1} is $\adj A$, which exists always.} $A$, $B$. (For completeness, recall also another useful and well known identity, used in our calculations: $\Tr \ln A = \ln \det A$.)

We start with the differentiation with respect to $\lambda$. The direct calculation reveals
\begin{eqnarray}
\frac{\delta \, V_\lambda}{\delta \, \lambda^\T} &=&
\frac{\delta \, V_{\Phi,\lambda}}{\delta \, \lambda^\T} \>=\>
\frac{1}{(2\pi)^4}\Big(G_\Phi-\sigma_1\,G_\Phi^\T\,\sigma_1\Big) \,.
\end{eqnarray}
Demanding that it vanishes we just obtain the constraint \eqref{chp:frm:NGcondprop} for $G_\Phi$ and through \eqref{chp:frm:PiDPhiGphi} also the constrain \eqref{chp:frm:NGcondpropPi} for $\boldsymbol{\Pi}$.

Let us proceed with the differentiation with respect to the propagator $G_\Phi$. The particular derivatives are
\begin{eqnarray}
\frac{\delta \, V_\Phi}{\delta \, G_\Phi^\T} &=&
-\I\frac{1}{2}\frac{1}{(2\pi)^4}\Big(D^{-1}-G_\Phi^{-1}\Big)\>=\>
-\I\frac{1}{2}\frac{1}{(2\pi)^4} \boldsymbol{\Pi}
\end{eqnarray}
and
\begin{eqnarray}
\frac{\delta \, V_{\Phi,\lambda}}{\delta \, G_\Phi^\T} &=&
\frac{1}{(2\pi)^4} \Big(\lambda-\sigma_1\,\lambda^\T\,\sigma_1\Big) \,.
\end{eqnarray}
Thus
\begin{eqnarray}
\frac{\delta \, V_{\lambda}}{\delta \, G_\Phi^\T} &=&
-\I\frac{1}{2}\frac{1}{(2\pi)^4} \boldsymbol{\Pi}
+\frac{1}{(2\pi)^4} \Big(\lambda-\sigma_1\,\lambda^\T\,\sigma_1\Big)
+\frac{\delta \, V_{2}}{\delta \, G_\Phi^\T} \,.
\end{eqnarray}
This must vanish, so that we can express $\boldsymbol{\Pi}$ from it as
\begin{eqnarray}
\label{chp:frm:Piexpress1}
-\I\,\boldsymbol{\Pi} &=& -2 \Big(\lambda-\sigma_1\,\lambda^\T\,\sigma_1\Big)
-2(2\pi)^4 \frac{\delta \, V_{2}}{\delta \, G_\Phi^\T} \,.
\end{eqnarray}
Now we must somehow eliminate the Lagrange multiplier. For this we make use of the equation \eqref{chp:frm:extrmz:lambda}, or more precisely, of its consequence \eqref{chp:frm:NGcondpropPi}:
\begin{subequations}
\label{chp:frm:Piexpress2}
\begin{eqnarray}
-\I\,\boldsymbol{\Pi} &=& -\I\,\sigma_1\,\boldsymbol{\Pi}^\T\,\sigma_1
\\ &=& +2 \Big(\lambda-\sigma_1\,\lambda^\T\,\sigma_1\Big)
-2(2\pi)^4 \sigma_1 \bigg(\frac{\delta \, V_{2}}{\delta \, G_\Phi^\T}\bigg)^\T \sigma_1 \,.
\end{eqnarray}
\end{subequations}
We can now sum the two equations \eqref{chp:frm:Piexpress1} and \eqref{chp:frm:Piexpress2} to eliminate the Lagrange multiplier $\lambda$ and to obtain the final scalar SD equation:
\begin{eqnarray}
\label{chp:frm:SDPiintrmd}
-\I\,\boldsymbol{\Pi} &=&
-(2\pi)^4 \left[\frac{\delta \, V_{2}}{\delta \, G_\Phi^\T} + \sigma_1 \bigg(\frac{\delta \, V_{2}}{\delta \, G_\Phi^\T}\bigg)^\T \sigma_1 \right] \,.
\end{eqnarray}
From this equation it is manifestly evident that $\boldsymbol{\Pi}$ will indeed satisfy the condition \eqref{chp:frm:NGcondpropPi}.

The fermions are much easier, since $\psi_i$ are complex (i.e., Dirac) fermions and therefore there is no special constraint on the form of their propagators. (For real, i.e., Majorana fermions there would be constraint $\psi = \psi^\C$; this situation will in fact arise in chapter~\ref{chp:ewdyn}, where we will in the context of electroweak interactions discuss the neutrinos.) The fermion SD equation is therefore given simply by
\begin{eqnarray}
\frac{\delta \, V}{\delta \, G_{\psi_i}^\T} &=& \frac{\delta \, V_\lambda}{\delta \, G_{\psi_i}^\T} \>=\> 0 \,.
\end{eqnarray}
Some algebra reveals
\begin{eqnarray}
\frac{\delta \, V_{\psi_i}}{\delta \, G_{\psi_i}^\T} &=&
\I\frac{1}{(2\pi)^4}\Big(S_i^{-1}-G_{\psi_i}^{-1}\Big)\>=\>
\I\frac{1}{(2\pi)^4} \boldsymbol{\Sigma}_i
\end{eqnarray}
and the SD equation consequently reads
\begin{eqnarray}
\label{chp:frm:SDSigmaintrmd}
-\I \, \boldsymbol{\Sigma}_i &=&    (2\pi)^4 \frac{\delta \, V_{2}}{\delta \, G_{\psi_i}^\T} \,.
\end{eqnarray}

\subsection{Hartree--Fock approximation}

\begin{figure}[t]
\begin{center}
\includegraphics[width=0.55\textwidth]{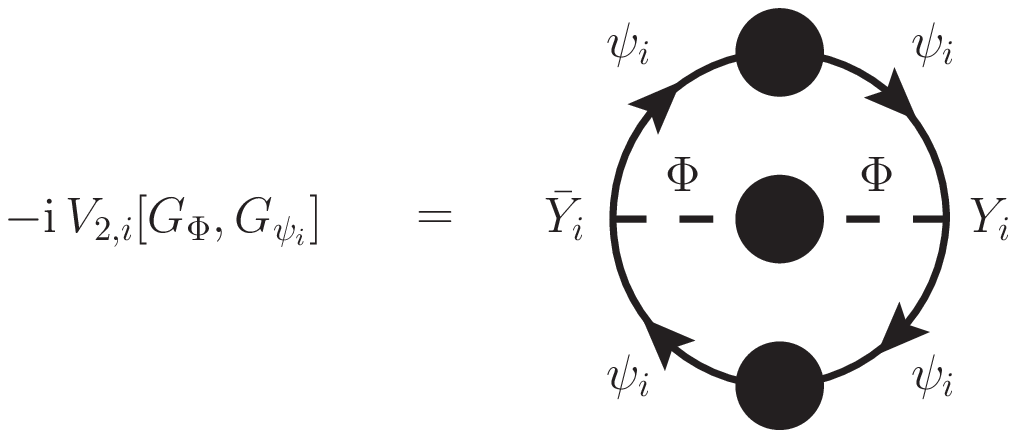}
\caption[Hartree--Fock approximation of $V_2$.]{Diagrammatical representation of $V_{2,i}$, Eq.~\eqref{chp:frm:V2i}.}
\label{fig:form:V2}
\end{center}
\end{figure}

To proceed further, we have to specify $V_2$. In principle, it is an infinite sum of all 2PI diagrams, as can be seen in Fig.~\ref{chp:frm:fig:bubbles}. We truncate this series at the simplest possible diagrams, i.e., we consider only the very first one in Fig.~\ref{chp:frm:fig:bubbles}. This particular choice of $V_2$ is called the Hartree--Fock approximation. Since the Yukawa interactions do not mix the fermion species (i.e., there are no interactions of the type $\bar \psi_1 \psi_2 \phi$) and in the Hartree--Fock approximation there is only one fermion loop in each diagram, $V_2$ can be written as a sum of independent contributions from the two fermion species,
\begin{eqnarray}
V_2[G_\Phi,G_{\psi_1},G_{\psi_2}] &=& \sum_{i=1,2} V_{2,i}[G_\Phi,G_{\psi_i}] \,,
\end{eqnarray}
where the particular terms on the right-hand side are given as
\begin{eqnarray}
\label{chp:frm:V2i}
-\I \, V_{2,i}[G_\Phi,G_{\psi_i}] &=& - \frac{1}{2} \I^5
\int\!\frac{\d^4 k}{(2\pi)^4}\,\frac{\d^4 p}{(2\pi)^4} \,
\Tr \! \Big\{ Y_{i} \, G_{\psi_i}(k) \, \bar Y_{i} \, G_{\psi_i}(p) \, G_\Phi(k-p)  \Big\} \,.
\end{eqnarray}
This expression is easily understood according to Fig.~\ref{fig:form:V2}. The minus sign on the right side is for the fermion loop, while the factor of $1/2$ is a combinatorial factor. The trace in \eqref{chp:frm:V2i} is over both the fermion and the Nambu--Gorkov scalar space. However, since the couplings constants $Y_i$, $\bar Y_i$ are rectangular matrices, one has to be careful when applying the rule about the cyclicity of the trace. E.g., one can move $Y_i$ from the beginning of the trace to its end; then, however, the trace is only over the fermion space; the trace over the Nambu--Gorkov space would be in such a case already effectively implemented by the matrix multiplication $\bar Y_{i} \, G_\Phi \, Y_i$.

The functional derivatives of $V_2$ relevant for the SD equations \eqref{chp:frm:SDPiintrmd}, \eqref{chp:frm:SDSigmaintrmd} are
\begin{subequations}
\label{chp:frm:SDtraces}
\begin{eqnarray}
\frac{\delta \, V_{2}}{\delta \, G_{\Phi}^\T(p)}   &=& \sum_{i=1,2} \frac{1}{2} \frac{1}{(2\pi)^4}
\int \! \frac{\d^4 k}{(2\pi)^4} \,
\Tr_\psi \! \Big\{ Y_{i} \, G_{\psi_i}(k) \, \bar Y_{i} \, G_{\psi_i}(k-p)  \Big\} \,,
\\
\frac{\delta \, V_{2}}{\delta \, G_{\psi_i}^\T(p)} &=& \frac{1}{(2\pi)^4}
\int \! \frac{\d^4 k}{(2\pi)^4} \,
\Tr_\Phi \! \Big\{ Y_{i} \, G_{\psi_i}(k) \, \bar Y_{i} \, G_{\Phi}(k-p)  \Big\} \,.
\end{eqnarray}
\end{subequations}
Notice that the particular traces in \eqref{chp:frm:SDtraces} are only over the indicated space. Therefore the rule about cyclicity of the trace (in terms of the quantities $Y_i$, $\bar Y_i$, $G_{\psi_i}$, $G_\Phi$) is no longer applicable.

One can now verify that the following identity holds:
\begin{eqnarray}
\label{chp:frm:identT}
\frac{\delta \, V_{2}}{\delta \, G_\Phi^\T} &=& \sigma_1 \bigg(\frac{\delta \, V_{2}}{\delta \,  G_\Phi^\T}\bigg)^\T \sigma_1 \,.
\end{eqnarray}
It holds due to the property of the Yukawa coupling constants $Y_i$, \eqref{chp:frm:Y1Y2def},
\begin{eqnarray}
Y_i &=& \sigma_1 \, \bar Y_i^{\T_\Phi} \,,
\end{eqnarray}
with the transpose $^{\T_\Phi}$ being understood only in the two-dimensional Nambu--Gorkov space. We stress, however, that the identity \eqref{chp:frm:identT} holds only within our special form \eqref{chp:frm:V2i} of $V_2$; in another than the Hartee--Fock approximation \eqref{chp:frm:V2i} may no longer be true.

\begin{figure}[t]
\begin{center}
\includegraphics[width=0.65\textwidth]{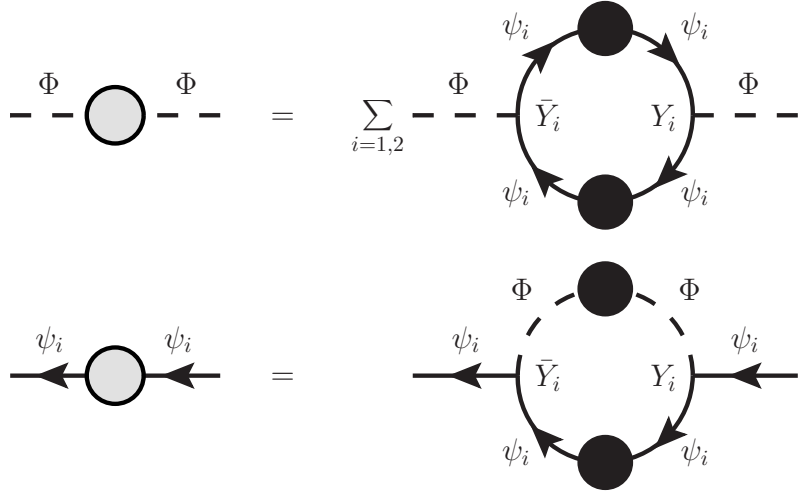}
\caption[SD equations in the Hartree--Fock approximation.]{Diagrammatical representation of the SD equations \eqref{chp:frm:SDnoans} in the Hartree--Fock approximation.}
\label{chp:frm:fig:SDHF}
\end{center}
\end{figure}

Now we can plug the functional derivatives \eqref{chp:frm:SDtraces} of $V_2$ into the general forms \eqref{chp:frm:SDPiintrmd}, \eqref{chp:frm:SDSigmaintrmd} of the SD equations and with the help of the identity \eqref{chp:frm:identT} we obtain
\begin{subequations}
\label{chp:frm:SDnoans}
\begin{eqnarray}
-\I \, \boldsymbol{\Pi}(p)      &=& - \sum_{i=1,2}
\int \! \frac{\d^4 k}{(2\pi)^4} \,
\Tr_\psi \! \Big\{ Y_{i} \, G_{\psi_i}(k) \, \bar Y_{i} \, G_{\psi_i}(k-p)  \Big\} \,,
\\
-\I \, \boldsymbol{\Sigma}_i(p) &=&
\int \! \frac{\d^4 k}{(2\pi)^4} \,
\Tr_\Phi \! \Big\{ Y_{i} \, G_{\psi_i}(k) \, \bar Y_{i} \, G_{\Phi}(k-p)  \Big\} \,,
\end{eqnarray}
\end{subequations}
see Fig.~\ref{chp:frm:fig:SDHF}.

\subsection{Employing the Ansatz}

The SD equations \eqref{chp:frm:SDnoans} hold generally, for any form of the propagators, unconstrained by any Ansatz. Now it is time to put our Ansatz \eqref{chp:frm:Ansatz} back into game.

If we just mechanically plug the Ansatz \eqref{chp:frm:Ansatz} into the SD equations \eqref{chp:frm:SDnoans}, we obtain\footnote{We indicate here the momentum arguments for the sake of brevity by subscripts, i.e., $\Pi_p \equiv \Pi(p^2)$, $\Sigma_{ip} \equiv \Sigma_i(p^2)$. This notation will be used repeatedly throughout the text.}
\begin{subequations}
\begin{eqnarray}
-\I\left(\begin{array}{cc} 0 & \Pi_p \\ \Pi_p^* & 0 \end{array}\right)
&=& -2\int\!\frac{\d^4 k}{(2\pi)^4}\, \frac{1}{k^2-|\Sigma_{1k}|^2}\frac{1}{\ell^2-|\Sigma_{1\ell}|^2}
\left(\begin{array}{cc}
|y_1|^2 (k \cdot \ell)                    &  y_1^{*2}\,\Sigma_{1k}\,\Sigma_{1\ell} \\
y_1^{2}\,\Sigma_{1k}^*\,\Sigma_{1\ell}^*  &  |y_1|^2 (k \cdot \ell)
\end{array}\right)
\nonumber \\ &&
-2\int\!\frac{\d^4 k}{(2\pi)^4}\, \frac{1}{k^2-|\Sigma_{2k}|^2}\frac{1}{\ell^2-|\Sigma_{2\ell}|^2}
\left(\begin{array}{cc}
|y_2|^2 (k \cdot \ell)                &  y_2^{*2}\,\Sigma_{2k}^*\,\Sigma_{2\ell}^* \\
y_2^{2}\,\Sigma_{2k}\,\Sigma_{2\ell}  &  |y_2|^2 (k \cdot \ell)
\end{array}\right) \,,
\nonumber \\ && \\
-\I \big(\Sigma_{1p}^*\,P_L + \Sigma_{1p}\,P_R\big)
&=& \int\!\frac{\d^4 k}{(2\pi)^4}\, \frac{1}{k^2-|\Sigma_{1k}|^2}\frac{1}{(\ell^2-M^2)^2-|\Pi_\ell|^2}
\nonumber \\ &&
\phantom{\int\!\frac{\d^4 k}{(2\pi)^4}}
{}\times
\bigg[
|y_1|^2\slashed{k}(\ell^2-M^2) + y_1^{*2}\,\Sigma_{1k}\,\Pi_\ell^*\,P_L + y_1^{2}\,\Sigma_{1k}^*\,\Pi_\ell\,P_R
\bigg] \,,
\nonumber \\ &&
\\
-\I \big(\Sigma_{2p}^*\,P_L + \Sigma_{2p}\,P_R\big)
&=& \int\!\frac{\d^4 k}{(2\pi)^4}\, \frac{1}{k^2-|\Sigma_{2k}|^2}\frac{1}{(\ell^2-M^2)^2-|\Pi_\ell|^2}
\nonumber \\ &&
\phantom{\int\!\frac{\d^4 k}{(2\pi)^4}}
{}\times
\bigg[
|y_2|^2\slashed{k}(\ell^2-M^2) + y_2^{2}\,\Sigma_{2k}\,\Pi_\ell\,P_L + y_2^{*2}\,\Sigma_{2k}^*\,\Pi_\ell^*\,P_R
\bigg] \,,
\nonumber \\ &&
\end{eqnarray}
\end{subequations}
where we denoted
\begin{eqnarray}
\ell &\equiv& k-p \,.
\end{eqnarray}
Each of these three matrix equations comprise in fact two independent (not related by the complex conjugation) scalar equations. Of these altogether six scalar equations let us first discuss the following three ones:
\begin{subequations}
\label{chp:frm:SDpatholog}
\begin{eqnarray}
0
&=& -2|y_1|^2 \int\!\frac{\d^4 k}{(2\pi)^4}\,
\frac{k_\alpha}{k^2-|\Sigma_{1k}|^2}
\frac{\ell^\alpha}{\ell^2-|\Sigma_{1\ell}|^2}
-2|y_2|^2 \int\!\frac{\d^4 k}{(2\pi)^4}\,
\frac{k_\alpha}{k^2-|\Sigma_{2k}|^2}
\frac{\ell^\alpha}{\ell^2-|\Sigma_{2\ell}|^2} \,,\qquad
\nonumber \\ &&
\\
0
&=& |y_1|^2 \int\!\frac{\d^4 k}{(2\pi)^4}\,
\frac{\slashed{k}}{k^2-|\Sigma_{1k}|^2}
\frac{\ell^2-M^2}{(\ell^2-M^2)^2-|\Pi_\ell|^2} \,,
\\
0
&=& |y_2|^2 \int\!\frac{\d^4 k}{(2\pi)^4}\,
\frac{\slashed{k}}{k^2-|\Sigma_{2k}|^2}
\frac{\ell^2-M^2}{(\ell^2-M^2)^2-|\Pi_\ell|^2} \,.
\end{eqnarray}
\end{subequations}
(The last two equations are strictly speaking not scalar, because as they contain $\slashed{k}$, they are proportional to $\slashed{p}$. Nevertheless, the true scalar equations can be easily projected out; effectively it suffices to make the replacement $\slashed{k} \rightarrow (k \cdot p)/p^2$.) Of course, these equations have to be discarded, as they do not comply with the Ansatz \eqref{chp:frm:Ansatz}. This is after all manifested in the pathological fact that their left-hand sides are vanishing. Nevertheless, it is useful to take a quick look at their right-hand sides. The integrals in all three equations \eqref{chp:frm:SDpatholog} are UV-divergent for any decreasing or constant self-energies $\Sigma_i$, $\Pi$. Since we assume that the symmetry-breaking self-energies $\Sigma_i$, $\Pi$ must be UV-finite, i.e., decreasing (see the discussion at the end of this section), we conclude that even if we included the symmetry-preserving self-energies into our Ansatz (so that the left-hand sides of \eqref{chp:frm:SDpatholog} would not be vanishing), they would come out necessarily UV-divergent. This is because symmetry-preserving self-energies (as well as any other symmetry-preserving Green's functions) contain, apart from possible non-perturbative parts, also perturbative parts, i.e., the parts calculable within the usual perturbation theory using the symmetry-preserving interactions from the (symmetric) Lagrangian.

\begin{figure}[t]
\begin{center}
\includegraphics[width=1\textwidth]{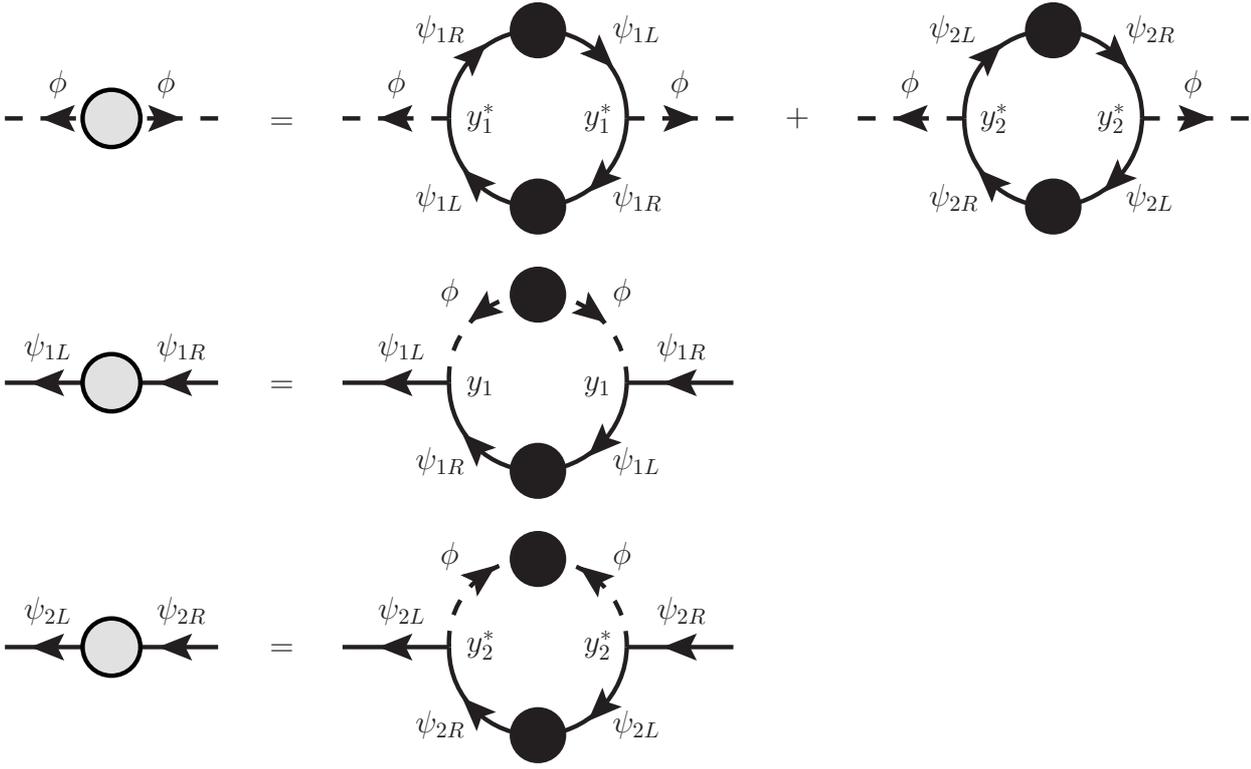}
\caption[SD equations for the symmetry-breaking parts of the propagators.]{Diagrammatical representation of the SD equations \eqref{chp:frm:SD} for the symmetry-breaking parts of the propagators.}
\label{fig:SDabelform}
\end{center}
\end{figure}

We are thus left with the remaining three equations:
\begin{subequations}
\label{chp:frm:SD}
\begin{eqnarray}
-\I \, \Pi_p
&=&
- 2 y_1^{*2} \int \! \frac{\d^4 k}{(2\pi)^4} \,
\frac{\Sigma_{1k}}{k^2-|\Sigma_{1k}|^2}\frac{\Sigma_{1\ell}}{\ell^2-|\Sigma_{1\ell}|^2}
- 2 y_2^{*2} \int \! \frac{\d^4 k}{(2\pi)^4} \,
\frac{\Sigma_{2k}^*}{k^2-|\Sigma_{2k}|^2}\frac{\Sigma_{2\ell}^*}{\ell^2-|\Sigma_{2\ell}|^2} \,,
\nonumber \\ &&
\\
-\I \, \Sigma_{1p}
&=&
y_1^{2} \int \! \frac{\d^4 k}{(2\pi)^4} \,
\frac{\Sigma_{1k}^*}{k^2-|\Sigma_{1k}|^2}
\frac{\Pi_\ell}{(\ell^2-M^2)^2-|\Pi_\ell|^2} \,,
\\
-\I \, \Sigma_{2p}
&=&
y_2^{*2} \int \! \frac{\d^4 \ell}{(2\pi)^4} \,
\frac{\Sigma_{2k}^*}{k^2-|\Sigma_{2k}|^2}
\frac{\Pi_\ell^*}{(\ell^2-M^2)^2-|\Pi_\ell|^2} \,,
\end{eqnarray}
\end{subequations}
depicted also in Fig.~\ref{fig:SDabelform}. Mathematically, this is a set of three non-linear integral equations for three unknown functions $\Sigma_1(p^2)$, $\Sigma_2(p^2)$ and $\Pi(p^2)$. These equations are homogenous, i.e., they have the trivial solution $\Sigma_1(p^2) = \Sigma_2(p^2) = \Pi(p^2) \equiv 0$, corresponding to no SSB. Our task is to find a non-trivial solution, which would indicate the occurence of the SSB.

Very little can be said about the possible non-trivial solutions. Even their number is in principle unknown. However, note that the equations \eqref{chp:frm:SD} are basically the same as those \eqref{chp:prelim:SD}, derived diagrammatically in the previous chapter (up to the different number of fermion equations). Recall that we have concluded, on the basis of the convergence properties of the integrals in the equations, that the resulting self-energies must be UV-finite, i.e., decreasing. Indeed, the form of the equations is clearly consistent with this assumption.

In fact, this assumption can be supported by another, more physical argument. If the self-energies were UV-divergent, appropriate counterterms would have to be added to the Lagrangian. However, as the self-energies are symmetry-breaking, so would have to be also the counterterms themselves. But the Lagrangian must be symmetric, which prohibits such counterterms. Consequently, as there is no possibility to add the symmetry-breaking counterterms to the Lagrangian, the symmetry-breaking self-energies (as well as any other symmetry-breaking Green's functions) must be necessarily UV-finite.

\section{Numerics}

As there is virtually no hope to solve the SD equations \eqref{chp:frm:SD} analytically, one has to resort to some kind of numerical approach. In this section the results of the numerical solution of SD equations are presented, together with a brief description of the numerical procedure itself.

\subsection{Approximations}
\label{chp:frm:ssec:Appr}

The SD equations in the form \eqref{chp:frm:SD}, yet being a result of numerous approximations, are still quite difficult to be solved even numerically, so further approximations have to be done. The most serious problem is the existence of the poles in the propagators. While vital for the very mass generation, these poles are extremely difficult to integrate numerically. Thus, we get rid of them by switching form the Minkowski to the Euclidean metric via the Wick rotation. Effectively, in the propagators the Wick rotation consists of changing $p^2 \rightarrow -p_{\mathrm{E}}^2$, with $p_{\mathrm{E}}^2$ being always non-negative. By this we remove the poles in the fermion propagators. In the scalar propagator the situation is more complicated, the pole still remains, only its position is changed. After wick rotation it is given by equation
\begin{eqnarray}
\label{chp:frm:polesclr}
p_{\mathrm{E}}^2+M^2 - |\Pi(-p_{\mathrm{E}}^2)|&=& 0 \,,
\end{eqnarray}
which, depending on $\Pi$, can still have a solution for some positive $p_{\mathrm{E}}^2$. This problem is \qm{solved} by considering in the numerical analysis only those $\Pi$ for which the pole equation \eqref{chp:frm:polesclr} has no (real and positive) solution.

Moreover, in order to reduce the number of independent equations to be solved, we deliberately consider both the fermion and scalar self-energies to be real. In fact, this approximation is consistent with the removing of the poles discussed above, as now there is no $\I\epsilon$-prescription to bring any imaginary parts (provided, of course, that the coupling constants $y_1$, $y_2$ are set real too).

As a net result, we solve the following set of equations for the unknown real functions $\Sigma_1(p^2)$, $\Sigma_2(p^2)$ and $\Pi(p^2)$:
\begin{subequations}
\label{chp:frm:SDwr}
\begin{eqnarray}
\Pi_p &=& \sum_{i=1,2}2 y_i^2\int\!\frac{\d^4k}{(2\pi)^4}\,
\frac{\Sigma_{ik}}{k^2+\Sigma_{ik}^2} \frac{\Sigma_{i\ell}}{\ell^2+\Sigma_{i\ell}^2} \,,
\label{chp:frm:SDwrSgm1}
\\
\Sigma_{1p} &=& y_1^2\int\!\frac{\d^4k}{(2\pi)^4}\,
\frac{\Sigma_{1k}}{k^2+\Sigma_{1k}^2} \frac{\Pi_{\ell}}{(\ell^2+M^2)^2-\Pi_{\ell}^2} \,,
\label{chp:frm:SDwrSgm2}
\\
\Sigma_{2p} &=& y_2^2\int\!\frac{\d^4k}{(2\pi)^4}\,
\frac{\Sigma_{2k}}{k^2+\Sigma_{2k}^2} \frac{\Pi_{\ell}}{(\ell^2+M^2)^2-\Pi_{\ell}^2} \,,
\end{eqnarray}
\end{subequations}
where $p^2=p_0^2+p_1^2+p_2^2+p_3^2\geq 0$ and $y_1,y_2 \in \mathbb{R}$. Notice that from now on we omit the subscript $_{\mathrm{E}}$ and redefine the self-energies like, e.g., $\Sigma_{i}(-p^2) \rightarrow \Sigma_{i}(p^2) \equiv \Sigma_{ip}$, and similarly for $\Pi$.

The system of equations \eqref{chp:frm:SDwr} has apparently three free parameters $y_1$, $y_2$ and $M$. In fact, the parameter $M$, as being the only parameter with dimension of mass in the theory, serves just as a scale parameter for the self-energies and momenta. Therefore it can be set to any value without the loss of generality. Hence we are left with only two relevant parameters, the Yukawa coupling constants $y_1$ and $y_2$, according to which the solutions will be classified.

Now the task is to solve the equations \eqref{chp:frm:SDwr} possibly for each pair $y_1$, $y_2$. However, as the equations \eqref{chp:frm:SDwr} depend on the \emph{squares} of $y_1$, $y_2$, one does not need to scan the full $(y_1,y_2)$ space, it suffices to check only one quadrant. Say, the one where both $y_1$, $y_2$ are positive. Moreover, even this quadrant need not to be probed full, due to the symmetricity of the equations \eqref{chp:frm:SDwr} under exchange of the fermion species, $1 \leftrightarrow 2$. If one knows a solution for at some point $(y_1,y_2)$, one automatically knows also the solution at the point $(y_2,y_1)$.

\subsection{Numerical procedure}
\label{chp:frm:ssec:Numpr}

The system of equations \eqref{chp:frm:SDwr} can be formally written as
\begin{subequations}
\label{chp:frm:SDnumform}
\begin{eqnarray}
\Pi        &=& G[\Sigma_{1},\Sigma_{2}] \,, \\
\Sigma_{1} &=& F_1[\Sigma_{1},\Pi] \,, \\
\Sigma_{2} &=& F_2[\Sigma_{2},\Pi] \,,
\end{eqnarray}
\end{subequations}
where the functionals $G$, $F_i$ are given by the integrals on the right-hand sides of \eqref{chp:frm:SDwr}. These integrals are four-dimensional over the full $\mathbb{R}^4$. Upon rewriting the integrals to the hyperspherical coordinates, two of the three angular integrals can be performed analytically and one is finally left with only two-dimensional integrals. There remain two integrals that cannot be in general solved analytically: One angular over the interval $[0,\pi]$ and one radial (momentum squared) over $[0,\infty)$.

Since the self-energies are presumably approaching zero at large momenta, so do the corresponding integrands in the integrals \eqref{chp:frm:SDwr}. Therefore introducing a sufficiently high momentum cut-off in the integral should not alter the solutions substantially. Such a cut-off corresponds to replacing the infinite interval of the radial integral by a finite one.

The next step is \emph{discretizing} the self-energies. That is to say, instead of computing the self-energies as the functions of \emph{all} momenta (eventually up to the cut-off introduced in the previous paragraph), we compute the self-energies only in the finite, but sufficiently large number of fixed discrete momenta, appropriately (i.e., not necessarily equidistantly) distributed between the zero and the cut-off. Choosing such a discretization, the integrals can be naturally substituted by sums by means of some quadrature rule for numerical integrating. To be concrete, we have used the Simpson's rule for radial integral and the Gauss--Chebyshev quadrature formula for the angular integral.

To summarize, we have traded the system of non-linear \emph{integral} equations for unknown \emph{functions} (i.e., the self-energies) by a system of non-linear \emph{algebraic} equations for finite set of unknown \emph{numbers} (i.e., the discretized self-energies). It has actually the same structure \eqref{chp:frm:SDnumform} as the original set of integral equations, only the symbols $\Pi$, $\Sigma_i$ have to be understood as vectors of finite dimensions, rather than functions, and the symbols $G$, $F_i$ are some complicated multivariable vector-valued functions rather than functionals.

Such an algebraic system is already directly amenable to the numerical treatment. The standard (and in fact the only) method for solving it is the method of iterations. It consists roughly of the following: One chooses some initial Ansatz, or \qm{zeroth} iteration, for the fermion self-energies:
\begin{equation}
\label{chp:frm:numAnz}
\Sigma_{1}^{(0)} \,,\quad \Sigma_{2}^{(0)} \,.
\end{equation}
Consequently, one can also calculate the \qm{zeroth} iteration of the scalar self-energy:
\begin{eqnarray}
\Pi^{(0)} &=& G[\Sigma_{1}^{(0)},\Sigma_{2}^{(0)}] \,.
\end{eqnarray}
By this the iteration process is established: the $(n+1)$-th iteration is calculated from the $n$-th iteration as
\begin{subequations}
\label{chp:frm:SDiterative}
\begin{eqnarray}
\Sigma_{1}^{(n+1)} &=& F_1[\Sigma_{1}^{(n)},\Pi^{(n)}] \,, \\
\Sigma_{2}^{(n+1)} &=& F_2[\Sigma_{2}^{(n)},\Pi^{(n)}] \,, \\
\Pi^{(n+1)}        &=& G[\Sigma_{1}^{(n)},\Sigma_{2}^{(n)}] \,,
\end{eqnarray}
\end{subequations}
with $n\geq0$. Clearly, if this procedure converges, then its limit is the solution to the discretized equations \eqref{chp:frm:SDnumform}. The convergence of the iteration process \eqref{chp:frm:SDiterative} can be controlled, e.g., by the quantities
\begin{eqnarray}
I_X^{(n)} &=& \frac{\int X^{(n)}}{\int X^{(n-1)}}  \,,
\end{eqnarray}
where $X=\Pi,\Sigma_1,\Sigma_2$. The advantage of the quantities $I_X^{(n)}$ is that they constitute only three scalar quantities, not vectors like the self-energies, so that their convergence can be controlled much easier that the convergence of the self-energies. Obviously, if a self-energy converges to some non-trivial fixed point, then the corresponding $I_X^{(n)}$ converges to $1$, and if the self-energy converges to zero, $I_X^{(n)}$ converges to some $I$ in the interval $0 \leq I < 1$. The opposite implications may not hold in general. However, in our case it turned out that they do hold, due to a \qm{good} behavior\footnote{The particular iterations of the self-energies turn out to be positive and monotonically decreasing functions. Or, loosely speaking, their shapes are similar, the only difference between the iterations is in their \qm{size}.} of the iterations. Thus, the quantities $I_X^{(n)}$ can indeed be used for controlling the convergence of the iteration process.

Usual behavior of such a nonlinear system in the case of only \emph{one equation for one unknown function} is such that for (almost) \emph{any} initial Ansatz the iteration procedure converges to a (trivial or non-trivial) solution. Example of such an equation is the equation \eqref{chp:prelim:SDpsi}, i.e., the SD equation for one $\Sigma$ (there is no subscript $_i$ in such a case) with $\Pi$ set to be a constant. Whether the solution $\Sigma$ is trivial or non-trivial depends typically on whether $y < y_\mathrm{crit.}$ or $y > y_\mathrm{crit.}$, respectively, for some critical value $y_\mathrm{crit.}$.

In our case of more \emph{coupled} equations the situation turns out to be, however, different. First, if there is only the trivial solution, the situation is the same as before: For any initial Ansatz \eqref{chp:frm:numAnz} the $I_X^{(n)}$ converges to some $0 \leq I < 1$. However, the existence of a non-trivial solution manifests differently than before: For the initial Ansatz \eqref{chp:frm:numAnz} too \qm{small} the $I_X^{(n)}$ behaves exactly like if there was only the non-trivial solution. However, for the Ansatz being sufficiently \qm{big} (and, needless to say, the same $y_1$, $y_2$) the iteration procedure blows up, i.e., $I_X^{(n)}$ converges to some $I > 1$ or even diverges.\footnote{In fact, this picture, as presented here, is somewhat simplified. In reality there are some additional complications due to the existence of the scalar pole \eqref{chp:frm:polesclr}.}

Let us specify the loose notions \qm{small} and \qm{big} more precisely. We can choose the initial Ansatz \eqref{chp:frm:numAnz} to be
\begin{eqnarray}
\label{chp:frm:numAnzx}
\Sigma_{1}^{(0)}(p^2) \;=\, \Sigma_{2}^{(0)}(p^2) &=& x \, f(p^2) \,,
\end{eqnarray}
where $f(p^2)$ is some fixed decreasing function, its concrete form turns out not to be very important. More important is the real parameter $x$, by setting of which we can manage the iteration process to converge to the trivial solution (which exists in any case) or to blow up (presumably in the case of existence of a non-trivial solution). Not surprisingly, the former is achieved by setting $x$ small enough, while the latter corresponds to $x$ large enough.

There must exist a limiting value of $x_{\mathrm{lim.}}$ between the two r\'{e}gimes. Since we can for any $x$ determine, according to the behavior of the iteration procedure, whether $x<x_{\mathrm{lim.}}$ or $x>x_{\mathrm{lim.}}$, the value $x_{\mathrm{lim.}}$ can be approximatively determined, with arbitrary accuracy, by means of the bisection method.

The behavior of the iterations process for $x$ close to $x_{\mathrm{lim.}}$ is rather peculiar: There exist some $n_0$ such that for $n<n_0$ the iteration process seems to converge to the non-trivial solution, but for $n>n_0$ it starts to go to the trivial solution or blows up (depending on whether $x<x_{\mathrm{lim.}}$ or $x>x_{\mathrm{lim.}}$, respectively). The point is that the closer is $x$ to $x_{\mathrm{lim.}}$, the larger is $n_0$. One can deduce that ideally, for $x=x_{\mathrm{lim.}}$, the $n_0$ would be infinite. Or in other words, for the Ansatz \eqref{chp:frm:numAnzx} with $x=x_{\mathrm{lim.}}$ the iteration procedure would converge to the non-trivial solution.

The procedure of finding a sufficiently accurate numerical solution (or, more precisely, a sufficiently accurate \emph{approximation} of the solution) of \eqref{chp:frm:SDnumform} therefore schematically consists of:
\begin{enumerate}
  \item Getting $x$ as close to $x_{\mathrm{lim.}}$ as possible. As this is numerically the most demanding part, the achieved proximity of $x$ to $x_{\mathrm{lim.}}$ is ultimately only a question of the available time and computer capacities.
  \item Finding the corresponding $n_0$, until which the iteration process (seemingly) converges.
  \item Taking the $n_0$-th iteration, i.e., $\Pi^{(n_0)}$, $\Sigma_{1}^{(n_0)}$, $\Sigma_{2}^{(n_0)}$, as the solution of \eqref{chp:frm:SDnumform}.
\end{enumerate}

Since the numerical procedure, as described above, has clearly many ambiguities, a special care was taken whether these ambiguities do not influence substantially the obtained results. In other words, the stability of the numerical algorithm was tested. Three main variations of the algorithm were considered:
\begin{description}
    \item[Class of Ans\"{a}tze] Several types of the decreasing function $f(x)$ in the Ansatz \eqref{chp:frm:numAnzx} were considered. For some of them (some very rapidly decreasing exponentials) the iteration procedure converged for \emph{any} $x$ \emph{only} to the trivial solution. However, if $f(p^2)$ was such that the iteration procedure converged to a non-trivial solution, then the non-trivial solution was always the same and hence presumably unique.

    \item[Integration method] There is an arbitrariness in the choice of the numerical integration method for the two of four integrals in each of the equations \eqref{chp:frm:SDwr} that have to be performed numerically. For the purpose of probing this arbitrariness we employed consecutively two methods: the trapezoidal rule and the Simpson's rule. On top of these, for the angular integration we have used also the Gauss--Chebyshev quadrature formula (using the Chebyshev polynomials of the second kind). The final results for all integration methods agreed, the differences were only in the speed of convergence.

    \item[Step-size] As a remnant of the numerical integration there is necessarily a step-size dependence of the results. The important question is how this dependence behaves for arbitrarily small step-sizes. If there is no sensible (i.e., finite) limit of the integral as the step-sizes are going to zero (the continuum limit), the results of the numerical integration have no meaning. We checked that this limit does exist and that all interesting phenomena (especially the strong $y_{1,2}$-dependence of the fermion masses, presented thereinafter) are present in it.

\end{description}

Moreover, in order to check the consistency of our numerical method by a comparison with an independent result, we calculated the equation for $\Sigma_i$ (either of \eqref{chp:frm:SDwrSgm1} or \eqref{chp:frm:SDwrSgm2}), with $\Pi$ set to be a constant (up to our knowledge, there are no independent calculations of the full set of the coupled equations \eqref{chp:frm:SDwr} we could compare with) and compared our result with the results of Ref.~\cite{Sauli:2006ba} (Eq.~(2) and Fig.~2 therein). They coincided.

\subsection{Numerical results}



\begin{figure}[t]
\begin{center}

\begingroup
  \makeatletter
  \providecommand\color[2][]{%
    \GenericError{(gnuplot) \space\space\space\@spaces}{%
      Package color not loaded in conjunction with
      terminal option `colourtext'%
    }{See the gnuplot documentation for explanation.%
    }{Either use 'blacktext' in gnuplot or load the package
      color.sty in LaTeX.}%
    \renewcommand\color[2][]{}%
  }%
  \providecommand\includegraphics[2][]{%
    \GenericError{(gnuplot) \space\space\space\@spaces}{%
      Package graphicx or graphics not loaded%
    }{See the gnuplot documentation for explanation.%
    }{The gnuplot epslatex terminal needs graphicx.sty or graphics.sty.}%
    \renewcommand\includegraphics[2][]{}%
  }%
  \providecommand\rotatebox[2]{#2}%
  \@ifundefined{ifGPcolor}{%
    \newif\ifGPcolor
    \GPcolorfalse
  }{}%
  \@ifundefined{ifGPblacktext}{%
    \newif\ifGPblacktext
    \GPblacktexttrue
  }{}%
  \let\gplgaddtomacro\g@addto@macro
  \gdef\gplbacktext{}%
  \gdef\gplfronttext{}%
  \makeatother
  \ifGPblacktext
    \def\colorrgb#1{}%
    \def\colorgray#1{}%
  \else
    \ifGPcolor
      \def\colorrgb#1{\color[rgb]{#1}}%
      \def\colorgray#1{\color[gray]{#1}}%
      \expandafter\def\csname LTw\endcsname{\color{white}}%
      \expandafter\def\csname LTb\endcsname{\color{black}}%
      \expandafter\def\csname LTa\endcsname{\color{black}}%
      \expandafter\def\csname LT0\endcsname{\color[rgb]{1,0,0}}%
      \expandafter\def\csname LT1\endcsname{\color[rgb]{0,1,0}}%
      \expandafter\def\csname LT2\endcsname{\color[rgb]{0,0,1}}%
      \expandafter\def\csname LT3\endcsname{\color[rgb]{1,0,1}}%
      \expandafter\def\csname LT4\endcsname{\color[rgb]{0,1,1}}%
      \expandafter\def\csname LT5\endcsname{\color[rgb]{1,1,0}}%
      \expandafter\def\csname LT6\endcsname{\color[rgb]{0,0,0}}%
      \expandafter\def\csname LT7\endcsname{\color[rgb]{1,0.3,0}}%
      \expandafter\def\csname LT8\endcsname{\color[rgb]{0.5,0.5,0.5}}%
    \else
      \def\colorrgb#1{\color{black}}%
      \def\colorgray#1{\color[gray]{#1}}%
      \expandafter\def\csname LTw\endcsname{\color{white}}%
      \expandafter\def\csname LTb\endcsname{\color{black}}%
      \expandafter\def\csname LTa\endcsname{\color{black}}%
      \expandafter\def\csname LT0\endcsname{\color{black}}%
      \expandafter\def\csname LT1\endcsname{\color{black}}%
      \expandafter\def\csname LT2\endcsname{\color{black}}%
      \expandafter\def\csname LT3\endcsname{\color{black}}%
      \expandafter\def\csname LT4\endcsname{\color{black}}%
      \expandafter\def\csname LT5\endcsname{\color{black}}%
      \expandafter\def\csname LT6\endcsname{\color{black}}%
      \expandafter\def\csname LT7\endcsname{\color{black}}%
      \expandafter\def\csname LT8\endcsname{\color{black}}%
    \fi
  \fi
  \setlength{\unitlength}{0.0500bp}%
  \begin{picture}(9360.00,6552.00)%
    \gplgaddtomacro\gplbacktext{%
      \csname LTb\endcsname%
      \put(2108,660){\makebox(0,0)[r]{\strut{} 0}}%
      \put(2108,1598){\makebox(0,0)[r]{\strut{} 20}}%
      \put(2108,2536){\makebox(0,0)[r]{\strut{} 40}}%
      \put(2108,3474){\makebox(0,0)[r]{\strut{} 60}}%
      \put(2108,4412){\makebox(0,0)[r]{\strut{} 80}}%
      \put(2108,5350){\makebox(0,0)[r]{\strut{} 100}}%
      \put(2108,6288){\makebox(0,0)[r]{\strut{} 120}}%
      \put(2240,440){\makebox(0,0){\strut{} 0}}%
      \put(3178,440){\makebox(0,0){\strut{} 20}}%
      \put(4116,440){\makebox(0,0){\strut{} 40}}%
      \put(5054,440){\makebox(0,0){\strut{} 60}}%
      \put(5992,440){\makebox(0,0){\strut{} 80}}%
      \put(6930,440){\makebox(0,0){\strut{} 100}}%
      \put(7868,440){\makebox(0,0){\strut{} 120}}%
      \put(1338,3474){\rotatebox{90}{\makebox(0,0){\strut{}$y_2$}}}%
      \put(5054,110){\makebox(0,0){\strut{}$y_1$}}%
      \put(3882,5116){\makebox(0,0){\strut{}$\mathbb{(I)}$}}%
      \put(3882,4553){\makebox(0,0){\strut{}$\Sigma_1 = 0 \quad \Sigma_2 \neq 0$}}%
      \put(6461,2442){\makebox(0,0){\strut{}$\mathbb{(III)}$}}%
      \put(6461,1879){\makebox(0,0){\strut{}$\Sigma_1 \neq 0 \quad \Sigma_2 = 0$}}%
      \put(7493,5913){\makebox(0,0){\strut{}$\mathbb{(II)}$}}%
      \put(7024,5444){\makebox(0,0){\strut{}$\Sigma_1 \neq 0$}}%
      \put(6555,4975){\makebox(0,0){\strut{}$\Sigma_2 \neq 0$}}%
      \put(3413,2161){\makebox(0,0){\strut{}$\mathbb{(IV)}$}}%
      \put(3413,1598){\makebox(0,0){\footnotesize{\textsc{hic svnt leones}}}}%
    }%
    \gplgaddtomacro\gplfronttext{%
    }%
    \gplbacktext


    \setlength\fboxsep{5bp}
    \setlength\fboxrule{0.5pt}
    \put(660,0){\fbox{\includegraphics[trim = 38bp 0bp 35bp 0bp, clip]{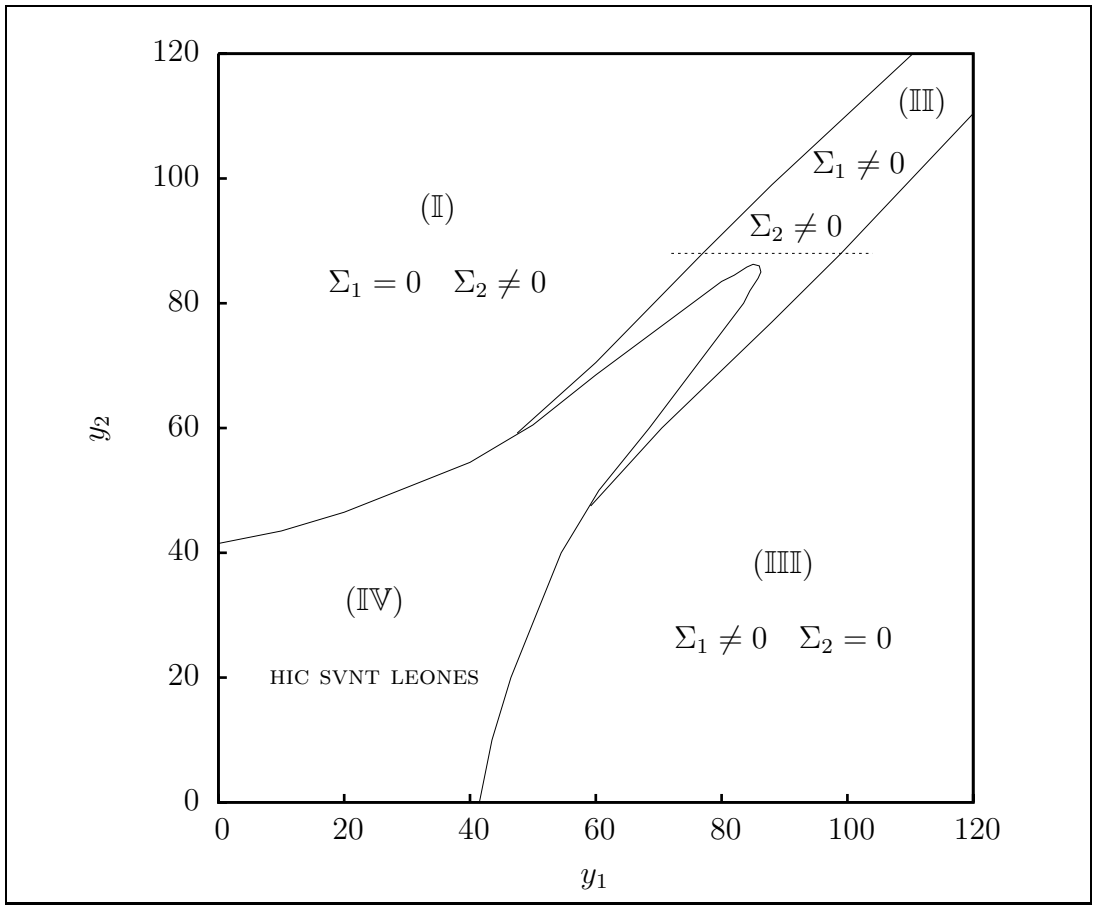}}}%
    \gplfronttext
  \end{picture}%
\endgroup

\end{center}
\caption[Qualitative behavior of the solutions to the SD equations, depending on the position in the $(y_1,y_2)$ plane.]{A quadrant of the $(y_1,y_2)$ plane with indicated areas of different behavior of the system of equations \eqref{chp:frm:SDwr}. According to the resulting fermion self-energies there are three main areas: first where $\Sigma_1 \equiv 0$ and $\Sigma_2 \neq 0$, second where $\Sigma_1 \neq 0$ and $\Sigma_2 \neq 0$ and third where $\Sigma_1 \neq 0$ and $\Sigma_2 \equiv 0$, denoted as $(\mathbb{I})$, $(\mathbb{II})$ and $(\mathbb{III})$, respectively. There is also the area, denoted as $(\mathbb{IV})$, where the pole \eqref{chp:frm:polesclr} in the scalar propagator prohibited us from finding solutions. The dashed line, going from $y_1=72$ and $y_2=88$ to $y_1=104$ and $y_2=88$, shows where the dependence of the spectrum on the Yukawa coupling constants was probed -- see Fig.~\ref{plot_Ms} and Fig.~\ref{plot_Mf}.}
\label{y_plane_pic}
\end{figure}


Using the numerical procedure described above, a part of the quadrant $y_1,y_2>0$ was probed and non-trivial solutions were found. Moreover, as far as we were able to check, all non-trivial solutions seem to be unique.

There are three types of the non-trivial solutions, according to whether only $\Sigma_1$, only $\Sigma_2$ or both $\Sigma_1$, $\Sigma_2$ are non-trivial. The locations of the three types of solutions in the $(y_1,y_2)$ plane are depicted in Fig.~\ref{y_plane_pic}. While for most of the values of $y_1$, $y_2$ the solutions were found, there is a region around the origin in the $(y_1,y_2)$ plane where the numerical analysis failed due to the existence of the scalar pole \eqref{chp:frm:polesclr}. Thus, we cannot say anything about the solutions of the equations \eqref{chp:frm:SDwr} for $y_1$, $y_2$ being simultaneously small.


\begin{figure}[t]
\begin{center}

\begin{picture}(13,0)%
\setlength\fboxsep{5bp}
\setlength\fboxrule{0.5pt}
\fbox{\includegraphics{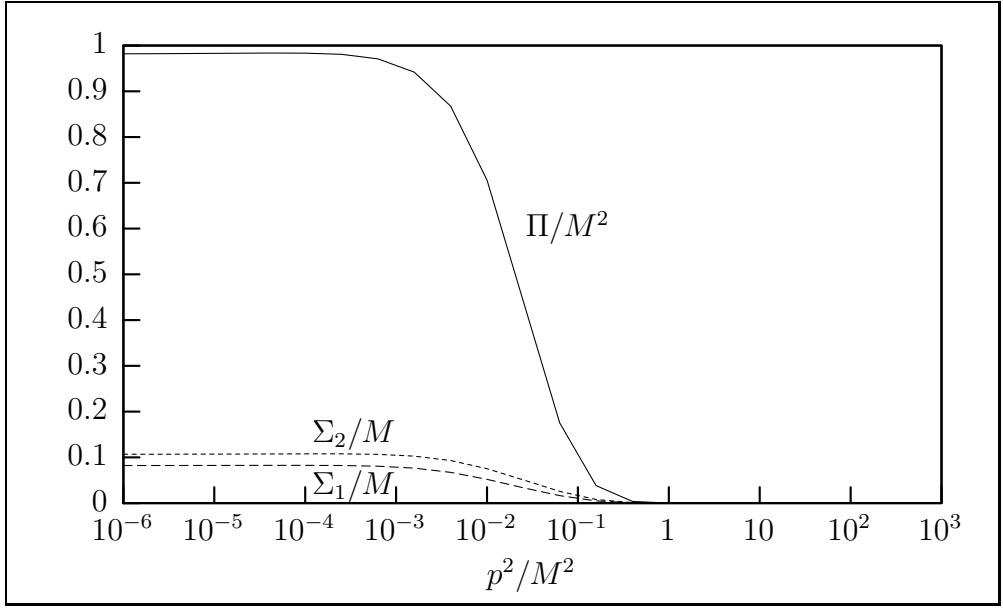}}%
\end{picture}%
\begingroup
\setlength{\unitlength}{0.0200bp}%
\begin{picture}(18000,10800)(0,0)%
\put(1650,1650){\makebox(0,0)[r]{\strut{} 0}}%
\put(1650,2510){\makebox(0,0)[r]{\strut{} 0.1}}%
\put(1650,3370){\makebox(0,0)[r]{\strut{} 0.2}}%
\put(1650,4230){\makebox(0,0)[r]{\strut{} 0.3}}%
\put(1650,5090){\makebox(0,0)[r]{\strut{} 0.4}}%
\put(1650,5950){\makebox(0,0)[r]{\strut{} 0.5}}%
\put(1650,6810){\makebox(0,0)[r]{\strut{} 0.6}}%
\put(1650,7670){\makebox(0,0)[r]{\strut{} 0.7}}%
\put(1650,8530){\makebox(0,0)[r]{\strut{} 0.8}}%
\put(1650,9390){\makebox(0,0)[r]{\strut{} 0.9}}%
\put(1650,10250){\makebox(0,0)[r]{\strut{} 1}}%
\put(17175,1100){\makebox(0,0){\strut{}$10^{3}$}}%
\put(15481,1100){\makebox(0,0){\strut{}$10^{2}$}}%
\put(13786,1100){\makebox(0,0){\strut{}$10$}}%
\put(12092,1100){\makebox(0,0){\strut{}$1$}}%
\put(10397,1100){\makebox(0,0){\strut{}$10^{-1}$}}%
\put(8703,1100){\makebox(0,0){\strut{}$10^{-2}$}}%
\put(7008,1100){\makebox(0,0){\strut{}$10^{-3}$}}%
\put(5314,1100){\makebox(0,0){\strut{}$10^{-4}$}}%
\put(3619,1100){\makebox(0,0){\strut{}$10^{-5}$}}%
\put(1925,1100){\makebox(0,0){\strut{}$10^{-6}$}}%
\put(9550,275){\makebox(0,0){\strut{}$p^2/M^2$}}%
\put(10233,6810){\makebox(0,0){\strut{}$\Pi/M^2$}}%
\put(6236,1960){\makebox(0,0){\strut{}$\Sigma_1/M$}}%
\put(6236,2940){\makebox(0,0){\strut{}$\Sigma_2/M$}}%
\end{picture}%
\endgroup

\end{center}
\caption[Typical shapes of the scalar and fermion self-energies.]{Typical shape of the solutions $\Sigma_1(p^2)$, $\Sigma_2(p^2)$ and $\Pi(p^2)$ to the system of equations \eqref{chp:frm:SDwr}, computed here for $y_1=83$ and $y_2=88$. Note the saturation of the self-energies at low momenta and fast decrease at high momenta.}
\label{self_energies_pic}
\end{figure}

The typical shape of the resulting non-trivial self-energies is depicted in Fig.~\ref{self_energies_pic}. They are saturated at low momenta and fall down rapidly at high momenta so that the integrals \eqref{chp:frm:SDwr} are indeed finite.


\begin{figure}[t]
\begin{center}

\begin{picture}(13,0)%
\setlength\fboxsep{5bp}
\setlength\fboxrule{0.5pt}
\fbox{\includegraphics{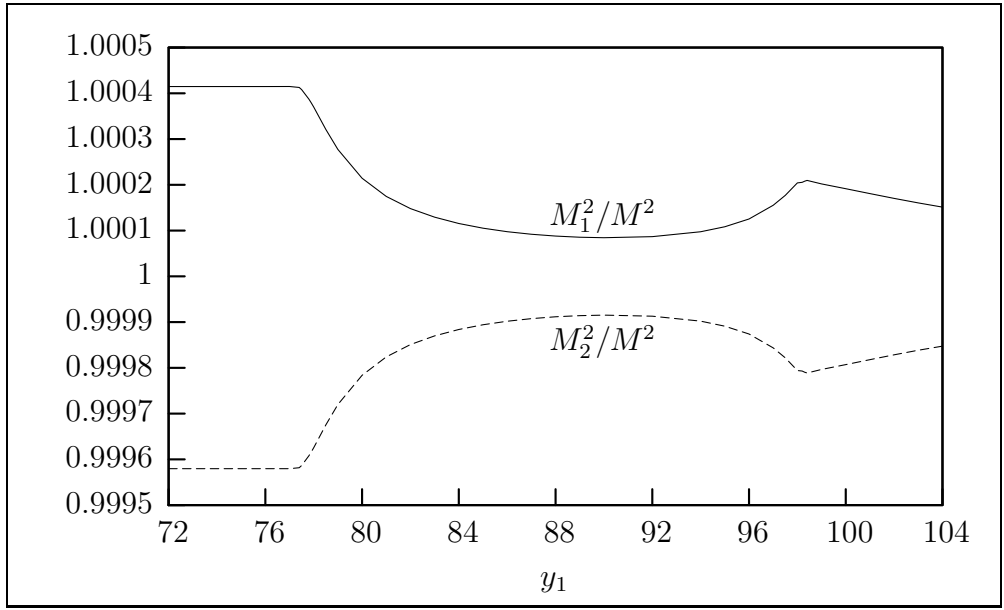}}%
\end{picture}%
\begingroup
\setlength{\unitlength}{0.0200bp}%
\begin{picture}(18000,10800)(0,0)%
\put(2475,1650){\makebox(0,0)[r]{\strut{} 0.9995}}%
\put(2475,2510){\makebox(0,0)[r]{\strut{} 0.9996}}%
\put(2475,3370){\makebox(0,0)[r]{\strut{} 0.9997}}%
\put(2475,4230){\makebox(0,0)[r]{\strut{} 0.9998}}%
\put(2475,5090){\makebox(0,0)[r]{\strut{} 0.9999}}%
\put(2475,5950){\makebox(0,0)[r]{\strut{} 1}}%
\put(2475,6810){\makebox(0,0)[r]{\strut{} 1.0001}}%
\put(2475,7670){\makebox(0,0)[r]{\strut{} 1.0002}}%
\put(2475,8530){\makebox(0,0)[r]{\strut{} 1.0003}}%
\put(2475,9390){\makebox(0,0)[r]{\strut{} 1.0004}}%
\put(2475,10250){\makebox(0,0)[r]{\strut{} 1.0005}}%
\put(2750,1100){\makebox(0,0){\strut{} 72}}%
\put(4553,1100){\makebox(0,0){\strut{} 76}}%
\put(6356,1100){\makebox(0,0){\strut{} 80}}%
\put(8159,1100){\makebox(0,0){\strut{} 84}}%
\put(9963,1100){\makebox(0,0){\strut{} 88}}%
\put(11766,1100){\makebox(0,0){\strut{} 92}}%
\put(13569,1100){\makebox(0,0){\strut{} 96}}%
\put(15372,1100){\makebox(0,0){\strut{} 100}}%
\put(17175,1100){\makebox(0,0){\strut{} 104}}%
\put(9962,275){\makebox(0,0){\strut{}$y_1$}}%
\put(10864,7068){\makebox(0,0){\strut{}$M_{1}^2/M^2$}}%
\put(10864,4703){\makebox(0,0){\strut{}$M_{2}^2/M^2$}}%
\end{picture}%
\endgroup

\end{center}
\caption[Scalar masses dependence on the Yukawa coupling constant.]{The $y_1$-dependence of the scalar masses $M^2_{1,2}$ with fixed $y_2=88$.}
\label{plot_Ms}
\end{figure}


\begin{figure}[t]
\begin{center}

\begin{picture}(13,0)%
\setlength\fboxsep{5bp}
\setlength\fboxrule{0.5pt}
\fbox{\includegraphics{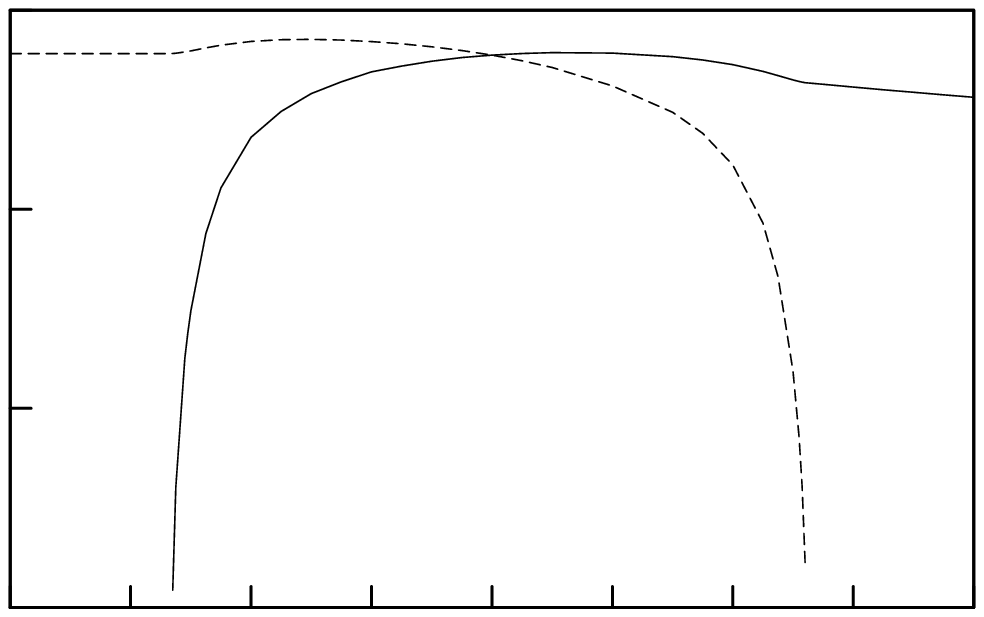}}%
\end{picture}%
\begingroup
\setlength{\unitlength}{0.0200bp}%
\begin{picture}(18000,10800)(0,0)%
\put(3025,1650){\makebox(0,0)[r]{\strut{}$10^{-5}$}}%
\put(3025,4517){\makebox(0,0)[r]{\strut{}$10^{-4}$}}%
\put(3025,7383){\makebox(0,0)[r]{\strut{}$10^{-3}$}}%
\put(3025,10250){\makebox(0,0)[r]{\strut{}$10^{-2}$}}%
\put(3300,1100){\makebox(0,0){\strut{} 72}}%
\put(5034,1100){\makebox(0,0){\strut{} 76}}%
\put(6769,1100){\makebox(0,0){\strut{} 80}}%
\put(8503,1100){\makebox(0,0){\strut{} 84}}%
\put(10238,1100){\makebox(0,0){\strut{} 88}}%
\put(11972,1100){\makebox(0,0){\strut{} 92}}%
\put(13706,1100){\makebox(0,0){\strut{} 96}}%
\put(15441,1100){\makebox(0,0){\strut{} 100}}%
\put(17175,1100){\makebox(0,0){\strut{} 104}}%
\put(10237,275){\makebox(0,0){\strut{}$y_1$}}%
\put(6986,8246){\makebox(0,0)[l]{\strut{}$m_{1}^2/M^2$}}%
\put(10888,8246){\makebox(0,0)[l]{\strut{}$m_{2}^2/M^2$}}%
\end{picture}%
\endgroup

\end{center}
\caption[Fermion masses dependence on the Yukawa coupling constant.]{The $y_1$-dependence of the fermion masses $m^2_{1,2}$ with fixed $y_2=88$.}
\label{plot_Mf}
\end{figure}

Our aim was to find the dependence of the spectrum -- the masses of the fermions and the scalars -- on the Yukawa coupling constants $y_1$, $y_2$. For the calculation of masses we have used the Minkowski-metric equations \eqref{chp:frm:eqsSpectrum}. We have probed the $y_{1,2}$-dependence along the cut depicted in Fig.~\ref{y_plane_pic}, since it connects all the three main areas $(\mathbb{I})$, $(\mathbb{II})$ and $(\mathbb{III})$ and therefore the resulting $y_{1,2}$-dependence of the spectrum can be regarded as quite typical. The results are depicted in Figs.~\ref{plot_Ms} and \ref{plot_Mf}. Note how the critical lines between the areas are evident in the $y_{1,2}$-dependence of the spectrum. The most significant result -- the behavior of the fermion spectrum -- can be seen in Fig.~\ref{plot_Mf}. As $y_1$ approaches the critical line between $(\mathbb{II})$ and $(\mathbb{I})$ (or $(\mathbb{III})$, respectively) \emph{in the direction} from $(\mathbb{II})$ to $(\mathbb{I})$ ($(\mathbb{III})$), the ratio $m^2_2/m^2_1$ becomes arbitrarily high (low)!

\section{Summary}

In this chapter we have redone the previous chapter's analysis in a more rigorous way:
\begin{itemize}
\item When defining the model, we have made sure this time, by introducing two fermion species with judiciously chosen axial charges, that the theory was free of the axial anomaly.
\item While in the previous chapter we have \qm{derived} the SD equations in a mere diagrammatical way, now we have derived them using the elaborate CJT formalism.
\item Moreover, we have derived the SD equations first for arbitrary self-energies and only then we have restricted them only on the properly chosen parts of the self-energies, sufficient for the task of \emph{demonstrating} the presence of SSB of the axial symmetry.
\item We have given a numerical evidence of viability of the present scheme by finding non-trivial UV-finite solutions to the SD equations with the following properties:
\begin{itemize}
\item They seem to be unique.
\item They exhibit a critical behavior in the sense that they exist only for Yukawa coupling constants large enough.
\item They allow for arbitrary amplification of fermion masses ratio.
\end{itemize}

\end{itemize}


\part{Electroweak interactions}
\label{part:ew}

\chapter{The model}
\label{chp:ew1}


\intro{In this part we apply the ideas from the previous one on a realistic theory of electroweak interactions. This chapter is dedicated merely to defining of the model and only in the subsequent chapters~\ref{chp:ewa} and \ref{chp:ewdyn} we will discuss the very possibility of breaking the electroweak symmetry by the Yukawa dynamics. Apart from sole definition of the model by means of its particle content and its symmetries, most of this chapter is dedicated to convenient reparameterization of the theory (i.e., the fields, the symmetry generators and the gauge and Yukawa interactions) in terms of the Nambu--Gorkov formalism. This time, in contrast to part~\ref{part:abel}, this reparameterization will apply not only to scalars, but also to fermions, namely to leptons.}

\intro{This chapter, as well as the whole part~\ref{part:ew}, is a technically-oriented extension of Refs.~\cite{Brauner:2004kg,Benes:2008ir}.}

\section{The Lagrangian}

\subsection{Particle content}
\label{ew1:ssec:PartCont}

We consider an $\group{SU}(2)_{\mathrm{L}} \times \group{U}(1)_{\mathrm{Y}}$ gauge-invariant theory equipped with the usual SM fermion content. That is to say, we consider the quark and lepton left-handed $\group{SU}(2)_{\mathrm{L}}$ doublets $\bigl(\begin{smallmatrix} u_{aL} \\ d_{aL} \end{smallmatrix} \bigr)$ and $\bigl(\begin{smallmatrix} \nu_{aL} \\ e_{aL} \end{smallmatrix} \bigr)$, respectively, together with the charged fermion right-handed singlets $u_{aR}$, $d_{aR}$, $e_{aR}$. We assume $n$ generations: $a=1,\ldots,n$. Moreover, on top of this SM particle content we introduce also $m$ right-handed neutrino singlets $\nu_{aR}$. Their number $m$ may be in principle arbitrary, unrelated to the number of generations, so in general we have to assume $a=0,\ldots,m \neq n$. (Notice that the numbers of the left-handed doublets and of the charged right-handed singlets are constrained to be the same due to the requirement of anomaly freedom.) Since the fields of each type with different values of the index $a$ are not distinguished by the quantum numbers of the $\group{SU}(2)_{\mathrm{L}} \times \group{U}(1)_{\mathrm{Y}}$ symmetry, we can call them the \emph{flavors}\footnote{See for example the very first sentence of Ref.~\cite{Perez:2009xj}: \qm{Flavors are just replication of states with identical quantum numbers.}}.

In the following it will be useful to suppress the flavor indices and adopt more compact notation. We will therefore denote the particular left-handed fields of all generations collectively as
\begin{eqnarray}
\label{ew1:fL}
f_{L} &\equiv& \left( \begin{array}{c} f_{1L} \\ \vdots \\ f_{nL} \end{array} \right)
\,,\qquad\qquad f=\nu,e,u,d,
\end{eqnarray}
and similarly the right-handed \emph{charged} fields as
\begin{eqnarray}
\label{ew1:fR}
f_{R} &\equiv& \left( \begin{array}{c} f_{1R} \\ \vdots \\ f_{nR} \end{array} \right)
\,,\qquad\qquad f=e,u,d.
\end{eqnarray}
The definition of the right-handed neutrino field $\nu_R$ is analogous and differs of course only formally by its different dimension ($m$ instead of $n$):
\begin{eqnarray}
\label{ew1:nuR}
\nu_{R} &\equiv& \left( \begin{array}{c} \nu_{1R} \\ \vdots \\ \nu_{mR} \end{array} \right) \,.
\end{eqnarray}
The $\group{SU}(2)_{\mathrm{L}}$ doublets are now given in terms of the left-handed fields \eqref{ew1:fL} as
\begin{equation}
\label{ew1:ellLqL}
\ell_{L} \equiv \left( \begin{array}{c} \nu_{L} \\ e_{L} \end{array} \right) \,,
\quad\quad
q_{L} \equiv \left( \begin{array}{c} u_{L} \\ d_{L} \end{array} \right) \,.
\end{equation}

Furthermore, we consider two scalar $\group{SU}(2)_{\mathrm{L}}$ doublets\footnote{The denotations \qm{$S$} and \qm{$N$} stand respectively for \qm{southern} and \qm{northern}, since, as we will see later, the $S$ doublet will contribute primarily to the masses of the down-type fermions, while the $N$ doublet primarily to the masses of the up-type fermions.}
\begin{subequations}
\label{ew1:SN}
\begin{eqnarray}
S &\equiv& \left(\begin{array}{c} S^{(+)} \\ S^{(0)} \end{array}\right) \,, \\
N &\equiv& \left(\begin{array}{c} N^{(0)} \\ N^{(-)} \end{array}\right) \,,
\end{eqnarray}
\end{subequations}
with bare masses $M_S$ and $M_N$, respectively. For definiteness, their Lagrangian thus reads
\begin{eqnarray}
\label{ew1:eL:sclr}
\eL_{\mathrm{scalar}} &=& (\partial_\mu S)^\dag (\partial^\mu S) + (\partial_\mu N)^\dag (\partial^\mu N)
- M_S^2 S^\dag S - M_N^2 N^\dag N \,.
\end{eqnarray}
We stress that $M_S^2,M_N^2 > 0$, so that, likewise in the previous part, we do not need to consider the scalar self-interactions.

Since the right-handed neutrinos are singlets under the whole $\group{SU}(2)_{\mathrm{L}} \times \group{U}(1)_{\mathrm{Y}}$, there is no protection for the their Majorana mass terms. In other words, the requirement of the electroweak symmetry $\group{SU}(2)_{\mathrm{L}} \times \group{U}(1)_{\mathrm{Y}}$ \emph{alone} is consistent with the assumption of the existence of hard mass term in the Lagrangian of the form
\begin{eqnarray}
\label{ew1:eL:MnuR}
\eL_{\mathrm{mass}} &=& {}-\frac{1}{2}(\bar\nu_R)^\C M_{\nu R}\nu_R -\frac{1}{2}\bar\nu_R M_{\nu R}^\dag(\nu_R)^\C \,,
\end{eqnarray}
where $M_{\nu R}$ is an $m \times m$ symmetric\footnote{See equation \eqref{appfrm:mR} in appendix~\ref{app:fermi propag}.} mass matrix and $(\nu_R)^\C$ denotes the charge conjugation \eqref{symbols:psiC}, discussed in more detail in appendix~\ref{app:charge}.

On the other hand, one can define the \emph{lepton number} symmetry $\group{U}(1)_\ell$ as follows: On the lepton fields $\ell_L$, $e_R$, $\nu_R$ it acts non-trivially as
\begin{eqnarray}
\label{ew1:leptonNum}
\group{U}(1)_\ell\,:\qquad \{\ell_L,e_R,\nu_R\}  &\TransformsTo&
{[\{\ell_L,e_R,\nu_R\}]}^\prime \>=\> \e^{\I Q_\ell \theta} \{\ell_L,e_R,\nu_R\}
\end{eqnarray}
(with $Q_\ell$ being the $\group{U}(1)_\ell$ charge) and leaves all other fields (i.e., the quark fields, the scalars and the gauge bosons) invariant. Clearly, the theory is invariant under \eqref{ew1:leptonNum}, except for the Lagrangian \eqref{ew1:eL:MnuR}, which breaks it explicitly. For $M_{\nu R}=0$ the lepton number symmetry $\group{U}(1)_\ell$ would be exact. Nevertheless, we will for definiteness assume $M_{\nu R} \neq 0$; implications of the case $M_{\nu R}=0$ will be discussed only occasionally.

\subsection{Yukawa interactions}
\label{ew1:ssec:Yukawa}


By assumption responsible for the eventual dynamical EWSB, the Yukawa interactions are for us of key importance. We postulate
\begin{eqnarray}
\label{eq:Yukawa}
\eL_{\mathrm{Yukawa}} &=& \eL_{\mathrm{Yukawa},q} + \eL_{\mathrm{Yukawa},\ell} \,,
\end{eqnarray}
where
\begin{subequations}
\label{ew1:eL:YqellSN}
\begin{eqnarray}
\eL_{\mathrm{Yukawa},q} &=& \bar q_{L}y_{d}d_{R} S + \bar q_{L}y_{u}u_{R} N + \hc \,,
\label{ew1:eL:Ykw:q}
\\
\eL_{\mathrm{Yukawa},\ell} &=& \bar \ell_{L}y_{e}e_{R} S + \bar \ell_{L}y_{\nu}\nu_{R} N + \hc
\label{ew1:eL:Ykw:ell}
\end{eqnarray}
\end{subequations}
The Yukawa coupling constants $y_{u}$, $y_{d}$, $y_{\nu}$, $y_{e}$ are in principle arbitrary complex matrices. For the sake of later references let us decompose \eqref{ew1:eL:YqellSN} also as
\begin{subequations}
\begin{eqnarray}
\eL_{\mathrm{Yukawa},q} &=& \eL_{\mathrm{Yukawa},q}^{(0)} + \eL_{\mathrm{Yukawa},q}^{(\pm)} \,,
\\
\eL_{\mathrm{Yukawa},\ell} &=& \eL_{\mathrm{Yukawa},\ell}^{(0)} + \eL_{\mathrm{Yukawa},\ell}^{(\pm)} \,,
\end{eqnarray}
\end{subequations}
i.e., into the interactions of the neutral scalars:
\begin{subequations}
\label{eq:Yukawa_neutral}
\begin{eqnarray}
\eL_{\mathrm{Yukawa},q}^{(0)} &=&
\bar d_{L} y_{d} d_{R} S^{(0)} + \bar d_{R} y_{d}^\dag d_{L} S^{(0)\dag} +
\bar u_{L} y_{u} u_{R} N^{(0)} + \bar u_{R} y_{u}^\dag u_{L} N^{(0)\dag} \,,
\\
\eL_{\mathrm{Yukawa},\ell}^{(0)} &=&
\bar   e_{L} y_{e}     e_{R} S^{(0)} + \bar   e_{R} y_{e}^\dag     e_{L} S^{(0)\dag} +
\bar \nu_{L} y_{\nu} \nu_{R} N^{(0)} + \bar \nu_{R} y_{\nu}^\dag \nu_{L} N^{(0)\dag}
\end{eqnarray}
\end{subequations}
and the interactions of the charged scalars:
\begin{subequations}
\label{eq:Yukawa_charged}
\begin{eqnarray}
\eL_{\mathrm{Yukawa},q}^{(\pm)} &=&
\bar u_{L} y_{d} d_{R} S^{(+)} + \bar d_{R} y_{d}^\dag u_{L} S^{(+)\dag} +
\bar d_{L} y_{u} u_{R} N^{(-)} + \bar u_{R} y_{u}^\dag d_{L} N^{(-)\dag} \,,
\\
\eL_{\mathrm{Yukawa},\ell}^{(\pm)} &=&
\bar \nu_{L} y_{e} e_{R} S^{(+)} + \bar e_{R} y_{e}^\dag \nu_{L} S^{(+)\dag} +
\bar e_{L} y_{\nu} \nu_{R} N^{(-)} + \bar \nu_{R} y_{\nu}^\dag e_{L} N^{(-)\dag} \,.
\end{eqnarray}
\end{subequations}

Notice that we do not consider the interactions
\begin{eqnarray}
\label{eq:Yukawa_tilde}
\tilde \eL_{\mathrm{Yukawa}} &=& \tilde \eL_{\mathrm{Yukawa},q} + \tilde \eL_{\mathrm{Yukawa},\ell} \,,
\end{eqnarray}
with
\begin{subequations}
\begin{eqnarray}
\tilde \eL_{\mathrm{Yukawa},q} &=& \bar q_{L} \tilde y_{d} d_{R} \tilde N + \bar q_{L} \tilde y_{u} u_{R} \tilde S + \hc \,,
\label{ew1:eL:tldYkw:q}
\\
\tilde \eL_{\mathrm{Yukawa},\ell} &=& \bar \ell_{L} \tilde y_{e} e_{R} \tilde N + \bar \ell_{L} \tilde y_{\nu} \nu_{R} \tilde S + \hc \,,
\label{ew1:eL:tldYkw:ell}
\end{eqnarray}
\end{subequations}
whose neutral and charged parts read
\begin{subequations}
\label{eq:Yukawa_neutral_tilde}
\begin{eqnarray}
\tilde \eL_{\mathrm{Yukawa},q}^{(0)} &=&
{}- \bar d_{L} \tilde y_{d} d_{R} N^{(0)\dag} - \bar d_{R} \tilde y_{d}^\dag d_{L} N^{(0)}
+ \bar u_{L} \tilde y_{u} u_{R} S^{(0)\dag} + \bar u_{R} \tilde y_{u}^\dag u_{L} S^{(0)} \,,
\\
\tilde \eL_{\mathrm{Yukawa},\ell}^{(0)} &=&
{}- \bar e_{L} \tilde y_{e} e_{R} N^{(0)\dag} - \bar e_{R} \tilde y_{e}^\dag e_{L} N^{(0)}
+ \bar \nu_{L} \tilde y_{\nu} \nu_{R} S^{(0)\dag} + \bar \nu_{R} \tilde y_{\nu}^\dag \nu_{L} S^{(0)}
\end{eqnarray}
\end{subequations}
and
\begin{subequations}
\label{eq:Yukawa_charged_tilde}
\begin{eqnarray}
\tilde \eL_{\mathrm{Yukawa},q}^{(\pm)} &=&
\bar u_{L} \tilde y_{d} d_{R} N^{(-)\dag} + \bar d_{R} \tilde y_{d}^\dag u_{L} N^{(-)}
-\bar d_{L} \tilde y_{u} u_{R} S^{(+)\dag} - \bar u_{R} \tilde y_{u}^\dag d_{L} S^{(+)} \,,
\\
\tilde \eL_{\mathrm{Yukawa},\ell}^{(\pm)} &=&
\bar \nu_{L} \tilde y_{e} e_{R} N^{(-)\dag} + \bar e_{R} \tilde y_{e}^\dag \nu_{L} N^{(-)}
-\bar e_{L} \tilde y_{\nu} \nu_{R} S^{(+)\dag} - \bar \nu_{R} \tilde y_{\nu}^\dag e_{L} S^{(+)} \,,
\end{eqnarray}
\end{subequations}
respectively. In fact, these interactions are also permitted by the underlying $\group{SU}(2)_{\mathrm{L}} \times \group{U}(1)_{\mathrm{Y}}$ symmetry: The doublets $\tilde S$, $\tilde N$ are defined in terms of $S$, $N$ as\footnote{The charge conjugates $S^\C$, $N^\C$ are defined in Eq.~\eqref{app:sclr:phiC} in appendix~\ref{app:sclr}.}

\begin{subequations}
\begin{eqnarray}
\tilde S &\equiv& \I \sigma_2 S^\C
\>=\> \left(\begin{array}{c} S^{(0)\dag} \\ -S^{(+)\dag} \end{array}\right) \,,
\\
\tilde N &\equiv& \I \sigma_2 N^\C
\>=\> \left(\begin{array}{c} N^{(-)\dag} \\ -N^{(0)\dag} \end{array}\right) \,,
\end{eqnarray}
\end{subequations}
i.e., they are basically just the charge conjugates of $S$, $N$ (and thus having opposite hypercharge), only rotated by the antisymmetric matrix $\I \sigma_2 = \bigl(\begin{smallmatrix} \phantom{-}0 & 1 \\ -1 & 0 \end{smallmatrix} \bigr)$, so that they are valid $\group{SU}(2)_{\mathrm{L}}$ doublets.

Dismissing of the interactions \eqref{eq:Yukawa_tilde} is in fact justified by assuming that there is a discrete symmetry, called $\mathcal{P}_{\mathrm{down}}$, acting non-trivially only on $e_R$, $d_R$, $S$ as
\begin{eqnarray}
\label{ew1:Pdown}
\mathcal{P}_{\mathrm{down}}\,:\qquad \{e_R,d_R,S\}  &\TransformsTo&
{[\{e_R,d_R,S\}]}^\prime \>=\> - \{e_R,d_R,S\}
\end{eqnarray}
and leaving all other fields invariant. Clearly, this symmetry forbids the interactions \eqref{eq:Yukawa_tilde}. Its most obvious advantage is at this moment the reduction of the number of the Yukawa coupling constants. Further reasons for imposing $\mathcal{P}_{\mathrm{down}}$ will be discussed at the end of Sec.~\ref{ssec:beyond1loop}.

At this point one may notice that the Yukawa Lagrangian \eqref{eq:Yukawa} is basically the same as in the SM (apart from the Yukawa interactions of $\nu_{R}$), with $S$ and $N$ playing the r\^{o}le of the Higgs doublet $\phi$ and $\tilde \phi = \I \sigma_2 \phi^\C$, respectively. One may then ask why not to consider only one scalar doublet like in the SM, instead of two distinct doublets $S$, $N$. This question will be discussed in Sec.~\ref{ssec:whytwo}, after introduction of the SD equations.

\subsection{Gauge interactions}

\subsubsection{Notation for the gauge basis}

The theory is invariant under the $\group{SU}(2)_{\mathrm{L}} \times \group{U}(1)_{\mathrm{Y}}$ gauge symmetry. We denote the generators of the respective subgroups as
\begin{subequations}
\begin{eqnarray}
\group{SU}(2)_{\mathrm{L}}\,: &&  t_{a=1,2,3} \,,   \\
\group{U}(1)_{\mathrm{Y}}\,: &&  t_{a=4}
\end{eqnarray}
\end{subequations}
and the corresponding gauge fields as
\begin{subequations}
\begin{eqnarray}
\group{SU}(2)_{\mathrm{L}}\,: &&  A^\mu_{a=1,2,3} \,,  \\
\group{U}(1)_{\mathrm{Y}}\,: &&  A^\mu_{a=4} \,.
\end{eqnarray}
\end{subequations}

\subsubsection{Another gauge basis}

However, for various reasons the gauge boson basis $A^\mu_{1}$, $A^\mu_{2}$, $A^\mu_{3}$, $A^\mu_{4}$ is not always convenient. Thus, we are now going to introduce another basis.

The generators $t_3$, $t_4$ can be rotated as
\begin{eqnarray}
\label{ew1:rot:t34-Zew}
\left(\begin{array}{c} t_{Z} \\ t_{\mathrm{em}} \end{array}\right) &\equiv& O_{\mathrm{W}}
\left(\begin{array}{c} t_3 \\ t_4    \end{array}\right) \,,
\end{eqnarray}
which corresponds to the rotation of the gauge bosons $A^\mu_3$, $A^\mu_4$
\begin{eqnarray}
\label{ew1:rot:A34-Zew}
\left(\begin{array}{c} A^\mu_{Z} \\ A^\mu_{\mathrm{em}} \end{array}\right) &\equiv& O_{\mathrm{W}}
\left(\begin{array}{c} A^\mu_3 \\ A^\mu_4    \end{array}\right) \,.
\end{eqnarray}
We defined here the orthogonal matrix $O_{\mathrm{W}}$ in terms of the gauge coupling constants $g$, $g^\prime$ (corresponding to the respective subgroups $\group{SU}(2)_{\mathrm{L}}$ and $\group{U}(1)_{\mathrm{Y}}$) as
\begin{eqnarray}
\label{ew1:Wangle}
O_{\mathrm{W}} &\equiv&
\left(\begin{array}{rr}
\cos\theta_{\mathrm{W}}  & -\sin\theta_{\mathrm{W}} \\
\sin\theta_{\mathrm{W}}  &  \cos\theta_{\mathrm{W}}
\end{array}\right)
\ \equiv \ \frac{1}{\sqrt{g^2+g^{\prime2}}}
\left(\begin{array}{ll} g & -g^\prime \\ g^\prime & \phantom{-}g \end{array}\right) \,,
\end{eqnarray}
where $\theta_{\mathrm{W}}$ is the Weinberg (or weak mixing) angle. Now the new generator $t_{\mathrm{em}}$ corresponds to the unbroken subgroup $\group{U}(1)_{\mathrm{em}}$:
\begin{eqnarray}
\group{U}(1)_{\mathrm{em}}\,: &&  t_{\mathrm{em}} \,.
\end{eqnarray}
Of course, the fields $A^\mu_{Z}$ and $A^\mu_{\mathrm{em}}$ correspond to the $Z$ boson and $\gamma$ (photon), respectively, i.e., to the gauge boson mass eigenstates after the eventual spontaneous breakdown of $\group{SU}(2)_{\mathrm{L}} \times \group{U}(1)_{\mathrm{Y}}$ down to $\group{U}(1)_{\mathrm{em}}$, as we will show in detail in chapter~\ref{ewM}.

Similarly we can rotate the gauge fields $A^\mu_1$, $A^\mu_2$ as
\begin{eqnarray}
\label{ew1:rot:A12-Wpm}
\left(\begin{array}{c} A^\mu_{{W}^+} \\ A^\mu_{{W}^-} \end{array}\right) &\equiv& U_{\mathrm{W}}
\left(\begin{array}{c} A^\mu_1 \\ A^\mu_2    \end{array}\right) \,,
\end{eqnarray}
where
\begin{eqnarray}
U_{\mathrm{W}} &\equiv& \frac{1}{\sqrt{2}}\left(\begin{array}{rr} 1 & -\I \\ 1 & \I \end{array}\right)
\end{eqnarray}
is a unitary matrix. The fields $A^\mu_{{W}^\pm}$ correspond to the $W^\pm$ bosons (being charge conjugation eigenstates) and satisfy
\begin{eqnarray}
(A^\mu_{{W}^\pm})^\dag &=& A^\mu_{{W}^\mp} \,.
\end{eqnarray}

\section{Reparameterization of the Lagrangian}

The theory has been so far formulated in terms of the fields $S$, $N$, $q_L$, $u_R$, $d_R$, $\ell_L$, $\nu_R$, $e_R$. This is convenient for the formulation of the model from the gauge principle, since all the mentioned fields are in fact directly the $\group{SU}(2)_{\mathrm{L}} \times \group{U}(1)_{\mathrm{Y}}$ irreducible representations. However, for practical calculations it will be more useful to reparameterize the theory in terms of new degrees of freedom $\Phi$, $q$, $\Psi_\ell$, defined in terms of the original ones as
\begin{equation}
\label{ew1:PhiqPsiell}
\Phi \>\equiv\> \left(\begin{array}{c} S^{(+)} \\ N^{(-)\dag} \\ S^{(+)\dag} \\ N^{(-)} \\ S^{(0)} \\ S^{(0)\dag} \\ N^{(0)}   \\ N^{(0)\dag}  \end{array}\right)
\,,\qquad
q \>\equiv\> \left(\begin{array}{c} u_L+u_R \\ d_L+d_R  \end{array}\right)
\,,\qquad
\Psi_\ell \>\equiv\> \left(\begin{array}{c} \nu_L+(\nu_L)^\C \\ \nu_R+(\nu_R)^\C \\ e_L+(e_L)^\C \\ e_R+(e_R)^\C  \end{array}\right)
\,.
\end{equation}
In the following sections we comment closer on the motivations for introducing this, at first sight unnecessarily complicated notation and give a more detailed technical treatment of each of the fields $\Phi$, $q$, $\Psi_\ell$. At this moment let us just say that the main motivation is the same as before within the Abelian toy model in part~\ref{part:abel}, where it was more convenient to work with the Nambu--Gorkov field $\Phi = \bigl(\begin{smallmatrix} \phi \\ \phi^\dag \end{smallmatrix} \bigr)$ than with $\phi$ due to the employed mechanism and pattern of the SSB. Similar argumentation will be used also in the present context, especially for scalars and leptons. Last but not least, the notation \eqref{ew1:PhiqPsiell} will allow us to write some (but not all) formul{\ae} in much more compact and elegant way. For instance, as a consequence of having only three independent fields $\Phi$, $q$, $\Psi_\ell$ there will be also only three (though matrix) SD equations.

The notation for fermions, introduced in this section, will be utilized also later in chapter \ref{ewM} when discussing masses of the electroweak gauge bosons.

\subsection{Scalars}

\subsubsection{Reparameterization of the fields}

Recall in the Abelian toy model, where the symmetry was broken by the two-point functions $\langle \phi\phi \rangle$, $\langle \phi^\dag\phi^\dag \rangle$, it was more convenient to work with the Nambu--Gorkov field $\Phi=\bigl(\begin{smallmatrix} \phi \\ \phi^\dag \end{smallmatrix} \bigr)$ instead of the single complex field $\phi$. Analogously, we now assume that the electroweak symmetry will be broken by formation of the two-point functions of the type, e.g., $\langle S^{(0)}S^{(0)} \rangle$, $\langle N^{(0)}N^{(0)} \rangle$, etc. Therefore, instead of working with the two scalar doublet $S$, $N$, organized perhaps in a single field $\phi$,
\begin{eqnarray}
\label{ew1:phi}
\phi &\equiv& \left(\begin{array}{c} S \\ N \end{array}\right)
\>=\> \left(\begin{array}{c} S^{(+)} \\ S^{(0)} \\ N^{(0)} \\ N^{(-)} \end{array}\right) \,,
\end{eqnarray}
it will be more convenient to work with the corresponding Nambu--Gorkov field, whose matrix propagator, in contrast to propagator of the field $\phi$, naturally incorporates the symmetry-breaking propagators of the desired type. I.e., we can, in accordance with appendix~\ref{app:sclr}, define:
\begin{eqnarray}
\label{ew1:Phiprime}
\Phi^\prime &\equiv& \left(\begin{array}{c} \phi \\ \phi^\C \end{array}\right)
\>=\>
\left(\begin{array}{c} S^{(+)} \\ S^{(0)} \\ N^{(0)} \\ N^{(-)} \\ S^{(+)\dag} \\ S^{(0)\dag} \\ N^{(0)\dag} \\ N^{(-)\dag} \end{array}\right)
\,.
\end{eqnarray}
Here the scalar charge conjugation $\phi^\C$ is defined by \eqref{app:sclr:phiC} in appendix~\ref{app:sclr}.

However, as indicated by the prime, the basis $\Phi^\prime$ is in fact still not the most convenient one (and consequently also not the one we will eventually use for actual calculations). The reason for this is that the electrically neutral and the electrically charged components are in $\Phi^\prime$ distributed in a rather inconvenient way. The following choice proves to be more (or perhaps the most) convenient:
\begin{eqnarray}
\label{ew1:Phi}
\Phi &\equiv&
\left(\begin{array}{c}
S^{(+)} \\ N^{(-)\dag} \\ S^{(+)\dag} \\ N^{(-)} \\ S^{(0)} \\ S^{(0)\dag} \\ N^{(0)} \\ N^{(0)\dag}
\end{array}\right) \,.
\end{eqnarray}
One can appreciate better this choice by noting that it has the generic structure
\begin{eqnarray}
\label{ew1:Phi+0}
\Phi &=& \left(\begin{array}{c} \Phi^{(+)} \\ \Phi^{(0)} \end{array}\right) \,,
\end{eqnarray}
where $\Phi^{(+)}$ and $\Phi^{(0)}$ are made exclusively of the charged and neutral scalars, respectively:
 \begin{equation}
\label{ew1:Phi+0def}
\Phi^{(+)} \>\equiv\> \left(\begin{array}{c} S^{(+)} \\ N^{(-)\dag} \\ S^{(+)\dag} \\ N^{(-)} \end{array}\right)
\,, \qquad\qquad
\Phi^{(0)} \>\equiv\> \left(\begin{array}{c} S^{(0)} \\ S^{(0)\dag} \\ N^{(0)} \\ N^{(0)\dag} \end{array}\right)
\,.
\end{equation}
Thus, due to the conservation of the electric charge the propagator $\langle \Phi\Phi^\dag \rangle$ will have in the basis \eqref{ew1:Phi+0} a block diagonal form. Moreover, the structure of \eqref{ew1:Phi+0} resembles the structure of an $\group{SU}(2)_{\mathrm{L}}$ doublet with the electroweak hypercharge $Y=1$, consequently the structure of the $\group{SU}(2)_{\mathrm{L}} \times \group{U}(1)_{\mathrm{Y}}$ generators in the $\Phi$ basis will be rather familiar.

In fact, $\Phi$ is related to $\Phi^\prime$ by a simple linear transformation
\begin{eqnarray}
\label{ew1:Phiprime-Phi}
\Phi &=& U \, \Phi^\prime \,,
\end{eqnarray}
where the unitary matrix $U$ is explicitly given as
\begin{eqnarray}
U &\equiv& \left(\begin{array}{cc|cc|cc|cc}
1 & 0 & 0 & 0 & 0 & 0 & 0 & 0 \\
0 & 0 & 0 & 0 & 0 & 0 & 0 & 1 \\ \hline
0 & 0 & 0 & 0 & 1 & 0 & 0 & 0 \\
0 & 0 & 0 & 1 & 0 & 0 & 0 & 0 \\ \hline
0 & 1 & 0 & 0 & 0 & 0 & 0 & 0 \\
0 & 0 & 0 & 0 & 0 & 1 & 0 & 0 \\ \hline
0 & 0 & 1 & 0 & 0 & 0 & 0 & 0 \\
0 & 0 & 0 & 0 & 0 & 0 & 1 & 0 \\
\end{array}\right)
\>=\>
\frac{1}{2}\left(\begin{array}{cccc}
1+\sigma_3          & 0          & 0          & 1-\sigma_3           \\
0                   & 1-\sigma_3 & 1+\sigma_3 & 0                    \\
\sigma_1+\I\sigma_2 & 0          & 1-\sigma_3 & 0                    \\
0                   & 1+\sigma_3 & 0          & \sigma_1-\I\sigma_2  \\
\end{array}\right) \,.
\nonumber \\ &&
\end{eqnarray}

In order to establish the link with the literature, we also introduce the notation
\begin{subequations}
\label{ew1:PhiSN-S-N}
\begin{eqnarray}
\Phi_{SN} &\equiv& \left(\begin{array}{c} S^{(+)} \\ N^{(-)\dag} \end{array}\right) \,, \\
\Phi_{S} &\equiv& \left(\begin{array}{c} S^{(0)} \\ S^{(0)\dag} \end{array}\right) \,, \\
\Phi_{N} &\equiv& \left(\begin{array}{c} N^{(0)} \\ N^{(0)\dag} \end{array}\right) \,,
\end{eqnarray}
\end{subequations}
which is used in Ref.~\cite{Benes:2008ir}. We will also use this notation later on when discussing the Ansatz for the scalar 1PI propagator. In this notation we clearly have
\begin{subequations}
\begin{eqnarray}
\Phi^{(+)} &=& \left(\begin{array}{c} \Phi_{SN} \\ \Phi_{SN}^\C \end{array}\right) \,, \\
\Phi^{(0)} &=& \left(\begin{array}{c} \Phi_{S} \\ \Phi_{N} \end{array}\right) \,,
\end{eqnarray}
\end{subequations}
with $\Phi^{(+)}$, $\Phi^{(0)}$ defined in \eqref{ew1:Phi+0def}.

Both Nambu--Gorkov fields $\Phi^\prime$ and $\Phi$ are real fields, since their charge conjugates are just their linear transforms. For the primed field the linear relation between $\Phi^\prime$ and $\Phi^{\prime\C}$ has the \qm{canonical} form (see~\eqref{app:sclr:NGcond})
\begin{eqnarray}
\label{ew1:NGcondprime}
\Phi^{\prime\C} &=& \sigma_1 \, \Phi^\prime \,,
\end{eqnarray}
with the Pauli matrix $\sigma_1$ operating in the Nambu--Gorkov doublet space of $\Phi^\prime$ (i.e., rotating $\phi$ and $\phi^\C$). In the unprimed basis this condition translates as
\begin{eqnarray}
\label{ew1:NGcond}
\Phi^\C &=& \Sigma_1 \, \Phi \,,
\end{eqnarray}
where
\begin{subequations}
\label{ew1:Sigma1}
\begin{eqnarray}
\Sigma_1 &\equiv& U\,\sigma_1\,U
\\ &=& \left(\begin{array}{cccc}
0 & \unitmatrix & 0        & 0         \\
\unitmatrix & 0 & 0        & 0         \\
0 & 0 & \sigma_1 & 0         \\
0 & 0 & 0        & \sigma_1  \\
\end{array}\right) \,.
\end{eqnarray}
\end{subequations}
The reality conditions \eqref{ew1:NGcondprime} and \eqref{ew1:NGcond} have impact on the form of the corresponding propagators: The full propagators in both bases
\begin{eqnarray}
\I\,G_{\Phi^\prime} &=& \langle \Phi^\prime \Phi^{\prime\dag} \rangle \,, \\
\I\,G_{\Phi} &=& \langle \Phi \Phi^{\dag} \rangle
\end{eqnarray}
must non-trivially satisfy
\begin{eqnarray}
G_{\Phi^\prime} &=& \sigma_1 \, G_{\Phi^\prime}^{\T} \, \sigma_1 \,, \\
G_{\Phi} &=& \Sigma_1 \, G_{\Phi}^{\T} \, \Sigma_1 \,. \label{ew1:GPhiNGcond}
\end{eqnarray}

\subsubsection{Rewriting of the gauge interactions}

It will be later useful to know explicitly the $\group{SU}(2)_{\mathrm{L}} \times \group{U}(1)_{\mathrm{Y}}$ generators in the $\Phi$ basis. In order to find them we now rewrite the gauge interaction of the scalars from the original basis $S$, $N$ to the $\Phi$ one. In terms of $S$, $N$ the scalar gauge interactions read\footnote{The Lagrangian \eqref{ew1:el:gaugeSN} includes also the scalar kinetic terms, entering the Lagrangian \eqref{ew1:eL:sclr}.}
\begin{eqnarray}
\label{ew1:el:gaugeSN}
\eL_{\mathrm{scalar,gauge}} &=& \sum_{X=S,N}\big(\D_\mu X\big)^\dag \big(\D^\mu X\big) \,,
\end{eqnarray}
where the covariant derivatives read
\begin{eqnarray}
\D^\mu &=& \partial^\mu - \I g \frac{\sigma_a}{2} A_a^\mu - \I g^\prime \frac{Y_{X}}{2} A_4^\mu \,.
\end{eqnarray}
The weak hypercharge $Y$ is, in general, related to the electric charge $Q$ and the third component of the weak isospin $t_3$ of the corresponding $\group{SU}(2)_L \times \group{U}(1)_Y$ irreducible representation by the Gell-Mann--Nishijima formula
\begin{eqnarray}
\label{ew1:GellMannNishijima}
Y &=& 2(Q-t_3) \,.
\end{eqnarray}
I.e., in our case of the electroweak doublets $S$, $N$, \eqref{ew1:SN}, we have numerically
\begin{subequations}
\label{ew1:YSYN}
\begin{eqnarray}
Y_{S} &=& +1 \,, \\
Y_{N} &=& -1 \,.
\end{eqnarray}
\end{subequations}

We can write the Lagrangian \eqref{ew1:el:gaugeSN} more compactly in terms of the field $\phi$, \eqref{ew1:phi}, as
\begin{eqnarray}
\label{ew1:el:gaugephi}
\eL_{\mathrm{scalar,gauge}} &=& \big(\D_\mu \phi\big)^\dag \big(\D^\mu \phi\big) \,,
\end{eqnarray}
with the covariant derivative given by\footnote{In order not to overload the notation, we denote the covariant derivative always as $\D_\mu$, irrespective of which basis ($\phi$, $\Phi^\prime$ or $\Phi$) it is written in.}
\begin{eqnarray}
\D^\mu &=& \partial^\mu - \I T_{\phi,a} A_a^\mu \,.
\end{eqnarray}
This time already $a=1,\ldots,4$. The generators $T_{\phi,a}$ are defined with the gauge coupling constants $g$, $g^\prime$ deliberately included in:
\begin{subequations}
\label{ew1:Tphi}
\begin{eqnarray}
T_{\phi,a=1,2,3} &\equiv& \frac{1}{2} g        \left(\begin{array}{cc} \sigma_a & 0 \\ 0 & \sigma_a \end{array}\right) \,,  \\
T_{\phi,a=4}     &\equiv& \frac{1}{2} g^\prime \left(\begin{array}{cc} Y_S & 0 \\ 0 & Y_N \end{array}\right) \,.
\end{eqnarray}
\end{subequations}

We will now reparameterize the model in terms of the field $\Phi$ instead of the field $\phi$. In order to do so, it turns out to be useful to utilize the primed basis $\Phi^\prime$ as an intermediate stage, i.e., to parameterize the model first in terms of the primed field $\Phi^\prime$ and only then to move on to the parameterization in terms of the field $\Phi$.

In terms of the primed Nambu--Gorkov basis $\Phi^\prime$ the gauge interaction Lagrangian \eqref{ew1:el:gaugephi} recasts as
\begin{eqnarray}
\label{ew1:el:gaugePhiprime}
\eL_{\mathrm{scalar,gauge}} &=& \frac{1}{2}\big(\D_\mu \Phi^\prime\big)^\dag \big(\D^\mu \Phi^\prime\big) \,,
\end{eqnarray}
where this time
\begin{eqnarray}
\D^\mu &=& \partial^\mu - \I T_{\Phi^\prime\!,a} A_a^\mu \,.
\end{eqnarray}
The generators $T_{\Phi^\prime\!,a}$ in the $\Phi^\prime$ basis can be expressed in terms of those $T_{\phi,a}$ in the $\phi$ basis, \eqref{ew1:Tphi}, as
\begin{eqnarray}
\label{ew1:TPhiprime}
T_{\Phi^\prime\!,a} &=& \left(\begin{array}{cc} T_{\phi,a} & 0 \\ 0 & -T_{\phi,a}^\T \end{array}\right) \,.
\end{eqnarray}

Now we can rewrite the gauge interaction Lagrangian \eqref{ew1:el:gaugePhiprime} and the symmetry generators \eqref{ew1:TPhiprime} into the unprimed Nambu--Gorkov basis $\Phi$. The Lagrangian has again the same form
\begin{eqnarray}
\label{ew1:el:gaugePhi}
\eL_{\mathrm{scalar,gauge}} &=& \frac{1}{2}\big(\D_\mu \Phi\big)^\dag \big(\D^\mu \Phi\big) \,,
\end{eqnarray}
with
\begin{eqnarray}
\D^\mu &=& \partial^\mu - \I T_{\Phi,a} A_a^\mu \,.
\end{eqnarray}
Using the relation \eqref{ew1:Phiprime-Phi} between the bases $\Phi^\prime$ and $\Phi$, the generators $T_{\Phi,a}$ can be given in terms of $T_{\Phi^\prime\!,a}$, \eqref{ew1:TPhiprime}, as
\begin{eqnarray}
T_{\Phi,a} &=& U \, T_{\Phi^\prime\!,a} \, U \,.
\end{eqnarray}
The generators $T_{\Phi,a}$ ($8 \times 8$ matrices) can be now for $a=1,2$ written in terms of $4 \times 4$ blocks as
\begin{subequations}
\label{ew1:TPhi}
\begin{eqnarray}
T_{\Phi,a=1,2} &=&
\left(\begin{array}{cc}
0                  &  \mathcal{T}_a  \\
\mathcal{T}_a^\dag &              0
\end{array}\right)
\label{ew1:TPhi12}
\end{eqnarray}
and for $a=3,4$ in terms of $2 \times 2$ blocks as
\begin{eqnarray}
T_{\Phi,a=3} &=& \frac{1}{2}g
\left(\begin{array}{cccc}
\unitmatrix &  0           &         0 &        0 \\
0           & -\unitmatrix &         0 &        0 \\
0           &  0           & -\sigma_3 &        0 \\
0           &  0           &         0 & \sigma_3
\end{array}\right) \,,
\\
T_{\Phi,a=4} &=& \frac{1}{2}g^\prime
\left(\begin{array}{cccc}
\unitmatrix &  0           &        0 &         0 \\
0           & -\unitmatrix &        0 &         0 \\
0           &  0           & \sigma_3 &         0 \\
0           &  0           &        0 & -\sigma_3
\end{array}\right) \,.
\end{eqnarray}
\end{subequations}
The blocks $\mathcal{T}_a$ are given by
\begin{subequations}
\label{ew1:TPhimathcalT}
\begin{eqnarray}
\mathcal{T}_1 &=& \frac{1}{2}g
\left(\begin{array}{cccc}
1 &   0 &  0 &   0 \\
0 &   0 &  0 &  -1 \\
0 &  -1 &  0 &   0 \\
0 &   0 &  1 &   0
\end{array}\right)
\ = \ \frac{1}{4}g \left(\begin{array}{cc}
(1+\sigma_3) &  -(1-\sigma_3)  \\
-(\sigma_1+\I\sigma_2) & (\sigma_1-\I\sigma_2)
\end{array}\right)
\,,
\\
\mathcal{T}_2 &=& \frac{1}{2}g
\left(\begin{array}{cccc}
-\I &    0 &  0 &   0 \\
0   &    0 &  0 &  \I \\
0   &  -\I &  0 &   0 \\
0   &    0 & \I &   0
\end{array}\right)
\ = \ \frac{1}{4}g \left(\begin{array}{cc}
-\I(1+\sigma_3) &  \I(1-\sigma_3)  \\
-\I(\sigma_1+\I\sigma_2) & \I(\sigma_1-\I\sigma_2)
\end{array}\right)
\,,
\end{eqnarray}
\end{subequations}
and for practical calculation it is useful to note that they are related to each other by
\begin{eqnarray}
\label{ew1:TPhimathcalTrel}
\mathcal{T}_2 &=& -\I\sigma_3\mathcal{T}_1 \,,
\end{eqnarray}
with $\sigma_3$ operating in the space of the indicated blocks of $\mathcal{T}_1$, $\mathcal{T}_2$. Notice also that in expressing $T_{\Phi,4}$ we have already used the numerical values \eqref{ew1:YSYN} of the hypercharges $Y_S$, $Y_N$.

Since we will break spontaneously the $\group{SU}(2)_{\mathrm{L}} \times \group{U}(1)_{\mathrm{Y}}$ symmetry down to the \emph{non-trivial} subgroup $\group{U}(1)_{\mathrm{em}}$, rather than completely to the trivial group, it will be later on useful to know explicitly the generators corresponding to broken symmetries and the generator of the unbroken $\group{U}(1)_{\mathrm{em}}$. The generators $T_{\Phi,1}$, $T_{\Phi,2}$ already correspond to the fully broken symmetries, so it suffices to rotate the generators $T_{\Phi,3}$, $T_{\Phi,4}$ according to \eqref{ew1:rot:t34-Zew} in order to find the completely broken generator $T_{\Phi,Z}$ and the conserved generator $T_{\Phi,\mathrm{em}}$. The resulting generators can be, regarding later applications, written in the form
\begin{subequations}
\begin{eqnarray}
T_{\Phi,\mathrm{em}} &=&
\left(\begin{array}{cc} T_{\Phi^{(+)}\!,\mathrm{em}} & 0 \\ 0 & T_{\Phi^{(0)}\!,\mathrm{em}} \end{array}\right)
\,,
\label{ew1:TPhiem}
\\
T_{\Phi,Z} &=& \left(\begin{array}{cc} T_{\Phi^{(+)}\!,Z} & 0 \\ 0 & T_{\Phi^{(0)}\!,Z} \end{array}\right) \,,
\label{ew1:TPhiZ}
\end{eqnarray}
\end{subequations}
with the components of $T_{\Phi,\mathrm{em}}$ given by
\begin{subequations}
\begin{eqnarray}
T_{\Phi^{(+)}\!,\mathrm{em}} &=& \frac{g g^\prime}{\sqrt{g^2+g^{\prime2}}}
\left(\begin{array}{cc} \unitmatrix & 0 \\ 0 & -\unitmatrix \end{array}\right) \,,
\label{ew1:TPhi+em}
\\
T_{\Phi^{(0)}\!,\mathrm{em}} &=& 0
\end{eqnarray}
\end{subequations}
and the components of $T_{\Phi,Z}$ by
\begin{subequations}
\label{ew1:TPhi+0Z}
\begin{eqnarray}
T_{\Phi^{(+)}\!,Z} &=& \frac{1}{2} \frac{g^2-g^{\prime2}}{\sqrt{g^2+g^{\prime2}}}
\left(\begin{array}{cc} \unitmatrix & 0 \\ 0 & -\unitmatrix \end{array}\right) \,,
\label{ew1:TPhi+Z}
\\
T_{\Phi^{(0)}\!,Z} &=& \frac{1}{2}\sqrt{g^2+g^{\prime2}}
\left(\begin{array}{cc} -\sigma_3 & 0 \\ 0 & \sigma_3 \end{array}\right) \,,
\label{ew1:TPhi0Z}
\end{eqnarray}
\end{subequations}
respectively.

\subsection{Quarks}

\subsubsection{Reparameterization of the fields}

Recall the definition of the left-handed quark doublet in terms of the left-handed quark fields $u_L$ and $d_L$, \eqref{ew1:ellLqL}:
\begin{subequations}
\label{ew1:qchiraldoublets}
\begin{eqnarray}
q_L &=& \left(\begin{array}{c} u_L \\ d_L \end{array}\right) \,.
\end{eqnarray}
It is convenient to organize analogously the right-handed singlet fields $u_R$ and $d_R$, \eqref{ew1:fR}, into the right-handed quark doublet
\begin{eqnarray}
q_R &\equiv& \left(\begin{array}{c} u_R \\ d_R \end{array}\right) \,.
\end{eqnarray}
\end{subequations}
Even more compact notation can be, however, achieved by combining the chiral doublets \eqref{ew1:qchiraldoublets} in the obvious way as
\begin{eqnarray}
\label{ew1:q}
q &\equiv& q_L + q_R
\>=\> \left(\begin{array}{c} u_L + u_R \\ d_L + d_R \end{array}\right)
\>\equiv\> \left(\begin{array}{c} u \\ d \end{array}\right) \,.
\end{eqnarray}
The field $q$ (and occasionally also the fields $u$, $d$) will be the most convenient to use and we will therefore rewrite the relevant interactions in its terms.

\subsubsection{Rewriting of the gauge interactions}

We start the rewriting of the theory in terms of $q$ with the gauge interactions, aiming mainly at the form of the electroweak generators in the basis $q$. In terms of the left- and right-handed doublets $q_L$, $q_R$, \eqref{ew1:qchiraldoublets}, the gauge interactions can be written as
\begin{eqnarray}
\label{ew1:eL:gaugeqLqR}
\eL_{\mathrm{quark,gauge}} & = &
\bar q_L \Big(g \frac{1}{2}\sigma^a A^\mu_a + g^\prime \frac{1}{2} Y_q A^\mu_4\Big) \gamma_\mu q_L +
\bar q_R g^\prime \frac{1}{2}
\left(\begin{array}{cc}Y_u & 0 \\ 0 & Y_d \end{array}\right) A^\mu_4 \gamma_\mu q_R \,.
\end{eqnarray}
The hyper-charges $Y_q$, $Y_u$, $Y_d$ correspond to the left-handed doublet $q_L$ and to the right-handed singlets $u_R$, $d_R$, respectively. They are given in terms of the electric charges $Q_f$,
\begin{subequations}
\begin{eqnarray}
Q_u  &=& + \frac{2}{3} \,, \\
Q_d  &=& - \frac{1}{3} \,,
\end{eqnarray}
\end{subequations}
and in terms of the third components of the weak isospin $t_{3f}$,
\begin{subequations}
\label{ew1:t3ut3d}
\begin{eqnarray}
t_{3u} &=& +\frac{1}{2} \,, \\
t_{3d} &=& -\frac{1}{2} \,,
\end{eqnarray}
\end{subequations}
by the general formula \eqref{ew1:GellMannNishijima}. Specifically, we have
\begin{subequations}
\label{ew1:GMNq}
\begin{eqnarray}
Y_q &=& 2(Q_f-t_{3f}) \,, \\
Y_f &=& 2 Q_f \,,
\end{eqnarray}
\end{subequations}
so that the numerical values are
\begin{subequations}
\begin{eqnarray}
Y_q  &=& + \frac{1}{3} \,, \\
Y_u  &=& + \frac{4}{3} \,,\\
Y_d  &=& - \frac{2}{3} \,.
\end{eqnarray}
\end{subequations}

In the basis $q$ the gauge interaction Lagrangian \eqref{ew1:eL:gaugeqLqR} recasts as
\begin{eqnarray}
\eL &=& \bar q \gamma^\mu T_{q,a} q A^a_\mu \,,
\end{eqnarray}
where the generators are defined again with the gauge coupling constants included in as
\begin{subequations}
\label{ew1:Tqa}
\begin{eqnarray}
T_{q,a=1,2,3}             &=& \phantom{-}g        \frac{\sigma_a}{2} P_L \,,
\label{ew1:Tq123} \\
T_{q,a=4\hphantom{,2,3}}  &=& \phantom{-}g^\prime \frac{Y_q}{2}      P_L
+ g^\prime\frac{1}{2} \left(\begin{array}{cc} Y_u & 0 \\ 0 & Y_d \end{array}\right) P_R \\
&=& -g^\prime \frac{\sigma_3}{2} P_L + g^\prime \left(\begin{array}{cc} Q_u & 0 \\ 0 & Q_d \end{array}\right) \,.
\label{ew1:Tq4druhy}
\end{eqnarray}
\end{subequations}
In expressing the generator $T_{q,4}$ in the form \eqref{ew1:Tq4druhy} we have used the relations \eqref{ew1:GMNq} and the explicit values \eqref{ew1:t3ut3d} of $t_{3f}$.

Again, it is useful to know the form of the generator $T_{q,\mathrm{em}}$, corresponding to the conserved $\group{U}(1)_{\mathrm{em}}$ subgroup, together with the orthogonal generator $T_{q,Z}$. Using the relations \eqref{ew1:rot:t34-Zew} we arrive at
\begin{subequations}
\begin{eqnarray}
T_{q,\mathrm{em}}  &=& \phantom{-}\frac{gg^\prime}{\sqrt{g^2+g^{\prime2}}}
\left(\begin{array}{cc} Q_u & 0 \\ 0 & Q_d \end{array}\right) \,,
\label{ew1:Tqem}
\\
T_{q,Z}  &=& -\frac{g^{\prime2}}{\sqrt{g^2+g^{\prime2}}}
\left(\begin{array}{cc} Q_u & 0 \\ 0 & Q_d \end{array}\right)
+\sqrt{g^2+g^{\prime2}} \frac{\sigma_3}{2} P_L \,.
\label{ew1:TqZ}
\end{eqnarray}
\end{subequations}

Just for the sake of later references we also mark explicitly the block-diagonal form of the generators $T_{q,3}$, $T_{q,4}$:
\begin{eqnarray}
\label{ew1:Tq34diag}
T_{q,a=3,4} &=& \left(\begin{array}{cc} T_{u,a} & 0 \\ 0 & T_{d,a} \end{array}\right) \,,
\end{eqnarray}
where
\begin{subequations}
\label{ew1:qTf34}
\begin{eqnarray}
T_{f,3} &\equiv& \phantom{-}g \, t_{3f} P_L \,, \\
T_{f,4} &\equiv& \phantom{-}g^\prime \frac{1}{2} \big( Y_q P_L + Y_f P_R \big) \\
&=& - g^\prime t_{3f} P_L + g^\prime Q_f \,,
\end{eqnarray}
\end{subequations}
and naturally also of the generators $T_{q,\mathrm{em}}$, $T_{q,Z}$:
\begin{subequations}
\begin{eqnarray}
T_{q,\mathrm{em}} &=& \left(\begin{array}{cc} T_{u,\mathrm{em}} & 0 \\ 0 & T_{d,\mathrm{em}} \end{array}\right) \,, \\
T_{q,Z}  &=& \left(\begin{array}{cc} T_{u,Z}  & 0 \\ 0 & T_{d,Z}  \end{array}\right) \,,
\end{eqnarray}
\end{subequations}
where
\begin{subequations}
\label{ew1:qTfemZ}
\begin{eqnarray}
T_{f,\mathrm{em}} &\equiv& \phantom{-}\frac{gg^\prime}{\sqrt{g^2+g^{\prime2}}} Q_f \,,
\label{ew1:qTfem}
\\
T_{f,Z}  &\equiv& -\frac{g^{\prime2}}{\sqrt{g^2+g^{\prime2}}} Q_f + \sqrt{g^2+g^{\prime2}} \, t_{3f} P_L \,.
\label{ew1:qTfZ}
\end{eqnarray}
\end{subequations}

\subsubsection{Rewriting of the Yukawa interactions}

Now when we have, in addition to the scalar part, reparameterized also the quark part of the theory, we can finally rewrite the Yukawa interactions \eqref{ew1:eL:Ykw:q} in terms of the new degrees of freedom $\Phi$ and $q$:
\begin{subequations}
\label{ew1:eL:Ykw:Phiq}
\begin{eqnarray}
\eL_{\mathrm{Yukawa},q} &=& \bar q \, \bar Y_q  \, q \, \Phi
\\ &=& \Phi^\dag \, \bar q \, Y_q \, q \,.
\end{eqnarray}
\end{subequations}
Here the coupling constant $Y_q$ is a complicated rectangular matrix, incorporating all of the particular Yukawa coupling constants from the original Lagrangian \eqref{ew1:eL:Ykw:q}, together with the chiral projectors $P_{L,R}$. It is a column with eight entries, corresponding to eight entries of the scalar field $\Phi$. Each of the eight entries is a $2 \times 2$ matrix in the space of the quark doublet $q$. Having in mind the expression of $\Phi$ in terms of the fields \eqref{ew1:PhiSN-S-N}, the coupling constant $Y_q$ can be expressed in the block form as
\begin{eqnarray}
Y_q &=& \left(\begin{array}{c} Y_{q,SN} \\ \bar Y_{q,SN}^{\T_\Phi} \\ Y_{q,S} \\ Y_{q,N} \end{array}\right) \,.
\end{eqnarray}
The block $Y_{q,SN}$ is explicitly given as
\begin{subequations}
\label{ew1:Yq}
\begin{align}
\hspace{2cm}
Y_{q,SN} & \ =\ \left(\begin{array}{c}
\left(\begin{array}{cc} 0 & 0 \\ y_d^\dag P_L - \tilde y_u P_R & 0 \end{array}\right)  \\
\\
\left(\begin{array}{cc} 0 & 0 \\ y_u P_R + \tilde y_d^\dag P_L & 0 \end{array}\right)
\end{array}\right) \,,
\intertext{so that $\bar Y_{q,SN}^{\T_\Phi}$, with $^{\T_\Phi}$ denoting the transpose only in the eight-dimensional space of the scalar field $\Phi$, has the form}
\bar Y_{q,SN}^{\T_\Phi} & \ =\ \left(\begin{array}{c}
\left(\begin{array}{cc} 0 & y_d P_R - \tilde y_u^\dag P_L \\ 0 & 0 \end{array}\right)  \\
\\
\left(\begin{array}{cc} 0 & y_u^\dag P_L + \tilde y_d P_R \\ 0 & 0 \end{array}\right)
\end{array}\right) \,.
\intertext{The blocks $Y_{q,S}$ and $Y_{q,N}$ read}
Y_{q,S} & \ =\ \left(\begin{array}{c}
\left(\begin{array}{cc} \tilde y_u P_R & 0 \\ 0 & y_d^\dag P_L \end{array}\right)  \\
\\
\left(\begin{array}{cc} \tilde y_u^\dag P_L & 0 \\ 0 & y_d P_R \end{array}\right)
\end{array}\right) \,,
\\ \nonumber && \\
Y_{q,N} & \ =\ \left(\begin{array}{c}
\left(\begin{array}{cc} y_u^\dag P_L & 0 \\ 0 & - \tilde y_d P_R \end{array}\right)  \\
\\
\left(\begin{array}{cc} y_u P_R & 0 \\ 0 & - \tilde y_d^\dag P_L \end{array}\right)
\end{array}\right)
\end{align}
\end{subequations}
and one can observe that they satisfy
\begin{subequations}
\begin{eqnarray}
Y_{q,S} &=& \sigma_1 \, \bar Y_{q,S}^{\T_\Phi} \,, \\
Y_{q,N} &=& \sigma_1 \, \bar Y_{q,N}^{\T_\Phi} \,.
\end{eqnarray}
\end{subequations}
Consequently, the whole $Y_q$ satisfies
\begin{eqnarray}
\label{ew1:Yqcond}
Y_q &=& \Sigma_1 \, \bar Y_q^{\T_\Phi} \,.
\end{eqnarray}
Notice that we have included in the expression \eqref{ew1:Yq} of $Y_q$, just for completeness, also the \qm{tilded} Yukawa coupling constants $\tilde y_u$ and $\tilde y_d$, corresponding to the interactions \eqref{ew1:eL:tldYkw:q}, forbidden by the discrete symmetry \eqref{ew1:Pdown}. In fact, we assume of course $\tilde y_u = \tilde y_d = 0$.

\subsection{Leptons}

\subsubsection{Reparameterization of the fields}

Likewise in the case of quarks, we have two $n$-plets $\nu_L$ and $e_L$ of the left-handed fields, organized into the $\group{SU}(2)_L$ doublet \eqref{ew1:ellLqL},
\begin{subequations}
\label{ew1:ellLellR}
\begin{eqnarray}
\label{ew1:ellL}
\ell_L & = & \left(\begin{array}{c} \nu_L \\ e_L \end{array}\right) \,,
\end{eqnarray}
and one $n$-plet $e_R$ of the right-handed charged lepton singlets. This time, in contrast to quarks, there are  $m$ (in general different from $n$) right-handed neutrino fields, organized into the $m$-plet $\nu_R$, \eqref{ew1:nuR}. Nevertheless, it is again convenient to introduce the right-handed lepton doublet field $\ell_R$:
\begin{eqnarray}
\label{ew1:ellR}
\ell_R & \equiv & \left(\begin{array}{c} \nu_R \\ e_R \end{array}\right) \,.
\end{eqnarray}
\end{subequations}

Since this time the dimensions of the two doublets $\ell_L$ and $\ell_R$ differ, we cannot sum them and define this way the doublet $\ell = \ell_L + \ell_R$, as we did with quarks (see Eq.~\eqref{ew1:q}). We can, however, organize them in the following way
\begin{eqnarray}
\label{ew1:Psiprime}
\Psi_{\ell}^\prime &\equiv& \left(\begin{array}{c} \ell_L + (\ell_L)^\C \\ \ell_R + (\ell_R)^\C \end{array}\right)
\>=\>
\left(\begin{array}{c}
\nu_L + (\nu_L)^\C \\
e_L + (e_L)^\C \\
\nu_R + (\nu_R)^\C \\
e_R + (e_R)^\C \\
\end{array}\right) \,.
\end{eqnarray}
We call it a Nambu--Gorkov field, since it is in a sense the same object as the scalar Nambu--Gorkov field $\Phi$. In particular, it is also a real or Majorana field, since it satisfies the Majorana condition:
\begin{eqnarray}
\label{ew1:majcondprime}
\Psi_{\ell}^{\prime\C} &=& \Psi_{\ell}^{\prime} \,.
\end{eqnarray}
The properties of the fermion Majorana field \eqref{ew1:Psiprime} are discussed in more detail in appendix~\ref{app:fermi propag}, Sec.~\ref{app:frm:Maj}.

The main advantage of the Nambu--Gorkov formalism \eqref{ew1:Psiprime} and actually the reason why we use it (apart from $m \neq n$, which could be after all overcome in a simpler way than by defining $\Psi_{\ell}^\prime$) is that its matrix propagator naturally contains the Majorana propagators, i.e., the propagators of the type $\langle \psi\bar\psi^\C \rangle$, $\langle \psi^\C\bar\psi \rangle$. Such propagators can be clearly in principle generated for the neutrinos without breaking the sacred electromagnetic invariance.

However, we are now at the same situation as before with scalars, or more precisely, with the field $\Phi^\prime$. The prime at $\Psi_{\ell}^\prime$ indicates that this basis is not the most convenient one. The reason for that is again the same: In $\Psi_{\ell}^\prime$ the neutral and charged components are mixed. There is a better basis $\Psi_\ell$ of the Nambu--Gorkov doublet, which can be constructed as follows: First we define the Nambu--Gorkov doublets separately for both types of leptons:
\begin{subequations}
\label{ew1:PsinuPsie}
\begin{eqnarray}
\Psi_\nu &\equiv&  \left(\begin{array}{c} \nu_L + (\nu_L)^\C \\ \nu_R + (\nu_R)^\C \end{array}\right) \,,
\label{ew1:Psinu}
\\ \nonumber && \\
\Psi_e &\equiv&  \left(\begin{array}{c} e_L + (e_L)^\C \\ e_R + (e_R)^\C \end{array}\right) \,,
\label{ew1:Psie}
\end{eqnarray}
\end{subequations}
and then we make out of them the doublet $\Psi_\ell$:
\begin{eqnarray}
\label{ew1:Psi}
\Psi_{\ell} &\equiv& \left(\begin{array}{c} \Psi_{\nu} \\ \Psi_{e} \end{array}\right)
\>=\>
\left(\begin{array}{c}
\nu_L + (\nu_L)^\C \\
\nu_R + (\nu_R)^\C \\
e_L + (e_L)^\C \\
e_R + (e_R)^\C \\
\end{array}\right) \,.
\end{eqnarray}
Notice that this field is again Majorana:
\begin{eqnarray}
\label{ew1:majcond}
\Psi_{\ell}^{\C} &=& \Psi_{\ell} \,.
\end{eqnarray}
The Nambu--Gorkov doublet in this basis is more convenient, because has the same natural form as the electroweak doublet \eqref{ew1:ellLellR} and consequently certain quantities (e.g., the propagators and the generators) will have more natural and familiar block forms.

Notice that the fields $\Psi_{\ell}$ and $\Psi_{\ell}^\prime$ are related to each other by a simple linear transformation:
\begin{eqnarray}
\label{ew1:PsiUPriprime}
\Psi_{\ell} &=& U \Psi_{\ell}^\prime \,,
\end{eqnarray}
where
\begin{eqnarray}
U &\equiv& \left(\begin{array}{cccc}
\unitmatrix_{n \times n} & 0                        & 0                        & 0                        \\
0                        & 0                        & \unitmatrix_{m \times m} & 0                        \\
0                        & \unitmatrix_{n \times n} & 0                        & 0                        \\
0                        & 0                        & 0                        & \unitmatrix_{n \times n}
\end{array}\right) \,.
\end{eqnarray}
The matrix $U$ satisfies $U^\dag U = U^\T U = 1$.

To complete the present discussion of the Nambu--Gorkov formalism, let us state the consequence of the Majorana character of the field $\Psi_\ell$ for its propagator
\begin{eqnarray}
\I\,G_{\Psi_\ell} &=& \langle \Psi_\ell \bar \Psi_\ell \rangle \,.
\end{eqnarray}
The Majorana condition \eqref{ew1:majcond} implies the following symmetry of the propagator:
\begin{eqnarray}
\label{ew1:majcondG}
G_{\Psi_\ell}(p) &=& G_{\Psi_\ell}^\C(-p) \,,
\end{eqnarray}
where $G_{\Psi_\ell}^\C$ (generally, a \qm{charge conjugation} of matrix in Dirac space) is defined in \eqref{symbols:AC}. (The same relation as \eqref{ew1:majcondG} holds also for the propagators of the primed Nambu--Gorkov field $\Psi_\ell^\prime$, as it satisfies the same Majorana condition \eqref{ew1:majcondprime}.) More details are to be found in appendix~\ref{app:fermi propag}.

\subsubsection{Rewriting of the gauge interactions}

We will now rewrite the gauge interaction into the basis $\Psi_\ell$. Likewise in the case of scalars, we will use the basis $\Psi_\ell^\prime$ as a convenient intermediate step, using the results from appendix~\ref{app:fermi propag}. In terms of the left- and right-handed doublets \eqref{ew1:ellLellR} the gauge interaction can be written as
\begin{subequations}
\label{ew1:eL:gaugeellLellR}
\begin{eqnarray}
\label{ew1:eL:gaugeellLellRnorm}
\eL_{\mathrm{lepton,gauge}} &=&
\bar \ell_L \Big(g \frac{1}{2}\sigma^a A^\mu_a + g^\prime \frac{1}{2} Y_\ell A^\mu_4\Big) \gamma_\mu \ell_L +
\bar \ell_R g^\prime \frac{1}{2}
\left(\begin{array}{cc}Y_\nu & 0 \\ 0 & Y_e \end{array}\right) A^\mu_4 \gamma_\mu \ell_R
\,,
\end{eqnarray}
or more compactly as
\begin{eqnarray}
\label{ew1:eL:gaugeellLellRcomp}
\eL_{\mathrm{lepton,gauge}} &=& \bar\ell_L \gamma_\mu T_{{\ell_L},a} \ell_L A^\mu_a + \bar\ell_R \gamma_\mu T_{{\ell_R},a} \ell_R A^\mu_a \,,
\end{eqnarray}
\end{subequations}
with $T_{{\ell_L},a}$, $T_{{\ell_R},a}$ defined for $a=1,2,3$ as
\begin{subequations}
\label{ew1:TellLTellR}
\begin{eqnarray}
T_{{\ell_L},a=1,2,3} &\equiv& g \frac{1}{2}\sigma^a \,, \\
T_{{\ell_R},a=1,2,3} &\equiv& 0
\end{eqnarray}
and for $a=4$ as
\begin{eqnarray}
T_{{\ell_L},a=4} &\equiv& g^\prime \frac{1}{2} Y_\ell \,, \\
T_{{\ell_R,}a=4} &\equiv& g^\prime \frac{1}{2} \left(\begin{array}{cc}Y_\nu & 0 \\ 0 & Y_e \end{array}\right) \,.
\end{eqnarray}
\end{subequations}
The hyper-charges $Y_\ell$, $Y_\nu$, $Y_e$, corresponding to $\ell_L$, $\nu_R$ and $e_R$, respectively, are again related to the corresponding electric charges $Q_f$,
\begin{subequations}
\begin{eqnarray}
Q_\nu &=& \phantom{-} 0 \,, \\
Q_e   &=&          -  1 \,,
\end{eqnarray}
\end{subequations}
and to the third components of the isospin $t_{3f}$,
\begin{subequations}
\begin{eqnarray}
t_{3\nu} &=& +\frac{1}{2} \,, \\
t_{3e}   &=& -\frac{1}{2} \,,
\end{eqnarray}
\end{subequations}
by the formula \eqref{ew1:GellMannNishijima}, i.e.,
\begin{subequations}
\begin{eqnarray}
Y_\ell &=& 2(Q_f-t_{3f}) \,, \\
Y_f    &=& 2 Q_f \,.
\end{eqnarray}
\end{subequations}
The numerical values are
\begin{subequations}
\begin{eqnarray}
Y_\ell  &=& -1 \,, \\
Y_\nu   &=& \hphantom{-} 0 \,, \\
Y_e     &=& -2 \,.
\end{eqnarray}
\end{subequations}

As shown in appendix~\ref{app:fermi propag}, the Lagrangian \eqref{ew1:eL:gaugeellLellRcomp} acquires in the Nambu--Gorkov basis $\Psi_\ell^\prime$ the form
\begin{eqnarray}
\label{ew1:eL:gaugePsiprime}
\eL_{\mathrm{lepton,gauge}} &=&
\frac{1}{2} \bar\Psi_\ell^\prime \gamma_\mu T_{\Psi_\ell^\prime,a} \Psi_\ell^\prime A_a^\mu \,,
\end{eqnarray}
with the generators $T_{\Psi_\ell^\prime,a}$ expressed in terms of the original generators $T_{{\ell_L},a}$, $T_{{\ell_R},a}$, \eqref{ew1:TellLTellR}, as
\begin{eqnarray}
T_{\Psi_\ell^\prime,a} &=&
\left( \begin{array}{cc}
T_{{\ell_L},a} P_L - T_{{\ell_L},a}^\T P_R & 0                                          \\
0                                          & T_{{\ell_R},a} P_R - T_{{\ell_R},a}^\T P_L
\end{array}\right) \,.
\end{eqnarray}
(Cf.~Eq.~\eqref{app:frm:tPsi-tpsiLtpsiR}.) For transition from $\Psi_\ell^\prime$ to $\Psi_\ell$ we can now use the relation \eqref{ew1:PsiUPriprime}. In the basis $\Psi_\ell$ the Lagrangian \eqref{ew1:eL:gaugePsiprime} recasts as
\begin{eqnarray}
\eL_{\mathrm{lepton,gauge}} &=& \frac{1}{2} \bar\Psi_\ell \gamma_\mu T_{\Psi_\ell,a} \Psi_\ell A_a^\mu \,,
\end{eqnarray}
with the generators $T_{\Psi_\ell,a}$ given in terms of the generators $T_{\Psi_\ell^\prime,a}$ simply as
\begin{eqnarray}
T_{\Psi_\ell,a} &=& U \, T_{\Psi_\ell^\prime,a} \, U^\dag \,.
\end{eqnarray}
Explicitly we obtain:
\begin{subequations}
\label{ew1:TPsi}
\begin{eqnarray}
T_{\Psi_\ell,1} &=& -\gamma_5 \, g \frac{1}{2}
\left(\begin{array}{cc} 0 & P_+^\T \\ P_+ & 0 \end{array}\right) \,,
\label{ew1:Tpsi1}
\\
T_{\Psi_\ell,2} &=& \hphantom{-\gamma_5} \, g \frac{1}{2}
\left(\begin{array}{cc} 0 & -\I P_+^\T \\ \I P_+ & 0 \end{array}\right) \,,
\label{ew1:Tpsi2}
\\
T_{\Psi_\ell,3} &=& -\gamma_5 \, g \frac{1}{2}
\left(\begin{array}{cc} P_{+\nu} & 0 \\ 0 & -P_{+e} \end{array}\right) \,,
\\
T_{\Psi_\ell,4} &=& -\gamma_5 \, g^\prime \frac{1}{2}
\left(\begin{array}{cc} Y_\ell P_{+\nu} & 0 \\ 0 & Y_\ell P_{+e} \end{array}\right)
+ \gamma_5 \, g^\prime \frac{1}{2}
\left(\begin{array}{cc} Y_\nu P_{-\nu} & 0 \\ 0 & Y_e P_{-e} \end{array}\right)
\\ &=&
\hphantom{-} \gamma_5 \, g^\prime \frac{1}{2} \left(\begin{array}{cc} P_{+\nu} & 0 \\ 0 & -P_{+e} \end{array}\right)
- \gamma_5 g^\prime \left(\begin{array}{cc} Q_\nu\sigma_{3\nu} & 0 \\ 0 & Q_e\sigma_{3e} \end{array}\right) \,.
\end{eqnarray}
\end{subequations}

The three matrices $P_+$, $P_{+\nu}$, $P_{+e}$ in \eqref{ew1:TPsi} differ only in their dimensions: While the first is rectangular, the other two are square with different dimensions:
\begin{subequations}
\begin{eqnarray}
P_{+}    &\equiv& \left(\begin{array}{cc} \unitmatrix_{n \times n} & 0 \\ 0 & \zeromatrix_{n \times m} \end{array}\right) \,,
\label{ew1:P+} \\
P_{+\nu} &\equiv& \left(\begin{array}{cc} \unitmatrix_{n \times n} & 0 \\ 0 & \zeromatrix_{m \times m} \end{array}\right) \,,
\label{ew1:P+nu} \\
P_{+e}   &\equiv& \left(\begin{array}{cc} \unitmatrix_{n \times n} & 0 \\ 0 & \zeromatrix_{n \times n} \end{array}\right) \,.
\label{ew1:P+e}
\end{eqnarray}
\end{subequations}
Note that the three matrices coincide in the special case $m=n$. In practical calculations it is also useful to note that they are related by the formul{\ae}
\begin{subequations}
\begin{eqnarray}
P_{+}^{\phantom{\T}}P_{+}^\T   &=& P_{+e} \,, \\
P_{+}^\T P_{+}^{\vphantom{\T}} &=& P_{+\nu} \,.
\end{eqnarray}
\end{subequations}
Analogously, one can also define the matrices $P_{-\nu}$, $P_{-e}$ as
\begin{subequations}
\begin{eqnarray}
P_{-\nu} &\equiv& \left(\begin{array}{cc} \zeromatrix_{n \times n} & 0 \\ 0 & \unitmatrix_{m \times m} \end{array}\right) \,, \\
P_{-e}   &\equiv& \left(\begin{array}{cc} \zeromatrix_{n \times n} & 0 \\ 0 & \unitmatrix_{n \times n} \end{array}\right) \,.
\end{eqnarray}
\end{subequations}
Notice that each pair $P_{+f}$, $P_{-f}$ forms a complete set of projectors on the two-dimensional Nambu--Gorkov space of each particular $\Psi_{\!f}$, \eqref{ew1:PsinuPsie}, $f=\nu,e$:
\begin{subequations}
\begin{eqnarray}
P_{\pm f} \, P_{\pm f} &=& P_{\pm f} \,, \\
P_{\pm f} \, P_{\mp f} &=& 0 \,, \label{ew1:PpmPmpkrat}\\
P_{\pm f} + P_{\mp f}  &=& 1 \,, \label{ew1:PpmPmpplus}
\end{eqnarray}
\end{subequations}
with the right-hand sides of \eqref{ew1:PpmPmpkrat}, \eqref{ew1:PpmPmpplus} being square matrices of dimensions $n+m$ and $2n$ for the neutrinos and charged leptons, respectively. We can also \qm{generalize} the Pauli matrix $\sigma_3$:
\begin{eqnarray}
\sigma_{3f} &\equiv& P_{+f} - P_{-f} \,,
\end{eqnarray}
i.e.,
\begin{subequations}
\begin{eqnarray}
\sigma_{3\nu} &=& \left(\begin{array}{cc} \unitmatrix_{n \times n} & 0  \\ 0 & -\unitmatrix_{m \times m} \end{array}\right) \,, \\
\sigma_{3e}   &=& \left(\begin{array}{cc} \unitmatrix_{n \times n} & 0  \\ 0 & -\unitmatrix_{n \times n} \end{array}\right) \,.
\end{eqnarray}
\end{subequations}
The generators $T_{\Psi_\ell,\mathrm{em}}$ and $T_{\Psi_\ell,Z}$ are again given by the formul{\ae} \eqref{ew1:rot:t34-Zew} and explicitly come out as
\begin{subequations}
\begin{eqnarray}
T_{\Psi_\ell,\mathrm{em}} &=&
- \frac{gg^\prime}{\sqrt{g^2+g^{\prime2}}} \, \gamma_5
\left(\begin{array}{cc} Q_\nu\,\sigma_{3\nu} & 0 \\ 0 & Q_e\,\sigma_{3e} \end{array}\right) \,,
\label{ew1:TPsiellem}
\\
T_{\Psi_\ell,Z} &=& \phantom{-}
\frac{g^{\prime2}}{\sqrt{g^2+g^{\prime2}}} \, \gamma_5
\left(\begin{array}{cc} Q_\nu\,\sigma_{3\nu} & 0 \\ 0 & Q_e\,\sigma_{3e} \end{array}\right)
-\frac{1}{2} \sqrt{g^2+g^{\prime2}} \, \gamma_5
\left(\begin{array}{cc} P_{+\nu} & 0 \\ 0 & -P_{+e} \end{array}\right) \,.
\qquad\qquad
\label{ew1:TPsiellZ}
\end{eqnarray}
\end{subequations}


Like in the case of quarks, we again, for the sake of later references, mark explicitly the block-diagonal form of the generators $T_{\Psi_\ell,3}$, $T_{\Psi_\ell,4}$:
\begin{eqnarray}
\label{ew1:TPsiell34diag}
T_{\Psi_\ell,a=3,4} &=& \left(\begin{array}{cc} T_{\Psi_\nu,a} & 0 \\ 0 & T_{\Psi_e,a} \end{array}\right) \,,
\end{eqnarray}
where
\begin{subequations}
\label{ew1:TPsif34}
\begin{eqnarray}
T_{\Psi_{\!f},3} &\equiv& -g\,\gamma_5\,t_{3f}\,P_{+f} \,, \\
T_{\Psi_{\!f},4} &\equiv& -\gamma_5 \, g^\prime \frac{1}{2} Y_\ell P_{+f} + \gamma_5 \, g^\prime \frac{1}{2} Y_\nu P_{-f}
\\
&=& \phantom{-} g^\prime\gamma_5\,t_{3f}\,P_{+f} - g^\prime\gamma_5\,Q_f\,\sigma_{3f} \,,
\end{eqnarray}
\end{subequations}
and analogously of the generators $T_{\Psi_\ell,\mathrm{em}}$, $T_{\Psi_\ell,Z}$:
\begin{subequations}
\begin{eqnarray}
T_{\Psi_\ell,\mathrm{em}} &=&
\left(\begin{array}{cc} T_{\Psi_\nu,\mathrm{em}} & 0 \\ 0 & T_{\Psi_e,\mathrm{em}} \end{array}\right) \,, \\
T_{\Psi_\ell,Z}  &=&
\left(\begin{array}{cc} T_{\Psi_\nu,Z}  & 0 \\ 0 & T_{\Psi_e,Z}  \end{array}\right) \,,
\end{eqnarray}
\end{subequations}
where
\begin{subequations}
\begin{eqnarray}
T_{\Psi_{\!f},\mathrm{em}} &\equiv&  - \frac{gg^\prime}{\sqrt{g^2+g^{\prime2}}} \gamma_5\,Q_f\,\sigma_{3f} \,,
\label{ew1:TPsifem}
\\
T_{\Psi_{\!f},Z}  &\equiv& \phantom{-} \frac{g^{\prime2}}{\sqrt{g^2+g^{\prime2}}} \gamma_5\,Q_f\,\sigma_{3f}
- \sqrt{g^2+g^{\prime2}}\,\gamma_5\,t_{3f} P_{+f} \,.
\label{ew1:TPsifZ}
\end{eqnarray}
\end{subequations}

Consider now the charged lepton generators $T_{\Psi_e,a}$ with $a=3,4$ or $a=\mathrm{em},Z$. These generators operate on the space of the Nambu--Gorkov doublet $\Psi_e$, which is made of the left-handed and the right-handed charged lepton fields $e_L$ and $e_R$, respectively. Since the number of components of both $e_L$ and $e_R$ is the same (i.e., $n$), the fields $e_L$, $e_R$ can be represented, apart from the Nambu--Gorkov doublet $\Psi_e$, also by the field
\begin{eqnarray}
\label{ew1:e}
e &\equiv& e_L + e_R \,,
\end{eqnarray}
just like the quarks (cf.~formulae \eqref{ew1:q}). It is useful to know the generators $T_{e,a}$ (with $a=3,4$ or $a=\mathrm{em},Z$), i.e., the generators $T_{\Psi_e,a}$ rewritten in the basis $e$. For this we can make use of the result \eqref{app:frm:tpsi-tPsi} (and generally the results from Sec.~\ref{app:frm:Rel} of appendix~\ref{app:fermi propag}), stating that
\begin{eqnarray}
T_{e,a} &=& \Big(P_L,P_R\Big) \, T_{\Psi_e,a} \left(\begin{array}{c} P_L \\ P_R \end{array}\right)
\,, \qquad a=3,4 \mbox{ or } a=\mathrm{em},Z \,.
\end{eqnarray}
Not surprisingly, the resulting generators $T_{e,a}$ are of the same form as the quark generators $T_{f,a}$, $f=u,d$, see Eqs.~\eqref{ew1:qTf34} and \eqref{ew1:qTfemZ}. Only for the sake of later reference, let us state the generators $T_{e,a}$ explicitly. We have
\begin{subequations}
\begin{eqnarray}
T_{e,3} &=& \phantom{-}g \, t_{3e} P_L \,, \\
T_{e,4} &=& \phantom{-}g^\prime \frac{1}{2} \big( Y_\ell P_L + Y_e P_R \big) \\
&=& - g^\prime t_{3e} P_L + g^\prime Q_e
\end{eqnarray}
\end{subequations}
for the $a=3,4$ basis and
\begin{subequations}
\begin{eqnarray}
T_{e,\mathrm{em}} &=& \phantom{-}\frac{gg^\prime}{\sqrt{g^2+g^{\prime2}}} Q_e \,,
\label{ew1:Teem} \\
T_{e,Z}  &=& -\frac{g^{\prime2}}{\sqrt{g^2+g^{\prime2}}} Q_e + \sqrt{g^2+g^{\prime2}} \, t_{3e} P_L
\end{eqnarray}
\end{subequations}
for the $a=\mathrm{em},Z$ basis.

\subsubsection{Rewriting the lepton number symmetry}

The lepton number symmetry $\group{U}(1)_\ell$, \eqref{ew1:leptonNum}, is in the Nambu--Gorkov basis $\Psi_\ell$ translated as
\begin{eqnarray}
\label{ew1:leptonNumNG}
\group{U}(1)_\ell\,:\qquad \Psi_\ell  &\TransformsTo&
{[\Psi_\ell]}^\prime \>=\> \e^{\I T_{\Psi_\ell} \theta} \, \Psi_\ell \,.
\end{eqnarray}
We denote the generator $T_{\Psi_\ell}$ for the sake of later references as
\begin{eqnarray}
\label{ew1:TPsiell}
T_{\Psi_\ell} &=& \left(\begin{array}{cc} T_{\Psi_\nu} & 0 \\ 0 & T_{\Psi_\nu} \end{array}\right) \,,
\end{eqnarray}
where $T_{\Psi_\nu}$ and $T_{\Psi_e}$ are of course the same
\begin{subequations}
\label{ew1:TPsif}
\begin{eqnarray}
T_{\Psi_\nu} &=& -\gamma_5 \, \sigma_3 \, Q_\ell \,, \\
T_{\Psi_e}   &=& -\gamma_5 \, \sigma_3 \, Q_\ell \,.
\end{eqnarray}
\end{subequations}

\subsubsection{Rewriting of the Yukawa interactions}

The lepton part \eqref{ew1:eL:Ykw:ell} of the Yukawa interactions \eqref{eq:Yukawa} is rewritten in terms of $\Phi$ and $\Psi_\ell$ as
\begin{subequations}
\label{ew1:eL:Ykw:PhiPsiell}
\begin{eqnarray}
\eL_{\mathrm{Yukawa},\ell} &=& \frac{1}{2}\bar \Psi_\ell \, \bar Y_{\Psi_\ell} \, \Psi_\ell \, \Phi
\\ &=& \frac{1}{2}\Phi^\dag \, \bar \Psi_\ell \, Y_\ell \, \Psi_\ell \,,
\end{eqnarray}
\end{subequations}
with the coupling constant $Y_{\Psi_\ell}$ being again, similarly as the quark coupling constant $Y_q$, an $8$-plet, being contracted in \eqref{ew1:eL:Ykw:PhiPsiell} with the $8$-plet $\Phi$. In contrast to quarks, however, the entries of $Y_{\Psi_\ell}$ are this time not $2 \times 2$, but rather $4 \times 4$ matrices in the space of the field $\Psi_\ell$. We can write $Y_{\Psi_\ell}$ as
\begin{eqnarray}
Y_{\Psi_\ell} &=&
\left(\begin{array}{c} Y_{{\Psi_\ell},SN} \\ \bar Y_{{\Psi_\ell},SN}^{\T_\Phi} \\ Y_{{\Psi_\ell},S} \\ Y_{{\Psi_\ell},N} \end{array}\right) \,.
\end{eqnarray}
The block $Y_{{\Psi_\ell},SN}$ is given explicitly as
\begin{subequations}
\label{ew1:Yell}
\begin{align}
\hspace{1cm}
Y_{{\Psi_\ell},SN} & \ =\
\left(\begin{array}{c}
\left(\begin{array}{cc|cc}
0            & 0                 & 0                    & y_e^* P_L \\
0            & 0                 & -\tilde y_\nu^\T P_R & 0         \\ \hline
0            & -\tilde y_\nu P_R & 0                    & 0         \\
y_e^\dag P_L & 0                 & 0                    & 0
\end{array}\right)
\\ \\
\left(\begin{array}{cc|cc}
0                   & 0         & 0            & \tilde y_e^* P_L \\
0                   & 0         & y_\nu^\T P_R & 0                \\ \hline
0                   & y_\nu P_R & 0            & 0                \\
\tilde y_e^\dag P_L & 0         & 0            & 0
\end{array}\right)
\end{array}\right) \,,
\intertext{so that $Y_{{\Psi_\ell},SN}^{\T_\Phi}$ reads}
Y_{{\Psi_\ell},SN}^{\T_\Phi} & \ =\
\left(\begin{array}{c}
\left(\begin{array}{cc|cc}
0           & 0                   & 0                      & y_e P_R \\
0           & 0                   & -\tilde y_\nu^\dag P_L & 0       \\ \hline
0           & -\tilde y_\nu^* P_L & 0                      & 0       \\
y_e^\T P_R  & 0                   & 0                      & 0
\end{array}\right)
\\ \\
\left(\begin{array}{cc|cc}
0                 & 0           & 0              & \tilde y_e P_R \\
0                 & 0           & y_\nu^\dag P_L & 0              \\ \hline
0                 & y_\nu^* P_L & 0              & 0              \\
\tilde y_e^\T P_R & 0           & 0              & 0
\end{array}\right)
\end{array}\right) \,.
\intertext{The remaining blocks are given as}
Y_{{\Psi_\ell},S} & \ =\
\left(\begin{array}{c}
\left(\begin{array}{cc|cc}
0                   & \tilde y_\nu P_R & 0            & 0         \\
\tilde y_\nu^\T P_R & 0                & 0            & 0         \\ \hline
0                   & 0                & 0            & y_e^* P_L \\
0                   & 0                & y_e^\dag P_L & 0
\end{array}\right)
\\ \\
\left(\begin{array}{cc|cc}
0                     & \tilde y_\nu^* P_L & 0          & 0       \\
\tilde y_\nu^\dag P_L & 0                  & 0          & 0       \\ \hline
0                     & 0                  & 0          & y_e P_R \\
0                     & 0                  & y_e^\T P_R & 0
\end{array}\right)
\end{array}\right) \,,
\\ \nonumber && \\
Y_{{\Psi_\ell},N} & \ =\
\left(\begin{array}{c}
\left(\begin{array}{cc|cc}
0                   &  y_\nu^* P_L & 0            & 0         \\
y_\nu^\dag P_L & 0                & 0            & 0         \\ \hline
0                   & 0                & 0            & -\tilde y_e P_R \\
0                   & 0                & -\tilde y_e^\T P_R & 0
\end{array}\right)
\\ \\
\left(\begin{array}{cc|cc}
0                   &  y_\nu P_R & 0            & 0         \\
y_\nu^\T P_R & 0                & 0            & 0         \\ \hline
0                   & 0                & 0            & -\tilde y_e^* P_L \\
0                   & 0                & -\tilde y_e^\dag P_L & 0
\end{array}\right)
\end{array}\right)
\end{align}
\end{subequations}
and since they satisfy
\begin{subequations}
\begin{eqnarray}
Y_{{\Psi_\ell},S} &=& \sigma_1 \, \bar Y_{{\Psi_\ell},S}^{\T_\Phi} \,, \\
Y_{{\Psi_\ell},N} &=& \sigma_1 \, \bar Y_{{\Psi_\ell},N}^{\T_\Phi} \,,
\end{eqnarray}
\end{subequations}
the whole $Y_{\Psi_\ell}$ satisfies, similarly to the quark case, the relation
\begin{eqnarray}
\label{ew1:Yellcond}
Y_{\Psi_\ell} &=& \Sigma_1 \, \bar Y_{\Psi_\ell}^{\T_\Phi} \,.
\end{eqnarray}
In addition, there is also the relation
\begin{eqnarray}
\label{ew1:YellcondMaj}
Y_{\Psi_\ell} &=& Y_{\Psi_\ell}^{\C\T_\Phi} \,,
\end{eqnarray}
which is a consequence of the Majorana nature of $\Psi_\ell$. Therefore there is no analogue of this relation for the quarks. For the sake of completeness we have again included in \eqref{ew1:Yell} also the coupling constants $\tilde y_\nu$, $\tilde y_e$ from \eqref{ew1:eL:tldYkw:ell}, which we actually assume to be vanishing due to the symmetry \eqref{ew1:Pdown}: $\tilde y_\nu = 0$, $\tilde y_e = 0$.

\section{Summary}

We have considered an $\group{SU}(2)_{\mathrm{L}} \times \group{U}(1)_{\mathrm{Y}}$ gauge theory equipped with $n$ generations of the SM fermions (i.e., the left-handed quark and lepton doublets $q_L$ and $\ell_L$ and the right-handed quark and charged lepton singlets $u_R$, $d_R$ and $e_R$). We have enhanced this theory with $m$ right-handed neutrino singlets $\nu_R$, allowing for the gauge-invariant Majorana mass term \eqref{ew1:eL:MnuR}.

Moreover, we introduced two scalar doublets $S$ and $N$ with opposite hypercharges $\pm 1$. The bare masses squared of these scalars are assumed to be positive, in contrast to the usual Higgs scalar doublet. Also, again in contrast to the SM, we neglected the scalar self-interactions. However, the Yukawa interactions (though in the form somewhat constrained by the imposed discrete symmetry $\mathcal{P}_{\mathrm{down}}$, \eqref{ew1:Pdown}) were kept, as they will be of vital importance for the quest of spontaneous symmetry breaking in chapter~\ref{chp:ewdyn}.

Most of the chapter was dedicated to the reparameterization of the theory in terms of the new degrees of freedom. Namely, instead of using the scalar ($S$, $N$), quark ($q_L$, $u_R$, $d_R$) and lepton ($\ell_L$, $\nu_R$, $e_R$) irreducible representations of $\group{SU}(2)_{\mathrm{L}} \times \group{U}(1)_{\mathrm{Y}}$, we introduced new fields $\Phi$, $q$ and $\ell$, in terms of which we have rewritten the gauge interactions (i.e., basically the symmetry generators) and the Yukawa interactions.

\chapter{Ans\"{a}tze for propagators}
\label{chp:ewa}

\intro{Our strategy of demonstrating the SSB in the next chapter will be to find its manifestation in the sector of fermion and scalar propagators. In other words, we will look for symmetry-breaking parts of those propagators. For that purpose it will be sufficient to probe only a subset of all possible propagators, i.e., to restrict to some Ansatz for the propagators. This chapter is dedicated to finding such Ansatz.}


\section{Strategy}

In constructing the Ans\"{a}tze for propagators we will above all make sure carefully that it will not break the sacred electromagnetic invariance. Apart from this rather obligatory requirement we will follow also two optional criteria, whose aim is rather to simplify the calculations as much as possible while keeping present the most essential physical properties of our pattern of the SSB:

First, we will consider only those self-energies that break the symmetry and will neglect the symmetry-preserving self-energies. We can do this, since we wish only to demonstrate the viability of the SSB and do not pretend to make any phenomenological predictions. Moreover, from the technical point of view, it will be convenient to consider only the symmetry-breaking self-energies since by general arguments we know that they must be UV-finite.

Second, we will also neglect those self-energies (though symmetry-breaking) that renormalize the wave function. This is because we concentrate here mainly on the effects of the SSB on the particle spectrum. The renormalization of the kinetic terms, though finite, is therefore not of much interest from our adopted point of view. Nevertheless, we will, just for curiosity, separately write down explicitly the SD equations for such self-energies and show that they really come out finite, as they should.

The finial, rather minor guiding principle for determining the Ansatz will be anticipating the Hartree--Fock approximation of the SD equations, to be introduced only in the next chapter~\ref{chp:ewdyn}. This will in fact apply only to the scalars. It will turn out that some of the scalar self-energies, even though symmetry-breaking and not renormalizing the kinetic terms, will be vanishing in the one-loop, Hartree--Fock approximation. We will therefore set them to zero from the very beginning, just in order to make the intermediate formul{\ae} as simple and tractable as possible.

\section{Scalars}

\subsection{Notation for propagators}


The scalar self-energy $\boldsymbol{\Pi}_{\Phi}$ is defined as the difference
\begin{eqnarray}
\label{ewa:Pidef}
\boldsymbol{\Pi}_{\Phi} &\equiv& D_\Phi^{-1} - G_\Phi^{-1}
\end{eqnarray}
between the inverse free propagator $D_\Phi$ and the inverse full propagator $G_\Phi$,
\begin{eqnarray}
\I\,D_\Phi &=& \langle \Phi \Phi^\dag \rangle_{0} \,, \\
\I\,G_\Phi &=& \langle \Phi \Phi^\dag \rangle \,.
\end{eqnarray}
Since the original scalar doublets $S$, $N$ have hard masses $M_S$, $M_N$, respectively, the free propagator $D_\Phi$ is given by
\begin{eqnarray}
\label{ewa:DPhi}
D_\Phi &=& \diag(D_S^0,D_N^0,D_S^0,D_N^0,D_S^0,D_S^0,D_N^0,D_N^0) \,,
\end{eqnarray}
where
\begin{eqnarray}
\label{ewq:Dscalbare}
D_S^0 \ \equiv \ \frac{1}{p^2-M_S^2} \,, &\quad&
D_N^0 \ \equiv \ \frac{1}{p^2-M_N^2} \,.
\end{eqnarray}
We will now construct an appropriate Ansatz for $\boldsymbol{\Pi}_{\Phi}$ by following the philosophy outlined above.

\subsection{General form of the self-energy}
\label{ewa:ssec:Phigen}

First, recall that since the Nambu--Gorkov field $\Phi$ satisfies the condition \eqref{ew1:NGcond}, there is the non-trivial condition \eqref{ew1:GPhiNGcond} for the full propagator $G_\Phi$. Since the free propagator \eqref{ewa:DPhi} satisfies this condition too (it must, as being just a special case of the full propagator in the case of no interactions), the self-energy \eqref{ewa:Pidef} must satisfy it as well:
\begin{eqnarray}
\label{ewa:PicondNG}
\boldsymbol{\Pi}_{\Phi}^{\phantom{\T}} &=& \Sigma_1\,\boldsymbol{\Pi}_{\Phi}^\T\,\Sigma_1 \,,
\end{eqnarray}
with $\Sigma_1$ given by \eqref{ew1:Sigma1}. This is the first, most basic requirement on $\boldsymbol{\Pi}_{\Phi}$.

Further, we demand that the $\group{U}(1)_{\mathrm{em}}$ is preserved by the scalar self-energy $\boldsymbol{\Pi}_{\Phi}$. That is to say, we demand that\footnote{See discussion of the quantities \eqref{chp:frm:noninv}, measuring the symmetry breaking.}
\begin{eqnarray}
\label{ewa:Picondem}
[\boldsymbol{\Pi}_{\Phi},T_{\Phi,\mathrm{em}}] &=& 0 \,,
\end{eqnarray}
with the electromagnetic generator $T_{\Phi,\mathrm{em}}$ given by \eqref{ew1:TPhiem}.

Moreover, we want rather for technical reasons the self-energy $\boldsymbol{\Pi}_{\Phi}$ to be Hermitian:
\begin{eqnarray}
\label{ewa:Picondhc}
\boldsymbol{\Pi}_{\Phi}^{\phantom{\dag}} &=& \boldsymbol{\Pi}_{\Phi}^\dag \,.
\end{eqnarray}
Apart from obviously convenient reduction of the number of independent parts of $\boldsymbol{\Pi}_{\Phi}$ this condition will later on ensure that the masses squared of the scalars bosons will be real.


The three conditions \eqref{ewa:PicondNG}, \eqref{ewa:Picondem}, \eqref{ewa:Picondhc} constrain the $\boldsymbol{\Pi}_{\Phi}$ to have the form
\begin{eqnarray}
\boldsymbol{\Pi}_{\Phi} &=&
\left(\begin{array}{cc} \boldsymbol{\Pi}_{\Phi^{(+)}} & 0 \\ 0 & \boldsymbol{\Pi}_{\Phi^{(0)}} \end{array}\right) \,,
\end{eqnarray}
where
\begin{subequations}
\begin{eqnarray}
\boldsymbol{\Pi}_{\Phi^{(+)}} &\equiv& \left(\begin{array}{cc} A & 0 \\ 0      & A^\T \end{array}\right) \,,
\label{ewa:Pi+intermed} \\
\boldsymbol{\Pi}_{\Phi^{(0)}} &\equiv& \left(\begin{array}{cc} C & E \\ E^\dag & D    \end{array}\right) \,,
\label{ewa:Pi0intermed}
\end{eqnarray}
\end{subequations}
with
\begin{subequations}
\label{ewa:ACDE}
\begin{eqnarray}
A &\equiv& \left(\begin{array}{cc} A_1 & A_2 \\ A_2^* & A_3 \end{array}\right) \,, \\
C &\equiv& \left(\begin{array}{cc} C_1 & C_2 \\ C_2^* & C_1 \end{array}\right) \,, \\
D &\equiv& \left(\begin{array}{cc} D_1 & D_2 \\ D_2^* & D_1 \end{array}\right) \,, \\
E &\equiv& \left(\begin{array}{cc} E_1 & E_2 \\ E_2^* & E_1^* \end{array}\right) \,. \label{ewa:E}
\end{eqnarray}
\end{subequations}
Here the numbers $A_1$, $A_3$, $C_1$, $D_1$ and $A_2$, $C_2$, $D_2$, $E_1$, $E_2$ are real and complex functions of $p^2$, respectively.

\subsection{Symmetry constraints}
\label{ewa:ssec:Phisym}

Not all of the functions $A_i$, $C_i$, $D_i$, $E_i$, however, break the symmetry. Some of them (or some linear combination(s) of them) may preserve it. Let us now check it.

\subsubsection{The $T_{\Phi,Z}$ generator}

Let us start with the symmetry associated with the generator $T_{\Phi,Z}$, \eqref{ew1:TPhiZ}. Its breaking induced by the scalar self-energy $\boldsymbol{\Pi}_{\Phi}$ is measured by the commutator $[\boldsymbol{\Pi}_{\Phi},T_{\Phi,Z}]$. Due to the block-diagonal structure of $T_{\Phi,Z}$ we have
\begin{eqnarray}
[\boldsymbol{\Pi}_{\Phi},T_{\Phi,Z}] &=&
\left(\begin{array}{cc}
[\boldsymbol{\Pi}_{\Phi^{(+)}},T_{\Phi^{(+)}\!,Z}] &        0              \\
0                     & [\boldsymbol{\Pi}_{\Phi^{(0)}},T_{\Phi^{(0)}\!,Z}]
\end{array}\right) \,,
\end{eqnarray}
with $T_{\Phi^{(+)}\!,Z}$, $T_{\Phi^{(0)}\!,Z}$ given by \eqref{ew1:TPhi+0Z}.



For the first commutator $[\boldsymbol{\Pi}_{\Phi^{(+)}},T_{\Phi^{(+)}\!,Z}]$ we have immediately
\begin{eqnarray}
[\boldsymbol{\Pi}_{\Phi^{(+)}},T_{\Phi^{(+)}\!,Z}]  &=& 0 \,.
\end{eqnarray}
Vanishing of $[\boldsymbol{\Pi}_{\Phi^{(+)}},T_{\Phi^{(+)}\!,Z}]$ is in fact due to the requirement of the electromagnetic invariance, since the generator $T_{\Phi^{(+)}\!,Z}$, \eqref{ew1:TPhi+Z}, is proportional to its electromagnetic counterpart $T_{\Phi^{(+)}\!,\mathrm{em}}$, \eqref{ew1:TPhi+em}.

More interesting is the second commutator $[\boldsymbol{\Pi}_{\Phi^{(0)}},T_{\Phi^{(0)}\!,Z}]$. Explicit calculation reveals
\begin{eqnarray}
\label{ewa:PiTZ}
[\boldsymbol{\Pi}_{\Phi^{(0)}},T_{\Phi^{(0)}\!,Z}]  &=& \frac{1}{2}\sqrt{g^2+g^{\prime2}}
\left(\begin{array}{cc}
-[C,\sigma_3]        & \{E,\sigma_3\} \\
-\{E^\dag,\sigma_3\} & [D,\sigma_3]
\end{array}\right) \,,
\end{eqnarray}
with the particular (anti)commutators
\begin{subequations}
\begin{eqnarray}
{[C,\sigma_3]}  &=& 2 \left(\begin{array}{cc} 0 & -C_2 \\ C_2^* & 0 \end{array}\right) \,, \\
{[D,\sigma_3]}  &=& 2 \left(\begin{array}{cc} 0 & -D_2 \\ D_2^* & 0 \end{array}\right) \,, \\
\{E,\sigma_3\}  &=& 2 \left(\begin{array}{cc} E_1 & 0 \\ 0 & E_1^* \end{array}\right) \,.
\label{ewa:Esigma3}
\end{eqnarray}
\end{subequations}
Therefore, we conclude that from the nine self-energies $A_1$, $A_3$, $C_1$, $D_1$, $A_2$, $C_2$, $D_2$, $E_1$, $E_2$ only the three $C_2$, $D_2$, $E_1$ break the symmetry associated with the generator $T_{\Phi,Z}$.

\subsubsection{The $T_{\Phi,1}$, $T_{\Phi,2}$ generators}

We can calculate similarly also the commutators of $\boldsymbol{\Pi}_{\Phi}$ with the generators $T_{\Phi,1}$, $T_{\Phi,2}$, \eqref{ew1:TPhi12}, and arrive at
\begin{eqnarray}
[\boldsymbol{\Pi}_{\Phi},T_{\Phi,a=1,2}] &=& \left(\begin{array}{cc} 0 & X_a \\ -X_a^\dag & 0 \end{array}\right) \,,
\end{eqnarray}
with $X_a$ given by
\begin{eqnarray}
X_a &=& \boldsymbol{\Pi}_{\Phi^{(+)}}\,\mathcal{T}_a - \mathcal{T}_a\,\boldsymbol{\Pi}_{\Phi^{(0)}} \,.
\end{eqnarray}
Noting the definitions \eqref{ew1:TPhimathcalT} of $\mathcal{T}_a$, we find
\begin{subequations}
\begin{align}
X_1
& \ =\
\frac{1}{2}g\left(\begin{array}{cc|cc}
(A_1-C_1)   & -C_2       & -E_1        & -(A_2+E_2) \\
(A_2+E_2)^* & E_1        & D_2^*       & -(A_3-D_1) \\ \hline
C_2^*       & -(A_1-C_1) & (A_2+E_2)^* & E_1^*      \\
-E_1^*      & -(A_2+E_2) & (A_3-D_1)   & -D_2       \\
\end{array}\right)
\\
\intertext{and}
X_2 & \ =\  -\I\sigma_3 X_1 \,.
\label{ewa:PhiX2}
\end{align}
\end{subequations}
In expression \eqref{ewa:PhiX2} for $X_2$ we have used the relation \eqref{ew1:TPhimathcalTrel}. Thus, we conclude that the self-energies $C_2$, $D_2$, $E_1$ separately break the invariance. Of the remaining six self-energies $A_1$, $A_3$, $C_1$, $D_1$, $A_2$, $E_2$ only the combinations
\begin{subequations}
\begin{eqnarray}
&A_1-C_1 \,, & \label{ewa:A1-C1} \\
&A_3-D_1 \,, & \label{ewa:A3-D1} \\
&A_2+E_2 \phantom{\,,}    & \label{ewa:A2+E2}
\end{eqnarray}
\end{subequations}
break the generators $T_{\Phi,1}$, $T_{\Phi,2}$, while the combinations
\begin{subequations}
\begin{eqnarray}
&A_1+C_1 \,, & \label{ewa:A1+C1} \\
&A_3+D_1 \,, & \label{ewa:A3+D1} \\
&A_2-E_2 \phantom{\,,}    & \label{ewa:A2-E2}
\end{eqnarray}
\end{subequations}
leave them invariant.

\subsubsection{The discrete $\mathcal{P}_{\mathrm{down}}$ symmetry}

Furthermore, recall that apart from the continuous symmetry $\group{SU}(2)_{\mathrm{L}} \times \group{U}(1)_{\mathrm{Y}}$ there is also the discrete symmetry $\mathcal{P}_{\mathrm{down}}$, \eqref{ew1:Pdown}, acting on the scalar doublets $S$, $N$ as
\begin{subequations}
\begin{eqnarray}
\mathcal{P}_{\mathrm{down}}\,:\,\qquad S &\TransformsTo& [S]^\prime \,\>=\> -S \,,\\
\mathcal{P}_{\mathrm{down}}\,:\qquad N &\TransformsTo& [N]^\prime \>=\> +N \,.
\end{eqnarray}
\end{subequations}
For the individual blocks $A$, $C$, $D$, $E$ of $\boldsymbol{\Pi}$ we have therefore
\begin{subequations}
\label{ewa:Pdowncomm}
\begin{eqnarray}
{[A,\mathcal{P}_{\mathrm{down}}]} &=& 2 \left(\begin{array}{cc} 0 & A_2 \\ A_2^* & 0 \end{array}\right) \,, \\
{[C,\mathcal{P}_{\mathrm{down}}]} &=& 0 \,, \\
{[D,\mathcal{P}_{\mathrm{down}}]} &=& 0 \,, \\
{[E,\mathcal{P}_{\mathrm{down}}]} &=& 2 E \,.
\end{eqnarray}
\end{subequations}
The \qm{commutators} in \eqref{ewa:Pdowncomm} are defined as
\begin{eqnarray}
\label{ewa:XPdown}
[X,\mathcal{P}_{\mathrm{down}}] &\equiv& X - [X]^\prime \,,
\end{eqnarray}
where $[X]^\prime$ is transformation of $X$ under $\mathcal{P}_{\mathrm{down}}$. Therefore we see, in particular, that the functions $A_2$ and $E_2$ do break the $\mathcal{P}_{\mathrm{down}}$ symmetry \emph{separately}, which is to be compared with the previous result that only the combination $A_2+E_2$ breaks the $\group{SU}(2)_{\mathrm{L}} \times \group{U}(1)_{\mathrm{Y}}$ symmetry, while the combination $A_2-E_2$ preserves it.

\subsubsection{Elimination of $E$}

Now, as we know which of the functions $A_i$, $C_i$, $D_i$, $E_i$, or their linear combinations do break the symmetries of the Lagrangian and which not, we can proceed to the construction of the Ansatz. We have seen that all functions, but the combinations $A_1+C_1$ and $A_3+D_1$, break at least a part of the full symmetry $\group{SU}(2)_{\mathrm{L}} \times \group{U}(1)_{\mathrm{Y}} \times \mathcal{P}_{\mathrm{down}}$. Since the symmetry-preserving combinations contain the perturbative and hence potentially UV-divergent contributions, we will not consider them in our Ansatz and set
\begin{subequations}
\begin{eqnarray}
A_1+C_1 &=& 0 \,, \\
A_3+D_1 &=& 0 \,.
\end{eqnarray}
\end{subequations}

Now all of the other functions, as being symmetry-breaking and hence UV-finite, should be in principle included into the Ansatz. However, for various reason we will neglect also some of these symmetry-breaking functions. First of all, we set $E=0$:
\begin{subequations}
\begin{eqnarray}
E_1 &=& 0 \,, \\
E_2 &=& 0 \,.
\end{eqnarray}
\end{subequations}
There are two reason for doing that. The first reason is rather pragmatic: If $E=0$, then the matrix $\boldsymbol{\Pi}_{\Phi^{(0)}}$, \eqref{ewa:Pi0intermed}, has a block-diagonal form with the non-vanishing blocks $C$, $D$ being $2 \times 2$ matrices. Recall that the self-energy $\boldsymbol{\Pi}_{\Phi^{(+)}}$, \eqref{ewa:Pi+intermed}, has already the same block-diagonal form too. Now since the free propagator $D_\Phi$, \eqref{ewa:DPhi}, is a diagonal matrix, the full propagator $G_\Phi$ has consequently the same block structure as the self-energy $\boldsymbol{\Pi}_{\Phi}$, i.e., it consists of four $2 \times 2$ blocks on the diagonal. The point is that each of these blocks is calculated from the self-energy by taking inverse of a $2 \times 2$ matrix, which is of course much easier than taking inverse of a $4 \times 4$ matrix (which would be inevitable if $E \neq 0$).

Second reason for setting $E=0$ actually anticipates what we will discuss only in the next chapter. Likewise in chapter~\ref{chp:frm} on the Abelian toy model, we will also here study the dynamics using the SD equations, derived from the CJT effective potential. And also likewise in the Abelian toy model, we will approximate the CJT effective potential by the single one-loop diagram, i.e., we will use the Hartree--Fock approximation. However, as we will show explicitly later, it turns out that in this approximation the SD equations for $E$ are vanishing, and consequently in one loop we have indeed $E=0$. Only at two loops there would be a non-vanishing contribution to $E$.

\subsection{Refining the notation}



Now when we have set $E=0$, it is convenient to change slightly our denotations. We rename the non-vanishing self-energy blocks as
\begin{subequations}
\begin{eqnarray}
A &\equiv& \boldsymbol{\Pi}_{\Phi_{SN}} \,, \\
C &\equiv& \boldsymbol{\Pi}_{\Phi_{S}} \,, \\
D &\equiv& \boldsymbol{\Pi}_{\Phi_{N}} \,,
\end{eqnarray}
\end{subequations}
so that the self-energies $\boldsymbol{\Pi}_{\Phi^{(+)}}$ and $\boldsymbol{\Pi}_{\Phi^{(0)}}$ now read
\begin{subequations}
\begin{eqnarray}
\boldsymbol{\Pi}_{\Phi^{(+)}} &=&
\left(\begin{array}{cc} \boldsymbol{\Pi}_{\Phi_{SN}} & 0 \\ 0 & \boldsymbol{\Pi}_{\Phi_{SN}}^\T \end{array}\right) \,, \\
\boldsymbol{\Pi}_{\Phi^{(0)}} &=&
\left(\begin{array}{cc} \boldsymbol{\Pi}_{\Phi_{S}} & 0 \\ 0 & \boldsymbol{\Pi}_{\Phi_{N}} \end{array}\right) \,.
\end{eqnarray}
\end{subequations}

The full propagator $G_\Phi$ has the form
\begin{eqnarray}
G_\Phi &=& \left(\begin{array}{cc} G_{\Phi^{(+)}} & 0 \\ 0 & G_{\Phi^{(0)}} \end{array}\right) \,,
\end{eqnarray}
with
\begin{subequations}
\begin{eqnarray}
G_{\Phi^{(+)}} &=& \left(\begin{array}{cc} G_{\Phi_{SN}} & 0 \\ 0 & G_{\Phi_{SN}}^\T \end{array}\right) \,, \\
G_{\Phi^{(0)}} &=& \left(\begin{array}{cc} G_{\Phi_{S}}  & 0 \\ 0 & G_{\Phi_{N}}     \end{array}\right) \,.
\label{ewa:G0blockdiag}
\end{eqnarray}
\end{subequations}
The particular propagators $G_{\Phi_{SN}}$, $G_{\Phi_{S}}$, $G_{\Phi_{N}}$ are given in terms of the self-energies $\boldsymbol{\Pi}_{\Phi_{SN}}$, $\boldsymbol{\Pi}_{\Phi_{S}}$, $\boldsymbol{\Pi}_{\Phi_{N}}$ as
\begin{subequations}
\label{ewa:GPsiSNSNintermed}
\begin{eqnarray}
G_{\Phi_{SN}} &=&
\left[
\left(\begin{array}{cc} p^2-M_S^2 & 0 \\ 0 & p^2-M_N^2 \end{array}\right) - \boldsymbol{\Pi}_{\Phi_{SN}}
\right]^{-1}\,,
\\
G_{\Phi_{S}} &=&
\left[
\left(\begin{array}{cc} p^2-M_S^2 & 0 \\ 0 & p^2-M_S^2 \end{array}\right) - \boldsymbol{\Pi}_{\Phi_{S}}
\right]^{-1}\,,
\\
G_{\Phi_{N}} &=&
\left[
\left(\begin{array}{cc} p^2-M_N^2 & 0 \\ 0 & p^2-M_N^2 \end{array}\right) - \boldsymbol{\Pi}_{\Phi_{N}}
\right]^{-1}\,.
\end{eqnarray}
\end{subequations}
Here we can see explicitly what was mentioned above: If the block $E$ was not assumed to be vanishing, the propagator $G^{(0)}$ could not be written in the convenient block-diagonal form \eqref{ewa:G0blockdiag}.

Also, in order to be in accordance with Ref.~\cite{Benes:2008ir}, we rename the functions $A_2$, $C_2$, $D_2$ as
\begin{subequations}
\label{ewa:PiPhiSNSN}
\begin{eqnarray}
A_2 &\equiv& \Pi_{SN} \,, \\
C_2 &\equiv& \Pi_{S} \,, \\
D_2 &\equiv& \Pi_{N} \,.
\end{eqnarray}
\end{subequations}
Furthermore, we denote the symmetry-breaking combinations \eqref{ewa:A1-C1}, \eqref{ewa:A3-D1} as
\begin{subequations}
\label{ewa:ASANdef}
\begin{eqnarray}
A_1-C_1 &\equiv& 2 A_S \,, \\
A_3-D_1 &\equiv& 2 A_N \,,
\end{eqnarray}
\end{subequations}
The self-energies $\boldsymbol{\Pi}_{\Phi_{SN}}$, $\boldsymbol{\Pi}_{\Phi_{S}}$, $\boldsymbol{\Pi}_{\Phi_{N}}$ therefore read
\begin{subequations}
\label{ewa:PiSNSNintermed}
\begin{eqnarray}
\boldsymbol{\Pi}_{\Phi_{SN}} &=&
\left(\begin{array}{cc} A_S & \Pi_{SN} \\ \Pi_{SN}^* & A_N \end{array}\right) \,,
\\
\boldsymbol{\Pi}_{\Phi_{S}} &=&
\left(\begin{array}{cc} -A_S & \Pi_{S} \\ \Pi_{S}^* & -A_S \end{array}\right) \,,
\\
\boldsymbol{\Pi}_{\Phi_{N}} &=&
\left(\begin{array}{cc} -A_N & \Pi_{N} \\ \Pi_{N}^* & -A_N \end{array}\right) \,.
\end{eqnarray}
\end{subequations}
Plugging these self-energies into the expressions \eqref{ewa:GPsiSNSNintermed} for the full propagators, we arrive at
\begin{subequations}
\label{ewa:GPsiSNSNintermed2}
\begin{eqnarray}
G_{\Phi_{SN}} &=& \frac{1}{(p^2-M_S^2-A_S)(p^2-M_N^2-A_N)-|\Pi_{SN}|^2}
\left(\begin{array}{cc} p^2-M_N^2-A_N & \Pi_{SN} \\ \Pi_{SN}^* & p^2-M_S^2-A_S \end{array}\right) \,,
\nonumber \\ &&
\\
G_{\Phi_{S}} &=& \frac{1}{(p^2-M_S^2+A_S)^2-|\Pi_{S}|^2}
\left(\begin{array}{cc} p^2-M_S^2+A_S & \Pi_{S} \\ \Pi_{S}^* & p^2-M_S^2+A_S \end{array}\right) \,,
\\
G_{\Phi_{N}} &=& \frac{1}{(p^2-M_N^2+A_N)^2-|\Pi_{N}|^2}
\left(\begin{array}{cc} p^2-M_N^2+A_N & \Pi_{N} \\ \Pi_{N}^* & p^2-M_N^2+A_N \end{array}\right) \,.
\end{eqnarray}
\end{subequations}

\subsection{Wave function renormalization self-energies}

The self-energy Ansatz \eqref{ewa:PiSNSNintermed}, with five symmetry-breaking functions $\Pi_{SN}$, $\Pi_{S}$, $\Pi_{N}$, $A_S$, $A_N$, is still quite complicated. One could ask whether it is possible to simplify it by neglecting some of the five functions, while keeping present the most significant features of the resulting scalar spectrum.

Consider the scalar spectrum, which is obtained as poles of the full propagator. From the three particular propagators \eqref{ewa:GPsiSNSNintermed2} we have altogether six pole equations:
\begin{subequations}
\label{ewa:PhiPoleeqsintermed}
\begin{eqnarray}
p^2 &=& \frac{M_S^2+A_S}{2}+\frac{M_N^2+A_N}{2} \pm \sqrt{\bigg(\frac{M_S^2+A_S}{2}-\frac{M_N^2+A_N}{2}\bigg)^2+|\Pi_{SN}|^2} \,,
\label{ewa:GPsiSNSNintermed2SN}
\\
p^2 &=& M_S^2-A_S \pm |\Pi_S| \,,
\label{ewa:GPsiSNSNintermed2S}
\\
p^2 &=& M_N^2-A_N \pm |\Pi_N| \,.
\label{ewa:GPsiSNSNintermed2N}
\end{eqnarray}
\end{subequations}
Recall that all the quantities $\Pi_{SN}$, $\Pi_{S}$, $\Pi_{N}$, $A_S$, $A_N$ are functions of $p^2$.

The first equation \eqref{ewa:GPsiSNSNintermed2SN} (which is actually two equations, thanks to the \qm{$\pm$} option) says that as a result of the SSB the original two charged fields $S^{(+)}$ and $N^{(-)\dag}$ with the respective masses $M_S$ and $M_N$ mix into two new charged fields with masses given by the pole equation(s) \eqref{ewa:GPsiSNSNintermed2SN}.

More interesting are the other two equations, \eqref{ewa:GPsiSNSNintermed2S} and \eqref{ewa:GPsiSNSNintermed2N}. For the sake of definiteness let us focus on the former one, \eqref{ewa:GPsiSNSNintermed2S}, as the latter one, \eqref{ewa:GPsiSNSNintermed2N}, is completely analogous. As discussed already on a similar example in chapter~\ref{chp:frm}, the equation \eqref{ewa:GPsiSNSNintermed2S} (comprising again actually two equations) describes mixing between the two \emph{complex} fields $S^{(0)}$ and $S^{(0)\dag}$ with the \emph{same} bare masses $M_S$, resulting into two new \emph{real} fields with \emph{different} masses. Clearly, this mass splitting is proportional to $\Pi_S$, as for the case $\Pi_S = 0$ the two equations \eqref{ewa:GPsiSNSNintermed2S} would coincide. If on the other hand $A_S = 0$, the mass splitting is still present. In fact, $A_S$ serves only as a finite renormalization of the bare mass $M_S$, with no impact on the interesting phenomenon of mass splitting. Therefore we will neglect in our Ansatz the function $A_S$, as well as on the basis of the same arguments also the function $A_N$:
\begin{subequations}
\begin{eqnarray}
A_S &=& 0 \,, \\
A_N &=& 0 \,.
\end{eqnarray}
\end{subequations}

\subsection{Final form of the Ansatz}
\label{ewa:final:Phi}

We can now state the final form of the Ansatz. The scalar self-energy $\boldsymbol{\Pi}_{\Phi}$ is given by
\begin{eqnarray}
\label{ewa:PiPhi}
\boldsymbol{\Pi}_{\Phi} &=&
\left(\begin{array}{cccc}
\boldsymbol{\Pi}_{\Phi_{SN}} & 0 & 0 & 0 \\
0 & \boldsymbol{\Pi}_{\Phi_{SN}}^\T & 0 & 0 \\
0 & 0 & \boldsymbol{\Pi}_{\Phi_{S}} & 0 \\
0&0&0& \boldsymbol{\Pi}_{\Phi_{N}}
\end{array}\right) \,,
\end{eqnarray}
where
\begin{subequations}
\label{ewa:PhiSNSN}
\begin{eqnarray}
\boldsymbol{\Pi}_{\Phi_{SN}} &=&
\left(\begin{array}{cc} 0 & \Pi_{SN} \\ \Pi_{SN}^* & 0 \end{array}\right) \,, \\
\boldsymbol{\Pi}_{\Phi_{S}} &=&
\left(\begin{array}{cc} 0 & \Pi_{S} \\ \Pi_{S}^* & 0 \end{array}\right) \,, \\
\boldsymbol{\Pi}_{\Phi_{N}} &=&
\left(\begin{array}{cc} 0 & \Pi_{N} \\ \Pi_{N}^* & 0 \end{array}\right) \,.
\end{eqnarray}
\end{subequations}
This corresponds to the full propagator $G_\Phi$ of the form
\begin{eqnarray}
\label{ewa:GPhi}
G_{\Phi} &=&
\left(\begin{array}{cccc}
G_{\Phi_{SN}} & 0 & 0 & 0 \\
0 & G_{\Phi_{SN}}^\T & 0 & 0 \\
0 & 0 & G_{\Phi_{S}} & 0 \\
0&0&0& G_{\Phi_{N}}
\end{array}\right) \,,
\end{eqnarray}
where the particular propagators are given by
\begin{subequations}
\label{ewa:GPhiSNSN}
\begin{eqnarray}
G_{\Phi_{SN}} &=& \frac{1}{(p^2-M_S^2)(p^2-M_N^2)-|\Pi_{SN}|^2}
\left(\begin{array}{cc} p^2-M_N^2 & \Pi_{SN} \\ \Pi_{SN}^* & p^2-M_S^2 \end{array}\right) \,,
\\
G_{\Phi_{S}} &=& \frac{1}{(p^2-M_S^2)^2-|\Pi_{S}|^2}
\left(\begin{array}{cc} p^2-M_S^2 & \Pi_{S} \\ \Pi_{S}^* & p^2-M_S^2 \end{array}\right) \,,
\\
G_{\Phi_{N}} &=& \frac{1}{(p^2-M_N^2)^2-|\Pi_{N}|^2}
\left(\begin{array}{cc} p^2-M_N^2 & \Pi_{N} \\ \Pi_{N}^* & p^2-M_N^2 \end{array}\right) \,.
\end{eqnarray}
\end{subequations}
The pole equations \eqref{ewa:PhiPoleeqsintermed} reduce to
\begin{subequations}
\label{ewa:PhiPoleeqs}
\begin{eqnarray}
p^2 &=& \frac{M_S^2+M_N^2}{2} \pm \sqrt{\bigg(\frac{M_S^2-M_N^2}{2}\bigg)^2+|\Pi_{SN}|^2} \,,
\\
p^2 &=& M_S^2 \pm |\Pi_S| \,,
\\
p^2 &=& M_N^2 \pm |\Pi_N| \,.
\end{eqnarray}
\end{subequations}
Finally, in order to make formul{\ae} more compact, it is convenient to introduce the notation
\begin{subequations}
\begin{eqnarray}
D_{SN} &\equiv& \frac{1}{(p^2-M_S^2)(p^2-M_N^2)-|\Pi_{SN}|^2} \,,
\\
D_{S} &\equiv& \frac{1}{(p^2-M_S^2)^2-|\Pi_{S}|^2} \,,
\\
D_{N} &\equiv& \frac{1}{(p^2-M_N^2)^2-|\Pi_{N}|^2}
\end{eqnarray}
\end{subequations}
for the fractions figuring in the expressions \eqref{ewa:GPhiSNSN} for the full propagators.


For the sake of later references, we state here explicitly the Feynman rules for the propagators. The Feynman rules for the self-energies \eqref{ewa:PiPhiSNSN} read
\begin{subequations}
\begin{eqnarray}
\langle S^{(0)} S^{(0)} \rangle_{\mathrm{1PI}}
\ = \
\begin{array}{c}
\scalebox{0.85}{\includegraphics[trim = 10bp 12bp 14bp 11bp,clip]{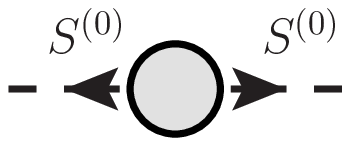}}
\end{array}
&=& -\I\,\Pi_S \,,
\\
\langle N^{(0)} N^{(0)} \rangle_{\mathrm{1PI}}
\ = \
\begin{array}{c}
\scalebox{0.85}{\includegraphics[trim = 10bp 12bp 14bp 11bp,clip]{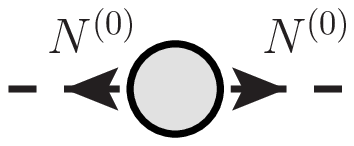}}
\end{array}
&=& -\I\,\Pi_N \,,
\\
\langle S^{(+)} N^{(-)} \rangle_{\mathrm{1PI}}
\ = \
\begin{array}{c}
\scalebox{0.85}{\includegraphics[trim = 10bp 12bp 14bp 11bp,clip]{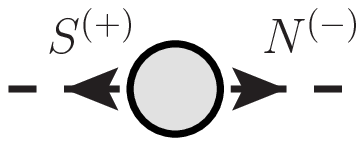}}
\end{array}
&=& -\I\,\Pi_{SN} \,.
\end{eqnarray}
\end{subequations}

The 1PI propagators with the opposite arrows differ from these only by complex conjugation of the corresponding function $\Pi_S$, $\Pi_N$, $\Pi_{SN}$, respectively.

The Feynman rules for the symmetry-breaking parts of the full propagators \eqref{ewa:GPhiSNSN} (the off-diagonal entries) read
\begin{subequations}
\begin{eqnarray}
\langle S^{(0)} S^{(0)} \rangle
\ = \
\begin{array}{c}
\scalebox{0.85}{\includegraphics[trim = 10bp 12bp 14bp 11bp,clip]{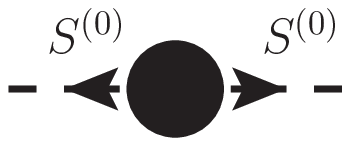}}
\end{array}
&=& \I\,\Pi_S\,D_S \,,
\\
\langle N^{(0)} N^{(0)} \rangle
\ = \
\begin{array}{c}
\scalebox{0.85}{\includegraphics[trim = 10bp 12bp 14bp 11bp,clip]{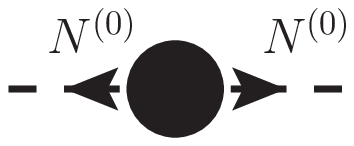}}
\end{array}
&=& \I\,\Pi_N\,D_N \,,
\\
\langle S^{(+)} N^{(-)} \rangle
\ = \
\begin{array}{c}
\scalebox{0.85}{\includegraphics[trim = 10bp 12bp 14bp 11bp,clip]{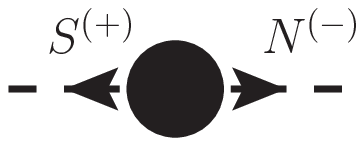}}
\end{array}
&=& \I\,\Pi_{SN}\,D_{SN} \,.
\end{eqnarray}
\end{subequations}
Again, the opposite arrows correspond to complex conjugation of the respective self-energy functions. The Feynman rules for the symmetry-preserving parts of the full propagators \eqref{ewa:GPhiSNSN} (the diagonal entries) are given by
\begin{subequations}
\begin{eqnarray}
\langle S^{(0)} S^{(0)\dag} \rangle
\ = \
\begin{array}{c}
\scalebox{0.85}{\includegraphics[trim = 10bp 12bp 14bp 11bp,clip]{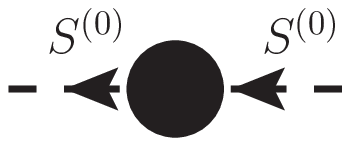}}
\end{array}
&=& \I\,(p^2-M_S^2)\,D_S \,,
\\
\langle N^{(0)} N^{(0)\dag} \rangle
\ = \
\begin{array}{c}
\scalebox{0.85}{\includegraphics[trim = 10bp 12bp 14bp 11bp,clip]{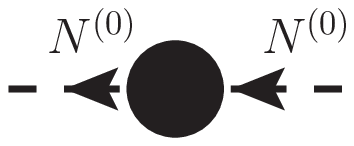}}
\end{array}
&=& \I\,(p^2-M_N^2)\,D_N \,,
\\
\langle S^{(+)} S^{(+)\dag} \rangle
\ = \
\begin{array}{c}
\scalebox{0.85}{\includegraphics[trim = 10bp 12bp 14bp 11bp,clip]{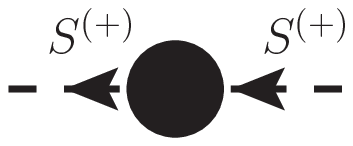}}
\end{array}
&=& \I\,(p^2-M_N^2)\,D_{SN} \,,
\label{ewa:S+S+full}
\\
\langle N^{(-)} N^{(-)\dag} \rangle
\ = \
\begin{array}{c}
\scalebox{0.85}{\includegraphics[trim = 10bp 12bp 14bp 11bp,clip]{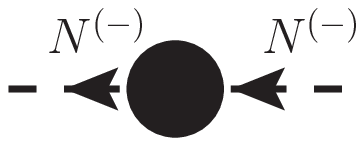}}
\end{array}
&=& \I\,(p^2-M_S^2)\,D_{SN} \,.
\end{eqnarray}
\end{subequations}

\section{Quarks}

Now we will construct an Ansatz for the quark self-energy. We will proceed basically in the same way as before with scalar: First we will demand the electromagnetic invariance together with the Hermiticity. Next we will investigate which parts of the self-energy break the symmetries of the model and which not, with the intention to keep in the Ansatz only those symmetry-breaking parts. And finally, we will argue that even some of the symmetry-breaking parts of the self-energy can be neglected without affecting the most significant impacts of the SSB on the particle spectrum, which will be this time the very generation of fermion masses (rather than the mass splitting in the scalar case).

\subsection{Notation for propagators}

The full scalar propagator $G_q$,
\begin{eqnarray}
\I\,G_q &=& \langle q \bar q \rangle \,,
\end{eqnarray}
is related to the free propagator $S_q$, whose inverse reads, due to the absence of quarks mass terms in the Lagrangian, simply
\begin{eqnarray}
\label{ewa:Sq}
S_q^{-1} &=& \slashed{p} \,,
\end{eqnarray}
by the self-energy $\boldsymbol{\Sigma}_{q}$:
\begin{eqnarray}
\label{ewa:Sgmqdef}
\boldsymbol{\Sigma}_{q} &\equiv& S_q^{-1} - G_q^{-1} \,.
\end{eqnarray}
We are now going to construct a suitable Ansatz for it.

\subsection{General form of the self-energy}
\label{ewa:ssec:qgen}

The requirement of the $\group{U}(1)_{\mathrm{em}}$ invariance for the quark self-energy $\boldsymbol{\Sigma}_{q}$ reads
\begin{eqnarray}
\label{ewa:qeminv}
\boldsymbol{\Sigma}_{q}\,T_{q,\mathrm{em}} - \bar T_{q,\mathrm{em}}\,\boldsymbol{\Sigma}_{q} &=& 0 \,.
\end{eqnarray}
Since the generator $T_{q,\mathrm{em}}$ has the block-diagonal form (see explicit form \eqref{ew1:Tqem} of $T_{q,\mathrm{em}}$), the self-energy $\boldsymbol{\Sigma}_{q}$ must have due to the condition \eqref{ewa:qeminv} a block-diagonal form too:
\begin{eqnarray}
\label{ewa:Sgmqdiag}
\boldsymbol{\Sigma}_{q} &=&
\left(\begin{array}{cc} \boldsymbol{\Sigma}_{u} & 0 \\ 0 & \boldsymbol{\Sigma}_{d} \end{array}\right) \,.
\end{eqnarray}
Here the self-energies $\boldsymbol{\Sigma}_{u}$, $\boldsymbol{\Sigma}_{d}$ are given in terms of the fields $u$, $d$, Eq.~\eqref{ew1:q}, as
\begin{subequations}
\begin{eqnarray}
-\I \, \boldsymbol{\Sigma}_{u} &=& \langle u \bar u \rangle_{\mathrm{1PI}} \,, \\
-\I \, \boldsymbol{\Sigma}_{d} &=& \langle d \bar d \rangle_{\mathrm{1PI}} \,.
\end{eqnarray}
\end{subequations}
Thus, the single condition \eqref{ewa:qeminv} of the electromagnetic invariance now decouples into two separate conditions
\begin{eqnarray}
\label{ewa:feminv}
\boldsymbol{\Sigma}_{f}\,T_{f,\mathrm{em}} - \bar T_{f,\mathrm{em}}\,\boldsymbol{\Sigma}_{f} &=& 0
\,, \quad\quad f=u,d \,.
\end{eqnarray}
However, since the particular generators $T_{f,\mathrm{em}}$, \eqref{ew1:qTfem}, are just pure real numbers, the conditions \eqref{ewa:feminv} are fulfilled automatically and the requirement of electromagnetic invariance gives us no further constraints on the particular quarks self-energies $\boldsymbol{\Sigma}_{u}$ and $\boldsymbol{\Sigma}_{d}$.

Further, we demand satisfaction of the condition
\begin{eqnarray}
\label{ewa:Sgmqhc}
\boldsymbol{\Sigma}_q &=& \boldsymbol{\bar\Sigma}_q
\end{eqnarray}
(recall that $\boldsymbol{\bar\Sigma}_q = \gamma_0\,\boldsymbol{\Sigma}_q^\dag\,\gamma_0$), which is just a direct analogue of the Hermiticity condition \eqref{ewa:Picondhc} for the scalar self-energy. Due to the block-diagonal form of \eqref{ewa:Sgmqdiag} of $\boldsymbol{\Sigma}_q$ the condition \eqref{ewa:Sgmqhc} implies $\boldsymbol{\Sigma}_f = \boldsymbol{\bar\Sigma}_f$ for both $f=u,d$. As a consequence we obtain the general form of both $\boldsymbol{\Sigma}_f$:
\begin{eqnarray}
\label{ewa:fSgmAnzintermed}
\boldsymbol{\Sigma}_f &=& \slashed{p}(A_{fL}\,P_L+A_{fR}\,P_R)+(\Sigma_f^\dag\,P_L + \Sigma_f\,P_R) \,,
\end{eqnarray}
where $A_{fL}$, $A_{fR}$, $\Sigma_f$ are $p^2$-dependent, complex $n \times n$ matrices and the matrices $A_{fL}$, $A_{fR}$ are moreover Hermitian. The Hermiticity condition \eqref{ewa:Sgmqhc} is technical in two senses: First, it reduces the number of independent parts of the quark self-energy, and second, it ensures that the resulting fermions spectrum will be real.\footnote{To be more precise, this is true only under the additional assumption, being made here implicitly, that the matrices $1-A_{fL}$ and $1-A_{fR}$ are positive definite. This is actually related to the positivity of the spectral function.}

\subsection{Symmetry constraints}

\subsubsection{The $T_{q,Z}$ and $T_{q,1}$, $T_{q,2}$ generators}

Let us now examine which parts of the quark self-energy do break the symmetry. For the symmetry associated with the generator $T_{q,Z}$ the relevant quantity is
\begin{eqnarray}
\label{ewa:qnoninvTqZ}
\boldsymbol{\Sigma}_q\,T_{q,Z} - \bar T_{q,Z}\,\boldsymbol{\Sigma}_q &=& -\frac{1}{2}\sqrt{g^2+g^{\prime2}}
\left(\begin{array}{cc} -\Sigma_u^\dag\,P_L+\Sigma_u\,P_R & 0 \\ 0 & \Sigma_d^\dag\,P_L-\Sigma_d\,P_R \end{array}\right) \,,
\end{eqnarray}
while for the generators $T_{q,1}$ and $T_{q,2}$ we have
\begin{subequations}
\label{ewa:qnoninvTq12}
\begin{eqnarray}
\boldsymbol{\Sigma}_q\,T_{q,1} - \bar T_{q,1}\,\boldsymbol{\Sigma}_q &=&
\nonumber \\ &&
\hspace{-2cm}
{}-\frac{1}{2}g \slashed{p}P_L (A_{uL}-A_{dL}) \left(\begin{array}{cc} 0 & -1 \\ 1 & 0 \end{array}\right)
{}-\frac{1}{2}g \left(\begin{array}{cc} 0 & \Sigma_d\,P_R - \Sigma_u^\dag\,P_L \\ \Sigma_u\,P_R - \Sigma_d^\dag\,P_L & 0 \end{array}\right) \,,
\nonumber \\ &&
\\
\boldsymbol{\Sigma}_q\,T_{q,2} - \bar T_{q,2}\,\boldsymbol{\Sigma}_q &=&
\nonumber \\ &&
\hspace{-2cm}
{}-\I\frac{1}{2}g \slashed{p}P_L (A_{uL}-A_{dL}) \left(\begin{array}{cc} 0 & 1 \\ 1 & 0 \end{array}\right)
{}-\I\frac{1}{2}g \left(\begin{array}{cc} 0 & -(\Sigma_d\,P_R - \Sigma_u^\dag\,P_L) \\ \Sigma_u\,P_R - \Sigma_d^\dag\,P_L & 0 \end{array}\right) \,.
\nonumber \\ &&
\end{eqnarray}
\end{subequations}

\subsubsection{The discrete $\mathcal{P}_{\mathrm{down}}$ symmetry}

Let us now check the invariance under the discrete symmetry $\mathcal{P}_{\mathrm{down}}$, \eqref{ew1:Pdown}. Its action quark fields can be written compactly as
\begin{subequations}
\label{ewa:Pdownq}
\begin{eqnarray}
\mathcal{P}_{\mathrm{down}}\,:\qquad u &\TransformsTo& [u]^\prime \>=\> \phantom{-\gamma_5} \, u \,,\\
\mathcal{P}_{\mathrm{down}}\,:\qquad d &\TransformsTo& [d]^\prime \>=\>          -\gamma_5  \, d \,.
\end{eqnarray}
\end{subequations}
It is clear that the up-type self-energy $\boldsymbol{\Sigma}_u$ stays intact under \eqref{ewa:Pdownq}:
\begin{subequations}
\begin{eqnarray}
{[\boldsymbol{\Sigma}_u,\mathcal{P}_{\mathrm{down}}]} &=& 0 \,.
\end{eqnarray}
(Recall the definition \eqref{ewa:XPdown} of this commutator.) On the other hand, the down-type self-energy $\boldsymbol{\Sigma}_d$, \eqref{ewa:fSgmAnzintermed}, does not commute with \eqref{ewa:Pdownq}, its chirality-changing part $\Sigma_d$ changes the sign under $\mathcal{P}_{\mathrm{down}}$, so that
\begin{eqnarray}
\label{ewa:Pdownd}
{[\boldsymbol{\Sigma}_d,\mathcal{P}_{\mathrm{down}}]} &=& 2(\Sigma_d^\dag\,P_L + \Sigma_d\,P_R) \,.
\end{eqnarray}
\end{subequations}
We conclude that only the $\Sigma_d$ is non-invariant under $\mathcal{P}_{\mathrm{down}}$. However, as we saw a moment ago (Eqs.~\eqref{ewa:qnoninvTqZ} and \eqref{ewa:qnoninvTq12}), $\Sigma_d$ was non-invariant also under the $\group{SU}(2)_{\mathrm{L}} \times \group{U}(1)_{\mathrm{Y}}$. In this sense the behavior under $\mathcal{P}_{\mathrm{down}}$ tells us nothing new concerning the (non-)invariance of the quark self-energy under the symmetries of the Lagrangian. This is in contrast with the scalar self-energy, where due to the discrete symmetry $\mathcal{P}_{\mathrm{down}}$ the self-energy $\Pi_{SN}$ is symmetry-breaking and thus capable of being a part of the scalar self-energy Ansatz.

\subsubsection{Purely symmetry-breaking self-energy}

We have probed the behavior of the quark self-energy under all symmetries of the Lagrangian and thus we can continue at the very construction of the Ansatz. We have seen that both $\Sigma_u$ and $\Sigma_d$ did break the symmetry and hence they have to be included into the Ansatz. On the other hand, we have also seen that out of the four functions $A_{uL}$, $A_{uR}$, $A_{dL}$, $A_{dR}$ only the combination $A_{uL}-A_{dL}$ did break the symmetry and hence should be included into the Ansatz too. We therefore denote
\begin{eqnarray}
\label{ewa:Aqdef}
A_{uL}-A_{dL} &\equiv& 2 A_q
\end{eqnarray}
and set
\begin{eqnarray}
\label{ewa:AuL+AdL}
A_{uL}+A_{dL} &=& 0 \,,
\end{eqnarray}
together with
\begin{subequations}
\label{ewa:AuRAdR}
\begin{eqnarray}
A_{uR} &=& 0 \,, \\
A_{dR} &=& 0 \,.
\end{eqnarray}
\end{subequations}
The most general, purely symmetry-breaking Ansatz therefore has the form
\begin{subequations}
\label{ewa:fSgmAnzintermed2}
\begin{eqnarray}
\boldsymbol{\Sigma}_u &=& \phantom{-}\slashed{p}A_q\,P_L+(\Sigma_u^\dag\,P_L + \Sigma_u\,P_R) \,, \\
\boldsymbol{\Sigma}_d &=&          - \slashed{p}A_q\,P_L+(\Sigma_d^\dag\,P_L + \Sigma_d\,P_R) \,.
\end{eqnarray}
\end{subequations}

\subsection{Wave function renormalization self-energies}

Likewise in the case of scalars, we are now going to argue that not all of the three symmetry-breaking functions $\Sigma_u$, $\Sigma_u$, $A_q$ in \eqref{ewa:fSgmAnzintermed2} are necessary in the quest for the phenomenon of the dynamical generation of fermion masses. These can be obtained as poles of the full propagators corresponding to the self-energies \eqref{ewa:fSgmAnzintermed2}. Explicitly the pole equations read
\begin{subequations}
\begin{eqnarray}
\det\big[p^2-\Sigma_u^\dag(1-A_q)^{-1}\Sigma_u\big] &=& 0 \,, \\
\det\big[p^2-\Sigma_d^\dag(1+A_q)^{-1}\Sigma_d\big] &=& 0 \,.
\end{eqnarray}
\end{subequations}
We immediately see that in order to have non-vanishing fermion masses we must have non-vanishing chirality changing parts of the propagators, i.e., the self-energies $\Sigma_u$, $\Sigma_d$. On the other hand, the self-energy $A_q$ is obviously not essential in this respect. Recall that we are primarily interested in the very demonstration of the generation of fermion masses, without an ambition to make the phenomenological predictions. For this purpose considering $A_q$ is redundant. Therefore we will neglect it in the Ansatz and set
\begin{eqnarray}
A_q &=& 0 \,,
\end{eqnarray}
which completes construction of the Ansatz.

\subsection{Final form of the Ansatz}
\label{ewa:final:q}

Let us summarize for the sake of later references various formul{\ae} concerning the final form of the quark self-energy Ansatz.

\subsubsection{Propagators}

The final form of the Ansatz reads
\begin{eqnarray}
\label{ewa:Sgmqdiag}
\boldsymbol{\Sigma}_q &=&
\left(\begin{array}{cc} \boldsymbol{\Sigma}_u & 0 \\ 0 & \boldsymbol{\Sigma}_d \end{array}\right) \,,
\end{eqnarray}
where
\begin{eqnarray}
\label{ewa:Sgmf}
\boldsymbol{\Sigma}_f &=& \Sigma_f^\dag\,P_L + \Sigma_f\,P_R \,.
\end{eqnarray}
Notice that the definition \eqref{ewa:Sgmf} is correct for $f$ standing both for $q$ and for $u$, $d$; we will use this convention in the rest of this section.

We can now write down the explicit form of the full quark propagator. Since both the self-energy $\boldsymbol{\Sigma}_q$ and the free propagator $S_q$ are diagonal in the space of the quark doublet, so must be the propagator $G_q$:
\begin{eqnarray}
\label{ewa:Gqdiag}
G_q &=& \left(\begin{array}{cc} G_u & 0 \\ 0 & G_d \end{array}\right) \,.
\end{eqnarray}
Using the relation \eqref{ewa:Sgmqdef} and the form of the Ansatz \eqref{ewa:Sgmqdiag}, we can express the propagators $G_f$,
\begin{eqnarray}
\label{ewa:Gfinv}
G_f &=& \big(\slashed{p}-\boldsymbol{\Sigma}_f\big)^{-1} \,,
\end{eqnarray}
either in terms of $\Sigma_f$ as
\begin{eqnarray}
\label{ewa:Gf}
G_f &=&
\big(\slashed{p}+\Sigma_f\big)\big(p^2-\Sigma_f^\dag\,\Sigma_f^{\phantom\dag}\big)^{-1} P_L +
\big(\slashed{p}+\Sigma_f^\dag\big)\big(p^2-\Sigma_f^{\phantom\dag}\,\Sigma_f^\dag\big)^{-1} P_R
\,,
\end{eqnarray}
or in terms of $\boldsymbol{\Sigma}_f$ as
\begin{subequations}
\label{ewa:Gftucne}
\begin{eqnarray}
G_f  &=&
\big(\slashed{p}+\boldsymbol{\Sigma}_f^\dag\big) \big(p^2-\boldsymbol{\Sigma}_f^{\phantom\dag}\,\boldsymbol{\Sigma}_f^\dag\big)^{-1}
\\ &=&
\big(p^2-\boldsymbol{\Sigma}_f^\dag\,\boldsymbol{\Sigma}_f^{\phantom\dag}\big)^{-1}
\big(\slashed{p}+\boldsymbol{\Sigma}_f^\dag\big)
\,.
\end{eqnarray}
\end{subequations}

\subsubsection{Mass spectrum}

Since the two quantities $\Sigma_f^\dag\,\Sigma_f^{\phantom\dag}$ and $\Sigma_f^{\phantom\dag}\,\Sigma_f^\dag$ are \emph{different} matrices, so are the two \qm{denominators} in \eqref{ewa:Gf}. However, their pole structure is the same, as both matrices $\Sigma_f^\dag\,\Sigma_f^{\phantom\dag}$ and $\Sigma_f^{\phantom\dag}\,\Sigma_f^\dag$ have the same ($p^2$-dependent) spectrum. In other words, the two, apparently different, pole equations
\begin{subequations}
\label{ewa:dvepole}
\begin{eqnarray}
\det\big(p^2-\Sigma_f^\dag\,\Sigma_f^{\phantom\dag}\big) &=& 0 \,, \\
\det\big(p^2-\Sigma_f^{\phantom\dag}\,\Sigma_f^\dag\big) &=& 0
\end{eqnarray}
\end{subequations}
are identical. To see this recall that the self-energy $\Sigma_f$ can be diagonalized by means of the bi-unitary transformation (see also Eq.~\eqref{eq:definition_UV} in appendix~\ref{app:fermi propag}):
\begin{eqnarray}
\label{ewa:BiUni}
\Sigma_f &=& V_f^\dag\,M_f\,U_f \,,
\end{eqnarray}
where $U_f$, $V_f$ are unitary matrices and $M_f$ is a diagonal, real and non-negative matrix. Needless to say that all three matrices $U_f$, $V_f$, $M_f$ are functions of $p^2$; some consequences of this general momentum dependence will be discusses in chapter~\ref{chp:mx}. Plugging the expression \eqref{ewa:BiUni} into the pole equations \eqref{ewa:dvepole} and using the unitarity of $U_f$, $V_f$, we find that both pole equations \eqref{ewa:dvepole} are reexpressed by the same equation
\begin{eqnarray}
\det\big(p^2-M_f^2\big) &=& 0
\end{eqnarray}
(which can be understood, due to the diagonality of $M_f$, as a set of $n$ independent equations $p^2-M_{fi}^2=0$, $i=1,\ldots,n$, rather than as a single equation). We have thus shown that the two pole equations \eqref{ewa:dvepole} are really the same. Moreover, we have also explicitly shown that the quark mass spectrum is real and positive. However, the number of solutions of the pole equation remains undetermined, due to the undetermined momentum dependence of $\Sigma_f$. Only in the special case of constant, momentum-independent $\Sigma_f$ one knows that there are exactly $n$ solutions.

\subsubsection{Notation for \qm{denominators}}

Let us also introduce, in accordance with appendix~\ref{app:fermi propag}, some useful notation for the \qm{denominators} of the propagators:
\begin{subequations}
\label{ewa:DfLDfR}
\begin{eqnarray}
D_{fL} &=& \big(p^2-\Sigma_f^{\phantom{\dag}}\,\Sigma_f^{\dag}\big)^{-1} \,, \\
D_{fR} &=& \big(p^2-\Sigma_f^{\dag}\,\Sigma_f^{\phantom{\dag}}\big)^{-1} \,,
\end{eqnarray}
\end{subequations}
and
\begin{subequations}
\begin{eqnarray}
\boldsymbol{D}_{fL} &=& \big(p^2-\boldsymbol{\Sigma}_f^{\phantom{\dag}}\,\boldsymbol{\Sigma}_f^{\dag}\big)^{-1}
\ = \  D_{fL}\,P_L + D_{fR}\,P_R  \,, \\
\boldsymbol{D}_{fR} &=& \big(p^2-\boldsymbol{\Sigma}_f^{\dag}\,\boldsymbol{\Sigma}_f^{\phantom{\dag}}\big)^{-1}
\ = \  D_{fL}\,P_R + D_{fR}\,P_L  \,.
\end{eqnarray}
\end{subequations}
For the reader's convenience we also present the \qm{commutation} relations
\begin{eqnarray}
\label{ewa:SgmfDfr:DfLSgmf}
\Sigma_f\,D_{fR} &=& D_{fL}\,\Sigma_f
\end{eqnarray}
and
\begin{eqnarray}
\label{ewa:bldSgmfDfr:DfLSgmf}
\boldsymbol{\Sigma}_f\,\boldsymbol{D}_{fR} &=& \boldsymbol{D}_{fL}\,\boldsymbol{\Sigma}_f \,,
\end{eqnarray}
useful for practical calculations.

\section{Leptons}

We are now going to construct the Ansatz for the self-energy of the lepton field $\Psi_\ell$. We will follow the same guidelines as in the case of scalars and quarks.

\subsection{Notation for propagators}

Just in order to establish the notation, recall that the lepton self-energy $\boldsymbol{\Sigma}_{\Psi_\ell}$ is defined as the difference between the free propagator $S_{\Psi_\ell}$ and the full propagator $G_{\Psi_\ell}$:
\begin{eqnarray}
\label{ewa:SgmPsielldef}
\boldsymbol{\Sigma}_{\Psi_\ell} &\equiv& S_{\Psi_\ell}^{-1} - G_{\Psi_\ell}^{-1} \,.
\end{eqnarray}
At the same time, it can be computed as the 1PI propagator:
\begin{eqnarray}
-\I \boldsymbol{\Sigma}_{\Psi_\ell} &=& \langle \Psi_\ell \bar \Psi_\ell \rangle_{\mathrm{1PI}} \,.
\end{eqnarray}
The full propagator is expressed in terms of the Nambu--Gorkov field $\Psi_\ell$ as
\begin{eqnarray}
\label{ewa:GPsielldef}
\I\,G_{\Psi_\ell} &=& \langle \Psi_\ell \bar \Psi_\ell \rangle \,.
\end{eqnarray}
The inverse free propagator $S_{\Psi_\ell}$ has due to the existence of the non-vanishing bare right-handed Majorana neutrino masses \eqref{ew1:eL:MnuR} the non-trivial form
\begin{eqnarray}
\label{ewa:Sell}
S_{\Psi_\ell}^{-1} &=& \slashed{p} - \big(M_{\nu R}^\dag\,P_L + M_{\nu R}\,P_R\big)
\left(\begin{array}{cccc}
0 & 0 & 0 & 0 \\
0 & 1 & 0 & 0 \\
0 & 0 & 0 & 0 \\
0 & 0 & 0 & 0 \end{array}\right) \,,
\end{eqnarray}
which is to be compared with much simpler form \eqref{ewa:Sq} of the quark free propagator.

\subsection{General form of the self-energy}
\label{ewa:ssec:Psiellgen}

Recall that since the lepton field $\Psi_\ell$ satisfies the Majorana condition \eqref{ew1:majcond}, there is the constraint \eqref{ew1:majcondG} on its full propagator $G_{\Psi_\ell}$. As the free propagator \eqref{ewa:Sell} satisfies this constraint too (since the mass matrix $M_{\nu R}$ is symmetric), so must also the self-energy $\boldsymbol{\Sigma}_{\Psi_\ell}$:
\begin{eqnarray}
\label{ewa:SgmPsiellmaj}
\boldsymbol{\Sigma}_{\Psi_\ell}(p) &=& \boldsymbol{\Sigma}_{\Psi_\ell}^\C(-p) \,.
\end{eqnarray}

Apart from the \qm{obligatory} constraint \eqref{ewa:SgmPsiellmaj} we will also demand that $\boldsymbol{\Sigma}_{\Psi_\ell}$ satisfy the \qm{optional} technical constraint
\begin{eqnarray}
\label{ewa:SgmPsiellhc}
\boldsymbol{\Sigma}_{\Psi_\ell} &=& \boldsymbol{\bar\Sigma}_{\Psi_\ell} \,.
\end{eqnarray}
The reasons for imposing this condition are the same as in the case of quarks: reduction of the independent parts of $\boldsymbol{\Sigma}_{\Psi_\ell}$ and reality and non-negativity of the lepton spectrum.

We require invariance of $\boldsymbol{\Sigma}_{\Psi_\ell}$ under the electromagnetic $\group{U}(1)_{\mathrm{em}}$ symmetry. Namely, we demand
\begin{eqnarray}
\label{ewa:SgmPsielleminv}
\boldsymbol{\Sigma}_{\Psi_\ell}\,T_{\Psi_\ell,\mathrm{em}} - \bar T_{\Psi_\ell,\mathrm{em}}\,\boldsymbol{\Sigma}_{\Psi_\ell} &=& 0 \,.
\end{eqnarray}
Due to the block-diagonal form of $T_{\Psi_\ell,\mathrm{em}}$, \eqref{ew1:TPsiellem}, the self-energy $\boldsymbol{\Sigma}_{\Psi_\ell}$ must have a block-diagonal form too:
\begin{eqnarray}
\boldsymbol{\Sigma}_{\Psi_\ell} &=&
\left(\begin{array}{cc} \boldsymbol{\Sigma}_{\Psi_\nu} & 0 \\ 0 & \boldsymbol{\Sigma}_{\Psi_e} \end{array}\right) \,.
\end{eqnarray}
The single condition \eqref{ewa:SgmPsielleminv} on the electromagnetic invariance now decouples into two conditions
\begin{eqnarray}
\label{ewa:SgmPsifeminv}
\boldsymbol{\Sigma}_{\Psi_{\!f}}\,T_{\Psi_{\!f},\mathrm{em}} - \bar T_{\Psi_{\!f},\mathrm{em}}\,\boldsymbol{\Sigma}_{\Psi_{\!f}} &=& 0
\,, \quad\quad f=\nu,e \,,
\end{eqnarray}
being subject to independent treatments.

The self-energies $\boldsymbol{\Sigma}_{\Psi_\nu}$, $\boldsymbol{\Sigma}_{\Psi_e}$ are given in terms of the fields $\Psi_\nu$, $\Psi_e$, Eq.~\eqref{ew1:PsinuPsie}, as ($f=\nu,e$)
\begin{eqnarray}
-\I \boldsymbol{\Sigma}_{\Psi_{\!f}} &=& \langle \Psi_{\!f} \bar \Psi_{\!f} \rangle_{\mathrm{1PI}} \,.
\end{eqnarray}
Since the fields $\Psi_\nu$, $\Psi_e$ are Majorana too, we have for $\boldsymbol{\Sigma}_{\Psi_\nu}$, $\boldsymbol{\Sigma}_{\Psi_e}$ the same constraints as the \eqref{ewa:SgmPsiellmaj} for $\boldsymbol{\Sigma}_{\Psi_\ell}$:
\begin{eqnarray}
\label{ewa:SgmPsifmaj}
\boldsymbol{\Sigma}_{\Psi_{\!f}}(p) &=& \boldsymbol{\Sigma}_{\Psi_{\!f}}^\C(-p) \,.
\end{eqnarray}
Similarly, the Hermiticity condition \eqref{ewa:SgmPsiellhc} now translates into two conditions
\begin{eqnarray}
\label{ewa:SgmPsifhc}
\boldsymbol{\Sigma}_{\Psi_{\!f}} &=& \boldsymbol{\bar\Sigma}_{\Psi_{\!f}} \,.
\end{eqnarray}

\subsection{Symmetry constraints}

\subsubsection{Electromagnetic invariance for neutrinos}

We are now going to discuss the conditions \eqref{ewa:SgmPsifeminv} of electromagnetic invariance for each of the two lepton types separately. We start with the neutrinos, as they are easier. Recall that
\begin{eqnarray}
T_{\Psi_\nu,\mathrm{em}} &=& 0 \,,
\end{eqnarray}
as can be seen from \eqref{ew1:TPsifem} with $Q_\nu=0$. Therefore the equation \eqref{ewa:SgmPsifeminv} is for neutrinos satisfied trivially and hence it gives no constraint on $\boldsymbol{\Sigma}_{\Psi_\nu}$. Put another way, due to electrical neutrality of neutrinos their self-energy can be arbitrary (up to the constraints \eqref{ewa:SgmPsifmaj} and \eqref{ewa:SgmPsifhc}) without affecting the electric charge conservation. In particular, in contrast to charged fermions, it can contain also the components of the Majorana type. To be explicit, the two conditions \eqref{ewa:SgmPsifmaj} and \eqref{ewa:SgmPsifhc} constrain the self-energy $\boldsymbol{\Sigma}_{\Psi_\nu}$ to have the form
\begin{eqnarray}
\label{ewa:Sgmnuansintermed}
\boldsymbol{\Sigma}_{\Psi_\nu} &=& \slashed{p}(A_{\Psi_\nu}^\T\,P_L+A_{\Psi_\nu}\,P_R)+(\Sigma_{\Psi_\nu}^\dag\,P_L + \Sigma_{\Psi_\nu}\,P_R) \,,
\end{eqnarray}
where the matrices $A_{\Psi_\nu}$ and $\Sigma_{\Psi_\nu}$ are respectively Hermitian and symmetric, but otherwise completely arbitrary. However, as the left-handed and the right-handed neutrino components have different transformation properties under the symmetries of the model, it is convenient to take into account the doublet structure of the Nambu--Gorkov field $\Psi_\nu$,
\begin{eqnarray}
\Psi_\nu &=& \left(\begin{array}{c} \nu_L + (\nu_L)^\C \\ \nu_R + (\nu_R)^\C \end{array}\right) \,,
\end{eqnarray}
and introduce for the sake of later references a special denotation for the corresponding blocks in the matrices $A_{\Psi_\nu}$ and $\Sigma_{\Psi_\nu}$, entering \eqref{ewa:Sgmnuansintermed}:
\begin{subequations}
\label{ewa:APsinuSgmpsinu}
\begin{eqnarray}
A_{\Psi_\nu} &\equiv&
\left(\begin{array}{cc} A_{\nu L}^\T & A_{\nu M} \\ A_{\nu M}^\dag & A_{\nu R} \end{array}\right) \,,
\\
\Sigma_{\Psi_\nu} &\equiv&
\left(\begin{array}{cc} \Sigma_{\nu L} & \Sigma_{\nu D} \\ \Sigma_{\nu D}^\T & \Sigma_{\nu R} \end{array}\right) \,.
\end{eqnarray}
\end{subequations}
Here $A_{\nu L}$, $A_{\nu R}$ are Hermitian matrices, while $\Sigma_{\nu L}$, $\Sigma_{\nu R}$ are symmetric matrices. The matrices $A_{\nu M}$, $\Sigma_{\nu D}$ are arbitrary. In terms of these blocks the self-energy $\boldsymbol{\Sigma}_{\Psi_\nu}$, \eqref{ewa:Sgmnuansintermed}, has the form:
\begin{eqnarray}
\boldsymbol{\Sigma}_{\Psi_\nu} &=&
\slashed{p} \left(\begin{array}{cc}
A_{\nu L}\,P_L+A_{\nu L}^\T\,P_R & A_{\nu M}^*\,P_L+A_{\nu M}\,P_R \\
A_{\nu M}^\T\,P_L+A_{\nu M}^\dag\,P_R & A_{\nu R}^\T\,P_L+A_{\nu R}\,P_R
\end{array}\right)
\nonumber \\ &&  {}
+
\left(\begin{array}{cc}
\Sigma_{\nu L}^\dag\,P_L+\Sigma_{\nu L}\,P_R & \Sigma_{\nu D}^*\,P_L+\Sigma_{\nu D}\,P_R \\
\Sigma_{\nu D}^\dag\,P_L+\Sigma_{\nu D}^\T\,P_R & \Sigma_{\nu R}^\dag\,P_L+\Sigma_{\nu R}\,P_R
\end{array}\right) \,.
\end{eqnarray}

\subsubsection{Electromagnetic invariance for charged leptons}

For the charged leptons $\Psi_e$ the application of the conditions \eqref{ewa:SgmPsifmaj} and \eqref{ewa:SgmPsifhc} yields $\boldsymbol{\Sigma}_{\Psi_e}$ in the same form as before $\boldsymbol{\Sigma}_{\Psi_\nu}$, \eqref{ewa:Sgmnuansintermed}, i.e.,
\begin{eqnarray}
\label{ewa:Sgmeansintermed}
\boldsymbol{\Sigma}_{\Psi_e} &=&
\slashed{p}(A_{\Psi_e}^\T\,P_L+A_{\Psi_e}\,P_R)+(\Sigma_{\Psi_e}^\dag\,P_L + \Sigma_{\Psi_e}\,P_R) \,,
\end{eqnarray}
with $A_{\Psi_e}$ and $\Sigma_{\Psi_e}$ being again respectively Hermitian and symmetric matrices. However, since the corresponding electromagnetic generator $T_{\Psi_e,\mathrm{em}}$, \eqref{ew1:TPsifem}, is this time non-trivial ($Q_e = -1$), the application of the condition of electromagnetic invariance \eqref{ewa:SgmPsifeminv} constrains the matrices $A_{\Psi_e}$, $\Sigma_{\Psi_e}$ to have in the Nambu--Gorkov space \eqref{ew1:Psie} the special block forms
\begin{eqnarray}
A_{\Psi_e} &=& \left(\begin{array}{cc} A_{e L}^\T & 0 \\ 0 & A_{e R} \end{array}\right) \,,
\\
\Sigma_{\Psi_e} &=& \left(\begin{array}{cc} 0 & \Sigma_{e} \\ \Sigma_{e}^\T & 0 \end{array}\right) \,.
\end{eqnarray}
In this block form the self-energy $\boldsymbol{\Sigma}_{\Psi_e}$ has the form
\begin{eqnarray}
\label{ewa:SgmPsieprel}
\boldsymbol{\Sigma}_{\Psi_e} &=&
\slashed{p} \left(\begin{array}{cc} A_{eL}\,P_L+A_{eL}^\T\,P_R & 0 \\ 0 & A_{eR}\,P_R+A_{eR}^\T\,P_L \end{array}\right)
+
\left(\begin{array}{cc} 0 & \Sigma_e\,P_R+\Sigma_e^*\,P_L \\ \Sigma_e^\dag\,P_L+\Sigma_e^\T\,P_R & 0 \end{array}\right) \,.
\nonumber \\ &&
\end{eqnarray}
Recall that the matrices $A_{eL}$, $A_{eR}$ are Hermitian, while $\Sigma_e$ can be arbitrary.

One can take also another, more illuminating view on the Ansatz \eqref{ewa:Sgmeansintermed} for $\boldsymbol{\Sigma}_{\Psi_e}$. We can notice that the generator $T_{\Psi_e,\mathrm{em}}$ is proportional to $\gamma_5\,\sigma_3$. Therefore, as shown in appendix~\ref{app:fermi propag}, Eq.~\eqref{app:frm:phasePsi}, the transformation of the Namu--Gorkov field $\Psi_e$, generated by the generator $T_{\Psi_e,\mathrm{em}}$, is equivalent to the transformation of the Dirac field $e=e_L + e_R$, \eqref{ew1:e}, induced by the generator $T_{e,\mathrm{em}}$, \eqref{ew1:Teem}. Accordingly, the condition \eqref{ewa:SgmPsifeminv} is equivalent to the condition
\begin{eqnarray}
\boldsymbol{\Sigma}_{e}\,T_{e,\mathrm{em}} - \bar T_{e,\mathrm{em}}\,\boldsymbol{\Sigma}_{e} &=& 0 \,,
\end{eqnarray}
with
\begin{eqnarray}
-\I\,\boldsymbol{\Sigma}_e &=& \langle e \bar e \rangle_{\mathrm{1PI}} \,.
\end{eqnarray}
In other words, due to the invariance under the electromagnetic symmetry $\group{U}(1)_{\mathrm{em}}$ (and since $e_L$ and $e_R$ are of the same dimensionality) the description using the \emph{charged} Nambu--Gorkov field $\Psi_e$ is completely equivalent to the description using the Dirac field $e$. This is shown in more detail in Sec.~\ref{app:frm:Rel} of appendix~\ref{app:fermi propag}, together with more results concerning passing between the two equivalent descriptions $\Psi_e$ and $e$. Using these results it can be shown that the self-energy $\boldsymbol{\Sigma}_{\Psi_e}$, \eqref{ewa:Sgmeansintermed}, of the Nambu--Gorkov field $\Psi_e$ corresponds to the following form of the self-energy $\boldsymbol{\Sigma}_e$ of the Dirac field $e$:
\begin{eqnarray}
\boldsymbol{\Sigma}_e &=& \slashed{p}(A_{eL}\,P_L+A_{eR}\,P_R)+(\Sigma_e^\dag\,P_L + \Sigma_e\,P_R) \,.
\end{eqnarray}
Notice that $\boldsymbol{\Sigma}_e$ has the same form as the self-energies \eqref{ewa:fSgmAnzintermed} of the quark fields $u$, $d$. It is of course not surprising, since for the electrically charged quarks the condition of electromagnetic invariance is the same as for charged leptons (compare the $\group{U}(1)_{\mathrm{em}}$ generators \eqref{ew1:qTfem} for $u$, $d$ with the generator \eqref{ew1:Teem} for $e$).

Two comments are in order now. First, we have now \emph{a posteriori} justified our choice in the case of quarks to work directly from the very beginning within the Dirac basis $q=q_L+q_R$. Formally it would have been more correct to start with Nambu--Gorkov field $\Psi_q$ and only afterwards to show its equivalence to $q$ due to non-vanishing quarks' electric charges and due the same number of left-handed and the right-handed quarks.

Second, if the description using the field $e$ is equivalent to the description using the field $\Psi_e$, there is a question why to introduce $\Psi_e$ at all and why not to work exclusively with $e$. Certainly this would be possible. However, we choose to work rather in terms $\Psi_e$, since it seems to be convenient to treat the charged leptons and the neutrinos on the same footing as long as possible and only in the final results to take back into play their different nature.

\subsubsection{The $T_{\Psi_\ell,Z}$ generator}

Having established the $\group{U}(1)_{\mathrm{em}}$ invariant Ansatz, we can analyze its transformation properties under the complementary symmetries of the full $\group{SU}(2)_{\mathrm{L}} \times \group{U}(1)_{\mathrm{Y}}$ symmetry, i.e., under the symmetries induced by the generators $T_{\Psi_\ell,Z}$, $T_{\Psi_\ell,1}$, $T_{\Psi_\ell,2}$.

We start with the generator $T_{\Psi_\ell,Z}$, \eqref{ew1:TPsiellZ}. As it is block-diagonal, we can analyze two separate quantities $\boldsymbol{\Sigma}_{\Psi_{\!f}}\,T_{\Psi_{\!f},Z} - \bar T_{\Psi_{\!f},Z}\,\boldsymbol{\Sigma}_{\Psi_{\!f}}$ for the two lepton types $f=\nu,e$.

For the neutrinos with the generator $T_{\Psi_\nu,Z}$ given explicitly by \eqref{ew1:TPsifZ} (recall that $Q_\nu=0$) and with $\boldsymbol{\Sigma}_{\Psi_\nu}$ given by \eqref{ewa:Sgmnuansintermed} we therefore arrive explicitly at
\begin{eqnarray}
\boldsymbol{\Sigma}_{\Psi_\nu}\,T_{\Psi_\nu,Z} - \bar T_{\Psi_\nu,Z}\,\boldsymbol{\Sigma}_{\Psi_\nu} &=&
\phantom{+\,}
\frac{1}{2}\sqrt{g^2+g^{\prime2}} \slashed{p}\bigg(
P_L[P_{+\nu},A_{\Psi_\nu}]^\T + P_R[P_{+\nu},A_{\Psi_\nu}]\bigg)
\nonumber \\ &&
+\,\frac{1}{2}\sqrt{g^2+g^{\prime2}}\bigg(
P_L \{P_{+\nu},\Sigma_{\Psi_\nu}\}^\dag + P_R \{P_{+\nu},\Sigma_{\Psi_\nu}\}
\bigg) \,,
\qquad\qquad
\end{eqnarray}
with the relevant (anti)commutators given in terms of \eqref{ewa:APsinuSgmpsinu} by
\begin{subequations}
\begin{eqnarray}
{[P_{+\nu},A_{\Psi_\nu}]} &=& \left(\begin{array}{cc} 0 & A_{\nu M} \\ - A_{\nu M}^\dag & 0 \end{array}\right) \,, \\
\{P_{+\nu},\Sigma_{\Psi_\nu}\} &=& \left(\begin{array}{cc} 2 \Sigma_{\nu L} & \Sigma_{\nu D} \\ \Sigma_{\nu D}^\T & 0 \end{array}\right) \,.
\end{eqnarray}
\end{subequations}

For the charged leptons with the generator $T_{\Psi_e,Z}$ given explicitly by \eqref{ew1:TPsifZ} and $\boldsymbol{\Sigma}_{\Psi_e}$ given by \eqref{ewa:SgmPsieprel} we obtain
\begin{eqnarray}
\boldsymbol{\Sigma}_{\Psi_e}\,T_{\Psi_e,Z} - \bar T_{\Psi_e,Z}\,\boldsymbol{\Sigma}_{\Psi_e} &=&
\frac{1}{2}\sqrt{g^2+g^{\prime2}}
\left(\begin{array}{cc}
0 & \Sigma_{e}^\dag\,P_L+\Sigma_{e}^\T\,P_R \\
\Sigma_{e}^*\,P_L+\Sigma_{e}\,P_R & 0
\end{array}\right) \,.
\qquad
\end{eqnarray}

\subsubsection{The $T_{\Psi_\ell,1}$, $T_{\Psi_\ell,2}$ generators}

Similarly can be treated the generators $T_{\Psi_\ell,1}$ and $T_{\Psi_\ell,2}$, \eqref{ew1:Tpsi1} and \eqref{ew1:Tpsi2}, respectively. Taking into account the form of $\boldsymbol{\Sigma}_{\Psi_\ell}$ obtained so far, we find
\begin{eqnarray}
\boldsymbol{\Sigma}_{\Psi_\ell}\,T_{\Psi_\ell,a} - \bar T_{\Psi_\ell,a}\,\boldsymbol{\Sigma}_{\Psi_\ell} &=&
\left(\begin{array}{cc} 0 & -X_a \\ \bar X_a & 0 \end{array}\right) \,,
\end{eqnarray}
where
\begin{subequations}
\begin{eqnarray}
X_1 &\equiv& \phantom{\I}\frac{1}{2} g \left[ \phantom{+}
\slashed{p}\,P_L \left(\begin{array}{cc} (A_{\nu L}-A_{eL}) & A_{\nu M}^* \\ 0 & 0 \end{array}\right) +
\slashed{p}\,P_R \left(\begin{array}{cc} -(A_{\nu L}-A_{eL})^\T & -A_{\nu M} \\ 0 & 0 \end{array}\right)
\right. \nonumber \\ && \phantom{\phantom{\I}\frac{1}{2} g \left[\right.} \left.
+ P_L \left(\begin{array}{cc} -\Sigma_{\nu L}^\dag & -\Sigma_{\nu D}^* \\ -\Sigma_{e D}^\dag & 0 \end{array}\right)
+ P_R \left(\begin{array}{cc} \Sigma_{\nu L} & \Sigma_{\nu D} \\ \Sigma_{e D}^\T & 0 \end{array}\right)
\right] \,,
\\
&& \nonumber \\
X_2 &\equiv& \I\frac{1}{2} g \left[
\slashed{p}\,P_L \left(\begin{array}{cc} -(A_{\nu L}-A_{eL}) & -A_{\nu M}^* \\ 0 & 0 \end{array}\right) +
\slashed{p}\,P_R \left(\begin{array}{cc} -(A_{\nu L}-A_{eL})^\T & -A_{\nu M} \\ 0 & 0 \end{array}\right)
\right. \nonumber \\ && \phantom{\phantom{\I}\frac{1}{2} g \left[\right.} \left.
+ P_L \left(\begin{array}{cc} -\Sigma_{\nu L}^\dag & -\Sigma_{\nu D}^* \\ \Sigma_{e D}^\dag & 0 \end{array}\right)
+ P_R \left(\begin{array}{cc} -\Sigma_{\nu L} & -\Sigma_{\nu D} \\ \Sigma_{e D}^\T & 0 \end{array}\right)
\right] \,.
\end{eqnarray}
\end{subequations}

\subsubsection{The discrete $\mathcal{P}_{\mathrm{down}}$ symmetry}

And finally, there is the discrete symmetry $\mathcal{P}_{\mathrm{down}}$, \eqref{ew1:Pdown}. Clearly, it does not affect the neutrinos at all. The charged leptons are nevertheless affected. We can repeat the result for the down-type quarks \eqref{ewa:Pdownd} that non-invariance of $\boldsymbol{\Sigma}_e$ under $\mathcal{P}_{\mathrm{down}}$ is proportional to $\Sigma_e$:
\begin{eqnarray}
{[\boldsymbol{\Sigma}_e,\mathcal{P}_{\mathrm{down}}]} &=& 2(\Sigma_e^\dag\,P_L + \Sigma_e\,P_R) \,.
\end{eqnarray}

\subsubsection{Lepton number symmetry}

As discussed in Sec.~\ref{ew1:ssec:PartCont}, the lepton number symmetry is in fact broken explicitly by the non-vanishing right-handed Majorana neutrino mass terms \eqref{ew1:eL:MnuR}. It is nevertheless useful to see how this symmetry would be broken spontaneously (i.e., by the lepton self-energies) in the case of $M_{\nu R}=0$.

Recall that the lepton number symmetry $\group{U}(1)_\ell$ acts on the Nambu--Gorkov field $\Psi_\ell$ as \eqref{ew1:leptonNumNG}, with the corresponding generator $T_{\Psi_\ell}$ having the diagonal form \eqref{ew1:TPsiell}. Thus again, since both $T_{\Psi_\ell}$ and $\boldsymbol{\Sigma}_{\Psi_\ell}$ are diagonal, we can investigate the quantity $\boldsymbol{\Sigma}_{\Psi_\ell}\,T_{\Psi_\ell} -\bar T_{\Psi_\ell}\,\boldsymbol{\Sigma}_{\Psi_\ell}$ separately for neutrinos and charged leptons. We obtain
\begin{subequations}
\begin{eqnarray}
\boldsymbol{\Sigma}_{\Psi_e}\,T_{\Psi_e} - \bar T_{\Psi_e}\,\boldsymbol{\Sigma}_{\Psi_e} &=& 0 \,,
\\
\boldsymbol{\Sigma}_{\Psi_\nu}\,T_{\Psi_\nu} - \bar T_{\Psi_\nu}\,\boldsymbol{\Sigma}_{\Psi_\nu} &=&
2 Q_\ell
\left(\begin{array}{cc}
\Sigma_{\nu L}^\dag\,P_L-\Sigma_{\nu L}\,P_R & 0                                            \\
0                                             & -\Sigma_{\nu R}^\dag\,P_L+\Sigma_{\nu R}\,P_R
\end{array}\right) \,.
\end{eqnarray}
\end{subequations}
We can see the expected result that only the two Majorana-type self-energies $\Sigma_{\nu L}$, $\Sigma_{\nu R}$ break the lepton number symmetry.

\subsubsection{Purely symmetry-breaking self-energy}

We conclude that the symmetry-preserving components of the self-energies are $A_{eR}$, $A_{\nu R}$, $A_{\nu L} + A_{eL}$, plus the $\Sigma_{\nu R}$ as we assume the explicit violation \eqref{ew1:eL:MnuR} of the lepton number symmetry. (If we assumed the lepton number symmetry to be at the Lagrangian level exact, we would include $\Sigma_{\nu R}$ into the Ansatz too.) The symmetry-breaking self-energies are then $\Sigma_{\nu L}$, $\Sigma_{\nu D}$, $\Sigma_{e}$, $A_{\nu M}$, $A_{\nu L} - A_{eL}$. We therefore neglect the symmetry-preserving components of the Ansatz,
\begin{subequations}
\begin{eqnarray}
A_{\nu L} + A_{eL} &=& 0 \,, \\
A_{eR} &=& 0 \,, \\
A_{\nu R} &=& 0 \,, \\
\Sigma_{\nu R} &=& 0
\end{eqnarray}
\end{subequations}
and upon denoting
\begin{eqnarray}
A_{\nu L} - A_{eL} &\equiv& 2 A_\ell
\end{eqnarray}
find the most general electromagnetically invariant Ansatz consisting only of the symmetry-breaking and thus UV-finite parts to be the following:
\begin{subequations}
\label{ewa:AnsPsinuealm}
\begin{eqnarray}
\boldsymbol{\Sigma}_{\Psi_\nu} &=&
\slashed{p} \left(\begin{array}{cc}
A_{\ell}\,P_L+A_{\ell}^\T\,P_R & A_{\nu M}^*\,P_L+A_{\nu M}\,P_R \\
A_{\nu M}^\T\,P_L+A_{\nu M}^\dag\,P_R & 0
\end{array}\right)
\nonumber \\ && {}
+
\left(\begin{array}{cc}
\Sigma_{\nu L}^\dag\,P_L+\Sigma_{\nu L}\,P_R & \Sigma_{\nu D}^*\,P_L+\Sigma_{\nu D}\,P_R \\
\Sigma_{\nu D}^\dag\,P_L+\Sigma_{\nu D}^\T\,P_R & 0
\end{array}\right) \,,
\\
\boldsymbol{\Sigma}_{\Psi_e} &=&
\slashed{p} \left(\begin{array}{cc} -A_{\ell}\,P_L-A_{\ell}^\T\,P_R & 0 \\ 0 & 0 \end{array}\right)
+
\left(\begin{array}{cc} 0 & \Sigma_e\,P_R+\Sigma_e^*\,P_L \\ \Sigma_e^\dag\,P_L+\Sigma_e^\T\,P_R & 0 \end{array}\right) \,.
\qquad
\end{eqnarray}
\end{subequations}
Or in terms of the forms \eqref{ewa:Sgmnuansintermed}, \eqref{ewa:Sgmeansintermed} for $\boldsymbol{\Sigma}_{\Psi_\nu}$, $\boldsymbol{\Sigma}_{\Psi_e}$ we have for the relevant quantities  $A_{\Psi_\nu}$, $\Sigma_{\Psi_\nu}$ and $A_{\Psi_e}$, $\Sigma_{\Psi_e}$:
\begin{subequations}
\label{ewa:APsinuSgmpsinu}
\begin{eqnarray}
A_{\Psi_\nu} &=& \left(\begin{array}{cc} A_{\ell}^\T & A_{\nu M} \\ A_{\nu M}^\dag & 0 \end{array}\right) \,,
\\
\Sigma_{\Psi_\nu} &=& \left(\begin{array}{cc} \Sigma_{\nu L} & \Sigma_{\nu D} \\ \Sigma_{\nu D}^\T & 0 \end{array}\right)
\end{eqnarray}
\end{subequations}
and
\begin{subequations}
\label{ewa:APsieSgmpsie}
\begin{eqnarray}
A_{\Psi_e} &=& \left(\begin{array}{cc} -A_{\ell}^\T & 0 \\ 0 & 0 \end{array}\right) \,,
\\
\Sigma_{\Psi_e} &=& \left(\begin{array}{cc} 0 & \Sigma_{e} \\ \Sigma_{e}^\T & 0 \end{array}\right) \,.
\end{eqnarray}
\end{subequations}

\subsection{Wave function renormalization self-energies}

The procedure of refining the Ansatz now continues in the same way as with the quarks. The pole equations corresponding to the full propagators with the self-energies given by the Ansatz \eqref{ewa:APsinuSgmpsinu}, \eqref{ewa:APsieSgmpsie} read
\begin{subequations}
\begin{eqnarray}
\det \big[ p^2 - (1-A_{\Psi_\nu})^{-1/2} \Sigma_{\Psi_\nu,M}^\dag (1-A_{\Psi_\nu}^\T)^{-1} \Sigma_{\Psi_\nu,M}^{\phantom{\dag}} (1-A_{\Psi_\nu})^{-1/2} \big] &=& 0 \,, \\
\det \big[ p^2 - (1-A_{\Psi_e})^{-1/2} \Sigma_{\Psi_e}^\dag (1-A_{\Psi_e}^\T)^{-1} \Sigma_{\Psi_e}^{\phantom{\dag}} (1-A_{\Psi_e})^{-1/2} \big] &=& 0 \,,
\end{eqnarray}
\end{subequations}
where we denoted
\begin{eqnarray}
\Sigma_{\Psi_\nu,M} &\equiv& \Sigma_{\Psi_\nu} + \left(\begin{array}{cc} 0 & 0 \\ 0 & M_{\nu R} \end{array}\right)
\>=\>
\left(\begin{array}{cc} \Sigma_{\nu L} & \Sigma_{\nu D} \\ \Sigma_{\nu D}^\T & M_{\nu R} \end{array}\right) \,.
\end{eqnarray}
Clearly, while the chirality-changing parts of the self-energies, $\Sigma_{\Psi_\nu}$ and $\Sigma_{\Psi_e}$, are necessary for generation of the lepton masses, the chirality preserving parts, $A_{\Psi_\nu}$ and $A_{\Psi_e}$, are not and will be accordingly discarded from the Ansatz. That is to say, we set
\begin{subequations}
\label{ewa:Ajsou0}
\begin{eqnarray}
A_{\Psi_\nu} &=& 0 \,, \\
A_{\Psi_e} &=& 0 \,,
\end{eqnarray}
or in terms of the individual entries of $A_{\Psi_\nu}$, $A_{\Psi_e}$,
\begin{eqnarray}
A_\ell &=& 0 \,,\\
A_{\nu M} &=& 0 \,.
\end{eqnarray}
\end{subequations}

\subsection{Final form of the Ansatz}
\label{ewa:final:Psiell}

By setting \eqref{ewa:Ajsou0} we have completed the construction of the lepton self-energy Ansatz. In this section we summarize the obtained results and for the reader's convenience we also repeat some of the formul{\ae} presented already above.

\subsubsection{Self-energies}

The final form of the Ansatz \eqref{ewa:AnsPsinuealm} upon considering \eqref{ewa:Ajsou0} thus reads
\begin{eqnarray}
\label{ewa:SgmPsif}
\boldsymbol{\Sigma}_{\Psi_{\!f}} &=& \Sigma_{\Psi_{\!f}}^\dag\,P_L + \Sigma_{\Psi_{\!f}}\,P_R \,,
\end{eqnarray}
where, likewise in the case of quarks, the subscript $f$ can stand both for $\ell$ as well as for $\nu$, $e$. On basis of the previous discussion we have
\begin{subequations}
\label{ewa:SgmPsinuPsie}
\begin{eqnarray}
\Sigma_{\Psi_\nu} &=&
\left(\begin{array}{cc}
\Sigma_{\nu L} & \Sigma_{\nu D} \\
\Sigma_{\nu D}^\T & 0
\end{array}\right) \,,
\\
\Sigma_{\Psi_e} &=& \left(\begin{array}{cc} 0 & \Sigma_e \\ \Sigma_e^\T & 0 \end{array}\right) \,.
\label{ewa:SgmPsie}
\end{eqnarray}
\end{subequations}
Just for completeness recall that
\begin{eqnarray}
\label{ewa:SgmPsiell}
\boldsymbol{\Sigma}_{\Psi_\ell} &=&
\left(\begin{array}{cc} \boldsymbol{\Sigma}_{\Psi_\nu} & 0 \\ 0 & \boldsymbol{\Sigma}_{\Psi_e} \end{array}\right) \,.
\end{eqnarray}
so that
\begin{eqnarray}
\Sigma_{\Psi_\ell} &=& \left(\begin{array}{cc} \Sigma_{\Psi_\nu} & 0 \\ 0 & \Sigma_{\Psi_e} \end{array}\right) \,.
\end{eqnarray}

Since there is a non-vanishing neutrino bare mass in the Lagrangian (the right-handed Majorana mass term $M_{\nu R}$, Eq.~\eqref{ew1:eL:MnuR}), it is convenient to define
\begin{eqnarray}
\label{ewa:SgmPsinuM}
\Sigma_{\Psi_\nu,M} &\equiv& \Sigma_{\Psi_\nu} + \left(\begin{array}{cc} 0 & 0 \\ 0 & M_{\nu R} \end{array}\right)
\ = \  \left(\begin{array}{cc} \Sigma_{\nu L} & \Sigma_{\nu D} \\ \Sigma_{\nu D}^\T & M_{\nu R} \end{array}\right)
\end{eqnarray}
and correspondingly also
\begin{eqnarray}
\Sigma_{\Psi_\ell,M} &\equiv& \left(\begin{array}{cc} \Sigma_{\Psi_\nu,M} & 0 \\ 0 & \Sigma_{\Psi_e} \end{array}\right) \,,
\end{eqnarray}
so that naturally
\begin{eqnarray}
\boldsymbol{\Sigma}_{\Psi_\nu,M} &\equiv& \Sigma_{\Psi_\nu,M}^\dag\,P_L + \Sigma_{\Psi_\nu,M}\,P_R
\end{eqnarray}
and
\begin{eqnarray}
\boldsymbol{\Sigma}_{\Psi_\ell,M} &\equiv& \Sigma_{\Psi_\ell,M}^\dag\,P_L + \Sigma_{\Psi_\ell,M}\,P_R \,.
\end{eqnarray}

\subsubsection{Full propagators}

Having arrived at the definitive self-energy Ansatz, we can now finally express the full propagator $G_{\Psi_\ell}$, \eqref{ewa:GPsielldef}, using the formula \eqref{ewa:SgmPsielldef}. It has necessarily the diagonal form
\begin{eqnarray}
\label{ewa:Gpsiell}
G_{\Psi_\ell} &=& \left(\begin{array}{cc} G_{\Psi_\nu} & 0 \\ 0 & G_{\Psi_e} \end{array}\right) \,,
\end{eqnarray}
where
\begin{subequations}
\label{ewa:GPsinue}
\begin{eqnarray}
G_{\Psi_\nu} &=& \big(\slashed{p}-\boldsymbol{\Sigma}_{\Psi_\nu,M}\big)^{-1} \,,
\label{ewa:GPsinu} \\
G_{\Psi_e} &=& \big(\slashed{p}-\boldsymbol{\Sigma}_{\Psi_e}\big)^{-1} \,,
\label{ewa:GPsie}
\end{eqnarray}
\end{subequations}
and
\begin{eqnarray}
\label{ewa:GPsiell}
G_{\Psi_\ell} &=& \big(\slashed{p}-\boldsymbol{\Sigma}_{\Psi_\ell,M}\big)^{-1} \,,
\end{eqnarray}
We do not state here explicit forms of the inversions, as they would be the same as for the quarks. In any case, detailed formul{\ae} can be found in appendix~\ref{app:fermi propag}.

\subsubsection{Mass spectrum}

The pole equations corresponding to the propagators $G_{\Psi_\nu}$, $G_{\Psi_e}$ are
\begin{subequations}
\label{ewa:poleeqell}
\begin{eqnarray}
\det \big[ p^2 - \Sigma_{\Psi_\nu,M}^\dag \, \Sigma_{\Psi_\nu,M}^{\phantom{\dag}} \big] &=& 0 \,,
\label{ewa:poleeqnu} \\
\det \big[ p^2 - \Sigma_{\Psi_e}^\dag \,\Sigma_{\Psi_e}^{\phantom{\dag}} \big] &=& 0 \,.
\label{ewa:poleeqe}
\end{eqnarray}
\end{subequations}
Pole equations with interchanged $\Sigma_{\Psi_\nu,M}^\dag \leftrightarrow \Sigma_{\Psi_\nu,M}^{\phantom{\dag}}$ and $\Sigma_{\Psi_e}^\dag \leftrightarrow \Sigma_{\Psi_e}^{\phantom{\dag}}$ are equivalent to \eqref{ewa:poleeqell} as could be shown in an analogous way as in the case of quarks. Notice that the charged lepton pole equation \eqref{ewa:poleeqe} can be simplified due to the special form \eqref{ewa:SgmPsie} of $\Sigma_{\Psi_e}$ as
\begin{eqnarray}
\det \big[ p^2 - \Sigma_{e}^\dag \,\Sigma_{e}^{\phantom{\dag}} \big] &=& 0 \,.
\end{eqnarray}
Due to the same argument as in the case of quarks, the pole equations again predict real and positive lepton masses. Their number, however, is for general self-energies undetermined. Only in the special case of the momentum-independent self-energies we know that there will be $n+m$ solutions to the neutrino equation \eqref{ewa:poleeqnu} and $n$ solutions to the charged lepton equation \eqref{ewa:poleeqe}.

\subsubsection{Notation for \qm{denominators}}

Likewise in the case of quarks, it is now convenient to introduce some notation for the \qm{denominators} in the full propagators $G_{\Psi_\nu}$ and $G_{\Psi_e}$, \eqref{ewa:GPsinue}:
\begin{subequations}
\label{ewa:bldDPsinuPsie}
\begin{eqnarray}
D_{\Psi_\nu} &\equiv& \big(p^2-\Sigma_{\Psi_\nu,M}^{\phantom{\dag}}\,\Sigma_{\Psi_\nu,M}^{\dag}\big)^{-1}
\ \equiv \ \left(\begin{array}{cc} D_{\nu L} & D_{\nu M} \\ D_{\nu M}^\dag & D_{\nu R}^\T \end{array}\right) \,,
\\
D_{\Psi_e} &\equiv&\ \big(p^2-\Sigma_{\Psi_e}^{\phantom{\dag}}\,\Sigma_{\Psi_e}^{\dag}\big)^{-1}
\ \equiv \ \left(\begin{array}{cc} D_{e L} & 0 \\ 0 & D_{e R}^\T \end{array}\right) \,,
\label{ewa:DPsie}
\end{eqnarray}
\end{subequations}
as well as
\begin{subequations}
\begin{eqnarray}
\boldsymbol{D}_{\Psi_\nu} &\equiv& \big(p^2-\boldsymbol{\Sigma}_{\Psi_\nu,M}^{\phantom{\dag}}\,\boldsymbol{\Sigma}_{\Psi_\nu,M}^{\dag}\big)^{-1} \,,
\\
\boldsymbol{D}_{\Psi_e} &\equiv& \big(p^2-\boldsymbol{\Sigma}_{\Psi_e}^{\phantom{\dag}}\,\boldsymbol{\Sigma}_{\Psi_e}^{\dag}\big)^{-1}  \,.
\end{eqnarray}
\end{subequations}
Similarly we also define, for the sake of later references, the notation concerning the propagator $G_{\Psi_\ell}$, \eqref{ewa:GPsiell}:
\begin{eqnarray}
\label{ewa:DPsiell}
D_{\Psi_\ell} &\equiv& \big(p^2-\Sigma_{\Psi_\ell,M}^{\phantom{\dag}}\,\Sigma_{\Psi_\ell,M}^{\dag}\big)^{-1}
\ = \ \left(\begin{array}{cc} D_{\Psi_\nu} & 0 \\ 0 & D_{\Psi_e} \end{array}\right)
\end{eqnarray}
and
\begin{eqnarray}
\label{ewa:bldDPsiell}
\boldsymbol{D}_{\Psi_\ell} &\equiv& \big(p^2-\boldsymbol{\Sigma}_{\Psi_\ell,M}^{\phantom{\dag}}\,\boldsymbol{\Sigma}_{\Psi_\ell,M}^{\dag}\big)^{-1}
\ = \ \left(\begin{array}{cc} \boldsymbol{D}_{\Psi_\nu} & 0 \\ 0 & \boldsymbol{D}_{\Psi_e} \end{array}\right) \,.
\end{eqnarray}
(Cf.~appendix~\ref{app:fermi propag}.) Unlike in the case of quarks, we do not need to introduce a special denotation for the expressions with interchanged $\Sigma^\dag \leftrightarrow \Sigma^{\phantom{\dag}}$ and $\boldsymbol{\Sigma}^\dag \leftrightarrow \boldsymbol{\Sigma}^{\phantom{\dag}}$, since they are given just by the transposition and charge conjugation, respectively, see \eqref{app:frm:DPsiDpsiT} and \eqref{app:frm:DPsiDpsiC}. The charged lepton notation \eqref{ewa:DPsie} is consistent with the quark notation \eqref{ewa:DfLDfR}: For the charged leptons the definition \eqref{ewa:DPsie} equivalent to
\begin{eqnarray}
D_{eL} &=& \big(p^2-\Sigma_e^{\phantom{\dag}}\,\Sigma_e^{\dag}\big)^{-1} \\
D_{eR} &=& \big(p^2-\Sigma_e^{\dag}\,\Sigma_e^{\phantom{\dag}}\big)^{-1}
\end{eqnarray}
to be compared with the analogous quark definition \eqref{ewa:DfLDfR}.

The lepton analogues of the quark relations \eqref{ewa:SgmfDfr:DfLSgmf} and \eqref{ewa:bldSgmfDfr:DfLSgmf} read in the Nambu--Gorkov basis
\begin{subequations}
\begin{eqnarray}
\Sigma_{\Psi_\nu,M}\,D_{\Psi_\nu}^\T &=& D_{\Psi_\nu}\,\Sigma_{\Psi_\nu,M} \,, \\
\Sigma_{\Psi_e}\,D_{\Psi_e}^\T &=& D_{\Psi_e}\,\Sigma_{\Psi_e}
\end{eqnarray}
\end{subequations}
and
\begin{subequations}
\begin{eqnarray}
\boldsymbol{\Sigma}_{\Psi_\nu,M}\,\boldsymbol{D}_{\Psi_\nu}^\C &=& \boldsymbol{D}_{\Psi_\nu}\,\boldsymbol{\Sigma}_{\Psi_\nu,M} \,, \\
\boldsymbol{\Sigma}_{\Psi_e}\,\boldsymbol{D}_{\Psi_e}^\C &=& \boldsymbol{D}_{\Psi_e}\,\boldsymbol{\Sigma}_{\Psi_e} \,,
\end{eqnarray}
\end{subequations}
respectively (similarly for the $\Psi_\ell$-quantities \eqref{ewa:DPsiell}, \eqref{ewa:bldDPsiell}). In terms of the individual Nambu--Gorkov components it translates for the charged leptons simply as
\begin{eqnarray}
\Sigma_e\,D_{eR} &=& D_{eL}\,\Sigma_e \,,
\end{eqnarray}
while for the neutrinos we obtain slightly more complicated set of relations
\begin{eqnarray}
\Sigma_{\nu D}    \, D_{\nu R} + \Sigma_{\nu L} \, D_{\nu M}^*        &=&
D_{\nu L}         \, \Sigma_{\nu D} + D_{\nu M}      \, M_{\nu R} \,,    \\
\Sigma_{\nu L} \, D_{\nu L}^\T   + \Sigma_{\nu D}    \, D_{\nu M}^\T &=&
D_{\nu L}      \, \Sigma_{\nu L} + D_{\nu M}         \, \Sigma_{\nu D}^\T \,,      \\
M_{\nu R}      \, D_{\nu R}      + \Sigma_{\nu D}^\T \, D_{\nu M}^*    &=&
D_{\nu R}^\T   \, M_{\nu R}      + D_{\nu M}^\dag    \, \Sigma_{\nu D} \,.
\end{eqnarray}



\section{Summary}

We looked for the self-energy Ansatz separately for scalars ($\Phi$), quarks ($q$) and leptons ($\Psi_\ell$). The procedure was in each case basically the same: First we found the most general form of the Ansatz consistent with the requirement of Hermiticity and electromagnetical invariance, as well as with the constraints following from the eventual Nambu--Gorkov nature of the field in question. Then we checked the invariance of the Ansatz under the generators of the coset space $\group{SU}(2)_{\mathrm{L}} \times \group{U}(1)_{\mathrm{Y}} / \, \group{U}(1)_{\mathrm{em}}$ and kept in the Ansatz only the non-invariant parts. This was followed by discarding the wave function renormalization parts. Finally, in sections~\ref{ewa:final:Phi}, \ref{ewa:final:q} and \ref{ewa:final:Psiell}, we made a short summary of the formul{\ae} associated with the final form of the Ansatz, including the expressions and notations for the corresponding full propagators.

\chapter{Dynamics}
\label{chp:ewdyn}

\intro{In this chapter we will study the Yukawa dynamics of the presented model with the aim to show that it is capable of breaking the electroweak symmetry down to the electromagnetic one by means of formation of symmetry-breaking parts of the scalar and fermion propagators. We will proceed basically in the same way as we did in part~\ref{part:abel} within the Abelian toy model. We will first derive the SD equations at the Hartree--Fock approximation for arbitrary self-energies and only then we will restrict them on the self-energies of the form derived in the previous chapter. The solutions to these SD equations are assumed to be UV-finite, as the considered self-energies Ans\"{a}tze contain only symmetry-breaking parts. For the sake of comparison, we will present also the SD equations for the non-Ansatz parts of the self-energies and show that they indeed come out UV-finite or UV-divergent, depending on whether they are symmetry-breaking or symmetry-preserving. In some cases, however, we will have for that purpose to resort to two-loop considerations.}

\intro{Apart from mere formulation of the SD equations, we will also give some numerical evidence that they have the solutions of the desired properties: UV-finite and allowing, at least in principle, for the realistic fermion spectrum, with large observed hierarchies. Finally, we will also briefly comment on the compatibility of the obtained results with the electroweak observables.}

\section{Schwinger--Dyson equations}

In analyzing the dynamics of the model we will now proceed exactly in the same way as we did in the context of the Abelian toy model in chapter~\ref{chp:frm}. That is to say, we will employ the method of the SD equations, truncated at the level of two-point functions; all other functions, in particular the three-point functions, will be approximated by the bare ones. We will again derive the SD equations using the CJT formalism, with the CJT effective potential calculated in the lowest, i.e., in the Hartree--Fock approximation.

\subsection{SD equations in general}

We will now derive the SD equations using the CJT formalism and again under the deliberate (and unjustified) assumption of vanishing scalar VEVs.\footnote{See the discussion at the beginning of Sec.~\ref{sec:dynabel}} As the procedure will be almost completely analogous to what we did in chapter~\ref{chp:frm}, we will not go into much detail.

The CJT effective potential is given by
\begin{eqnarray}
\label{ewd:V}
V[G_\Phi,G_q,G_{\Psi_\ell}] &=& V_\Phi[G_\Phi] + V_q[G_q] + V_{\Psi_\ell}[G_{\Psi_\ell}] + V_2[G_\Phi,G_q,G_{\Psi_\ell}] \,.
\end{eqnarray}
The first three terms are defined standardly (cf.~\eqref{chp:frm:VpsiiVPhi} in the Abelian case) as
\begin{subequations}
\begin{eqnarray}
V_\Phi[G_\Phi] &=& \I \frac{1}{2} \int\!\frac{\d^4 k}{(2\pi)^4}\,
\Tr \Big\{ \ln(D_\Phi^{-1}G_\Phi^{\phantom{1}})-D_\Phi^{-1}G_\Phi^{\phantom{1}}+1 \Big\} \,,
\\
V_q[G_q] &=& -\I \int\!\frac{\d^4 k}{(2\pi)^4}\,
\Tr \Big\{ \ln(S_q^{-1}G_q^{\phantom{1}})-S_q^{-1}G_q^{\phantom{1}}+1 \Big\} \,,
\\
V_{\Psi_\ell}[G_{\Psi_\ell}] &=& -\I \frac{1}{2} \int\!\frac{\d^4 k}{(2\pi)^4}\,
\Tr \Big\{ \ln(S_{\Psi_\ell}^{-1}G_{\Psi_\ell}^{\phantom{1}})-S_{\Psi_\ell}^{-1}G_{\Psi_\ell}^{\phantom{1}}+1 \Big\} \,.
\end{eqnarray}
\end{subequations}
The factor of $1/2$ at $V_{\Psi_\ell}$ is due to the Majorana character of the field $\Psi_\ell$. The functional $V_2$ is again given by the sum of all 2PI diagrams.

The SD equations correspond to the condition for the stationary point of the effective potential $V$, \eqref{ewd:V}, with respect to the variations of the propagators $G_\Phi$, $G_q $, $G_{\Psi_\ell}$. However, similarly to the Abelian toy model, the space of the allowed variations is not arbitrary. Recall that while the quark field $q$ is unrestricted, yielding also no restriction on its propagator $G_q$, the scalar and lepton fields $\Phi$, $\Psi_{\ell}$ satisfy the conditions \eqref{ew1:NGcond}, \eqref{ew1:majcond}, implying the non-trivial conditions for their propagators \eqref{ew1:GPhiNGcond}, \eqref{ew1:majcondG}, respectively. Therefore in varying the propagators one has to take carefully into account these restrictions.

Technically the procedure of extremizing the effective potential $V$ under the constraints \eqref{ew1:GPhiNGcond}, \eqref{ew1:majcondG} is carried out again using the method of the Lagrange multipliers. Without going into the detail we only state the resulting SD equations:
\begin{subequations}
\label{ewd:SD:general}
\begin{eqnarray}
-\I\,\boldsymbol{\Pi}_\Phi(p) &=&
-(2\pi)^4 \left[\frac{\delta \, V_{2}}{\delta \, G_\Phi^\T(p)} + \Sigma_1 \bigg(\frac{\delta \, V_{2}}{\delta \, G_\Phi^\T(p)}\bigg)^\T \Sigma_1 \right] \,,
\label{ewd:SD:general:Phi}
\\
-\I \, \boldsymbol{\Sigma}_q(p) &=& (2\pi)^4 \frac{\delta \, V_{2}}{\delta \, G_q^\T(p)} \,,
\label{ewd:SD:general:q}
\\
-\I \, \boldsymbol{\Sigma}_{\Psi_\ell}(p) &=& (2\pi)^4
\left[\frac{\delta \, V_{2}}{\delta \, G_{\Psi_\ell}^\T(p)} + \bigg(\frac{\delta \, V_{2}}{\delta \, G_{\Psi_\ell}^\T(-p)}\bigg)^\C \right] \,.
\label{ewd:SD:general:Psiell}
\end{eqnarray}
\end{subequations}
Notice that the form of the equations for $\boldsymbol{\Pi_\Phi}$, $\boldsymbol{\Sigma}_{\Psi_\ell}$ indeed does guarantee the satisfaction of the respective constraints \eqref{ewa:PicondNG}, \eqref{ewa:SgmPsiellmaj}.

\subsection{Hartree--Fock approximation}

\begin{figure}[t]
\begin{center}
\includegraphics[width=0.55\textwidth]{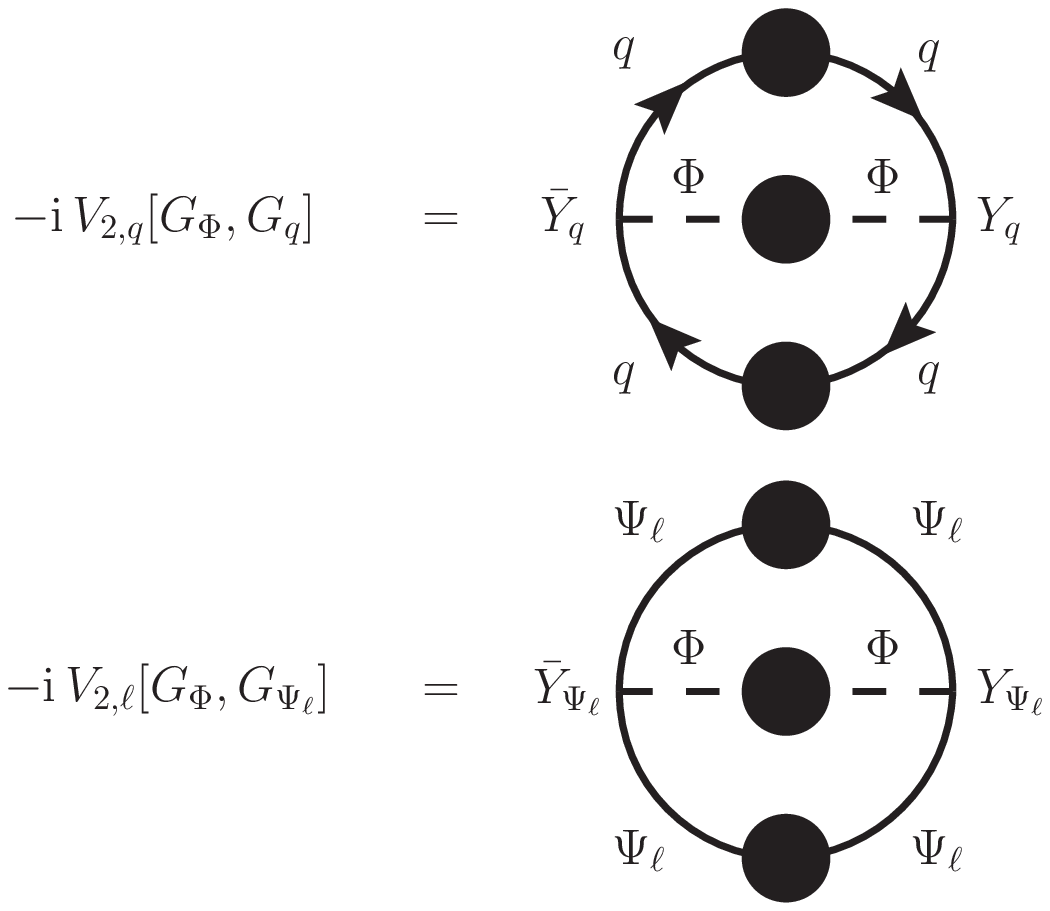}
\caption[Hartree--Fock approximation of $V_2$.]{Diagrammatical representation of $V_{2,q}$ and $V_{2,\ell}$, Eqs.~\eqref{ewd:V2qell}. Note the missing arrows at the lepton lines due to the Majorana (i.e., real) character of the corresponding field $\Psi_\ell$.}
\label{ewd:fig:V2}
\end{center}
\end{figure}

We approximate the functional $V_2$ in \eqref{ewd:V} by the Hartree--Fock approximation, used also in chapter~\ref{chp:frm}. Again, $V_2$ therefore decouples into the sum of the independent contributions from the quarks and the leptons:
\begin{eqnarray}
\label{ewd:V2}
V_2[G_\Phi,G_q,G_{\Psi_\ell}] &=& V_{2,q}[G_\Phi,G_q] + V_{2,\ell}[G_\Phi,G_{\Psi_\ell}] \,,
\end{eqnarray}
with the particular fermion contributions given according to Fig.~\ref{ewd:fig:V2} (and taking into account the Yukawa interactions \eqref{ew1:eL:Ykw:Phiq} and \eqref{ew1:eL:Ykw:PhiPsiell}) as
\begin{subequations}
\label{ewd:V2qell}
\begin{eqnarray}
-\I \, V_{2,q}[G_\Phi,G_q] &=& - \frac{1}{2} \I^5 N_c \int\!\frac{\d^4 k}{(2\pi)^4}\frac{\d^4 p}{(2\pi)^4}\,
\Tr \Big\{ Y_q \, G_q(k) \, \bar Y_q \, G_q(p) \, G_\Phi(k-p) \Big\} \,,
\\
-\I \, V_{2,\ell}[G_\Phi,G_{\Psi_\ell}] &=& - \frac{1}{4} \I^5 \int\!\frac{\d^4 k}{(2\pi)^4}\frac{\d^4 p}{(2\pi)^4}\,
\Tr \Big\{ Y_{\Psi_\ell} \, G_{\Psi_\ell}(k) \, \bar Y_{\Psi_\ell} \, G_{\Psi_\ell}(p) \, G_\Phi(k-p) \Big\} \,,
\end{eqnarray}
\end{subequations}
where $N_c = 3$ is the number of colors. One can compare this with the analogous expression \eqref{chp:frm:V2i} in the Abelian toy model, with the two fermion species $\psi_1$, $\psi_2$ being analogues of $q$, $\Psi_\ell$, respectively.

We can now plug the approximation \eqref{ewd:V2} of $V_2$ into the equations \eqref{ewd:SD:general} and calculate the corresponding functional derivatives in order to arrive at the final form of the SD equations (before taking into account the Ansatz for the propagators). Likewise in the Abelian case, computing of the SD equation \eqref{ewd:SD:general:Phi} for $\boldsymbol{\Pi}_\Phi$ can be somewhat simplified by noting that due to the properties \eqref{ew1:Yqcond}, \eqref{ew1:Yellcond} of the coupling constants $Y_q$, $Y_{\Psi_\ell}$, respectively, and also due to the using of the Hartree--Fock approximation \eqref{ewd:V2qell} there are the following identities:
\begin{subequations}
\begin{eqnarray}
\frac{\delta \, V_{2,q}}{\delta \, G_\Phi^\T} &=& \Sigma_1 \bigg(\frac{\delta \, V_{2,q}}{\delta \, G_\Phi^\T}\bigg)^\T \Sigma_1 \,,
\\
\frac{\delta \, V_{2,\ell}}{\delta \, G_\Phi^\T} &=& \Sigma_1 \bigg(\frac{\delta \, V_{2,\ell}}{\delta \, G_\Phi^\T}\bigg)^\T \Sigma_1 \,.
\end{eqnarray}
\end{subequations}
On top of this, there is also property \eqref{ew1:YellcondMaj} of the lepton coupling constant $Y_\ell$, following from the Majorana character of the field $\Psi_\ell$. It implies, again together with the Hartree--Fock approximation \eqref{ewd:V2qell}, the relation
\begin{eqnarray}
\frac{\delta \, V_{2,\ell}}{\delta \, G_{\Psi_\ell}^\T(p)} &=& \bigg(\frac{\delta \, V_{2,\ell}}{\delta \, G_{\Psi_\ell}^\T(-p)}\bigg)^\C \,,
\end{eqnarray}
allowing to simplify the calculation of the SD equation \eqref{ewd:SD:general:Psiell} for $\boldsymbol{\Sigma}_{\Psi_\ell}$.

\begin{figure}[t]
\begin{center}
\includegraphics[width=1\textwidth]{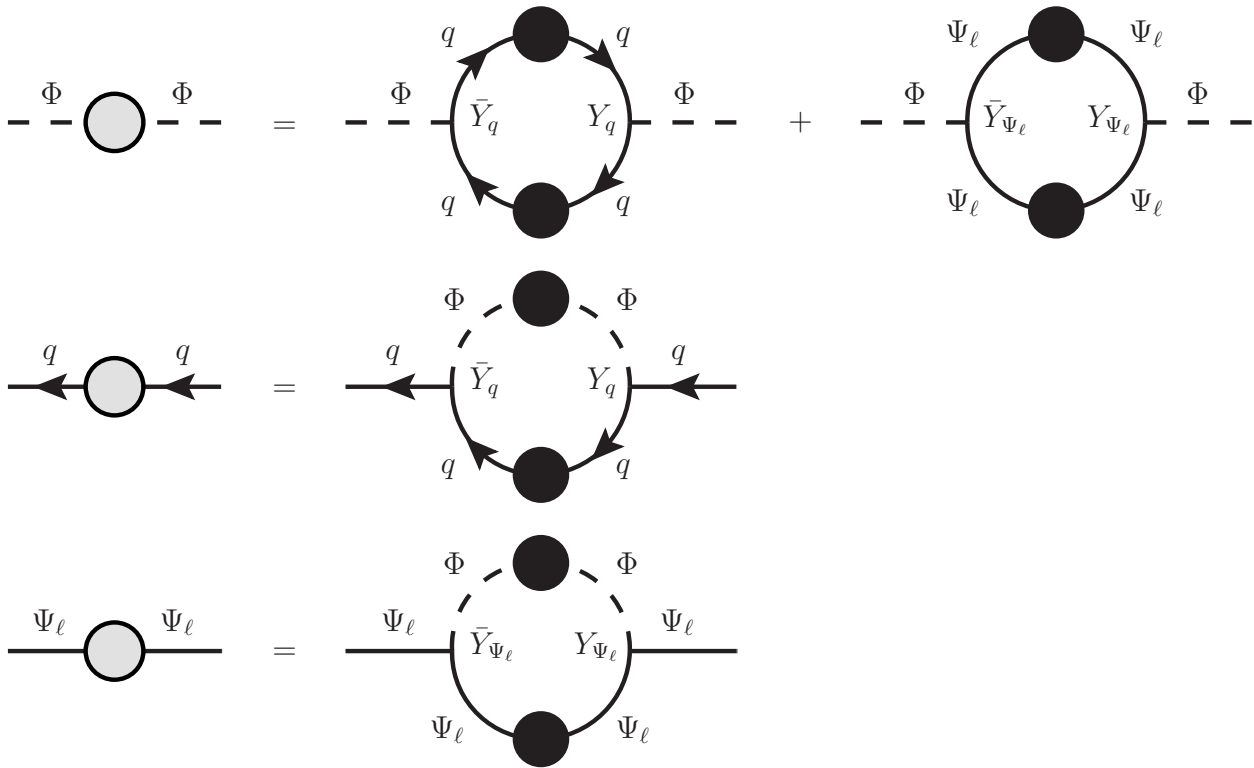}
\caption[SD equations in the Hartree--Fock approximation.]{The SD equations \eqref{ewd:SD:HF} for $\boldsymbol{\Pi}_\Phi$, $\boldsymbol{\Sigma}_q$, $\boldsymbol{\Sigma}_{\Psi_\ell}$ in the Hartree--Fock approximation, yet without employment of a specific Ansatz for the self-energies.}
\label{ewd:fig:SDHF}
\end{center}
\end{figure}

As a net result, we arrive at the following set of SD equations:\footnote{We do not indicate the momentum dependencies in the SD equations (here as well as elsewhere in this chapter) as they can be easily revealed from the corresponding Feynman diagrams.}
\begin{subequations}
\label{ewd:SD:HF}
\begin{eqnarray}
-\I\,\boldsymbol{\Pi}_\Phi &=&
-N_c\int\!\frac{\d^4 k}{(2\pi)^4}\, \Tr_\psi \Big\{ \, Y_q \, G_q \, \bar Y_q \, G_q \Big\}
-\frac{1}{2}\int\!\frac{\d^4 k}{(2\pi)^4}\, \Tr_\psi \Big\{ \, Y_{\Psi_\ell} \, G_{\Psi_\ell} \, \bar Y_{\Psi_\ell} \, G_{\Psi_\ell} \Big\} \,,
\qquad
\label{ewd:SD:HF:Phi} \\
-\I\,\boldsymbol{\Sigma}_q &=& \int\!\frac{\d^4 k}{(2\pi)^4}\, \Tr_\Phi \Big\{ \, Y_q \, G_q \, \bar Y_q \, G_\Phi \Big\} \,,
\label{ewd:SD:HF:q} \\
-\I\,\boldsymbol{\Sigma}_{\Psi_\ell} &=& \int\!\frac{\d^4 k}{(2\pi)^4}\, \Tr_\Phi \Big\{ \, Y_{\Psi_\ell} \, G_{\Psi_\ell} \, \bar Y_{\Psi_\ell} \, G_\Phi \Big\} \,,
\label{ewd:SD:HF:Psiell}
\end{eqnarray}
\end{subequations}
see Fig.~\ref{ewd:fig:SDHF}.

\subsection{Employing the Ansatz}


We now restrict the general SD equations \eqref{ewd:SD:HF} to the self-energy Ans\"{a}tze derived in the previous chapter. This means two things:
\begin{enumerate}
  \item We substitute the full propagators $G_\Phi$, $G_q$, $G_{\Psi_\ell}$ in the SD equations \eqref{ewd:SD:HF} by the expressions \eqref{ewa:GPhi}, \eqref{ewa:Gqdiag}, \eqref{ewa:Gpsiell}, corresponding to the self-energies Ans\"{a}tze \eqref{ewa:PiPhi}, \eqref{ewa:Sgmqdiag}, \eqref{ewa:SgmPsiell}. \label{ewd:enumSD:rhs}
  \item We keep only those equations, which contribute to the parts of $\boldsymbol{\Pi}_\Phi$, $\boldsymbol{\Sigma}_q$, $\boldsymbol{\Sigma}_{\Psi_\ell}$, consistent with the corresponding Ansatz. \label{ewd:enumSD:lhs}
\end{enumerate}
In other words, we plug the Ansatz derived in chapter~\ref{chp:ewa} to both sides of the SD equations \eqref{ewd:SD:HF} and keep only those equations with non-vanishing left-hand side.


\begin{figure}[t]
\begin{center}
\includegraphics[width=0.92\textwidth]{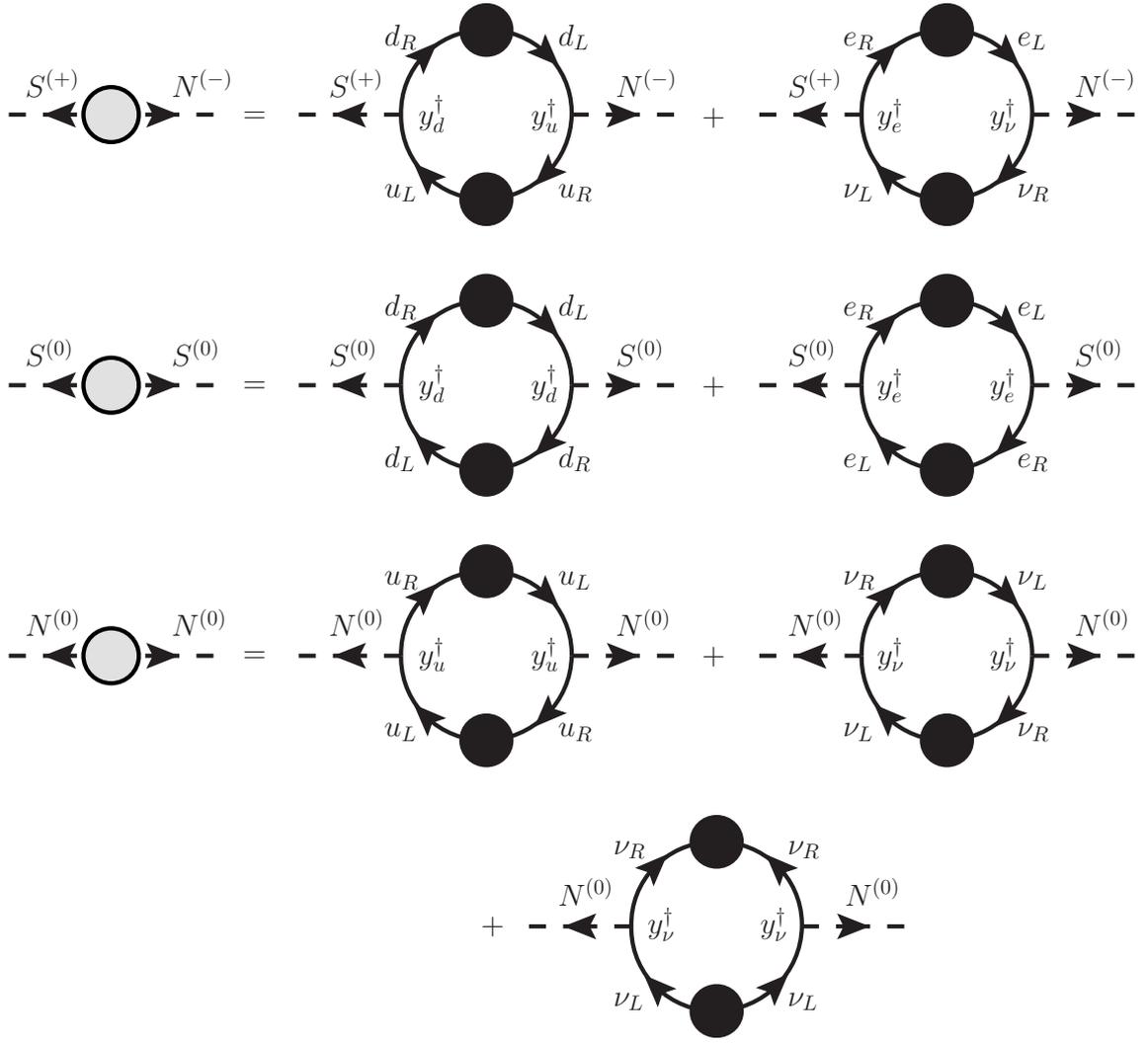}
\caption[SD equations for the symmetry-breaking parts of the scalar propagators.]{The SD equations \eqref{ewd:SD:PiSNSN} for the scalar self-energies $\Pi_{SN}$, $\Pi_{S}$, $\Pi_{N}$, respectively.}
\label{ewd:fig:SD:PiSNSN}
\end{center}
\end{figure}

The Ansatz \eqref{ewa:PiPhi} for the scalar self-energy $\boldsymbol{\Pi}_\Phi$ consists of the three independent components $\Pi_{SN}$, $\Pi_{S}$, $\Pi_{N}$, see \eqref{ewa:PhiSNSN}. The single matrix equation \eqref{ewd:SD:HF:Phi} for $\boldsymbol{\Pi}_\Phi$ thus yields the following three non-matrix SD equations:
\begin{subequations}
\label{ewd:SD:PiSNSN}
\begin{eqnarray}
-\I\,\Pi_{SN} &=&
-2N_c\int\!\frac{\d^4 k}{(2\pi)^4}\, \Tr\Big\{ y_d^\dag\,\Sigma_u\,D_{uR}\,y_u^\dag\,\Sigma_d\,D_{dR}\Big\}
\nonumber \\ &&
-2\int\!\frac{\d^4 k}{(2\pi)^4}\, \Tr\Big\{ (\Sigma_{\nu L}\,D_{\nu M}^*+\Sigma_{\nu D}\,D_{\nu R}) y_\nu^\dag\,\Sigma_{e}\,D_{eR}\,y_e^\dag\Big\} \,,
\\
-\I\,\Pi_{S} &=&
-2N_c\int\!\frac{\d^4 k}{(2\pi)^4}\, \Tr\Big\{ y_d^\dag\,\Sigma_d\,D_{dR}\,y_d^\dag\,\Sigma_d\,D_{dR}\Big\}
\nonumber \\ &&
-2\int\!\frac{\d^4 k}{(2\pi)^4}\, \Tr\Big\{ y_e^\dag\,\Sigma_{e}\,D_{eR}\,y_e^\dag\,\Sigma_{e}\,D_{eR}\Big\} \,,
\\
-\I\,\Pi_{N} &=&
-2N_c\int\!\frac{\d^4 k}{(2\pi)^4}\, \Tr\Big\{ y_u^\dag\,\Sigma_u\,D_{uR}\,y_u^\dag\,\Sigma_u\,D_{uR}\Big\}
\nonumber \\ &&
-2\int\!\frac{\d^4 k}{(2\pi)^4}\, \Tr\Big\{ y_\nu^\dag (\Sigma_{\nu L}\,D_{\nu M}^*+\Sigma_{\nu D}\,D_{\nu R}) y_\nu^\dag (\Sigma_{\nu L}\,D_{\nu M}^*+\Sigma_{\nu D}\,D_{\nu R})\Big\}
\nonumber \\ &&
-2\int\!\frac{\d^4 k}{(2\pi)^4}\, \Tr\Big\{ y_\nu^\dag (\Sigma_{\nu L}\,D_{\nu L}^\T+\Sigma_{\nu D}\,D_{\nu M}^\T) y_\nu^* (\Sigma_{\nu D}^\T\,D_{\nu M}^*+M_{\nu R}\,D_{\nu R})\Big\} \,,
\qquad
\end{eqnarray}
\end{subequations}
see Fig.~\ref{ewd:fig:SD:PiSNSN}. Notice the employment of the notation for the propagator \qm{denominators}, introduced in sections~\ref{ewa:final:Phi}, \ref{ewa:final:q}, \ref{ewa:final:Psiell}

\begin{figure}[t]
\begin{center}
\includegraphics[width=0.92\textwidth]{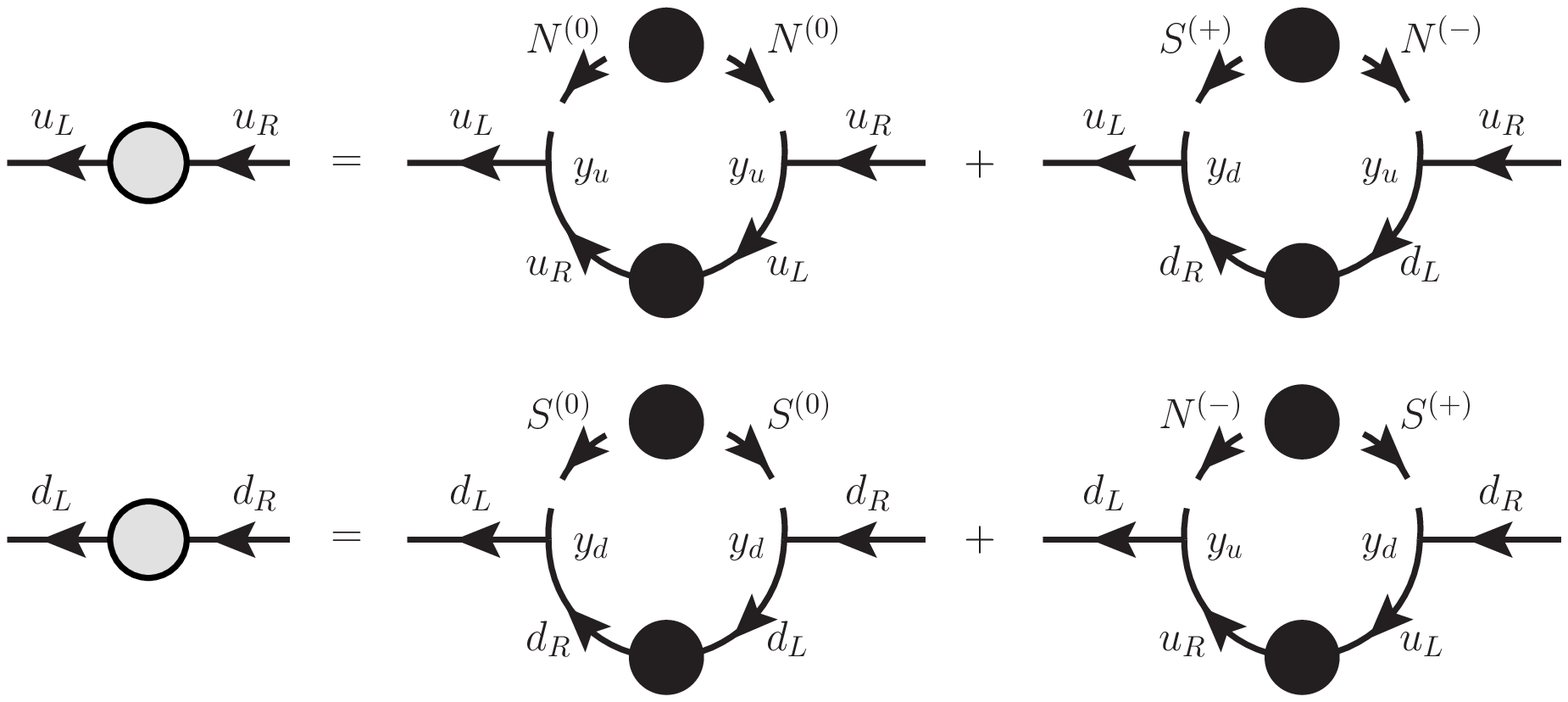}
\caption[SD equations for the symmetry-breaking parts of the quark propagators.]{The SD equations \eqref{ewd:SD:Sgmud} (up to the missing overall factor of $P_R$ in \eqref{ewd:SD:Sgmud}) for the quark self-energies $\Sigma_u$ and $\Sigma_d$, respectively.}
\label{ewd:fig:SD:Sgmud}
\end{center}
\end{figure}

The Ansatz \eqref{ewa:Sgmqdiag} for the quark self-energy $\boldsymbol{\Sigma}_q$ consists of the two flavor-matrix functions $\Sigma_u$, $\Sigma_d$. Correspondingly, the equation \eqref{ewd:SD:HF:q} for $\boldsymbol{\Sigma}_q$ gives rise to the two SD equations
\begin{subequations}
\label{ewd:SD:Sgmud}
\begin{eqnarray}
-\I\,\Sigma_u &=&
\int\!\frac{\d^4 k}{(2\pi)^4}\, y_u\,D_{uR}\,\Sigma_u^\dag\,y_u \,\Pi_{N}\,D_{N} +
\int\!\frac{\d^4 k}{(2\pi)^4}\, y_d\,D_{dR}\,\Sigma_d^\dag\,y_u \,\Pi_{SN}\,D_{SN} \,,
\\
-\I\,\Sigma_d &=&
\int\!\frac{\d^4 k}{(2\pi)^4}\, y_d\,D_{dR}\,\Sigma_d^\dag\,y_d \,\Pi_{S}\,D_{S} +
\int\!\frac{\d^4 k}{(2\pi)^4}\, y_u\,D_{uR}\,\Sigma_u^\dag\,y_d \,\Pi_{SN}\,D_{SN} \,,
\end{eqnarray}
\end{subequations}
see Fig.~\ref{ewd:fig:SD:Sgmud}.

\begin{figure}[t]
\begin{center}
\includegraphics[width=0.92\textwidth]{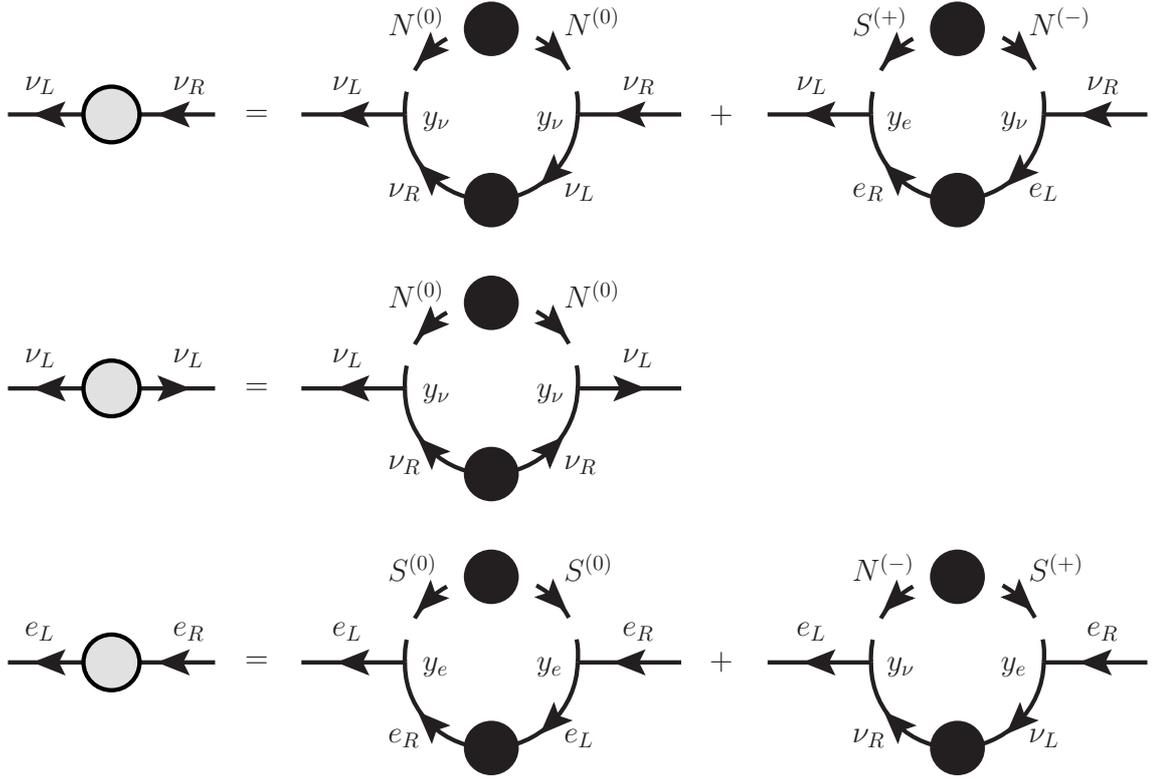}
\caption[SD equations for the symmetry-breaking parts of the lepton propagators.]{The SD equations \eqref{ewd:SD:Sgmnue} (up to the missing overall factor of $P_R$ in \eqref{ewd:SD:Sgmnue}) for the lepton self-energies $\Sigma_{\nu D}$, $\Sigma_{\nu L}$ and $\Sigma_e$, respectively.}
\label{ewd:fig:SD:Sgmnue}
\end{center}
\end{figure}

And finally, the Ansatz \eqref{ewa:SgmPsiell} for the lepton self-energy $\boldsymbol{\Sigma}_{\Psi_\ell}$ consists again of two flavor-matrix functions $\Sigma_{\Psi_\nu}$, $\Sigma_{\Psi_e}$. This time, however, the self-energies $\Sigma_{\Psi_\nu}$, $\Sigma_{\Psi_e}$ are \qm{reducible} in the sense that each of them contains mutually dependent parts. To see this explicitly, consider their respective matrix forms \eqref{ewa:SgmPsinuPsie}: the Dirac parts $\Sigma_{\nu D}$, $\Sigma_e$ are contained twice in each $\Sigma_{\Psi_\nu}$, $\Sigma_{\Psi_e}$. Moreover, some of the blocks of $\Sigma_{\Psi_\nu}$, $\Sigma_{\Psi_e}$ are vanishing. Thus, it is convenient to decompose $\Sigma_{\Psi_\nu}$, $\Sigma_{\Psi_e}$ into the block in Nambu--Gorkov space and consider only those independent and non-vanishing, i.e., the mentioned Dirac parts $\Sigma_{\nu D}$, $\Sigma_e$ and the Neutrino left-handed Majorana part $\Sigma_{\nu L}$. We have therefore the following three independent equations:
\begin{subequations}
\label{ewd:SD:Sgmnue}
\begin{eqnarray}
-\I\,\Sigma_{\nu D} &=&
\int\!\frac{\d^4 k}{(2\pi)^4}\, y_\nu (D_{\nu R}\,\Sigma_{\nu D}^\dag+D_{\nu M}^\T\,\Sigma_{\nu L}^\dag) y_\nu \,\Pi_{N}\,D_{N} +
\int\!\frac{\d^4 k}{(2\pi)^4}\, y_e\,D_{eR}\,\Sigma_e^\dag\,y_\nu \,\Pi_{SN}\,D_{SN} \,,
\nonumber \\ &&
\\
-\I\,\Sigma_{\nu L} &=&
\int\!\frac{\d^4 k}{(2\pi)^4}\, y_\nu (D_{\nu M}^\T\,\Sigma_{\nu D}^*+D_{\nu R}\,M_{\nu R}^\dag) y_\nu^\T \,\Pi_{N}\,D_{N} \,,
\label{ewd:SD:Sgmnue:L}
\nonumber \\ &&
\\
-\I\,\Sigma_e &=&
\int\!\frac{\d^4 k}{(2\pi)^4}\, y_e\,D_{eR}\,\Sigma_e^\dag\,y_e \,\Pi_{S}\,D_{S} +
\int\!\frac{\d^4 k}{(2\pi)^4}\, y_\nu (D_{\nu R}\,\Sigma_{\nu D}^\dag+D_{\nu M}^\T\,\Sigma_{\nu L}^\dag) y_e \,\Pi_{SN}\,D_{SN} \,,
\nonumber \\ &&
\end{eqnarray}
\end{subequations}
see Fig.~\ref{ewd:fig:SD:Sgmnue}. (In fact, the matrix equation for $\Sigma_{\nu L}$ still includes some mutually dependent equations, since $\Sigma_{\nu L} = \Sigma_{\nu L}^\T$.)

\subsection{Why two scalar doublets?}
\label{ssec:whytwo}

Now it is time to comment on why we considered two scalar doublets $S$ and $N$ with opposite hypercharges, instead of only one, like in the SM. Thus, assume for a moment that the scalar doublet $N$ is missing and the only scalar doublet in the theory is $S$. Relax also the requirement of invariance under the discrete symmetry $\mathcal{P}_{\mathrm{down}}$. Then the Yukawa interactions of both $S$ and $\tilde S$ are present and have the same form as those in the SM. Consequently, the doublet $S$ can be in such a case regarded as a direct analogue of the SM Higgs doublet.

\begin{figure}[t]
\begin{center}
\includegraphics[width=0.92\textwidth]{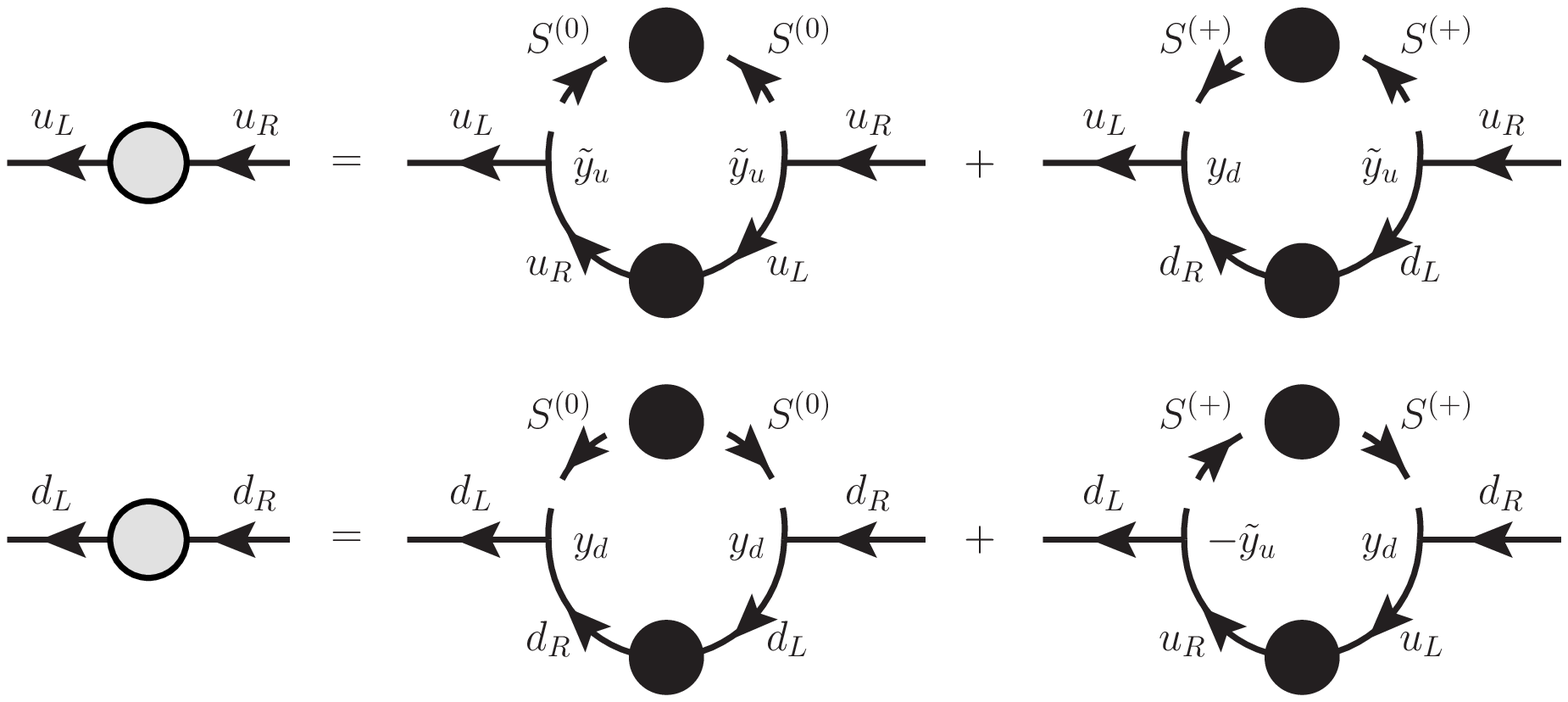}
\caption[SD equations for the symmetry-breaking parts of the quark propagators in the case of only the scalar doublet $S$.]{The SD equations \eqref{ewd:SD:Sgmud_S} for the quark self-energies $\Sigma_u$ and $\Sigma_d$, respectively, in the case of only one scalar doublet $S$ and without the discrete symmetry $\mathcal{P}_{\mathrm{down}}$.}
\label{ewd:fig:SD:Sgmud_S}
\end{center}
\end{figure}

The SD equations \eqref{ewd:SD:Sgmud} for the quark self-energies $\Sigma_u$, $\Sigma_d$ (we consider for simplicity only the quarks, as the case of neutrinos would be analogous) then modify as
\begin{subequations}
\label{ewd:SD:Sgmud_S}
\begin{eqnarray}
-\I\,\Sigma_u &=&
\int\!\frac{\d^4 k}{(2\pi)^4}\, \tilde y_u\,D_{uR}\,\Sigma_u^\dag\, \tilde y_u \,\Pi_{S}^*\,D_{S} +
\int\!\frac{\d^4 k}{(2\pi)^4}\, y_d\,D_{dR}\,\Sigma_d^\dag\,\tilde y_u \,D_{S}^0 \,,
\\
-\I\,\Sigma_d &=&
\int\!\frac{\d^4 k}{(2\pi)^4}\, y_d\,D_{dR}\,\Sigma_d^\dag\,y_d \,\Pi_{S}\,D_{S} -
\int\!\frac{\d^4 k}{(2\pi)^4}\, \tilde y_u\,D_{uR}\,\Sigma_u^\dag\,y_d \,D_{S}^0 \,.
\end{eqnarray}
\end{subequations}
In understanding these equations and their relation to the corresponding diagrams in Fig.~\ref{ewd:fig:SD:Sgmud_S} it is useful to take into account that in the case of missing $N$ we have effectively $M_N = \Pi_{SN} = 0$, so that $(p^2-M_N^2)\,D_{SN} = D_{S}^0$ and accordingly the expression \eqref{ewa:S+S+full} for the propagator $\langle S^{(+)} S^{(+)\dag} \rangle$ reduces to
\begin{eqnarray}
\label{ewd:S+S+}
\langle S^{(+)} S^{(+)\dag} \rangle
\ = \
\begin{array}{c}
\scalebox{0.85}{\includegraphics[trim = 10bp 12bp 14bp 11bp,clip]{fig_ew/scalar_S+S+.eps}}
\end{array}
&=& \I\,D_{S}^0 \,,
\end{eqnarray}
where $D_{S}^0$ is the bare propagator \eqref{ewq:Dscalbare}.


First of all we see that in the case of only one scalar doublet $S$ we must not postulate the discrete symmetry $\mathcal{P}_{\mathrm{down}}$, as otherwise we would have $\tilde y_u = \tilde y_d = 0$ and consequently only the down-type quarks would become massive, while the up-type quarks would remain massless.

What is important, however, is the presence of the symmetry-preserving propagator $\langle S^{(+)} S^{(+)\dag} \rangle = \I\,D_{S}^0$, \eqref{ewd:S+S+}, which behaves asymptotically as $1/k^2$. This is to be compared with the symmetry-breaking scalar propagator $\langle S^{(+)} N^{(-)} \rangle = \I\,\Pi_{SN}\,D_{SN}$ in equations \eqref{ewd:SD:Sgmud}, which behaves asymptotically\footnote{Recall that this asymptotic behavior was interpreted in chapter \ref{chp:inf} to be due to difference of two propagators, corresponding to the scalar mass eigenstates.} as $1/k^4$. In other words, the integrals in \eqref{ewd:SD:Sgmud_S} have worse asymptotic behavior than the integrals in \eqref{ewd:SD:Sgmud}. This is the very reason why we considered two scalar doublets instead of one, since the better asymptotic behavior of integrands in the SD equations \eqref{ewd:SD:Sgmud} makes the existence of UV-finite solutions more probable and accordingly the whole proposed mechanism of dynamical EWSB more viable.

\subsection{Non-Ansatz SD equations}

This section is just an informational aside with the aim to show explicitly that:
\begin{itemize}
  \item Those symmetry-breaking parts of the self-energies, not included in our regular self-energies Ans\"{a}tze and consequently also not in the SD equations presented in the previous section, are indeed UV-finite. This applies namely to the wave function renormalization parts of the self-energies, i.e., to the scalar functions $A_S$, $A_N$, the quark function $A_q$ and the lepton functions $A_\ell$, $A_{\nu M}$.
  \item The symmetry-preserving parts of the self-energies are UV-divergent.
\end{itemize}
For the sake of simplicity we do not present here the SD equations for the non-Ansatz parts of the self-energies in a self-consistent way, but rather investigate how the loops with the Ansatz propagators (i.e., as presented in sections~\ref{ewa:final:Phi}, \ref{ewa:final:q}, \ref{ewa:final:Psiell}) contribute to them. Put more formally, we consider the right-hand sides of the SD equations \eqref{ewd:SD:HF} as described in point~\ref{ewd:enumSD:rhs}.~on page~\pageref{ewd:enumSD:lhs}, but, in contrast to what is described in subsequent point~\ref{ewd:enumSD:lhs}., we consider for the left-hand sides the self-energies of the general Hermitian and electromagnetically invariant forms, i.e., as presented at the ends of sections~\ref{ewa:ssec:Phigen}, \ref{ewa:ssec:qgen}, \ref{ewa:ssec:Psiellgen}.

\subsubsection{Scalar self-energies}

Let us start with the scalars. The SD equations for the relevant individual functions $A_1$, $A_3$, $C_1$, $D_1$ (see \eqref{ewa:ACDE}), obtained from \eqref{ewd:SD:HF:Phi} by the procedure described above, then read\footnote{We consider here for simplicity only the quark contributions; for the leptons the argument would be essentially the same, only unnecessarily obscured due to the technicalities connected with the Nambu--Gorkov formalism.\label{ftnt:onlyquarks}}
\begin{subequations}
\label{ewd:SD:A1A3C1D1}
\begin{eqnarray}
-\I\,A_1 &=&
-2\int\!\frac{\d^4 k}{(2\pi)^4}\, (k \cdot \ell) \Tr \Big\{ y_d^\dag \, D_{uL} \, y_d \, D_{dR} \Big\} \,, \\
-\I\,A_3 &=&
-2\int\!\frac{\d^4 k}{(2\pi)^4}\, (k \cdot \ell) \Tr \Big\{ y_u^\dag \, D_{dL} \, y_u \, D_{uR} \Big\} \,, \\
-\I\,C_1 &=&
-2\int\!\frac{\d^4 k}{(2\pi)^4}\, (k \cdot \ell) \Tr \Big\{ y_d^\dag \, D_{dL} \, y_d \, D_{dR} \Big\} \,, \\
-\I\,D_1 &=&
-2\int\!\frac{\d^4 k}{(2\pi)^4}\, (k \cdot \ell) \Tr \Big\{ y_u^\dag \, D_{uL} \, y_u \, D_{uR} \Big\} \,,
\end{eqnarray}
\end{subequations}
where $\ell \equiv p-k$ and $p$ is the external momentum. We see that each of the four integrals in \eqref{ewd:SD:A1A3C1D1} is separately divergent, since the propagators $D_{uL}$, $D_{dL}$, $D_{uR}$, $D_{dR}$ are not suppressed by any self-energy and hence behave as $1/k^2$ for large $k^2$.

However, we know that the combinations $2 A_S = A_1-C_1$, $2 A_N = A_3-D_1$ (Eq.~\eqref{ewa:ASANdef}) should be UV-finite, since they break the symmetry. Indeed, it is the case:
\begin{subequations}
\label{ewd:SD:ASAN}
\begin{eqnarray}
-\I\,A_S &=&
-\int\!\frac{\d^4 k}{(2\pi)^4}\, (k \cdot \ell) \Tr \Big\{ y_d^\dag \, (D_{uL}-D_{dL}) \, y_d \, D_{dR} \Big\} \,, \\
-\I\,A_N &=&
+\int\!\frac{\d^4 k}{(2\pi)^4}\, (k \cdot \ell) \Tr \Big\{ y_u^\dag \, (D_{uL}-D_{dL}) \, y_u \, D_{uR} \Big\} \,.
\end{eqnarray}
\end{subequations}
The UV-finiteness is consequence of the fact that the \emph{difference} of the two propagators $D_{uL}$, $D_{dL}$ is already suppressed by fermion symmetry-breaking and consequently also decreasing, UV-finite self-energies $\Sigma_u$, $\Sigma_d$:
\begin{eqnarray}
D_{uL}-D_{dL} &=&
\Sigma_u^{\phantom{\dag}}\,\Sigma_u^{\dag}\,D_{uL}\,D_{dL} - D_{uL}\,D_{dL}\,\Sigma_d^{\phantom{\dag}}\,\Sigma_d^{\dag} \,.
\end{eqnarray}
Correspondingly $D_{uL}-D_{dL}$ falls faster than $1/k^2$, rendering the integrals \eqref{ewd:SD:ASAN} UV-finite. Obviously, the symmetry-preserving combinations $A_1+C_1$, $A_3+D_1$ (Eqs.~\eqref{ewa:A1+C1}, \eqref{ewa:A3+D1}) remain divergent, as the sum $D_{uL}+D_{dL}$ still behaves like $1/k^2$.

\subsubsection{Fermion Dirac self-energies}

Let us consider only the case of quarks, for similar reasons as mentioned in footnote~\ref{ftnt:onlyquarks} on page~\pageref{ftnt:onlyquarks}. The integrals for the particular functions $A_{uL}$, $A_{uR}$, $A_{dL}$, $A_{dR}$, Eq.~\eqref{ewa:fSgmAnzintermed}, read
\begin{subequations}
\begin{eqnarray}
-\I\slashed{p}\,A_{uL} &=& \int\!\frac{\d^4 k}{(2\pi)^4}\, \slashed{k}
\Big[ y_d\,D_{dR}\,y_d^\dag\,(\ell^2-M_N^2)\,D_{SN} + y_u\,D_{uR}\,y_u^\dag\,(\ell^2-M_N^2)\,D_{N} \Big] \,,
\qquad
\\
-\I\slashed{p}\,A_{uR} &=& \int\!\frac{\d^4 k}{(2\pi)^4}\, \slashed{k}
\Big[ y_u^\dag\,D_{dL}\,y_u\,(\ell^2-M_S^2)\,D_{SN} + y_u^\dag\,D_{uL}\,y_u\,(\ell^2-M_N^2)\,D_{N} \Big] \,,
\\
-\I\slashed{p}\,A_{dL} &=& \int\!\frac{\d^4 k}{(2\pi)^4}\, \slashed{k}
\Big[ y_u\,D_{uR}\,y_u^\dag\,(\ell^2-M_S^2)\,D_{SN} + y_d\,D_{dR}\,y_d^\dag\,(\ell^2-M_S^2)\,D_{S} \Big] \,,
\\
-\I\slashed{p}\,A_{dR} &=& \int\!\frac{\d^4 k}{(2\pi)^4}\, \slashed{k}
\Big[ y_d^\dag\,D_{uL}\,y_d\,(\ell^2-M_N^2)\,D_{SN} + y_d^\dag\,D_{dL}\,y_d\,(\ell^2-M_S^2)\,D_{S} \Big] \,.
\end{eqnarray}
\end{subequations}
Clearly, all the four self-energies $A_{uL}$, $A_{uR}$, $A_{dL}$, $A_{dR}$ are separately divergent. However, it is again easy to see that the symmetry-breaking combination $2 A_q = A_{uL}-A_{dL}$ (Eq.~\eqref{ewa:Aqdef}) is UV-finite, as it must be:
\begin{eqnarray}
\label{ewd:SD:Aq}
-\I\slashed{p}\,A_q &=&
\phantom{-\,}
\frac{1}{2}\int\!\frac{\d^4 k}{(2\pi)^4}\, \slashed{k} \,
y_d\,D_{dR}\,y_d^\dag \Big[(\ell^2-M_N^2)\,D_{SN}-(\ell^2-M_S^2)\,D_{S}\Big]
\nonumber \\ &&
- \,\frac{1}{2}\int\!\frac{\d^4 k}{(2\pi)^4}\, \slashed{k} \,
y_u\,D_{uR}\,y_u^\dag \Big[(\ell^2-M_S^2)\,D_{SN}-(\ell^2-M_N^2)\,D_{N}\Big] \,.
\end{eqnarray}
Again, crucial are the differences of the scalar propagators in the square brackets:
\begin{subequations}
\begin{eqnarray}
(\ell^2-M_N^2)\,D_{SN} - (\ell^2-M_S^2)\,D_{S}
&=&
\Big[(\ell^2-M_S^2)|\Pi_{SN}|^2-(\ell^2-M_N^2)|\Pi_{S}|^2\Big] D_{SN}\,D_{S} \,,\qquad
\\
(\ell^2-M_S^2)\,D_{SN} - (\ell^2-M_N^2)\,D_{N}
&=&
\Big[(\ell^2-M_N^2)|\Pi_{SN}|^2-(\ell^2-M_S^2)|\Pi_{N}|^2\Big] D_{SN}\,D_{N} \,.
\qquad\qquad
\end{eqnarray}
\end{subequations}
Since these quantities are suppressed by the presumably decreasing self-energies $\Pi_{SN}$, $\Pi_{S}$, $\Pi_{N}$, the integrals \eqref{ewd:SD:Aq} for the symmetry-breaking self-energy $A_q$ are indeed UV-finite. Needless to say that the symmetry-preserving self-energies $A_{uL}+A_{dL}$, $A_{uR}$, $A_{dR}$ remain UV-divergent.


\subsubsection{Fermion Majorana self-energy $A_{\nu M}$}

Finally we mention the neutrino Majorana self-energy $A_{\nu M}$. Its SD equation reads
\begin{eqnarray}
\label{ewd:SD:AnuM}
-\I\slashed{p}\,A_{\nu M} &=&
\int\!\frac{\d^4 k}{(2\pi)^4} \, \slashed{k} \, y_\nu^* \, D_{\nu M}^\dag \, y_\nu (\ell^2-M_N^2) D_{N} \,.
\end{eqnarray}
Recall that $A_{\nu M}$ is symmetry-breaking and as such it should be also UV-finite. This is actually the case: The necessary suppression of the integrand in \eqref{ewd:SD:AnuM} is this time maintained by the propagator $D_{\nu M}$, which is proportional to (and thus suppressed by) the self-energies $\Sigma_{\nu L}$ and $\Sigma_{\nu D}$, as shown in \eqref{app:frm:DMpropto} in appendix~\ref{app:fermi propag}.

\subsection{Beyond one loop}
\label{ssec:beyond1loop}

In this section we make two unsystematic remarks connected in some way with the three-loop effective potential, corresponding to two-loop SD equations.

The Hartree--Fock (i.e., two-loop) approximation of the effective potential \eqref{ewd:V2} leads to the one-loop SD equations \eqref{ewd:SD:HF}. It turns out that by neglecting higher-loop contributions some self-energies obtain rather special accidental properties, which are not protected once the diagrams with more loops are taken into account. This applies namely to the right-handed neutrino Majorana self-energy $\Sigma_{\nu R}$, which is at one loop UV-finite, and to the scalar self-energy $E$, which is at one loop vanishing.

\subsubsection{Right-handed neutrino Majorana self-energy $\Sigma_{\nu R}$}


Consider the right-handed neutrino Majorana self-energy $\Sigma_{\nu R}$:
\begin{eqnarray}
-\I \Sigma_{\nu R}\,P_R &=& \langle (\nu_R)^\C \bar\nu_R \rangle_{\mathrm{1PI}} \,.
\end{eqnarray}
Recall that since we have broken explicitly the lepton number by including into the Lagrangian the right-handed Majorana mass terms \eqref{ew1:eL:MnuR}, this self-energy does not break any of the symmetries of the theory and hence can be in general UV-divergent. For this reason we have not included it in section~\ref{ewa:final:Psiell} in our self-energy Ansatz.

Nevertheless, pretend for a moment that we did include $\Sigma_{\nu R}$ into our Ansatz. The corresponding SD equation then reads
\begin{eqnarray}
\label{ewd:SD:SgmnuR}
-\I\Sigma_{\nu R} &=&
\int\!\frac{\d^4 k}{(2\pi)^4}\, y_\nu^\T \big(D_{\nu L}^\T\,\Sigma_{\nu L}^\dag+D_{\nu M}\,\Sigma_{\nu d}^\dag\big) y_\nu \,\Pi_{N}\,D_{N} \,.
\end{eqnarray}
We see that $\Sigma_{\nu R}$ comes out from \eqref{ewd:SD:SgmnuR} as UV-finite!

The point is that the UV-divergent part of $\Sigma_{\nu R}$ should be calculable perturbatively, i.e., using only the bare propagators and vertices, defined by the Lagrangian. However, it turns out that in one-loop approximation (i.e., in the second order in the expansion in the Yukawa coupling constants) there are actually no perturbative corrections to $\Sigma_{\nu R}$. Therefore, since our SD equations \eqref{ewd:SD:HF} are only one-loop (which corresponds to the two-loop Hartree--Fock approximation \eqref{ewd:V2} of the effective potential), they do not include the perturbative contributions and thus the equation \eqref{ewd:SD:SgmnuR} for $\Sigma_{\nu R}$ is coincidentally UV-finite.

\subsubsection{The scalar self-energy $E$}

Consider the scalar self-energy $E$, \eqref{ewa:E},
\begin{eqnarray}
E &=& \left(\begin{array}{cc} E_1 & E_2 \\ E_2^* & E_1^* \end{array}\right) \,,
\end{eqnarray}
where
\begin{subequations}
\begin{eqnarray}
-\I\,E_1 &=& \langle N^{(0)\dag} S^{(0)} \rangle_{\mathrm{1PI}} \,, \\
-\I\,E_2 &=& \langle N^{(0)}     S^{(0)} \rangle_{\mathrm{1PI}} \,.
\end{eqnarray}
\end{subequations}
This self-energy is symmetry-breaking and hence UV-finite. Recall that we have not included it in the Ansatz \eqref{ewa:PiPhi}, arguing that in the Hartree--Fock approximation there are no contributions to it anyway. Let us now discuss this issue in more details.

Consider the Yukawa interactions \eqref{eq:Yukawa} of the scalar doublets $S$, $N$. In particular, we are interested only in the part describing interactions of the neutral components $S^{(0)}$, $N^{(0)}$, i.e., in \eqref{eq:Yukawa_neutral}. Notice that $S^{(0)}$ interacts at tree-level only with the down-type fermions ($d$, $e$), while $N^{(0)}$ only with the up-type fermions ($u$, $\nu$). Therefore it is clear that it is impossible to make a one-loop correction to the propagators of the type $\langle N^{(0)\dag} S^{(0)} \rangle$, $\langle N^{(0)} S^{(0)} \rangle$, simply as one cannot bilinearly connect propagators of different fermions.

\begin{figure}[t]
\begin{center}
\includegraphics[width=1\textwidth]{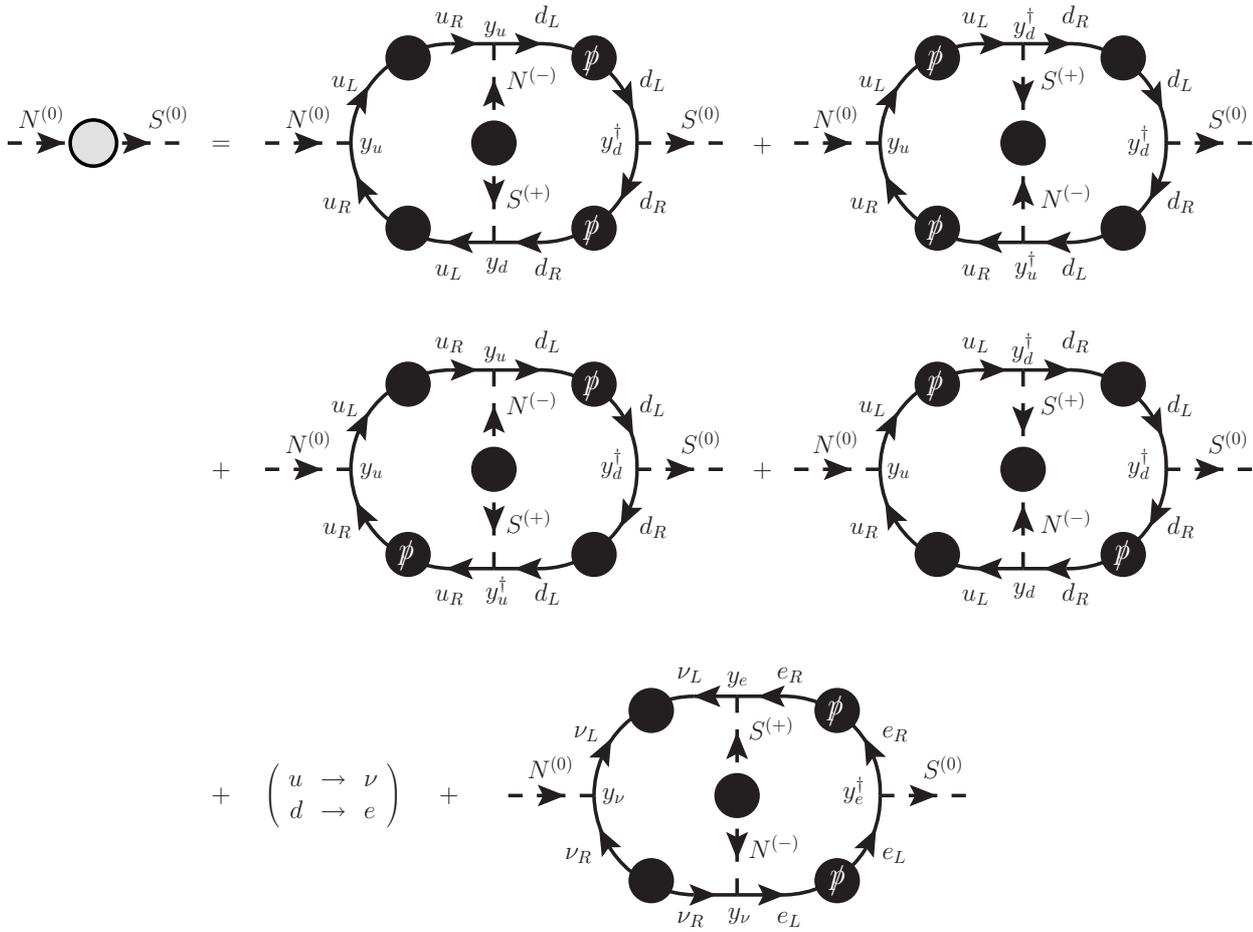}
\caption[Two-loop contributions to $\langle N^{(0)\dag} S^{(0)} \rangle_{\mathrm{1PI}} = -\I E_1$.]{Two-loop contributions to $-\I E_1 = \langle N^{(0)\dag} S^{(0)} \rangle_{\mathrm{1PI}}$. (The $\slashed{p}$'s at some of the fermion lines just schematically indicate that the corresponding full propagators are odd functions of the momentum, whose name need not be necessarily $p$. Cf.~also the notation \eqref{app:frm:slashedp} in appendix~\ref{app:fermi propag}.)}
\label{ewd:fig:SD:E1}
\end{center}
\end{figure}

\begin{figure}[t]
\begin{center}
\includegraphics[width=1\textwidth]{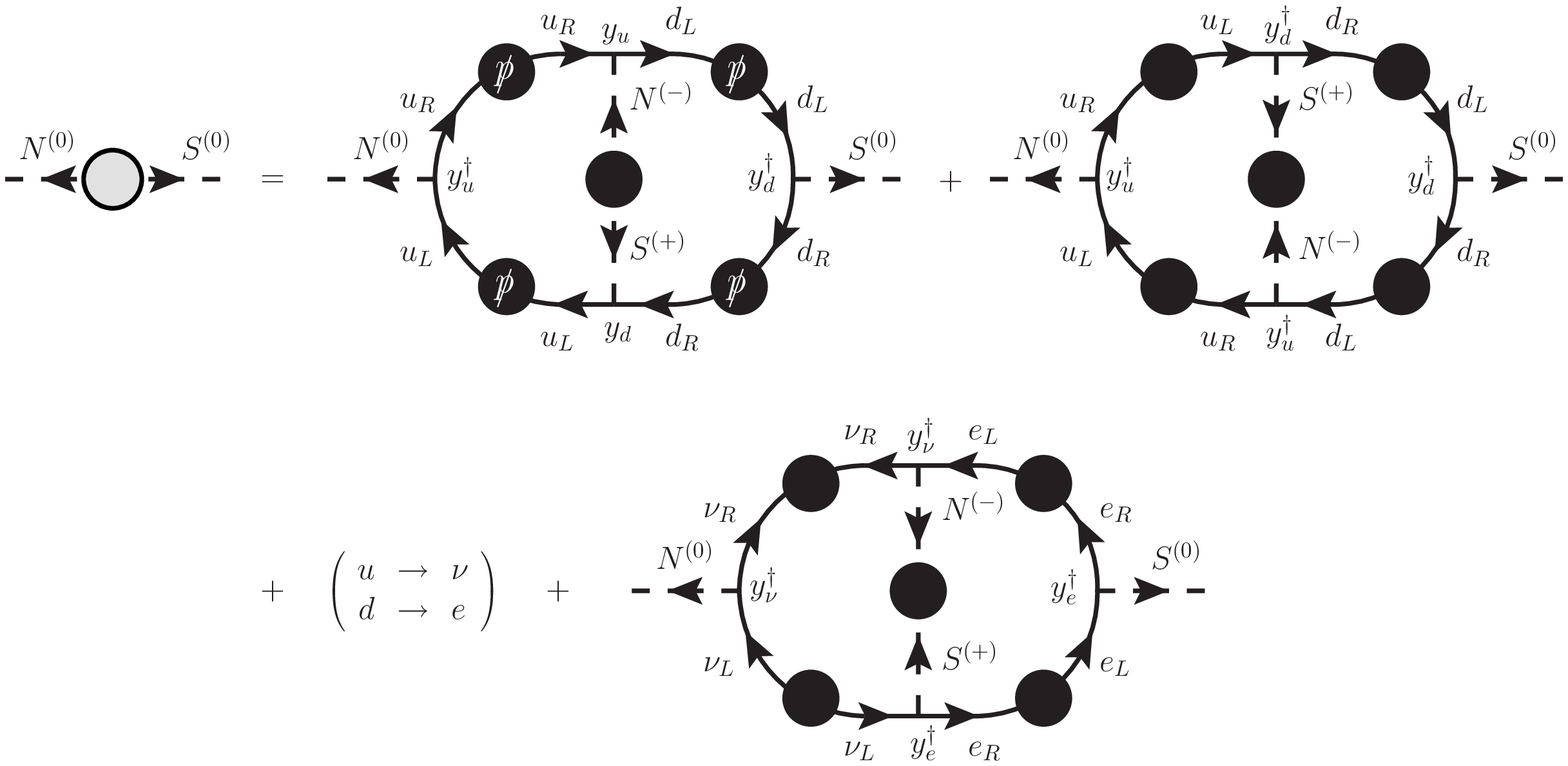}
\caption[Two-loop contributions to $\langle N^{(0)} S^{(0)} \rangle_{\mathrm{1PI}} = -\I E_2$.]{Two-loop contributions to $-\I E_2 = \langle N^{(0)} S^{(0)} \rangle_{\mathrm{1PI}}$.}
\label{ewd:fig:SD:E2}
\end{center}
\end{figure}

However, at two loops there are already non-trivial contributions to $E$. Recall that there are the charged scalars $S^{(+)}$, $N^{(-)}$, capable of changing an up-type fermion to the down-type and vice versa, see \eqref{eq:Yukawa_charged}. Thus, adding a charged scalar internal line inside the fermion loop allows to overcome the problems described in the previous paragraph and draw non-vanishing contributions to $E_1$, $E_2$, see Figs.~\ref{ewd:fig:SD:E1},~\ref{ewd:fig:SD:E2}.

\begin{figure}[t]
\begin{center}
\includegraphics[width=1\textwidth]{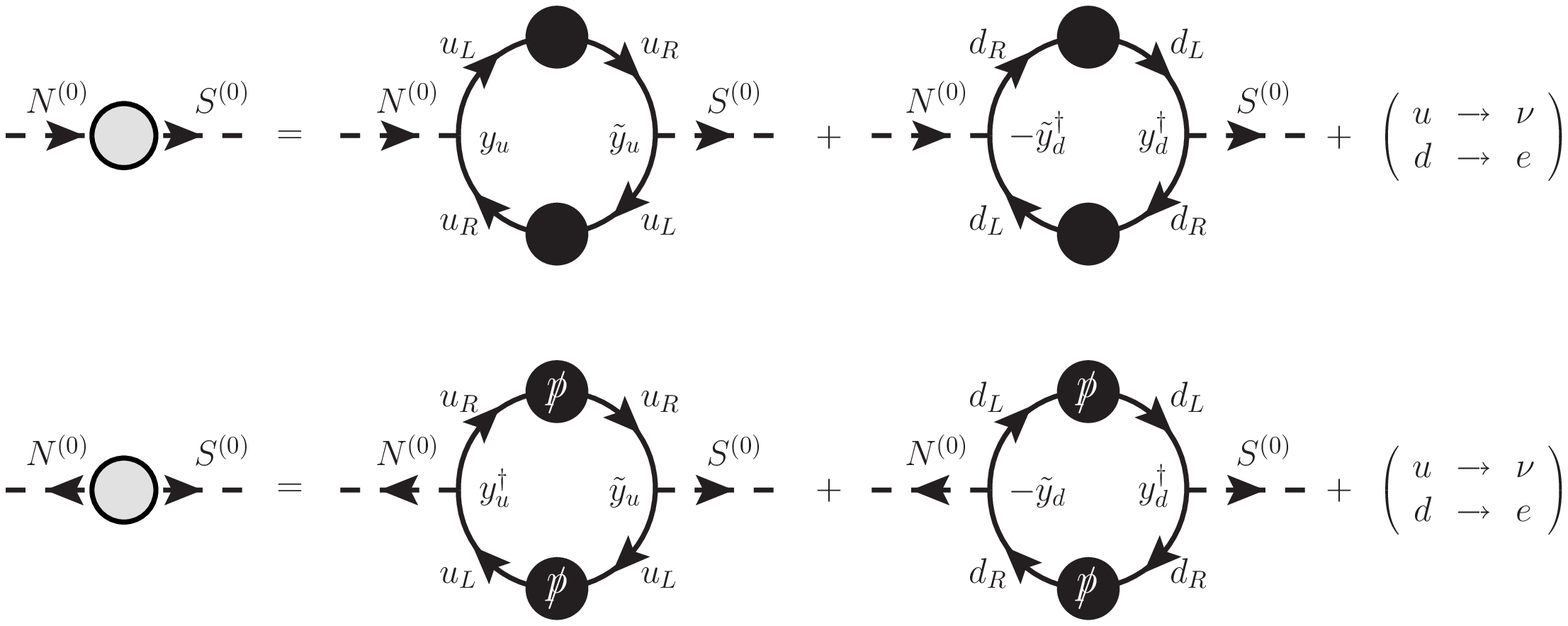}
\caption[One-loop contributions to $E_1$ and $E_2$ in the case of no $\mathcal{P}_{\mathrm{down}}$ symmetry.]{One-loop contributions to $E$ in the case of no $\mathcal{P}_{\mathrm{down}}$ symmetry. Notice that $E_1$ (the first line) and $E_2$ (the second line) are respectively UV-finite and UV-divergent.}
\label{ewd:fig:SD:Eoneloop}
\end{center}
\end{figure}

The situation would be however different if we have not postulated the discrete symmetry $\mathcal{P}_{\mathrm{down}}$, as in such a case there would be present also the Yukawa interactions \eqref{eq:Yukawa_tilde} of the charge conjugated scalar doublets $\tilde S$, $\tilde N$. The point is that considering \emph{both} Lagrangians \eqref{eq:Yukawa} and \eqref{eq:Yukawa_tilde} there would be contributions to $E_1$, $E_2$ already at one-loop level, as can be seen in Fig.~\ref{ewd:fig:SD:Eoneloop}. One can also notice that the loop integrals for $E_1$ and $E_2$ in Fig.~\ref{ewd:fig:SD:Eoneloop} are respectively UV-finite and UV-divergent. This is not a coincidence. Recall that even though the $\mathcal{P}_{\mathrm{down}}$ symmetry is broken, the self-energy $E_1$ is still symmetry-breaking (see Eqs.~\eqref{ewa:PiTZ} and \eqref{ewa:Esigma3}) and hence necessarily UV-finite. On the other hand, in the case of broken $\mathcal{P}_{\mathrm{down}}$ symmetry the self-energy $E_2$ contains symmetry-preserving and hence potentially (and also actually) UV-divergent part, as can be seen from \eqref{ewa:A2+E2} and \eqref{ewa:A2-E2}.

Recall at this point, however, the very reason why we assumed the $\mathcal{P}_{\mathrm{down}}$ symmetry: The actual vanishing of $E$ in the one-loop approximation makes the construction of the scalar self-energy Ansatz more tractable. That is because in such a case it suffices, when expressing the full scalar propagator, to invert only $2 \times 2$ matrices and not $4 \times 4$ matrices. Therefore, on top of the reducing of the free parameters of the Lagrangian, mentioned in Sec.~\ref{ew1:ssec:Yukawa}, the postulation of $\mathcal{P}_{\mathrm{down}}$ is desirable also from this practical reason.


\section{Numerical results}
\label{chp:ewdyn:ssec:num}

In order to make the numerical treatment of the model more tractable, some simplifications of the Lagrangian were made:
\begin{itemize}
  \item We considered $M_{\nu R}=0$.
  \item We considered only one generation of the charged fermions, i.e., $n=1$.
  \item We considered only one right-handed neutrino, i.e., $m=1$.
\end{itemize}

Let us comment the assumption $M_{\nu R}=0$. As already discussed in Sec.~\ref{ew1:ssec:PartCont}, in this case lepton number symmetry $\group{U}(1)_\ell$ is exact at the level of Lagrangian and one should include both $\Sigma_{\nu L}$ and $\Sigma_{\nu R}$ into the Ansatz, as both are $\group{U}(1)_\ell$ symmetry-breaking and thus UV-finite. On the other hand, one can also restrict oneself only to the solutions of the SD equations, which preserve $\group{U}(1)_\ell$, or, in other words, one can assume $\Sigma_{\nu L}=0$ and $\Sigma_{\nu R}=0$ from the beginning. This is exactly what we did in the numerical analysis: We considered only the Dirac-type neutrino self-energy $\Sigma_{\nu D}$, while the Majorana-type $\Sigma_{\nu L}$, $\Sigma_{\nu R}$ were neglected.

Under these assumptions the SD equation \eqref{ewd:SD:Sgmnue:L} for $\Sigma_{\nu L}$ is dismissed and the set \eqref{ewd:SD:Sgmnue} of lepton SD equations is formally the same as the set \eqref{ewd:SD:Sgmud} of quark SD equations. We emphasize that all fermion self-energies are then Dirac and they have no non-trivial matrix structure in the flavor space.

Further approximations of the SD equations consist of:
\begin{itemize}
  \item Considering all Yukawa coupling constants real.
  \item Considering all self-energies real.
\end{itemize}
These approximations are the same as before in Sec.~\ref{chp:frm:ssec:Appr} within the Abelian toy model. The same is also the very numerical procedure, see Sec.~\ref{chp:frm:ssec:Numpr}.

Recall that in the Abelian toy model the parametric space to be scanned was essentially two-dimensional: The true free parameters were the two (real) Yukawa coupling constants $y_1$ and $y_2$; the bare scalar mass $M$, as the only dimensional-full parameter of the model, served only as a scale parameter for the self-energies and momenta. Therefore it was possible to scan the parameter space at least to the extent of being able to decide whether for a given pair $y_1$, $y_2$ the solution is trivial or not. As a result of this scanning we obtained Fig.~\ref{y_plane_pic}.

This time the situation is considerably more complicated. We have four (real) Yukawa coupling constants, $y_u$, $y_d$, $y_\nu$, $y_e$, and two bare scalar masses $M_S$ and $M_N$. Of the two masses $M_S$, $M_N$ only one can be considered as a free parameter, the other serves again merely as a scaling parameter for the dimension-full quantities. Thus, we have altogether five free parameters and consequently five-dimensional space to be scanned. However, a \emph{systematic} scanning of such a vast parametric space was not possible. We have therefore checked the solutions only in some, rather randomly selected points in the parametric space in order to get some feeling about the general features and behavior of the solutions.

Thus, even though the parametric space was not scanned systematically, it was found that the above described SD equations have a similar behavior as the SD equations of the Abelian toy model:
\begin{enumerate}
  \item Non-trivial, UV-finite solutions do exist.
  \item The solutions are found only for relatively large values of the Yukawa coupling constants (of order of tens).
  \item Large ratios of fermion masses can be accommodated while having the corresponding Yukawa coupling constants of the same order of magnitude.\label{enum:point:Large}
\end{enumerate}
The point~\ref{enum:point:Large}.~above is promising in the quest for realistic fermion mass hierarchy. Because of the large parameter space which needs to be scanned this has not been accomplished. However, some achievements, which suggest that it should be possible, have been made. First, we accommodated the hierarchy between the lepton and quark doublets. For $y_\nu=63$, $y_e\approx84$, $y_u=65$, $y_d=90$ (and $M_S^2=2$, $M_N^2=1$) we found $m_\nu \apprle m_e=\mathcal{O}(10^{-4})$ and $m_u  \apprle m_d=\mathcal{O}(10^{-2})$. (Note that all masses are expressed in the units of $M_N$.) Second, we managed to generate a large hierarchy within one doublet. Considering only the leptons and neglecting the quarks ($y_u=y_d=0$), we found $m_e/m_\nu=\mathcal{O}(10^{2})$, calculated for $y_\nu\approx50$, $y_e=80$ (and again $M_S^2=2$, $M_N^2=1$). Nevertheless, it should be emphasized that this lepton mass ratio would be presumably significantly enhanced by the seesaw mechanism upon taking the Majorana right-handed neutrino mass term into account.

\section{Compatibility with electroweak observables}

\subsection{$\rho$-parameter}

While the realistic fermion spectrum together with the Yukawa coupling constants not vastly different can be presumably accommodated, it brings on the other hand the problem how to keep the $\rho$-parameter
\begin{eqnarray}
\label{ewd:rho}
\rho &\equiv& \frac{M_W^2}{M_Z^2\cos^2\theta_{\mathrm{W}}}
\end{eqnarray}
close to $1$. Note that in the case of exact custodial symmetry of the Lagrangian, i.e., when $n=m$, $y_\nu=y_e$, $y_u=y_d$, $M_S=M_N$ and $M_{\nu R} = 0$, one expects $\rho=1$ exactly. In chapter~\ref{ewM}, after calculating the explicit form of the fermion contribution to the $W^\pm$, $Z$ masses, it will be possible to see this for fermions explicitly.

Of course, in reality the custodial symmetry is in any case broken at least by fermions, since $y_\nu\neq y_e$, $y_u \neq y_d$. However, there is a possibility that if the scalar sector is (at least reasonably approximately) custodially symmetric and remains so even after the SSB, then the scalars can render $\rho$ to be close to $1$, provided they are heavy enough so that they can overcome the effect of the custodial symmetry breaking in the fermion sector.


\subsection{Flavor-changing neutral currents}


The new scalars must be heavy enough in order to avoid constraints from FCNC. We can make in this respect a rough, order-of-magnitude estimate. Consider, for instance, the decay $\mu \rightarrow e + S^{(0)}$. The virtual heavy scalar can subsequently decay as $S^{(0)} \rightarrow \bar e + e$. The Yukawa interactions will therefore induce the flavor-changing muon decay, $\mu \rightarrow e + \bar e + e $, with the amplitude being roughly given by $y^2/M_S^2$. (We assume that in the absence of fine tuning, all Yukawa couplings, including the flavor-changing ones, will be of the same order of magnitude.) The dominant muon decay channel, with branching ratio close to $100\%$, is $\mu \rightarrow e + \bar\nu_e + \nu_\mu$, whose amplitude is analogously proportional to $G_\mathrm{F}$. From here we infer the estimate $\BR(\mu \rightarrow e + \bar e + e)\sim(y^2/G_\mathrm{F}M_S^2)^2$. Taking the current experimental limit \cite{Amsler:2008zzb}, $\BR(\mu \rightarrow e + \bar e + e)<10^{-12}$, we find $M_S/y\apprge10^{2.5}\,\mathrm{TeV}$.

\subsection{$S$-parameter}


\begin{figure}[t]
\begin{center}

\begin{picture}(13,0)%
\setlength\fboxsep{5bp}
\setlength\fboxrule{0.5pt}
\fbox{\includegraphics{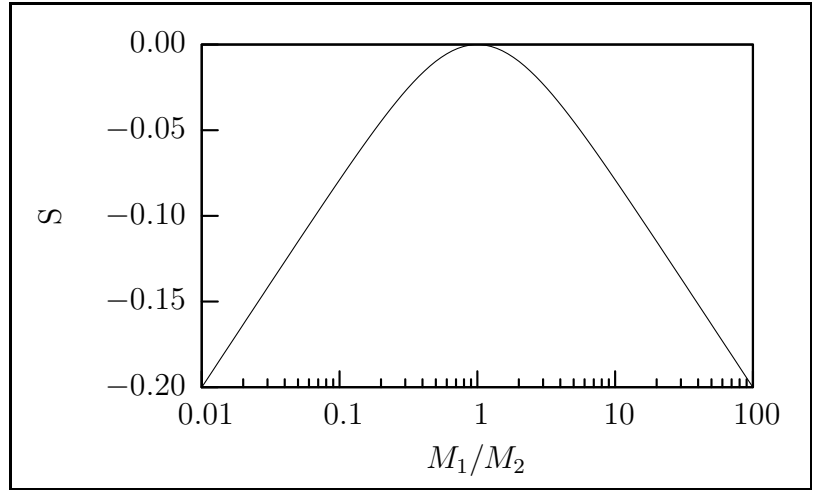}}%
\end{picture}%
\begingroup
\setlength{\unitlength}{0.0200bp}%
\begin{picture}(14400,8640)(0,0)%
\put(3025,1650){\makebox(0,0)[r]{\strut{}$-0.20$}}%
\put(3025,3260){\makebox(0,0)[r]{\strut{}$-0.15$}}%
\put(3025,4870){\makebox(0,0)[r]{\strut{}$-0.10$}}%
\put(3025,6480){\makebox(0,0)[r]{\strut{}$-0.05$}}%
\put(3025,8090){\makebox(0,0)[r]{\strut{}$0.00$}}%
\put(3300,1100){\makebox(0,0){\strut{} 0.01}}%
\put(5869,1100){\makebox(0,0){\strut{} 0.1}}%
\put(8438,1100){\makebox(0,0){\strut{} 1}}%
\put(11006,1100){\makebox(0,0){\strut{} 10}}%
\put(13575,1100){\makebox(0,0){\strut{} 100}}%
\put(550,4870){\rotatebox{90}{\makebox(0,0){\strut{}S}}}%
\put(8437,275){\makebox(0,0){\strut{}$M_{1}/M_{2}$}}%
\end{picture}%
\endgroup

\end{center}
\caption[The $S$-parameter.]{The $S$-parameter \eqref{ewd:Sparam} plotted for the special case $M_{S\pm} = M_{N\pm} = M_{SN\pm} \equiv M_{\pm}$. Note that, according to the Particle Data Group \cite{Amsler:2008zzb}, $S = -0.10 \pm 0.10 $.}
\label{fig_plot_S}
\end{figure}

The very introduction of new scalars also affects the Peskin--Takeuchi $S$-parameter \cite{Peskin:1991sw}. In order to estimate the scalar contribution to it, we set for simplicity the scalar self-energies $\Pi_{S}$, $\Pi_{N}$, $\Pi_{SN}$ to be constant. The spectrum (see the pole equations \eqref{ewa:PhiPoleeqs}) is then given simply by
\begin{subequations}
\begin{eqnarray}
M_{SN\pm}^2 &=& \frac{M_S^2+M_N^2}{2} \pm \sqrt{\bigg(\frac{M_S^2-M_N^2}{2}\bigg)^2+|\Pi_{SN}|^2} \,,
\\
M_{S\pm}^2 &=& M_S^2 \pm |\Pi_S| \,,
\\
M_{N\pm}^2 &=& M_N^2 \pm |\Pi_N| \,.
\end{eqnarray}
\end{subequations}
The resulting $S$-parameter then can be written as
\begin{eqnarray}
\label{ewd:Sparam}
S & = &  S_{S} + S_{N} + S_{SN} \,,
\end{eqnarray}
where
\begin{subequations}
\begin{eqnarray}
S_{S} &\equiv&  \frac{1}{12 \pi}
\Bigg[
\frac{5}{6} - \frac{2M_{S+}^2 M_{S-}^2}{\big(M_{S+}^2-M_{S-}^2\big)^2}
-\frac{1}{2}\ln\frac{M_{S+}^2 M_{S-}^2}{\mu^4}
\nonumber \\ &&
\phantom{\frac{1}{12 \pi}\Bigg[\frac{5}{6}}
{}-\frac{1}{2} \frac{M_{S+}^6+M_{S-}^6 - 3 M_{S+}^2M_{S-}^2
(M_{S+}^2+M_{S-}^2)}{\big(M_{S+}^2-M_{S-}^2\big)^3}\ln\frac{M_{S+}^2}{M_{S-}^2}
\Bigg] \,,
\\
S_{N} &\equiv&  \frac{1}{12 \pi}
\Bigg[
\frac{5}{6} - \frac{2M_{N+}^2 M_{N-}^2}{\big(M_{N+}^2-M_{N-}^2\big)^2}
-\frac{1}{2}\ln\frac{M_{N+}^2 M_{N-}^2}{\mu^4}
\nonumber \\ &&
\phantom{\frac{1}{12 \pi}\Bigg[\frac{5}{6}}
{}-\frac{1}{2} \frac{M_{N+}^6+M_{N-}^6 - 3 M_{N+}^2M_{N-}^2
(M_{N+}^2+M_{N-}^2)}{\big(M_{N+}^2-M_{N-}^2\big)^3}\ln\frac{M_{N+}^2}{M_{N-}^2}
\Bigg] \,, \qquad
\\
S_{SN} &\equiv&  \frac{1}{12 \pi}
\ln \frac{M_{SN+}^2 M_{SN-}^2}{\mu^4} \,.
\end{eqnarray}
\end{subequations}
(The $\mu$ is just an arbitrary mass scale introduced for {\ae}sthetic reasons; the total $S$-parameter \eqref{ewd:Sparam} is independent of it.) Taking into account the previous discussion of the $\rho$-parameter and the scalar masses, we plotted the $S$-parameter for the special case $M_{S\pm} = M_{N\pm} = M_{SN\pm} \equiv M_{\pm}$. The resulting $S$-parameter, which is then function only of the mass ratio $M_+/M_-$, is plotted in Fig.~\ref{fig_plot_S}. When in this special case the ratio $M_+/M_-$ is far from one, the $S$-parameter is well approximated by the simple formula $S=\frac{1}{6\pi}\bigl(\frac56-\ln\bigl|\frac{M_+}{M_-}\bigr|\bigr)$. On the other hand, for $M_+/M_-$ close to one the $S$-parameter behaves like $\frac{-1}{15 \pi} \bigl(1-\frac{M_+}{M_-}\bigr)^2$. From Fig.~\ref{fig_plot_S} one can see that the $S$-parameter meets the experimental bounds for any value of $M_+/M_-$ from $0.01$ up to $100$.

\section{Summary}

We have derived, within the Yukawa dynamics considered in chapter~\ref{chp:ew1} and using the CJT formalism, the SD equations for the scalar and fermion self-energies. They were derived for general self-energies and eventually restricted to the Ansatz introduced in the previous chapter~\ref{chp:ewa}. The form of these SD equations suggests the expected UV-finiteness of the solutions.

Due to huge parametric space of the set of SD equations (even in the oversimplified case of only one generation of charged fermions, one Dirac neutrino and real Yukawa coupling constants), the systematic numerical survey for the solutions was not possible, in contrast to the Abelian toy model in Sec.~\ref{chp:frm:ssec:Numpr}. Nevertheless, an unsystematic (i.e., rather random) scanning of the parametric space revealed some promising points (one of which was presented in Sec.~\ref{chp:ewdyn:ssec:num}), suggesting the possibility of obtaining a realistic fermion spectrum.

Finally, we commented on the compatibility with the electroweak observables. While the sole fermion sector pushes the $\rho$-parameter away from $1$, the scalar sector can render it close to $1$, provided the scalars are heavy enough, as they after all tend to be, as the numerical analysis suggests. This presumable heaviness of scalars is also consistent with the desired suppression of the FCNC. Finally, we showed that even the $S$-parameter remains in norm, provided the scalar masses are mutually not too different.


\part{Flavor mixing}
\label{part:mx}

\chapter{Quark flavor mixing}
\label{chp:mx}

\intro{We will now discuss the implications of the very momentum-dependence of the dynamically generated quark self-energies for the mixing of the physical quarks, i.e., of the quark mass eigenstates. We will show that, unlike in models (e.g., the MCS and particularly the SM) with mass matrices (i.e., momentum-independent self-energies), the resulting effective Cabibbo--Kobayashi--Maskawa (CKM) matrix \cite{Cabibbo:1963yz,Kobayashi:1973fv} is in general non-unitary and the FCNC are present already at lowest (tree) order in the gauge coupling constant. We will not present the case of leptons, as the discussion would be essentially the same, with the resulting Pontecorvo--Maki--Nakagawa--Sakata (PMNS) matrix \cite{Pontecorvo:1957cp,Maki:1962mu} being non-unitary as well.}

\intro{We stress that the following discussion applies to any model of EWSB generating the quark self-energies with a momentum dependency. In this respect the model presented in part~\ref{part:ew} can be regarded as a particular example of such class of models.}

\intro{This chapter is a concise version of paper \cite{Benes:2009iz}, where more details can be found.}

\section{Gauge interactions in the interaction eigenstate basis}

We start off by a slight change of notation, to be applied exclusively in this chapter. Instead of the denotation $u$, $d$, introduced in section~\ref{ew1:ssec:PartCont} and used throughout part~\ref{part:ew}, we will use the primed denotation $u^\prime$, $d^\prime$ and call it the \emph{(weak) interaction eigenstate basis}. The denotation $u$, $d$ will be reserved for the so-called \emph{mass eigenstate basis}, to be introduced thereinafter.

In the interaction eigenstate basis $u^\prime$, $d^\prime$ and in the basis \eqref{ew1:rot:A34-Zew} and \eqref{ew1:rot:A12-Wpm} of the EW gauge fields the gauge interactions \eqref{ew1:eL:gaugeqLqR} read
\begin{eqnarray}
\eL_{\mathrm{quark,qauge}} &=& \eL_{\mathrm{cc}}(u^\prime,d^\prime) + \eL_{\mathrm{nc}}(u^\prime,d^\prime) + \eL_{\mathrm{em}}(u^\prime,d^\prime) \,,
\end{eqnarray}
where
\begin{subequations}
\label{eq:L_gauge}
\begin{eqnarray}
\eL_{\mathrm{cc}}(u^\prime,d^\prime) &=& \frac{g}{\sqrt{2}}\bar u^\prime \gamma_\mu P_L d^\prime A_{W^+}^\mu + \hc \,,
\label{eq:L_gauge_cc}
\\
\eL_{\mathrm{nc}}(u^\prime,d^\prime) &=& \frac{g}{2 \cos \theta_{\mathrm{W}}}
\sum_{f=u,d}
\bar f^\prime \gamma_\mu (v_f-a_f\gamma_5) f^\prime A_{Z}^\mu \,,
\\
\eL_{\mathrm{em}}(u^\prime,d^\prime) &=& \sum_{f=u,d} e Q_f
\bar f^\prime \gamma_\mu f^\prime A_{\mathrm{em}}^\mu \,.
\end{eqnarray}
\end{subequations}
We use the standard notation \cite{Horejsi:2002ew}
\begin{subequations}
\begin{eqnarray}
v_f &\equiv& t_{3f}-2Q_f \sin^2\theta_{\mathrm{W}} \,,
\\
a_f &\equiv& t_{3f} \,.
\end{eqnarray}
\end{subequations}

\section{Momentum-independent self-energies}
\label{sec:SM}

Before investigating the general case of momentum-dependent self-energies in the next section, we revise in this section how the fermion flavor mixing is treated in the special case of constant self-energies. In other words, we review here the SM (and generally any MCS, the most prominent representative of which the SM is).\footnote{We will thus use the superscript $\mathrm{SM}$ for the quantities calculated using the assumption of constant self-energies.} Nevertheless, we present it here, in order to establish some notation and to make the text reasonably self-contained. The primary reason is, however, that the case of constant self-energies provides a natural reference point when discussing in the next section some novel consequences stemming from self-energies' momentum dependence.

Since the quark self-energies $\boldsymbol{\Sigma}_u$, $\boldsymbol{\Sigma}_d$ are by assumption momentum-independent, they can be regarded as mass matrices sitting in the Lagrangian:
\begin{subequations}
\label{mx:eL:mass:SM}
\begin{eqnarray}
\eL_{\mathrm{mass}}^{\mathrm{(SM)}}(u^\prime,d^\prime) &=&
{}- \bar u^\prime \boldsymbol{\Sigma}_u u^\prime - \bar d^\prime \boldsymbol{\Sigma}_d d^\prime
\\ &=& {}- \bar u_L^\prime \Sigma_u u_R^\prime - \bar d_L^\prime \Sigma_d d_R^\prime + \hc \,,
\end{eqnarray}
\end{subequations}
where we took into account the form $\boldsymbol{\Sigma}_f = \Sigma_f^\dag\,P_L+\Sigma_f\,P_R$ ($f=u,d$), Eq.~\eqref{ewa:Sgmf}. The component $\Sigma_f$ can be diagonalized via the bi-unitary transformation\footnote{The following diagonalization of mass matrices is a special case of more general analysis for the momentum-dependent self-energies, presented in section \eqref{frm:sec:diag} of appendix~\ref{app:fermi propag}.}
\begin{eqnarray}
\label{mx:definitionUV}
\Sigma_f &=& V_f^\dag \, M_f \, U_f \,,
\end{eqnarray}
where $U_f$, $V_f$ are some unitary matrices and $M_f$ is a diagonal, real, non-negative matrix:
\begin{eqnarray}
\label{mx:Mf}
M_f &\equiv& \diag\big(m_{f_1},m_{f_2}, \ldots, m_{f_n}\big) \,.
\end{eqnarray}
More compact notation can be achieved by defining the unitary matrix $X_f$ as
\begin{eqnarray}
X_f &\equiv& V_f^\dag\, P_L + U_f^\dag\, P_R \,,
\end{eqnarray}
so that the full $\boldsymbol{\Sigma}_f$ can be written as
\begin{eqnarray}
\boldsymbol{\Sigma}_f &=& \bar X_f^\dag \, M_f \, X_f^\dag \,.
\end{eqnarray}
We can now define new fields as
\begin{subequations}
\label{eq:def:quark_unitary_redef}
\begin{eqnarray}
u &=& X_u^\dag u^\prime \,, \\
d &=& X_d^\dag d^\prime \,,
\end{eqnarray}
which can be rewritten in terms of the original chiral components as
\begin{align}
u_L & \ =\  V_u \, u_L^\prime\,, &
d_L & \ =\  V_d \, d_L^\prime\,,
\\
u_R & \ =\  U_u \, u_R^\prime\,, &
d_R & \ =\  U_d \, d_R^\prime\,.
\end{align}
\end{subequations}
Thus, the Lagrangian \eqref{mx:eL:mass:SM}, expressed in terms of $u$, $d$ (or their chiral components) is mass-diagonal:
\begin{subequations}
\label{mx:eL:mass:SMdiag}
\begin{eqnarray}
\eL_{\mathrm{mass}}^{\mathrm{(SM)}}(u,d) &=&
{}- \bar u M_u u - \bar d M_d d \\
&=&
{}- \bar u_L M_u u_R - \bar d_L M_d d_R + \hc
\end{eqnarray}
\end{subequations}
I.e., particular components $u_i$, $d_j$ ($i,j=1,\ldots,n$) of the fields $u$, $d$ have now straightforward interpretation as operators creating the states $\ket{u_i}$, $\ket{d_j}$ with definite masses $m_{u_i}$, $m_{d_j}$ (see \eqref{mx:Mf}) from the vacuum and we are allowed to call the operators $u$, $d$ the \emph{mass eigenstate basis}.


The redefinitions \eqref{eq:def:quark_unitary_redef} apply also for the rest of the Lagrangian, in particular for the gauge interactions \eqref{eq:L_gauge}. One obtains
\begin{subequations}
\label{eq:L_gauge_SM}
\begin{eqnarray}
\eL_{\mathrm{cc}}^{\mathrm{(SM)}}(u,d) &=& \frac{g}{\sqrt{2}}\bar u \gamma_\mu P_L V_{\mathrm{CKM}} d A_{W^+}^\mu + \hc \,,
\label{eq:L_gauge_SM_cc}
\\
\eL_{\mathrm{nc}}^{\mathrm{(SM)}}(u,d) &=& \frac{g}{2\cos\theta_{\mathrm{W}}} \sum_{f=u,d} \bar f \gamma_\mu (v_f-a_f\gamma_5) f A_{Z}^\mu \,,
\label{eq:L_gauge_SM_nc}
\\
\eL_{\mathrm{em}}^{\mathrm{(SM)}}(u,d) &=& \sum_{f=u,d} e Q_f \bar f \gamma_\mu f A_{\mathrm{em}}^\mu \,.
\label{eq:L_gauge_SM_em}
\end{eqnarray}
\end{subequations}
We see that in contrast to the Lagrangian \eqref{eq:L_gauge}, the charged current interactions $\eL_{\mathrm{cc}}^{\mathrm{(SM)}}$ are no longer flavor-diagonal, but rather exhibit the flavor mixing parameterized by the celebrated Cabibbo--Kobayashi--Maskawa (CKM) matrix \cite{Cabibbo:1963yz,Kobayashi:1973fv}, which is expressed in terms of the matrices $V_u$, $V_d$, \eqref{mx:definitionUV}, as
\begin{eqnarray}
\label{eq:CKM_SM}
V_{\mathrm{CKM}} &\equiv& V_{u\vphantom{d}}^{\vphantom{\dag}} V_d^\dag \,.
\end{eqnarray}
Note that $V_{\mathrm{CKM}}$ is unitary\footnote{An $n \times n$ unitary matrix has $n^2$ real parameters. Of these, $\frac{1}{2}n(n-1)$ are angles and $\frac{1}{2}n(n+1)$ are complex phases. However, for $V_{\mathrm{CKM}}$ the number of free parameters can be further reduced, since one column and one row can be made real by appropriate redefinitions of quark fields. This amounts to $2n-1$ redundant phases, so that in $V_{\mathrm{CKM}}$ there are only $\frac{1}{2}(n-1)(n-2)$ physical, $\mathcal{CP}$-violating phases.} due to the unitarity of matrices $V_u$, $V_d$. On the other hand, the electromagnetic and neutral current interactions remain diagonal, which is again a consequence of the unitarity of the matrices $X_u$, $X_d$.

Consider now, for the sake of later references, the decay process $W^+ \rightarrow u_i + \bar d_j$ and its $S$-matrix element
\begin{subequations}
\label{eq:definition:Mfi}
\begin{eqnarray}
S_{fi} &=& \bra{u_i,\bar d_j}S\ket{W^+} \\
&=& \delta_{fi}+(2\pi)^4\,\delta^4(p+k-q)N_p N_k N_q\,\I\mathcal{M}_{fi}\,,
\end{eqnarray}
\end{subequations}
where the factors $N_p$, $N_k$, $N_q$ are defined in Eq.~\eqref{app:charge:Np} (we assign the external momenta as $W^+(q) \rightarrow u_i(p) + \bar d_j(k)$). This is the simplest process in which the effect of the CKM matrix takes place. Within the SM-like Lagrangian \eqref{eq:L_gauge_SM_cc} we have in the lowest order in the gauge coupling constant $g$ for the corresponding amplitude immediately
\begin{eqnarray}
\label{eq:Mfi_SM}
\mathcal{M}_{fi}^{\mathrm{(SM)}} &=&
\frac{g}{\sqrt{2}}\,\bar u_{u_i}\!(p) \, \gamma^\mu P_L (V_{\mathrm{CKM}})_{ij} \, v_{d_j}\!(k) \, \varepsilon_\mu(q) \,.
\end{eqnarray}

\section{Momentum-dependent self-energies}
\label{sec:general_selfenergies}


Let us now relax the requirement of the self-energies' momentum-independence and allow them to depend on momentum in a general way.\footnote{Problem of extracting physical information from matrix-like momentum-dependent self-energies has been already discussed (although in the different context of perturbative radiative corrections), e.g., in Refs.~\cite{Machet:2004rv,Duret:2006wk,Barroso:2000is,Diener:2001qt,Zhou:2003te,Kniehl:2006rc}.} In this situation the self-energies cannot be any longer interpreted as mass matrices and there is no obvious way how to reexpress the Lagrangian from the interaction eigenstate basis into the mass eigenstate basis. We will show that even in this situation one can define the mass eigenstate basis, though in an effective sense, together with the effective CKM matrix.

The crucial observation is that although we do not have the theory expressed in terms of the mass eigenstate basis $u$, $d$ (i.e., in terms of the fields that create the quarks with definite mass), it is still possible to calculate the amplitudes of the processes involving the mass eigenstates $\ket{u_i}$, $\ket{d_j}$, with the masses $m_{u_i}$, $m_{d_j}$ given by the pole equations \eqref{ewa:dvepole}. This is allowed by the Lehmann--Symanzik--Zimmermann (LSZ) reduction formula \cite{Lehmann:1954rq}, which states that the amplitude of a given process involving the mass eigenstates $u$, $d$ can be calculated (up to the polarization vectors and possible sign due to the fermionic nature of involved particles) as a residue of the appropriate (momentum space) connected Green's function for the external momenta going on their mass-shell. The point is that the Green's function need \emph{not} be calculated in terms of eventual operators $u$, $d$ of the mass eigenstates, but rather in terms of the original interaction eigenstate basis operators $u^\prime$, $d^\prime$, which have no direct connection to the mass eigenstates (possibly even up to any unitary redefinition, as we will see later). Note that the Green's functions are easily calculated: One can apply the usual perturbation theory given by the Lagrangian \eqref{eq:L_gauge}, with the additional Feynman rule that the fermion lines in the diagrams are given by the full quark propagators
\begin{eqnarray}
\I G_f &=& \langle f^\prime \bar f^\prime \rangle \,,
\end{eqnarray}
which are expressed in terms of the self-energies $\boldsymbol{\Sigma}_f$ as
\begin{eqnarray}
\label{mx:Gf}
G_f &=& (\slashed{p}-\boldsymbol{\Sigma}_f)^{-1} \,,
\end{eqnarray}
cf.~Eq.~\eqref{ewa:Gfinv}.

The possibility of calculating processes involving the mass eigenstates, as sketched in the previous paragraph, opens the way to investigating the fermion flavor mixing in the case of momentum-dependent self-energies. We explain it in more detail in the following section on the example of flavor mixing in the charged current sector. Next, in the subsequent section, we state (without detailed derivation) the analogous results for the electromagnetic and neutral current sectors.

\subsection{Charged current interactions}
\label{subsec:cc}

\subsubsection{Effective CKM matrix}

The idea is simple and can be roughly stated as follows: First, we calculate (using the approach described above) the $S$-matrix element for the process $W^+ \rightarrow u_i + \bar d_j$ in the lowest order in the gauge coupling constant. Second, we demand that the obtained amplitude has the same form as the amplitude \eqref{eq:Mfi_SM} calculated within the SM (Sec.~\ref{sec:SM}) and define this way the effective CKM matrix. This effective CKM matrix is eventually interpreted to be a part of the effective Lagrangian of the SM form \eqref{eq:L_gauge_SM}.

\begin{figure}[t]
\begin{center}
\includegraphics[width=0.71\textwidth]{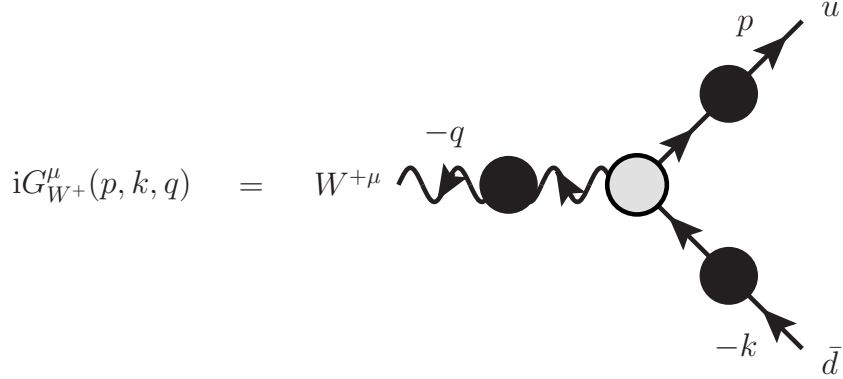}
\caption[Connected Green's function $\langle u^\prime \bar d^\prime A_{W^+}^\mu \rangle = \I G_{W^+}^\mu(p,k,q)$.]{Diagrammatical representation and momenta assignment of the connected Green's function $\I G_{W^+}^\mu(p,k,q)$, Eq.~\eqref{mw:GWp}. The shaded blob denotes its 1PI part, $\I \Gamma_{W^+}^\mu(p,-k)$, while the dark blobs represent the full propagators. (Notice the arrows on the boson line: We conventionally define the $W^+$ as an antiparticle.)}
\label{fig:vertex}
\end{center}
\end{figure}

Let us work out the idea in detail. Consider the connected Green's function $\langle u^\prime \bar d^\prime A_{W^+}^\mu \rangle$ and define its Fourier transform $\I G_{W^+}^\mu(p,k,q)$ as
\begin{eqnarray}
\label{eq:Gamma_general}
\int \! \d^4 x \, \d^4 y \, \d^4 z \,
\e^{\I p\cdot x}\,\e^{\I k\cdot y}\,\e^{-\I q\cdot z}\,
\bra{0}T \, u^\prime(x)\,\bar d^\prime(y)\, A_{W^+}^\mu(z)\ket{0}
&=&
(2\pi)^4\,\delta^4(p+k-q)\,\I G_{W^+}^\mu(p,k,q) \,.
\nonumber \\ &&
\end{eqnarray}
For the assignment of the momenta see Fig.~\ref{fig:vertex}. Recall that a connected Green's function is generally calculated as a proper (1PI) Green's function with full propagators at the external lines:
\begin{eqnarray}
\label{mw:GWp}
\I G_{W^+}^\mu(p,k,q)
&=&
\I G_u(p) \, \I\Gamma_{W^+}^\nu(p,-k) \, \I G_d(-k) \, \I G^{\mu}_{\hphantom{\mu}\nu}(q) \,.
\end{eqnarray}
For the external fermion lines we consider the full propagators $G_u(p)$, $G_d(p)$, as defined by Eq.~\eqref{mx:Gf}. The $W^\pm$ propagator $G_{\mu\nu}(q)$ is taken at this moment to be just the bare propagator of a massive vector field \cite{Horejsi:2002ew} with hard mass $M_W$ (the mass $M_W$ will be discussed in more detail in chapter~\ref{ewM}). Similarly, the proper vertex $\Gamma_{W^+}^\mu(p,-k)$ is taken to be the tree one, determined by the charged current Lagrangian $\eL_{\mathrm{cc}}(u^\prime,d^\prime)$, \eqref{eq:L_gauge_cc}, i.e.,
\begin{eqnarray}
\label{eq:vertex_bare}
\Gamma_{W^+}^\mu(p,-k) &=& \frac{g}{\sqrt{2}}\gamma^\mu P_L \,.
\end{eqnarray}
Thus, we have at the leading order in the gauge coupling constant $g$ immediately
\begin{eqnarray}
\label{eq:Gamma_explicit}
\I G_{W^+}^\mu(p,k,q)
&=&
\I G_u(p) \, \I\frac{g}{\sqrt{2}}\gamma^\nu P_L \, \I G_d(-k) \, \I G^\mu_{\phantom{\mu}\nu}(q) \,.
\end{eqnarray}

We are now ready to apply the LSZ reduction formula. (In the following we will rely on the results from section \eqref{frm:sec:diag} of appendix~\ref{app:fermi propag}, concerning the diagonalization of momentum-dependent self-energies, as well as of the corresponding full propagators.) Recall that upon taking the limit $p^2 \rightarrow m_{u_i}^2$, $k^2 \rightarrow m_{d_j}^2$, $q^2 \rightarrow M_{W}^2$ in the Green's function $\I G_{W^+}^\mu(p,k,q)$, the residue of the leading divergent term (i.e., the one with the triple pole) is (up to polarization vectors and a sign) the desired amplitude $\mathcal{M}_{fi}$ of the process $W^+ \rightarrow u_i + \bar d_j$, \eqref{eq:definition:Mfi}. Taking into account the explicit form \eqref{eq:Gamma_explicit} of $\I G_{W^+}^\mu(p,k,q)$ and applying the asymptotic formul{\ae} \eqref{eq:S_asymptotics} for the propagators $G_u(p)$, $G_d(p)$, we arrive straightforwardly at the result
\begin{eqnarray}
\label{eq:LSZ}
\I G_{W^+}^\mu(p,k,q) &
\xrightarrow[\begin{subarray}{c}
p^2 \rightarrow m_{u_i}^2 \\
k^2 \rightarrow m_{d_j}^2 \\
q^2 \rightarrow M_{W}^2 \\
\vphantom{x}
\end{subarray}]{}
&
-
\frac{\I\mathcal{U}_{u_i}(p)}{p^2-m_{u_i}^2}
\frac{\I\mathcal{\bar V}_{d_j}(k)}{k^2-m_{d_j}^2}
\frac{\I\varepsilon^{\mu*}(q)}{q^2-M_{W}^2}
\,\I\mathcal{M}_{fi}
\>\> + \> \ldots \,,
\end{eqnarray}
where the ellipsis represents less divergent terms (i.e., the terms with double and single poles and regular terms). The amplitude $\mathcal{M}_{fi}$ in \eqref{eq:LSZ} comes out as
\begin{eqnarray}
\label{eq:Mfi_exact}
\mathcal{M}_{fi}
&=&
\frac{g}{\sqrt{2}} \, \bar u_{u_i}\!(p) \, \big(\tilde V_{u\vphantom{d}}^{\vphantom{\dag}} \tilde V_d^\dag \big)_{ij} \gamma^\mu P_L \, v_{d_j}\!(k)  \, \varepsilon_\mu(q)
\,.
\end{eqnarray}
For the precise definition of the matrices $\tilde V_u$, $\tilde V_d$ see Eq.~\eqref{eq:def:U_V_tilde}. Now let just say that the matrices $\tilde V_u$, $\tilde V_d$ are in general non-unitary and their definition is related to the diagonalization of the (momentum-dependent) self-energies $\boldsymbol{\Sigma}_u$, $\boldsymbol{\Sigma}_d$ in a similar manner as the definition \eqref{mx:definitionUV} of $V_u$, $V_d$. In fact, the both pairs of matrices coincides in the limit of momentum-independent self-energies.

We are now going to compare the amplitude $\mathcal{M}_{fi}$, \eqref{eq:Mfi_exact} with the amplitude $\mathcal{M}_{fi}^{\mathrm{(SM)}}$, \eqref{eq:Mfi_SM}), calculated within the SM for the same process $W^+ \rightarrow u_i + \bar d_j$ and in the same (lowest) order in the gauge coupling constant. Demanding that both amplitudes have the same form, we conclude that the effective CKM matrix is given by
\begin{eqnarray}
\label{eq:CKM_eff}
V_{\mathrm{CKM}}^{\mathrm{(eff)}} &\equiv& \tilde V_{u\vphantom{d}}^{\vphantom{\dag}} \tilde V_d^\dag \,.
\end{eqnarray}
This effective CKM matrix has the striking feature of being in general non-unitary, in contrast to the CKM matrix \eqref{eq:CKM_SM} in the SM, thanks to the mentioned non-unitarity of $\tilde V_u$, $\tilde V_d$. Note however, that in the special case of constant self-energies $\boldsymbol{\Sigma}_u$, $\boldsymbol{\Sigma}_d$ the two expressions \eqref{eq:CKM_SM} and \eqref{eq:CKM_eff} coincide and the unitarity of CKM matrix is restored.

\subsubsection{Effective Lagrangian}

Let us now proceed to the definition of the effective Lagrangian. The CKM matrix in the SM occurs not only in the matrix elements of the type \eqref{eq:Mfi_SM} (in the same way as our effective CKM matrix \eqref{eq:CKM_eff} does), but it also lives in the charged current Lagrangian \eqref{eq:L_gauge_SM_cc}, written in terms of the mass-diagonalized quark fields $u$, $d$. The natural question arises whether and to what extent it is analogously possible to reexpress the Lagrangian in terms of the mass eigenstate basis $u$, $d$ also in the present case of momentum-dependent self-energies and how to incorporate this way the effective CKM matrix obtained above. The answer is that it is possible merely in an effective sense to be specified below.

We define the effective Lagrangian $\eL^{\mathrm{(eff)}}(u,d)$ in the following way: We \emph{postulate} the mass eigenstate basis operators $u$, $d$ in such a way that they are operators creating the quarks with the masses given by the momentum-dependent self-energies $\boldsymbol{\Sigma}_u$, $\boldsymbol{\Sigma}_d$ via the pole equations \eqref{ewa:dvepole}. More precisely, $\eL^{\mathrm{(eff)}}(u,d)$ contains, on top of the fermion kinetic terms, the mass Lagrangian $\eL^{\mathrm{(eff)}}_{\mathrm{mass}}(u,d)$ of the form \eqref{mx:eL:mass:SMdiag}, i.e.,
\begin{eqnarray}
\label{eq:L_eff_mass}
\eL^{\mathrm{(eff)}}_{\mathrm{mass}}(u,d) &=& {}- \bar u M_u u - \bar d M_d d \,.
\end{eqnarray}
Here the mass matrices $M_u$, $M_d$ are of the form \eqref{mx:Mf}:
\begin{subequations}
\begin{eqnarray}
M_u &=& \diag(m_{u_1},m_{u_2},\ldots,m_{u_n}) \,,
\\
M_d &=& \diag(m_{d_1},m_{d_2},\ldots,m_{d_n}) \,,
\end{eqnarray}
\end{subequations}
with the entries determined by the poles of the full propagators $G_u(p)$, $G_d(p)$. Let the effective Lagrangian $\eL^{\mathrm{(eff)}}(u,d)$ contain also the kinetic terms of the gauge bosons $W^\pm$, $Z$, $\gamma$ and the corresponding mass terms. Since $\eL^{\mathrm{(eff)}}(u,d)$ is written in terms of massive fields, it is capable of describing processes like $W^+ \rightarrow u_i + \bar d_j$ directly, without employing the LSZ reduction formula. Indeed, postulating that $\eL^{\mathrm{(eff)}}(u,d)$ contains the SM-like charged current interactions of the form
\begin{eqnarray}
\label{eq:L_eff_cc}
\eL^{\mathrm{(eff)}}_{\mathrm{cc}}(u,d) &=& \frac{g}{\sqrt{2}}\bar u \gamma_\mu P_L \tilde V_{u\vphantom{d}}^{\vphantom{\dag}} \tilde V_d^\dag d A_{W^+}^\mu + \hc \,,
\end{eqnarray}
it is straightforward to see that this leads to the same matrix element as the one \eqref{eq:Mfi_exact} obtained using the LSZ formula. As expected, comparing this effective charged current interaction Lagrangian with that of the SM \eqref{eq:L_gauge_SM_cc}, we are again led to the definition \eqref{eq:CKM_eff} of the effective CKM matrix.

\subsection{Electromagnetic and neutral current interactions}
\label{subsec:nc_em}

In the same way as we probed in the previous section the charged current sector, it is possible to investigate the flavor mixing also in the electromagnetic and neutral current sectors. Since the procedure is technically completely analogous, we merely state the results. Considering the decay processes $Z \rightarrow f_i + \bar f_j$ and $\gamma \rightarrow f_i + \bar f_j$, $f=u,d$, we arrive at the corresponding effective interaction Lagrangians (to be part of $\eL^{\mathrm{(eff)}}(u,d)$)
\begin{subequations}
\label{eq:L_eff_nc_em}
\begin{eqnarray}
\eL_{\mathrm{nc}}^{\mathrm{(eff)}}(u,d) &=& \frac{g}{2 \cos \theta_{\mathrm{W}}}
\sum_{f=u,d} \bar f \gamma_\mu \Big[ (v_f+a_f) \tilde V_f^{} \tilde V_f^\dag P_L + (v_f-a_f) \tilde U_f^{} \tilde U_f^\dag P_R \Big] f A_{Z}^\mu \,,
\qquad
\label{ew:eL:eff:nc}
\\
\eL_{\mathrm{em}}^{\mathrm{(eff)}}(u,d) &=& \sum_{f=u,d} e Q_f
\bar f \gamma_\mu \Big( \tilde V_f^{} \tilde V_f^\dag P_L + \tilde U_f^{} \tilde U_f^\dag P_R \Big) f A_{\mathrm{em}}^\mu \,.
\label{ew:eL:eff:em}
\end{eqnarray}
\end{subequations}
The matrices $\tilde U_u$, $\tilde U_d$ are defined in \eqref{eq:def:U_V_tilde}. Again, they are in general non-unitary, but in the special case of constant self-energies they reduce to the unitary matrices $U_u$, $U_d$, \eqref{mx:definitionUV}.

We see that, in contrast to their SM counterparts \eqref{eq:L_gauge_SM_nc}, \eqref{eq:L_gauge_SM_em}, the effective Lagrangians \eqref{ew:eL:eff:nc}, \eqref{ew:eL:eff:em} exhibit non-trivial flavor mixing. However, as expected, they reduce to those \eqref{eq:L_gauge_SM_nc}, \eqref{eq:L_gauge_SM_em} of the SM with no flavor mixing in the special case of constant self-energies, since then the matrices $\tilde V_f$, $\tilde U_f$ are unitary.

\section{Discussion}
\label{sec:summary_discussion}

First, a few comments are in order concerning the effectiveness of $\eL^{\mathrm{(eff)}}(u,d)$.
It is not effective in the usual sense as being a low energy approximation of the full theory.\footnote{By \qm{full theory} one can in the narrower sense understand the theory described in part~\ref{part:ew}. In the wider sense, though, as the discussion of this chapter can be applied to \emph{any} EWSB dynamics, generating momentum-dependent quarks self-energies of the type \eqref{ewa:Sgmf}, the term \qm{full theory} can be understood just as the $\group{SU}(2)_{\mathrm{L}} \times \group{U}(1)_{\mathrm{Y}}$ gauge-invariant theory of quarks \emph{plus} the symmetry-breaking quark self-energies \eqref{ewa:Sgmf}, without specification of the precise mechanism of their generation.} Rather, it is by construction effective in the sense that it reproduces predictions of the full theory, but only for a very limited set of processes (and only at the tree level). Namely, on top of reproducing the quark mass spectrum, only the processes $W^+ \rightarrow u_i + \bar d_j$ and $Z / \gamma \rightarrow q_i + \bar q_j$, $q=u,d$, modulo crossing symmetry, are computed correctly (i.e., in accordance with the full theory). If one calculates any more complicated process (e.g., $W^+ + W^- \rightarrow q_i + \bar q_j$) within this effective theory, one obtains an answer differing from the answer obtained within the full theory. Clearly, we have lost some amount of the physical information contained in the full theory when passing to the effective one. However, this makes sense, since the self-energies as the momentum-dependent matrix \emph{functions} (in the full theory) contain \qm{much more} physical information than the \emph{constants} like the masses and the flavor mixing matrices (in the effective theory).

There is a significant exception, though. In the case of constant self-energies the amount of physical information remains the same while going from the full theory to the effective one. Recall that in this case the effective Lagrangian $\eL^{\mathrm{(eff)}}(u,d)$ (Eqs.~\eqref{eq:L_eff_cc}, \eqref{eq:L_eff_nc_em}) reduces precisely to the SM Lagrangian $\eL^{\mathrm{(SM)}}(u,d)$ (Eq.~\eqref{eq:L_gauge_SM}), which is indeed fully physically equivalent to the full theory, since the two are related by the unitary transformation \eqref{eq:def:quark_unitary_redef}.

This leads us to another substantial difference between the two cases. We are accustomed from the SM that the interaction eigenstate basis ($u^\prime$, $d^\prime$) and mass eigenstate basis ($u$, $d$) are related to each other by the unitary transformation \eqref{eq:def:quark_unitary_redef} and working in either of them is merely a matter of taste. This is clearly not the case in the more general situation of momentum-dependent self-energies: Here the interaction eigenstate basis operators $u^\prime$, $d^\prime$ are the fundamental ones and there is no way to obtain from them the mass eigenstate basis operators $u$, $d$ by a suitable unitary transformation. This is of course related to the effective nature of the corresponding Lagrangian $\eL^{\mathrm{(eff)}}(u,d)$, since the operators $u$, $d$ are nothing more that merely postulated, effective fields.

The comparison with Refs.~\cite{Machet:2004rv,Duret:2006wk} is in order now. We have confirmed the phenomenological results concerning non-unitarity of the effective CKM matrix and occurence of the flavor changing electromagnetic and neutral currents. In particular, we recovered the explicit formula \eqref{eq:CKM_eff} for the former. What is new in our treatment is that we provided also explicit formul{\ae} \eqref{eq:L_eff_nc_em} for the flavor mixing in the electromagnetic and neutral current sectors. Moreover, we found out that the corresponding mixing matrices are only effective ones: They allow to compute the processes in the lowest order in the gauge coupling constants, but if one wants to go to higher orders of the perturbation theory, it is necessary to come back to the self-energies and consider their full momentum dependence.

We also contributed to the discussion of the relation between the interaction eigenstate basis $u^\prime$, $d^\prime$ and the mass eigenstate basis $u$, $d$. We confirmed that both bases cannot be related by a unitary transformation. The authors of Refs.~\cite{Machet:2004rv,Duret:2006wk} showed, however, that the two bases can be related by the \emph{non-unitary} transformation
\begin{equation}
\label{eq:def:quark_unitary_redef_tilde}
u = \tilde X_u^\dag u^\prime \,,
\quad\quad\quad
d = \tilde X_d^\dag d^\prime \,,
\end{equation}
(cf.~Eq.~\eqref{eq:def:quark_unitary_redef}) with non-unitary $\tilde X$'s defined by Eq.~\eqref{eq:def:X_tilde}. (The resulting non-diagonality of quark kinetic terms due to non-unitarity of matrices $\tilde X$ can be cured by adding appropriate finite counterterms to Lagrangian \cite{Shabalin:1978rs,Duret:2008md,Duret:2008st}.) This is in accordance with our result: Using the non-unitary redefinitions \eqref{eq:def:quark_unitary_redef_tilde} in the Lagrangian \eqref{eq:L_gauge_SM} (and neglecting impacts on kinetic terms), we arrive precisely at our effective Lagrangian \eqref{eq:L_eff_cc}, \eqref{eq:L_eff_nc_em}. Since we argued, however, that any Lagrangian, written in the mass eigenstate basis, should be (at least in principle) regarded as an effective one, in the sense described above, we conclude that the non-unitary transformations \eqref{eq:def:quark_unitary_redef_tilde} should be regarded effective as well.

\section{Summary}

We have investigated some of the implications of the non-trivial momentum dependence of the quark self-energies. We concentrated on the mixing between the quark mass eigenstates in the charge current, as well as in the neutral current and electromagnetic sector. We found that, depending on the details of the momentum-dependency of the self-energies, the resulting CKM matrix can be, in general, non-unitary and the neutral and electromagnetic currents can change flavor already at the tree level.

These results were expressed by the interaction Lagrangians \eqref{eq:L_eff_cc}, \eqref{eq:L_eff_nc_em} in terms of the mass eigenstate basis, i.e., with the operators creating (and annihilating) quark states with definite mass. We argued that the mass eigenstate basis cannot be, in general, related to the original interaction eigenstate basis by a unitary transformation of the type \eqref{eq:def:quark_unitary_redef}. In this sense the mentioned interaction Lagrangians were considered only as effective ones.


\part{Gauge boson masses}
\label{part:gauge}

\chapter{Preliminaries}
\label{chp:gbp}

\intro{The spontaneous symmetry breaking of a gauged symmetry leads to the generation of masses of (at least some of) the gauge bosons. This part is dedicated to the problem how to calculate these masses under the assumption that the symmetry is broken by fermion propagators. The scalar contribution will not be considered for the reasons briefly discussed at the end of Sec.~\ref{gbm:ssec:symm} in the next chapter.}

\intro{In this and in the following chapter we will discuss the problem in as general way as possible and only in chapters~\ref{chp:ablgg} and \ref{ewM} we will apply the obtained results to the gauged Abelian toy model (from part~\ref{part:abel}) and to the electroweak interactions (from part~\ref{part:ew}). Thus, this chapter is dedicated to mere setting the stage, i.e., to introducing the notation and stating the assumptions under which we will in the next chapter~\ref{chp:gbm} derive the very formula for the gauge boson mass matrix. Although this chapter can be therefore perhaps omitted at first reading, we do not put it in appendices, as it provides an organic introduction to the subsequent chapters.}

\intro{This chapter is organized deliberately into two main sections. In the first section~\ref{chp:gbp:sec:glob}, \qm{Global symmetry}, we discuss the issues, which are not related to eventual gauging. In particular, we introduce the fermion content together with the assumed (global) symmetries and derive the Ward--Takahashi (WT) identity for the corresponding Green's function $\langle j_a^\mu \psi \bar\psi \rangle$. In the subsequent section~\ref{chp:gbp:sec:loc}, \qm{Local symmetry}, we gauge the theory and discuss various properties of gauge bosons and their propagator and also derive, using the result from the preceding section, the WT identity for the Green's function $\langle A_a^\mu \psi \bar\psi \rangle$.}

\section{Global symmetry}
\label{chp:gbp:sec:glob}

\subsection{Fermion content}

\subsubsection{General}

Assume that we have a theory with $n$ left-handed fermion fields $\psi_{Li}$, $i=1,\ldots,n$, and with $m$ right-handed fermion fields $\psi_{Rj}$, $j=1,\ldots,m$. We organize these fields into the left-handed $n$-plet $\psi_L$ and the right-handed $m$-plet $\psi_R$, respectively:
\begin{eqnarray}
\label{gbp:ferms}
\psi_L \>\equiv\> \left(\begin{array}{c} \psi_{L1} \\ \vdots \\ \psi_{Ln}\end{array} \right)
\,, &\qquad&
\psi_R \>\equiv\> \left(\begin{array}{c} \psi_{R1} \\ \vdots \\ \psi_{Rm}\end{array} \right)
\,,
\end{eqnarray}
and denote the corresponding Lagrangian as $\eL(\psi)$. Its kinetic part is
\begin{subequations}
\label{gbp:eL:kinpsi}
\begin{eqnarray}
\eL_{\mathrm{kinetic}}(\psi) &=&
\sum_{i=1}^{n} \bar\psi_{Li} \I\slashed{\partial} \psi_{Li} +
\sum_{j=1}^{m} \bar\psi_{Rj} \I\slashed{\partial} \psi_{Rj}
\\
&=& \bar\psi_L \I\slashed{\partial} \psi_L + \bar\psi_R \I\slashed{\partial} \psi_R \,.
\end{eqnarray}
\end{subequations}

Assume further that the theory possesses a \emph{global} symmetry with some Lie group $\group{G}$, which can be possibly non-Abelian. The fields $\psi_L$ and $\psi_R$, \eqref{gbp:ferms}, transform under $\group{G}$ as
\begin{subequations}
\label{gbp:psiLRG}
\begin{eqnarray}
\group{G}\,:\qquad \psi_L  & \TransformsTo & {[\psi_L]}^\prime \>=\> \e^{\I\theta_a t_{La}} \, \psi_L \,,
\\
\group{G}\,:\qquad \psi_R  & \TransformsTo & {[\psi_R]}^\prime \>=\> \e^{\I\theta_a t_{Ra}} \, \psi_R \,,
\end{eqnarray}
\end{subequations}
where $\theta_a$ are the parameters of the transformation and the generators $t_{La}$, $t_{Ra}$ are Hermitian matrices with the dimensions $n \times n$, $m \times m$, respectively, forming some representations of $\group{G}$, which need not be necessarily irreducible. The range of the gauge index $a=1,\ldots,N_{\group{G}}$ is given by the dimension of $\group{G}$.

The right-hand sides of the transformations \eqref{gbp:psiLRG} are infinitesimally given by
\begin{subequations}
\begin{eqnarray}
{[\psi_L]}^\prime &=& \psi_L + \theta_a \, \delta_a \psi_L + \mathcal{O}(\theta_a^2) \,, \\
{[\psi_R]}^\prime &=& \psi_R + \theta_a \, \delta_a \psi_R + \mathcal{O}(\theta_a^2) \,,
\end{eqnarray}
\end{subequations}
where
\begin{subequations}
\begin{eqnarray}
\delta_a \psi_L &\equiv& \I \, t_{La} \psi_L \,, \\
\delta_a \psi_R &\equiv& \I \, t_{Ra} \psi_R \,.
\end{eqnarray}
\end{subequations}
Thus, the Noether current $j_a^\mu$ corresponding to the transformations \eqref{gbp:psiLRG} is defined as\footnote{We assume that there are no other fields than $\psi_L$, $\psi_R$, transforming non-trivially under $\group{G}$. Otherwise such fields would contribute to $j_a^\mu$ as well.}
\begin{eqnarray}
j_a^\mu &=&
- \frac{\partial\eL(\psi)}{\partial(\partial_\mu\psi_L)}\delta_a \psi_L
- \frac{\partial\eL(\psi)}{\partial(\partial_\mu\psi_R)}\delta_a \psi_R
\end{eqnarray}
and explicitly reads
\begin{eqnarray}
\label{gbp:jamu:psiLR}
j_a^\mu &=& \bar \psi_L \gamma^\mu t_{La} \psi_L + \bar \psi_R \gamma^\mu t_{Ra} \psi_R \,.
\end{eqnarray}
(We assume, of course, that in $\eL(\psi)$ there are no other derivatives of the fermion fields than those in the kinetic terms \eqref{gbp:eL:kinpsi}.) Recall the crucial property of $j_a^\mu$ of being conserved:\footnote{We neglect the possibility of anomalous non-conservation of the current.}
\begin{eqnarray}
\label{gbp:currcons}
\partial_\mu j_a^\mu &=& 0 \,,
\end{eqnarray}
as can be seen by taking into account the corresponding equations of motion.

And finally and most importantly, we assume that there is some dynamics in the theory. We leave this dynamics unspecified in order to make present discussion as general as possible and also because we actually do not need to specify it in much detail. The only thing we assume is that the dynamics spontaneously breaks the symmetry $\group{G}$ down to a subgroup $\group{H} \subseteq \group{G}$:
\begin{eqnarray}
\label{gbp:SSBpatt}
\group{G} &\longrightarrow& \group{H} \subseteq \group{G} \,.
\end{eqnarray}
Operational meaning of this assumption will be specified in the following sections; in nutshell, we will only assume that the dynamics provides us with symmetry-breaking fermion self-energies of the type discussed in the previous chapters.

Likewise we denoted the number of generators of $\group{G}$ as $N_{\group{G}}$, we will denote the number of generators of $\group{H}$ as $N_{\group{H}}$.

\subsubsection{Dirac case}

The picture introduced so far is quite general in the sense that it does not assume anything special concerning the fermion content \eqref{gbp:ferms}, the symmetry group $\group{G}$ and the pattern \eqref{gbp:SSBpatt} of the eventual SSB. However, we will from now assume for simplicity the following:
\begin{description}
  \item[A1] The numbers of the left-handed and the right-handed fermions are the same:
\begin{eqnarray}
n &=& m \,. \label{gbp:n=m}
\end{eqnarray}
  \item[A2] The symmetry group $\group{G}$ has a $\group{U}(1)$ subgroup:
\begin{eqnarray}
\group{U}(1) &\subseteq& \group{G} \,.
\end{eqnarray}
  \item[A3] The dynamics is such that the $\group{U}(1)$ subgroup, mentioned in \textbf{A2}, remains unbroken:
\begin{eqnarray}
\group{U}(1) &\subseteq& \group{H} \,.
\end{eqnarray}
\end{description}

The consequences of the assumptions \textbf{A1}--\textbf{A3} are discussed in more detail in appendix~\ref{app:fermi propag}, now let us state only the main points. The assumption \textbf{A1} implies that since the multiplets \eqref{gbp:ferms} have the same dimensions, one can define the field
\begin{eqnarray}
\label{gbp:psi}
\psi &\equiv& \psi_L + \psi_R \,,
\end{eqnarray}
allowing for more compact formalism. The assumption \textbf{A2} implies that there are no Majorana mass terms in the free Lagrangian and the bare fermion propagator can be consequently expressed just as $\langle \psi\bar \psi\rangle_0$ (i.e., there is no necessity for introducing the Nambu--Gorkov formalism \eqref{frm:Psidef} in order to incorporate the Majorana propagators of the type $\langle (\psi)^\C\bar \psi\rangle$ etc.). And finally, the assumption \textbf{A3} implies that even though the dynamics is switched on, still no Majorana self-energies are generated and the full fermion propagator can be expressed as $\langle \psi\bar \psi\rangle$ too.

The assumptions \textbf{A1}--\textbf{A3} are by no means necessary and we make them here only for simplicity. Any of them can be violated and in fact in the case of neutrinos it is violated, as we saw on in the previous chapters. In such a case, when the assumptions \textbf{A1}--\textbf{A3} are not fulfilled, one can work with the Nambu--Gorkov field
\begin{eqnarray}
\Psi &\equiv& \left(\begin{array}{c}\psi_L + (\psi_L)^\C \\ \psi_R + (\psi_R)^\C \end{array}\right)
\end{eqnarray}
instead of $\psi$ (for details see appendix~\ref{app:fermi propag}). In fact, all considerations that we will do from now with quantities (like propagators, symmetry generators, vertices, etc.) expressed in the basis $\psi$, can be equally well done with the same quantities expressed in the basis $\Psi$. Therefore we will not lose any generality by assuming \textbf{A1}--\textbf{A3}.

Having accepted the assumptions \textbf{A1}--\textbf{A3}, let us now rewrite the formul{\ae} above from the bases $\psi_L$, $\psi_R$ into the single basis $\psi$. The kinetic terms \eqref{gbp:eL:kinpsi} recast as
\begin{eqnarray}
\label{gbp:eL:kinpsii}
\eL_{\mathrm{kinetic}}(\psi) &=& \bar\psi \I\slashed{\partial} \psi \,.
\end{eqnarray}
The transformation \eqref{gbp:psiLRG} in terms of $\psi$ is
\begin{subequations}
\label{gbp:psiG}
\begin{eqnarray}
\group{G}\,:\qquad \psi  & \TransformsTo & {[\psi]}^\prime \>=\> \e^{\I\theta_a t_a} \, \psi \,.
\end{eqnarray}
Let us state this time explicitly also the corresponding transformation of $\bar \psi$:
\begin{eqnarray}
\group{G}\,:\qquad \bar\psi  & \TransformsTo & {[\bar\psi]}^\prime \>=\> \bar\psi \, \e^{-\I\theta_a \bar t_a} \,.
\end{eqnarray}
\end{subequations}
Notice that while the transformation of $\psi$ is generated by the generator $t_a$, defined as
\begin{subequations}
\label{gbp:taLR}
\begin{eqnarray}
t_a &\equiv& t_{La} P_L + t_{Ra} P_R
\end{eqnarray}
(needless to say that due to \eqref{gbp:n=m} the matrices $t_{La}$, $t_{Ra}$ are of the same dimension and hence it is correct to add them up), the transformation of $\bar\psi$ is generated rather by $\bar t_a$, defined standardly (cf.~\eqref{symbols:barA}) as $\bar t_a \equiv \gamma_0 t_a^\dag \gamma_0$, i.e., having the form
\begin{eqnarray}
\bar t_a &=& t_{La} P_R + t_{Ra} P_L \,.
\end{eqnarray}
\end{subequations}
Notice that the generators $t_a$ and $\bar t_a$ differ only by the sign at the $\gamma_5$ matrix.

In some applications later it will prove to be more convenient to parameterize the generators $t_a$, $\bar t_a$ not as \eqref{gbp:taLR}, i.e., as a linear combination of the chiral projectors $P_L$, $P_R$, but rather as a linear combination of $1$ and $\gamma_5$:
\begin{subequations}
\label{gbp:taVA}
\begin{eqnarray}
t_a &=& t_{Va} + t_{Aa} \gamma_5 \,,
\\
\bar t_a &=& t_{Va} - t_{Aa} \gamma_5 \,,
\end{eqnarray}
\end{subequations}
where the new generators $t_{Va}$, $t_{Aa}$ are again Hermitian and again do not contain of course any $\gamma_5$ matrices. The two bases $t_{La}$, $t_{Ra}$ and $t_{Va}$, $t_{Aa}$ are related to each other by obvious relations
\begin{subequations}
\begin{eqnarray}
t_{Va} &=& \frac{1}{2}\big(t_{Ra}+t_{La}\big) \,, \\
t_{Aa} &=& \frac{1}{2}\big(t_{Ra}-t_{La}\big)
\end{eqnarray}
\end{subequations}
and
\begin{subequations}
\begin{eqnarray}
t_{Ra} &=& t_{Va}+t_{Aa} \,, \\
t_{La} &=& t_{Va}-t_{Aa} \,.
\end{eqnarray}
\end{subequations}

The Noether current $j_a^\mu$ in terms of the generators $t_a$ reads
\begin{eqnarray}
\label{gbp:jamu:psi}
j_a^\mu &=& \bar \psi \gamma^\mu t_a \psi \,.
\end{eqnarray}
Notice that now, in contrast to the expression \eqref{gbp:jamu:psiLR} for $j_a^\mu$, the order of $\gamma^\mu$ and $t_a$ matters, since in general $\gamma^\mu t_a = \bar t_a \gamma^\mu \neq t_a \gamma^\mu$ due to presence of $\gamma_5$ in $t_a$.

\subsubsection{Fermion propagator}

Consider the full fermion propagator $\I\,G = \langle \psi \bar\psi \rangle$. For the sake of later references, we state here the proper definition of its Fourier transform:
\begin{eqnarray}
\label{gbp:GpsiFourier}
\bra{0} T \big[\psi(x) \bar \psi(y) \big] \ket{0}
&=& \int\!\frac{\d^4 p}{(2\pi)^4}\,\I G(p) \, \e^{-\I p \cdot (x-y)} \,.
\end{eqnarray}
The full propagator $G$ has general form
\begin{eqnarray}
G^{-1} &=& S^{-1}-\boldsymbol{\Sigma}
\end{eqnarray}
where $S$ is the free (bare) propagator, defined by the part of the Lagrangian quadratic in the fermion fields, i.e., with the interactions neglected. The $\boldsymbol{\Sigma}$ is the (proper) self-energy, i.e., the 1PI part of the propagator: $-\I \boldsymbol{\Sigma} = \langle \psi \bar\psi \rangle_{\mathrm{1PI}}$.

The full propagator $G$ transforms under $\group{G}$ as
\begin{eqnarray}
\label{gbp:Gtranf}
\group{G}\,:\qquad G  & \TransformsTo & [G]^\prime \>=\>  \e^{\I \theta \cdot t} \, G \, \e^{-\I \theta \cdot \bar t} \,,
\end{eqnarray}
as can be seen by applying the transformation \eqref{gbp:psiG} on the definition \eqref{gbp:GpsiFourier} of $G$. From this we can deduce the transformation rule for $G^{-1}$ and consequently also for $\boldsymbol{\Sigma}$:
\begin{subequations}
\label{gbp:Sgm:trans}
\begin{eqnarray}
\group{G}\,:\qquad \boldsymbol{\Sigma}  \>\>\TransformsTo\>\> [\boldsymbol{\Sigma}]^\prime &=&  \e^{\I \theta \cdot \bar t} \, \boldsymbol{\Sigma} \, \e^{-\I \theta \cdot t}
\label{gbp:Sgm:transglob} \\
&=& \boldsymbol{\Sigma}
- \I \theta_a \big(\boldsymbol{\Sigma}\,t_a - \bar t_a\,\boldsymbol{\Sigma}\big)
+ \mathcal{O}(\theta^2) \,.
\label{gbp:Sgm:transinf}
\end{eqnarray}
\end{subequations}
In the infinitesimal form \eqref{gbp:Sgm:transinf} we can identify the quantity
\begin{eqnarray}
\label{gbp:Sgmnoninv}
\bbl \boldsymbol{\Sigma},t_a \bbr &\equiv& \boldsymbol{\Sigma}\,t_a - \bar t_a\,\boldsymbol{\Sigma} \,,
\end{eqnarray}
which measures the non-invariance of the self-energy $\boldsymbol{\Sigma}$ under action of $\group{G}$ generated by the generator $t_a$.

Likewise in most this text, also in this chapter we will not consider the fermion propagator in the most general form, but rather somewhat constrained. First of all, we will assume that it satisfies the Hermiticity condition
\begin{eqnarray}
\label{gbp:hermG}
G &=& \bar G \,.
\end{eqnarray}
Notice that, as shown in the appendix~\ref{app:fermi propag}, the free propagator $S$ (being actually only a \emph{special case} of $G$) satisfies this condition automatically, due to Hermiticity of the Lagrangian. Thus, the condition \eqref{gbp:hermG} for $G$ is in fact equivalent to the similar condition for $\boldsymbol{\Sigma}$:
\begin{eqnarray}
\label{gbp:hermSgm}
\boldsymbol{\Sigma} &=& \boldsymbol{\bar \Sigma} \,.
\end{eqnarray}

Furthermore, we will assume that the fermions are at the level of the Lagrangian massless, i.e., that the free propagator has the simple form
\begin{eqnarray}
\label{gbp:Sbare}
S^{-1} &=& \slashed{p}
\end{eqnarray}
and is therefore invariant under the transformation \eqref{gbp:Gtranf}, generated by the group $\group{G}$: $[S]^\prime = S$. We in fact do not lose any generality by making the assumption \eqref{gbp:Sbare}; its purpose is merely to have a simpler notation, since any potential hard masses can be included by redefinition of $\boldsymbol{\Sigma}$.

More crude assumption is made concerning the $\boldsymbol{\Sigma}$ itself, as we will assume it to be a function only of $p^2$, not $\slashed{p}$. This assumption can be compactly written as
\begin{eqnarray}
\label{gbp:Sgnnowfr}
{\big[\boldsymbol{\Sigma},\gamma_5\big]} &=& 0
\end{eqnarray}
and corresponds to the assumptions made within the Abelian toy model and the electroweak interactions. Thus, under the assumptions \eqref{gbp:hermSgm} and \eqref{gbp:Sgnnowfr} the self-energy $\boldsymbol{\Sigma}$ has the familiar form
\begin{eqnarray}
\label{gbp:Sgm:form}
\boldsymbol{\Sigma} &=& \Sigma^\dag P_L+\Sigma\,P_R \,,
\end{eqnarray}
where $\Sigma$ is a complex $n \times n$ matrix and is a function only of $p^2$. Concerning the momentum dependence we only assume that $\lim_{p^2 \rightarrow \infty} \Sigma(p^2) = 0$, in order that certain integrals, to be introduced thereinafter, be UV-finite. Apart from this, the form \eqref{gbp:Sgm:form} of the self-energy is no longer constrained, in particular, we do not assume any special commutation relations between the self-energy $\boldsymbol{\Sigma}$ and the symmetry generators $t_a$.


Finally, under the assumptions \eqref{gbp:Sbare} and \eqref{gbp:Sgm:form} the full propagator
\begin{eqnarray}
\label{gbp:Gfrm}
G &=& \big( \slashed{p} - \boldsymbol{\Sigma} \big)^{-1}
\end{eqnarray}
can be explicitly expressed as
\begin{eqnarray}
\label{gbp:Gfrminv}
G &=& \big(\slashed{p} + \boldsymbol{\Sigma}^\dag\big)\boldsymbol{D}_L
\ = \
\boldsymbol{D}_R\big(\slashed{p} + \boldsymbol{\Sigma}^\dag\big) \,,
\end{eqnarray}
where we denoted
\begin{subequations}
\label{gbp:DLDR}
\begin{eqnarray}
\boldsymbol{D}_L &\equiv& \big(p^2-\boldsymbol{\Sigma}\,\boldsymbol{\Sigma}^\dag\big)^{-1} \,, \\
\boldsymbol{D}_R &\equiv& \big(p^2-\boldsymbol{\Sigma}^\dag\boldsymbol{\Sigma}\big)^{-1} \,,
\end{eqnarray}
\end{subequations}
in accordance with definitions \eqref{frm:def:bldDRDL} in appendix \eqref{app:fermi propag}. Needless to say that in general $\boldsymbol{D}_{L} \neq \boldsymbol{D}_{R}$.

\subsubsection{$\mathcal{C}$, $\mathcal{P}$ and $\mathcal{CP}$ transformations}

The discrete symmetries $\mathcal{C}$ and $\mathcal{P}$ (i.e., the charge conjugation and the parity, respectively) act on the fermion field $\psi(x)$ as
\begin{eqnarray}
\mathcal{C}\,:\qquad \psi(x)  & \TransformsTo & {[\psi(x)]}^{\mathcal{C}} \>=\>  \psi^{\C}(x) \,,
\label{gbp:PsitrC} \\
\mathcal{P}\,:\qquad \psi(x)  & \TransformsTo & {[\psi(x)]}^{\mathcal{P}} \>=\>  \gamma_0 \, \psi(\tilde x) \,.
\label{gbp:PsitrP}
\end{eqnarray}
Here the charge conjugated field $\psi^{\C}$ is defined in \eqref{symbols:psiC} and more details can be found in appendix~\ref{app:charge}. The symbol $\tilde x$ denotes the parity-transformed $4$-vector $x$,
\begin{eqnarray}
\tilde x &\equiv& (x_0,-\threevector{x})
\end{eqnarray}
or
\begin{eqnarray}
\tilde x^\mu &\equiv& \mathcal{P}^\mu_{\phantom{\mu}\nu} \, x^\nu \,,
\end{eqnarray}
where $\mathcal{P}^\mu_{\phantom{\mu}\nu}$ is the Lorentz transformation corresponding to the space reflection, i.e.,
\begin{eqnarray}
\mathcal{P}^\mu_{\phantom{\mu}\nu} &\equiv&
\left(\begin{array}{rrrr}
1 &  0 &  0 &  0 \\
0 & -1 &  0 &  0 \\
0 &  0 & -1 &  0 \\
0 &  0 &  0 & -1
\end{array}\right) \,.
\end{eqnarray}
For the combined $\mathcal{CP}$ transformation (first is applied the charge conjugation $\mathcal{C}$ and then the parity $\mathcal{P}$) of $\psi(x)$ we therefore have
\begin{subequations}
\label{gbp:PsitrCP}
\begin{eqnarray}
\mathcal{CP}\,:\qquad \psi(x)  \>\>\TransformsTo\>\> {[\psi(x)]}^{\mathcal{CP}} &=&  \gamma_0 \, \psi^\C(\tilde x)
\\
&=& \gamma_0 \, C \, \bar \psi^\T(\tilde x) \,,
\end{eqnarray}
\end{subequations}
where $C$ is the matrix of charge conjugations, introduced in appendix~\ref{app:charge}.

The full fermion propagator $G(p)$ now transforms under $\mathcal{C}$ and $\mathcal{P}$ as
\begin{eqnarray}
{[G(p)]}^{\mathcal{C}} &=& C \, G^\T(-p) \, C^{-1} \,,
\\
{[G(p)]}^{\mathcal{P}} &=& \gamma_0 \, G(\tilde p) \, \gamma_0 \,,
\end{eqnarray}
where $\tilde p$ is defined analogously as $\tilde x$ above. The transformation under $\mathcal{CP}$ follows as
\begin{eqnarray}
[G(p)]^{\mathcal{CP}} &=& \gamma_0 \, C \, G^\T(-\tilde p) \, C^{-1} \,\gamma_0 \,.
\end{eqnarray}
The same transformation rules hold also for the inverse propagator $G^{-1}(p)$, i.e., in particular, also for the self-energy $\boldsymbol{\Sigma}(p)$.

Let us now return to the assumptions made above concerning the form of the propagator. Considering the bare propagator $S(p)$, \eqref{gbp:Sbare}, we easily observe that it is invariant under both $\mathcal{C}$ and $\mathcal{P}$, simply due to $-C \, \gamma_\mu^\T \, C^{-1} = \gamma_\mu$ and $\gamma_0\,\slashed{\tilde p}\,\gamma_0 = \slashed{p}$, respectively. Considering the Ansatz \eqref{gbp:Sgm:form} for the self-energy $\boldsymbol{\Sigma}$, we first note that as it depends only on $p^2$ (rather than on $\slashed{p}$), we can suppress the momentum argument due to $p^2 = (-\tilde p)^2$ and write the transformation rules simply as
\begin{eqnarray}
{[\boldsymbol{\Sigma}]}^{\mathcal{C}}   &=& C             \, \boldsymbol{\Sigma}^\T \, C^{-1}  \,,
\label{gbp:SgmC} \\
{[\boldsymbol{\Sigma}]}^{\mathcal{P}}   &=& \gamma_0      \, \boldsymbol{\Sigma}    \, \gamma_0 \,,
\label{gbp:SgmP} \\
{[\boldsymbol{\Sigma}]}^{\mathcal{CP}}  &=& \gamma_0 \, C \, \boldsymbol{\Sigma}^\T \, C^{-1} \,\gamma_0 \,.
\label{gbp:SgmCP}
\end{eqnarray}
For concreteness, in terms of the Ansatz \eqref{gbp:Sgm:form} these transformations explicitly read
\begin{eqnarray}
{[\boldsymbol{\Sigma}]}^{\mathcal{C}}  &=& \Sigma^*    P_L + \Sigma^\T P_R \,, \\
{[\boldsymbol{\Sigma}]}^{\mathcal{P}}  &=& \Sigma^\dag P_R + \Sigma    P_L \,, \\
{[\boldsymbol{\Sigma}]}^{\mathcal{CP}} &=& \Sigma^\T   P_L + \Sigma^*  P_R \,.
\end{eqnarray}
Put another way, in terms of $\Sigma$ the $\mathcal{C}$, $\mathcal{P}$ and $\mathcal{CP}$ transformations consist merely of transposition, Hermitian conjugation and complex conjugation, respectively:
\begin{eqnarray}
{[\Sigma]}^{\mathcal{C}}  &=& \Sigma^\T \,, \\
{[\Sigma]}^{\mathcal{P}}  &=& \Sigma^\dag \,, \\
{[\Sigma]}^{\mathcal{CP}} &=& \Sigma^* \,.
\end{eqnarray}

It will be also useful to know the transformation properties under $\mathcal{C}$, $\mathcal{P}$ and $\mathcal{CP}$ of the Noether current $j_a^\mu$, which we rename here for convenience as $j_{t_a}^\mu$,
\begin{eqnarray}
j^\mu_{t_a}(x) &\equiv& \bar\psi(x)\, \gamma^\mu \,t_a\, \psi(x) \,,
\end{eqnarray}
in order to mark explicitly its dependence on the symmetry generator $t_a$. The transformation rules under $\mathcal{C}$ and $\mathcal{P}$, induced solely by the corresponding transformation \eqref{gbp:PsitrC}, \eqref{gbp:PsitrP} of the fermion fields, are
\begin{eqnarray}
\big[j^{\mu}_{t_a}(x)\big]^{\mathcal{C}} &=& -j^\mu_{\bar t_a^\C}(x) \,,
\label{gbp:jamu:mathcalC}
\\
\big[j^{\mu}_{t_a}(x)\big]^{\mathcal{P}} &=& \mathcal{P}^\mu_{\hphantom{\mu}\nu} \, j^\nu_{\bar t_a}(\tilde x) \,,
\end{eqnarray}
where
\begin{eqnarray}
\label{gbp:tc}
t_a^\C &\equiv& C\,t_a^\T\,C^{-1} \,,
\end{eqnarray}
see \eqref{symbols:AC}. For the combined transformation $\mathcal{CP}$ we have
\begin{eqnarray}
\label{gbp:jamu:mathcalCP}
\big[j^{\mu}_{t_a}(x)\big]^{\mathcal{CP}} &=& -\mathcal{P}^\mu_{\hphantom{\mu}\nu} \, j^\nu_{t_a^\C}(\tilde x) \,.
\end{eqnarray}

At this point it is appropriate to introduce some notation to be used in the following sections. Assuming that $t_a^\C$ can be expressed as a linear combination of the generators $t_a$, we define the corresponding matrix $C_{ab}$ as\footnote{Do not confuse this matrix with the fermion matrix $C$ of charge conjugation.}
\begin{eqnarray}
\label{gbp:tclin}
t_a^\C &=& C_{ab} \, t_b \,.
\end{eqnarray}
It can be shown, using $\Tr[t_a^\C t_b^\C] = \Tr[t_a t_b] \sim \delta_{ab}$ and $(t_a^\C)^\C = t_a$, that $C_{ab}$ must be symmetric and orthogonal:
\begin{eqnarray}
\label{gbp:Cab:prop}
C &=& C^{\T} \ = \  C^{-1}
\end{eqnarray}
and also independent of $\gamma_5$.

It may be instructive to give some explicit examples of $C_{ab}$:
\begin{description}
\item[Group $\group{U}(1)$:] The only generator $t$ is in general given as $t = t_L P_L + t_R P_R$, where $t_L$, $t_R$ are arbitrary real numbers. Obviously:
\begin{eqnarray}
C &=& 1 \,.
\end{eqnarray}
\item[Group $\group{SU}(2)$:] The generators are given as $t_a = \tfrac{1}{2} \sigma_a$ in terms of the Pauli matrices $\sigma_a$, $a=1,2,3$. Recalling that $\sigma_1$ and $\sigma_3$ are symmetric, while $\sigma_2$ is antisymmetric, we obtain \cite{Peccei:1998jv}
\begin{eqnarray}
C &=& \diag(+1,-1,+1) \,.
\end{eqnarray}
\item[Group $\group{SU}(3)$:] The generators are given as $t_a = \tfrac{1}{2} \lambda_a$, where $\lambda_a$, $a=1,\ldots,8$, are the Gell-Mann matrices. This time we get \cite{Peccei:1998jv}
\begin{eqnarray}
C &=& \diag(+1,-1,+1,+1,-1,+1,-1,+1) \,.
\end{eqnarray}
\end{description}

Using the notion of the matrix $C_{ab}$ the transformation \eqref{gbp:jamu:mathcalC} of the current under $\mathcal{C}$ can be reexpressed as
\begin{eqnarray}
\big[j^{\mu}_{t_a}(x)\big]^{\mathcal{C}} &=& -C_{ab} \, j^\mu_{\bar t_b}(x) \,.
\end{eqnarray}
Considering the $\mathcal{CP}$ transformation of the current, we can return to the original notation \qm{$j_a^\mu$} (rather than \qm{$j_{t_a}^\mu$}) and rewrite \eqref{gbp:jamu:mathcalCP} as
\begin{eqnarray}
\big[j^{\mu}_{a}(x)\big]^{\mathcal{CP}} &=& -C_{ab} \, \mathcal{P}^\mu_{\hphantom{\mu}\nu} \, j^\nu_{b}(\tilde x) \,.
\end{eqnarray}

\subsection{Global Ward--Takahashi identity}
\label{gbp:globWT}

In this section we derive the Ward--Takahashi (WT) \cite{Ward:1950xp,Takahashi:1957xn} identity for the global symmetry $\group{G}$. Later, in section~\ref{gbp:locWT}, we will argue that the same result holds also once the symmetry is gauged.

%
%
%
%
%

\subsubsection{Preliminary calculation}

Before proceeding to the very derivation of the WT identity, we make some preliminary calculation, which we will use also later in section~\ref{gbp:locWT}.

Consider the $T$-product of the type $T \big[V^\mu \, \psi \, \bar \psi\big]$, where $V^\mu$ is a bosonic operator. The $T$-product is then given explicitly by
\begin{eqnarray}
\label{gbp:Tproddef}
T \big[ V^\mu(x) \, \psi(y) \, \bar \psi(z) \big]
&=&
\nonumber \\ &&
\hphantom{+\,}
\hspace{-3cm}
\theta(x_0-y_0)\,\theta(y_0-z_0) \, V^\mu(x)    \, \psi(y)     \, \bar\psi(z) -
\theta(z_0-y_0)\,\theta(y_0-x_0) \, \bar\psi(z) \, \psi(y)     \, V^\mu(x)
\nonumber \\ &&
\hspace{-3cm}
+\,
\theta(y_0-z_0)\,\theta(z_0-x_0) \, \psi(y)     \, \bar\psi(z) \, V^\mu(x)    -
\theta(x_0-z_0)\,\theta(z_0-y_0) \, V^\mu(x)    \, \bar\psi(z) \, \psi(y)
\nonumber \\ &&
\hspace{-3cm}
+\,
\theta(y_0-x_0)\,\theta(x_0-z_0) \, \psi(y)     \, V^\mu(x)    \, \bar\psi(z) -
\theta(z_0-x_0)\,\theta(x_0-y_0) \, \bar\psi(z) \, V^\mu(x)    \, \psi(y) \,.
\nonumber \\ &&
\end{eqnarray}
We now compute its derivative with respect to $x$, i.e., apply the operator $\partial_\alpha^x$. At doing so one must remember that not only $V^\mu$ itself is $x$-dependent, but so are also some of the Heawiside functions $\theta$ in the definition \eqref{gbp:Tproddef} of the $T$-product. Thus, taking this carefully into account and using the formula $\frac{\d}{\d x} \theta(x) = \delta(x)$, we arrive at
\begin{eqnarray}
\partial_\alpha^x \, T \big[ V^\mu(x) \, \psi(y) \, \bar \psi(z) \big]
&=&
T \big[ \big(\partial_\alpha^x V^\mu(x)\big) \, \psi(y) \, \bar \psi(z) \big]
\nonumber \\ && \hspace{-3cm}{}+
g_{\alpha 0}\,\delta(x_0-y_0)\Big(
\theta(y_0-z_0)\,             \big[V^\mu(x),\psi(y)\big]\,\bar\psi(z)      -
\theta(z_0-y_0)\,\bar\psi(z)\,\big[V^\mu(x),\psi(y)\big]              \Big)
\nonumber \\ && \hspace{-3cm}{}+
g_{\alpha 0}\,\delta(x_0-z_0)\Big(
\theta(y_0-z_0)\,\psi(y)\,\big[V^\mu(x),\bar\psi(z)\big]               -
\theta(z_0-y_0)\,         \big[V^\mu(x),\bar\psi(z)\big]\,\psi(y) \Big) \,.
\nonumber \\ &&
\label{gbp:Tprodder}
\end{eqnarray}
Here we have already rearranged the resulting terms in order to have them in the convenient form of the commutators. Notice that these commutators are in fact equal-time, due to the preceding delta-functions.

\subsubsection{Derivation of the WT identity}

Consider now the Green's function $\langle j_a^\mu \psi \bar \psi \rangle$. We will calculate its divergence with respect to $x$, i.e., the quantity $\partial_\mu^x\langle j_a^\mu \psi \bar \psi \rangle$. Recall that $\langle j_a^\mu \psi \bar \psi \rangle$ is shorthand for $\bra{0} T \big[ j_a^\mu(x) \, \psi(y) \, \bar\psi(z) \big] \ket{0}$. Thus, as it contains the $T$-product, we can use the result \eqref{gbp:Tprodder} with $V^\mu = j_a^\mu$. We obtain
\begin{eqnarray}
\partial_\mu^x \bra{0} T \big[ j_a^\mu(x) \, \psi(y) \, \bar \psi(z) \big] \ket{0}
&=&
\bra{0} T \big[ \big(\partial_\mu^x j_a^\mu(x)\big) \, \psi(y) \, \bar \psi(z) \big] \ket{0}
\nonumber \\ && \hspace{-4.5cm}{}+
\delta(x_0-y_0)\Big(
\theta(y_0-z_0)             \bra{0} \big[j_a^0(x),\psi(y)\big]\,\bar\psi(z)\ket{0}      -
\theta(z_0-y_0)\bra{0}\bar\psi(z)\,\big[j_a^0(x),\psi(y)\big]\ket{0}              \Big)
\nonumber \\ && \hspace{-4.5cm}{}+
\delta(x_0-z_0)\Big(
\theta(y_0-z_0)\bra{0}\psi(y)\,\big[j_a^0(x),\bar\psi(z)\big]\ket{0}               -
\theta(z_0-y_0)         \bra{0}\big[j_a^0(x),\bar\psi(z)\big]\,\psi(y)\ket{0} \Big) \,.
\nonumber \\ &&
\label{gbp:derJamupsibarpsi}
\end{eqnarray}
First of all, the first term, containing $\big(\partial_\mu^x j_a^\mu(x)\big)$, can be dismissed due to the conservation \eqref{gbp:currcons} of the current $j_a^\mu$. To proceed we have to calculate the commutators $\big[j_a^0(x),\psi(y)\big]$ and $\big[j_a^0(x),\bar\psi(z)\big]$. Invoking the form \eqref{gbp:jamu:psi} of the current $j_a^\mu$ we can use the simple matrix identity $[AB,C] = A\{B,C\}-\{A,C\}B$ to rewrite the commutators in terms of the anticommutators of the type $\{\psi,\psi\}$ and $\{\psi,\psi^\dag\}$. Recall that the commutators are equal-time, thus so are the anticommutators. However, they are therefore nothing else than the canonical anticommutators \eqref{app:quant:equaltime} of the fermion fields, stemming from the process of quantization, as shown in appendix~\ref{app:quant}. Using this fact we readily arrive at
\begin{subequations}
\begin{eqnarray}
\big[j_a^0(x),\psi(y)\big]_{\mathrm{e.t.}}     &=&
-\delta^3(\threevector{x}-\threevector{y}) \, t_a \, \psi(y) \,, \\
\big[j_a^0(x),\bar\psi(y)\big]_{\mathrm{e.t.}} &=&
\phantom{-}\delta^3(\threevector{x}-\threevector{y}) \, \bar\psi(y) \, \bar t_a \,.
\end{eqnarray}
\end{subequations}
We can now plug these results into \eqref{gbp:derJamupsibarpsi}. After factorizing the $\delta^3$ functions out of the round brackets we see that the contents of the round brackets have the form of $T$-products of fermion operators:
\begin{eqnarray}
\partial_{\mu}^x \bra{0} T \big[ j_a^\mu(x) \, \psi(y) \, \bar \psi(z) \big] \ket{0}
&=&
\nonumber \\ && \hspace{-2cm} {}
- \delta^4(x-y) \, t_a \, \bra{0} T \big[\psi(y) \, \bar \psi(z) \big] \ket{0}
+ \delta^4(x-z) \, \bra{0} T \big[\psi(y) \, \bar \psi(z) \big] \ket{0} \, \bar t_a \,.
\nonumber \\ &&
\label{gbp:WTglobx}
\end{eqnarray}
We recognize the quantities on the right-hand side as the fermion propagators $\I G = \langle \psi \bar\psi \rangle$. This equation is in fact the coveted WT identity, relating the three-point Green's function $\langle j_a^\mu \psi \bar \psi \rangle$ with the two-point Green's function $\langle \psi \bar \psi \rangle$.

Finally, it is useful to rewrite the WT identity \eqref{gbp:WTglobx} into the momentum space in terms of the 1PI function $\langle j_a^\mu \psi \bar \psi \rangle_{\mathrm{1PI}} = \gamma_a^\mu$. The Fourier transform of $\langle j_a^\mu \psi \bar \psi \rangle$ is defined as
\begin{eqnarray}
\label{gbp:brkt:jamu:ft}
\bra{0} T \big[ j_a^\mu(x) \psi(y) \bar \psi(z) \big] \ket{0}
&=&
\nonumber \\ && \hspace{-3cm}
\int \! \frac{\d^4 q}{(2\pi)^4} \frac{\d^4 p^\prime}{(2\pi)^4} \frac{\d^4 p}{(2\pi)^4}
\, \e^{\I q \cdot x} \, \e^{-\I p^\prime \cdot y} \, \e^{\I p \cdot z}
\, (2\pi)^4 \delta^4(p+q-p^\prime) \,
\, \I G(p^\prime) \, \gamma_a^\mu(p^\prime,p) \, \I G(p) \,,
\nonumber \\ &&
\end{eqnarray}
where we have already explicitly indicated its 1PI part $\gamma_a^\mu$. Taking this definition and the definition \eqref{gbp:GpsiFourier} of the Fourier transform of the fermion propagator $G$ into account we can rewrite the WT identity \eqref{gbp:WTglobx} in a more convenient and familiar form as
\begin{eqnarray}
\label{gbp:WTglob}
q_{\mu}\gamma_a^\mu(p^\prime,p) &=& G^{-1}(p^\prime)\,t_a - \bar t_a\,G^{-1}(p) \,,
\end{eqnarray}
where $p^\prime \equiv p+q$.

\section{Local symmetry}
\label{chp:gbp:sec:loc}

\subsection{Gauging of the theory}

\subsubsection{Lagrangian}

We now \qm{switch on} the gauge interactions. That it to say, we assume that the transformation \eqref{gbp:psiG} is local,\footnote{We assume here, again merely for simplicity, that the whole group $\group{G}$ is gauged, whereas in some applications this may be the case only for some its subgroup.} i.e., the parameters $\theta_a$ are now position-dependent. In order to maintain the invariance of the theory under such \emph{gauge transformation}, we are forced to introduce a set of \emph{gauge bosons} -- the spin-1 massless particles $A_a^\mu$, $a=1,\ldots,N_{\group{G}}$, coupled in a specific way to the fermions. At the Lagrangian level instead of $\eL(\psi)$ we have to deal now with its extension \cite{Peskin:1995ev}
\begin{eqnarray}
\label{gbp:eL:psiAamu}
\eL(\psi,A_a^\mu) &=& \eL(\psi) + g j_a^\mu A_{a\mu} - \frac{1}{4} F_a^{\mu\nu} F_{a \mu\nu} \,,
\end{eqnarray}
where $F_a^{\mu\nu}$ is the gauge boson \emph{field-strength tensor}, defined as
\begin{eqnarray}
\label{gbp:Fmunu}
F_a^{\mu\nu} &\equiv& \partial^\mu A_a^\nu - \partial^\nu A_a^\mu + g f_{abc} A_b^\mu A_c^\nu \,.
\end{eqnarray}
As there are derivatives in it, the last term in \eqref{gbp:eL:psiAamu}, proportional to $(F_a^{\mu\nu})^2$, thus contains the kinetic terms for the gauge bosons. Moreover, it potentially contains also the gauge boson self-interaction terms, proportional to the structure constants $f_{abc}$ of the group $\group{G}$. These are defined using the commutation relations of the generators of $\group{G}$ as
\begin{eqnarray}
[t_a,t_b] &=& \I f_{abc} \, t_c \,.
\end{eqnarray}

The term $g j_a^\mu A_{a\mu}$ in \eqref{gbp:eL:psiAamu} contains the coveted interactions of the fermions with the gauge bosons:
\begin{subequations}
\label{gbp:eL:int}
\begin{eqnarray}
\eL_{\mathrm{int.}} &=& g j_a^\mu A_{a\mu}
\\ &=& g \bar\psi \gamma_\mu t_{a} \psi A_a^\mu \,.
\end{eqnarray}
\end{subequations}
Another way of deriving them consists of trading the partial derivatives in the fermion kinetic term \eqref{gbp:eL:kinpsii} for the covariant derivatives, i.e.,
\begin{eqnarray}
\partial^\mu  \ \longrightarrow \ \D^\mu &\equiv& \partial^\mu - \I g \, t_{a} A_a^\mu \,,
\end{eqnarray}
so that the kinetic term \eqref{gbp:eL:kinpsii} modifies as
\begin{eqnarray}
\eL_{\mathrm{kinetic}}(\psi) \ = \ \bar\psi \I\slashed{\partial} \psi &\longrightarrow&  \bar\psi \I\slashed{\D} \psi
\ = \  \eL_{\mathrm{kinetic}}(\psi) + g j_a^\mu A_{a\mu} \,.
\end{eqnarray}

\subsubsection{Quantization}

The Lagrangian \eqref{gbp:eL:psiAamu} describes a classical theory. The process of its quantization entails effectively two modifications.

First, one must fix the gauge. By fixing the gauge we avoid the problem of overcounting the gauge boson degrees of freedom in the functional integral by counting in the gauge fields related by a gauge transformation (and hence physically equivalent). We will fix the gauge by adding the gauge-fixing term of the form
\begin{eqnarray}
\label{gbp:eL:gf}
\eL_{\mathrm{g.f.}} &=& - \frac{1}{2\xi}(\partial_\mu A_a^\mu)^2
\end{eqnarray}
to the Lagrangian \eqref{gbp:eL:psiAamu}.

Second, the quantization of the Lagrangian \eqref{gbp:eL:psiAamu} requires also introduction of the Faddeev--Popov ghosts $c_a$, $a=1,\ldots,N_{\group{G}}$, with the Lagrangian
\begin{subequations}
\begin{eqnarray}
\eL_{\mathrm{ghosts}} &=& -\bar c_a \partial_\mu \D^\mu_{ab} c_b
\\ &=& - \bar c_a \square c_a - g f_{abc} \bar c_a \partial_\mu (A_b^\mu c_c) \,.
\end{eqnarray}
\end{subequations}
These ghost fields, emerging in the process of the quantizing, are scalars obeying the Fermi--Dirac statistics. Thus, as being unphysical, they can appear in the Feynman diagrams only in the closed loops, with the aim to preserve the unitarity of the theory. We will however not need them in the present text.


\subsection{Simplifying assumptions about the gauge dynamics}

\subsubsection{Weak gauge dynamics}

Let us comment on the quantity $g$, the \emph{gauge coupling constant}. We make now the key assumption that the gauge dynamics is weak, i.e.,
\begin{eqnarray}
\label{gbp:gsmall}
g &\ll& 1 \,,
\end{eqnarray}
so that the perturbative expansions in $g$ is meaningful.

The assumption \eqref{gbp:gsmall} about the weakness of the gauge dynamics was in fact implicitly present already in applications considered in the previous parts. Recall the situation in the Abelian toy model and in the electroweak interactions: The symmetry is broken spontaneously by the strong Yukawa dynamics, through the formation of appropriate fermion and scalar self-energies, while the gauge dynamics is actually not considered at all. It is assumed to be merely a passive spectator and can be incorporated only perturbatively, provided it is weakly coupled, as we are going to show in this chapter.

\subsubsection{More gauge coupling constants}


At this point a remark concerning the nature of the symmetry group $\group{G}$ is in order. In gauging the theory we have so far implicitly assumed that the symmetry $\group{G}$ is simple, which allowed us to introduce only one coupling constant $g$. If $\group{G}$ was not simple but rather a product of some simple subgroups, we would need to introduce a special coupling constant for each such subgroup. In fact, in the applications of our interest the group $\group{G}$ is always of the form of a product of two or more simple groups. Recall that in the Abelian toy model we had $\group{G} = \group{U}(1)_{\mathrm{V}_{\!1}} \times \group{U}(1)_{\mathrm{V}_{\!2}} \times \group{U}(1)_{\mathrm{A}}$, while in the electroweak interactions we had $\group{G} = \group{SU}(2)_{\mathrm{L}} \times \group{U}(1)_{\mathrm{Y}}$.

This problem can be overcome as follows. Assume that the group in question, $\group{G}$, is of the form
\begin{eqnarray}
\label{gbp:nonsimpleG}
\group{G} &=& \group{G}_1 \times \group{G}_2 \times \ldots \times \group{G}_N \,,
\end{eqnarray}
where each subgroup $\group{G}_i$ ($i=1,\ldots,N$) is a simple group with $N_{\group{G}_i}$ generators and we attribute a gauge coupling constant $g_{\group{G}_i}$ to it. Now we can define the diagonal matrix
\begin{eqnarray}
\label{gbl:gmatrix}
g &\equiv& \diag\big(
\underbrace{g_{\group{G}_1},\ldots,g_{\group{G}_1}}_{N_{\group{G}_1}\mbox{ times}},
\underbrace{g_{\group{G}_2},\ldots,g_{\group{G}_2}}_{N_{\group{G}_2}\mbox{ times}},
\ldots,
\underbrace{g_{\group{G}_N},\ldots,g_{\group{G}_N}}_{N_{\group{G}_N}\mbox{ times}}
\big) \,.
\end{eqnarray}
The point is that the gauge coupling constants typically appear in formul{\ae} in combinations with the quantities carrying the gauge index. E.g., in the case of only one gauge coupling constant $g$, considered so far, we deal typically with quantities of the type
\begin{equation}
\label{gbp:gXa}
g X_a \,,
\end{equation}
where $X_a$ can stand for a generator $t_a$, symmetry current $j_a^\mu$, gauge field $A_a^\mu$, etc. In the case of $\group{G}$ given by \eqref{gbp:nonsimpleG} the expressions of the type \eqref{gbp:gXa} generalize straightforwardly as
\begin{equation}
\label{gbp:gabXb}
g_{ab} X_b \,,
\end{equation}
where $g_{ab}$ is given by \eqref{gbl:gmatrix}.

In this text, however, for the sake of notational simplicity, we will still use the notation of the type \eqref{gbp:gXa} (i.e., pretending that $\group{G}$ is simple) and will keep in mind that such a notation is in a more general case \eqref{gbp:nonsimpleG} merely a shorthand for \eqref{gbp:gabXb}. In fact, later, after introducing the notation \eqref{gbp:Tagta} (combining a gauge coupling constant and a symmetry generator into a single symbol) we will not need to deal with this issue anymore.

\subsubsection{More Abelian factors in $\group{G}$}

If the gauge group $\group{G}$ contains more that one Abelian factor of $\group{U}(1)$, another subtlety comes into play \cite{Holdom:1985ag,Babu:1997st,Fonseca:2011vn}. Recall that strictly non-Abelian field-strength tensor $F_a^{\mu\nu}$, \eqref{gbp:Fmunu}, is gauge-covariant, but not gauge-invariant. The only way how to make a (renormalizable and $\mathcal{CP}$-conserving) gauge-invariant quantity out of it is to consider its \qm{square} $F_{a\mu\nu}^{\vphantom{\mu}}F_a^{\mu\nu}$, i.e., the usual kinetic term, as in the Lagrangian \eqref{gbp:eL:psiAamu}. In particular, the \qm{off-diagonal} kinetic terms, contracting field-strength tensor of two different groups, at least one of them being non-Abelian, are forbidden. On the other hand, in Abelian gauge theories this need not be true, as the Abelian field-strength tensor alone is already gauge-invariant. For definiteness, consider the gauge group in question to be
\begin{eqnarray}
\group{G} &=& \prod_a \group{U}(1)_a \,.
\end{eqnarray}
Then since each $F_{a}^{\mu\nu}$ is gauge-invariant, one can write the most general kinetic term as
\begin{eqnarray}
\eL_{\mathrm{kinetic}} &=& -\frac{1}{4} \xi_{ab} F_a^{\mu\nu} F_{b \mu\nu} \,,
\end{eqnarray}
where $\xi_{ab}$ (not to be confused with the gauge-fixing parameter above) is in principle \emph{non-diagonal}, real and positive matrix, which can be also without loss of generality assumed to be symmetric. By an appropriate rotation of the gauge fields the matrix $\xi_{ab}$ can be transformed into the unit matrix $\delta_{ab}$, but the prize is that the matrix $g$, \eqref{gbl:gmatrix}, of the gauge couplings constants is no longer diagonal. Still, however, $g$ can be made symmetric by a specific choice of coordinates in the gauge space.


This should be in principle taken into account especially in chapter~\ref{chp:ablgg}, where the Abelian toy model with the group $\group{G} = \group{U}(1)_{\mathrm{V}_{\!1}} \times \group{U}(1)_{\mathrm{V}_{\!2}} \times \group{U}(1)_{\mathrm{A}}$ will be gauged. However we will for the sake of simplicity treat the subject in the usual way: We will consider diagonal kinetic terms, so that there will be no mixing in the free propagator of the gauge bosons, and we will also associate each $\group{U}(1)$ with just one gauge coupling constant.

\subsection{Properties of the gauge fields}

Let us discuss briefly, to the needed extent, the properties of the gauge fields.

\subsubsection{Transformation properties}

Not only the fermions, but also the gauge fields themselves transform non-trivially under $\group{G}$. Assuming that $t_a$ are generators of some representation of $\group{G}$, then the action of $\group{G}$ on the gauge fields can be written as\footnote{In order not to waste the indices, we use in the exponentials the shorthand notation $\theta_a t_a \equiv \theta \cdot t$.}
\begin{eqnarray}
\group{G}\,:\qquad t_a A_a^\mu  & \TransformsTo & t_a{[A_a^\mu]}^\prime \>=\> \e^{\I \theta \cdot t} \Big(t_a A^\mu_a + \frac{\I}{g}\partial^\mu\Big) \e^{-\I \theta \cdot t} \,,
\end{eqnarray}
or, more compactly, as
\begin{eqnarray}
\label{gbp:Aamutrans}
\group{G}\,:\qquad A_a^\mu  & \TransformsTo & {[A_a^\mu]}^\prime \>=\> X_{ab}(\theta) \, A^\mu_b + \frac{1}{g} Y^\mu_{a}(\theta) \,,
\end{eqnarray}
where we defined the quantities $X_{ab}(\theta)$, $Y^\mu_{a}(\theta)$ as
\begin{eqnarray}
t_{a} \, X_{ab}(\theta)    &\equiv& \e^{\I \theta \cdot t} \, t_b \, \e^{-\I \theta \cdot t} \,,
\label{gbp:taXabetbe} \\
t_{a} \, Y^\mu_{a}(\theta) &\equiv& \e^{\I \theta \cdot t} \, \I \partial^\mu \, \e^{-\I \theta \cdot t} \,.
\end{eqnarray}
In the lowest order in the transformation parameters $\theta_a$ and their derivatives $\partial^\mu \theta_a$ we have explicitly
\begin{eqnarray}
X_{ab}(\theta)    &=& \delta_{ab} + f_{abc}\theta_c + \mathcal{O}(\theta^2) \,, \\
Y^\mu_{a}(\theta) &=& \partial^\mu \theta_a + \mathcal{O}(\theta^2) \,.
\end{eqnarray}

In the following we will be concerned mostly with the matrix $X(\theta)$. It satisfies
\begin{eqnarray}
\label{gbp:Xproperties}
X(-\theta) &=&  X^{-1}(\theta) \ =\  X^{\T}(\theta) \ =\ X^\dag(\theta) \,,
\end{eqnarray}
as can be seen from its expression in the form
\begin{eqnarray}
\label{gbp:XeiT}
X(\theta) &=& \exp (\I \theta_a \mathcal{T}_a) \,,
\end{eqnarray}
where the matrices $\mathcal{T}_a$ are generators of the adjoint representation of $\group{G}$, i.e., their elements are given by
\begin{eqnarray}
\label{gbp:adjoint}
(\mathcal{T}_a)_{bc} &=& -\I f_{abc} \,,
\end{eqnarray}
so that $\mathcal{T}_a$ are antisymmetric:
\begin{eqnarray}
\mathcal{T}_a^\T &=& -\mathcal{T}_a \,.
\end{eqnarray}
Recall that the structure constants $f_{abc}$ are real and antisymmetric. In terms of \eqref{gbp:XeiT} the relation \eqref{gbp:taXabetbe} recasts as
\begin{eqnarray}
\label{gbp:taeabetbe}
t_{a} \, \big(\e^{\I\theta\cdot\mathcal{T}}\big)_{ab} &=&
\e^{\I \theta \cdot t} \, t_b \, \e^{-\I \theta \cdot t} \,.
\end{eqnarray}

For the sake of later references let us write again the transformation \eqref{gbp:Aamutrans} of $A_a^\mu$, this time under \emph{global} $\group{G}$, i.e., with $Y^\mu_{a}(\theta) \equiv 0$:
\begin{subequations}
\label{gbp:Aamutransglob}
\begin{eqnarray}
\group{G}\,:\qquad A_a^\mu  \>\>\TransformsTo\>\> {[A_a^\mu]}^\prime
&=& \big(\e^{\I\theta\cdot\mathcal{T}}\big)_{ab} \, A^\mu_b \\
&=& A^{\mu}_a + \theta_b \, \delta_b A^{\mu}_a + \mathcal{O}(\theta^2) \,,
\label{gbp:Aamutransglobinf}
\end{eqnarray}
\end{subequations}
where
\begin{subequations}
\label{gbp:deltaaAmub}
\begin{eqnarray}
\delta_a A^{\mu}_b &\equiv& \I (\mathcal{T}_a)_{bc} A_c^\mu \\ &=& f_{abc} A_c^{\mu} \,.
\end{eqnarray}
\end{subequations}

\subsubsection{Equations of motion}

The equations of motion of the gauge bosons $A_a^\mu$, following from the Lagrangian \eqref{gbp:eL:psiAamu} by means of the standard Euler--Lagrange procedure, are
\begin{eqnarray}
\partial_\mu F_a^{\mu\nu} &=& - g J_{a}^\nu \,.
\end{eqnarray}
The quantity $J_a^\mu$ is the Noether current associated with the global symmetry $\group{G}$ of the Lagrangian \eqref{gbp:eL:psiAamu}, i.e., given by
\begin{subequations}
\label{gbp:Jamu:psiA}
\begin{eqnarray}
J_a^\mu &=&
- \frac{\partial\eL}{\partial(\partial_\mu\psi)} \, \delta_a \psi
- \frac{\partial\eL}{\partial(\partial_\mu A_b^\nu)} \, \delta_a A_b^\nu
\\
&=& j_{a}^\mu +  f_{abc} F_{b}^{\mu\nu} A_{c\nu} \,,
\end{eqnarray}
\end{subequations}
where $j_a^\mu$ is the given by \eqref{gbp:jamu:psi} and for $\delta_a A_b^\nu$ we used \eqref{gbp:deltaaAmub}. The current $J_{a}^\mu$ is conserved:
\begin{eqnarray}
\partial_\mu J_{a}^\mu &=& 0 \,,
\end{eqnarray}
implying that the current $j_{a}^\mu$ is no longer conserved as in \eqref{gbp:currcons}, its divergence is now proportional to $f_{abc}$:
\begin{eqnarray}
\label{gbp:currconsloc}
\partial_\mu j_{a}^\mu &=& \mathcal{O}(f_{abc}) \,.
\end{eqnarray}

Adding the gauge fixing term \eqref{gbp:eL:gf} to the Lagrangian \eqref{gbp:eL:psiAamu} the equations of motion modifiy as
\begin{eqnarray}
\partial_\mu F_a^{\mu\nu} &=& - g J_{a}^\nu - \frac{1}{\xi}\partial^\nu(\partial_\mu A_a^\mu) \,.
\end{eqnarray}
Employing the explicit form of $J_a^\mu$, \eqref{gbp:Jamu:psiA}, the equations of motion can be rewritten in a more convenient way as
\begin{eqnarray}
\label{gbl:gaugeEOM}
(\mathcal{D}^{-1})^{\mu\nu}_{ab} A_{b\nu} &=& - g j_a^\mu + \mathcal{O}(f_{abc})
\end{eqnarray}
Here we have introduced the differential operator $(\mathcal{D}^{-1})^{\mu\nu}_{ab}$, defined as
\begin{eqnarray}
\label{gbp:mathcalDm1munuab}
(\mathcal{D}^{-1})^{\mu\nu}_{ab} &\equiv&
\Big[ \partial^2 g^{\mu\nu} - \Big(1-\frac{1}{\xi}\Big) \partial^\mu \partial^\nu \Big]\unitmatrix_{ab} \,.
\end{eqnarray}

\subsubsection{Propagators}

Consider first the free propagator of the gauge bosons $A_a^\mu$, denoted as
\begin{eqnarray}
\I \, D_{ab}^{\mu\nu} &=& \langle A_a^\mu A_b^\nu\rangle_0 \,.
\end{eqnarray}
The free part of the Lagrangian $\eL(\psi,A_a^\mu)$, \eqref{gbp:eL:psiAamu}, i.e., the part quadratic in $A_a^\mu$, is given explicitly by
\begin{subequations}
\begin{eqnarray}
\eL_{\mathrm{gauge,quadratic}}(\psi,A_a^\mu) &=&
-\frac{1}{4} \big(\partial^\mu A_a^\nu - \partial^\nu A_a^\mu\big)^2 - \frac{1}{2\xi}\big(\partial_\mu A_a^\mu\big)^2
\\ &=& \frac{1}{2}
A_{a\mu} \, (\mathcal{D}^{-1})^{\mu\nu}_{ab} \, A_{b\nu}
+ \partial_\mu V^\mu(A,\partial A) \,,
\end{eqnarray}
\end{subequations}
where the differential operator $(\mathcal{D}^{-1})^{\mu\nu}_{ab}$ is given by \eqref{gbp:mathcalDm1munuab} and the four-vector $V^\mu(A,\partial A)$ is certain function of the gauge fields $A_a^\mu$ and their derivatives. Assuming that the surface term $\partial_\mu V^\mu$ can be neglected when computing the action, the Fourier transform of $(\mathcal{D}^{-1})^{\mu\nu}_{ab}$ defines the momentum space inverse free propagator $(D^{-1})_{ab}^{\mu\nu}$:
\begin{subequations}
\label{gbp:Dm1munuab}
\begin{eqnarray}
\big(D^{-1}\big)^{\mu\nu}_{ab} &=&
\int \! \d^4 x \, (\mathcal{D}^{-1})^{\mu\nu}_{ab} \, \e^{\I q \cdot x}
\label{gbp:Dm1munuabpoprve}
\\
&=&
- \Big[ q^2 g^{\mu\nu} - \Big(1-\frac{1}{\xi}\Big) q^\mu q^\nu \Big] \unitmatrix_{ab}
\\
&=& -\Big(g^{\mu\nu}-\frac{q^\mu q^\nu}{q^2}\Big) q^2 \unitmatrix_{ab} -\frac{1}{\xi} \frac{q^\mu q^\nu}{q^2} q^2 \unitmatrix_{ab} \,.
\end{eqnarray}
\end{subequations}
The full propagator $D_{ab}^{\mu\nu}$ is obtained by inverting \eqref{gbp:Dm1munuab},
\begin{eqnarray}
\big(D^{-1}D\big)^{\mu\nu}_{ab} \ \equiv \  (D^{-1})_{ac\,\rho}^{\mu} \, D_{cb}^{\rho\nu}  &=& g^{\mu\nu} \unitmatrix_{ab} \,,
\end{eqnarray}
so that we arrive at
\begin{eqnarray}
\label{gbp:Dmunuab}
D^{\mu\nu}_{ab} &=& -\Big(g^{\mu\nu}-\frac{q^\mu q^\nu}{q^2}\Big) \frac{\unitmatrix_{ab}}{q^2} -\xi \frac{q^\mu q^\nu}{q^2} \frac{\unitmatrix_{ab}}{q^2} \,.
\end{eqnarray}
Notice here the r\^{o}le of the gauge-fixing parameter $1/\xi$: If it was missing, the inverse propagator $(D^{-1})_{ab}^{\mu\nu}$ would be proportional to the projector $g^{\mu\nu}-q^\mu q^\nu/q^2$, so that it would be a singular matrix without a meaningful inversion. Only by addition of the projector $q^\mu q^\nu/q^2$ (proportional to $1/\xi$), which makes together with $g^{\mu\nu}-q^\mu q^\nu/q^2$ a complete set of projectors, we obtain a regular matrix suitable for inversion.

The full propagator $G^{\mu\nu}_{ab}$ of the gauge bosons,
\begin{eqnarray}
\label{gbp:Gabmunudef}
\I \, G_{ab}^{\mu\nu} &=& \langle A_a^\mu A_b^\nu\rangle \,,
\end{eqnarray}
is given in terms of the free one as (we suppress the Lorentz and gauge indices)
\begin{eqnarray}
G^{-1} &=& D^{-1} + \Pi \,,
\end{eqnarray}
where the polarization tensor $\Pi_{ab}^{\mu\nu}$ is the gauge boson self-energy:
\begin{eqnarray}
\label{gbp:Piabmunudef}
\I \Pi_{ab}^{\mu\nu} &=& \langle A_a^\mu A_b^\nu\rangle_{\mathrm{1PI}} \,.
\end{eqnarray}

It can be proved (see, e.g., \cite{Pokorski:1987ed}) that as a consequence of the symmetry of the Lagrangian the polarization tensor $\Pi_{ab}^{\mu\nu}$ must be transversal,
\begin{eqnarray}
\label{gbp:Piabmunucontract}
q_\mu \Pi_{ab}^{\mu\nu}(q) &=& 0 \,,
\end{eqnarray}
i.e., it is proportional to the transversal projector:
\begin{eqnarray}
\label{gbp:Piabmunutrans}
\Pi_{ab}^{\mu\nu}(q) &=& \Big(g^{\mu\nu}q^2-q^\mu q^\nu\Big) \Pi_{ab}(q^2) \,,
\end{eqnarray}
where the form factor $\Pi_{ab}$ (being a function of $q^2$ due to Lorentz invariance) is symmetric in the gauge indices $a$, $b$. The full propagator $G^{\mu\nu}_{ab}$ has consequently the form
\begin{eqnarray}
\label{gbp:Gabmunuexpl}
G^{\mu\nu}_{ab} &=&
-\Big(g^{\mu\nu}-\frac{q^\mu q^\nu}{q^2}\Big)\big[(q^2-q^2\Pi)^{-1}\big]_{ab}
-\xi \frac{q^\mu q^\nu}{q^2} \frac{\unitmatrix_{ab}}{q^2} \,.
\end{eqnarray}
Notice that only the transversal part of the full propagator $G^{\mu\nu}_{ab}$ gets renormalized, whereas the part proportional to the gauge-fixing parameter $\xi$ stays intact and is identical to its counterpart in the free propagator $D^{\mu\nu}_{ab}$, \eqref{gbp:Dmunuab}. This is in fact due to the transversality \eqref{gbp:Piabmunucontract} of the polarization tensor $\Pi^{\mu\nu}_{ab}$: Assuming an additional term $q^\mu q^\nu \Pi_{ab}^{(L)}$ in \eqref{gbp:Piabmunutrans}, the term proportional to $\xi$ in the full propagator $G^{\mu\nu}_{ab}$ would be modified as $\unitmatrix_{ab}/q^2 \rightarrow \big[(q^2-q^2 \Pi^{(L)})^{-1}\big]_{ab}$.

\subsubsection{Transformation under $\group{G}$}

The transformation rule for the full propagator $G^{\mu\nu}_{ab}$ under the \emph{global} symmetry $\group{G}$, i.e., under \eqref{gbp:Aamutransglob}, is
\begin{subequations}
\label{gbp:trnsf:Gabmunu}
\begin{align}
\group{G}\,:\qquad G_{ab}^{\mu\nu}
&\quad\TransformsTo\quad
{[G_{ab}^{\mu\nu}]}^\prime
\>=\> \big(\e^{\I\theta\cdot\mathcal{T}}\big)_{ac} \, G_{cd}^{\mu\nu} \, \big(\e^{-\I\theta\cdot\mathcal{T}}\big)_{db} \,,
\\
\intertext{or, by suppression the gauge indices, in a more compact matrix form}
\group{G}\,:\qquad G^{\mu\nu}
&\quad\TransformsTo\quad
{[G^{\mu\nu}]}^\prime
\>=\> \e^{\I\theta\cdot\mathcal{T}} \, G^{\mu\nu} \, \e^{-\I\theta\cdot\mathcal{T}} \,.
\end{align}
\end{subequations}
The transformation rule for the polarization tensor follows:
\begin{subequations}
\label{gbp:trnsf:Piabmunu}
\begin{align}
\group{G}\,:\qquad \Pi_{ab}^{\mu\nu}
&\quad\TransformsTo\quad
{[\Pi_{ab}^{\mu\nu}]}^\prime
\>=\> \big(\e^{\I\theta\cdot\mathcal{T}}\big)_{ac} \, \Pi_{cd}^{\mu\nu} \, \big(\e^{-\I\theta\cdot\mathcal{T}}\big)_{db} \,,
\\
\intertext{or, in the matrix form,}
\group{G}\,:\qquad \Pi^{\mu\nu}
&\quad\TransformsTo\quad {[\Pi^{\mu\nu}]}^\prime
\>=\> \e^{\I\theta\cdot\mathcal{T}} \, \Pi^{\mu\nu} \, \e^{-\I\theta\cdot\mathcal{T}} \,.
\end{align}
\end{subequations}
One can immediately see that the free propagator \eqref{gbp:Dmunuab} is invariant under $\group{G}$:
\begin{eqnarray}
\group{G}\,:\qquad D_{ab}^{\mu\nu}  & \TransformsTo & {[D_{ab}^{\mu\nu}]}^\prime \>=\> D_{ab}^{\mu\nu} \,.
\end{eqnarray}

As the transformations of the gauge propagators under $\group{G}$ do not touch the Lorentz indices, it is in particular clear that the form factor $\Pi_{ab}$, \eqref{gbp:Piabmunutrans} transforms in the same way as $\Pi_{ab}^{\mu\nu}$. I.e., one can use the transformation rule \eqref{gbp:trnsf:Piabmunu}, just with the Lorentz indices missing. Nevertheless, let us write, only for the sake of later references, the transformation rule of $\Pi_{ab}$ together with its infinitesimal form:
\begin{subequations}
\label{gbp:Pitransf}
\begin{eqnarray}
\group{G}\,:\qquad \Pi \>\>\TransformsTo\>\>  {[\Pi]}^\prime &=&
\e^{\I\theta\cdot\mathcal{T}} \, \Pi \, \e^{-\I\theta\cdot\mathcal{T}} \\
&=& \Pi + \I \theta_a \big[\mathcal{T}_a,\Pi\big] + \mathcal{O}(\theta^2) \,.
\end{eqnarray}
\end{subequations}

\subsubsection{$\mathcal{C}$, $\mathcal{P}$ and $\mathcal{CP}$ transformations}

Consider first the behavior of the gauge field $A_a^\mu$ under the charge conjugation $\mathcal{C}$. It transforms in such a way that the following relation holds \cite{Pokorski:1987ed}:
\begin{eqnarray}
{[A_a^\mu(x)]}^{\mathcal{C}} (-t_a^\C) &=& A_a^\mu(x) \, t_a \,.
\end{eqnarray}
Here $-t_a^\C$ can be recognized as the generators of the conjugate representation of $\group{G}$; recall in this respect also the definition \eqref{gbp:tc} of $t_a^\C$. It follows that the field-strength tensor $F_a^{\mu\nu}$, \eqref{gbp:Fmunu}, transforms in the same way
\begin{eqnarray}
{[F_a^{\mu\nu}(x)]}^{\mathcal{C}} (-t_a^\C) &=& F_a^{\mu\nu}(x) \, t_a \,,
\end{eqnarray}
so that the Yang--Mills Lagrangian $\eL = -\tfrac{1}{4} F_a^{\mu\nu} F_{a\mu\nu}$ stays invariant under $\mathcal{C}$.

In order to find a more compact expression for ${[A_a^\mu(x)]}^{\mathcal{C}}$, we recall that $t_a^\C$ can be expressed as a linear combination of the generators $t_a$. Using the corresponding relation \eqref{gbp:tclin} we arrive at more compact expression for the transformation rule of $A_a^\mu$ under $\mathcal{C}$ \cite{Peccei:1998jv}:
\begin{eqnarray}
{[A_a^\mu(x)]}^{\mathcal{C}} &=& - C_{ab} A_b^\mu(x) \,,
\end{eqnarray}
where we also used the properties \eqref{gbp:Cab:prop} of the matrix $C_{ab}$.

The transformation of the gauge field $A_a^\mu$ under parity $\mathcal{P}$ is straightforward:
\begin{eqnarray}
{[A_a^\mu(x)]}^{\mathcal{P}} &=& \mathcal{P}^\mu_{\phantom{\mu}\nu} \, A_a^\nu(\tilde x) \,,
\end{eqnarray}
so that we can readily write the transformation law for $A_a^\mu$ under the combined parity $\mathcal{CP}$:
\begin{eqnarray}
{[A_a^\mu(x)]}^{\mathcal{CP}} &=& - C_{ab} \, \mathcal{P}^\mu_{\phantom{\mu}\nu} \, A_b^\nu(\tilde x) \,.
\end{eqnarray}

The $\mathcal{C}$ and $\mathcal{P}$ transformations of the full propagator $G_{ab}^{\mu\nu}$, \eqref{gbp:Gabmunudef}, are given by
\begin{eqnarray}
{[G^{\mu\nu}_{ab}(q)]}^{\mathcal{C}} &=& C_{ac}\,C_{bd}\, G^{\mu\nu}_{cd}(q) \,,
\\
{[G^{\mu\nu}_{ab}(q)]}^{\mathcal{P}} &=& \mathcal{P}^\mu_{\phantom{\mu}\alpha} \, \mathcal{P}^\nu_{\phantom{\nu}\beta} \, G^{\alpha\beta}_{ab}(\tilde q) \,,
\end{eqnarray}
and the combined $\mathcal{CP}$ transformation is consequently
\begin{eqnarray}
{[G^{\mu\nu}_{ab}(q)]}^{\mathcal{CP}} &=& C_{ac}\,C_{bd}\,
\mathcal{P}^\mu_{\phantom{\mu}\alpha} \, \mathcal{P}^\nu_{\phantom{\nu}\beta} \, G^{\alpha\beta}_{cd}(\tilde q) \,.
\end{eqnarray}
The same transformation rules as for $G_{ab}^{\mu\nu}$ hold also for the polarization tensor $\Pi_{ab}^{\mu\nu}$, \eqref{gbp:Piabmunudef}. Noting the explicit forms \eqref{gbp:Gabmunuexpl} and \eqref{gbp:Piabmunutrans} of $G_{ab}^{\mu\nu}$ and $\Pi_{ab}^{\mu\nu}$, respectively, we see that the propagators are in fact invariant under parity: ${[G^{\mu\nu}_{ab}(q)]}^{\mathcal{P}} = G^{\mu\nu}_{ab}(q)$, so that effectively ${[G^{\mu\nu}_{ab}(q)]}^{\mathcal{CP}} = {[G^{\mu\nu}_{ab}(q)]}^{\mathcal{C}}$; analogously for $\Pi_{ab}^{\mu\nu}$. Thus, the $\mathcal{CP}$ transformation of $G_{ab}^{\mu\nu}$ and $\Pi_{ab}^{\mu\nu}$ manifests itself only by its effect on the form factor $\Pi_{ab}$:
\begin{eqnarray}
{[\Pi_{ab}]}^{\mathcal{CP}} &=& C_{ac}\,C_{bd}\, \Pi_{cd} \,.
\end{eqnarray}
From this we can in particular see that the free propagators $D_{ab}^{\mu\nu}$, \eqref{gbp:Dmunuab}, as well as the gauge-fixing part of the full propagator $G_{ab}^{\mu\nu}$, \eqref{gbp:Gabmunuexpl}, is invariant under $\mathcal{CP}$, since $C_{ac}\,C_{bd}\,\delta_{cd} = \delta_{ab}$

Consider finally the Lagrangian describing the interactions of the gauge fields with fermions:
\begin{eqnarray}
\label{gbp:eL:jamuAamu}
\eL(x) &=& g \, j_{t_a}^\mu\!(x) \, A_{a\mu}(x) \,.
\end{eqnarray}
Using the transformation rules for the current and for the gauge fields, we straightforwardly obtain
\begin{eqnarray}
\label{gbp:eL:jamuAamuCandP}
{[\eL(x)]}^{\mathcal{C}} \ = \ {[\eL(x)]}^{\mathcal{P}}
&=& g \, j_{\bar t_a}^\mu\!(\tilde x) \, A_{a\mu}(\tilde x) \,.
\end{eqnarray}
We therefore see that $\mathcal{C}$ and $\mathcal{P}$ are not good symmetries of the gauge interactions, unless  $\bar t_a = t_a$ for all $a$, i.e., unless the generators $t_a$ do not contain any $\gamma_5$ matrices. On the other hand, consider the transformation of the interaction Lagrangian \eqref{gbp:eL:jamuAamu} under the combined symmetry $\mathcal{CP}$. Using the results \eqref{gbp:eL:jamuAamuCandP} we readily observe
\begin{subequations}
\label{gbp:eL:jamuAamuCP}
\begin{eqnarray}
{[\eL(x)]}^{\mathcal{CP}} &=& g \, j_{t_a}^\mu\!(x) \, A_{a\mu}(x)
\\ &=& \eL(x) \,,
\end{eqnarray}
\end{subequations}
since $\tilde{\tilde x} = x$ and $\bar{\bar t}_a = t_a$. I.e., the gauge interactions are always invariant under $\mathcal{CP}$.\footnote{We mean here \emph{unbroken} gauge interactions. If the gauge symmetry is broken, the gauge sector can be $\mathcal{CP}$-violating. This happens, e.g., in the charged current interactions in the SM. Even in this case, however, it is not the gauge sector itself, but rather the Yukawa sector (or generally a sector, generating the fermion masses), which is ultimately responsible for the $\mathcal{CP}$-violation.}

\subsection{Gauge boson masses}
\label{gbp:ssec:gbmasses}

The gauge boson mass spectrum is given by the poles of their propagator. The free propagator $D_{ab}^{\mu\nu}$ has its only pole in $q^2=0$, which reflects the fact that the gauge bosons are at the Lagrangian level massless. However, the dynamics may be such that once the bare propagator is corrected by the polarization tensor $\Pi_{ab}^{\mu\nu}$, the resulting full propagator $G_{ab}^{\mu\nu}$ already has poles at some $q^2 \neq 0$, corresponding to massive gauge bosons. Clearly, the inspection of the full propagator $G_{ab}^{\mu\nu}$ reveals that the source of non-vanishing poles can be only the $\big(q^2 - q^2 \Pi\big)^{-1}$ part. Thus, the poles are given by the equation
\begin{eqnarray}
\label{gbp:gaugepole}
\det \big(q^2 - q^2 \Pi(q^2)\big) &=& 0 \,.
\end{eqnarray}

Let us investigate the conditions for the pole equation \eqref{gbp:gaugepole} to have a non-vanishing solution. In order to simplify the problem it is convenient first to diagonalize the symmetric matrix $\Pi(q^2)$ via the orthogonal transformation as
\begin{eqnarray}
\label{gbp:Pi=OpiO}
\Pi(q^2) &=& O(q^2) \, \pi(q^2) \, O^\T(q^2) \,,
\end{eqnarray}
where $\pi(q^2)$ is a diagonal matrix:
\begin{eqnarray}
\label{gbp:pi}
\pi(q^2) &=& \diag\big(\pi_1(q^2),\ldots,\pi_{N_{\group{G}}}(q^2)\big)
\end{eqnarray}
and $O(q^2)$ is an orthogonal matrix:
\begin{eqnarray}
\label{gbp:Oorthogon}
O(q^2) \, O^\T(q^2) &=& 1 \,.
\end{eqnarray}
We demand that $O(q^2)$ is orthogonal for all $q^2$, which implies that $O(q^2)$ is also \emph{regular} for all $q^2$, i.e., it has, in particular, no pole at $q^2=0$ and therefore is expressible in the form
\begin{eqnarray}
\label{gbp:Otaylor}
O(q^2) &=& \sum_{n=0}^{\infty} (q^2)^n \, O_{n} \,,
\end{eqnarray}
where $O_{n}$ are some momentum-independent matrix coefficients. The orthogonality condition \eqref{gbp:Oorthogon} is in terms of the first few coefficients $O_n$ expressed as
\begin{subequations}
\begin{eqnarray}
O_0^{\phantom{\T}} \, O_0^\T &=& 1 \,,
\label{gbp:Onulaorth} \\
O_0^{\phantom{\T}} \, O_1^\T + O_1^{\phantom{\T}} \, O_0^\T &=& 0 \,, \\
O_1^{\phantom{\T}} \, O_1^\T + O_0^{\phantom{\T}} \, O_2^\T + O_2^{\phantom{\T}} \, O_0^\T &=& 0 \,, \\
&\vdots& \nonumber
\end{eqnarray}
\end{subequations}

Using the orthogonal transformation \eqref{gbp:Pi=OpiO} the determinant in \eqref{gbp:gaugepole} simplifies as
\begin{eqnarray}
\det\big(q^2-q^2\Pi(q^2)\big) &=&  \prod_{a=1}^{N_{\group{G}}} \big(q^2-q^2\pi_a(q^2)\big) \,,
\end{eqnarray}
so that instead of the single pole equation \eqref{gbp:gaugepole} we have now a separate equation
\begin{eqnarray}
\label{gbp:pipole}
q^2-q^2\pi_a(q^2) &=& 0
\end{eqnarray}
for each $a=1,\ldots,{N_{\group{G}}}$.

Let us now discuss possibilities of the analytic structure of $\pi_a(q^2)$ (for some fixed $a$). Assume first that it is regular at $q^2=0$ (i.e., it has no pole at $q^2=0$). In such a case the pole equation \eqref{gbp:pipole} has clearly the vanishing solution $q^2=0$, which corresponds to the massless gauge boson. Now assume on contrary that $\pi_a(q^2)$ has a simple pole, i.e., a pole of the type $1/q^2$. Then the term $q^2 \pi_a(q^2)$ in the pole equation \eqref{gbp:pipole} contains a constant part, given by the residue of the pole $1/q^2$ of $\pi_a(q^2)$, and correspondingly $q^2=0$ cannot be a solution of the equation \eqref{gbp:pipole}. In other words, a pole of the type $1/q^2$ in $\pi_a(q^2)$ with a non-vanishing residue guarantees that the gauge boson acquires a non-vanishing mass. Finally, to complete the argument, one might also in principle assume that the pole in $\pi_a(q^2)$ is not simple (i.e., that it is, for instance, of the type $1/q^4$). However, this cannot happen, since any pole of a Green's function should be physically interpretable as a propagator of some intermediate particle. In the present case the allowed pole of the type $1/q^2$ corresponds to a massless scalar particle, the NG boson, coupled bilinearly to the gauge boson. We will discuss this interpretation more closely later in section~\ref{sec:NGboson}.

Thus, the most general form of $\pi_a(q^2)$ is
\begin{eqnarray}
\label{gbp:piaLaurent}
\pi_a(q^2) &=& \frac{1}{q^2} m_a^2 + \sum_{n=0}^\infty (q^2)^n \, \pi_{n,a} \,,
\end{eqnarray}
where $m_a^2$ and $\pi_{n,a}$ are some coefficients independent of $q^2$, and the full diagonal $\pi(q^2)$, \eqref{gbp:pi}, has the form
\begin{eqnarray}
\label{gbp:pilaurent}
\pi(q^2) &=& \frac{1}{q^2} m^2 + \sum_{n=0}^\infty (q^2)^n \, \pi_{n} \,,
\end{eqnarray}
where
\begin{eqnarray}
m^2     &\equiv& \diag\big(m_1^2,    \ldots,m_{N_{\group{G}}}^2    \big) \,, \\
\pi_{n} &\equiv& \diag\big(\pi_{n,1},\ldots,\pi_{n,{N_{\group{G}}}}\big) \,.
\end{eqnarray}

We showed that each particular $\pi(q^2)$ has the pole of the type $1/q^2$. On the other hand, recall that $O(q^2)$ has no pole at $q^2=0$, \eqref{gbp:Otaylor}. Thus, when applied to $\pi(q^2)$ to obtain $\Pi(q^2)$ via \eqref{gbp:Pi=OpiO}, $O(q^2)$ protects the pole structure \eqref{gbp:pilaurent} so that $\Pi(q^2)$ can be written in the same form as $\pi(q^2)$:
\begin{eqnarray}
\label{gbp:Pi:Laurent}
\Pi(q^2) &=& \frac{1}{q^2} M^2 + \sum_{n=0}^\infty (q^2)^n \, \Pi_{n} \,.
\end{eqnarray}
Here the coefficients $M^2$, $\Pi_{n}$ are some symmetric matrices, in principle non-diagonal. For the sake of later references we state here explicit relations between $M^2$, $\Pi_{n}$ and $m^2$, $\pi_{n}$ for the first few terms:
\begin{subequations}
\begin{eqnarray}
M^2 &=& O_{0}^{\phantom{\T}} \, m^2 \,  O_{0}^\T \,,
\label{gbp:MOmO0} \\
\Pi_{0} &=& O_{0}^{\phantom{\T}} \, \pi_{0}^{\phantom{\T}} \, O_{0}^\T +
O_{1}^{\phantom{\T}} \, m^2 \, O_{0}^\T + O_{0}^{\phantom{\T}} \, m^2 \, O_{1}^\T \,, \\
\Pi_{1} &=& O_{0}^{\phantom{\T}} \, \pi_{1}^{\phantom{\T}} \, O_{0}^\T +
O_{1}^{\phantom{\T}} \, \pi_{0}^{\phantom{\T}} \, O_{0}^\T + O_{0}^{\phantom{\T}} \, \pi_{0}^{\phantom{\T}} \, O_{1}^\T + O_{1}^{\phantom{\T}} \, m^2 \, O_{1}^\T +
O_{2}^{\phantom{\T}} \, m^2 \, O_{0}^\T + O_{0}^{\phantom{\T}} \, m^2 \, O_{2}^\T \,,
\qquad\qquad
\\ &\vdots& \nonumber
\end{eqnarray}
\end{subequations}

\subsection{Three-point function}
\label{gbp:3point}

The three-point function $\langle A_a^\mu \psi \bar\psi \rangle$ and especially its 1PI part will be subject of the most of the next chapter. We will now state some of its properties and derive the WT identity for it.

\subsubsection{Definitions}

\begin{figure}[t]
\begin{center}
\includegraphics[width=0.55\textwidth]{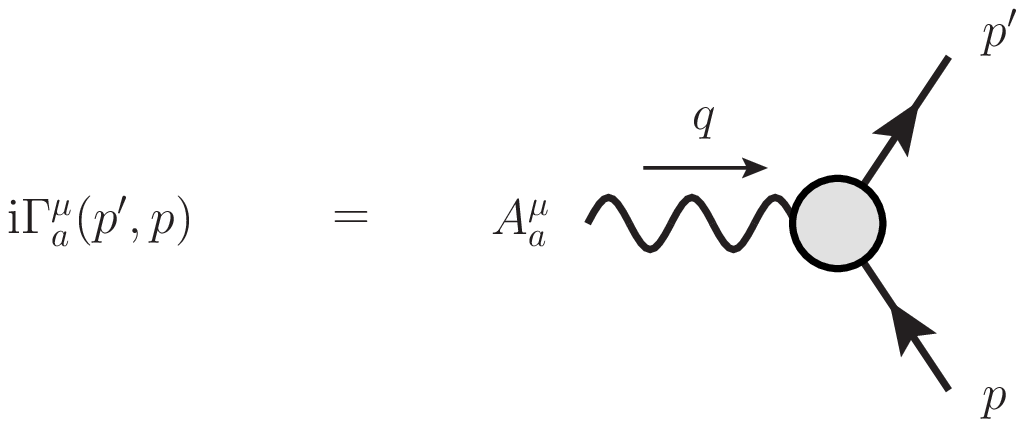}
\caption[Proper vertex $\langle A_a^\mu \psi \bar\psi \rangle_{\mathrm{1PI}} = \I\Gamma_a^\mu(p^\prime,p)$.]{Assignment of momenta of the proper vertex $\Gamma_a^\mu(p^\prime,p)$, \eqref{gbp:Gamupprimep}. Momentum conservation $q = p^\prime - p$ is implied.}
\end{center}
\end{figure}

Consider the three-point Green's function $\langle A_a^\mu \psi \bar\psi \rangle$,
\begin{eqnarray}
\label{gbp:Gamuxyzdef}
\I G_a^\mu(x,y,z) &=& \bra{0} T \big[ A_a^\mu(x) \psi(y) \bar \psi(z) \big] \ket{0} \,,
\end{eqnarray}
and its momentum representation
\begin{eqnarray}
\label{gbp:Gamuxyz}
G_a^\mu(x,y,z)
&=& \int\! \frac{\d^4 q}{(2\pi)^4} \frac{\d^4 p^\prime}{(2\pi)^4} \frac{\d^4 p}{(2\pi)^4}
\, \e^{\I q \cdot x} \, \e^{-\I p^\prime \cdot y} \, \e^{\I p \cdot z}
\, (2\pi)^4 \delta^4(p+q-p^\prime) \, G_a^\mu(p^\prime,p) \,.
\qquad
\end{eqnarray}
Since it is a \emph{full} Green's function, it can be written as the 1PI function $\I\Gamma_a^\mu = \langle A_a^\mu \psi \bar\psi \rangle_{\mathrm{1PI}}$ times the full propagators attached to the external legs:
\begin{eqnarray}
\label{gbp:Gamupprimep}
\I G_a^\mu(p^\prime,p) &=& \I G_{ab\nu}^\mu(q) \, \I G(p^\prime) \, \I \Gamma_b^\nu(p^\prime,p) \, \I G(p) \,,
\end{eqnarray}
where $q=p^\prime-p$.

Note that according to the Lagrangian \eqref{gbp:eL:int} the bare (tree) proper vertex $\Gamma_a^\mu(p^\prime,p)$ reads
\begin{subequations}
\label{gbp:Gmm:bare}
\begin{eqnarray}
\Gamma_a^\mu(p^\prime,p)\big|_{\mathrm{bare}} &=& g \, \gamma^\mu t_a \\ &=& \gamma^\mu T_a \,.
\end{eqnarray}
\end{subequations}
Here we have introduced the notation
\begin{subequations}
\label{gbp:Tagta}
\begin{eqnarray}
\label{gbp:Tagtasimple}
T_a &\equiv& g \, t_a \,,
\end{eqnarray}
to be used in the following extensively. Recall that, as discussed above, the definition \eqref{gbp:Tagtasimple} generalizes for $\group{G}$ given by \eqref{gbp:nonsimpleG} naturally as
\begin{eqnarray}
T_a &\equiv& g_{ab} \, t_b \,,
\end{eqnarray}
\end{subequations}
where $g_{ab}$ is given by \eqref{gbl:gmatrix}. We will call the quantity $T_a$ a \qm{generator} too.

\subsubsection{Hermiticity}

Since the fermion self-energy $\boldsymbol{\Sigma}$ is assumed to satisfy the Hermiticity condition \eqref{gbp:hermSgm}, it is only natural to assume that the vertex $\Gamma_a^\mu(p^\prime,p)$ satisfies an analogous condition too:
\begin{eqnarray}
\label{gbm:hermGmm}
\Gamma_a^\mu(p^\prime,p) &=& \bar \Gamma_a^{\mu}(p,p^\prime) \,.
\end{eqnarray}
The two Hermiticity conditions \eqref{gbp:hermSgm} and \eqref{gbm:hermGmm} will eventually, in the next chapter, ensure that the polarization tensor will be real.

\subsubsection{Transformation under $\group{G}$}

Let us now check the transformation properties. According to the transformation rules \eqref{gbp:psiG} and \eqref{gbp:Aamutransglob} for fermions and gauge bosons, respectively, we observe that the full (i.e., not 1PI) three-point function $\langle A_a^\mu \psi \bar\psi \rangle$, \eqref{gbp:Gamuxyzdef}, must transform under the \emph{global} symmetry $\group{G}$ as
\begin{eqnarray}
\group{G}\,:\qquad G_a^\mu(x,y,z)  & \TransformsTo & {[G_a^\mu(x,y,z)]}^\prime \>=\>
\e^{\I \theta \cdot t} \, \big(\e^{\I\theta\cdot\mathcal{T}}\big)_{ab} \, G_a^\mu(x,y,z) \, \e^{-\I \theta \cdot \bar t} \,.
\end{eqnarray}
The same transformation rule must, due to \eqref{gbp:Gamuxyz}, hold also in the momentum representation:
\begin{eqnarray}
\label{gbp:Gprimepoprve}
\group{G}\,:\qquad G_a^{\mu}(p^\prime,p)  & \TransformsTo & {[G_a^{\mu}(p^\prime,p)]}^\prime \>=\>
\e^{\I \theta \cdot t} \, \big(\e^{\I\theta\cdot\mathcal{T}}\big)_{ab} \, G_b^\mu(p^\prime,p) \, \e^{-\I \theta \cdot \bar t} \,.
\end{eqnarray}
Later we will be however interested rather in the transformation rule of the 1PI function $\I\Gamma_a^\mu = \langle A_a^\mu \psi \bar\psi \rangle_{\mathrm{1PI}}$. Recall that in the momentum representation the full and 1PI vertices are related by \eqref{gbp:Gamupprimep}. The transformation of $\langle A_a^\mu \psi \bar\psi \rangle$ under $\group{G}$ must be therefore induced by the transformations of its particular components expressed in \eqref{gbp:Gamupprimep}, i.e.,
\begin{eqnarray}
\label{gbp:Gprimepoddruhe}
\group{G}\,:\qquad \I G_a^{\mu}(p^\prime,p)  & \TransformsTo & {[\I G_a^{\mu}(p^\prime,p)]}^\prime \>=\>
[G_{ab\nu}^{\mu}(q)]^\prime \, [G(p^\prime)]^\prime \, [\Gamma_b^{\nu}(p^\prime,p)]^\prime \, [G(p)]^\prime \,.
\end{eqnarray}
We can now plug the expression \eqref{gbp:Gamupprimep} for $G_a^{\mu}(p^\prime,p)$ into \eqref{gbp:Gprimepoprve} and compare the resulting form of the expression for ${[G_a^{\mu}(p^\prime,p)]}^\prime$ with the other expression \eqref{gbp:Gprimepoddruhe} for the same quantity to obtain the equation
\begin{eqnarray}
\label{gbp:Gammaprimeeq}
\e^{\I \theta \cdot t} \, \big(\e^{\I\theta\cdot\mathcal{T}}\big)_{ab} \,
G_{ab\nu}^\mu(q) \, G(p^\prime) \, \Gamma_b^\nu(p^\prime,p) \, G(p)
\, \e^{-\I \theta \cdot \bar t}
&=&
[G_{ab\nu}^{\mu}(q)]^\prime \, [G(p^\prime)]^\prime \, [\Gamma_b^{\nu}(p^\prime,p)]^\prime \, [G(p)]^\prime \,.
\qquad\qquad
\end{eqnarray}
Notice that we know the transformation rules \eqref{gbp:Gtranf} and \eqref{gbp:trnsf:Gabmunu} for the fermion and gauge boson propagator, respectively, entering the right-hand side of \eqref{gbp:Gammaprimeeq}. Using these we can finally extract from the equation \eqref{gbp:Gammaprimeeq} the desired transformation rule for the proper vertex $\Gamma_{a}^{\mu}(p^\prime,p)$:
\begin{eqnarray}
\label{gbp:Gmm:transf}
\group{G}\,:\qquad \Gamma_{a}^{\mu}(p^\prime,p)  & \TransformsTo & {[\Gamma_{a}^{\mu}(p^\prime,p)]}^\prime \>=\>
\e^{\I \theta \cdot \bar t} \, \big(\e^{\I\theta\cdot\mathcal{T}}\big)_{ab} \, \Gamma_{b}^{\mu}(p^\prime,p) \, \e^{-\I \theta \cdot t} \,.
\end{eqnarray}
It is easy to show that the bare vertex \eqref{gbp:Gmm:bare} is invariant under global $\group{G}$, as it after all must be:
\begin{eqnarray}
\group{G}\,:\qquad \Gamma_{a}^{\mu}(p^\prime,p)\big|_{\mathrm{bare}}  & \TransformsTo &
{\big[\Gamma_{a}^{\mu}(p^\prime,p)\big|_{\mathrm{bare}}\big]}^\prime \>=\>
\Gamma_{a}^{\mu}(p^\prime,p)\big|_{\mathrm{bare}} \,.
\end{eqnarray}

\subsubsection{Transformation under $\mathcal{C}$, $\mathcal{P}$ and $\mathcal{CP}$}

The transformation of the three-point function $G_{a}^{\mu}(p^\prime,p)$ under $\mathcal{C}$, $\mathcal{P}$ and $\mathcal{CP}$ is induced by the corresponding transformations of the gauge bosons and fermions. I.e., schematically
\begin{eqnarray}
[\I G_{a}^{\mu}(p^\prime,p)]^{\mathcal{X}} &=&
\langle [A_a^\mu]^{\mathcal{X}} [\psi]^{\mathcal{X}} [\bar\psi]^{\mathcal{X}} \rangle \,,
\end{eqnarray}
where $\mathcal{X} = \mathcal{C}, \mathcal{P}, \mathcal{CP}$. Using the transformation rules for individual fields we obtain this way the transformation rules for the three-point function $G_{a}^{\mu}(p^\prime,p)$:
\begin{subequations}
\label{gbl:Gamudiscpoprve}
\begin{eqnarray}
{[G_{a}^{\mu}(p^\prime,p)]}^{\mathcal{C}\phantom{\mathcal{P}}} &=&
- C_{ab} \, C \, G_b^{\mu\T}(-p,-p^\prime) \, C^{-1} \,,
\\
{[G_{a}^{\mu}(p^\prime,p)]}^{\mathcal{P}\phantom{\mathcal{C}}} &=&
\mathcal{P}^\mu_{\phantom{\mu}\nu} \, \gamma_0 \, G_a^{\nu}(\tilde p^\prime,\tilde p) \, \gamma_0 \,,
\\
{[G_{a}^{\mu}(p^\prime,p)]}^{\mathcal{CP}} &=&
- \mathcal{P}^\mu_{\phantom{\mu}\nu} \, C_{ab} \, \gamma_0 \, C \, G_b^{\nu\T}(-\tilde p,-\tilde p^\prime) \, C^{-1} \, \gamma_0 \,.
\end{eqnarray}
\end{subequations}

In order to find the transformation rules for the proper vertex $\Gamma_{a}^{\mu}(p^\prime,p)$, we proceed exactly in the same way as above when probing the transformation properties under $\group{G}$: We note that since the full Green's function $G_{a}^{\mu}(p^\prime,p)$ is of the form \eqref{gbp:Gamupprimep}, its transformations under $\mathcal{C}$, $\mathcal{P}$ and $\mathcal{CP}$ must be induced also by the corresponding transformations of the propagators and the proper vertex $\Gamma_{a}^{\mu}(p^\prime,p)$:
\begin{eqnarray}
\label{gbl:Gamudiscpodruhe}
[\I G_{a}^{\mu}(p^\prime,p)]^{\mathcal{X}} &=&
[G_{ab\nu}^\mu(q)]^{\mathcal{X}} \, [G(p^\prime)]^{\mathcal{X}} \, [\Gamma_b^\nu(p^\prime,p)]^{\mathcal{X}} \, [G(p)]^{\mathcal{X}} \,.
\end{eqnarray}
Thus, noting that both expressions \eqref{gbl:Gamudiscpoprve} and \eqref{gbl:Gamudiscpodruhe} must be the same and taking into account the known transformation rules for the propagators on the right-hand side of \eqref{gbl:Gamudiscpodruhe}, we arrive at the transformation rules for the proper vertex $\Gamma_{a}^{\mu}(p^\prime,p)$:
\begin{eqnarray}
{[\Gamma_{a}^{\mu}(p^\prime,p)]}^{\mathcal{C}\phantom{\mathcal{P}}} &=&
- C_{ab} \, C \, \Gamma_b^{\mu\T}(-p,-p^\prime) \, C^{-1} \,,
\label{gbp:Gmm:C}
\\
{[\Gamma_{a}^{\mu}(p^\prime,p)]}^{\mathcal{P}\phantom{\mathcal{C}}} &=&
\mathcal{P}^\mu_{\phantom{\mu}\nu} \, \gamma_0 \, \Gamma_a^{\nu}(\tilde p^\prime,\tilde p) \, \gamma_0 \,,
\label{gbp:Gmm:P}
\\
{[\Gamma_{a}^{\mu}(p^\prime,p)]}^{\mathcal{CP}} &=&
- \mathcal{P}^\mu_{\phantom{\mu}\nu} \, C_{ab} \, \gamma_0 \, C \, \Gamma_b^{\nu\T}(-\tilde p,-\tilde p^\prime) \, C^{-1} \, \gamma_0 \,,
\label{gbp:Gmm:CP}
\end{eqnarray}
which are the same as those \eqref{gbl:Gamudiscpoprve} for $G_{a}^{\mu}(p^\prime,p)$.

Consider now the free vertex $\Gamma_{a}^{\mu}(p^\prime,p)|_{\mathrm{bare}} = \gamma^\mu T_a$, \eqref{gbp:Gmm:bare}. Transforming it under $\mathcal{C}$ and $\mathcal{P}$ via \eqref{gbp:Gmm:C} and \eqref{gbp:Gmm:P}, respectively, we obtain
\begin{eqnarray}
{\big[\Gamma_{a}^{\mu}(p^\prime,p)\big|_{\mathrm{bare}}\big]}^{\mathcal{C}} \ = \
{\big[\Gamma_{a}^{\mu}(p^\prime,p)\big|_{\mathrm{bare}}\big]}^{\mathcal{P}} &=&
\gamma^\mu \bar T_a \,.
\end{eqnarray}
We see that the non-invariance of the free vertex under $\mathcal{C}$ and $\mathcal{P}$ applied separately is attributed to the presence of $\gamma_5$ in the generators $T_a$. On the other hand, under combined transformation $\mathcal{CP}$ the free vertex remains invariant:
\begin{subequations}
\begin{eqnarray}
{\big[\Gamma_{a}^{\mu}(p^\prime,p)\big|_{\mathrm{bare}}\big]}^{\mathcal{CP}} &=& \gamma^\mu T_a \\
&=& \Gamma_{a}^{\mu}(p^\prime,p)\big|_{\mathrm{bare}} \,.
\end{eqnarray}
\end{subequations}
Of course, this is nothing else than mere rephrasing of the above discussion of the (non-)invariance of the Lagrangian $\eL = g j_a^\mu A_{a\mu}$, \eqref{gbp:eL:jamuAamu}, under $\mathcal{C}$, $\mathcal{P}$ and $\mathcal{CP}$; compare with equations \eqref{gbp:eL:jamuAamuCandP}, \eqref{gbp:eL:jamuAamuCP}.

\subsection{Local Ward--Takahashi identity}
\label{gbp:locWT}

Now we are going to derive the Abelian approximation of WT identity for the Green's function $\langle A_a^\mu \psi \bar\psi \rangle$, or, more precisely, for its 1PI part $\I\Gamma_a^\mu = \langle A_a^\mu \psi \bar\psi \rangle_{\mathrm{1PI}}$. We will use for this purpose the results from section~\ref{gbp:globWT} concerning the WT identity for the Green's function $\gamma_a^\mu = \langle j_a^\mu \psi \bar\psi \rangle_{\mathrm{1PI}}$.

\subsubsection{Relation between $\Gamma_a^\mu$ and $\gamma_a^\mu$}

We start by finding the relation between $\Gamma_a^\mu(p^\prime,p)$ and $\gamma_a^\mu(p^\prime,p)$. By applying the operator $(\mathcal{D}^{-1})_{ab}^{\mu\nu}$, \eqref{gbp:mathcalDm1munuab}, on $\langle A_a^\mu \psi \bar\psi \rangle$ we obtain
\begin{subequations}
\label{gbp:divBracket}
\begin{eqnarray}
(\mathcal{D}^{-1}_x)_{ab\,\nu}^{\mu}
\bra{0} T \big[ A_{b}^{\nu}(x) \, \psi(y) \, \bar\psi(z) \big] \ket{0}
&=&
\bra{0} T \big[ \big((\mathcal{D}^{-1}_x)_{ab\,\nu}^{\mu}\,A_{b}^{\nu}(x)\big) \, \psi(y) \, \bar\psi(z) \big] \ket{0}
\label{gbp:divBracket1}
\\
&=& -g
\bra{0} T \big[ j_a^\mu(x) \, \psi(y) \, \bar\psi(z) \big] \ket{0} + \mathcal{O}(f_{abc}) \,.
\qquad
\label{gbp:divBracket2}
\end{eqnarray}
\end{subequations}
The commuting of $(\mathcal{D}^{-1})_{ab}^{\mu\nu}$ through the $T$-product in the first equality, \eqref{gbp:divBracket1}, is done by applying twice (recall that $(\mathcal{D}^{-1})_{ab}^{\mu\nu}$ is a differential operator of the \emph{second} order) the formula \eqref{gbp:Tprodder}, first time with $V^\mu = A_a^\mu$ and second time with $V^\mu = \partial_\alpha A_a^\mu$, and noting that
\begin{subequations}
\begin{eqnarray}
\big[A_a^\mu(x),    \psi(y)\big] &=& 0 \,, \\
\big[A_a^\mu(x),\bar\psi(y)\big] &=& 0 \,,
\end{eqnarray}
\end{subequations}
as well as
\begin{subequations}
\begin{eqnarray}
\big[\partial_\alpha^x A_a^\mu(x),    \psi(y)\big] &=& 0 \,, \\
\big[\partial_\alpha^x A_a^\mu(x),\bar\psi(y)\big] &=& 0 \,.
\end{eqnarray}
\end{subequations}
The second equality, \eqref{gbp:divBracket2}, is just an application of the equations of motion \eqref{gbl:gaugeEOM}.

Using the momentum representations \eqref{gbp:Gamuxyz} and \eqref{gbp:brkt:jamu:ft} of $\langle A_a^\mu \psi \bar\psi \rangle$ and $\langle j_a^\mu \psi \bar\psi \rangle$, respectively, we obtain the Fourier transform of the equation \eqref{gbp:divBracket}:
\begin{eqnarray}
\label{gbp:divBracketmoment}
\big[D^{-1}(q)\,\I G(q)\big]^{\mu}_{ab\,\nu} \, \I G(p^\prime) \, \I\Gamma_b^\nu(p^\prime,p) \, \I G(p)
&=& -g \, \I G(p^\prime) \, \gamma_a^\mu(p^\prime,p) \, \I G(p) + \mathcal{O}(f_{abc}) \,.
\qquad
\end{eqnarray}
We also used here the fact that $(\mathcal{D}^{-1})_{ab}^{\mu\nu}$ is a Fourier transform of the inverse free propagator $(D^{-1})_{ab}^{\mu\nu}$, \eqref{gbp:Dm1munuabpoprve}. After some manipulations with propagators in \eqref{gbp:divBracketmoment} we finally express $\Gamma_a^\mu(p^\prime,p)$ in terms of $\gamma_a^\mu(p^\prime,p)$ as
\begin{eqnarray}
\label{gbl:Gmmgmm}
\Gamma_a^\mu(p^\prime,p) &=&
g \big[G^{-1}(q)\,D(q)\big]^{\mu}_{ab\,\nu} \, \gamma_b^\nu(p^\prime,p) + \mathcal{O}(f_{abc}) \,.
\end{eqnarray}

\subsubsection{Derivation of the WT identity}

Now we want to calculate $q_\mu \Gamma_a^\mu(p^\prime,p)$ by contracting \eqref{gbl:Gmmgmm} with $q_\mu$. In doing so an awkward quantity $q_\mu \big[G^{-1}\,D\big]^{\mu}_{ab\,\nu}$ appears on the right-hand side. It can be, however, significantly simplified. From the explicit forms \eqref{gbp:Gabmunuexpl} and \eqref{gbp:Dmunuab} of $G_{ab}^{\mu\nu}$ and $D_{ab}^{\mu\nu}$, respectively, it follows
\begin{eqnarray}
\big[G^{-1}\,D\big]^{\mu}_{ab\,\nu} &=&
(\unitmatrix_{ab}-\Pi_{ab})\Big(g^{\mu}_{\phantom{\mu}\nu}-\frac{q^\mu q_\nu}{q^2}\Big) +
\frac{q^\mu q_\nu}{q^2} \unitmatrix_{ab} \,,
\end{eqnarray}
so that
\begin{eqnarray}
q_\mu \big[G^{-1}\,D\big]^{\mu}_{ab\,\nu} &=& q_\nu \unitmatrix_{ab}
\end{eqnarray}
and consequently
\begin{eqnarray}
\label{gbl:Gmmgmmcontr}
q_\mu \Gamma_a^\mu(p^\prime,p) &=& g q_\mu \gamma_a^\mu(p^\prime,p) + \mathcal{O}(f_{abc}) \,.
\end{eqnarray}
However, the contraction $q_\mu \gamma_a^\mu(p^\prime,p)$ has been already calculated in Sec.~\ref{gbp:globWT}, it is just the WT identity \eqref{gbp:WTglob}.\footnote{Recall that in Sec.~\ref{gbp:globWT} we derived the contraction $q_\mu \gamma_a^\mu(p^\prime,p)$, where $\gamma_a^\mu = \langle j_a^\mu \psi \bar\psi \rangle_{\mathrm{1PI}}$, under the assumption $\partial_\mu j_a^\mu = 0$, \eqref{gbp:currcons}. This assumption is not fulfilled now, see \eqref{gbp:currconsloc}. However, $\partial_\mu j_a^\mu$ is proportional to $\mathcal{O}(f_{abc})$, which is exactly the order in which we are interested in, see \eqref{gbl:Gmmgmmcontr}.} We therefore arrive at the analogous WT identity\footnote{A remark concerning the terminology is in order now. The identity \eqref{gbp:WTloc} for $\Gamma_a^\mu(p^\prime,p)$ is usually in the case of non-vanishing $\mathcal{O}(f_{abc})$ referred to as the \emph{Slavnov--Taylor identity} \cite{Taylor:1971ff,Slavnov:1972fg} and only in the case of $\mathcal{O}(f_{abc})=0$ as the \emph{WT identity}. However, since we will in the following neglect the terms $\mathcal{O}(f_{abc})$, we use in this text the term \emph{WT identity} exclusively, for both cases.} for $\Gamma_a^\mu(p^\prime,p)$:
\begin{eqnarray}
\label{gbp:WTloc}
q_{\mu}\Gamma_a^\mu(p^\prime,p) &=& G^{-1}(p^\prime)\,T_a - \bar T_a\,G^{-1}(p) + \mathcal{O}(f_{abc}) \,,
\end{eqnarray}
differing from the WT identity \eqref{gbp:WTglob} for $\gamma_a^\mu(p^\prime,p)$ basically only by the overall factor of $g$ (recall the definition \eqref{gbp:Tagta} of $T_a$) and by the presence of the non-Abelian terms $\mathcal{O}(f_{abc})$.

The WT identity, as derived in \eqref{gbp:WTloc}, is ambiguous due to the presence of the undetermined terms $\mathcal{O}(f_{abc})$. In fact, these terms can be determined as well and the resulting identity is called the Slavnov--Taylor identity. For our purposes, however, the simple derivation of the ambiguous WT identity \eqref{gbp:WTloc} will turn out to be sufficient, because later on we will show that neglecting of the ambiguous terms $\mathcal{O}(f_{abc})$ will be consistent with our approximations of the polarization tensor.

We finally note that the bare (tree) proper vertex $\Gamma_a^\mu(p^\prime,p)|_{\mathrm{bare}}$, \eqref{gbp:Gmm:bare}, does satisfy the WT identity \eqref{gbp:WTloc} (actually with vanishing $\mathcal{O}(f_{abc})$) provided one takes as the fermion propagators $G^{-1}(p)$ the bare ones $S^{-1}(p) = \slashed{p}$. Indeed, the WT identity, reading in such case
\begin{eqnarray}
q_{\mu}\Gamma_a^\mu(p^\prime,p)\big|_{\mathrm{bare}} &=& S^{-1}(p^\prime)\,T_a - \bar T_a\,S^{-1}(p) \,,
\end{eqnarray}
reduces to the simple identity
\begin{eqnarray}
\slashed{q}\, T_a &=& \slashed{p}^\prime\,T_a - \bar T_a\,\slashed{p} \,,
\end{eqnarray}
holding due to $\bar T_a\,\slashed{p} = \slashed{p} \, T_a$ and $p^\prime = p+q$.

\section{Summary}

Major part of this chapter consisted of reviewing some textbook, as well as some less-textbook (but still rather straightforward) facts, concerning a gauge field theory with fermions. This was for the sake of later references accompanied by introducing the corresponding notation. Most importantly, we have stated, among other things, the transformation rules of fermion and gauge boson propagators, as well as of the three-point function $\langle A_a^\mu \psi \bar\psi \rangle$, both full and 1PI, under continuous and discrete symmetries. We also derived WT identity for the three-point function.

On top of mere reviewing \emph{facts}, we also stated some \emph{assumptions} under which we would work in the following chapters:
\begin{itemize}
  \item There is the same number of the left-handed and the right-handed fermions $\psi_L$ and $\psi_R$, respectively, and their common fermion number symmetry remains unbroken. Hence we can work with the field $\psi = \psi_L + \psi_R$, \eqref{gbp:psi}.
  \item The theory is at the Lagrangian level massless, so that the free propagator of the fermion field $\psi$ is given simply by $S^{-1} = \slashed{p}$, \eqref{gbp:Sbare}.
  \item The self-energy $\boldsymbol{\Sigma}$ of $\psi$ contains no $\slashed{p}$, \eqref{gbp:Sgnnowfr}, and satisfies the Hermiticity condition $\boldsymbol{\Sigma} = \boldsymbol{\bar\Sigma}$, \eqref{gbp:hermSgm}. It has consequently the form $\boldsymbol{\Sigma} = \Sigma^\dag P_L + \Sigma \, P_R$, \eqref{gbp:Sgm:form}, considered already in the previous chapters.
  \item The symmetry group $\group{G}$ is broken down to $\group{H} \subseteq \group{G}$ by the self-energy $\boldsymbol{\Sigma}$. Operationally:
\begin{subequations}
\begin{eqnarray}
\bbl \boldsymbol{\Sigma}, t_a \bbr &=&    0 \quad\mbox{for } t_a \in \algebra{h} \,, \\
\bbl \boldsymbol{\Sigma}, t_a \bbr &\neq& 0 \quad\mbox{for } t_a \in \algebra{g} \backslash \algebra{h} \,,
\end{eqnarray}
\end{subequations}
      where $\algebra{g}$ and $\algebra{h}$ are Lie algebras corresponding to the groups $\group{G}$ and $\group{H}$, respectively.
  \item The gauge dynamics is weak: $g \ll 1$, \eqref{gbp:gsmall}.
  \item There are no \qm{off-diagonal} gauge coupling constants in the case of more Abelian factors in $\group{G}$.
\end{itemize}

\chapter{Gauge boson mass matrix formula}
\label{chp:gbm}

\intro{This chapter is the very heart of the part~\ref{part:gauge}. We will derive here, on the basis of assumptions and formalism developed in the previous chapter, as well as within certain additional assumptions made in this chapter, the formula for the gauge boson mass matrix. However, as will be discussed in detail, the resulting formula will not be applicable for an arbitrary spontaneously broken gauge theory with fermions, but merely to a specific subclass of such theories, satisfying certain condition. Luckily enough, this subclass contains the models discussed in parts~\ref{part:abel} and \ref{part:ew}, so that we will be able to apply the gauge boson mass matrix formula to them in the following two chapters.}

\intro{Massiveness of a gauge boson manifests itself also by existence of its longitudinal polarization. This new degree of freedom can be physically interpreted as a (\qm{would-be}) NG boson, associated with the spontaneous breakdown of the (gauged) symmetry. Thus, besides mere calculating the gauge bosons masses, we will in this chapter in section \ref{sec:NGboson} also occupy ourselves with their interpretation in terms of the NG bosons.}

\section{Strategy}

\subsection{Pole approximation of the polarization tensor}
\label{gbl:ssec:Poleapp}

\subsubsection{Why the pole approximation}

The gauge boson spectrum (as well as any other spectrum) is given by poles of their full propagator. The key quantity is here the polarization tensor $\Pi_{ab}^{\mu\nu}(q)$, or, due to its transversality \eqref{gbp:Piabmunutrans}, the form factor $\Pi_{ab}(q^2)$. We have already mentioned in section~\ref{gbp:ssec:gbmasses} that the sufficient (though not necessary) condition for the gauge bosons to become massive is existence of a pole of the type $1/q^2$ in the Laurent expansion of $\Pi_{ab}(q^2)$.

In fact, our approach will be to approximate the $\Pi(q^2)$ \emph{only} by its pole part. That is to say, we will focus only on the residue $M^2$ and neglect all the coefficients $\Pi_{n}$, $n \geq 0$, in the Laurent expansion \eqref{gbp:Pi:Laurent} of $\Pi(q^2)$:
\begin{eqnarray}
\label{gbm:PoleAppr}
\Pi(q^2) &\doteq& \frac{1}{q^2} M^2 \,.
\end{eqnarray}
Thus, using this \emph{pole approximation} of $\Pi(q^2)$ the pole equation \eqref{gbp:gaugepole} reduces to
\begin{eqnarray}
\det \big(q^2 - M^2 \big) &=& 0 \,,
\end{eqnarray}
i.e., to a simpler problem of finding the eigenvalues of the matrix $M^2$, which can be accordingly interpreted as the gauge boson \emph{mass matrix}.

The pole approximation \eqref{gbm:PoleAppr} is motivated and justified by the assumption $g \ll 1$, \eqref{gbp:gsmall}, concerning the weakness of the gauge dynamics, i.e., by the fact that perturbative calculations in $g$ are possible. As we are going to prove below, it turns out non-trivially that if one wants to calculate, by solving the pole equation \eqref{gbp:gaugepole}, the gauge boson mass spectrum in the lowest (second) order in $g$, then it is sufficient to consider only the pole term of $\Pi(q^2)$, since the higher terms in Laurent expansion of $\Pi(q^2)$ happen to contribute only to higher terms in the $g$-expansion of the gauge boson masses.




\subsubsection{Proof}

In order to prove the statement made in the previous paragraph, we are now going to investigate the perturbative expansion in $g$ of solutions of the pole equation \eqref{gbp:gaugepole}. For that purpose it is convenient to make use of the diagonalization \eqref{gbp:Pi=OpiO} of the self-energy $\Pi(q^2)$ and investigate first in this respect only one particular pole equation \eqref{gbp:pipole} and only then generalize the result to the full non-diagonal $\Pi(q^2)$.

We start with the observation that the polarization tensor is in any case at least of the second order in $g$. This is a consequence of the fact that any interaction of $A_a^\mu$ with any other field (including the fermions, ghost, as well as the gauge bosons themselves), is proportional to $g$; see the Lagrangian $\eL(\psi,A_a^\mu)$, \eqref{gbp:eL:psiAamu}.

Let us diagonalize $\Pi(q^2)$ via \eqref{gbp:Pi=OpiO} to obtain the diagonal $\pi(q^2)$ with the elements $\pi_a(q^2)$, \eqref{gbp:pi}. Since $\Pi(q^2)$ is of the second order in $g$, so must be also each $\pi_a(q^2)$. Recalling the general form \eqref{gbp:piaLaurent} of $\pi_a(q^2)$, it is therefore convenient to factorize $g^2$ out of the corresponding pole residue $m_a^2$ as
\begin{eqnarray}
\label{gbm:mag2mu}
m_a^2 &\equiv& g^2 \mu_a^2 \,,
\end{eqnarray}
where $\mu_a^2$ is now of zeroth order in $g$. Similarly can be treated the coefficients $\pi_{n,a}$:
\begin{eqnarray}
\pi_{n,a} &\equiv& g^2 a_{n,a} (\mu_a^2)^{-n} \,,
\end{eqnarray}
where we have also utilized the dimension-full coefficient $\mu_a^2$ from \eqref{gbm:mag2mu} to carry the mass dimension of each $\pi_{n,a}$ (assuming, of course, that $\mu_a^2 \neq 0$). Consequently the coefficients $a_{n,a}$ are dimensionless and again of order $g^0$, due to explicit factorization of $g^2$. Using these definitions the expression \eqref{gbp:piaLaurent} for $\pi_a(q^2)$ recasts as
\begin{eqnarray}
\pi_a(q^2) &=& g^2
\bigg[ \Big(\frac{q^2}{\mu_a^2}\Big)^{-1} + \sum_{n=0}^\infty a_{n,a} \Big(\frac{q^2}{\mu_a^2}\Big)^{n} \bigg]
\,.
\end{eqnarray}
Finally, it is also convenient to introduce the dimensionless quantity $x_a$,
\begin{eqnarray}
x_a &\equiv& \frac{q^2}{\mu_a^2} \,,
\end{eqnarray}
to be used in the following.

Now recall that $\pi_a(q^2)$ enters the pole equation \eqref{gbp:pipole} for the unknown $q^2$. Using the definitions above, this pole equation for the unknown $q^2$ transforms as
\begin{eqnarray}
\label{gbm:poleeqxa}
x_a - g^2\bigg[1 + \sum_{n=0}^\infty a_{n,a} x_a^{n+1}\bigg] &=& 0
\end{eqnarray}
and turning thus into an equation for the unknown $x_a$.

If $g^2 = 0$, the equation \eqref{gbm:poleeqxa} has the solution $x_a=0$. We can therefore expect that for $g^2 \neq 0$ the solution $x_a$ will be proportional\footnote{We insist that $x_a$ be an analytic function of $g^2$, i.e., not, for instance, proportional to $1/g^2$. To see that such situation can easily happen, it is instructive to consider the case when all $a_{n,a}$, except $a_{0,a}$ and $a_{1,a}$, vanish. Then the equation \eqref{gbm:poleeqxa} for $x_a$ is quadratic and it is straightforward to show that while the first of its two solutions is indeed proportional to $g^2$, the second solution is proportional to $1/g^2$ and thus non-analytic.\label{footnote:analytic}} to $g^2$ and hence without loss of generality expressible in the form
\begin{eqnarray}
\label{gbm:xaepsilon}
x_a &=& g^2 (1 + \epsilon_a) \,,
\end{eqnarray}
where $\epsilon_a$ is some function of $g^2$. We are now going to argue that $\epsilon_a$ is proportional to $g^2$. By plugging the Ansatz \eqref{gbm:xaepsilon} for $x_a$ into the equation \eqref{gbm:poleeqxa} we obtain the equation for $\epsilon_a$:
\begin{eqnarray}
\label{gbm:poleeqepsilon}
\epsilon_a - g^2 \sum_{n=0}^\infty a_{n,a} g^{2n} (1+\epsilon_a)^{n+1} &=& 0 \,.
\end{eqnarray}
Now the argument is the same as above with $x_a$: If $g^2=0$, the equation \eqref{gbm:poleeqepsilon} has the solution $\epsilon_a = 0$, from which we conclude that $\epsilon_a$ must be really proportional\footnote{Cf.~footnote~\ref{footnote:analytic}.} to $g^2$, as proposed. Put in the original formalisms, we see that the solution of the pole equation \eqref{gbp:pipole}, i.e., the gauge boson mass, is given in the lowest order in the gauge coupling constant $g$ as
\begin{subequations}
\begin{eqnarray}
q^2 &=& g^2 \mu_a^2 (1 + \epsilon_a) \\ &=& g^2 \mu_a^2 (1 + \mathcal{O}(g^2)) \,,
\end{eqnarray}
\end{subequations}
where we recall that $\mu_a^2$, given by \eqref{gbm:mag2mu}, is of zeroth order in $g^2$. In other words, the residue $m_a^2$ in the $q^2$-expansion of $\pi_a(q^2)$, \eqref{gbp:piaLaurent}, is just the lowest order of the $g$-expansion of the gauge boson mass.

We have shown that in the case of diagonal $\pi(q^2)$, \eqref{gbp:pi}, the spectrum obtained considering only the residue $m^2$ is the lowest approximation of the $g$-expansion of the full spectrum, obtained from the full $\pi(q^2)$ with higher orders in $q^2$ properly included. Now we are going to generalize this result to the case of non-diagonal $\Pi(q^2)$.

Obviously, $\pi(q^2)$ and $\Pi(q^2)$, related by the orthogonal transformation \eqref{gbp:Pi=OpiO}, must yield the same spectrum. Therefore it remains to prove that the orthogonal transformation \eqref{gbp:Pi=OpiO} preserves also the spectrum obtained by the pole approximations of $\pi(q^2)$ and $\Pi(q^2)$, i.e., that the residues $m^2$ and $M^2$ of both the diagonal $\pi(q^2)$ and non-diagonal $\Pi(q^2)$, respectively, have the same eigenvalues. To see this we recall that the residues $m^2$ and $M^2$ are related by the transformation \eqref{gbp:MOmO0}: $M^2 = O_0 \, m^2 \, O_0^\T $. However, this transformation itself is also orthogonal, see \eqref{gbp:Onulaorth}. Thus, $m^2$ and $M^2$ indeed must have the same eigenvalues, which completes the proof.

\subsubsection{Structure of the gauge boson mass matrix}

Let us also, for the sake of later references, investigate the structure of the gauge boson mass matrix.

Recall that the mass matrix $M^2$, considered in the pole approximation \eqref{gbm:PoleAppr}, is a (symmetric) $N_{\group{G}} \times N_{\group{G}}$ matrix and thus its rank can be at most $N_{\group{G}}$. Assume therefore that its rank is\footnote{The rank $N^\prime$ of the gauge boson mass matrix is of course equal to the number of \qm{broken generators}: $N^\prime = N_{\group{G}} - N_{\group{H}}$.} $N^\prime \leq N_{\group{G}}$. We are now going to show for the sake of later purposes that the matrix $M^2$ can be written as
\begin{eqnarray}
\label{gbm:M2=FFT}
M^2 &=& F F^\T \,,
\end{eqnarray}
where the matrix $F$ is rectangular, of dimension $N_{\group{G}} \times N^\prime$. Needless to say that the rank of the matrix $F$ must be the maximal possible, i.e., $\min(N^\prime,N_{\group{G}}) = N^\prime$, so that the rank of the matrix $M^2$ is $N^\prime$ too, as supposed.

The matrix $M^2$, as being symmetric, can be transformed via the orthogonal transformation as
\begin{eqnarray}
\label{gbp:MOmO0podruhe}
M^2 &=& O \, m^2 \, O^\T \,,
\end{eqnarray}
where $O$ is an orthogonal matrix and $m^2$ is a matrix of the block form
\begin{eqnarray}
\label{gbm:m2block}
m^2 &=& \left(\begin{array}{cc} m_R^2 & 0 \\ 0 & 0 \end{array}\right) \,,
\end{eqnarray}
where $m_R^2$ is a symmetric matrix, which is \emph{regular}, i.e., of the dimension $N^\prime \times N^\prime$. (Notice that although the two orthogonal transformations \eqref{gbp:MOmO0podruhe} and \eqref{gbp:MOmO0} look similar, they are not the same. While the former assumes the special block structure \eqref{gbm:m2block}, with $m_R^2$ being regular, but not necessarily diagonal, the latter does not assume any special block structure, but on the other hand it insists on the diagonality .)

We may write the orthogonal matrix $O$ in a block form too:
\begin{eqnarray}
\label{gbm:O0block}
O &=& \left(\begin{array}{cc} A & B \\ C & D \end{array}\right) \,,
\end{eqnarray}
where the dimensions of the blocks $A$, $B$, $C$, $D$ are the same as the dimension of the corresponding blocks in $m^2$, \eqref{gbm:m2block}. If we plug the block forms \eqref{gbm:m2block} and \eqref{gbm:O0block} of $m^2$ and $O$, respectively, into \eqref{gbp:MOmO0podruhe}, we find the corresponding block form of $M^2$:
\begin{eqnarray}
\label{gbm:M2block}
M^2 &=& \left(\begin{array}{cc} A\,m_R^2\,A^\T & A\,m_R^2\,C^\T \\ C\,m_R^2\,A^\T & C\,m_R^2\,C^\T \end{array}\right) \,.
\end{eqnarray}

We now assert that the coveted matrix $F$ is given by
\begin{eqnarray}
\label{gbm:Fdef}
F &\equiv& \left(\begin{array}{c} A \\ C \end{array}\right) m_R \,.
\end{eqnarray}
Let us first check that $F$, defined as \eqref{gbm:Fdef}, really does satisfy the basic equation \eqref{gbm:M2=FFT}:
\begin{subequations}
\begin{eqnarray}
F F^\T
&=& \left(\begin{array}{c} A \\ C \end{array}\right) m_R^2 \Big( A^\T, C^\T\Big) \\
&=& \left(\begin{array}{cc} A\,m_R^2\,A^\T & A\,m_R^2\,C^\T \\ C\,m_R^2\,A^\T & C\,m_R^2\,C^\T \end{array}\right) \\
&=& M^2 \,.
\end{eqnarray}
\end{subequations}
Thus, the equation \eqref{gbm:M2=FFT} is satisfied. Furthermore, the dimension of the matrix $F$, \eqref{gbm:Fdef}, is clearly $N_{\group{G}} \times N^\prime$. This completes the proof that the matrix $F$ with desired properties exists.

We also note that while the product $F F^\T$ is given by the defining relation \eqref{gbm:M2=FFT}, for the product $F^\T F$ we obtain
\begin{eqnarray}
F^\T F &=& m_R^2 \,.
\end{eqnarray}
In deriving this relation one has to take into account that $A^\T A+C^\T C = 1$, which follows from the fact that the matrix $O$, \eqref{gbm:O0block}, is orthogonal: $O^\T O = 1$.

We finally briefly discuss the ambiguity in determining $F$. It is determined by the above requirements uniquely up to the orthogonal rotation
\begin{eqnarray}
F^\prime &=& F\,O_F \,,
\end{eqnarray}
where $O_F$ is an orthogonal matrix of the dimension $N^\prime \times N^\prime$. This orthogonal rotation of $F$ corresponds to the orthogonal rotation of $m_R^2$ as
\begin{eqnarray}
m_R^{2\prime} &=& O_F^\T\,m_R^2\,O_F \,.
\end{eqnarray}
From this we can in particular see that the matrix $F^\T F = m_R^2$ is always regular, irrespectively of the basis.

\subsection{Loop integral for the polarization tensor}

\begin{figure}[t]
\begin{center}
\includegraphics[width=1\textwidth]{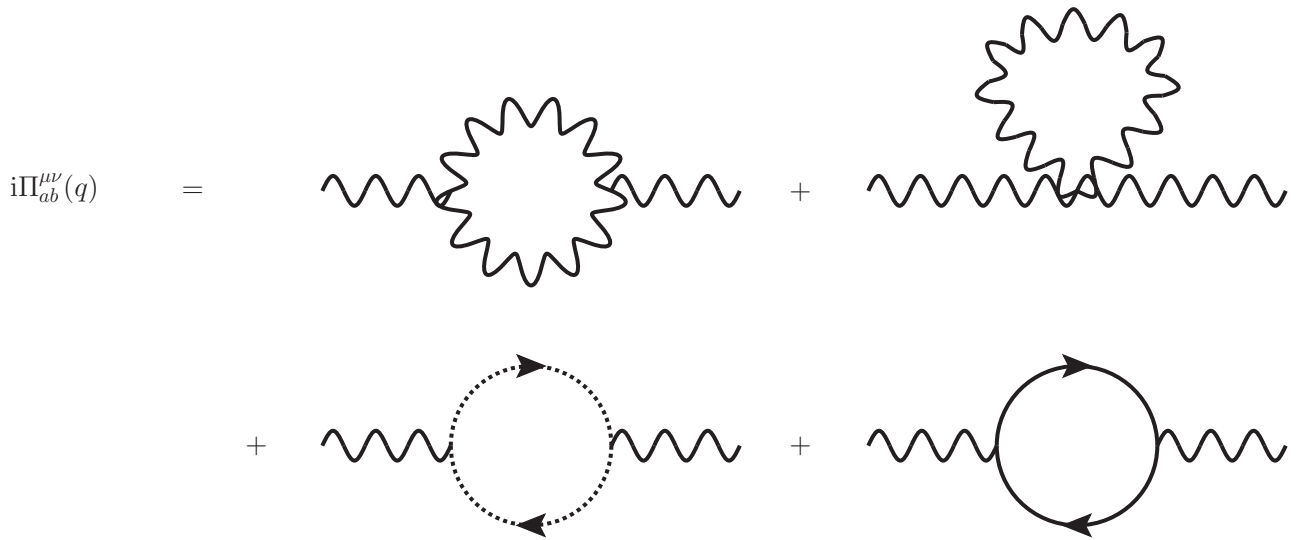}
\caption[Leading order contributions to $\Pi_{ab}^{\mu\nu}(q)$.]{Schematic diagrammatical representation of the lowest order (i.e., $g^2$) diagrams \eqref{gbm:symmdiagrams} of purely gauge (the first three diagrams) and fermion (the last diagram) origin, contributing to the polarization tensor $\Pi_{ab}^{\mu\nu}(q)$.}
\label{gbm:fig:symmdiagrams}
\end{center}
\end{figure}

Since we assume that the gauge dynamics is perturbative, we will calculate the gauge boson mass spectrum in the lowest order in the gauge coupling constant $g$, i.e., in the order $g^2$. In the previous section we showed that for that purpose it suffices to calculate only the pole part \eqref{gbm:PoleAppr} of the polarization tensor.

We will also assume that the symmetry group $\group{G}$ is broken spontaneously down to a subgroup $\group{H}$ by the fermion self-energies, as in the parts~\ref{part:abel} and  \ref{part:ew}. Since these self-energies are of course non-perturbative phenomena, we will therefore calculate the gauge boson spectrum in a mixed way: perturbatively in the gauge dynamics and at the same time non-perturbatively in the symmetry-breaking dynamics.

Let us start by reviewing how the polarization tensor is calculated in the symmetric (i.e., not spontaneously broken) theory. The perturbative contributions to the polarization tensor in the order $g^2$ are one-loop and can be divided into two groups as
\begin{eqnarray}
\label{gbm:symmdiagrams}
\I \Pi_{ab}^{\mu\nu}(q) &=&
\I \Pi_{ab}^{\mu\nu}(q)\big|_{\mathrm{gauge}} + \I \Pi_{ab}^{\mu\nu}(q)\big|_{\mathrm{fermions}} \,,
\end{eqnarray}
according to whether they are of purely gauge origin (including the ghost contribution) or whether they come from the fermion loop. The corresponding diagrams are depicted in Fig.~\ref{gbm:fig:symmdiagrams}.

Let us first discuss the pure gauge diagrams $\I \Pi_{ab}^{\mu\nu}(q)|_{\mathrm{gauge}}$. It is a textbook fact \cite{Peskin:1995ev} that their sum is transversal and that they do not contribute to the pole part of the polarization tensor. This remains true even once the symmetry is broken, since the SSB is, by assumption, triggered only by fermion propagators, which do not enter the pure gauge diagrams. We can therefore safely discard them.

We are thus left with the fermion contribution $\I \Pi_{ab}^{\mu\nu}(q)|_{\mathrm{fermions}}$, only which can potentially contribute to the pole \eqref{gbm:PoleAppr} of the polarization tensor. Clearly, for that purpose one has to include the symmetry-breaking dressed fermion propagators, since the symmetry-preserving propagators do not contribute to the pole of the polarization tensor \cite{Peskin:1995ev}. However, it turns out that in such a case the fermion loop diagram, as depicted in Fig.~\ref{gbm:fig:symmdiagrams}, is not correct. The point is that the bare vertices in the loop do not satisfy the correct WT identity \eqref{gbp:WTloc}, once the fermion propagators in the loop are considered to be not bare, but rather dressed, of the form \eqref{gbp:Gfrm}. As a result the polarization tensor is not transversal. As we are going to show in detail below, the most direct and easy way to cure this problem is to exchange one of the two bare vertices in Fig.~\ref{gbm:fig:symmdiagrams} by a dressed one, satisfying the WT identity.

\begin{figure}[t]
\begin{center}
\includegraphics[width=.7\textwidth]{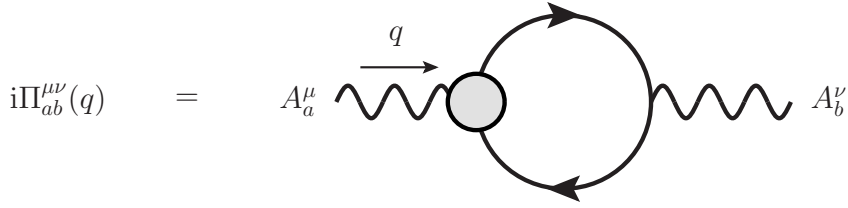}
\caption[Fermion one-loop contribution to $\Pi_{ab}^{\mu\nu}(q)$.]{The polarization tensor $\Pi_{ab}^{\mu\nu}(q)$, given by Eq.~\eqref{gbm:Pimunuab}. The fermion lines are not bare propagators, but rather the full ones (unlike in the previous chapters, in this chapter we will not provide them with the black blobs).}
\label{gbm:fig:Pi}
\end{center}
\end{figure}

We will therefore consider the polarization tensor with only the one-loop fermion contribution to be explicitly given by\footnote{Notice that there is no combinatorial factor due to the Dirac (i.e., complex) character of the field $\psi$, Eq.~\eqref{gbp:psi}. When dealing with a Majorana (i.e., real) field, there would be a combinatorial factor of $1/2$ in front of the integral \eqref{gbm:Pimunuab}.} \cite{Jackiw:1973tr,Cornwall:1973ts}
\begin{eqnarray}
\label{gbm:Pimunuab}
\I \Pi_{ab}^{\mu\nu}(q) &=& - \int\!\frac{\d^d p}{(2\pi)^d}\,
\Tr \Big\{ \Gamma^\mu_a(p+q,p) \, G(p) \, \gamma^\nu T_b \, G(p+q) \Big\}
\end{eqnarray}
and diagrammatically depicted in Fig.~\ref{gbm:fig:Pi}.\footnote{The symbol $d$ is the space-time dimension and has nothing to do with the dimensional regularization, we will always assume $d=4$. We prefer this symbolic denotation, because the space-time dimension will emerge explicitly in various formulae and by using the symbol $d$ rather than $4$ we will prevent at least some of the numerical factors (like, e.g., $1/(d-1) = 1/3$) from looking unnecessarily mysteriously.} Notice that while one of the two vertices is bare, $\gamma^\nu T_b$, the other is the full proper vertex $\Gamma_a^\mu(p^\prime,p)$, introduced in Sec.~\ref{gbp:3point}. As already mentioned, the presence of $\Gamma_a^\mu(p^\prime,p)$ is in fact necessary in order to maintain the transversality \eqref{gbp:Piabmunutrans} of $\Pi_{ab}^{\mu\nu}$, as we are going to show below, provided that it satisfies the WT identity \eqref{gbp:WTloc} in the \emph{Abelian approximation}:
\begin{eqnarray}
\label{gbm:WT}
q_\mu \Gamma_a^\mu(p^\prime,p) &=& G^{-1}(p^\prime)\,T_a - \bar T_a\,G^{-1}(p) \,,
\end{eqnarray}
i.e., with the structure constants $f_{abc}$ set to zero. Such approximation is consistent with neglecting the pure gauge diagrams $\I \Pi_{ab}^{\mu\nu}(q)|_{\mathrm{gauge}}$, which are proportional to $f_{abc}$ as well.


\section{Properties of the polarization tensor}

Let us now investigate some of the most important properties of the polarization tensor given by the loop integral \eqref{gbm:Pimunuab}.

\subsection{Transversality}
\label{gbm:ssec:Transv}

Let us check the transversality of the polarization tensor \eqref{gbm:Pimunuab} under the assumption \eqref{gbm:WT}. Explicit calculation of $q_\mu \Pi_{ab}^{\mu\nu}(q)$ reveals
\begin{subequations}
\label{gbm:Pitrpoprve}
\begin{eqnarray}
q_\mu \Pi_{ab}^{\mu\nu}(q)
&=& \I \int\!\frac{\d^d p}{(2\pi)^d}\,
\Tr \Big\{ \big[q_\mu \Gamma^\mu_a(p+q,p)\big] \, G(p) \, \gamma^\nu T_b \, G(p+q) \Big\}
\label{gbm:Pitrpoprve1}
\\ &=& \I \int\!\frac{\d^d p}{(2\pi)^d}\,
\Tr \Big\{ \big[G^{-1}(p+q)\,T_a - \bar T_a\,G^{-1}(p)\big] \, G(p) \, \gamma^\nu T_b \, G(p+q) \Big\}
\label{gbm:Pitrpoprve2}
\qquad
\\ &=& \I \int\!\frac{\d^d p}{(2\pi)^d}\,
\Tr \Big\{ G(p) \, \gamma^\nu \big[T_b,T_a\big] \Big\}
\label{gbm:Pitrpoprve3}
\\ &=& 0 \,.
\label{gbm:Pitrpoprve4}
\end{eqnarray}
\end{subequations}
Here we have in \eqref{gbm:Pitrpoprve2} used the WT identity \eqref{gbm:WT}, in \eqref{gbm:Pitrpoprve3} we shifted the integration variable of one of the two resulting integrals and using the cyclicity of the trace made some rearrangements of the integrand and finally in \eqref{gbm:Pitrpoprve4} we used the fact that the integral independent of any other momentum than the integration momentum and carrying at the same time a Lorentz index must vanish due to the Lorentz invariance.

Since $\Pi_{ab}^{\mu\nu}(q)$ depends only on $q$, it must be, by Lorenz invariance, a linear combination of $g^{\mu\nu}$ and $q^\mu q^\nu$. Therefore if $q_\mu \Pi_{ab}^{\mu\nu}(q)$ is vanishing, so must be also $q_\nu \Pi_{ab}^{\mu\nu}(q)$. It may be however instructive to check it explicitly:
\begin{subequations}
\label{gbm:Pitrpodruhe}
\begin{eqnarray}
q_\nu \Pi_{ab}^{\mu\nu}(q) &=& \I \int\!\frac{\d^d p}{(2\pi)^d}\,
\Tr \Big\{ \Gamma^\mu_a(p+q,p) \, G(p) \, \slashed{q} T_b \, G(p+q) \Big\}
\label{gbm:Pitrpodruhe1}
\\ &=& \I \frac{q^\mu}{q^2}\int\!\frac{\d^d p}{(2\pi)^d}\,
\Tr \Big\{ \big[q_\alpha \Gamma^\alpha_a(p+q,p)\big] \, G(p) \, \slashed{q} T_b \, G(p+q) \Big\}
\label{gbm:Pitrpodruhe2}
\\ &=& \I \frac{q^\mu}{q^2}\int\!\frac{\d^d p}{(2\pi)^d}\,
\Tr \Big\{ \big[G^{-1}(p+q)\,T_a - \bar T_a\,G^{-1}(p)\big] \, G(p) \, \slashed{q} T_b \, G(p+q) \Big\}
\label{gbm:Pitrpodruhe3}
\qquad
\\ &=& \I \frac{q^\mu}{q^2}q_\nu\int\!\frac{\d^d p}{(2\pi)^d}\,
\Tr \Big\{ G(p) \, \gamma^\nu \big[T_b,T_a\big] \Big\}
\label{gbm:Pitrpodruhe4}
\\ &=& 0 \,.
\label{gbm:Pitrpodruhe5}
\end{eqnarray}
\end{subequations}
Most of the steps here are the same as before in \eqref{gbm:Pitrpoprve}, only in \eqref{gbm:Pitrpodruhe2} we used the fact that due to the Lorentz invariance it holds
\begin{eqnarray}
\int \! \d^d p \, f^\mu(p,q) &=& \frac{q^\mu}{q^2} \int \! \d^d p \; q_\alpha f^\alpha(p,q)
\end{eqnarray}
for any function $f(p,q)$, depending, apart from the integration momentum $p$, on the \emph{only} external momentum $q$.

Thus, the polarization tensor is indeed transversal and therefore of the form \eqref{gbp:Piabmunutrans}. The corresponding form factor $\Pi_{ab}(q^2)$ can be explicitly written as
\begin{eqnarray}
\Pi_{ab}(q^2) &=& \I \frac{1}{d-1} \frac{1}{q^2} \int\!\frac{\d^d p}{(2\pi)^d}\,
\Tr \Big\{ \Gamma^{\mu}_a(p+q,p) \, G(p) \, \gamma_\mu T_b \, G(p+q)\Big\} \,.
\end{eqnarray}

\subsection{Reality}
\label{gbm:ssec:real}

In the previous chapter we made the assumption that both the fermion propagator $G(p)$ and the vertex $\Gamma^{\mu}_a(p^\prime,p)$ satisfy the respective Hermiticity conditions \eqref{gbp:hermG} and \eqref{gbm:hermGmm}. We have suggested in Sec.~\ref{gbp:3point} that the two Hermiticity conditions will eventually ensure the reality of the polarization tensor $\Pi_{ab}^{\mu\nu}(q)$ (and consequently, within the pole approximation \eqref{gbm:PoleAppr}, also the reality of the gauge boson mass matrix $M_{ab}^2$).

Indeed, consider the expression \eqref{gbm:Pimunuab} for the polarization tensor. Using the Hermiticity conditions \eqref{gbp:hermG}, \eqref{gbm:hermGmm} and the cyclicity of the trace, shifting appropriately the integration momentum and taking into account the identity $\Tr A^\dag = \Tr^*\!A$, one can rewrite \eqref{gbm:Pimunuab} as
\begin{eqnarray}
\I \Pi_{ab}^{\mu\nu}(q) &=&
-\int\!\frac{\d^d p}{(2\pi)^d}\,\Tr^* \Big\{ \Gamma^\mu_a(p+q,p) \, G(p) \, \gamma^\nu T_b \, G(p+q) \Big\} \,.
\end{eqnarray}
Comparing this expression with the original expressions \eqref{gbm:Pimunuab} for the polarization tensor and recalling that via the Wick rotation an extra factor of $\I$ appears in the measure $\d^d p$, one concludes that $\Pi_{ab}^{\mu\nu}(q)$ (and consequently also the mass matrix $M_{ab}^2$) must be really real.

\subsection{Transformation properties}
\label{gbm:ssec:tranf}

Consider first the transformation of the polarization tensor \eqref{gbm:Pimunuab} under the continuous symmetry $\group{G}$. It must be induced by the corresponding transformations of $G(p)$, $G(p+q)$ and $\Gamma^\mu_a(p+q,p)$ under $\group{G}$. Assuming that these transform as they should, i.e., according to \eqref{gbp:Gtranf} and \eqref{gbp:Gmm:transf}, respectively, we obtain
\begin{subequations}
\label{gbm:Piabmunutransproof}
\begin{eqnarray}
\group{G}\,:\qquad \I \Pi_{ab}^{\mu\nu}(q)  \>\>\TransformsTo\>\>
&& \nonumber \\
{[\I \Pi_{ab}^{\mu\nu}(q)]}^\prime &=& - \int\!\frac{\d^d p}{(2\pi)^d}\,
\Tr \Big\{ [\Gamma^\mu_a(p+q,p)]^\prime \, [G(p)]^\prime \, \gamma^\nu T_b \, [G(p+q)]^\prime \Big\}
\label{gbm:Piabmunutransproof1}
\\ &=&
- \big(\e^{\I\theta\cdot\mathcal{T}}\big)_{ac} \int\!\frac{\d^d p}{(2\pi)^d}\,
\Tr \Big\{ \Gamma^\mu_c(p+q,p) \, G(p) \, \gamma^\nu \underbrace{\e^{-\I\theta\cdot t} \, T_b \, \e^{\I\theta\cdot t}}_{T_d\,(\e^{-\I\theta\cdot\mathcal{T}})_{db}} \, G(p+q) \Big\}
\label{gbm:Piabmunutransproof2}
\nonumber \\ &&
\\ &=&
\big(\e^{\I\theta\cdot\mathcal{T}}\big)_{ac} \, \I \Pi_{cd}^{\mu\nu}(q) \, \big(\e^{-\I\theta\cdot\mathcal{T}}\big)_{db} \,,
\label{gbm:Piabmunutransproof3}
\end{eqnarray}
\end{subequations}
where we used in the line \eqref{gbm:Piabmunutransproof2} the relation \eqref{gbp:taeabetbe}. Thus we see that provided the fermion propagators and the vertex transform correctly under $\group{G}$, so does $\Pi_{ab}^{\mu\nu}(q)$, since the result \eqref{gbm:Piabmunutransproof3} corresponds to the correct transformation rule \eqref{gbp:trnsf:Piabmunu}.

Transformation properties of the polarization tensor \eqref{gbm:Pimunuab} under the discrete symmetries $\mathcal{C}$, $\mathcal{P}$ can be treated in the same manner. Again, it could be shown that provided the propagators and the vertex in the integral \eqref{gbm:Pimunuab} transform correctly according to their respective transformation rules, so does the resulting polarization tensor.

\section{The vertex}
\label{gbm:ssec:vertexprel}

The integral \eqref{gbm:Pimunuab} for the polarization tensor contains the fermion propagators $G(p)$, $G(p+q)$ and the dressed vertex $\Gamma^{\mu}_a(p^\prime,p)$. While the propagators are known by assumption, the vertex will have to be somehow \emph{constructed} in the following section. In order not to make that task unnecessarily difficult, it is worth observing which part of $\Gamma^{\mu}_a(p^\prime,p)$ are actually needed for the purpose of calculating the gauge boson mass spectrum under the pole approximation \eqref{gbm:PoleAppr}.

\subsection{Momentum expansions}
\label{gbm:ssec:MmntExp}

Let us start by investigating the analytic structure of $\Gamma^{\mu}_a(p^\prime,p)$. Consider the right-hand side of its WT identity \eqref{gbm:WT} for $q=0$ (recall that $p^\prime=p+q$):
\begin{subequations}
\label{gbm:WTrhs}
\begin{eqnarray}
G^{-1}(p)\,T_a - \bar T_a\,G^{-1}(p) &=& -\boldsymbol{\Sigma}\,T_a + \bar T_a\boldsymbol{\Sigma}
\\ &=& - g \, \bbl  \boldsymbol{\Sigma},t_a \bbr \,,
\end{eqnarray}
\end{subequations}
where we plugged in the Ansatz \eqref{gbp:Gfrm} for the fermion propagator $G(p)$. Notice that it is proportional to the quantity \eqref{gbp:Sgmnoninv}, measuring the non-invariance of the propagator under the symmetry generated by $t_a$. Since we assume that the symmetry $\group{G}$ is actually broken, \eqref{gbm:WTrhs} must be considered in general non-vanishing. Consequently, since the left-hand side of the WT identity is proportional to $q$, the vertex $\Gamma^{\mu}_a(p^\prime,p)$ must have a pole of the type $1/q$, if the WT identity is to be satisfied.

Any pole in a Green's function can be attributed only to a propagator of an intermediate particle. In our case, since the vertex $\Gamma^{\mu}_a(p^\prime,p)$ is by construction 1PI, it cannot contain propagators neither of the fermions nor of the gauge bosons. Any possible pole of $\Gamma^{\mu}_a(p^\prime,p)$ can therefore occur only due to some dynamically generated composite particles. But such particles are in fact there: They are the NG bosons, associated with the SSB of the group $\group{G}$ down to $\group{H}$. Notice that the NG bosons can be in this respect understood as \qm{compensating fields}, serving to ensure the satisfaction of the WT identity even if the symmetry of the Lagrangian is broken by the ground state (or by the Green's functions).\footnote{Recall that the WT identity is a consequence of the symmetry of the Lagrangian, not of the ground state.}

We will discuss the interpretation of the vertex in terms of the NG bosons closer in Sec.~\ref{sec:NGboson}. For the moment it suffices to note that the NG bosons couple bilinearly to the gauge bosons and are massless, hence the pole of the type $1/q$, dictated by the WT identity, must be of the form $q^\mu/q^2$, where $q$ is the momentum carried by the gauge boson. The vertex $\Gamma^{\mu}_a(p^\prime,p)$ must have therefore the general form
\begin{subequations}
\label{gbm:Gmm:decomp}
\begin{eqnarray}
\label{gbm:Gmm:decompgen}
\Gamma^{\mu}_a(p^\prime,p) &=& \Gamma^\mu_a(p^\prime,p)\big|_{\mathrm{NG}} + \Gamma^{\mu}_a(p^\prime,p)\big|_{\mathrm{reg.}} \,,
\end{eqnarray}
where NG part $\Gamma^\mu_a(p^\prime,p)|_{\mathrm{NG}}$ has the form
\begin{eqnarray}
\label{gbm:Gmm:decompNG}
\Gamma^\mu_a(p^\prime,p)\big|_{\mathrm{NG}} &=& \frac{q^\mu}{q^2}\Gamma_a(p^\prime,p)\big|_{\mathrm{NG}} \,,
\end{eqnarray}
\end{subequations}
and where both $\Gamma^{\mu}_a(p^\prime,p)|_{\mathrm{reg.}}$ and $\Gamma_a(p^\prime,p)|_{\mathrm{NG}}$ are \emph{regular} for all $p^\prime$ and $p$.

The vertex $\Gamma^{\mu}_a(p^\prime,p)$ of the form \eqref{gbm:Gmm:decomp} can be expanded into the Laurent series in $q = p^\prime - p$ about $q=0$ as
\begin{eqnarray}
\label{gbm:Gmmexp}
\Gamma^{\mu}_a(p^\prime,p) &=&
\frac{q^\mu}{q^2} A_a(p) + \frac{q^\mu}{q^2} q_\alpha B_a^\alpha(p) + C_a^\mu(p) + \mathcal{O}(q) \,,
\end{eqnarray}
where $A_a(p)$, $B_a^\alpha(p)$, $C_a^\mu(p)$ are some functions only of $p$. Note that we can \emph{uniquely} identify
\begin{subequations}
\label{gbm:Gmmseparation}
\begin{eqnarray}
\Gamma_a^\mu(p^\prime,p)\big|_{\mathrm{NG}} &=& \frac{q^\mu}{q^2}\bigg[A_a(p) + q_\alpha B_a^\alpha(p)\bigg] + \mathcal{O}(q) \,,
\label{gbm:GmmNGexp}
\\
\Gamma_a^\mu(p^\prime,p)\big|_{\mathrm{reg.}} &=& C_a^\mu(p) + \mathcal{O}(q) \,.
\label{gbm:Gmmregexp}
\end{eqnarray}
\end{subequations}
This uniqueness is actually possible only in the lowest orders in $q$. Already for the terms linear in $q$ the separation \eqref{gbm:Gmmseparation} of the expansion \eqref{gbm:Gmmexp} into the NG part and the non-NG part is ambiguous: Assume, e.g., that there is a term $q^2 D_a(p)$ in the square bracket of the expansion \eqref{gbm:GmmNGexp} of the NG part $\Gamma_a^\mu(p^\prime,p)|_{\mathrm{NG}}$. Clearly, the $q^2$ in it can be canceled with the NG pole $1/q^2$ and thus the same term can be equally well considered, in the form $q^\mu D_a(p)$, as a part of expansion \eqref{gbm:Gmmregexp} of the regular part $\Gamma_a^\mu(p^\prime,p)|_{\mathrm{reg.}}$. We will come across this problem in Sec.~\ref{gbm:vertexNGinterp}, where we will see that this ambiguity can be parameterized, within our approximation scheme, by one real parameter.

We can similarly expand the fermion propagator $G(p+q)$ about $q=0$:
\begin{eqnarray}
G(p+q) &=& G(p) + q_\alpha G^\alpha(p) + \mathcal{O}(q^2) \,,
\end{eqnarray}
where
\begin{eqnarray}
\label{gbm:Gexp}
G^\alpha(p) &=& \partial^\alpha G(p) \\ &=& - G(p) \big(\partial^\alpha G^{-1}(p)\big) G(p) \,.
\end{eqnarray}
E.g., for the fermion propagator of the form \eqref{gbp:Gfrm} (or better \eqref{gbp:Gfrminv}) we have explicitly
\begin{eqnarray}
G^\alpha(p) &=&
\gamma^\alpha \boldsymbol{D}_L
+ 2 p^\alpha (\slashed{p} + \boldsymbol{\Sigma}^\dag) \boldsymbol{D}_L^\prime
+ 2 p^\alpha \boldsymbol{\Sigma}^{\dag\prime} \boldsymbol{D}_L \,,
\end{eqnarray}
where the prime denotes the derivative with respect to $p^2$. Straightforward plugging of the expansions \eqref{gbm:Gmmexp} and \eqref{gbm:Gexp} into the basic expression \eqref{gbm:Pimunuab} for the polarization tensor $\Pi_{ab}^{\mu\nu}(q)$ yields
\begin{eqnarray}
\label{gbm:Piminuabexpinterm}
\Pi_{ab}^{\mu\nu}(q) &=& \I \frac{q^\mu}{q^2} \int\!\frac{\d^d p}{(2\pi)^d}\,
\Tr \Big\{ A_a(p) \, G(p) \, \gamma^\nu T_b \, G(p) \Big\}
\nonumber \\ &&
{}+ \I \int\!\frac{\d^d p}{(2\pi)^d}\,\Tr \Big\{ C_a^\mu(p) \, G(p) \, \gamma^\nu T_b \, G(p) \Big\}
\nonumber \\ &&
{}+ \I \frac{q^\mu}{q^2} q_\alpha \int\!\frac{\d^d p}{(2\pi)^d}\,
\Tr \Big\{ A_a(p) \, G(p) \, \gamma^\nu T_b \, G^\alpha(p) + B_a^\alpha(p) \, G(p) \, \gamma^\nu T_b \, G(p) \Big\}
\nonumber \\ && {}+ \mathcal{O}(q) \,.
\end{eqnarray}
The integral in the first line actually vanishes. This can be seen already from the Lorentz invariance, technically it is maintained by a symmetric integration. On the basis of a similar argument there will survive only terms even in $q$ in the expansion \eqref{gbm:Piminuabexpinterm}, so that the terms $\mathcal{O}(q)$ are actually $\mathcal{O}(q^2)$. We make further simplifications by noting that under integral we can make the substitution
\begin{eqnarray}
\int \! \d^d p \, f^{\mu\nu}(p) &=& \frac{1}{d} g^{\mu\nu} \int \! \d^d p \, f^{\alpha}_{\phantom{\alpha}\alpha}(p) \,,
\end{eqnarray}
provided $f^{\mu\nu}(p)$ does not depend on any other four-vector than $p$. This allows to make the Lorentz structure of \eqref{gbm:Piminuabexpinterm} explicit:
\begin{eqnarray}
\label{gbm:Piminuabexp}
\Pi_{ab}^{\mu\nu}(q) &=&
\I \frac{1}{d} g^{\mu\nu} \int\!\frac{\d^d p}{(2\pi)^d}\,\Tr \Big\{ C_a^\alpha(p) \, G(p) \, \gamma_\alpha T_b \, G(p) \Big\}
\nonumber \\ &&
{}+ \I \frac{1}{d} \frac{q^\mu q^\nu}{q^2} \int\!\frac{\d^d p}{(2\pi)^d}\,
\Tr \Big\{ A_a(p) \, G(p) \, \gamma^\alpha T_b \, G_\alpha(p) + B_a^\alpha(p) \, G(p) \, \gamma_\alpha T_b \, G(p) \Big\}
\nonumber \\ && {}+ \mathcal{O}(q^2) \,.
\end{eqnarray}

\subsection{Preliminary expression for the mass matrix}

Under the pole approximation \eqref{gbm:PoleAppr} of $\Pi_{ab}(q^2)$ the polarization tensor $\Pi_{ab}^{\mu\nu}(q)$ has the form
\begin{eqnarray}
\label{gbm:PimunuabM2}
\Pi_{ab}^{\mu\nu}(q) &=& \bigg(g^{\mu\nu}-\frac{q^\mu q^\nu}{q^2}\bigg)M_{ab}^2 \,.
\end{eqnarray}
Assuming that the vertex $\Gamma_a^\mu(p^\prime,p)$ satisfies the WT identity \eqref{gbm:WT}, the expression \eqref{gbm:Piminuabexp} must be transversal. Thus, by comparing it with \eqref{gbm:PimunuabM2}, we arrive at two seemingly different explicit expressions for the gauge boson mass matrix:
\begin{subequations}
\label{gbm:Mab:prel}
\begin{eqnarray}
M_{ab}^2 &=&
\phantom{-}\I \frac{1}{d} \int\!\frac{\d^d p}{(2\pi)^d}\,\Tr \Big\{ C_a^\alpha(p) \, G(p) \, \gamma_\alpha T_b \, G(p) \Big\}
\label{gbm:Mab:prelpoprve}
\\ &=&
- \I \frac{1}{d} \int\!\frac{\d^d p}{(2\pi)^d}\,
\Tr \Big\{ A_a(p) \, G(p) \, \gamma^\alpha T_b \, G_\alpha(p) + B_a^\alpha(p) \, G(p) \, \gamma_\alpha T_b \, G(p) \Big\} \,.
\label{gbm:Mab:prelpodruhe}
\end{eqnarray}
\end{subequations}
Both expression \eqref{gbm:Mab:prelpoprve} and \eqref{gbm:Mab:prelpodruhe} must be of course the same due to the WT identity \eqref{gbm:WT}, which relates the vertex and the fermion propagator to each other.

\subsection{Recapitulation}

Consider now the contraction $q_\mu \Gamma^{\mu}_a(p^\prime,p)$ of the expansion \eqref{gbm:Gmmexp} of the vertex $\Gamma^{\mu}_a(p^\prime,p)$:
\begin{eqnarray}
q_\mu \Gamma^{\mu}_a(p^\prime,p) &=&
A_a(p) + q_\mu\big( B_a^\mu(p) + C_a^\mu(p)\big) + \mathcal{O}(q^2) \,.
\end{eqnarray}
Recall that due to the WT identity \eqref{gbm:WT} this expression must be equal to $G^{-1}(p^\prime)\,T_a - \bar T_a\,G^{-1}(p)$. We can see that while $A_a$ (i.e., the leading part of the NG part \eqref{gbm:GmmNGexp} of $\Gamma^{\mu}_a(p^\prime,p)$) is determined by the WT uniquely as
\begin{eqnarray}
\label{eq:vertex_leadingNGpart}
A_a(p) &=& G^{-1}(p)\,T_a - \bar T_a\,G^{-1}(p) \,,
\end{eqnarray}
for the functions $B_a^\alpha$ and $C_a^\mu$ the WT identity determines only their sum.

Recall that, as we observed above, in order to compute the gauge boson mass matrix via the expression(s) \eqref{gbm:Mab:prel}, one has to know either $A_a$, $B_a^\alpha$, or $C_a^\mu$. In particular, at least one of the functions $B_a^\alpha$ and $C_a^\mu$ must be necessarily known. However, as we have just seen, on the basis of the WT identity we can determine only $B_a^\mu + C_a^\mu$, which is clearly insufficient for our purposes. Thus, the following section will be dedicated to the task of inventing some additional well motivated requirements on the vertex $\Gamma^{\mu}_a(p^\prime,p)$, allowing to determine separately each of the functions $B_a^\alpha$ and $C_a^\mu$ uniquely.

\section{Construction of the vertex}
\label{gbm:sec:construction}

We already mentioned that the dressed fermion propagator is known, e.g., by solving the corresponding SD equations like in parts~\ref{part:abel} and \ref{part:ew}. In principle, in order to be entirely consistent, the vertex should be calculated in the same way and in the same time as the fermion propagators: by means of solving the corresponding SD (or Bethe--Salpeter) equation for the vertex. However, guided by the applications, we assume that it is not. We have therefore to approximate it somehow, or in other words, we have to choose or construct a suitable Ansatz for it.

This section is dedicated to the construction of such an Ansatz. We will first state the minimal reasonable form of the vertex and then constrain it by imposing various additional requirements. These requirements will be of two kinds: First, we will require correct transformation behavior under various symmetries. Second, we will require that the vertex be consistent with the underlying NG boson interpretation. This way we will finally end up with (almost) uniquely determined vertex.\footnote{The vertex will be determined up to certain terms which do not contribute to the gauge boson mass matrix in the pole approximation and hence we will not need to worry much about them.\label{ftnt:almost}}

\subsection{General form of the Ansatz}
\label{gbm:ssec:general}

We start the construction of the vertex Ansatz by stating some general assumptions about its form.

Recall first that the SSB is by assumption driven by symmetry-breaking parts of the fermion self-energy $\boldsymbol{\Sigma}$. Let us then assume that $\boldsymbol{\Sigma}$ actually contains \emph{only} those symmetry-breaking (and consequently UV-finite) parts and is free of any symmetry-preserving (and potentially UV-divergent) parts. (This assumption of course corresponds to how we have constructed Ans\"{a}tze for $\boldsymbol{\Sigma}$ in parts~\ref{part:abel} and \ref{part:ew}.) It follows that in the case of no SSB the self-energy $\boldsymbol{\Sigma}$ vanishes.

We can then make the natural requirement that in the case of no SSB (i.e., when $\boldsymbol{\Sigma} = 0$) the vertex reduces to the bare on. This can be accomplished if the vertex is written as the bare one plus \qm{something} proportional to $\boldsymbol{\Sigma}$. Put more formally, we assume the vertex $\Gamma_{a}^{\mu}(p^\prime,p)$ to be of the form
\begin{eqnarray}
\label{gbm:Gmm:gen}
\Gamma_{a}^{\mu}(p^\prime,p) &=& \gamma^\mu T_a + \Gamma_{a}^{\mu}(p^\prime,p)\big|_{\mathrm{cor.}} \,,
\end{eqnarray}
where the \emph{correction} $\Gamma_{a}^{\mu}(p^\prime,p)|_{\mathrm{cor.}}$ to the bare vertex $\gamma^\mu T_a$, \eqref{gbp:Gmm:bare}, is proportional to the self-energies $\boldsymbol{\Sigma}$ and hence vanishing in the limit $\boldsymbol{\Sigma}=0$.

Just proportionality to $\boldsymbol{\Sigma}$ is however still quite general. Following the philosophy of making a \emph{minimal reasonable} Ansatz we impose the following simplifying restrictions: We assume that $\Gamma_{a}^{\mu}(p^\prime,p)|_{\mathrm{cor.}}$
\begin{itemize}
  \item is \emph{linear} in the self-energies $\boldsymbol{\Sigma}$,
  \item contains the self-energies $\boldsymbol{\Sigma}$ evaluated only in $p^\prime$ and $p$.
\end{itemize}
Moreover, recall that since we calculate the gauge boson mass matrix in the pole approximation, it is sufficient to calculate the polarization tensor in the order $g^2$. We are therefore interested only in the part of the vertex linear in $g$, which leads us to assume that $\Gamma_{a}^{\mu}(p^\prime,p)|_{\mathrm{cor.}}$
\begin{itemize}
  \item is \emph{linear} in the generators $T_a$ and $\bar T_a$.
\end{itemize}
(Recall that $T_a$ is linear in $g$, due to the definition \eqref{gbp:Tagta}.) The three conditions imply that $\Gamma_{a}^{\mu}(p^\prime,p)|_{\mathrm{cor.}}$ is a linear combination of the eight terms
\begin{equation}
\label{gbm:8terms}
\boldsymbol{\Sigma}_p\,T_a \,,\quad
\boldsymbol{\Sigma}_{p^\prime}\,T_a \,,\quad
\bar T_a\,\boldsymbol{\Sigma}_p \,,\quad
\bar T_a\,\boldsymbol{\Sigma}_{p^\prime} \,,\quad
T_a\,\boldsymbol{\Sigma}_p \,,\quad
T_a\,\boldsymbol{\Sigma}_{p^\prime} \,,\quad
\boldsymbol{\Sigma}_p\,\bar T_a \,,\quad
\boldsymbol{\Sigma}_{p^\prime}\,\bar T_a \,,\quad
\end{equation}
with the coefficients of the linear combination being only some functions of the two available momenta $p^\prime$ and $p$ and of the gamma matrices.

To conclude, we are led to assume that the vertex $\Gamma_{a}^\mu(p^\prime,p)$ has the form
\begin{eqnarray}
\label{gbm:Gmmgen}
\Gamma_{a}^\mu(p^\prime,p)  &=& \gamma^\mu T_a
+ \Big[f^\mu (\threevector{a}) \, T_a + \bar T_a \, f^\mu(\threevector{b}) +
T_a \, f^\mu(\threevector{c}) + f^\mu(\threevector{d}) \, \bar T_a\Big]
\nonumber \\ &&
\phantom{\gamma^\mu T_a}
{}+ \gamma_5 \Big[f^\mu (\threevector{e}) \, T_a + \bar T_a \, f^\mu(\threevector{f}) +
T_a \, f^\mu(\threevector{g}) + f^\mu(\threevector{h}) \, \bar T_a\Big] \,,
\end{eqnarray}
where each $f^\mu(\threevector{x})$, $\threevector{x} = \threevector{a},\threevector{b},\ldots,\threevector{h}$, is a linear combination of $\Sigma_{p^\prime}$ and $\Sigma_{p}$ and $\threevector{x}$ is some \qm{multiindex} parameterizing each linear combination. In the following section we will, under certain assumptions, find the general momentum and Lorentz structure of $f^\mu(\threevector{x})$ and show that $\threevector{x}$ is in fact a \emph{finite} set of complex numbers.

\subsection{Momentum and Lorentz structure}
\label{gbm:ssec:analytic}

The main guiding principle in determining the analytic structure of $\Gamma_{a}^{\mu}(p^\prime,p)$ will be to insist on its good interpretability in terms of the NG bosons. We have already encountered it in Sec.~\ref{gbm:ssec:vertexprel} when we assumed the vertex to be of the form \eqref{gbm:Gmm:decomp}. Let us now rephrase that assumption in another way. The vertex is a function of two independent momenta $p^\prime$ and $p$. Their linear combination $q = p^\prime - p$ is special in the sense that it is the momentum carried by the gauge boson and also by the eventual NG boson, bilinearly coupled to it. Then the correct NG interpretability of the vertex $\Gamma_{a}^{\mu}(p^\prime,p)$  technically means imposing the following conditions on the vertex:
\begin{itemize}
\item The poles of the types $1/\ell^2$ and $1/(\ell \cdot q)$, where $\ell$ is some linear combination of $p^\prime$ and $p$, being \emph{linearly independent} of $q$, are forbidden.
\item Any pole of the type $1/q^2$ can be only simple and must be multiplied by $q^\mu$.
\end{itemize}
These conditions hold of course in general, independently of the special vertex form \eqref{gbm:Gmmgen}, assumed in the previous section. Nevertheless we will use them now for determining $f^\mu(\threevector{x})$.

Since the vertex $\Gamma_{a}^{\mu}(p^\prime,p)$ is dimensionless (in the units of mass), so must be also $f^\mu(\threevector{x})$. Recall that $f^\mu(\threevector{x})$ is a linear combination of
\begin{equation}
\label{gbm:Sgmpprimep}
\boldsymbol{\Sigma}_{p^\prime} \,,\quad \boldsymbol{\Sigma}_{p} \,.
\end{equation}
But the self-energies $\boldsymbol{\Sigma}$ are of the dimension $+1$. Therefore we have to find the coefficients of the linear combination of \eqref{gbm:Sgmpprimep} with the negative dimension $-1$ in order to have dimensionless $f^\mu(\threevector{x})$. It seems, on the basis of the requirements made in the previous paragraph, that the only possibility for the coefficient is
\begin{equation}
\frac{q^\mu}{q^2}
\end{equation}
times a complex factor.

However, it turns out that this does not in fact exhaust all possibilities. Consider the following linear combination of $\boldsymbol{\Sigma}_{p^\prime}$ and $\boldsymbol{\Sigma}_{p}$:
\begin{equation}
\label{gbm:Sgmder}
\frac{\boldsymbol{\Sigma}_{p^\prime}-\boldsymbol{\Sigma}_{p}}{p^{\prime2}-p^2} \,.
\end{equation}
The crucial fact is that this quantity is regular for all values $p^\prime$ and $p$.\footnote{Unless the self-energy itself has a pole at some $p^2$. We do not take this possibility into account.} In the worst case, when $p^\prime \rightarrow p$ (i.e., when $q \rightarrow 0$), it just converges to the derivative of $\boldsymbol{\Sigma}_{p}$:
\begin{eqnarray}
\label{gbm:Sgmderexp}
\frac{\boldsymbol{\Sigma}_{p^\prime}-\boldsymbol{\Sigma}_{p}}{p^{\prime2}-p^2}
&=& \boldsymbol{\Sigma}^\prime_{p} + \mathcal{O}(q) \,,
\end{eqnarray}
where the prime denotes the derivative with respect to $p^2$. Thus, due to \eqref{gbm:Sgmderexp}, we can use \eqref{gbm:Sgmder} as a building block for $f^\mu(\threevector{x})$ as well, without introducing any unwanted kinematic singularity. Since it has the dimension $-1$, is suffices to multiply it by something of dimension $+1$ and carrying the Lorentz index. Taking into account the conditions above, the only possibilities turn out to be
\begin{equation}
\label{gbm:LortzCarry}
q^\mu \,,\quad
p^{\mu} \,,\quad
[\gamma^\mu,\slashed{q}] \,,\quad
[\gamma^\mu,\slashed{p}] \,,\quad
\frac{q^\mu}{q^2} [\slashed{q},\slashed{p}] \,,\quad
\frac{q^\mu}{q^2} p^2 \,,\quad
\frac{q^\mu}{q^2} (q \cdot p) \,,
\end{equation}
again up to complex factors.

We conclude from the previous discussion that $f^\mu(\threevector{x})$ is an element of the complex eight-dimensional vector space, spanned by the basis
\begin{equation}
\label{gbm:basispq}
\frac{q^\mu}{q^2} \bigg\{ \boldsymbol{\Sigma}_{p^\prime} \oplus \boldsymbol{\Sigma}_{p} \bigg\}
\oplus
\frac{\boldsymbol{\Sigma}_{p^\prime}-\boldsymbol{\Sigma}_{p}}{p^{\prime2}-p^2}
\bigg\{q^\mu \oplus
p^{\mu} \oplus
[\gamma^\mu,\slashed{q}] \oplus
[\gamma^\mu,\slashed{p}] \oplus
\frac{q^\mu}{q^2} [\slashed{q},\slashed{p}] \oplus
\frac{q^\mu}{q^2} p^2
\bigg\} \,.
\end{equation}
Notice that we have not included here the last term from \eqref{gbm:LortzCarry} (proportional to $(q \cdot p)$), since it depends linearly on the other terms.

The basis \eqref{gbm:basispq} is however not the most convenient one. Recall that transformation law \eqref{gbp:Gmm:C} of $\Gamma_{a}^\mu(p^\prime,p)$ under $\mathcal{C}$, as well as the Hermiticity condition \eqref{gbm:hermGmm} include the exchanges
\begin{eqnarray}
\label{gbm:pprimepexch}
p^\prime &\leftrightarrow& p \,.
\end{eqnarray}
As we will eventually apply the conditions \eqref{gbp:Gmm:C}, \eqref{gbm:hermGmm} to the vertex, it will prove convenient to have expressed $f^\mu(\threevector{x})$ in terms of a basis made of eigenstates of \eqref{gbm:pprimepexch}. Thus, instead of being a linear combination of $\boldsymbol{\Sigma}_{p^\prime}$ and $\boldsymbol{\Sigma}_{p}$, we will use the linear combination of $\boldsymbol{\Sigma}_{+}$ and $\boldsymbol{\Sigma}_{-}$, defined as
\begin{eqnarray}
\label{gbm:Sgmpm}
\boldsymbol{\Sigma}_{\pm} &\equiv& \boldsymbol{\Sigma}_{p^\prime} \pm \boldsymbol{\Sigma}_{p} \,.
\end{eqnarray}
Clearly, $\boldsymbol{\Sigma}_{\pm}$ are eigenstates of \eqref{gbm:pprimepexch} with the eigenvalues $\pm 1$. Similarly, we will express everything in terms of the two linearly independent momenta $q$ and $q^\prime$:
\begin{subequations}
\label{gbm:qprimeq}
\begin{eqnarray}
q        &\equiv& p^\prime - p \,, \\
q^\prime &\equiv& p^\prime + p \,,
\end{eqnarray}
\end{subequations}
which are eigenstates of \eqref{gbm:pprimepexch} with the eigenvalues $-1$ and $+1$, respectively. Notice also that the denominator of \eqref{gbm:Sgmder} can be expressed in terms of $q$ and $q^\prime$ conveniently as
\begin{eqnarray}
p^{\prime2}-p^2 &=& (q \cdot q^\prime) \,.
\end{eqnarray}
We can now rewrite the basis \eqref{gbm:basispq} in terms of the self-energies $\boldsymbol{\Sigma}_{\pm}$ and the momenta $q$, $q^\prime$ and arrive finally at $f^\mu(\threevector{x})$ of the form
\begin{eqnarray}
\label{gbm:fmux}
f^\mu(\threevector{x}) &\equiv& \phantom{+\,}
x_1 \frac{q^\mu}{q^2} \boldsymbol{\Sigma}_{+} +
x_2 \frac{q^\mu}{q^2} \boldsymbol{\Sigma}_{-} +
x_3 \frac{q^{\prime\mu}}{(q \cdot q^\prime)} \boldsymbol{\Sigma}_{-} +
x_4 \frac{q^\mu}{q^2} \frac{[\slashed{q},\slashed{q}^\prime]}{(q \cdot q^\prime)} \boldsymbol{\Sigma}_{-}
\nonumber \\ &&
+\,
x_5 \frac{[\gamma^\mu,\slashed{q}^\prime]}{(q \cdot q^\prime)} \boldsymbol{\Sigma}_{-} +
x_6 \frac{[\gamma^\mu,\slashed{q}]}{(q \cdot q^\prime)} \boldsymbol{\Sigma}_{-} +
x_7 \frac{q^\mu}{(q \cdot q^\prime)} \boldsymbol{\Sigma}_{-} +
x_8 \frac{q^\mu}{q^2} \frac{q^{\prime\mu}}{(q \cdot q^\prime)} \boldsymbol{\Sigma}_{-} \,,
\end{eqnarray}
where $\threevector{x} = (x_1,\ldots,x_8)$ is a vector of eight complex numbers. The Ansatz \eqref{gbm:Gmmgen} is thus parameterized altogether by $8 \times 8 = 64$, at this moment completely arbitrary complex numbers. In the following sections we will, step by step, determine almost\footnote{Cf.~footnote~\ref{ftnt:almost} on page~\pageref{ftnt:almost}.} all of them.

\subsection{WT identity}

The most basic condition that must be satisfied by the vertex $\Gamma_{a}^\mu(p^\prime,p)$ is certainly the WT identity \eqref{gbm:WT}, since it ensures the transversality of the polarization tensor \eqref{gbm:Pimunuab}, as shown in section~\ref{gbm:ssec:Transv}. Using the form \eqref{gbp:Gfrm} for the fermion propagator the WT identity has the form
\begin{eqnarray}
\label{gbm:WTansSgmpprimep}
q_\mu \Gamma_{a}^\mu(p^\prime,p) &=&
\slashed{q} T_a - \boldsymbol{\Sigma}_{p^\prime}\,T_a + \bar T_a\,\boldsymbol{\Sigma}_{p} \,.
\end{eqnarray}
The requirement that the Ansatz \eqref{gbm:Gmmgen}, with $f^\mu(\threevector{x})$ given by \eqref{gbm:fmux}, satisfies this WT identity will now enable us to determine more than half of the $64$ parameters of the Ansatz.

Let us rewrite the WT identity \eqref{gbm:WTansSgmpprimep} in terms of $\boldsymbol{\Sigma}_{\pm}$,
\begin{eqnarray}
\label{gbm:WTansSgmpm}
q_\mu \Gamma_{a}^\mu(p^\prime,p)&=&
\slashed{q} T_a - \frac{1}{2} (\boldsymbol{\Sigma}_{+}+\boldsymbol{\Sigma}_{-})\,T_a + \frac{1}{2} \bar T_a\,(\boldsymbol{\Sigma}_{+}-\boldsymbol{\Sigma}_{-}) \,,
\end{eqnarray}
and consider the contraction $q_\mu f^\mu(\threevector{x})$:
\begin{eqnarray}
q_\mu f^\mu(\threevector{x}) &=&
x_1 \Sigma_{+} + (x_2 + x_3) \boldsymbol{\Sigma}_{-} +
(x_4 + x_5) \frac{[\slashed{q},\slashed{q}^\prime]}{(q \cdot q^\prime)} \boldsymbol{\Sigma}_{-} +
x_7 \frac{q^2}{(q \cdot q^\prime)} \boldsymbol{\Sigma}_{-} +
x_8 \frac{q^{\prime2}}{(q \cdot q^\prime)} \boldsymbol{\Sigma}_{-} \,.
\nonumber \\ &&
\end{eqnarray}
By imposing WT identity \eqref{gbm:WTansSgmpm} on the Ansatz \eqref{gbm:Gmmgen} (with $f^\mu(\threevector{x})$ given by \eqref{gbm:fmux}) we can, using the contraction $q_\mu f^\mu(\threevector{x})$, readily read off the constraints on the free parameters $\threevector{x}$. For $x_1$ we obtain the constraints
\begin{eqnarray}
a_1 &=& -\frac{1}{2} \,, \\
b_1 &=& +\frac{1}{2} \,, \\
x_1 &=& \phantom{-} 0 \qquad \mbox{for} \quad x \neq a,b \,.
\end{eqnarray}
For the parameters $x_2$, $x_3$ we find
\begin{alignat}{2}
x_2 + x_3 &\quad=\quad          -  \frac{1}{2} &\qquad& \mbox{for} \quad x = a,b    \,, \\
x_2 + x_3 &\quad=\quad \phantom{-} 0           &\qquad& \mbox{for} \quad x \neq a,b \,,
\end{alignat}
enabling us to eliminate, say, $x_3$ in favor of $x_2$. For the rest we have
\begin{eqnarray}
x_4+x_5  &=& 0 \,, \\
x_7      &=& 0 \,, \\
x_8      &=& 0 \,,
\end{eqnarray}
for all $x = a,\ldots,h$, which enables us again to eliminate, e.g., $x_5$ in favor of $x_4$. On the other hand, note that $x_6$ (again for all $x = a,\ldots,h$) remains unconstrained by the WT identity.

Thus, applying the WT identity the vertex \eqref{gbm:Gmmgen} reduces to
\begin{eqnarray}
\label{gbm:Gmm:ansWT}
\Gamma_{a}^\mu(p^\prime,p) &=&
\gamma^\mu T_a
- \frac{1}{2}\frac{q^\mu}{q^2} \big( \boldsymbol{\Sigma}_{+}\,T_a - \bar T_a\,\boldsymbol{\Sigma}_{+} \big)
- \frac{1}{2}\frac{q^{\prime\mu}}{q \cdot q^\prime} \big( \boldsymbol{\Sigma}_{-}\,T_a + \bar T_a\,\boldsymbol{\Sigma}_{-} \big)
\nonumber \\ && +\,\bigg(\frac{q^\mu}{q^2}-\frac{q^{\prime\mu}}{q \cdot q^\prime}\bigg)
\bigg[
\phantom{+\gamma_5}\hspace{1.9mm}
\big(a_2\,\boldsymbol{\Sigma}_{-}\,T_a + b_2\,\bar T_a\,\boldsymbol{\Sigma}_{-} \big) +
\big(c_2\,T_a\,\boldsymbol{\Sigma}_{-} + d_2\,\boldsymbol{\Sigma}_{-}\,\bar T_a \big)
\nonumber \\ &&
\phantom{+\,\bigg(\frac{q^\mu}{q^2}-\frac{q^{\prime\mu}}{q \cdot q^\prime}\bigg)\bigg[}
+
\gamma_5 \big(e_2\,\boldsymbol{\Sigma}_{-}\,T_a + f_2\,\bar T_a\,\boldsymbol{\Sigma}_{-} \big) +
\gamma_5 \big(g_2\,T_a\,\boldsymbol{\Sigma}_{-} + h_2\,\boldsymbol{\Sigma}_{-}\,\bar T_a \big)
\bigg]
\nonumber \\ && +\,
\bigg(\frac{q^\mu}{q^2}\frac{[\slashed{q},\slashed{q}^\prime]}{q \cdot q^\prime} - \frac{[\gamma^\mu,\slashed{q}^\prime]}{q \cdot q^\prime} \bigg)
\bigg[
\phantom{+\gamma_5}\hspace{1.9mm}
\big(a_4\,\boldsymbol{\Sigma}_{-}\,T_a + b_4\,\bar T_a\,\boldsymbol{\Sigma}_{-} \big) +
\big(c_4\,T_a\,\boldsymbol{\Sigma}_{-} + d_4\,\boldsymbol{\Sigma}_{-}\,\bar T_a \big)
\nonumber \\ &&
\phantom{+\,\bigg(\frac{q^\mu}{q^2}\frac{[\slashed{q},\slashed{q}^\prime]}{q \cdot q^\prime} - \frac{[\gamma^\mu,\slashed{q}^\prime]}{q \cdot q^\prime} \bigg)
\bigg[}
+
\gamma_5 \big(e_4\,\boldsymbol{\Sigma}_{-}\,T_a + f_4\,\bar T_a\,\boldsymbol{\Sigma}_{-} \big) +
\gamma_5 \big(g_4\,T_a\,\boldsymbol{\Sigma}_{-} + h_4\,\boldsymbol{\Sigma}_{-}\,\bar T_a \big)
\bigg]
\nonumber \\ && + \,\frac{[\gamma^\mu,\slashed{q}]}{(q \cdot q^\prime)}\bigg[
\phantom{+\gamma_5}\hspace{1.9mm}
\big(a_6\,\boldsymbol{\Sigma}_{-}\,T_a + b_6\, \bar T_a \,\boldsymbol{\Sigma}_{-}\big) + \big(c_6\, T_a \,\boldsymbol{\Sigma}_{-} + d_6\,\boldsymbol{\Sigma}_{-}\,\bar T_a \big)
\nonumber \\ &&
\phantom{+ \,\frac{[\gamma^\mu,\slashed{q}]}{(q \cdot q^\prime)}\bigg[}
+
\gamma_5 \big(e_6\,\boldsymbol{\Sigma}_{-}\,T_a + f_6\, \bar T_a \,\boldsymbol{\Sigma}_{-}\big) + \gamma_5\big(g_6\, T_a \,\boldsymbol{\Sigma}_{-} + h_6\,\boldsymbol{\Sigma}_{-}\,\bar T_a \big)
\bigg] \,.
\end{eqnarray}
Notice that by imposing the WT identity we have reduced the number of free complex parameters from $64$ to $24$.

\subsection{Transformation under $\group{G}$}

\subsubsection{Correct transformation behavior under full $\group{G}$}

We continue by recalling that the vertex must transform properly under the global symmetry $\group{G}$, i.e., as \eqref{gbp:Gmm:transf}, in order to guarantee the correct transformation property of the polarization tensor \eqref{gbm:Pimunuab} (see Sec.~\ref{gbm:ssec:tranf}). If we suppress the momentum arguments, which do not play any substantial r\^{o}le in the present considerations, we can write the vertex \eqref{gbm:Gmm:ansWT} in a schematic form
\begin{eqnarray}
\label{gbm:Gmm:ansTrnsfSchm}
\Gamma_{a}^{\mu} &=&
v_1^{\mu} \, \boldsymbol{\Sigma} \, T_a + v_2^{\mu} \, \bar T_a \, \boldsymbol{\Sigma} +
v_3^{\mu} \, \boldsymbol{\Sigma} \, \bar T_a + v_4^{\mu} \, T_a \, \boldsymbol{\Sigma} \,.
\end{eqnarray}
The transformation of $\Gamma_{a}^{\mu}$ under $\group{G}$ must be induced by the corresponding transformation of $\boldsymbol{\Sigma}$:
\begin{subequations}
\label{gbm:Gmm:ansTrnsfProof}
\begin{eqnarray}
\group{G}\,:\qquad \Gamma_{a}^{\mu}  \>\>\TransformsTo\>\> {[\Gamma_{a}^{\mu}]}^\prime &=& v_1^{\mu} \, [\boldsymbol{\Sigma}]^\prime \, T_a + v_2^{\mu} \, \bar T_a \, [\boldsymbol{\Sigma}]^\prime +
v_3^{\mu} \, [\boldsymbol{\Sigma}]^\prime \, \bar T_a + v_4^{\mu} \, T_a \, [\boldsymbol{\Sigma}]^\prime
\label{gbm:Gmm:ansTrnsfProof1}
\\ &=&
\e^{\I \theta \cdot \bar t} \Big[ \phantom{+\,}
v_1^{\mu} \, \boldsymbol{\Sigma} \big(\e^{-\I \theta \cdot t}\,T_a\,\e^{\I \theta \cdot t}\big) +
v_2^{\mu} \big(\e^{-\I \theta \cdot \bar t}\,\bar T_a\,\e^{\I \theta \cdot \bar t}\big) \boldsymbol{\Sigma}
\nonumber \\ && \phantom{\e^{\I \theta \cdot \bar t} \Big[}
+\,
v_3^{\mu} \, \boldsymbol{\Sigma} \big(\e^{-\I \theta \cdot t}\,\bar T_a\,\e^{\I \theta \cdot t}\big) +
v_4^{\mu} \big(\e^{-\I \theta \cdot \bar t}\,T_a\,\e^{\I \theta \cdot \bar t}\big) \boldsymbol{\Sigma}
\Big]\e^{-\I \theta \cdot t} \,,
\qquad\qquad
\label{gbm:Gmm:ansTrnsfProof2}
\end{eqnarray}
\end{subequations}
where we have already used the transformation rule \eqref{gbp:Sgm:transglob} for $\boldsymbol{\Sigma}$.

Let us first check the first two terms (proportional to $v_1^\mu$, $v_2^\mu$). Using \eqref{gbp:taeabetbe} we find that the round brackets can be expressed as
\begin{subequations}
\begin{eqnarray}
\e^{-\I \theta \cdot t}\,T_a\,\e^{\I \theta \cdot t}
&=& \big(\e^{\I\theta\cdot\mathcal{T}}\big)_{ab}\,T_b \,, \\
\e^{-\I \theta \cdot \bar t}\,\bar T_a\,\e^{\I \theta \cdot \bar t}
&=& \big(\e^{\I\theta\cdot\mathcal{T}}\big)_{ab}\,\bar T_b \,.
\end{eqnarray}
\end{subequations}
Plugging these expressions into \eqref{gbm:Gmm:ansTrnsfProof2}, we find that the first two terms of the vertex \eqref{gbm:Gmm:ansTrnsfSchm} do transform correctly according to the rule \eqref{gbp:Gmm:transf}.

On the other hand, it turns out that the third and fourth term (proportional to $v_3^\mu$, $v_4^\mu$) do not transform properly, since in general
\begin{subequations}
\begin{eqnarray}
\e^{-\I \theta \cdot t}\,\bar T_a\,\e^{\I \theta \cdot t}
&\neq& \big(\e^{\I\theta\cdot\mathcal{T}}\big)_{ab}\,\bar T_b \,, \\
\e^{-\I \theta \cdot \bar t}\,T_a\,\e^{\I \theta \cdot \bar t}
&\neq& \big(\e^{\I\theta\cdot\mathcal{T}}\big)_{ab}\,T_b \,.
\end{eqnarray}
\end{subequations}

In other words, we found that only terms of the type $\boldsymbol{\Sigma} \, T_a$ and $\bar T_a \, \boldsymbol{\Sigma}$ are allowed, while the terms of the type $\boldsymbol{\Sigma} \, \bar T_a$ and $T_a \, \boldsymbol{\Sigma}$ are forbidden by the requirement that the vertex must transform under $\group{G}$ according to \eqref{gbp:Gmm:transf}. In terms of the free parameters of the vertex \eqref{gbm:Gmm:ansWT} we therefore must set
\begin{eqnarray}
x_2 \ = \ x_4 \ = \ x_6 &=& 0 \qquad \mbox{for} \quad x=c,d,g,h\,,
\end{eqnarray}
so that the vertex $\Gamma_{a}^\mu(p^\prime,p)$ now acquires the form
\begin{eqnarray}
\label{gbm:Gmm:ansTrnsfG}
\Gamma_{a}^\mu(p^\prime,p) &=&
\gamma^\mu T_a
- \frac{1}{2}\frac{q^\mu}{q^2} \big( \boldsymbol{\Sigma}_{+}\,T_a - \bar T_a\,\boldsymbol{\Sigma}_{+} \big)
- \frac{1}{2}\frac{q^{\prime\mu}}{q \cdot q^\prime} \big( \boldsymbol{\Sigma}_{-}\,T_a + \bar T_a\,\boldsymbol{\Sigma}_{-} \big)
\nonumber \\ &&
{}+\bigg(\frac{q^\mu}{q^2}-\frac{q^{\prime\mu}}{q \cdot q^\prime}\bigg)
\bigg[
\big(a_2\,\boldsymbol{\Sigma}_{-}\,T_a + b_2\,\bar T_a\,\boldsymbol{\Sigma}_{-} \big)
+
\gamma_5 \big(e_2\,\boldsymbol{\Sigma}_{-}\,T_a + f_2\,\bar T_a\,\boldsymbol{\Sigma}_{-} \big)
\bigg]
\nonumber \\ &&
{}+ \bigg(\frac{q^\mu}{q^2}\frac{[\slashed{q},\slashed{q}^\prime]}{q \cdot q^\prime} - \frac{[\gamma^\mu,\slashed{q}^\prime]}{q \cdot q^\prime} \bigg)
\bigg[
\big(a_4\,\boldsymbol{\Sigma}_{-}\,T_a + b_4\,\bar T_a\,\boldsymbol{\Sigma}_{-} \big)
+
\gamma_5 \big(e_4\,\boldsymbol{\Sigma}_{-}\,T_a + f_4\,\bar T_a\,\boldsymbol{\Sigma}_{-} \big)
\bigg]
\nonumber \\ &&
{}+ \frac{[\gamma^\mu,\slashed{q}]}{(q \cdot q^\prime)}\bigg[
\big(a_6\,\boldsymbol{\Sigma}_{-}\,T_a + b_6\, \bar T_a \,\boldsymbol{\Sigma}_{-}\big)
+
\gamma_5 \big(e_6\,\boldsymbol{\Sigma}_{-}\,T_a + f_6\, \bar T_a \,\boldsymbol{\Sigma}_{-}\big)
\bigg] \,,
\end{eqnarray}
which contains half as many complex free parameters as \eqref{gbm:Gmm:ansWT}, i.e., $12$.

\subsubsection{Invariance under unbroken $\group{H} \subseteq \group{G}$}

The next natural requirement is to demand the vertex $\Gamma_{a}^\mu(p^\prime,p)$ to be invariant under unbroken $\group{H} \subseteq \group{G}$:
\begin{eqnarray}
\label{gbm:Gmm:Hinv}
\group{H}\,:\qquad \Gamma_{a}^\mu(p^\prime,p)  & \TransformsTo & {[\Gamma_{a}^\mu(p^\prime,p)]}^\prime \>=\> \Gamma_{a}^\mu(p^\prime,p) \,.
\end{eqnarray}
However, it is easy to see that this invariance is already automatically \emph{guaranteed} due to the correct transformation behavior of $\Gamma_{a}^\mu(p^\prime,p)$ under the full symmetry group $\group{G}$, ensured above. To see it, let us first remind the r\^{o}le of the self-energy $\boldsymbol{\Sigma}$ here: It is this $\boldsymbol{\Sigma}$ which is assumed to break $\group{G}$ down to $\group{H}$. In other words, it is by definition invariant under $\group{H}$:
\begin{eqnarray}
\group{H}\,:\qquad \boldsymbol{\Sigma}  & \TransformsTo & [\boldsymbol{\Sigma}]^\prime \>=\> \boldsymbol{\Sigma} \,.
\end{eqnarray}
Second, note that the only part of $\Gamma_{a}^\mu(p^\prime,p)$, transforming non-trivially under $\group{G}$, is the self-energy $\boldsymbol{\Sigma}$, see \eqref{gbm:Gmm:ansTrnsfProof1}. Therefore, since $\boldsymbol{\Sigma}$ stays invariant under $\group{H}$, so must $\Gamma_{a}^\mu(p^\prime,p)$, which completes the proof of \eqref{gbm:Gmm:Hinv}.

\subsection{Transformation under $\mathcal{C}$, $\mathcal{P}$ and $\mathcal{CP}$}

\subsubsection{Separate $\mathcal{C}$ and $\mathcal{P}$ invariance}

Consider now the transformations of the vertex under the discrete symmetries $\mathcal{C}$ and $\mathcal{P}$. Applying the corresponding transformation rules \eqref{gbp:Gmm:C} and \eqref{gbp:Gmm:P}, respectively, on the vertex \eqref{gbm:Gmm:ansTrnsfG}, we obtain
\begin{eqnarray}
{[\Gamma_{a}^\mu(p^\prime,p)]}^{\mathcal{C}} &=&
\gamma^\mu \bar T_a -
\frac{1}{2}\frac{q^\mu}{q^2} \Big( [\boldsymbol{\Sigma}_{+}]^{\mathcal{C}}\,\bar T_a - T_a\,[\boldsymbol{\Sigma}_{+}]^{\mathcal{C}} \Big) -
\frac{1}{2}\frac{q^{\prime\mu}}{q \cdot q^\prime} \Big( [\boldsymbol{\Sigma}_{-}]^{\mathcal{C}}\,\bar T_a + T_a\,[\boldsymbol{\Sigma}_{-}]^{\mathcal{C}} \Big)
\nonumber \\ &&
+\,\bigg(\frac{q^\mu}{q^2}-\frac{q^{\prime\mu}}{q \cdot q^\prime}\bigg)
\bigg[     \Big(b_2  \,[\boldsymbol{\Sigma}_{-}]^{\mathcal{C}}\,\bar T_a + a_2  \,T_a\,[\boldsymbol{\Sigma}_{-}]^{\mathcal{C}} \Big)
\nonumber \\ &&
\phantom{
+\,\bigg(\frac{q^\mu}{q^2}-\frac{q^{\prime\mu}}{q \cdot q^\prime}\bigg)
\bigg[
}
+ \gamma_5 \Big(f_2\,[\boldsymbol{\Sigma}_{-}]^{\mathcal{C}}\,\bar T_a + e_2\,T_a\,[\boldsymbol{\Sigma}_{-}]^{\mathcal{C}} \Big)
\bigg]
\nonumber \\ &&
+\,\bigg(\frac{q^\mu}{q^2}\frac{[\slashed{q},\slashed{q}^\prime]}{q \cdot q^\prime} - \frac{[\gamma^\mu,\slashed{q}^\prime]}{q \cdot q^\prime} \bigg)
\bigg[   -\Big(b_4\,[\boldsymbol{\Sigma}_{-}]^{\mathcal{C}}\,\bar T_a + a_4\,T_a\,[\boldsymbol{\Sigma}_{-}]^{\mathcal{C}} \Big)
\nonumber \\ &&
\phantom{
+\,\bigg(\frac{q^\mu}{q^2}\frac{[\slashed{q},\slashed{q}^\prime]}{q \cdot q^\prime} - \frac{[\gamma^\mu,\slashed{q}^\prime]}{q \cdot q^\prime} \bigg)
\bigg[
}
{}-\gamma_5 \Big(f_4\,[\boldsymbol{\Sigma}_{-}]^{\mathcal{C}}\,\bar T_a + e_4\,T_a\,[\boldsymbol{\Sigma}_{-}]^{\mathcal{C}} \Big)
\bigg]
\nonumber \\ &&
{}+ \frac{[\gamma^\mu,\slashed{q}]}{(q \cdot q^\prime)}\bigg[
\Big(b_6\,[\boldsymbol{\Sigma}_{-}]^{\mathcal{C}}\,\bar T_a + a_6\,T_a \,[\boldsymbol{\Sigma}_{-}]^{\mathcal{C}}\Big)
+
\gamma_5 \Big(f_6\,[\boldsymbol{\Sigma}_{-}]^{\mathcal{C}}\,\bar T_a + e_6\, T_a \,[\boldsymbol{\Sigma}_{-}]^{\mathcal{C}}\Big)
\bigg]\,,
\nonumber \\ &&
\label{gbm:Gmm:ansC}
\\
{[\Gamma_{a}^\mu(p^\prime,p)]}^{\mathcal{P}} &=&
\gamma^\mu \bar T_a -
\frac{1}{2}\frac{q^\mu}{q^2} \Big( [\boldsymbol{\Sigma}_{+}]^{\mathcal{P}}\,\bar T_a - T_a\,[\boldsymbol{\Sigma}_{+}]^{\mathcal{P}} \Big) -
\frac{1}{2}\frac{q^{\prime\mu}}{q \cdot q^\prime} \Big( [\boldsymbol{\Sigma}_{-}]^{\mathcal{P}}\,\bar T_a + T_a\,[\boldsymbol{\Sigma}_{-}]^{\mathcal{P}} \Big)
\nonumber \\ &&
+\,\bigg(\frac{q^\mu}{q^2}-\frac{q^{\prime\mu}}{q \cdot q^\prime}\bigg)
\bigg[
\Big(a_2  \,[\boldsymbol{\Sigma}_{-}]^{\mathcal{P}}\,\bar T_a + b_2  \,T_a\,[\boldsymbol{\Sigma}_{-}]^{\mathcal{P}} \Big)
\nonumber \\ &&
\phantom{
+\,\bigg(\frac{q^\mu}{q^2}-\frac{q^{\prime\mu}}{q \cdot q^\prime}\bigg)
\bigg[
}
-\,\gamma_5 \Big(e_2\,[\boldsymbol{\Sigma}_{-}]^{\mathcal{P}}\,\bar T_a + f_2\,T_a\,[\boldsymbol{\Sigma}_{-}]^{\mathcal{P}} \Big)
\bigg]
\nonumber \\ &&
+\,\bigg(\frac{q^\mu}{q^2}\frac{[\slashed{q},\slashed{q}^\prime]}{q \cdot q^\prime} - \frac{[\gamma^\mu,\slashed{q}^\prime]}{q \cdot q^\prime} \bigg)
\bigg[     \Big(a_4\,[\boldsymbol{\Sigma}_{-}]^{\mathcal{P}}\,\bar T_a + b_4\,T_a\,[\boldsymbol{\Sigma}_{-}]^{\mathcal{P}} \Big)
\nonumber \\ &&
\phantom{
+\,\bigg(\frac{q^\mu}{q^2}\frac{[\slashed{q},\slashed{q}^\prime]}{q \cdot q^\prime} - \frac{[\gamma^\mu,\slashed{q}^\prime]}{q \cdot q^\prime} \bigg)
\bigg[
}
{}- \gamma_5 \Big(e_4\,[\boldsymbol{\Sigma}_{-}]^{\mathcal{P}}\,\bar T_a + f_4\,T_a\,[\boldsymbol{\Sigma}_{-}]^{\mathcal{P}} \Big)
\bigg]
\nonumber \\ &&
{}+ \frac{[\gamma^\mu,\slashed{q}]}{(q \cdot q^\prime)}\bigg[
\Big(a_6\,[\boldsymbol{\Sigma}_{-}]^{\mathcal{P}}\,\bar T_a + b_6\,T_a \,[\boldsymbol{\Sigma}_{-}]^{\mathcal{P}}\Big)
-
\gamma_5 \Big(e_6\,[\boldsymbol{\Sigma}_{-}]^{\mathcal{P}}\,\bar T_a + f_6\,T_a \,[\boldsymbol{\Sigma}_{-}]^{\mathcal{P}}\Big)
\bigg]\,,
\nonumber \\&&
\label{gbm:Gmm:ansP}
\end{eqnarray}
where the transformation $[\boldsymbol{\Sigma}]^{\mathcal{C}}$ and $[\boldsymbol{\Sigma}]^{\mathcal{P}}$ are given by \eqref{gbp:SgmC} and \eqref{gbp:SgmP}, respectively.

We can now impose two natural requirements on the vertex:
\begin{description}

\item[$\mathcal{C}$ invariance:] Recall that the non-invariance of the gauge interactions under $\mathcal{C}$ is attributed solely to the presence of $\gamma_5$ in the symmetry generators $T_a$, i.e., to $T_a \neq \bar T_a$. We can therefore require that in the case of $[\boldsymbol{\Sigma}]^{\mathcal{C}} = \boldsymbol{\Sigma}$ the only source of non-invariance of the vertex under $\mathcal{C}$ should be also the generators $T_a$. Rephrased more formally, we require
\begin{eqnarray}
\label{gbm:invC}
\Big(T_a = \bar T_a \quad \mbox{and} \quad  [\boldsymbol{\Sigma}]^{\mathcal{C}} = \boldsymbol{\Sigma} \Big)
&\quad\Longrightarrow\quad&
{[\Gamma_{a}^\mu(p^\prime,p)]}^{\mathcal{C}} = \Gamma_{a}^\mu(p^\prime,p) \,.
\end{eqnarray}
Taking into account the explicit form \eqref{gbm:Gmm:ansC} of ${[\Gamma_{a}^\mu(p^\prime,p)]}^{\mathcal{C}}$, we are forced to set
\begin{subequations}
\label{gbm:constrC}
\begin{eqnarray}
b_2 &=& a_2 \,, \\
f_2 &=& e_2 \,, \\
b_4 &=& -a_4 \,, \\
f_4 &=& -e_4 \,, \\
b_6 &=& a_6 \,, \\
f_6 &=& e_6 \,.
\end{eqnarray}
\end{subequations}

\item[$\mathcal{P}$ invariance:] Similarly can be treated the non-invariance under the $\mathcal{P}$ symmetry, whose only source in the gauge interactions is also the presence of $\gamma_5$ in the generators $T_a$. We can therefore analogously require that if $[\boldsymbol{\Sigma}]^{\mathcal{P}} = \boldsymbol{\Sigma}$, then the only source of parity violation of the vertex should be also the generators $T_a$:
\begin{eqnarray}
\label{gbm:invP}
\Big(T_a = \bar T_a \quad \mbox{and} \quad  [\boldsymbol{\Sigma}]^{\mathcal{P}} = \boldsymbol{\Sigma} \Big)
&\quad\Longrightarrow\quad&
{[\Gamma_{a}^\mu(p^\prime,p)]}^{\mathcal{P}} = \Gamma_{a}^\mu(p^\prime,p) \,.
\end{eqnarray}
This time we obtain, using ${[\Gamma_{a}^\mu(p^\prime,p)]}^{\mathcal{P}}$, \eqref{gbm:Gmm:ansP}, the constraints
\begin{eqnarray}
\label{gbm:constrP}
e_2 \ = \ f_2 \ = \ e_4 \ = \ f_4 \ = \ e_6 \ = \ f_6 &=& 0 \,,
\end{eqnarray}
leading to the absence of $\gamma_5$ in the vertex, elsewhere than in the generators $T_a$ and in the self-energy $\boldsymbol{\Sigma}$.

\end{description}

By putting the two constraints \eqref{gbm:constrC} and \eqref{gbm:constrP} together we obtain the vertex in the form
\begin{eqnarray}
\label{gbm:Gmm:ansDiscSym}
\Gamma_{a}^\mu(p^\prime,p) &=&
\gamma^\mu T_a
- \frac{1}{2}\frac{q^\mu}{q^2} \big( \boldsymbol{\Sigma}_{+}\,T_a - \bar T_a\,\boldsymbol{\Sigma}_{+} \big)
- \frac{1}{2}\frac{q^{\prime\mu}}{q \cdot q^\prime} \big( \boldsymbol{\Sigma}_{-}\,T_a + \bar T_a\,\boldsymbol{\Sigma}_{-} \big)
\nonumber \\ &&
{}+a_2\bigg(\frac{q^\mu}{q^2}-\frac{q^{\prime\mu}}{q \cdot q^\prime}\bigg)
\big(\boldsymbol{\Sigma}_{-}\,T_a + \bar T_a\,\boldsymbol{\Sigma}_{-} \big)
\nonumber \\ &&
{}+ a_4\bigg(\frac{q^\mu}{q^2}\frac{[\slashed{q},\slashed{q}^\prime]}{q \cdot q^\prime} - \frac{[\gamma^\mu,\slashed{q}^\prime]}{q \cdot q^\prime} \bigg)
\big(\boldsymbol{\Sigma}_{-}\,T_a - \bar T_a\,\boldsymbol{\Sigma}_{-} \big)
\nonumber \\ &&
{}+ a_6 \frac{[\gamma^\mu,\slashed{q}]}{(q \cdot q^\prime)}
\big(\boldsymbol{\Sigma}_{-}\,T_a + \bar T_a \,\boldsymbol{\Sigma}_{-}\big) \,,
\end{eqnarray}
i.e., now with only $3$ free complex parameters $a_2$, $a_4$, $a_6$.

\subsubsection{Combined $\mathcal{CP}$ invariance}

Consider now the $\mathcal{CP}$ transformation of the vertex $\Gamma_{a}^\mu(p^\prime,p)$. Since the gauge interactions are always invariant under $\mathcal{CP}$, any $\mathcal{CP}$ violation of the vertex should be attributed only to the $\mathcal{CP}$ violation of $\boldsymbol{\Sigma}$:
\begin{eqnarray}
\label{gbm:invCP}
[\boldsymbol{\Sigma}]^{\mathcal{CP}} = \boldsymbol{\Sigma}
&\quad\Longrightarrow\quad&
{[\Gamma_{a}^\mu(p^\prime,p)]}^{\mathcal{CP}} = \Gamma_{a}^\mu(p^\prime,p) \,.
\end{eqnarray}
Not surprisingly, this condition is now satisfied by the vertex \eqref{gbm:Gmm:ansDiscSym} automatically, since we have already imposed the two conditions \eqref{gbm:invC} and \eqref{gbm:invP}. Indeed, using \eqref{gbp:Gmm:CP} we find the $\mathcal{CP}$ transformation of \eqref{gbm:Gmm:ansDiscSym} to be
\begin{eqnarray}
{[\Gamma_{a}^\mu(p^\prime,p)]}^{\mathcal{CP}} &=&
\gamma^\mu T_a
- \frac{1}{2}\frac{q^\mu}{q^2} \Big( [\boldsymbol{\Sigma}_{-}]^{\mathcal{CP}}\,T_a - \bar T_a\,[\boldsymbol{\Sigma}_{-}]^{\mathcal{CP}} \Big)
- \frac{1}{2}\frac{q^{\prime\mu}}{q \cdot q^\prime} \Big( [\boldsymbol{\Sigma}_{-}]^{\mathcal{CP}}\,T_a + \bar T_a\,[\boldsymbol{\Sigma}_{-}]^{\mathcal{CP}} \Big)
\nonumber \\ &&
{}+a_2\bigg(\frac{q^\mu}{q^2}-\frac{q^{\prime\mu}}{q \cdot q^\prime}\bigg)
\Big([\boldsymbol{\Sigma}_{-}]^{\mathcal{CP}}\,T_a + \bar T_a\,[\boldsymbol{\Sigma}_{-}]^{\mathcal{CP}} \Big)
\nonumber \\ &&
{}+ a_4\bigg(\frac{q^\mu}{q^2}\frac{[\slashed{q},\slashed{q}^\prime]}{q \cdot q^\prime} - \frac{[\gamma^\mu,\slashed{q}^\prime]}{q \cdot q^\prime} \bigg)
\Big([\boldsymbol{\Sigma}_{-}]^{\mathcal{CP}}\,T_a - \bar T_a\,[\boldsymbol{\Sigma}_{-}]^{\mathcal{CP}} \Big)
\nonumber \\ &&
{}+ a_6 \frac{[\gamma^\mu,\slashed{q}]}{(q \cdot q^\prime)}
\Big([\boldsymbol{\Sigma}_{-}]^{\mathcal{CP}}\,T_a + \bar T_a \,[\boldsymbol{\Sigma}_{-}]^{\mathcal{CP}}\Big)
\end{eqnarray}
(where $[\boldsymbol{\Sigma}]^{\mathcal{CP}}$ is given by \eqref{gbp:SgmCP}), which clearly satisfies the condition \eqref{gbm:invCP}.

\subsection{Hermiticity}

We have shown in Sec.~\ref{gbm:ssec:real} that in order to arrive at a real gauge boson mass matrix, the vertex must satisfy the Hermiticity condition $\Gamma_{a}^\mu(p^\prime,p) = \bar \Gamma_{a}^\mu(p,p^\prime)$, \eqref{gbm:hermGmm}. For the vertex of the form \eqref{gbm:Gmm:ansDiscSym} we have for $\bar \Gamma_{a}^\mu(p,p^\prime)$ explicitly
\begin{eqnarray}
\bar \Gamma_{a}^\mu(p,p^\prime) &=&
\gamma^\mu T_a
- \frac{1}{2}\frac{q^\mu}{q^2} \big( \boldsymbol{\Sigma}_{+}\,T_a - \bar T_a\,\boldsymbol{\Sigma}_{+} \big)
- \frac{1}{2}\frac{q^{\prime\mu}}{q \cdot q^\prime} \big( \boldsymbol{\Sigma}_{-}\,T_a + \bar T_a\,\boldsymbol{\Sigma}_{-} \big)
\nonumber \\ &&
{}+a_2^* \bigg(\frac{q^\mu}{q^2}-\frac{q^{\prime\mu}}{q \cdot q^\prime}\bigg)
\big(\boldsymbol{\Sigma}_{-}\,T_a + \bar T_a\,\boldsymbol{\Sigma}_{-} \big)
\nonumber \\ &&
{}+ a_4^* \bigg(\frac{q^\mu}{q^2}\frac{[\slashed{q},\slashed{q}^\prime]}{q \cdot q^\prime} - \frac{[\gamma^\mu,\slashed{q}^\prime]}{q \cdot q^\prime} \bigg)
\big(\boldsymbol{\Sigma}_{-}\,T_a - \bar T_a\,\boldsymbol{\Sigma}_{-} \big)
\nonumber \\ &&
{}+ a_6^* \frac{[\gamma^\mu,\slashed{q}]}{(q \cdot q^\prime)}
\big(\boldsymbol{\Sigma}_{-}\,T_a + \bar T_a \,\boldsymbol{\Sigma}_{-}\big) \,.
\end{eqnarray}
Comparing this with \eqref{gbm:Gmm:ansDiscSym} the requirement \eqref{gbm:hermGmm} of Hermiticity leads to the conclusion that the three free parameters $a_2$, $a_4$, $a_6$ in \eqref{gbm:Gmm:ansDiscSym} must be real:
\begin{eqnarray}
a_2\,, \,a_4\,, \,a_6 &\in& \mathbb{R} \,.
\end{eqnarray}

\subsection{The NG interpretation}
\label{gbm:vertexNGinterp}

\subsubsection{Ambiguity in extracting the NG part}

Let us now come back to the separation \eqref{gbm:Gmm:decomp} of the vertex $\Gamma_{a}^\mu(p^\prime,p)$ into the NG and regular part. Taking into account the form \eqref{gbm:Gmm:ansDiscSym} of the vertex we can write
\begin{eqnarray}
\Gamma_{a}^\mu(p^\prime,p)\big|_{\mathrm{NG}} &=& \frac{q^\mu}{q^2}\bigg[
- \frac{1}{2}\big( \boldsymbol{\Sigma}_{+}\,T_a - \bar T_a\,\boldsymbol{\Sigma}_{+} \big)
+ a_2 \big(\boldsymbol{\Sigma}_{-}\,T_a + \bar T_a\,\boldsymbol{\Sigma}_{-} \big)
+ a_4 \frac{[\slashed{q},\slashed{q}^\prime]}{q \cdot q^\prime} \big(\boldsymbol{\Sigma}_{-}\,T_a - \bar T_a\,\boldsymbol{\Sigma}_{-} \big)
\nonumber \\ &&
\phantom{\frac{q^\mu}{q^2}\bigg\{}
+ q^2 \Gamma_a(p^\prime,p)\big|_{\mathrm{amb.}}
\bigg] \,,
\\
\Gamma_{a}^\mu(p^\prime,p)\big|_{\mathrm{reg.}} &=&
\gamma^\mu T_a
-\Big(\frac{1}{2}+a_2\Big)\frac{q^{\prime\mu}}{q \cdot q^\prime} \big( \boldsymbol{\Sigma}_{-}\,T_a + \bar T_a\,\boldsymbol{\Sigma}_{-} \big)
- a_4 \frac{[\gamma^\mu,\slashed{q}^\prime]}{q \cdot q^\prime}
\big(\boldsymbol{\Sigma}_{-}\,T_a - \bar T_a\,\boldsymbol{\Sigma}_{-} \big)
\nonumber \\ && {}
+ a_6 \frac{[\gamma^\mu,\slashed{q}]}{q \cdot q^\prime}
\big(\boldsymbol{\Sigma}_{-}\,T_a + \bar T_a \,\boldsymbol{\Sigma}_{-}\big)
{} - q^\mu \Gamma_a(p^\prime,p)\big|_{\mathrm{amb.}} \,.
\end{eqnarray}
Notice the presence of the terms proportional to the quantity $\Gamma_a(p^\prime,p)|_{\mathrm{amb.}}$. Its aim is to parameterize the ambiguity in the identification of the NG part of the vertex, which we have already discussed in Sec.~\ref{gbm:ssec:MmntExp}. In principle, $\Gamma_a(p^\prime,p)|_{\mathrm{amb.}}$ can be apparently arbitrary, as the whole vertex $\Gamma_{a}^\mu(p^\prime,p)$ is independent of it.

However, we can determine the $\Gamma_a(p^\prime,p)|_{\mathrm{amb.}}$ in the same spirit as we have determined (so far) the whole vertex $\Gamma^\mu_a(p^\prime,p)$. That is to say, on top of the natural requirement that $\Gamma_a(p^\prime,p)|_{\mathrm{amb.}}$ is free of any kinematical singularities, we can assume it to be a linear combination of the terms \eqref{gbm:8terms}, to transform under $\group{G}$ according to \eqref{gbp:Gmm:transf}, to be invariant under $\mathcal{C}$ and $\mathcal{P}$ in the sense of \eqref{gbm:invC} and \eqref{gbm:invP}, respectively, and to satisfy the Hermiticity condition \eqref{gbm:hermGmm}. As a result we find $\Gamma_a(p^\prime,p)|_{\mathrm{amb.}}$ to be given by
\begin{eqnarray}
\Gamma_a(p^\prime,p)\big|_{\mathrm{amb.}}  &=& b \frac{1}{q \cdot q^\prime} \big(\boldsymbol{\Sigma}_{-}\,T_a - \bar T_a\,\boldsymbol{\Sigma}_{-} \big) \,,
\end{eqnarray}
where $b \in \mathbb{R}$ is its only free parameter. Thus, as advertised, the ambiguity of the separation \eqref{gbm:Gmm:decomp} is parameterized by only one real number.

\subsubsection{Unbroken gauge index}

Let us now focus on the NG part of the vertex, which we found to be
\begin{eqnarray}
\label{gbm:GmmNGpartPrel}
\Gamma_{a}^\mu(p^\prime,p)\big|_{\mathrm{NG}} &=& \frac{q^\mu}{q^2}\bigg[
- \frac{1}{2}\big( \boldsymbol{\Sigma}_{+}\,T_a - \bar T_a\,\boldsymbol{\Sigma}_{+} \big)
+ a_2 \big(\boldsymbol{\Sigma}_{-}\,T_a + \bar T_a\,\boldsymbol{\Sigma}_{-} \big)
\nonumber \\ &&
\phantom{\frac{q^\mu}{q^2}\bigg[}
{}+ \bigg(b \frac{q^2}{q \cdot q^\prime} + a_4 \frac{[\slashed{q},\slashed{q}^\prime]}{q \cdot q^\prime} \bigg)
\big(\boldsymbol{\Sigma}_{-}\,T_a - \bar T_a\,\boldsymbol{\Sigma}_{-} \big)
\bigg] \,.
\end{eqnarray}
The $\Gamma_{a}^\mu(p^\prime,p)|_{\mathrm{NG}}$ contains the bilinear coupling between the gauge boson $A_a^\mu$ and some linear combination of the NG bosons (this will be investigated in more detail in Sec.~\ref{sec:NGboson}). However, this linear combination is non-trivial if and only if the generator corresponding to $A_a^\mu$ is broken, i.e., when the quantity \eqref{gbp:Sgmnoninv} vanishes. Therefore we demand
\begin{eqnarray}
\label{gbm:noNGcuplDmnd}
\boldsymbol{\Sigma}\,T_a - \bar T_a\,\boldsymbol{\Sigma} \ = \ 0
&\quad\Longrightarrow\quad&  \Gamma_{a}^\mu(p^\prime,p)\big|_{\mathrm{NG}} \ = \ 0 \,.
\end{eqnarray}
If we did not demand this, it could be possible to generate masses for the gauge bosons corresponding to an unbroken subgroup. E.g., the photon would come out massive. That is why the requirement \eqref{gbm:noNGcuplDmnd} is so crucial. Upon its application to the NG part \eqref{gbm:GmmNGpartPrel} of the vertex we immediately find
\begin{eqnarray}
a_2  &=& 0 \,.
\end{eqnarray}
For the NG and the regular part of the vertex we thus have
\begin{subequations}
\begin{eqnarray}
\Gamma_{a}^\mu(p^\prime,p)\big|_{\mathrm{NG}} &=& \frac{q^\mu}{q^2}\bigg[
- \frac{1}{2}\big( \boldsymbol{\Sigma}_{+}\,T_a - \bar T_a\,\boldsymbol{\Sigma}_{+} \big)
+ \bigg(b \frac{q^2}{q \cdot q^\prime} + a_4 \frac{[\slashed{q},\slashed{q}^\prime]}{q \cdot q^\prime} \bigg)
\big(\boldsymbol{\Sigma}_{-}\,T_a - \bar T_a\,\boldsymbol{\Sigma}_{-} \big)
\bigg] \,,
\qquad\quad
\\
\Gamma_{a}^\mu(p^\prime,p)\big|_{\mathrm{reg.}} &=&
\gamma^\mu T_a
-\bigg(\frac{1}{2}\frac{q^{\prime\mu}}{q \cdot q^\prime}
- a_6 \frac{[\gamma^\mu,\slashed{q}]}{q \cdot q^\prime}\bigg)
\big(\boldsymbol{\Sigma}_{-}\,T_a + \bar T_a \,\boldsymbol{\Sigma}_{-}\big)
\nonumber \\ && {}
\phantom{\gamma^\mu T_a}
{}- \bigg(b \frac{q^\mu}{q \cdot q^\prime} + a_4 \frac{[\gamma^\mu,\slashed{q}^\prime]}{q \cdot q^\prime}\bigg)
\big(\boldsymbol{\Sigma}_{-}\,T_a - \bar T_a\,\boldsymbol{\Sigma}_{-} \big) \,,
\end{eqnarray}
\end{subequations}
and the full vertex reads
\begin{eqnarray}
\label{gbm:Gmm:ansNG}
\Gamma_{a}^\mu(p^\prime,p) &=& \gamma^\mu T_a
- \frac{1}{2}\frac{q^\mu}{q^2}\big( \boldsymbol{\Sigma}_{+}\,T_a - \bar T_a\,\boldsymbol{\Sigma}_{+} \big)
-\bigg(\frac{1}{2}\frac{q^{\prime\mu}}{q \cdot q^\prime}
- a_6 \frac{[\gamma^\mu,\slashed{q}]}{q \cdot q^\prime}\bigg)
\big(\boldsymbol{\Sigma}_{-}\,T_a + \bar T_a \,\boldsymbol{\Sigma}_{-}\big)
\nonumber \\ && {}
+ a_4 \bigg(\frac{q^\mu}{q^2}\frac{[\slashed{q},\slashed{q}^\prime]}{q \cdot q^\prime} - \frac{[\gamma^\mu,\slashed{q}^\prime]}{q \cdot q^\prime} \bigg)
\big(\boldsymbol{\Sigma}_{-}\,T_a - \bar T_a\,\boldsymbol{\Sigma}_{-} \big) \,.
\end{eqnarray}
Thus, the vertex $\Gamma_{a}^\mu(p^\prime,p)$, \eqref{gbm:Gmm:ansNG}, has now only two free real parameters.

\section{Gauge boson mass matrix}

The vertex of the form \eqref{gbm:Gmm:ansNG} is the best what we can obtain by imposing various requirements on the vertex alone. Now we will return to our ultimate task of calculating the gauge boson spectrum and things will soon start to be less elegant.

\subsection{Intermediate formula}

We are now going to calculate the gauge bosons mass matrix using the expression \eqref{gbm:Mab:prel}. The needed coefficients of the expansion \eqref{gbm:Gmmexp} of the vertex \eqref{gbm:Gmm:ansNG} read explicitly
\begin{subequations}
\label{gbm:coefABC}
\begin{eqnarray}
A_a(p) &=& -\boldsymbol{\Sigma}_{p}\,T_a + \bar T_a\,\boldsymbol{\Sigma}_{p} \,,
\\
B_a^\alpha(p) &=&
\Big(p^\alpha - 2 a_4 [\gamma^\alpha,\slashed{p}]\Big) \Big(-\boldsymbol{\Sigma}_{p}^\prime\,T_a + \bar T_a\,\boldsymbol{\Sigma}_{p}^\prime\Big) \,,
\\
C_a^\mu(p) &=& \gamma^\mu T_a
+ 2 a_4 [\gamma^\mu,\slashed{p}] \Big(-\boldsymbol{\Sigma}_{p}^\prime\,T_a + \bar T_a\,\boldsymbol{\Sigma}_{p}^\prime\Big)
- p^\mu \Big(\boldsymbol{\Sigma}_{p}^\prime\,T_a + \bar T_a\,\boldsymbol{\Sigma}_{p}^\prime\Big) \,.
\end{eqnarray}
\end{subequations}
Notice that of the two parameters $a_4$, $a_6$ of the vertex \eqref{gbm:Gmm:ansNG} the $a_6$ does not enter here, as it would enter only terms linear and higher in $q$ in the expansion \eqref{gbm:Gmmexp} of the vertex \eqref{gbm:Gmm:ansNG}. Thus, the gauge boson mass matrix in the pole approximation \eqref{gbm:PoleAppr} will depend only on $a_4$.

Upon plugging the coefficients \eqref{gbm:coefABC} into the formula \eqref{gbm:Mab:prel} and making some algebra we arrive at the gauge boson mass matrix $M^2_{ab}$ of the form\footnote{We will now use extensively the notation \eqref{symbols:comAB} in order to make the formul{\ae} more compact}
\begin{eqnarray}
\label{gbm:M2ab:intermed}
M^2_{ab} &=& \I \frac{1}{2} \int\!\frac{\d^d p}{(2\pi)^d}\,\Tr\bigg\{
2\boldsymbol{D}_L \, \bbl \boldsymbol{\Sigma}, T_a  \bbr \, \boldsymbol{D}_R \, \bbl \boldsymbol{\Sigma}^\dag, \bar T_b \bbr
\nonumber \\ &&
\phantom{\I \frac{1}{2} \int\!\frac{\d^d p}{(2\pi)^d}\,\Tr\bigg\{}
+ 2 \frac{2}{d}p^2 \Big( \boldsymbol{D}_R^\prime \, \bbl \boldsymbol{\Sigma}^\dag, \bar T_a  \bbr \, \boldsymbol{D}_L \, \bbl \boldsymbol{\Sigma}, T_b  \bbr + \boldsymbol{D}_R^\prime \, \bbl \boldsymbol{\Sigma}^\dag, \bar T_b  \bbr \, \boldsymbol{D}_L \, \bbl \boldsymbol{\Sigma}, T_a  \bbr \Big)
\nonumber \\ &&
\phantom{\I \frac{1}{2} \int\!\frac{\d^d p}{(2\pi)^d}\,\Tr\bigg\{}
+ \frac{1}{2}\big(1+4(d-1)a_4\big)  \frac{4}{d}p^2   \boldsymbol{D}_L \, \bbl \boldsymbol{\Sigma}^\prime, T_a  \bbr \, \boldsymbol{D}_R \, \bbl \boldsymbol{\Sigma}^\dag, \bar T_b  \bbr
\nonumber \\ &&
\phantom{\I \frac{1}{2} \int\!\frac{\d^d p}{(2\pi)^d}\,\Tr\bigg\{}
+ \frac{4}{d}p^2  \bbl \boldsymbol{\Sigma}, T_a  \bbr
\bbl \boldsymbol{D}_R \, T_b \, \boldsymbol{\Sigma}^\dag , \boldsymbol{D}_L^\prime \bbr
\bigg\} \,,
\end{eqnarray}
where $\boldsymbol{D}_L$, $\boldsymbol{D}_R$ are given by \eqref{gbp:DLDR}. This can be expressed as a sum of the symmetric and antisymmetric part:
\begin{eqnarray}
M^2_{ab} &=& M^2_{ab}\big|_{\mathrm{S}} + M^2_{ab}\big|_{\mathrm{A}} \,,
\end{eqnarray}
where
\begin{eqnarray}
M^2_{ab}\big|_{\mathrm{S}} &=& \I \frac{1}{2} \int\!\frac{\d^d p}{(2\pi)^d}\,\Tr\bigg\{
2\boldsymbol{D}_L \, \bbl \boldsymbol{\Sigma}, T_a  \bbr \, \boldsymbol{D}_R \, \bbl \boldsymbol{\Sigma}^\dag, \bar T_b \bbr
\nonumber \\ &&
\hspace{0.5em}
{}+ \frac{1}{2}\big(1+4(d-1)a_4\big)  \frac{2}{d}p^2 \Big(  \boldsymbol{D}_L \, \bbl \boldsymbol{\Sigma}^\prime, T_a  \bbr \, \boldsymbol{D}_R \, \bbl \boldsymbol{\Sigma}^\dag, \bar T_b  \bbr + \boldsymbol{D}_L \, \bbl \boldsymbol{\Sigma}^\prime, T_b  \bbr \, \boldsymbol{D}_R \, \bbl \boldsymbol{\Sigma}^\dag, \bar T_a  \bbr \Big)
\nonumber \\ &&
\hspace{7.84em}
{}+2 \frac{2}{d}p^2 \Big( \boldsymbol{D}_R^\prime \, \bbl \boldsymbol{\Sigma}^\dag, \bar T_a  \bbr \, \boldsymbol{D}_L \, \bbl \boldsymbol{\Sigma}, T_b  \bbr + \boldsymbol{D}_R^\prime \, \bbl \boldsymbol{\Sigma}^\dag, \bar T_b  \bbr \, \boldsymbol{D}_L \, \bbl \boldsymbol{\Sigma}, T_a  \bbr \Big)
\nonumber \\ &&
\hspace{0.5em}
{}+ \frac{2}{d}p^2 \Big[ \bbl \boldsymbol{\Sigma}, T_a  \bbr
\bbl \boldsymbol{D}_R \, T_b \, \boldsymbol{\Sigma}^\dag , \boldsymbol{D}_L^\prime \bbr
+ \bbl \boldsymbol{\Sigma}, T_b  \bbr
\bbl \boldsymbol{D}_R \, T_a \, \boldsymbol{\Sigma}^\dag , \boldsymbol{D}_L^\prime \bbr \Big]
\bigg\}
\,,
\label{gbm:M2abS}
\\
M^2_{ab}\big|_{\mathrm{A}} &=& \I \frac{1}{2} \int\!\frac{\d^d p}{(2\pi)^d}\,\Tr\bigg\{
\frac{2}{d}p^2 \Big[ \bbl \boldsymbol{\Sigma}, T_a  \bbr
\bbl \boldsymbol{D}_R \, T_b \, \boldsymbol{\Sigma}^\dag , \boldsymbol{D}_L^\prime \bbr
- \bbl \boldsymbol{\Sigma}, T_b  \bbr
\bbl \boldsymbol{D}_R \, T_a \, \boldsymbol{\Sigma}^\dag , \boldsymbol{D}_L^\prime \bbr \Big]
\nonumber \\ &&
\hspace{0.5em}
{}+
\frac{1}{2}\big(1+4(d-1)a_4\big)  \frac{2}{d}p^2 \Big(  \boldsymbol{D}_L \, \bbl \boldsymbol{\Sigma}^\prime, T_a  \bbr \, \boldsymbol{D}_R \, \bbl \boldsymbol{\Sigma}^\dag, \bar T_b  \bbr - \boldsymbol{D}_L \, \bbl \boldsymbol{\Sigma}^\prime, T_b  \bbr \, \boldsymbol{D}_R \, \bbl \boldsymbol{\Sigma}^\dag, \bar T_a  \bbr \Big)
\bigg\} \,.
\nonumber \\ &&
\label{gbm:M2abA}
\end{eqnarray}

\subsection{Requirement of symmetricity}
\label{gbm:ssec:symm}

We require that the gauge boson mass matrix be symmetric:
\begin{eqnarray}
\label{gbm:Masym=0}
M^2_{ab}\big|_{\mathrm{A}} &=& 0 \,.
\end{eqnarray}
Notice that the mass matrix $M^2_{ab}$ and, in particular, also the antisymmetric part $M^2_{ab}|_{\mathrm{A}}$ depend on the free parameter $a_4$. Being experienced from the previous process of deriving the vertex, one might think that now it suffices just to set $a_4$ to some suitable value in order to fulfil \eqref{gbm:Masym=0}.

However, it turns out that for general\footnote{Of course, we do not want $a_4$ to depend on particular details of the theory (i.e., the gauge group $\group{G}$ and the fermion representations and self-energies), but rather to have the same value of $a_4$ for all possible theories. Otherwise it would be certainly possible to make $M^2_{ab}|_{\mathrm{A}}$ vanishing by tuning $a_4$ appropriately.} fermion setting (given by the self-energy $\boldsymbol{\Sigma}$ and the generators $T_a$, $\bar T_a$) it is just not possible to find such a value of $a_4$. It is a pathological feature of the present scheme (defined both by the vertex Ansatz \eqref{gbm:Gmm:ansNG} and by the loop diagram \eqref{gbm:Pimunuab} for the polarization tensor) that the resulting gauge boson mass matrix does not come out symmetric.

Nevertheless, there is a good news. It turns out that in all of the applications of interest the quantity on the first line of the expression \eqref{gbm:M2abA} for $M^2_{ab}|_{\mathrm{A}}$ \qm{miraculously} vanishes:
\begin{eqnarray}
\label{gbm:Aab=0}
A_{ab} &=& 0 \,,
\end{eqnarray}
where we introduced for the sake of later references the denotation
\begin{subequations}
\label{gbm:Aamdef}
\begin{eqnarray}
A_{ab} &\equiv& \Tr \bigg\{
\bbl \boldsymbol{\Sigma}, T_a  \bbr
\bbl \boldsymbol{D}_R \, T_b \, \boldsymbol{\Sigma}^\dag , \boldsymbol{D}_L^\prime \bbr
- \bbl \boldsymbol{\Sigma}, T_b  \bbr
\bbl \boldsymbol{D}_R \, T_a \, \boldsymbol{\Sigma}^\dag , \boldsymbol{D}_L^\prime \bbr
\bigg\}
\label{gbm:Aamdefkratsi}
\\
&=& \Tr \bigg\{
\phantom{+\,}
T_a\,\boldsymbol{\Sigma}^\dag\,\boldsymbol{D}_L^\prime\,\bar T_b\,\boldsymbol{\Sigma}\,\boldsymbol{D}_R -
T_a\,\boldsymbol{\Sigma}^\dag\,\boldsymbol{D}_L\,\bar T_b\,\boldsymbol{\Sigma}\,\boldsymbol{D}_R^\prime
\nonumber \\ &&
\phantom{\Tr \bigg\{}
\hspace{-0.45em}
+\,
\bar T_a\,\boldsymbol{D}_L\,\boldsymbol{\Sigma}\,T_b\,\boldsymbol{D}_R^\prime\,\boldsymbol{\Sigma}^\dag -
\bar T_a\,\boldsymbol{D}_L^\prime\,\boldsymbol{\Sigma}\,T_b\,\boldsymbol{D}_R\,\boldsymbol{\Sigma}^\dag
\bigg\} \,.
\label{gbm:Aamdefdelsi}
\end{eqnarray}
\end{subequations}
In particular, this happens in both the Abelian toy model and the electroweak interactions; we will show it in detail in the respective chapters~\ref{chp:ablgg} and \ref{ewM} when discussing the gauge boson masses in these models. Thus, we will from now assume that the condition \eqref{gbm:Aab=0} does hold.

Now, when we assume the condition \eqref{gbm:Aab=0}, the situation greatly improves. It obviously suffices to set
\begin{eqnarray}
\label{gbm:a4}
a_4 &=& - \frac{1}{4} \frac{1}{d-1}
\end{eqnarray}
in order to fulfil the condition \eqref{gbm:Masym=0} by eliminating the term in $M^2_{ab}|_{\mathrm{A}}$, \eqref{gbm:M2abA}, proportional to $1+4(d-1)a_4$. In fact, this elimination by setting \eqref{gbm:a4} is in fact necessary, as the term in question does not vanish in some applications of interest, unlike the term $A_{ab}$, \eqref{gbm:Aamdef}.


At this point we can finally briefly comment on why we have not considered the scalar contribution to the gauge boson mass matrix. If we considered the scalars, we would arrive at the vertex of the same form as the fermion vertex \eqref{gbm:Gmm:ansNG}, but this time without gamma matrix structure, which is in \eqref{gbm:Gmm:ansNG} parameterized by the parameters $a_4$ $a_6$. By properly adjusting one of these parameters ($a_4$) we were able to make the fermion contribution to the gauge boson mass matrix symmetric (at least in the cases satisfying the condition \eqref{gbm:Masym=0}). However, for the scalars this is not possible, simply because the scalar vertex is free of any free tunable parameters. This inability of making the scalar contribution to the gauge boson mass matrix symmetric is the reason why we neglect the scalars.

\subsection{The final formula}

Thus, under setting \eqref{gbm:a4} of $a_4$ and under the assumption \eqref{gbm:Aab=0} we arrive at the final expression for the gauge boson mass matrix:
\begin{eqnarray}
\label{gbm:M2ab}
M^2_{ab} &=& \I \frac{1}{2} \int\!\frac{\d^d p}{(2\pi)^d}\,\Tr\bigg\{
2\boldsymbol{D}_L \, \bbl \boldsymbol{\Sigma}, T_a  \bbr \, \boldsymbol{D}_R \, \bbl \boldsymbol{\Sigma}^\dag, \bar T_b \bbr
\nonumber \\ &&
\phantom{\I \frac{1}{2} \int\!\frac{\d^d p}{(2\pi)^d}\,\Tr\bigg\{}
+ 2 \frac{2}{d}p^2 \Big( \boldsymbol{D}_R^\prime \, \bbl \boldsymbol{\Sigma}^\dag, \bar T_a  \bbr \, \boldsymbol{D}_L \, \bbl \boldsymbol{\Sigma}, T_b  \bbr + \boldsymbol{D}_R^\prime \, \bbl \boldsymbol{\Sigma}^\dag, \bar T_b  \bbr \, \boldsymbol{D}_L \, \bbl \boldsymbol{\Sigma}, T_a  \bbr \Big)
\nonumber \\ &&
\phantom{\I \frac{1}{2} \int\!\frac{\d^d p}{(2\pi)^d}\,\Tr\bigg\{}
+ \frac{2}{d}p^2 \Big( 
\bbl \boldsymbol{\Sigma}, T_a  \bbr
\bbl \boldsymbol{D}_R \, T_b \, \boldsymbol{\Sigma}^\dag , \boldsymbol{D}_L^\prime \bbr
+ \bbl \boldsymbol{\Sigma}, T_b  \bbr
\bbl \boldsymbol{D}_R \, T_a \, \boldsymbol{\Sigma}^\dag , \boldsymbol{D}_L^\prime \bbr  \Big)
\bigg\} \,.
\nonumber \\ &&
\end{eqnarray}
Let us summarize some of the features of $M^2_{ab}$, given by \eqref{gbm:M2ab}:
\begin{itemize}
  \item It is real.
  \item It is symmetric.
  \item Its signature can be, depending on the self-energy $\boldsymbol{\Sigma}$, virtually arbitrary. (I.e., in particular the positive definiteness is not guaranteed.)
  \item The element $M^2_{ab}$ vanishes if at least one of the generators $T_a$ and $T_b$ is unbroken in the sense of \eqref{gbp:Sgmnoninv} (in the first two lines in \eqref{gbm:M2ab} it can be seen directly, while for the last line one has to utilize the condition \eqref{gbm:Aab=0}).
  \item It is free of any undetermined parameters and thus in this sense unique.
  \item It is UV-finite, as long as the self-energy $\boldsymbol{\Sigma}$ is suppressed at high momenta.
\end{itemize}
Also recall that the formula \eqref{gbm:M2ab} for $M^2_{ab}$ is applicable only under the condition \eqref{gbm:Aab=0}.

For the sake of later references let us also state the vertex $\Gamma_{a}^\mu(p^\prime,p)$ with the parameter $a_4$ determined as \eqref{gbm:a4}:
\begin{eqnarray}
\label{gbm:Gmm:ans}
\Gamma_{a}^\mu(p^\prime,p) &=& \gamma^\mu T_a
- \frac{1}{2}\frac{q^\mu}{q^2}\big( \boldsymbol{\Sigma}_{+}\,T_a - \bar T_a\,\boldsymbol{\Sigma}_{+} \big)
-\bigg(\frac{1}{2}\frac{q^{\prime\mu}}{q \cdot q^\prime}
- a_6 \frac{[\gamma^\mu,\slashed{q}]}{q \cdot q^\prime}\bigg)
\big(\boldsymbol{\Sigma}_{-}\,T_a + \bar T_a \,\boldsymbol{\Sigma}_{-}\big)
\nonumber \\ && {}
- \frac{1}{4} \frac{1}{d-1} \bigg(\frac{q^\mu}{q^2}\frac{[\slashed{q},\slashed{q}^\prime]}{q \cdot q^\prime} - \frac{[\gamma^\mu,\slashed{q}^\prime]}{q \cdot q^\prime} \bigg)
\big(\boldsymbol{\Sigma}_{-}\,T_a - \bar T_a\,\boldsymbol{\Sigma}_{-} \big) \,.
\end{eqnarray}
Notice that it still depends on one real parameter, $a_6$, which we nevertheless leave undetermined.\footnote{Cf.~footnote~\ref{ftnt:almost} on page~\pageref{ftnt:almost}.} It could be presumably determined in an analogous way as $a_4$, i.e., by requiring that the whole $\Pi_{ab}(q^2)$, not only the lowest order of its Laurent series (i.e., the $M_{ab}^2$), be symmetric (under the condition \eqref{gbm:Aab=0}). The vertex \eqref{gbm:Gmm:ans} can be divided into the NG part and the regular part as
\begin{subequations}
\begin{eqnarray}
\Gamma_{a}^\mu(p^\prime,p)\big|_{\mathrm{NG}} &=& \frac{q^\mu}{q^2}\bigg[
- \frac{1}{2}\big( \boldsymbol{\Sigma}_{+}\,T_a - \bar T_a\,\boldsymbol{\Sigma}_{+} \big)
+ \bigg(b \frac{q^2}{q \cdot q^\prime} - \frac{1}{4} \frac{1}{d-1} \frac{[\slashed{q},\slashed{q}^\prime]}{q \cdot q^\prime} \bigg)
\big(\boldsymbol{\Sigma}_{-}\,T_a - \bar T_a\,\boldsymbol{\Sigma}_{-} \big)
\bigg] \,,
\nonumber \\ &&
\label{gbm:Gmm:ans:NG}
\\
\Gamma_{a}^\mu(p^\prime,p)\big|_{\mathrm{reg.}} &=&
\gamma^\mu T_a
-\bigg(\frac{1}{2}\frac{q^{\prime\mu}}{q \cdot q^\prime}
- a_6 \frac{[\gamma^\mu,\slashed{q}]}{q \cdot q^\prime}\bigg)
\big(\boldsymbol{\Sigma}_{-}\,T_a + \bar T_a \,\boldsymbol{\Sigma}_{-}\big)
\nonumber \\ &&
\phantom{\gamma^\mu T_a}
- \bigg(b \frac{q^\mu}{q \cdot q^\prime} - \frac{1}{4} \frac{1}{d-1} \frac{[\gamma^\mu,\slashed{q}^\prime]}{q \cdot q^\prime}\bigg)
\big(\boldsymbol{\Sigma}_{-}\,T_a - \bar T_a\,\boldsymbol{\Sigma}_{-} \big) \,.
\end{eqnarray}
\end{subequations}
Again, the real free parameter $b$, parameterizing this separation, could be presumably determined by insisting  on the symmetricity of the contribution of only the NG bosons to the polarization tensor for all $q$ (see Eq.~\eqref{gbm:PiNGcontrib} below).

\section{Nambu--Goldstone boson interpretation}
\label{sec:NGboson}

\subsection{Introduction}


Since the symmetry $\group{G}$ is by assumption spontaneously broken to some subgroup $\group{H} \subseteq \group{G}$, we expect appearance of the corresponding NG bosons -- composite spin-$0$ massless particles. The number of the NG bosons is given as $N_{\mathrm{NG}} = N_{\group{G}} - N_{\group{H}}$, where $N_{\group{G}}$, $N_{\group{H}}$ are dimensions (numbers of generators) of $\group{G}$, $\group{H}$, respectively. We will denote the NG bosons as $\pi_A$, $A=1,\ldots,N_{\mathrm{NG}}$.

\begin{figure}[t]
\begin{center}
\includegraphics[width=0.55\textwidth]{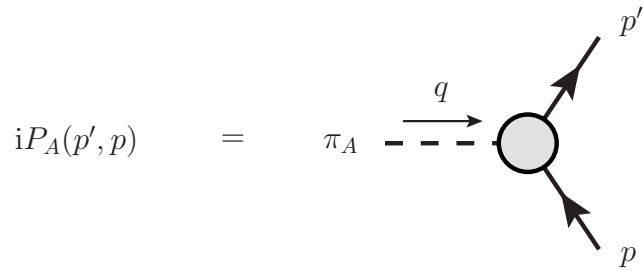}
\caption[NG boson effective vertex $\langle \pi_A \psi \bar\psi \rangle_{\mathrm{1PI}} = \I P_A(p^\prime,p)$.]{Assignment of momenta of the effective vertex $P_A(p^\prime,p)$, connecting the NG boson $\pi_A$ with fermions. Momentum conservation $q = p^\prime - p$ is implied.}
\label{gbm:fig:PA}
\end{center}
\end{figure}

As the current \eqref{gbp:jamu:psi}, corresponding to the broken symmetry $\group{G}$, is made of the fermion fields, the NG bosons are composites of the fermions and there will a direct coupling between the NG bosons and the fermion--antifermion pairs. These couplings can be parameterized by an effective vertex $P_A(p^\prime,p)$, see Fig.~\ref{gbm:fig:PA} for assignment of the momenta.

\begin{figure}[t]
\begin{center}
\includegraphics[width=0.65\textwidth]{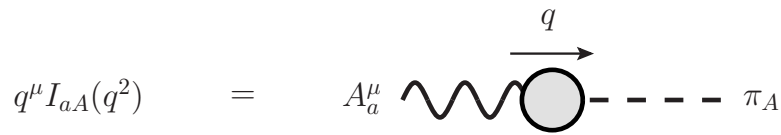}
\caption[NG boson bilinear coupling $\langle A_a^\mu \pi_A \rangle_{\mathrm{1PI}} = q^\mu I_{aA}(q^2)$.]{The bilinear coupling $q^\mu I_{aA}(q^2)$ between the gauge boson $A_a^\mu$ and the NG boson $\pi_A$.}
\label{gbm:fig:IaA}
\end{center}
\end{figure}

\begin{figure}[t]
\begin{center}
\includegraphics[width=0.7\textwidth]{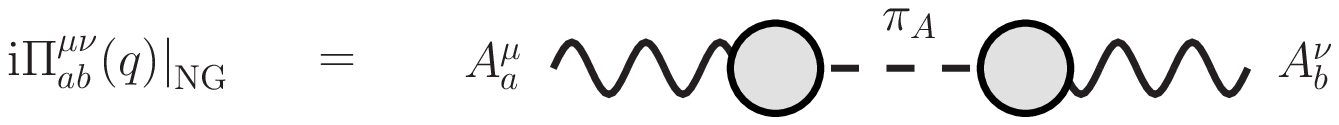}
\caption[NG boson contribution to $\Pi_{ab}^{\mu\nu}(q)$.]{The contribution $\Pi_{ab}^{\mu\nu}(q)|_{\mathrm{NG}}$ of the NG bosons to the polarization tensor, induced by the bilinear couplings $I_{aA}(q^2)$.}
\label{gbm:PiNG}
\end{center}
\end{figure}

If the broken symmetry were global, the NG bosons would be seen in the spectrum as normal particles (asymptotic states) and they would interact with fermions via the vertices $P_A(p^\prime,p)$. However, even if the symmetry is gauged and the NG bosons decouple from the spectrum, the couplings $P_A(p^\prime,p)$ play an important r\^{o}le: They induce the necessary bilinear couplings of the gauge bosons and NG bosons, which can be only a loop effect, since the two types of bosons do not couple directly. Due to Lorentz invariance the bilinear coupling of the NG bosons $\pi_A$ and the gauge bosons $A_a^\mu$ has the most general form $q^\mu I_{aA}(q^2)$, see Fig.~\eqref{gbm:fig:IaA}. The bilinear couplings are of course crucial for decoupling of the NG boson from the spectrum and generating the gauge boson masses. In fact, the contribution of the NG bosons to the polarization tensor is in terms of the bilinear coupling $q^\mu I_{aA}(q^2)$ given by
\begin{eqnarray}
\label{gbm:PiNGcontrib}
\I \Pi_{ab}^{\mu\nu}(q) \big|_{\mathrm{NG}} &=& - \I \frac{q^\mu q^\nu}{q^2} I_{aA}(q^2)\,I_{bA}(q^2) \,,
\end{eqnarray}
see Fig.~\ref{gbm:PiNG}, i.e., the NG bosons contribute only to the longitudinal part of the polarization tensor.


\subsection{Decomposition of $\Gamma_{a}^\mu(p^\prime,p)$}


The starting point of our analysis is the full vertex $\Gamma_{a}^\mu(p^\prime,p)$. As already briefly discussed in previous sections, among other contributions to it there are also contributions from the NG bosons and the vertex $\Gamma_{a}^\mu(p^\prime,p)$ can be therefore decomposed as \eqref{gbm:Gmm:decompgen}. We have already stated the general form \eqref{gbm:Gmm:decompNG} of the NG part $\Gamma_{a}^\mu(p^\prime,p)|_{\mathrm{NG}}$,
\begin{eqnarray}
\label{gbm:GmmNGbasic}
\Gamma_{a}^\mu(p^\prime,p)\big|_{\mathrm{NG}}
&=& \frac{q^\mu}{q^2} \Gamma_{a}(p^\prime,p)\big|_{\mathrm{NG}} \,,
\end{eqnarray}
which was in this form so far sufficient for our purposes. Recall that under our approximation scheme we have explicitly found
\begin{eqnarray}
\label{gbm:Gmm:NG}
\Gamma_{a}(p^\prime,p)\big|_{\mathrm{NG}} &=&
- \frac{1}{2}\big( \boldsymbol{\Sigma}_{+}\,T_a - \bar T_a\,\boldsymbol{\Sigma}_{+} \big)
+ \bigg(b \frac{q^2}{q \cdot q^\prime} - \frac{1}{4} \frac{1}{d-1} \frac{[\slashed{q},\slashed{q}^\prime]}{q \cdot q^\prime} \bigg)
\big(\boldsymbol{\Sigma}_{-}\,T_a - \bar T_a\,\boldsymbol{\Sigma}_{-} \big) \,,
\qquad\qquad
\end{eqnarray}
where $b$ is some real undetermined constant, see \eqref{gbm:Gmm:ans:NG}.

\begin{figure}[t]
\begin{center}
\includegraphics[width=0.65\textwidth]{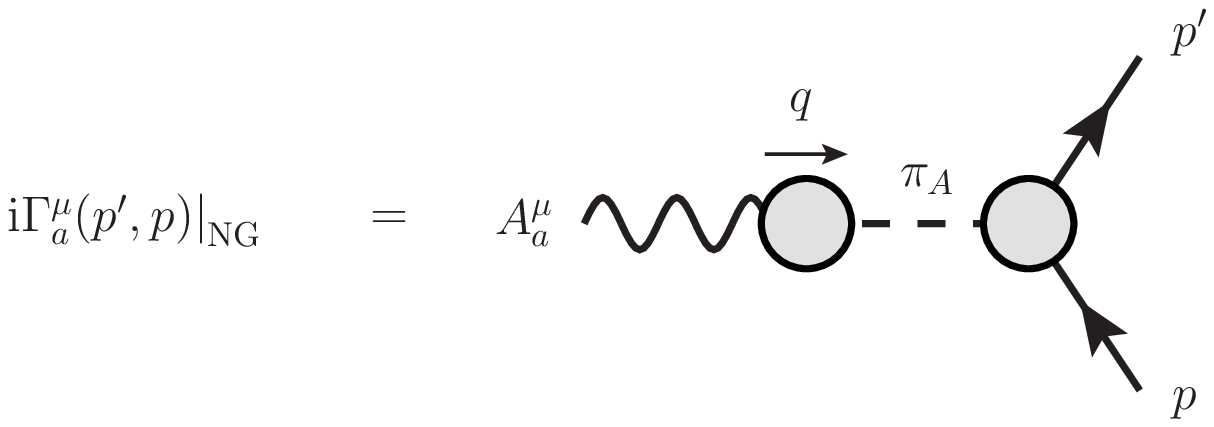}
\caption[NG boson part of $\Gamma_{a}^\mu(p^\prime,p)$.]{The \qm{anatomy} \eqref{gbm:GmmNGanat} of the NG part $\Gamma_{a}^\mu(p^\prime,p)|_{\mathrm{NG}}$ of the vertex.}
\label{gbm:fig:GmmNGanat}
\end{center}
\end{figure}

Now, when we have introduced the bilinear coupling of the NG and gauge bosons, we can investigate the \qm{anatomy} of $\Gamma_{a}^\mu(p^\prime,p)|_{\mathrm{NG}}$ in more detail. It turns out that $\Gamma_{a}^\mu(p^\prime,p)|_{\mathrm{NG}}$ can be expressed as (see also Fig.~\ref{gbm:fig:GmmNGanat})
\begin{eqnarray}
\label{gbm:GmmNGanat}
\Gamma_{a}^\mu(p^\prime,p)\big|_{\mathrm{NG}}
&=& q^\mu I_{aA}(q^2) \frac{\I\delta_{AB}}{q^2}\,P_B(p^\prime,p) \,,
\end{eqnarray}
where $q^\mu I_{aA}(q^2)$ is the bilinear coupling of the gauge boson $A^\mu_a$ to the NG boson $\pi_A$, $\I\delta_{AB}/q^2 = \langle \pi_A \pi_B \rangle$ is the propagator of the NG bosons and finally $P_B(p^\prime,p)$ is the coupling of the NG boson $\pi_B$ to the fermions.

\subsection{Expression for $P_A(p^\prime,p)$}

We will now investigate how to express the NG effective vertex $P_A(p^\prime,p)$ in terms of $\Gamma_{a}^\mu(p^\prime,p)|_{\mathrm{NG}}$ and $I_{aA}(q^2)$, which may be useful in situations when the latter two quantities are known.

Comparing the two expressions \eqref{gbm:GmmNGbasic} and \eqref{gbm:GmmNGanat} for $\Gamma_{a}^\mu(p^\prime,p)|_{\mathrm{NG}}$, we obtain the equation
\begin{eqnarray}
\label{gbm:Ngeqindices}
I_{aA}(q^2) \, P_A(p^\prime,p) &=& -\I\,\Gamma_{a}(p^\prime,p)\big|_{\mathrm{NG}} \,.
\end{eqnarray}
In the following it will be more convenient to suppress the gauge boson ($a$) as well as the NG boson ($A$) indices and utilize instead the matrix form. The equation \eqref{gbm:Ngeqindices} in the matrix formalism then reads
\begin{eqnarray}
\label{gbm:Ngeqmatrix}
I(q^2) \, P(p^\prime,p) &=& -\I\,\Gamma(p^\prime,p)\big|_{\mathrm{NG}} \,.
\end{eqnarray}
Now we would like to extract $P(p^\prime,p)$ from this equation. However, we cannot multiply the equation \eqref{gbm:Ngeqmatrix} by the inverse matrix of $I(q^2)$, simply because it may not in general exist: Recall that $I(q^2)$ is after all, in general, a rectangular matrix and hence singular. However, we can do the following: We can multiply the equation \eqref{gbm:Ngeqmatrix} from left with $I^\T(q^2)$ (i.e., contract the equation \eqref{gbm:Ngeqindices} with $I_{aB}(q^2)$) to arrive at
\begin{eqnarray}
I^\T(q^2) \, I(q^2) \, P(p^\prime,p) &=& -\I\,I^\T(q^2) \, \Gamma(p^\prime,p)\big|_{\mathrm{NG}} \,.
\end{eqnarray}
Recall that the matrix $I(q^2)$ is $N_{\group{G}} \times N_{\mathrm{NG}}$, with $N_{\mathrm{NG}} \leq N_{\group{G}}$. Assume now that the rank of $I(q^2)$ is the maximal possible, i.e., $N_{\mathrm{NG}}$, and assume this \emph{for all} $q^2$. Then the matrix $I^\T(q^2)\,I(q^2)$, which is $N_{\mathrm{NG}} \times N_{\mathrm{NG}}$, has the rank $N_{\mathrm{NG}}$ and therefore is regular and invertible. We can therefore finally express the effective NG vertex $P_A(p^\prime,p)$ as
\begin{eqnarray}
\label{gbm:PNG}
P(p^\prime,p)
&=& -\I \big[ I^\T(q^2) \, I(q^2) \big]^{-1} I^\T(q^2) \, \Gamma(p^\prime,p)\big|_{\mathrm{NG}} \,,
\end{eqnarray}
or in components,
\begin{eqnarray}
P_A(p^\prime,p)
&=& -\I \Big(\big[ I^\T(q^2) \, I(q^2) \big]^{-1}\Big)_{\!AB} I_{aB}(q^2) \, \Gamma_a(p^\prime,p)\big|_{\mathrm{NG}} \,.
\end{eqnarray}


\subsection{Loop expression for $I_{aA}(q^2)$}

\begin{figure}[t]
\begin{center}
\includegraphics[width=0.78\textwidth]{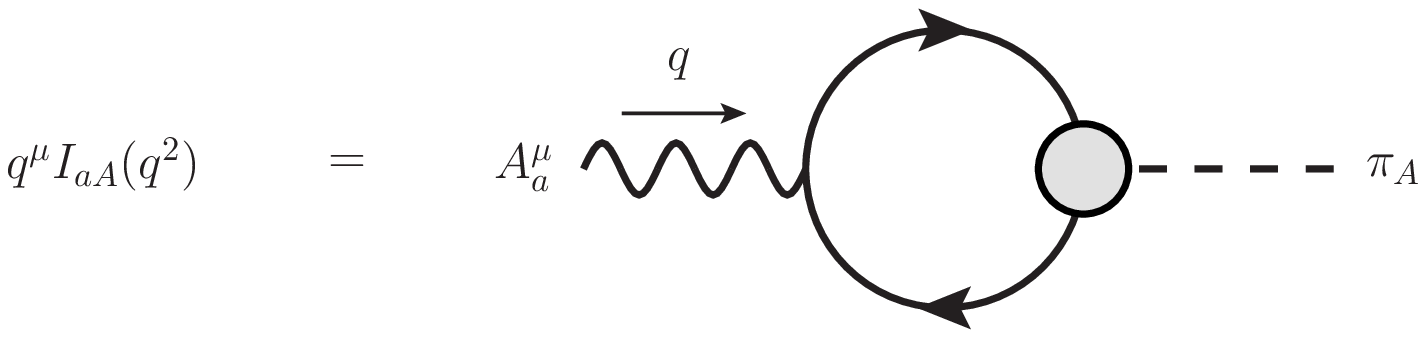}
\caption[One-loop fermion contribution to $q^\mu I_{aA}(q^2)$.]{The one-loop contribution \eqref{gbm:IaAoneloop} to the bilinear coupling $q^\mu I_{aA}(q^2)$ between the gauge boson $A_a^\mu$ and the NG boson $\pi_A$.}
\label{gbm:fig:IaAoneloop}
\end{center}
\end{figure}

We asserted above that the bilinear coupling $I_{aA}(q^2)$ between the gauge and the NG bosons is a loop effect. Let us now check this issue closer. Recall that the NG contribution to the polarization tensor is in terms of $I_{aA}(q^2)$ given by \eqref{gbm:PiNGcontrib}. Taking into account the loop integral \eqref{gbm:Pimunuab} for the polarization tensor we see that the same NG contribution should be given by the NG part of the vertex, $\Gamma^\mu_a(p+q,p)|_{\mathrm{NG}}$:
\begin{eqnarray}
\I \Pi_{ab}^{\mu\nu}(q)\big|_{\mathrm{NG}} &=& - \int\!\frac{\d^d p}{(2\pi)^d}\,
\Tr \Big\{ \Gamma^\mu_a(p+q,p)\big|_{\mathrm{NG}} \, G(p) \, \gamma^\nu T_b \, G(p+q) \Big\} \,.
\end{eqnarray}
If we substitute the expression \eqref{gbm:GmmNGanat} for $\Gamma^\mu_a(p+q,p)|_{\mathrm{NG}}$ in it and compare the resulting integral with the other expression \eqref{gbm:PiNGcontrib} for $\I \Pi_{ab}^{\mu\nu}(q)|_{\mathrm{NG}}$, we obtain an equation containing only the two NG vertices $I_{aA}(q^2)$ and $P_A(p^\prime,p)$. After some manipulation (involving the multiplication of the equation by $I^\T(q^2)$ from left in order to obtain the invertible matrix $I^\T(q^2)\,I(q^2)$ as above) we arrive at the explicit expression for $I_{aA}(q^2)$ in terms of $P_A(p^\prime,p)$:
\begin{eqnarray}
\label{gbm:IaAoneloop}
q^\mu I_{aA}(q^2) &=& -\int\!\frac{\d^d p}{(2\pi)^d} \, \Tr\!\Big\{ \gamma^\mu T_a \, G(p-q) \, P_A(p-q,p) \, G(p) \Big\} \,.
\end{eqnarray}
This is diagrammatically depicted in Fig.~\ref{gbm:fig:IaAoneloop}.

\subsection{Lowest orders in $q$}

Recall now the expansion \eqref{gbm:Gmmexp} of the vertex $\Gamma_{a}^\mu(p^\prime,p)$ in the lowest powers of $q$ and the identification \eqref{gbm:GmmNGexp} of $\Gamma^\mu_a(p^\prime,p)|_{\mathrm{NG}}$, corresponding to
\begin{eqnarray}
\label{gbm:Gmm:NGexp}
\Gamma_{a}(p^\prime,p)\big|_{\mathrm{NG}} &=& A_a(p) + q_\alpha B_a^\alpha(p) + \mathcal{O}(q^2) \,.
\end{eqnarray}
The explicit form \eqref{gbm:Gmm:NG} of $\Gamma_a(p^\prime,p)|_{\mathrm{NG}}$ yields
\begin{subequations}
\begin{eqnarray}
A_a(p) &=& - \big( \boldsymbol{\Sigma}\,T_a - \bar T_a\,\boldsymbol{\Sigma} \big) \,,
\\
B_a^\alpha(p) &=& - v^\alpha \big(\boldsymbol{\Sigma}^\prime\,T_a - \bar T_a\,\boldsymbol{\Sigma}^\prime \big) \,,
\end{eqnarray}
\end{subequations}
where we defined
\begin{eqnarray}
\label{gbm:valpha}
v^\alpha &\equiv& p^\alpha + \frac{1}{2} \frac{1}{d-1} [\gamma^\alpha,\slashed{p}] \,.
\end{eqnarray}
From \eqref{gbm:PNG} we get the corresponding expansion of the effective NG vertex $P_A(p^\prime,p)$:
\begin{eqnarray}
\label{gbm:PNGexp}
P(p^\prime,p) &=&
-\I \big[ I^\T(0) \, I(0) \big]^{-1} I^\T(0) \Big( A(p) + q_\alpha B^\alpha(p) \Big) + \mathcal{O}(q^2) \,.
\end{eqnarray}
Note that only $I(0)$ enters here, high terms in $q^2$ of $I(q^2)$ are dispensable in the given order of the expansion \eqref{gbm:PNGexp}.

Let us now take a closer look at $I(0)$. We saw that the NG part $\Gamma_{a}^\mu(p^\prime,p)|_{\mathrm{NG}}$ does contribute only to the longitudinal part of the polarization tensor. Of course, the regular part $\Gamma_{a}^\mu(p^\prime,p)|_{\mathrm{reg.}}$ can in principle contribute to it as well. However, in the considered lowest orders in $q$ it is only $\Gamma_{a}^\mu(p^\prime,p)|_{\mathrm{NG}}$ which actually does contribute to it, see the expansion \eqref{gbm:Piminuabexp} of $\Pi_{ab}^{\mu\nu}(q)$. We can thus, due to the expression \eqref{gbm:PiNGcontrib} for $\Pi_{ab}^{\mu\nu}(q)|_{\mathrm{NG}}$ (and taking into account the transversality of $\Pi_{ab}^{\mu\nu}(q)$), write
\begin{eqnarray}
\Pi_{ab}(q^2)  &=& \frac{1}{q^2} I_{aA}(0)\,I_{bA}(0) + \mathcal{O}(q^0) \,.
\end{eqnarray}
I.e., the gauge boson mass matrix is within the pole approximation \eqref{gbm:PoleAppr} given by
\begin{eqnarray}
M_{ab}^2 &=& I_{aA}(0)\,I_{bA}(0) \,,
\end{eqnarray}
or in the matrix form,
\begin{eqnarray}
M^2 &=& I(0)\,I^\T(0) \,.
\end{eqnarray}
We can therefore on the basis of \eqref{gbm:M2=FFT} identify
\begin{eqnarray}
I(0) &=& F \,,
\end{eqnarray}
as discussed above in Sec.~\ref{gbl:ssec:Poleapp}, and rewrite cosmetically the expression \eqref{gbm:PNGexp} for $P(p^\prime,p)$ as
\begin{eqnarray}
\label{gbm:Ppprimep}
P(p^\prime,p) &=&
-\I \big[ F^\T F \big]^{-1} F^\T \Big( A(p) + q_\alpha B^\alpha(p) \Big) + \mathcal{O}(q^2) \,.
\end{eqnarray}

\section{Summary}

Let us summarize the main points of this chapter. Before arriving at the final formula \eqref{gbm:M2ab} for the gauge boson mass matrix, several simplifying assumptions have been made.

First of all, we specified the way of treating the polarization tensor. We argued that for the purpose of computing the gauge boson masses in the lowest (second) order in the gauge coupling constant it is sufficient to consider only the pole part of the form factor $\Pi_{ab}(q^2)$ of the polarization tensor. For its very computation we chose the one-loop diagram \eqref{gbm:Pimunuab}, Fig.~\ref{gbm:fig:Pi}, with the fermion lines given by the symmetry-breaking propagators of the form and properties considered in the previous chapter.

In order to arrive at transversal polarization tensor, one of the two vertices in \eqref{gbm:Pimunuab} had to be dressed and satisfying the WT identity (while the other had to be the bare one and thus not satisfying the WT identity). In principle, if the symmetry breaking dynamics yields the dressed fermion propagators, it should be capable of yielding in the same manner also the dressed vertex. However, we assumed, motivated primarily by the models presented in parts~\ref{part:abel} and \ref{part:ew}, that due to the used approximation scheme (i.e., in our case the truncation scheme of the SD equations) we did not have the dressed vertex at disposal. Hence, the only way to arrive at the dressed vertex, satisfying the WT identity, was  to construct it.


A great part of the chapter was dedicated to this construction, which would not be possible without making some non-trivial assumptions concerning the general form of the vertex. Namely, we imposed in Sec.~\ref{gbm:ssec:general} linearity in $\boldsymbol{\Sigma}_{p^\prime}$, $\boldsymbol{\Sigma}_{p}$, as well as in $T_a$, $\bar T_a$, and in Sec.~\ref{gbm:ssec:analytic} the correct analytic structure. The rest was rather straightforward (though somewhat tedious): We imposed the indispensable WT identity, correct transformation behavior under both the continuous and discrete symmetries and Hermiticity. Finally, we exploited one more time the NG boson interpretation of the vertex, already used before for constraining the analytic structure of the vertex.

The vertex developed this way still contained two real free parameters. In order to determine them we returned to our ultimate task of calculating the gauge boson mass matrix and required it to be symmetric. This requirement actually applied only to one of those free parameters, as the other did contribute only to the higher orders of the Laurent expansion of the polarization tensor.

At this point the troubles appeared. It turned out that there is no single value of the mentioned free parameter of the vertex, ensuring the symmetricity of the gauge boson mass matrix for arbitrary gauge theory with arbitrary setting of fermion multiplets, unless the quantity $A_{ab}$, \eqref{gbm:Aamdef}, vanishes. Thus, we could not choose but assuming $A_{ab}=0$ and write down the final unique formula \eqref{gbm:M2ab} for the gauge boson mass matrix under this assumption.

Such result is of course not satisfactory in general. However, for our purposes of calculating the gauge boson masses within the Abelian toy model and the electroweak interactions in the following two chapters the formula \eqref{gbm:M2ab} will be in fact sufficient, as the condition $A_{ab}=0$ will be in both cases satisfied.

\chapter{Application to the Abelian toy model}
\label{chp:ablgg}

\intro{In this chapter we apply the results from the previous chapter to the Abelian toy model, discussed in part~\ref{part:abel}. However, we will not directly jump into the Abelian model (to be discussed only in section~\ref{chp:ablgg:sec:abl}), but rather gradually, step by step, decrease the level of generality considered in the previous two chapters. This will eventually allow us to use some of the results from this chapter also in the following chapter, dedicated to the gauge boson masses in electroweak interactions.}


\section{Some special cases}
\label{chp:ablgg:sec:spec}

\subsection{Assumption $[\boldsymbol{\Sigma},T_a] = 0$}
\label{gba:ssec:cmt}

We start with the assumption that the self-energy $\boldsymbol{\Sigma}$ commutes with the generators $T_a$ (for all $a$):
\begin{eqnarray}
\label{gba:commut}
{[\boldsymbol{\Sigma},T_a]}  &=& 0 \,.
\end{eqnarray}
(Needless to say that this is, in general, not the same as the seemingly similar condition $\bbl \boldsymbol{\Sigma},T_a \bbr$, \eqref{gbp:Sgmnoninv}, for the generator $T_a$ to be unbroken.) Notice that the analogous condition $[\boldsymbol{\Sigma},\bar T_a] = 0$ already follows automatically from this one due to the Hermiticity condition $\boldsymbol{\Sigma} = \boldsymbol{\bar\Sigma}$, \eqref{gbp:hermSgm}.
Under this assumption the vertex \eqref{gbm:Gmm:ans}, derived in the previous chapter, simplifies as
\begin{eqnarray}
\label{gba:Gmm:commut}
\Gamma_a^\mu(p^\prime,p) &=& \gamma^\mu T_a
-\frac{1}{2} \frac{q^\mu}{q^2} \big(\boldsymbol{\Sigma}_{p^\prime}+\boldsymbol{\Sigma}_{p}\big) \big(  T_a - \bar T_a \big)
-\bigg(\frac{1}{2} \frac{q^{\prime \mu}}{q \cdot q^\prime} - a_6 \frac{[\gamma^\mu,\slashed{q}]}{q \cdot q^\prime}\bigg)
\big(\boldsymbol{\Sigma}_{p^\prime}-\boldsymbol{\Sigma}_{p}\big) \big(  T_a + \bar T_a \big)
\nonumber \\ && \phantom{\gamma^\mu T_a}
-\frac{1}{4} \frac{1}{d-1}
\bigg(\frac{q^\mu}{q^2}\frac{[\slashed{q},\slashed{q}^\prime]}{q \cdot q^\prime} - \frac{[\gamma^\mu,\slashed{q}^\prime]}{q \cdot q^\prime} \bigg)
\big(\boldsymbol{\Sigma}_{p^\prime}-\boldsymbol{\Sigma}_{p}\big)
\big( T_a - \bar T_a \big) \,.
\end{eqnarray}

Let us now check the crucial condition \eqref{gbm:Aab=0}, whose fulfilment is necessary for being able to calculate the gauge boson masses using the formula \eqref{gbm:M2ab}. Direct calculation reveals
\begin{eqnarray}
\Tr \bigg\{
\bbl \boldsymbol{\Sigma}, T_a  \bbr
\bbl \boldsymbol{D}_R \, T_b \, \boldsymbol{\Sigma}^\dag , \boldsymbol{D}_L^\prime \bbr
\bigg\} &=&
\frac{1}{2} \Tr \bigg\{
\big(T_a-\bar T_a\big) \big(T_b-\bar T_b\big) \boldsymbol{\Sigma}\boldsymbol{\Sigma}^\dag \boldsymbol{D}_L^{2\prime}
\bigg\} \,,
\end{eqnarray}
which is symmetric in the gauge indices $a$, $b$, so that (recall the definition \eqref{gbm:Aamdefkratsi} of $A_{ab}$)
\begin{subequations}
\begin{eqnarray}
A_{ab} &=&
\Tr \bigg\{
\bbl \boldsymbol{\Sigma}, T_a  \bbr
\bbl \boldsymbol{D}_R \, T_b \, \boldsymbol{\Sigma}^\dag , \boldsymbol{D}_L^\prime \bbr
\bigg\} - (a \leftrightarrow b)
\\ &=& 0 \,.
\end{eqnarray}
\end{subequations}
I.e., the condition \eqref{gbm:Aab=0} is indeed fulfilled. We can therefore safely use the expression \eqref{gbm:M2ab} for the gauge boson matrix and we find
\begin{subequations}
\label{gba:Mab}
\begin{eqnarray}
\label{gba:Mabcomm}
M_{ab}^2 &=& -\I \frac{1}{2} \int\!\frac{\d^d p}{(2\pi)^d}\, \Tr \bigg\{
\big(T_a-\bar T_a\big) \big(T_b-\bar T_b\big)
\Big[\boldsymbol{\Sigma}\boldsymbol{\Sigma}^\dag - \frac{2}{d}p^2 \big(\boldsymbol{\Sigma}\boldsymbol{\Sigma}^\dag\big)^\prime\Big]
\big(p^2-\boldsymbol{\Sigma}\boldsymbol{\Sigma}^\dag\big)^{-2}
\bigg\} \,.
\qquad\qquad
\end{eqnarray}
This result can be further simplified by eliminating $\boldsymbol{\Sigma}$ in favor of $\Sigma$ (recall that $\boldsymbol{\Sigma} = \Sigma^\dag P_L + \Sigma\,P_R$). For this purpose one can use the fact that the commutation relations \eqref{gba:commut} hold not only for $\boldsymbol{\Sigma}$, but also for $\Sigma$. This, together with the cyclicity of the trace and the commutation relations \eqref{frm:SgmDDR=DLSgmD} leads to simplification of the mass matrix \eqref{gba:Mabcomm} as
\begin{eqnarray}
\label{gba:Mabcommnog5}
M_{ab}^2 &=& -\I\frac{1}{2} \int\!\frac{\d^d p}{(2\pi)^d}\, \Tr \bigg\{
\big(T_a-\bar T_a\big) \big(T_b-\bar T_b\big)
\Big[{\Sigma}{\Sigma}^\dag - \frac{2}{d}p^2 \big({\Sigma}{\Sigma}^\dag\big)^\prime\Big]
\big(p^2-{\Sigma}{\Sigma}^\dag\big)^{-2}
\bigg\} \,.
\qquad\qquad
\end{eqnarray}
\end{subequations}

Recall that in Sec.~\ref{gbm:ssec:symm} we have set the parameter $a_4$ of the vertex \eqref{gbm:Gmm:ansNG} to the non-trivial value \eqref{gbm:a4} to ensure that the resulting gauge boson mass matrix is indeed symmetric, once the condition \eqref{gbm:Aab=0} is satisfied. Nevertheless, let us, just for curiosity, calculate the gauge boson mass matrix using the formula \eqref{gbm:M2ab:intermed} with arbitrary $a_4$:
\begin{subequations}
\label{gba:M2aba4}
\begin{eqnarray}
M_{ab}^2 &=&
\nonumber \\ && \hspace{-2.2em}
-\I \frac{1}{2} \int\!\frac{\d^d p}{(2\pi)^d}\,
\Tr \bigg\{
\big(T_a-\bar T_a\big) \big(T_b-\bar T_b\big)
\Big[\boldsymbol{\Sigma}\boldsymbol{\Sigma}^\dag - \frac{1}{d}\big(1+4(d-1)a_4\big)p^2 \big(\boldsymbol{\Sigma}\boldsymbol{\Sigma}^\dag\big)^\prime\Big]
\big(p^2-\boldsymbol{\Sigma}\boldsymbol{\Sigma}^\dag\big)^{-2}
\bigg\}
\nonumber \\ && \\
&=&
\nonumber \\ && \hspace{-2.2em}
-\I \frac{1}{2} \int\!\frac{\d^d p}{(2\pi)^d}\,
\Tr \bigg\{
\big(T_a-\bar T_a\big) \big(T_b-\bar T_b\big)
\Big[{\Sigma}{\Sigma}^\dag - \frac{1}{d}\big(1+4(d-1)a_4\big)p^2 \big({\Sigma}{\Sigma}^\dag\big)^\prime\Big]
\big(p^2-{\Sigma}{\Sigma}^\dag\big)^{-2}
\bigg\} \,.
\nonumber \\ &&
\label{gba:M2aba4nonbold}
\end{eqnarray}
\end{subequations}
Thus, incidentally, we see that in the present special case (defined by \eqref{gba:commut}) the gauge boson mass matrix is actually symmetric for any value of $a_4$. Nevertheless, in the following we will keep the special value \eqref{gbm:a4} of $a_4$, as it follows from the general requirement that the formula \eqref{gbm:M2ab} be applicable, upon fulfilling the condition \eqref{gbm:Aab=0}, for \emph{any} theory (e.g., the electroweak interactions), in which the gauge boson mass matrix may not be symmetric for arbitrary $a_4$ like in the present case.

\subsection{Case of $\group{U}(1)^N$}
\label{gba:ssec:U1N}

Moreover, let us now assume, on top of the assumption \eqref{gba:commut}, that the group $\group{G}$ is Abelian. More precisely, we assume that the generators $T_a$ commute with each other, as well as with $\bar T_a$:
\begin{subequations}
\begin{eqnarray}
{[T_a,     T_b]} &=& 0 \,, \\
{[T_a,\bar T_b]} &=& 0 \,.
\end{eqnarray}
\end{subequations}
For our purposes it will be moreover sufficient to assume that all fermions sit in the same representation of $\group{G}$. This implies that in the expressions
\begin{subequations}
\label{gba:Tabases}
\begin{eqnarray}
T_a &=& T_{La} \, P_L + T_{Ra} \, P_R  \\  &=& T_{Va} \, \unitmatrix + T_{Aa} \, \gamma_5
\end{eqnarray}
\end{subequations}
for $T_a$ (cf.~Eqs.~\eqref{gbp:taLR}, \eqref{gbp:taVA}) the components $T_{La}$, $T_{Ra}$, as well as $T_{Va}$, $T_{Aa}$, can be considered as Hermitian matrices $1 \times 1$, i.e., mere real numbers.


\subsubsection{Gauge boson mass matrix}

Note that the generators $T_a$, \eqref{gba:Tabases}, as being only some real linear combinations of $P_L$ and $P_R$, now act only in the Dirac space, while the self-energy $\Sigma$ operates only in the flavor space. Thus, since $T_a$ and $\Sigma$ operates in different spaces, their product is a Kronecker product and the trace of their product can be written as a product of their traces: $\Tr[T_a \Sigma] = \Tr[T_a]\Tr[\Sigma]$. The gauge boson mass matrix \eqref{gba:Mabcommnog5} can be therefore written as
\begin{subequations}
\label{gba:M2abUN}
\begin{eqnarray}
M_{ab}^2 &=& \frac{1}{16} \Tr\!\Big\{\big(T_a-\bar T_a\big)\big(T_b-\bar T_b\big)\Big\} \mu^2
\\ &=& T_{Aa} T_{Ab} \, \mu^2 \,,
\end{eqnarray}
\end{subequations}
where we denoted
\begin{eqnarray}
\label{gba:mu2}
\mu^2 &\equiv& -8\I \int\!\frac{\d^d p}{(2\pi)^d}\, \Tr \bigg\{
\Big[{\Sigma}{\Sigma}^\dag - \frac{2}{d}p^2 \big({\Sigma}{\Sigma}^\dag\big)^\prime\Big]
\big(p^2-{\Sigma}{\Sigma}^\dag\big)^{-2}
\bigg\} \,.
\end{eqnarray}

After Wick rotation this expression for $\mu^2$ becomes
\begin{eqnarray}
\label{gba:mu2wick}
\mu^2 &=& 8 \int\!\frac{\d^d p}{(2\pi)^d}\, \Tr \bigg\{
\Big[{\Sigma}{\Sigma}^\dag - \frac{2}{d}p^2 \big({\Sigma}{\Sigma}^\dag\big)^\prime\Big]
\big(p^2+{\Sigma}{\Sigma}^\dag\big)^{-2}
\bigg\} \,.
\end{eqnarray}
From this expression one can in particular see that without additional assumptions about the behavior of the matrix function $\Sigma(p^2)$ the positivity of $\mu^2$ (and consequently the positivity of the mass squared of the gauge boson corresponding to the broken subgroup) is indeed not automatically guaranteed, as advertised above.

The mass matrix \eqref{gba:M2abUN} is of the expected form $M^2 = FF^\T$, \eqref{gbm:M2=FFT}, with $F$ being identified as the vector (i.e., matrix $N_{\group{G}} \times 1$)
\begin{eqnarray}
\label{gba:F}
F &\equiv&  \left(\begin{array}{c} T_{A1} \, \mu  \\ \vdots \\ T_{AN_{\group{G}}} \, \mu \end{array}\right) \,.
\end{eqnarray}
Thus, the mass matrix is singular, with rank $1$, and the only non-vanishing eigenvalue
\begin{subequations}
\label{gba:M2AUN}
\begin{eqnarray}
M_{\mathrm{A}}^2 &\equiv& \Tr FF^\T \ = \  F^\T F \\ &=& \mu^2 \sum_{a=1}^{N_{\group{G}}} T_{Aa}^{2}
\end{eqnarray}
\end{subequations}
expresses the mass squared of the gauge boson, corresponding to the spontaneously broken axial subgroup $\group{U}(1)_{\mathrm{A}}$. The remaining $N_{\group{G}}-1$ gauge bosons stay massless.

\subsubsection{NG boson coupling}

Consider now the expansion \eqref{gbm:Gmm:NGexp} of the NG part $\Gamma_a^\mu(p^\prime,p)|_{\mathrm{NG}}$. For the simplified vertex \eqref{gba:Gmm:commut} we have
\begin{subequations}
\begin{eqnarray}
A_a(p) &=& - \boldsymbol{\Sigma} \big(T_a - \bar T_a\big) \,, \\
B_a^\alpha(p) &=& - v^\alpha \boldsymbol{\Sigma}^\prime \big(T_a - \bar T_a \big) \,,
\end{eqnarray}
\end{subequations}
where $v^\alpha$ is defined in \eqref{gbm:valpha}. It can be, upon noting that $T_a - \bar T_a = 2 T_{Aa} \gamma_5$, rewritten as
\begin{subequations}
\begin{eqnarray}
A_a(p) &=& - 2 T_{aA} \boldsymbol{\Sigma} \gamma_5 \,, \\
B_a^\alpha(p) &=& - 2 T_{aA} v^\alpha \boldsymbol{\Sigma}^\prime \gamma_5 \,.
\end{eqnarray}
\end{subequations}
Recalling the expression \eqref{gba:F} for $F$, we thus find
\begin{eqnarray}
F^\T \Big( A(p) + q_\alpha B^\alpha(p) \Big) &=& - 2 \mu \Big( \boldsymbol{\Sigma} + (q \cdot v) \boldsymbol{\Sigma}^\prime \Big) \gamma_5 \sum_a T_{Aa}^2 \,.
\end{eqnarray}
Upon substituting this expression, together with the expression \eqref{gba:M2AUN} for $F^\T F$, into the formula \eqref{gbm:Ppprimep} for the NG coupling $P(p^\prime,p)$, the quantities $\sum_a T_{Aa}^2$ cancel and we finally obtain
\begin{eqnarray}
\label{gba:PpprimepUN}
P(p^\prime,p) &=& 2\I \frac{1}{\mu}\Big(\boldsymbol{\Sigma}+(q \cdot v)\boldsymbol{\Sigma}^\prime\Big)\gamma_5 + \mathcal{O}(q^2) \,.
\end{eqnarray}

Note that although our treatment of the NG boson was based on the gauge boson polarization tensor and in intermediate stages of calculation the gauge coupling constant $g$ appeared implicitly (through its presence in generators $T_a$), the final expression \eqref{gba:PpprimepUN} for the NG boson coupling $P(p^\prime,p)$ is independent of $g$, due to cancelation of $\sum_a T_{Aa}^2$. This correctly suggests that the result \eqref{gba:PpprimepUN} holds irrespective of whether the spontaneously broken symmetry is gauged or not.

\subsection{Comparison with the Pagels--Stokar formula}
\label{gba:ssec:PS}

Assume now for simplicity that $\Sigma$ is just a real scalar function, without any non-trivial matrix structure (i.e., the number of fermion flavors is one). Then the Wick-rotated expression \eqref{gba:mu2wick} for $\mu^2$ reads
\begin{eqnarray}
\label{gba:mu2w}
\mu^2 &=& 8 \int\!\frac{\d^4 p}{(2\pi)^4}\,
\frac{\Sigma^2 - \frac{1}{2}p^2 \big(\Sigma^2\big)^\prime}{\big(p^2+\Sigma^2\big)^{2}} \,,
\end{eqnarray}
where we have also explicitly set $d=4$.

A similar expression has already been derived in the literature: It is the Pagels--Stokar (PS) formula \cite{Pagels:1979hd}, which can be in the present context for the sake of comparison recast as
\begin{eqnarray}
\label{gba:mu2wPS}
\mu^2_{\mathrm{PS}} &=& 8 \int\!\frac{\d^4 p}{(2\pi)^4}\,
\frac{\Sigma^2 - \frac{1}{4}p^2 \big(\Sigma^2\big)^\prime}{\big(p^2+\Sigma^2\big)^{2}} \,.
\end{eqnarray}
We can see that there is a slight difference between the two formul{\ae} \eqref{gba:mu2w} and \eqref{gba:mu2wPS}: The coefficient at the term $(\Sigma^2)^\prime$ in our formula \eqref{gba:mu2w} is twice as large as in the PS formula \eqref{gba:mu2wPS}.

Origin of this discrepancy is easily revealed. It is the different value of the parameter $a_4$ in the expression \eqref{gbm:Gmm:ansNG} for the vertex $\Gamma_{a}^\mu(p^\prime,p)$ which makes the difference: For a general $a_4$ our expression \eqref{gba:mu2w} for $\mu^2$ would be modified as
\begin{eqnarray}
\mu^2 &=& 8 \int\!\frac{\d^4 p}{(2\pi)^4}\,
\frac{\Sigma^2 - \frac{1}{4}\big(1-12a_4\big)p^2 \big(\Sigma^2\big)^\prime}{\big(p^2+\Sigma^2\big)^{2}} \,.
\end{eqnarray}
as can be seen from \eqref{gba:M2aba4}. Clearly, while our formula \eqref{gba:mu2w} corresponds to non-vanishing value \eqref{gbm:a4} of $a_4$,
\begin{eqnarray}
a_4 &=& - \frac{1}{12} \,,
\end{eqnarray}
the PS formula \eqref{gba:mu2wPS} corresponds simply to
\begin{eqnarray}
a_4 &=& 0 \,.
\end{eqnarray}

Pagels and Stokar introduced in Ref.~\cite{Pagels:1979hd} the \qm{dynamical perturbation theory}, which, upon adapting on the present context, states roughly the same as what we did in Sec.~\ref{gbm:ssec:general}: One keeps in the vertex only the terms linear in the gauge coupling constant $g$. This actually implies the form \eqref{gbm:Gmm:gen}, i.e., the vertex must have the form of the bare vertex plus something which vanishes in the case of no SSB, since at order $g$ there are no perturbative corrections to the vertex. However, Pagels and Stokar moreover assumed that the only non-perturbative correction to the bare vertex is the pole term $q^\mu/q^2$ and overlooked the possibility that there can be also non-perturbative contributions regular in $q=0$. Recall now that the present discussion concerns about an axial symmetry $\group{U}(1)_{\mathrm{A}}$, whose generator is $T_a = g \gamma_5 \tau_a$, with $\tau_a$ being a real number, so that the vertex \eqref{gbm:Gmm:ansNG} within the simplifying assumptions about $\boldsymbol{\Sigma}$ made in this section reads
\begin{eqnarray}
\Gamma_{a}^\mu(p^\prime,p) &=& g\gamma^\mu \gamma_5 \tau_a
- g\frac{q^\mu}{q^2}{\Sigma}_{+}\gamma_5 \tau_a
+ 2ga_4 \bigg(\frac{q^\mu}{q^2}\frac{[\slashed{q},\slashed{q}^\prime]}{q \cdot q^\prime} - \frac{[\gamma^\mu,\slashed{q}^\prime]}{q \cdot q^\prime} \bigg) {\Sigma}_{-} \gamma_5 \tau_a \,.
\end{eqnarray}
We can see clearly that insisting that only the pole term can be non-perturbative (i.e., proportional to ${\Sigma}$) indeed effectively corresponds to setting $a_4=0$.

Notice for the sake of completeness that there also exists in the literature an improved version of the PS formula, introduced in \cite{Barducci:1997jh}:
\begin{eqnarray}
\label{gba:mu2wimpr}
\mu^2_{\mathrm{improved}} &=& 8 \int\!\frac{\d^4 p}{(2\pi)^4}\,\bigg\{
\frac{\Sigma^2 - \frac{1}{2}p^2 \big(\Sigma^2\big)^\prime}{\big(p^2+\Sigma^2\big)^{2}}
+\frac{1}{2}p^2 \big(\Sigma^\prime\big)^2 \frac{p^2-\Sigma^2}{\big(p^2+\Sigma^2\big)^{2}}
\bigg\} \,.
\end{eqnarray}
Although our formula \eqref{gba:mu2w} is not identical to this improved one, it reduces to it if one neglects the terms proportional to $(\Sigma^\prime)^2$.

Pagels and Stokar have used for the fermion self-energy a rather crude Ansatz $\Sigma = 4 m_D^3/p^2$ (with $p^2$ being in Minkowski metric), where $m_D$ is the \qm{dynamical quark mass}, to estimate value of the pion decay constant $f_\pi$, related to $\mu^2$ as
\begin{eqnarray}
f_\pi^2 &=& \frac{N_c}{2} \mu^2 \,,
\end{eqnarray}
where $N_c=3$ is the number of colors. Using the value $m_D=244 \MeV$ from \cite{Hagiwara:1978ns} they were surprised to obtain from their formula \eqref{gba:mu2wPS} the estimate $f_\pi = 83 \MeV$ (the same value is actually obtained also using the improved PS formula \eqref{gba:mu2wimpr}), which is rather close to the experimental value $f_\pi = 93 \MeV$. Interestingly enough, had they used rather the expression \eqref{gba:mu2w} for $\mu^2$, instead of \eqref{gba:mu2wPS}, they would obtain $f_\pi = 96 \MeV$, i.e., the agreement would be even better.

\subsection{Mixing in $\group{U}(1)^2$}

Let us now discuss in some detail a special case from the previous sections with $N=2$. Recall that the gauge boson mass matrix \eqref{gba:M2abUN} has the explicit form
\begin{eqnarray}
M^2 &=&
\left(\begin{array}{cc}
T_{A1}^{2}     &  T_{A1} \, T_{A2} \\
T_{A1} \, T_{A2}  &  T_{A2}^{2}
\end{array}\right)  \mu^2 \,,
\end{eqnarray}
with the two eigenvalues
\begin{subequations}
\begin{eqnarray}
M^2_{\mathrm{V}} &=& 0 \,, \\
M^2_{\mathrm{A}} &=& \mu^2 (T_{A1}^2 + T_{A2}^2) \,,
\end{eqnarray}
\end{subequations}
corresponding to the masses squared of the mass-diagonal gauge fields, denoted as $A_{\mathrm{V}}^\mu$, $A_{\mathrm{A}}^\mu$, respectively. They are given by an orthogonal rotation of the original gauge fields $A_1^\mu$, $A_2^\mu$:
\begin{eqnarray}
\label{gba:A1A2rot}
\left(\begin{array}{c} A_{\mathrm{V}}^\mu \\ A_{\mathrm{A}}^\mu \end{array}\right)
&=&
\left(\begin{array}{cr}
\cos\theta  & -\sin\theta \\
\sin\theta  &  \cos\theta
\end{array}\right)
\left(\begin{array}{c} A_1^\mu \\ A_2^\mu \end{array}\right) \,.
\end{eqnarray}
For the mixing angle $\theta$ there is the identity
\begin{eqnarray}
\tan \theta &=& \frac{T_{A1}}{T_{A2}} \,,
\end{eqnarray}
i.e.,
\begin{subequations}
\begin{eqnarray}
\sin \theta &=& \frac{T_{A1}}{\sqrt{T_{A1}^{2}+T_{A2}^{2}}} \,, \\
\cos \theta &=& \frac{T_{A2}}{\sqrt{T_{A1}^{2}+T_{A2}^{2}}} \,.
\end{eqnarray}
\end{subequations}

Consider now the interaction Lagrangian, reading in terms of the original fields $A_1^\mu$, $A_2^\mu$
\begin{eqnarray}
\eL &=& \bar\psi \gamma_\mu T_1 \psi A_{1}^{\mu} + \bar\psi \gamma_\mu T_2 \psi A_{2}^{\mu} \,.
\end{eqnarray}
In terms of the mass-diagonal fields $A_{\mathrm{V}}^\mu$, $A_{\mathrm{A}}^\mu$ we have
\begin{eqnarray}
\eL &=& \bar\psi \gamma_\mu T_{\mathrm{V}} \psi A_{\mathrm{V}}^\mu + \bar\psi \gamma_\mu T_{\mathrm{A}} \psi A_{\mathrm{A}}^\mu \,,
\end{eqnarray}
where the new generators $T_{\mathrm{V}}$, $T_{\mathrm{A}}$ are given by the rotation of the original generators $T_1$, $T_2$ in much the same way as the fields themselves, \eqref{gba:A1A2rot}, i.e., as
\begin{eqnarray}
\left(\begin{array}{c} T_{\mathrm{V}} \\ T_{\mathrm{A}} \end{array}\right)
&=&
\left(\begin{array}{cr}
\cos\theta & -\sin\theta \\
\sin\theta &  \cos\theta \\
\end{array}\right)
\left(\begin{array}{c} T_1 \\ T_2 \end{array}\right) \,.
\end{eqnarray}
Explicitly in terms of components of the original generators $T_1$, $T_2$ we have
\begin{subequations}
\label{gba:TVTA}
\begin{eqnarray}
\left(\begin{array}{c} T_{\mathrm{V}} \\ T_{\mathrm{A}} \end{array}\right)
&=&
\frac{1}{\sqrt{T_{A1}^{2}+T_{A2}^{2}}}
\left(\begin{array}{c}
T_{V1} T_{A2} - T_{V2} T_{A1} \\
T_{V1} T_{A1} + T_{V2} T_{A2} + \gamma_5 \Big(T_{A1}^{2}+T_{A2}^{2}\Big)
\end{array}\right)
\\ &=&
\frac{1}{\sqrt{T_{A1}^{2}+T_{A2}^{2}}}
\left(\begin{array}{c}
T_{V1} T_{A2} - T_{V2} T_{A1} \\
T_{V1} T_{A1} + T_{V2} T_{A2}
\end{array}\right)
+ \gamma_5
\frac{1}{\sqrt{T_{A1}^{2}+T_{A2}^{2}}} \left(\begin{array}{c} 0 \\ T_{A1}^{2}+T_{A2}^{2} \end{array}\right) \,.
\qquad\qquad
\end{eqnarray}
\end{subequations}
Note that $T_{\mathrm{V}}$ is non-vanishing due to the linear independency of the generators $T_1$, $T_2$, which is expressed by the condition
\begin{eqnarray}
\label{gba:detT1T2}
\det \left(\begin{array}{cc} T_{V1} & T_{A1} \\ T_{V2} & T_{A2} \end{array}\right)
\ = \ T_{V1} T_{A2} - T_{V2} T_{A1} &\neq& 0 \,.
\end{eqnarray}
Also note that $T_{\mathrm{V}}$ is purely vectorial (i.e., it does not contain the axial $\gamma_5$-component). On the other hand the generator $T_{\mathrm{A}}$ is not purely axial: While its $\gamma_5$-component is certainly non-vanishing (at least one of $T_{A1}$, $T_{A2}$ must be non-zero due to the condition \eqref{gba:detT1T2}), its vectorial component can be, in general, non-vanishing too. Evidently, the generator $T_{\mathrm{A}}$ is purely axial only if
\begin{eqnarray}
T_{V1} T_{A1} + T_{V2} T_{A2} &=& 0 \,.
\end{eqnarray}
Under this assumption the expression \eqref{gba:TVTA} for $T_{\mathrm{V}}$ and $T_{\mathrm{A}}$ simplifies significantly as
\begin{subequations}
\begin{eqnarray}
\left(\begin{array}{c} T_{\mathrm{V}} \\ T_{\mathrm{A}} \end{array}\right)
&=&
\sqrt{T_{A1}^{2}+T_{A2}^{2}} \left(\begin{array}{c} T_{V1} / T_{A2} \\ \gamma_5 \end{array}\right)
\\ &=&
\sqrt{T_{A1}^{2}+T_{A2}^{2}} \left(\begin{array}{c} -T_{V2} / T_{A1} \\ \gamma_5 \end{array}\right) \,.
\end{eqnarray}
\end{subequations}
Note that at least one of these two expressions makes sense, as at least one the two quantities $T_{A1}$, $T_{A2}$ is non-vanishing, due to \eqref{gba:detT1T2}.

\section{Abelian toy model}
\label{chp:ablgg:sec:abl}

\subsection{Mass spectrum}

We can now finally proceed to discussing the Abelian toy model, introduced in chapter~\ref{chp:frm}. Assume that its symmetry group $\group{G} = \group{U}(1)_{\mathrm{V}_{\!1}} \times \group{U}(1)_{\mathrm{V}_{\!2}} \times \group{U}(1)_{\mathrm{A}}$ is gauged. We denote the corresponding gauge bosons and coupling constants as $A_{\mathrm{V}_{\!1}}^\mu$, $A_{\mathrm{V}_{\!2}}^\mu$, $A_{\mathrm{A}}^\mu$ and $g_{\mathrm{V}_{\!1}}$, $g_{\mathrm{V}_{\!2}}$, $g_{\mathrm{A}}$, respectively. Recall that the axial subgroup $\group{U}(1)_{\mathrm{A}}$ was spontaneously broken by the fermion self-energies, while the vectorial subgroups $\group{U}(1)_{\mathrm{V}_{\!i}}$ remained unbroken. We thus expect that the gauge boson $A_{\mathrm{A}}^\mu$ will acquire a non-vanishing mass, while the other two gauge bosons $A_{\mathrm{V}_{\!i}}^\mu$ will remain massless.

In chapter~\ref{chp:gbp}, in the course of introducing the formalism for the quest of calculating the gauge boson masses, we assumed for convenience that all of the fermion fields present in the theory were organized in a single field $\psi$. In the present case the theory contains two fermion species $\psi_1$ and $\psi_2$, so we put them together as
\begin{eqnarray}
\psi &\equiv& \left(\begin{array}{c} \psi_1 \\ \psi_2 \end{array}\right) \,.
\end{eqnarray}
The corresponding representation of the symmetry generators then reads
\begin{subequations}
\label{gba:Ta}
\begin{eqnarray}
T_{\mathrm{V}_{\!i}} &=& \left(\begin{array}{cc} T_{1,\mathrm{V}_{\!i}} & 0 \\ 0 & T_{2,\mathrm{V}_{\!i}} \end{array}\right) \,, \\
T_{\mathrm{A}} &=& \left(\begin{array}{cc} T_{1,\mathrm{A}} & 0 \\ 0 & T_{2,\mathrm{A}} \end{array}\right) \,,
\end{eqnarray}
\end{subequations}
where $T_{i,\mathrm{V}_{\!j}}$, $T_{i,\mathrm{A}}$ are defined in terms of $t_{i,\mathrm{V}_{\!j}}$, $t_{i,\mathrm{A}}$, Eqs.~\eqref{chp:frm:U1Vipsigen}, \eqref{chp:frm:Apsigen}, respectively, as (no sum over $j$, cf.~footnote~\ref{footnote:nosum} on page~\pageref{footnote:nosum})
\begin{subequations}
\label{gba:Tia}
\begin{eqnarray}
T_{i,\mathrm{V}_{\!j}} &=& g_{\mathrm{V}_{\!j}} \, t_{i,\mathrm{V}_{\!j}} \,, \\
T_{i,\mathrm{A}} &=& g_{\mathrm{A}} \, t_{i,\mathrm{A}} \,,
\label{gba:TiA}
\end{eqnarray}
\end{subequations}
i.e., according to the definition \eqref{gbp:Tagta}, with the gauge coupling constants included in the generators.

Consider now the self-energy $\boldsymbol{\Sigma}$ of the field $\psi$. We assumed in chapter~\ref{chp:frm} that the vectorial symmetries $\group{U}(1)_{\mathrm{V}_{\!1}}$, $\group{U}(1)_{\mathrm{V}_{\!2}}$ remained unbroken, i.e., in particular that there was no mixing between the two fermion species $\psi_1$ and $\psi_2$. Therefore the off-diagonal elements of $\boldsymbol{\Sigma}$ must be vanishing:
\begin{eqnarray}
\label{gba:Sgmdiag}
\boldsymbol{\Sigma} &=& \left(\begin{array}{cc} \boldsymbol{\Sigma}_1 & 0 \\ 0 & \boldsymbol{\Sigma}_2 \end{array}\right) \,.
\end{eqnarray}
The assumption \eqref{gba:Sgmdiag} has two consequences. First, the self-energy $\boldsymbol{\Sigma}$ commutes with all generators:
\begin{eqnarray}
{[\boldsymbol{\Sigma},T_a]}  &=& 0 \quad\quad(a=\mathrm{V}_{\!1},\mathrm{V}_{\!2},\mathrm{A}) \,,
\end{eqnarray}
because the particular generators $T_{i,\mathrm{V}_{\!j}}$ and $T_{i,\mathrm{A}}$ have no non-trivial matrix structure (up to $\gamma_5$ in the case of $T_{i,\mathrm{A}}$, which commutes with $\boldsymbol{\Sigma}_i$ anyway). We can therefore use the results from Sec.~\ref{gba:ssec:cmt}, namely the satisfaction of the condition \eqref{gba:commut} and the simplified expression \eqref{gba:Mab} for the gauge boson mass matrix. The second consequence of the diagonal form \eqref{gba:Sgmdiag} of the self-energy $\boldsymbol{\Sigma}$ is that since the generators $T_a$, \eqref{gba:Ta}, are diagonal in the fermion species space as well, the one-loop expression \eqref{gba:Mab} for the gauge boson matrix decouples into the sum of independent contributions of the fermion species $\psi_1$, $\psi_2$:
\begin{eqnarray}
\label{gba:M2absep}
M_{ab}^2 &=& M_{ab}^2\big|_1 + M_{ab}^2\big|_2 \,.
\end{eqnarray}
Due to the already mentioned fact that the generators $T_{i,a}$, \eqref{gba:Tia}, are, up to some $\gamma_5$, just real numbers, we can use the results from Sec.~\ref{gba:ssec:U1N} and write $M_{ab}^2|_i$ in the form \eqref{gba:M2abUN}:
\begin{eqnarray}
M_{ab}^2\big|_i &=&  \frac{1}{16} \Tr\!\Big\{\big(T_{i,a}-\bar T_{i,a}\big) \big(T_{i,b}-\bar T_{i,b}\big)\Big\} \mu^2\big|_i \,,
\end{eqnarray}
where
\begin{eqnarray}
\label{gba:mu2i}
\mu^2\big|_i &=& -8\I \int\!\frac{\d^d p}{(2\pi)^d}\,
\frac{|\Sigma_i|^2 - \frac{2}{d}p^2 |\Sigma_i|^{2\prime}}{\big(p^2-|\Sigma_i|^2\big)^{2}} \,.
\end{eqnarray}
Notice that this expression for $\mu^2|_i$ is considerably simpler than the analogous general expression \eqref{gba:mu2} above, since $\Sigma_i$ are now just complex scalar functions without any non-trivial matrix structure in the flavor space.


Taking into account the forms of the generators $T_{i,a}$, namely the fact that $T_{i,\mathrm{A}}$ are proportional to $\gamma_5$, while $T_{i,\mathrm{V}_{\!j}}$ are proportional to $\unitmatrix$ (see definition \eqref{gba:Tia} of $T_{i,a}$ in terms of $t_{i,a}$, Eqs.~\eqref{chp:frm:U1Vipsigen}, \eqref{chp:frm:Apsigen}), we find the contribution $M^2|_i$ of $\psi_i$ to gauge boson mass matrix to be explicitly given (in the basis $A_{\mathrm{V}_{\!1}}^\mu$, $A_{\mathrm{V}_{\!2}}^\mu$, $A_{\mathrm{A}}^\mu$) as
\begin{eqnarray}
M^2\big|_i &=&
\left(\begin{array}{ccc}
0 & 0 & 0 \\
0 & 0 & 0 \\
0 & 0 & 1 \\
\end{array}\right) g_{\mathrm{A}}^2 \, Q_{i,\mathrm{A}}^2 \, \mu^2\big|_i \,.
\end{eqnarray}

We therefore arrive at the final result that the gauge bosons $A_{\mathrm{V}_{\!1}}^\mu$, $A_{\mathrm{V}_{\!2}}^\mu$, corresponding to the unbroken vectorial subgroup $\group{U}(1)_{\mathrm{V}_{\!1}} \times \group{U}(1)_{\mathrm{V}_{\!2}}$, remain massless, while the gauge boson $A_{\mathrm{A}}^\mu$, corresponding to the spontaneously broken axial subgroup $\group{U}(1)_{\mathrm{A}}$, acquires non-vanishing mass, which is proportional to the symmetry-breaking fermion self-energies $\Sigma_1$ and $\Sigma_2$. Explicitly thus the gauge boson mass spectrum reads
\begin{subequations}
\begin{eqnarray}
M^2_{\mathrm{V}_{\!1}} &=& 0 \,, \\
M^2_{\mathrm{V}_{\!2}} &=& 0 \,, \\
M^2_{\mathrm{A}} &=& g_{\mathrm{A}}^2 \, Q_{1,\mathrm{A}}^2 \Big(\mu^2\big|_1 + \mu^2\big|_2\Big)
\label{gba:MA}
\end{eqnarray}
\end{subequations}
(recall that $Q_{1,\mathrm{A}}^2 = Q_{2,\mathrm{A}}^2$, due to \eqref{chp:frm:anomalyfree}). Finally, the coupling of $\psi_i$ to the (would-be) NG boson is
\begin{eqnarray}
P_i(p^\prime,p) &=&
2\I \frac{1}{\mu|_i}\Big(\boldsymbol{\Sigma}_i+(q \cdot v)\boldsymbol{\Sigma}_i^\prime\Big)\gamma_5
+ \mathcal{O}(q^2) \,,
\end{eqnarray}
see \eqref{gba:PpprimepUN}.

\subsection{Effective trilinear gauge boson self-coupling}

Spontaneous breakdown of the axial symmetry $\group{U}(1)_{\mathrm{A}}$ manifests itself in the sector of the corresponding gauge boson $A^\mu_{\mathrm{A}}$ not only by giving the non-vanishing mass $M_{\mathrm{A}}^2$, \eqref{gba:MA}, to it, but also by generating various Green's functions, non-invariant under $\group{U}(1)_{\mathrm{A}}$. In particular, the three-point function $\langle A^\mu_{\mathrm{A}} A^\nu_{\mathrm{A}} A^\rho_{\mathrm{A}} \rangle$ can emerge.

\begin{figure}[t]
\begin{center}
\includegraphics[width=0.58\textwidth]{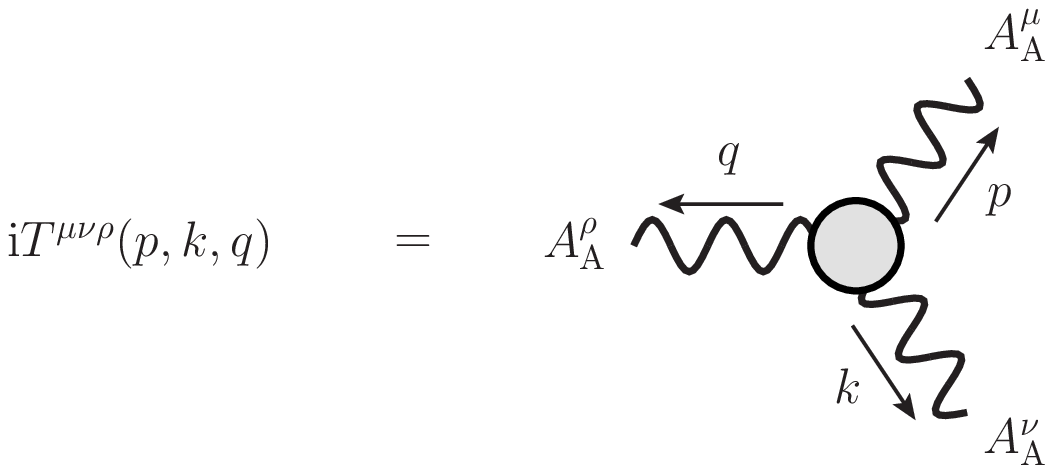}
\caption[Proper vertex $\langle A^\mu_{\mathrm{A}} A^\nu_{\mathrm{A}} A^\rho_{\mathrm{A}} \rangle_{\mathrm{1PI}} = \I T^{\mu\nu\rho}(p,k,q)$.]{Diagrammatical representation and assignment of momenta of the amplitude $\I T^{\mu\nu\rho}(p,k,q) = \langle A^\mu_{\mathrm{A}} A^\nu_{\mathrm{A}} A^\rho_{\mathrm{A}} \rangle_{\mathrm{1PI}}$, \eqref{gba:T}. Momentum conservation is assumed: $p+k+q=0$.}
\label{fig:T}
\end{center}
\end{figure}

\begin{figure}[t]
\begin{center}
\includegraphics[width=0.7\textwidth]{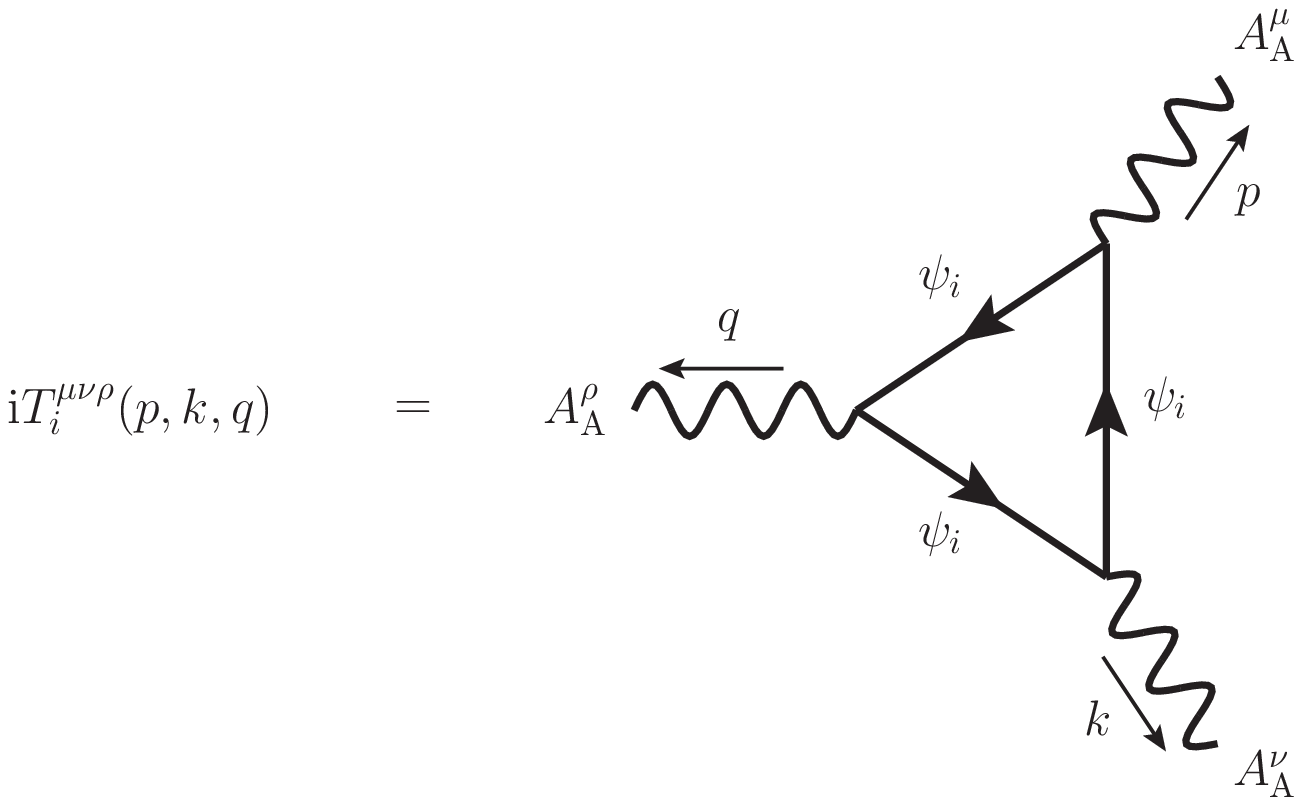}
\caption[One-loop fermion contribution to $T^{\mu\nu\rho}(p,k,q)$.]{Diagrammatical representation of the particular amplitude $\I T^{\mu\nu\rho}_i(p,k,q)$, \eqref{gba:Ti}.}
\label{fig:Ti}
\end{center}
\end{figure}

This function was analyzed in some detail in Ref.~\cite{Benes:2006ny}. If we denote its 1PI part as $\I T^{\mu\nu\rho}(p,k,q)$ (see Fig.~\ref{fig:T} for the assignment of momenta), it can be calculated in the lowest (third) order in the gauge coupling constant $g_{\mathrm{A}}$ as\footnote{Likewise we did not consider the scalar contribution to the gauge boson mass $M_{\mathrm{A}}^2$, we do not consider their contribution to $\I T^{\mu\nu\rho}(p,k,q)$ either.}
\begin{eqnarray}
\label{gba:T}
\I T^{\mu\nu\rho}(p,k,q) &=&
\sum_{i=1,2} \Big[ \I T^{\mu\nu\rho}_{i}(p,k,q) + \I T^{\nu\mu\rho}_{i}(k,p,q) \Big] \,,
\end{eqnarray}
where each particular $\I T^{\mu\nu\rho}_{i}(p,k,q)$ is given by the diagram in Fig.~\ref{fig:Ti}, with the fermion lines given by the symmetry-breaking propagator \eqref{chp:frm:Gpsii} and the vertices given as $\I g_{\mathrm{A}} \gamma^\mu t_{i,\mathrm{A}}$, where the $\group{U}(1)_{\mathrm{A}}$ generators $t_{i,\mathrm{A}}$ have been defined in \eqref{chp:frm:Apsigen}. I.e., we explicitly have
\begin{eqnarray}
\label{gba:Ti}
\I T^{\mu\nu\rho}_{i}(p,k,q) &=& g_{\mathrm{A}}^3 Q_{i,\mathrm{A}}^3 \int\!\frac{\d^4 \ell}{(2\pi)^4}\, \frac{1}{ \big[\ell^2-\Sigma_{i,\ell}^2\big] \big[(\ell+p)^2-\Sigma_{i,\ell+p}^2\big] \big[(\ell-k)^2-\Sigma_{i,\ell-k}^2 \big]}
\nonumber \\ &&
\hspace{-3cm}
\times\,\Tr\Big\{  \gamma^\mu\slashed{\ell}\gamma^\nu(\slashed{\ell}-\slashed{k})\gamma^\rho(\slashed{\ell}+\slashed{p})\gamma_5
\nonumber \\ &&
\hspace{-3cm}
\phantom{\times\Tr\Big\{}
-\gamma^\mu\slashed{\ell}\gamma^\nu\gamma^\rho\gamma_5\boldsymbol{\Sigma}_{i,\ell-k}^{\phantom{\dag}}\boldsymbol{\Sigma}_{i,\ell+p}^\dag
+\gamma^\mu\gamma^\nu(\slashed{\ell}-\slashed{k})\gamma^\rho\gamma_5\boldsymbol{\Sigma}_{i,\ell}^{\phantom{\dag}}\boldsymbol{\Sigma}_{i,\ell+p}^\dag
-\gamma^\mu\gamma^\nu\gamma^\rho(\slashed{\ell}+\slashed{p})\gamma_5\boldsymbol{\Sigma}_{i,\ell}^{\phantom{\dag}}\boldsymbol{\Sigma}_{i,\ell-k}^\dag
\Big\} \,.
\nonumber \\ &&
\end{eqnarray}
Notice that each $\I T^{\mu\nu\rho}_{i}(p,k,q)$ is logarithmically divergent. However, the full $\I T^{\mu\nu\rho}(p,k,q)$, \eqref{gba:T}, is UV-finite (provided the self-energies $\boldsymbol{\Sigma}_i$ are non-increasing functions of momentum), as the logarithmical divergencies cancel due to the anomaly-free condition $Q_{1,\mathrm{A}} + Q_{2,\mathrm{A}} = 0$, \eqref{chp:frm:anomalyfree}. On the basis of the same argument one can see that in the case of no SSB, i.e., when $\boldsymbol{\Sigma}_{1} = \boldsymbol{\Sigma}_{2} = 0$, the amplitude $\I T^{\mu\nu\rho}_{i}(p,k,q)$, given by \eqref{gba:T}, indeed vanishes.


\begin{figure}[t]
\begin{center}

\begin{picture}(13,0)%
\setlength\fboxsep{5bp}
\setlength\fboxrule{0.5pt}
\fbox{\includegraphics{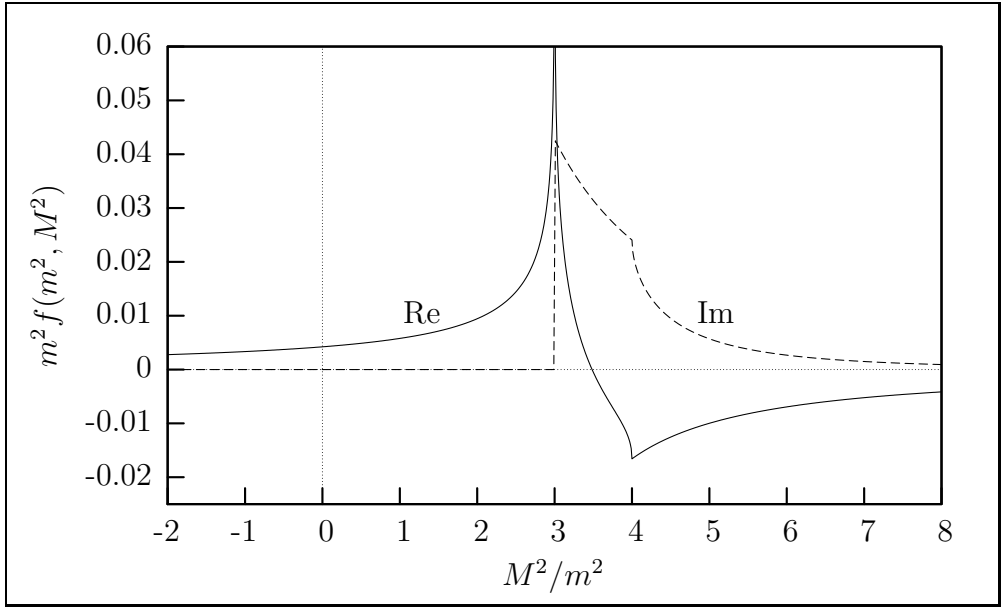}}%
\end{picture}%
\begingroup
\setlength{\unitlength}{0.0200bp}%
\begin{picture}(18000,10800)(0,0)%
\put(2475,2156){\makebox(0,0)[r]{\strut{}-0.02}}%
\put(2475,3168){\makebox(0,0)[r]{\strut{}-0.01}}%
\put(2475,4179){\makebox(0,0)[r]{\strut{} 0}}%
\put(2475,5191){\makebox(0,0)[r]{\strut{} 0.01}}%
\put(2475,6203){\makebox(0,0)[r]{\strut{} 0.02}}%
\put(2475,7215){\makebox(0,0)[r]{\strut{} 0.03}}%
\put(2475,8226){\makebox(0,0)[r]{\strut{} 0.04}}%
\put(2475,9238){\makebox(0,0)[r]{\strut{} 0.05}}%
\put(2475,10250){\makebox(0,0)[r]{\strut{} 0.06}}%
\put(2750,1100){\makebox(0,0){\strut{}-2}}%
\put(4193,1100){\makebox(0,0){\strut{}-1}}%
\put(5635,1100){\makebox(0,0){\strut{} 0}}%
\put(7078,1100){\makebox(0,0){\strut{} 1}}%
\put(8520,1100){\makebox(0,0){\strut{} 2}}%
\put(9963,1100){\makebox(0,0){\strut{} 3}}%
\put(11405,1100){\makebox(0,0){\strut{} 4}}%
\put(12848,1100){\makebox(0,0){\strut{} 5}}%
\put(14290,1100){\makebox(0,0){\strut{} 6}}%
\put(15733,1100){\makebox(0,0){\strut{} 7}}%
\put(17175,1100){\makebox(0,0){\strut{} 8}}%
\put(550,5950){\rotatebox{90}{\makebox(0,0){\strut{}$m^2 f(m^2,M^2)$}}}%
\put(9962,275){\makebox(0,0){\strut{}$M^2 / m^2$}}%
\put(7510,5191){\makebox(0,0){\strut{}$\Re$}}%
\put(12992,5191){\makebox(0,0){\strut{}$\Im$}}%
\end{picture}%
\endgroup

\end{center}
\caption[Function $f(m^2,M^2)$.]{The $M^2$-dependence of the function $f(m^2,M^2)$, \eqref{chp:frm:fmM}. Both quantities are normalized by $m^2$ to be dimensionless; note that $m^2 f(m^2,M^2)$ is only function of $M^2/m^2$. The cusps appear at $M^2=3m^2$ and $M^2=4m^2$. The former one also indicates the beginning of the imaginary part.}
\label{fig:plot_g}
\end{figure}

For illustration, let us now evaluate the amplitude $\I T^{\mu\nu\rho}(p,k,q)$ under certain approximations. First, we set the fermion self-energies to be constant, $\boldsymbol{\Sigma}_i = m_i$, where $m_i$ is a positive real number (i.e., in particular, it does not contain $\gamma_5$) and thus directly interpretable as the fermion's mass. Second, we put, for the sake of simplicity, all external momenta on their mass-shell: $p^2 = k^2 = q^2 = M_{\mathrm{A}}^2$. The momentum conservation $p+k+q=0$ enables us to easily compute the dot products of external momenta: $p\cdot k = p\cdot q = k\cdot q = -\frac{1}{2}M_{\mathrm{A}}^2$. The amplitude \eqref{gba:T} then simplifies as
\begin{eqnarray}
\I T^{\mu\nu\rho}(p,k,q) &=& G_{\mathrm{eff}} \Big[
(q^\mu k^\alpha - k^\mu q^\alpha)p^\beta \epsilon^{\nu\rho}_{\hphantom{\nu\rho}\alpha\beta} +
(p^\nu q^\alpha - q^\nu p^\alpha)k^\beta \epsilon^{\rho\mu}_{\hphantom{\rho\mu}\alpha\beta} +
(k^\rho p^\alpha - p^\rho k^\alpha)q^\beta \epsilon^{\mu\nu}_{\hphantom{\mu\nu}\alpha\beta}
\Big] \,,
\nonumber \\ &&
\end{eqnarray}
which corresponds to effective Lagrangian
\begin{eqnarray}
\eL_{\mathrm{eff}} &=& G_{\mathrm{eff}} \, \epsilon_{\alpha\beta\gamma\delta} \big(\partial_\sigma A_{\mathrm{A}}^\alpha\big)\big(\partial^\beta A_{\mathrm{A}}^\sigma\big)\big(\partial^\gamma A_{\mathrm{A}}^\delta\big) \,.
\end{eqnarray}
Here the effective coupling constant $G_{\mathrm{eff}}$ can be expressed as
\begin{eqnarray}
G_{\mathrm{eff}} &=& g_{\mathrm{A}}^3 \sum_{i=1,2} Q_{i,\mathrm{A}}^3 \, f(m_i^2,M_{\mathrm{A}}^2) \,.
\end{eqnarray}
The function $f(m^2,M^2)$ is defined by the integral
\begin{eqnarray}
\label{chp:frm:fmM}
f(m^2,M^2) &\equiv& \frac{2}{\pi^2 M^2}\int_0^1 \d x \frac{x(1-x)}{\sqrt{x(3x-4)+\frac{4m^2}{M^2}}} \arctan\frac{x}{\sqrt{x(3x-4)+\frac{4m^2}{M^2}}}
\end{eqnarray}
(here $m^2$ should be replaced by $m^2-\I 0^+$ whenever the correct branch choice of a multivalued analytic function is in question), which can be calculated analytically in some special cases:
\begin{subequations}
\begin{eqnarray}
f(m^2,0) &=& \frac{1}{24 \pi^2 m^2} \,, \\
f(0,M^2) &=& \frac{-1}{6 \pi^2 M^2} \,.
\end{eqnarray}
\end{subequations}
More information about the shape of $f(m^2,M^2)$ can be extracted numerically, see Fig.~\ref{fig:plot_g}.

\section{Summary}

We did not directly assumed an Abelian symmetry (i.e., the mutually commuting generators), but rather started with generators commuting with the fermion self-energy. Already this simple assumption (allowing for satisfaction of the \emph{sine qua non} condition \eqref{gbm:Aab=0}) led to a significant simplification of the formula for the gauge boson mass matrix, derived in the previous chapter. Only after this we continued with the very assumption of Abelianity, due to the specific form of the generators assumed to be $1 \times 1$ matrices in the flavor space.

At this point we were ready to compare the results obtained so far with the corresponding results in the literature, namely with the Pagels--Stokar formula. We found a small discrepancy and associated it with the specific form of our vertex, ensuring the symmetricity of the gauge boson mass matrix (under the condition \eqref{gbm:Aab=0}) also for other than Abelian theories (for which the gauge boson mass matrix is incidentally symmetric in any case).

The results mentioned in the first paragraph made the following computation of the gauge boson masses within the gauged Abelian toy model from part~\ref{part:abel}, chapter~\ref{chp:frm}, fairly easy. We arrived at the expected result that of the three gauge bosons only one (corresponding to the broken subgroup $\group{U}(1)_{\mathrm{A}}$) obtained mass, while the other two (corresponding the unbroken subgroup $\group{U}(1)_{\mathrm{V}_{\!1}} \times \group{U}(1)_{\mathrm{V}_{\!2}}$) remained massless.

\chapter{Application to the electroweak interactions}
\label{ewM}

\intro{In this chapter we will calculate, using the procedure introduced in chapter~\ref{chp:gbm}, the gauge boson mass matrix for the electroweak theory in terms of the symmetry-breaking parts of the fermion propagators, considered in detail in part~\ref{part:ew}, and will show that the spectrum will be the expected one, i.e., containing one massless gauge boson (the photon) and three massive gauge bosons, two of which ($W^\pm$) with the same mass.}


\section{Expected form the of gauge boson mass matrix}

The electroweak symmetry $\group{SU}(2)_{\mathrm{L}} \times \group{U}(1)_{\mathrm{Y}}$ is assumed to be spontaneously broken by the fermion propagators down to the subgroup $\group{U}(1)_{\mathrm{em}}$:
\begin{eqnarray}
\group{SU}(2)_{\mathrm{L}} \times \group{U}(1)_{\mathrm{Y}} &\longrightarrow& \group{U}(1)_{\mathrm{em}} \,.
\end{eqnarray}
Thus, we expect that three of the four gauge bosons corresponding to the full group $\group{SU}(2)_{\mathrm{L}} \times \group{U}(1)_{\mathrm{Y}}$ will acquire a non-vanishing mass, while the fourth one (the photon), corresponding to unbroken subgroup $\group{U}(1)_{\mathrm{em}}$, will remain massless.

Before starting the very calculation, we will first investigate more precisely what form of the gauge boson mass matrix we expect to obtain. We will employ for this purpose three mutually independent assumptions: electromagnetic invariance of the gauge boson mass matrix, masslessness of the photon and symmetricity of gauge boson mass matrix.

\subsection{Electromagnetic invariance}

Let us first find the form of the electromagnetic generator $t_{\mathrm{em}}$, \eqref{ew1:rot:t34-Zew}, in the adjoint representation \eqref{gbp:adjoint}. Its matrix elements are given by
\begin{eqnarray}
\big(\mathcal{T}_{\mathrm{em}}\big)_{ab} &=& -\I f_{ab} \,,
\end{eqnarray}
where the coefficients $f_{ab}$ are defined as
\begin{eqnarray}
{[t_{\mathrm{em}},t_a]} &\equiv& \I f_{ab} \, t_b \,.
\end{eqnarray}
The particular commutators can be easily calculated, e.g., by taking $t_{a=1,2,3}$ to be given by the Pauli matrices, $t_{a=1,2,3}=\sigma_a$, and $t_{a=4}$ to be the unit matrix, $t_{a=4} = \unitmatrix_{2 \times 2}$, so that $t_{\mathrm{em}}$ has the form $t_{\mathrm{em}} = \sigma_3 \sin\theta_{\mathrm{W}} + \unitmatrix \cos\theta_{\mathrm{W}}$. One obtains
\begin{subequations}
\begin{eqnarray}
{[t_{\mathrm{em}},t_1]} &=& \phantom{-}2\I\sin\theta_{\mathrm{W}}\,t_2 \,, \\
{[t_{\mathrm{em}},t_2]} &=& -2\I\sin\theta_{\mathrm{W}}\,t_1 \,, \\
{[t_{\mathrm{em}},t_3]} &=& \phantom{-}0 \,, \\
{[t_{\mathrm{em}},t_4]} &=& \phantom{-}0 \,.
\end{eqnarray}
\end{subequations}
We therefore find
\begin{eqnarray}
\label{gbew:mathcalTem}
\mathcal{T}_{\mathrm{em}} &=& 2\I\sin\theta_{\mathrm{W}}
\left(\begin{array}{rrrr}
0 & -1 & 0 & 0 \\
1 &  0 & 0 & 0 \\
0 &  0 & 0 & 0 \\
0 &  0 & 0 & 0
\end{array}\right) \,.
\end{eqnarray}

Since the subgroup $\group{U}(1)_{\mathrm{em}}$ is unbroken, the polarization tensor, and consequently also the gauge boson mass matrix $M^2$, must be invariant under it. Operationally it means that $M^2$ must commute with $\mathcal{T}_{\mathrm{em}}$:
\begin{eqnarray}
\label{gbew:emInv}
\big[M^2,\mathcal{T}_{\mathrm{em}}\big] &=& 0 \,,
\end{eqnarray}
as can be inferred from \eqref{gbp:Pitransf}. If we now, upon taking into account the explicit form \eqref{gbew:mathcalTem} of $\mathcal{T}_{\mathrm{em}}$, apply this condition on the matrix $M^2$, we find that $M^2$ must have the block diagonal form
\begin{eqnarray}
\label{gbew:M2gen}
M^2 &=& \left(\begin{array}{cc} M_{W^\pm}^2 & 0 \\ 0 & M_{Z\gamma}^2 \end{array}\right) \,,
\end{eqnarray}
where $M_{W^\pm}^2$ is a $2 \times 2$ matrix of the special form
\begin{eqnarray}
\label{gbew:M2Wpm:gen}
M_{W^\pm}^2 &=& \left(\begin{array}{rr} A & B \\ -B & A \end{array}\right) \,,
\end{eqnarray}
while $M_{Z\gamma}^2$ is an arbitrary $2 \times 2$ matrix, i.e., of the general form
\begin{eqnarray}
\label{gbew:M2zg:gen}
M_{Z\gamma}^2 &=& \left(\begin{array}{rr} C & D \\ E & F \end{array}\right) \,,
\end{eqnarray}
and $A$, $B$, $C$, $D$, $E$, $F$ are arbitrary complex numbers.

\subsection{Masslessness of the photon}

We have found the most general form of the mass matrix $M^2$, consistent with the requirement \eqref{gbew:emInv} of the electromagnetic invariance. However, the fact that the subgroup $\group{U}(1)_{\mathrm{em}}$ is unbroken implies not only such invariance of $M^2$, but also vanishing of the mass of the gauge boson $A_{\mathrm{em}}^\mu$ (photon), corresponding to $\group{U}(1)_{\mathrm{em}}$. In fact, the masslessness of the photon is \emph{not guaranteed} by the electromagnetic invariance \eqref{gbew:emInv} of $M^2$: Notice, that the mass matrix $M_{Z\gamma}^2$ of photon and $Z$ boson, \eqref{gbew:M2zg:gen}, is electromagnetically invariant, but still its spectrum is virtually arbitrary and in particular it admits a massive photon. The masslessness of the photon must be therefore assumed independently. We are now going to show how it constrains the form of the matrix $M_{Z\gamma}^2$.

Let us first note that $M_{Z\gamma}^2$, \eqref{gbew:M2zg:gen}, can be without loss of generality expressed also in the form
\begin{eqnarray}
\label{gbew:M2zg:gen2}
M_{Z\gamma}^2 &=&
\left(\begin{array}{cc} g^2\,a & -gg^\prime\,b \\ -gg^\prime\,c & g^{\prime2}\,d \end{array}\right) \,,
\end{eqnarray}
where the coefficients $a$, $b$, $c$, $d$ are regular for $g = g^\prime = 0$. It can be understood by noting that the polarization tensor is calculated using the one-loop integral \eqref{gbm:Pimunuab} (cf.~also Fig.~\ref{gbm:fig:Pi}) with two vertices, where the vertex connected to $A_3^\mu$ or $A_4^\mu$ is proportional to $g$ or $g^\prime$, respectively.

The matrix $M_{Z\gamma}^2$, \eqref{gbew:M2zg:gen2}, is written in the basis $(A_3^\mu,A_4^\mu)$, hence we can denote it also as
\begin{eqnarray}
M_{Z\gamma}^2 &\equiv& M_{Z\gamma}^2\big|_{(3,4)}
\end{eqnarray}
to make the basis explicit. For the present considerations it will be however more convenient to have it expressed in the basis $(A_{Z}^\mu,A_{\mathrm{em}}^\mu)$, given in terms of the $(A_3^\mu,A_4^\mu)$ one by \eqref{ew1:rot:A34-Zew}. The matrix $M_{Z\gamma}^2$ can be therefore transformed from the basis $(A_3^\mu,A_4^\mu)$ to the basis $(A_Z^\mu,A_{\mathrm{em}}^\mu)$ as
\begin{eqnarray}
M_{Z\gamma}^2\big|_{(\mathrm{em},Z)} &=& O_{\mathrm{W}} \, M_{Z\gamma}^2\big|_{(3,4)} \, O_{\mathrm{W}}^\T \,.
\end{eqnarray}
Taking into account the explicit form \eqref{gbew:M2zg:gen2} of $M_{Z\gamma}^2|_{(3,4)}$, we find
\begin{eqnarray}
M_{Z\gamma}^2\big|_{(\mathrm{em},Z)} &=& \frac{1}{\sqrt{g^2+g^{\prime2}}}
\left(\begin{array}{cc}
g^4 a + g^2g^{\prime2}\big(b+c\big) + g^{\prime4} d & g^3g^\prime\big(a-b\big) + gg^{\prime3}\big(c-d\big) \\
g^3g^\prime\big(a-c\big) + gg^{\prime3}\big(b-d\big)  &  g^2g^{\prime2}\big(a-b-c+d\big)
\end{array}\right) \,.
\nonumber \\ &&
\end{eqnarray}

Now we can easily apply the assumption that the photon $A_{\mathrm{em}}^\mu$ must be massless. It implies that all $a$, $b$, $c$, $d$ must be the same:
\begin{equation}
a \ = \ b \ = \ c \ = \ d \,.
\end{equation}
(In fact, this results already from a less strong assumption that $M_{Z\gamma}^2|_{(\mathrm{em},Z)}$ is diagonal, i.e., that $A_{Z}^\mu$ and $A_{\mathrm{em}}^\mu$ are mass eigenstates. The masslessness of the photon then follows automatically.) Thus, upon defining
\begin{eqnarray}
\mu_{Z\gamma}^2 &\equiv& a \ = \ b \ = \ c \ = \ d \,,
\end{eqnarray}
the matrix $M_{Z\gamma}^2$ in the basis $(A_{Z}^\mu,A_{\mathrm{em}}^\mu)$ acquires the form
\begin{eqnarray}
\label{gbew:M2zg:genemZ}
M_{Z\gamma}^2\big|_{(\mathrm{em},Z)} &=&
\left(\begin{array}{cc} g^2+g^{\prime2} & 0 \\ 0 & 0 \end{array}\right)\mu_{Z\gamma}^2
\end{eqnarray}
and in the basis $(A_3^\mu,A_4^\mu)$ the form
\begin{eqnarray}
\label{gbew:M2zg:gen34}
M_{Z\gamma}^2\big|_{(3,4)} &=&
\left(\begin{array}{cc} g^2 & -gg^\prime \\ -gg^\prime & g^{\prime2} \end{array}\right) \mu_{Z\gamma}^2 \,.
\end{eqnarray}

\subsection{Symmetricity}

Finally, we assume that the mass matrix $M^2$ is symmetric:
\begin{eqnarray}
M^{2\T} &=& M^2 \,.
\end{eqnarray}
The part $M_{Z\gamma}^2$, \eqref{gbew:M2zg:gen34}, is already symmetric, hence this assumption applies non-trivially only on $M_{W^\pm}^2$, \eqref{gbew:M2Wpm:gen}, and yields
\begin{eqnarray}
B &=& 0 \,,
\end{eqnarray}
so that
\begin{eqnarray}
M_{W^\pm}^2 &=& \left(\begin{array}{cc} A & 0 \\ 0 & A \end{array}\right) \,.
\end{eqnarray}
We can now similarly as above argue that $A$ is proportional to $g^2$, so that $M_{W^\pm}^2$ can be expressed as
\begin{eqnarray}
\label{gbew:M2Wpm:gen12}
M_{W^\pm}^2 &\equiv& \left(\begin{array}{cc} g^2 & 0 \\ 0 & g^2 \end{array}\right) \mu_{W^\pm}^2 \,,
\end{eqnarray}
where $\mu_{W^\pm}^2$ is regular for $g=0$. Needless to stress that since $M_{W^\pm}^2$ is proportional to the unit matrix, it is invariant under any regular transformation and thus, in particular, it has the same form in both bases $(A^\mu_1,A^\mu_2)$ and $(A^\mu_{W^+},A^\mu_{W^-})$, related to each other by the unitary transformation \eqref{ew1:rot:A12-Wpm}.

\section{Quark contribution}
\label{gbew:sec:quarks}

Since the polarization tensor is calculated at one-loop level, it will be a sum of separate contributions from the quarks and the leptons:
\begin{eqnarray}
\Pi^{\mu\nu}_{ab}(q^2) &=& \Pi^{\mu\nu}_{ab}(q^2)\big|_{q} + \Pi^{\mu\nu}_{ab}(q^2)\big|_{\ell} \,.
\end{eqnarray}
This is in direct analogy with the expression \eqref{gba:M2absep} of the gauge boson mass matrix in the Abelian toy model as a sum of independent contributions from the two fermion species $\psi_1$ and $\psi_2$. In this section we will calculate the quark contribution $\Pi^{\mu\nu}_{ab}(q^2)|_{q}$, while the lepton contribution $\Pi^{\mu\nu}_{ab}(q^2)|_{\ell}$ is postponed to Sec.~\ref{gbew:sec:leptons}.

The quark contribution to the polarization tensor is given by
\begin{eqnarray}
\label{gbew:Piabmunu:q}
\I \Pi_{ab}^{\mu\nu}(q)\big|_{q} &=& - N_c \int\!\frac{\d^d p}{(2\pi)^d}\,
\Tr \Big\{ \Gamma^\mu_{q,a}(p+q,p) \, G_{q}(p) \, \gamma^\nu T_{q,b} \, G_{q}(p+q) \Big\} \,,
\end{eqnarray}
where $N_c=3$ is the number of colors. For the vertex we use the Ansatz \eqref{gbm:Gmm:ans} derived in Sec.~\ref{gbm:sec:construction}:
\begin{eqnarray}
\label{gbew:Gmm:q}
\Gamma_{q,a}^\mu(p^\prime,p) &=& \gamma^\mu T_{q,a}
-\frac{1}{2} \frac{q^\mu}{q^2} \Big[ \boldsymbol{\Sigma}_{q,+}\,T_{q,a} - \bar T_{q,a}\,\boldsymbol{\Sigma}_{q,+} \Big]
\nonumber \\ && \phantom{\gamma^\mu T_{q,a}}
-\bigg(\frac{1}{2} \frac{q^{\prime \mu}}{q \cdot q^\prime} - a_6 \frac{[\gamma^\mu,\slashed{q}]}{q \cdot q^\prime}\bigg)
\Big[ \boldsymbol{\Sigma}_{q,-}\,T_{q,a} + \bar T_{q,a} \,\boldsymbol{\Sigma}_{q,-} \Big]
\nonumber \\ && \phantom{\gamma^\mu T_{q,a}}
-\frac{1}{4} \frac{1}{d-1}
\bigg(\frac{q^\mu}{q^2}\frac{[\slashed{q},\slashed{q}^\prime]}{q \cdot q^\prime} - \frac{[\gamma^\mu,\slashed{q}^\prime]}{q \cdot q^\prime} \bigg)
\Big[ \boldsymbol{\Sigma}_{q,-}\,T_{q,a} - \bar T_{q,a}\,\boldsymbol{\Sigma}_{q,-} \Big] \,,
\end{eqnarray}
with the generators $T_{q,a}$ given by \eqref{ew1:Tqa}. We also use the notation
\begin{eqnarray}
\boldsymbol{\Sigma}_{q,\pm} &\equiv& \boldsymbol{\Sigma}_{q,p^\prime} \pm \boldsymbol{\Sigma}_{q,p} \,,
\end{eqnarray}
in agreement with \eqref{gbm:Sgmpm}. Recall that $a_6$ in \eqref{gbew:Gmm:q} is an undetermined parameter, which will nevertheless not enter the final formula for the mass matrix. All quantities (vertices and propagators) are here expressed in terms of the quark doublet field $q = \bigl(\begin{smallmatrix} u \\ d \end{smallmatrix}\bigr)$, Eq.~\eqref{ew1:q}. Since the vertex \eqref{gbew:Gmm:q} satisfies by constructions the WT identity
\begin{subequations}
\begin{eqnarray}
q_\mu \Gamma_{q,a}^\mu(p^\prime,p)
&=& G_q(p^\prime) \, T_{q,a} - \bar T_{q,a} \, G_q(p) \\
&=& \slashed{q} - \boldsymbol{\Sigma}_{q,p^\prime} \, T_{q,a} + \bar T_{q,a} \, \boldsymbol{\Sigma}_{q,p} \,,
\end{eqnarray}
\end{subequations}
the polarization tensor \eqref{gbew:Piabmunu:q} is transverse.

The polarization tensor $\Pi^{\mu\nu}_{ab}(q^2)|_{q}$ is a $4 \times 4$ matrix in the gauge space. It can be therefore considered as a $2 \times 2$ block matrix, with each block itself being also a $2 \times 2$ matrix. Now it is important to note that the off-diagonal blocks actually vanish, so that $\Pi^{\mu\nu}_{ab}(q^2)|_{q}$ can be written, upon suppressing the gauge indices, in the block matrix form
\begin{eqnarray}
\label{gbew:Piabmunu:q:blocks}
\Pi^{\mu\nu}(q^2)\big|_{q} &=&
\left(\begin{array}{cc}
\Pi^{\mu\nu}_{W^\pm}(q^2)\big|_{q} & 0 \\ 0 & \Pi^{\mu\nu}_{Z\gamma}(q^2)\big|_{q}
\end{array}\right) \,.
\end{eqnarray}
This is due to the fact that the trace over the two-dimensional electroweak space of a product of two generators in the integral \eqref{gbew:Piabmunu:q} (one sitting in the vertex $\Gamma^{\mu}_{q,a}(p+q,p)$ and the other being a part of the bare vertex $\gamma_\mu T_{q,b}$), with one being antisymmetric ($T_{q,1}$, $T_{q,2}$) and the other symmetric ($T_{q,3}$, $T_{q,4}$), is zero. There are also the fermion propagators (both full and 1PI), but as they are diagonal in the considered space, they do not affect the argument. The subscripts $W^\pm$ and $Z\gamma$ on the right-hand side of \eqref{gbew:Piabmunu:q:blocks} are to suggest that the corresponding quantities are the polarization tensors of the indicated gauge bosons, with no mixing between them. Since the whole $\Pi^{\mu\nu}_{ab}(q^2)|_{q}$ is transversal, so must be also the particular $\Pi^{\mu\nu}_{W^\pm,ab}(q^2)|_{q}$ and $\Pi^{\mu\nu}_{Z\gamma,ab}(q^2)|_{q}$.

In other words, the quark contribution to the gauge boson mass matrix is indeed of the expected form \eqref{gbew:M2gen}. We can therefore now treat the quark contribution to the masses of $W^\pm$ and $Z$ separately.

\subsection{Masses of $W^\pm$}

The quark contribution $M_{W^\pm}^2|_{q}$ to the mass matrix $M^2_{W^\pm}$, \eqref{gbew:M2Wpm:gen12}, of $W^\pm$ is now given by the polarization tensor \eqref{gbew:Piabmunu:q} with gauge indices restricted to $a,b=1,2$, by means of the pole approximation \eqref{gbm:PoleAppr} described in detail in chapter~\ref{chp:gbm}. In order to use the corresponding general formula \eqref{gbm:M2ab} for the gauge boson mass matrix, we have first to check satisfaction of the condition $A_{ab}=0$, \eqref{gbm:Aab=0}, where $A_{ab}$ is given by
\begin{eqnarray}
\label{gbew:Aab:q}
A_{ab} &=&
\Tr \bigg\{
\phantom{+\,}
T_{q,a}\,\boldsymbol{\Sigma}_{q}^\dag\,\boldsymbol{D}_{qL}^\prime\,\bar T_{q,b}\,\boldsymbol{\Sigma}_{q}\,\boldsymbol{D}_{qR}
-
T_{q,a}\,\boldsymbol{\Sigma}_{q}^\dag\,\boldsymbol{D}_{qL}\,\bar T_{q,b}\,\boldsymbol{\Sigma}_{q}\,\boldsymbol{D}_{qR}^\prime
\nonumber \\ &&
\phantom{\Tr \bigg\{}
\hspace{-0.45em}
+\,
\bar T_{q,a}\,\boldsymbol{D}_{qL}\,\boldsymbol{\Sigma}_{q}\, T_{q,b}\,\boldsymbol{D}_{qR}^\prime\,\boldsymbol{\Sigma}_{q}^\dag
-
\bar T_{q,a}\,\boldsymbol{D}_{qL}^\prime\,\boldsymbol{\Sigma}_{q}\, T_{q,b}\,\boldsymbol{D}_{qR}\,\boldsymbol{\Sigma}_{q}^\dag
\bigg\} \,.
\end{eqnarray}
Notice in particular that in each of the four terms in \eqref{gbew:Aab:q} there is one generator without the bar ($T_{q,a}$ or $T_{q,b}$) and one with the bar ($\bar T_{q,a}$ or $\bar T_{q,b}$). Recall now the form \eqref{ew1:Tq123} of the generators $T_{q,a}$ with $a=1,2$: $T_{q,a} = g \tfrac{\sigma_a}{2} P_L$, implying $\bar T_{q,a} = g \tfrac{\sigma_a}{2} P_R$. Now since the chiral projectors $P_L$, $P_R$ commutes with anything in \eqref{gbew:Aab:q} and due to their property $P_L \, P_R = 0$, we conclude that each of the four terms in \eqref{gbew:Aab:q} vanishes and the condition \eqref{gbm:Aab=0} is indeed fulfilled.

We can now plug the quark self-energy $\boldsymbol{\Sigma}_{q}$ (given by \eqref{ewa:Sgmqdiag} and \eqref{ewa:Sgmf}) and the generators $T_{q,1}$, $T_{q,2}$ into the formula \eqref{gbm:M2ab} and we arrive at the result of the expected form (cf.~\eqref{gbew:M2Wpm:gen12})
\begin{eqnarray}
M_{W^\pm}^2\big|_{q} &=&
\left(\begin{array}{cc} g^2 & 0 \\ 0 & g^2 \end{array}\right) \mu_{W^\pm}^2\big|_{q} \,,
\end{eqnarray}
where
\begin{eqnarray}
\label{gbew:mu2:Wpm:q}
\mu_{W^\pm}^2\big|_{q} &=&
\nonumber \\ &&
\hspace{-5em}
-\I\frac{1}{2} N_c \int\!\frac{\d^d p}{(2\pi)^d}\, \Tr\bigg\{
\Big[
\big(\Sigma_u^{\vphantom{\dag}}\,\Sigma_u^\dag\big) - \frac{2}{d} p^2
\big(\Sigma_u^{\vphantom{\dag}}\,\Sigma_u^\dag\big)^\prime
\Big]
D_{dL}\,D_{uL}
-\frac{2}{d} p^2 \big(\Sigma_u^{\vphantom{\dag}}\,\Sigma_u^\dag\big)
\Big[
D_{dL}^{\vphantom{\prime}}\,D_{uL}^{          \prime} -
D_{dL}^{          \prime} \,D_{uL}^{\vphantom{\prime}}
\Big]
\nonumber \\ && \hphantom{-\I\frac{1}{2} N_c \int\!\frac{\d^d p}{(2\pi)^d} \, \Tr\bigg\{}
\hspace{-5mm}
\hspace{-5em}
+
\Big[
\big(\Sigma_d^{\vphantom{\dag}}\,\Sigma_d^\dag\big) - \frac{2}{d} p^2
\big(\Sigma_d^{\vphantom{\dag}}\,\Sigma_d^\dag\big)^\prime
\Big]
D_{uL}\,D_{dL}
-\frac{2}{d} p^2 \big(\Sigma_d^{\vphantom{\dag}}\,\Sigma_d^\dag\big)
\Big[
D_{uL}^{\vphantom{\prime}}\,D_{dL}^{          \prime} -
D_{uL}^{          \prime} \,D_{dL}^{\vphantom{\prime}}
\Big]
\bigg\} \,.
\nonumber \\ &&
\end{eqnarray}

\subsection{Masses of $Z$ and $\gamma$}

Let us continue with the quark contribution $M_{Z\gamma}^2|_{q}$ to the $Z$ and $\gamma$ mass matrix \eqref{gbew:M2zg:gen34}, which is given by the polarization tensor \eqref{gbew:Piabmunu:q} with $a,b=3,4$. We can make the following observation: Since both the generators $T_{q,3}$, $T_{q,4}$, \eqref{ew1:Tq34diag}, and the self-energy $\boldsymbol{\Sigma}_q$, \eqref{ewa:Sgmqdiag}, are diagonal in the two-dimensional space of the quark species (up-type and down-type), so is for $a=3,4$ also the vertex $\Gamma_{q,a}^\mu(p^\prime,p)$, \eqref{gbew:Gmm:q}, itself:
\begin{eqnarray}
\Gamma_{q,a}^\mu(p^\prime,p) &\equiv&
\left(\begin{array}{cc} \Gamma_{u,a}^\mu(p^\prime,p) & 0 \\ 0 & \Gamma_{d,a}^\mu(p^\prime,p) \end{array}\right) \,.
\end{eqnarray}
Consequently, the contributions from up-type and down-type quarks to the gauge boson polarization tensor decouple and the polarization tensor can be written as
\begin{eqnarray}
\Pi^{\mu\nu}_{Z\gamma,ab}(q^2)\big|_{q} &=& \sum_{f=u,d} \Pi^{\mu\nu}_{Z\gamma,ab}(q^2)\big|_{f} \,.
\end{eqnarray}
Here each $\Pi_{Z\gamma,ab}^{\mu\nu}(q)|_{f}$ is given by
\begin{eqnarray}
\I \Pi_{Z\gamma,ab}^{\mu\nu}(q)\big|_{f} &=& - N_c \int\!\frac{\d^d p}{(2\pi)^d}\,
\Tr \Big\{ \Gamma^\mu_{f,a}(p+q,p) \, G_{f}(p) \, \gamma^\nu T_{f,b} \, G_{f}(p+q) \Big\} \,,
\end{eqnarray}
where $G_f(p)$ and $T_{f,b}$ are given by \eqref{ewa:Gfinv} and \eqref{ew1:qTf34}, respectively. The vertex $\Gamma^\mu_{f,a}(p+q,p)$ is given by the formula \eqref{gbew:Gmm:q} with the subscript $q$ changed to $f=u,d$ and since it satisfies the corresponding WT identity, $\Pi_{Z\gamma,ab}^{\mu\nu}(q)|_{f}$ is transversal.

Recall the explicit form \eqref{ew1:qTf34} of the generators $T_{f,3}$, $T_{f,4}$. They are just real linear combinations of $\unitmatrix$ and $\gamma_5$, i.e., they are of the same special form which was considered in section~\ref{gba:ssec:U1N}. Thus, using the results from there (including fulfilment of the condition \eqref{gbm:Aab=0}), we readily obtain the quark contribution $M_{Z\gamma}^2|_{q}$ in the form
\begin{eqnarray}
M_{Z\gamma}^2\big|_{q} &=&
\left(\begin{array}{cc}
g^2          &  -g g^\prime   \\
-g g^\prime  &  g^{\prime 2}
\end{array}\right)
\sum_{f=u,d}\mu_{Z\gamma}^2\big|_{f} \,,
\end{eqnarray}
in agreement with the desired general form \eqref{gbew:M2zg:gen34}, ensuring the masslessness of the photon. The parameters $\mu_{Z\gamma}^2|_{f}$ are
\begin{eqnarray}
\label{gbew:mu2:gZ:f}
\mu_{Z\gamma}^2\big|_{f} &=& -\I\frac{1}{2} N_c \int\!\frac{\d^d p}{(2\pi)^d}\,\Tr
\bigg\{
\Big[
\big(\Sigma_f^{\vphantom{\dag}}\,\Sigma_f^\dag\big) - \frac{2}{d} p^2
\big(\Sigma_f^{\vphantom{\dag}}\,\Sigma_f^\dag\big)^\prime
\Big]
D_{fL}^{2\vphantom{\dag}}
\bigg\} \,,
\end{eqnarray}
which is, up to a factor, of the same form as the analogous parameter $\mu^2|_i$, \eqref{gba:mu2i}, in the Abelian toy model.

\subsection{Comparison with the Pagels--Stokar formula}

The formul{\ae} similar to those \eqref{gbew:mu2:Wpm:q}, \eqref{gbew:mu2:gZ:f} for $\mu_{W^\pm}^2|_{q}$, $\mu_{Z\gamma}^2|_{f=u,d}$ have been already presented in the literature \cite{Miransky:1988xi}. They have been derived as a straightforward generalization of the Pagels--Stokar result \cite{Pagels:1979hd}.

We have already encountered similar situation in Sec.~\ref{gba:ssec:PS}, when we compared our result to the PS result \cite{Pagels:1979hd} and discussed the discrepancy between them. In fact, now the problem is exactly the same: Our results \eqref{gbew:mu2:Wpm:q}, \eqref{gbew:mu2:gZ:f} do not correspond to those from Ref.~\cite{Miransky:1988xi} and the reason lies again in the different choice of the parameter $a_4$ of the vertex \eqref{gbm:Gmm:ansNG}. While we set $a_4$ to the value \eqref{gbm:a4}, the results in \cite{Miransky:1988xi} correspond to the vanishing value $a_4=0$.

Let us discuss in more detail this issue for the coefficient $\mu_{W^\pm}^2|_{q}$, \eqref{gbew:mu2:Wpm:q}. (The discussion of the coefficient $\mu_{Z\gamma}^2|_{f=u,d}$, \eqref{gbew:mu2:gZ:f}, would be exactly the same as in Sec.~\ref{gbew:mu2:gZ:f}.) Assume now that $a_4$ in the Ansatz \eqref{gbm:Gmm:ansNG} has a general, undetermined value. Then the quark contribution $M_{W^\pm}^2|_{q}$ can be written as
\begin{eqnarray}
\label{gbew:M2qa4}
M_{W^\pm}^2\big|_{q} &=&  \left(\begin{array}{cc} g^2 & 0 \\ 0 & g^2 \end{array}\right) \big(\mu_{W^\pm}^2\big|_{q} + \mu_{W^\pm}^2\big|_{a_4/\mathrm{S}}\big) +
\left(\begin{array}{cc} 0 & g^2 \\ -g^2 & 0 \end{array}\right) \mu_{W^\pm}^2\big|_{a_4/\mathrm{A}} \,,
\end{eqnarray}
where $\mu_{W^\pm}^2|_{q}$ is the contribution already computed in \eqref{gbew:mu2:Wpm:q} and the coefficients $\mu_{W^\pm}^2|_{a_4/\mathrm{S}}$, $\mu_{W^\pm}^2|_{a_4/\mathrm{A}}$ of the symmetric and antisymmetric part of $M_{W^\pm}^2|_{q}$ are given as
\begin{subequations}
\begin{eqnarray}
\mu_{W^\pm}^2\big|_{a_4/\mathrm{S}} &=& -\I\frac{1}{2} N_c \big(1+4(d-1)a_4\big) \int\!\frac{\d^d p}{(2\pi)^d}\,\Tr\bigg\{
\frac{1}{d} p^2
\Big[\Sigma_u^{\vphantom{\dag}}\,\Sigma_u^{\dag\prime} + \Sigma_d^{\vphantom{\dag}\prime}\,\Sigma_d^{\dag}\Big]
D_{dL}\,D_{uL}
\nonumber \\ &&
\phantom{-\I\frac{1}{2} N_c \big(1+4(d-1)a_4\big) \int\!\frac{\d^d p}{(2\pi)^d}\,\Tr\bigg\{}
\hspace{-1.4em}
+\,
\frac{1}{d} p^2
\Big[\Sigma_u^{\vphantom{\dag}\prime}\,\Sigma_u^{\dag} + \Sigma_d^{\vphantom{\dag}}\,\Sigma_d^{\dag\prime}\Big]
D_{uL}\,D_{dL}
\bigg\} \,,
\\
\mu_{W^\pm}^2\big|_{a_4/\mathrm{A}} &=& -\I\frac{1}{2} N_c \big(1+4(d-1)a_4\big) \int\!\frac{\d^d p}{(2\pi)^d}\,\Tr\bigg\{
\I\frac{1}{d} p^2
\Big[\Sigma_u^{\vphantom{\dag}}\,\Sigma_u^{\dag\prime} + \Sigma_d^{\vphantom{\dag}\prime}\,\Sigma_d^{\dag}\Big]
D_{dL}\,D_{uL}
\nonumber \\ &&
\phantom{-\I\frac{1}{2} N_c \big(1+4(d-1)a_4\big) \int\!\frac{\d^d p}{(2\pi)^d}\,\Tr\bigg\{}
\hspace{-1.4em}
-\,
\I\frac{1}{d} p^2
\Big[\Sigma_u^{\vphantom{\dag}\prime}\,\Sigma_u^{\dag} + \Sigma_d^{\vphantom{\dag}}\,\Sigma_d^{\dag\prime}\Big]
D_{uL}\,D_{dL}
\bigg\} \,.\qquad\qquad
\end{eqnarray}
\end{subequations}
One can see that antisymmetric part of $M_{W^\pm}^2|_{q}$, proportional to $\mu_{W^\pm}^2|_{a_4/\mathrm{A}}$, is indeed in general non-vanishing, unless one sets $a_4$ as in \eqref{gbm:a4}.

Consider, however, the case of only one fermion generation. In such a case the self-energies $\Sigma_u$, $\Sigma_d$ (and consequently also $D_{uL}$, $D_{dL}$) are just complex numbers, without any matrix structure, and thus commuting. The coefficients $\mu_{W^\pm}^2|_{a_4/\mathrm{S}}$ and $\mu_{W^\pm}^2|_{a_4/\mathrm{A}}$ then simplify as
\begin{subequations}
\begin{eqnarray}
\mu_{W^\pm}^2\big|_{a_4/\mathrm{S}} &=& -\I\frac{1}{2} N_c \big(1+4(d-1)a_4\big) \int\!\frac{\d^d p}{(2\pi)^d} \,\frac{1}{d} p^2
\Big(|\Sigma_u|^{2\prime}+|\Sigma_d|^{2\prime}\Big) D_u D_d \,,
\\
\mu_{W^\pm}^2\big|_{a_4/\mathrm{A}} &=& -\I\frac{1}{2} N_c \big(1+4(d-1)a_4\big) \int\!\frac{\d^d p}{(2\pi)^d} \,\I\frac{1}{d} p^2 \Big(
\big[\Sigma_u^{\vphantom{*}}\,\Sigma_u^{*\prime} - \Sigma_u^{*}\,\Sigma_u^{\vphantom{*}\prime}\big]
\nonumber \\ && \hspace{15.7em}
+
\big[\Sigma_d^{*}\,\Sigma_d^{\vphantom{*}\prime} - \Sigma_d^{\vphantom{*}}\,\Sigma_d^{*\prime}\big]\Big) D_u D_d \,,
\qquad\qquad
\label{gbew:mu2:a4A}
\end{eqnarray}
\end{subequations}
where
\begin{subequations}
\begin{eqnarray}
D_u &\equiv& \frac{1}{p^2-|\Sigma_u|^2} \,, \\
D_d &\equiv& \frac{1}{p^2-|\Sigma_d|^2} \,.
\end{eqnarray}
\end{subequations}
Consider now further simplifying assumption of real self-energies:
\begin{subequations}
\begin{eqnarray}
\Sigma_u &=& \Sigma_u^* \,, \\ \Sigma_d &=& \Sigma_d^* \,.
\end{eqnarray}
\end{subequations}
Under this assumption each of the two square brackets in \eqref{gbew:mu2:a4A} vanishes and consequently $\mu_{W^\pm}^2|_{a_4/\mathrm{A}}$ vanishes too:
\begin{eqnarray}
\mu_{W^\pm}^2\big|_{a_4/\mathrm{A}} &=& 0 \,.
\end{eqnarray}
We stress that this happens for any value of $a_4$. The mass matrix $M_{W^\pm}^2|_{q}$, \eqref{gbew:M2qa4}, now acquires the symmetric form
\begin{eqnarray}
M_{W^\pm}^2\big|_{q} &=&  \left(\begin{array}{cc} g^2 & 0 \\ 0 & g^2 \end{array}\right) \bar\mu_{W^\pm}^2\big|_{q} \,,
\end{eqnarray}
where
\begin{subequations}
\begin{eqnarray}
\bar\mu_{W^\pm}^2\big|_{q} &\equiv& \mu_{W^\pm}^2\big|_{q} + \mu_{W^\pm}^2\big|_{a_4/\mathrm{S}}
\\ &=&
-\I\frac{1}{2} N_c \int\!\frac{\d^d p}{(2\pi)^d}\,\Tr\bigg\{
\Big[\big(\Sigma_u^2+\Sigma_d^2\big)-\frac{1}{d}\big(1-4(d-1)a_4\big)p^2\big(\Sigma_u^{2\prime}+\Sigma_d^{2\prime}\big)\Big] D_u D_d
\nonumber \\ &&
\phantom{-\I\frac{1}{2} N_c \int\!\frac{\d^d p}{(2\pi)^d}\,\Tr\bigg\{}
-\frac{2}{d}p^2\big(\Sigma_u^2-\Sigma_d^2\big)\big(D_u^\prime D_d - D_u D_d^\prime\big)
\bigg\} \,.
\end{eqnarray}
\end{subequations}
If we now set $a_4=0$, we reproduce the result from \cite{Miransky:1988xi}. We stress again that this result is not correct, as it holds (i.e., seems to be correct in the sense that the gauge boson mass matrix is symmetric) only in the very special case of $\Sigma_u$, $\Sigma_d$ being \emph{real numbers}. Once one considers a more general case of $\Sigma_u$, $\Sigma_d$ being either \emph{complex} or \emph{matrices} (or both), setting $a_4$ to be anything but the unique non-vanishing value \eqref{gbm:a4} gives wrong results (i.e., non-symmetric gauge boson mass matrix).

\section{Lepton contribution}
\label{gbew:sec:leptons}

The lepton contribution to the polarization tensor reads
\begin{eqnarray}
\label{gbew:Piabmunu:ell}
\I \Pi_{ab}^{\mu\nu}(q)\big|_{\ell} &=& - \frac{1}{2} \int\!\frac{\d^d p}{(2\pi)^d}\,
\Tr \Big\{ \Gamma^\mu_{\Psi_\ell,a}(p+q,p) \, G_{\Psi_\ell}(p) \, \gamma^\nu T_{\Psi_\ell,b} \, G_{\Psi_\ell}(p+q) \Big\} \,,
\end{eqnarray}
where the vertex is given by
\begin{eqnarray}
\label{gbew:Gmm:ell}
\Gamma_{\Psi_{\ell},a}^\mu(p^\prime,p) &=& \gamma^\mu T_{{\Psi_\ell},a}
-\frac{1}{2} \frac{q^\mu}{q^2}
\Big[                         \boldsymbol{\Sigma}_{{\Psi_\ell},+}\,T_{{\Psi_\ell},a}
-     \bar T_{{\Psi_\ell},a}\,\boldsymbol{\Sigma}_{{\Psi_\ell},+}                   \Big]
\nonumber \\ &&
\phantom{\gamma^\mu T_{{\Psi_\ell},a}}
-\bigg(\frac{1}{2} \frac{q^{\prime \mu}}{q \cdot q^\prime} - a_6 \frac{[\gamma^\mu,\slashed{q}]}{q \cdot q^\prime}\bigg)
\Big[        \boldsymbol{\Sigma}_{{\Psi_\ell},-}\,T_{{\Psi_\ell},a}
+   \bar T_{{\Psi_\ell},a}\,\boldsymbol{\Sigma}_{{\Psi_\ell},-}     \Big]
\nonumber \\ && \phantom{\gamma^\mu T_{{\Psi_\ell},a}}
-\frac{1}{4} \frac{1}{d-1}
\bigg(\frac{q^\mu}{q^2}\frac{[\slashed{q},\slashed{q}^\prime]}{q \cdot q^\prime} - \frac{[\gamma^\mu,\slashed{q}^\prime]}{q \cdot q^\prime} \bigg)
\Big[        \boldsymbol{\Sigma}_{{\Psi_\ell},-}\,T_{{\Psi_\ell},a}
-   \bar T_{{\Psi_\ell},a}\,\boldsymbol{\Sigma}_{{\Psi_\ell},-}     \Big]
\qquad\qquad
\end{eqnarray}
and where
\begin{eqnarray}
\boldsymbol{\Sigma}_{{\Psi_\ell},\pm} &\equiv& \boldsymbol{\Sigma}_{{\Psi_\ell},p^\prime}\pm\boldsymbol{\Sigma}_{{\Psi_\ell},p}
\end{eqnarray}
as usual. Notice that all quantities under the trace in \eqref{gbew:Piabmunu:ell} are written in the Nambu--Gorkov basis $\Psi_\ell$, \eqref{ew1:Psi}, which is a real field, therefore there is the extra factor of $1/2$ in \eqref{gbew:Piabmunu:ell}. Since the vertex \eqref{gbew:Gmm:ell} satisfies the WT identity, the polarization tensor $\Pi_{ab}^{\mu\nu}(q)|_{\ell}$ is transversal.

The generators $T_{{\Psi_\ell},1}$, $T_{{\Psi_\ell},2}$ are off-diagonal in the two-dimensional electroweak space, while $T_{{\Psi_\ell},3}$, $T_{{\Psi_\ell},4}$ are diagonal, see \eqref{ew1:TPsi}. Since the propagators are in this space diagonal too, we conclude, on the basis of the same arguments as in the quark case, that $\Pi_{ab}^{\mu\nu}(q^2)|_{\ell}$ has the block-diagonal form
\begin{eqnarray}
\Pi^{\mu\nu}(q^2)\big|_{\ell} &=&
\left(\begin{array}{cc} \Pi^{\mu\nu}_{W^\pm}(q^2)\big|_{\ell} & 0 \\ 0 & \Pi^{\mu\nu}_{Z\gamma}(q^2)\big|_{\ell} \end{array}\right) \,,
\end{eqnarray}
with each block being a $2 \times 2$ matrix in the gauge space. It follows that both particular polarization tensors $\Pi^{\mu\nu}_{ab,W^\pm}(q^2)|_{\ell}$ and $\Pi^{\mu\nu}_{ab,Z\gamma}(q^2)|_{\ell}$ are transversal. Again, the lepton contribution to the gauge boson mass matrix is therefore of the form \eqref{gbew:M2gen}.

\subsection{Masses of $W^\pm$}


The lepton contribution $M^2_{W^\pm}|_{\ell}$ to the $W^\pm$ mass matrix \eqref{gbew:M2Wpm:gen12} is calculated in completely the same way as the quark contribution $M^2_{W^\pm}|_{q}$ above, i.e., using the formula \eqref{gbm:M2ab} for the gauge boson mass matrix, with gauge indices $a,b=1,2$. Before using it, however, we have to check the condition \eqref{gbm:Aab=0}, i.e., vanishing of the quantity
\begin{eqnarray}
\label{gbew:Aab:ell}
A_{ab} &=& \Tr \bigg\{\phantom{+\,}
T_{\Psi_\ell,a}\,\boldsymbol{\Sigma}_{\Psi_\ell}^\dag\,\boldsymbol{D}_{\Psi_\ell}^\prime\,\bar T_{\Psi_\ell,b}\,\boldsymbol{\Sigma}_{\Psi_\ell}\,\boldsymbol{D}_{\Psi_\ell}^\C
-
T_{\Psi_\ell,a}\,\boldsymbol{\Sigma}_{\Psi_\ell}^\dag\,\boldsymbol{D}_{\Psi_\ell}\,\bar T_{\Psi_\ell,b}\,\boldsymbol{\Sigma}_{\Psi_\ell}\,\boldsymbol{D}_{\Psi_\ell}^{\C\prime}
\nonumber \\ &&
\phantom{\Tr \bigg\{}
\hspace{-0.45em}
+\,
\bar T_{\Psi_\ell,a}\,\boldsymbol{D}_{\Psi_\ell}\,\boldsymbol{\Sigma}_{\Psi_\ell}\, T_{\Psi_\ell,b}\,\boldsymbol{D}_{\Psi_\ell}^{\C\prime}\,\boldsymbol{\Sigma}_{\Psi_\ell}^\dag
-
\bar T_{\Psi_\ell,a}\,\boldsymbol{D}_{\Psi_\ell}^\prime\,\boldsymbol{\Sigma}_{\Psi_\ell}\, T_{\Psi_\ell,b}\,\boldsymbol{D}_{\Psi_\ell}^\C\,\boldsymbol{\Sigma}_{\Psi_\ell}^\dag
\bigg\} \,.
\qquad
\end{eqnarray}
Recall first the form of the relevant generators $T_{{\Psi_\ell},1}$, $T_{{\Psi_\ell},2}$:
\begin{eqnarray}
T_{{\Psi_\ell},a} &=&  \left(\begin{array}{cc} 0 & \tau_{{\Psi_\ell},a}^\dag P_{+}^\T \\ \tau_{{\Psi_\ell},a} P_{+} & 0 \end{array}\right) \,,
\end{eqnarray}
where $P_{+}$ is given by \eqref{ew1:P+} and where the $\tau_{{\Psi_\ell},a}$ are denotations for
\begin{subequations}
\begin{eqnarray}
\tau_{{\Psi_\ell},1} &\equiv& \,-\gamma_5\phantom{\I} g \frac{1}{2} \,,  \\
\tau_{{\Psi_\ell},2} &\equiv& \phantom{-\gamma_5} \I \, g \frac{1}{2} \,.
\end{eqnarray}
\end{subequations}
(See also definitions \eqref{ew1:Tpsi1} and \eqref{ew1:Tpsi2} of $T_{{\Psi_\ell},1}$ and $T_{{\Psi_\ell},2}$, respectively.) Recall also that $\boldsymbol{\Sigma}_{\Psi_\ell}$, \eqref{ewa:SgmPsiell}, is diagonal in the electroweak space and so is thus also $\boldsymbol{D}_{\Psi_\ell}$, \eqref{ewa:bldDPsiell}. The first term in \eqref{gbew:Aab:ell} can be therefore expanded as
\begin{eqnarray}
\Tr \bigg\{
T_{\Psi_\ell,a}\,\boldsymbol{\Sigma}_{\Psi_\ell}^\dag\,\boldsymbol{D}_{\Psi_\ell}^\prime\,\bar T_{\Psi_\ell,b}\,\boldsymbol{\Sigma}_{\Psi_\ell}\,\boldsymbol{D}_{\Psi_\ell}^\C
\bigg\}
&=&
\phantom{+\,}
\Tr \bigg\{
\bar \tau_{{\Psi_\ell},b} \, P_{+}^\T\, \boldsymbol{\Sigma}_{\Psi_e}\,\boldsymbol{D}_{\Psi_e}^\C\,
P_{+} \, \tau_{{\Psi_\ell},a} \, \boldsymbol{\Sigma}_{\Psi_\nu}^\dag\,\boldsymbol{D}_{\Psi_\nu}^\prime
\bigg\}
\nonumber \\ &&
+\,
\Tr \bigg\{
\tau_{{\Psi_\ell},a}^\dag \, P_{+}^\T\, \boldsymbol{\Sigma}_{\Psi_e}^\dag\,\boldsymbol{D}_{\Psi_e}^\prime\,
P_{+} \, \bar \tau_{{\Psi_\ell},b}^\dag \, \boldsymbol{\Sigma}_{\Psi_\nu}\,\boldsymbol{D}_{\Psi_\nu}^\C
\bigg\} \,.
\label{gbew:ell1stterm}
\qquad\qquad
\end{eqnarray}
Recall now that $\boldsymbol{\Sigma}_{\Psi_e}$, \eqref{ewa:SgmPsif} and \eqref{ewa:SgmPsie}, is off-diagonal in the 2-dimensional Nambu--Gorkov space of $\Psi_e$, \eqref{ew1:Psie}. Therefore $\boldsymbol{D}_{\Psi_e}^\C = \big(p^2-\boldsymbol{\Sigma}_{\Psi_e}^\dag\,\boldsymbol{\Sigma}_{\Psi_e}^{\vphantom{\dag}}\big)^{-1}$ is in this space diagonal and $\boldsymbol{\Sigma}_{\Psi_e}\,\boldsymbol{D}_{\Psi_e}^\C$ is again off-diagonal. Since the only non-vanishing block element of $P_{+}$ is the upper left one (we consider $P_{+}$ to be $2 \times 2$ block matrix in the Nambu--Gorkov space), we arrive at
\begin{subequations}
\begin{eqnarray}
P_{+}^\T\, \boldsymbol{\Sigma}_{\Psi_e}\,\boldsymbol{D}_{\Psi_e}^\C\, P_{+} &=&  0 \,.
\end{eqnarray}
One can find analogously
\begin{eqnarray}
P_{+}^\T\, \boldsymbol{\Sigma}_{\Psi_e}^\dag\,\boldsymbol{D}_{\Psi_e}^\prime\, P_{+} &=&  0 \,.
\end{eqnarray}
\end{subequations}
Thus the quantity \eqref{gbew:ell1stterm}, i.e., the first term in $A_{ab}$, \eqref{gbew:Aab:ell}, vanishes. Similarly can be treated also the remaining three terms in \eqref{gbew:Aab:ell} and shown to be vanishing as well. Thus, we conclude that the condition \eqref{gbm:Aab=0} is indeed fulfilled.

We can now freely use the formula \eqref{gbm:M2ab} to find
\begin{eqnarray}
M_{W^\pm}^2\big|_{\ell} &=&  \left(\begin{array}{cc} g^2 & 0 \\ 0 & g^2 \end{array}\right) \mu_{W^\pm}^2\big|_{\ell} \,,
\end{eqnarray}
where $\mu_{W^\pm}^2|_{\ell}$ can be expressed, e.g., as
\begin{subequations}
\label{gbew:mu2:Wpm:ell}
\begin{eqnarray}
\mu_{W^\pm}^2\big|_{\ell} &=&
\nonumber \\ &&
\hspace{-3em}
-\I\frac{1}{2} \int\!\frac{\d^d p}{(2\pi)^d}\,\Tr\bigg\{
\Big[
\big( \Sigma_{\Psi_\nu,M}^{\vphantom{\dag}}\,\Sigma_{\Psi_\nu,M}^{\dag} \big) - \frac{2}{d}p^2
\big( \Sigma_{\Psi_\nu,M}^{\vphantom{\dag}}\,\Sigma_{\Psi_\nu,M}^{\dag} \big)^\prime
\Big]
\big(P_+^\T\,D_{\Psi_e}\big)\big(P_+^{\vphantom{\T}}\,D_{\Psi_\nu}\big)
\nonumber \\ &&
\hspace{6em}
{}- \frac{2}{d}p^2
\big( \Sigma_{\Psi_\nu,M}^{\vphantom{\dag}}\,\Sigma_{\Psi_\nu,M}^{\dag} \big)
\Big[
\big(P_+^\T\,D_{\Psi_e}\big)         \big(P_+^{\vphantom{\T}}\,D_{\Psi_\nu}\big)^{\prime} -
\big(P_+^\T\,D_{\Psi_e}\big)^{\prime}\big(P_+^{\vphantom{\T}}\,D_{\Psi_\nu}\big)
\Big]
\nonumber \\ &&
\hspace{3.55em}
{}+
\Big[
\big( \Sigma_{\Psi_e}^{\vphantom{\dag}}\,\Sigma_{\Psi_e}^{\dag} \big) - \frac{2}{d}p^2
\big( \Sigma_{\Psi_e}^{\vphantom{\dag}}\,\Sigma_{\Psi_e}^{\dag} \big)^\prime
\Big]
\big(P_+^{\vphantom{\T}}\,D_{\Psi_\nu}\big)\big(P_+^\T \, D_{\Psi_e}\big)
\nonumber \\ &&
\hspace{6em}
{}- \frac{2}{d}p^2
\big( \Sigma_{\Psi_e}^{\vphantom{\dag}}\,\Sigma_{\Psi_e}^{\dag} \big)
\Big[
\big(P_+^{\vphantom{\T}}\,D_{\Psi_\nu}\big)         \big(P_+^\T \, D_{\Psi_e}\big)^{\prime} -
\big(P_+^{\vphantom{\T}}\,D_{\Psi_\nu}\big)^{\prime}\big(P_+^\T \, D_{\Psi_e}\big)
\Big]
\bigg\} \,.
\nonumber \\ &&
\end{eqnarray}
This form of $\mu_{W^\pm}^2|_{\ell}$ is relatively compact and elegant in the sense that the charged leptons and neutrinos are treated in it symmetrically, on the same footing. However, one can consider the Nambu--Gorkov components \eqref{ewa:SgmPsinuM}, \eqref{ewa:SgmPsie} and \eqref{ewa:bldDPsinuPsie} of $\Sigma_{\Psi_\nu,M}$, $\Sigma_{\Psi_e}$ and $D_{\Psi_\nu}$, $D_{\Psi_e}$, respectively, and express the $\mu_{W^\pm}^2|_{\ell}$ in a less compact form as
\begin{eqnarray}
\mu_{W^\pm}^2\big|_{\ell} &=&
\nonumber \\ &&
\hspace{-3em}
-\I\frac{1}{2} \int\!\frac{\d^d p}{(2\pi)^d}\,\Tr\bigg\{
\Big[
\big(\Sigma_{\nu D}^{\vphantom{\dag}}\,\Sigma_{\nu D}^\dag+\Sigma_{\nu L}^{\vphantom{\dag}}\,\Sigma_{\nu L}^\dag\big) - \frac{2}{d}p^2
\big(\Sigma_{\nu D}^{\vphantom{\dag}}\,\Sigma_{\nu D}^\dag+\Sigma_{\nu L}^{\vphantom{\dag}}\,\Sigma_{\nu L}^\dag\big)^\prime
\Big]
D_{eL}^{\vphantom{\dag}}\,D_{\nu L}^{\vphantom{\dag}}
\nonumber \\ &&
\hspace{7em}
{}- \frac{2}{d}p^2
\big(\Sigma_{\nu D}^{\vphantom{\dag}}\,\Sigma_{\nu D}^\dag+\Sigma_{\nu L}^{\vphantom{\dag}}\,\Sigma_{\nu L}^\dag\big)
\Big[
D_{eL}^{\vphantom{\prime}}\,D_{\nu L}^{\prime} - D_{eL}^{\prime}\,D_{\nu L}^{\vphantom{\prime}}
\Big]
\nonumber \\ &&
\hspace{3.55em}
{}+
\Big[
\big( \Sigma_{\nu D}^{\T\vphantom{\dag}}\,\Sigma_{\nu L}^\dag+M_{\nu R}^{\vphantom{\dag}}\,\Sigma_{\nu D}^\dag \big) - \frac{2}{d}p^2
\big( \Sigma_{\nu D}^{\T\vphantom{\dag}}\,\Sigma_{\nu L}^\dag+M_{\nu R}^{\vphantom{\dag}}\,\Sigma_{\nu D}^\dag \big)^\prime
\Big]
D_{eL}^{\vphantom{\dag}}\,D_{\nu M}^{\vphantom{\dag}}
\nonumber \\ &&
\hspace{7em}
{}- \frac{2}{d}p^2
\big( \Sigma_{\nu D}^{\T\vphantom{\dag}}\,\Sigma_{\nu L}^\dag+M_{\nu R}^{\vphantom{\dag}}\,\Sigma_{\nu D}^\dag \big)
\Big[
D_{eL}^{\vphantom{\prime}}\,D_{\nu M}^{\prime} - D_{eL}^{\prime}\,D_{\nu M}^{\vphantom{\prime}}
\Big]
\nonumber \\ &&
\hspace{3.55em}
{}+
\Big[
\big( \Sigma_{e}^{\vphantom{\dag}}\,\Sigma_{e}^\dag \big) - \frac{2}{d}p^2
\big( \Sigma_{e}^{\vphantom{\dag}}\,\Sigma_{e}^\dag \big)^\prime
\Big]
D_{\nu L}^{\vphantom{\dag}}\,D_{eL}^{\vphantom{\dag}}
- \frac{2}{d}p^2
\big( \Sigma_{e}^{\vphantom{\dag}}\,\Sigma_{e}^\dag \big)
\Big[
D_{\nu L}^{\vphantom{\prime}}\,D_{eL}^{\prime} - D_{\nu L}^{\prime}\,D_{eL}^{\vphantom{\prime}}
\Big]
\bigg\} \,.
\nonumber \\ &&
\end{eqnarray}
\end{subequations}
This latter form of $\mu_{W^\pm}^2|_{\ell}$ can be used for a crosscheck, since one can see from it more clearly that in the case of Dirac neutrinos ($\Sigma_{\nu L} = M_{\nu R} = 0$, implying also $D_{\nu M} = 0$) the lepton contribution $\mu_{W^\pm}^2|_{\ell}$ would be formally the same as the analogous quark contribution $\mu_{W^\pm}^2|_{q}$, expressed by \eqref{gbew:mu2:Wpm:q}.

\subsection{Masses of $Z$ and $\gamma$}

Again, as before with the quarks, the generators $T_{{\Psi_\ell},3}$, $T_{{\Psi_\ell},4}$, \eqref{ew1:TPsiell34diag}, are block-diagonal in the two-dimensional electroweak space. Thus, since the self-energy $\boldsymbol{\Sigma}_{\Psi_\ell}$, \eqref{ewa:SgmPsiell}, is also block diagonal, so is for $a=3,4$ also the vertex $\Gamma_{\Psi_\ell,a}^\mu(p^\prime,p)$, \eqref{gbew:Gmm:ell}:
\begin{eqnarray}
\Gamma_{\Psi_\ell,a}^\mu(p^\prime,p) &\equiv&
\left(\begin{array}{cc} \Gamma_{\Psi_\nu,a}^\mu(p^\prime,p) & 0 \\ 0 & \Gamma_{\Psi_e,a}^\mu(p^\prime,p) \end{array}\right) \,.
\end{eqnarray}
Consequently the contributions from the neutrinos and charged leptons to the polarization tensor decouple and we can write
\begin{eqnarray}
\Pi^{\mu\nu}_{Z\gamma,ab}(q^2)\big|_{\ell} &=& \sum_{f=\nu,e} \Pi^{\mu\nu}_{Z\gamma,ab}(q^2)\big|_{f} \,,
\end{eqnarray}
where each contribution $\Pi^{\mu\nu}_{Z\gamma,ab}(q^2)|_{f}$ is given by ($a,b=3,4$)
\begin{eqnarray}
\label{gbew:Piabmunu:f}
\I \Pi_{Z\gamma,ab}^{\mu\nu}(q)\big|_{f} &=& - \frac{1}{2}\int\!\frac{\d^d p}{(2\pi)^d}\,
\Tr \Big\{ \Gamma^\mu_{\Psi_{\!f},a}(p+q,p) \, G_{\Psi_{\!f}}(p) \, \gamma^\nu T_{\Psi_{\!f},b} \, G_{\Psi_{\!f}}(p+q) \Big\} \,.
\end{eqnarray}
The vertex $\Gamma^\mu_{\Psi_{\!f},a}(p+q,p)$ is of the same form as \eqref{gbew:Gmm:ell} satisfying the WT identity so that $\Pi_{Z\gamma,ab}^{\mu\nu}(q)|_{f}$ is transversal.

In contrast to the quarks where we calculated the contributions from the up-type and the down-type quarks at the same time, now this is not convenient due to substantial differences between the two types of leptons. We will therefore calculate the contributions from the charged leptons and neutrinos separately.

\subsubsection{Contribution of charged leptons}

We start with the charged leptons, as they are substantially easier than the neutrinos. Since the number of the left-handed and the right-handed charged leptons is the same (i.e., $n$) and since the Majorana components of the charged leptons' propagators vanish, the Dirac basis \eqref{ew1:e},
\begin{eqnarray}
e &=& e_L + e_R \,,
\end{eqnarray}
makes sense. The expression \eqref{gbew:Piabmunu:f} for the polarization tensor $\Pi_{Z\gamma,ab}^{\mu\nu}(q)|_{e}$ can be therefore rewritten from the basis $\Psi_e$ into the basis $e$ (using the results from appendix~\ref{app:fermi propag}, section~\ref{app:frm:Rel}) as
\begin{eqnarray}
\I \Pi_{Z\gamma,ab}^{\mu\nu}(q)\big|_{e} &=& - \int\!\frac{\d^d p}{(2\pi)^d}\,
\Tr \Big\{ \Gamma^\mu_{e,a}(p+q,p) \, G_{e}(p) \, \gamma^\nu T_{e,b} \, G_{e}(p+q) \Big\} \,,
\end{eqnarray}
where $a,b=3,4$. The point is that now we are in the same situation as before with quarks, since the basis $e$ corresponds to the quark bases $u$ and $d$. Therefore we can use the results from Sec.~\ref{gbew:sec:quarks} about quarks. We obtain the charged lepton contribution $M^2_{Z\gamma}|_{e}$ to the $Z$ and $\gamma$ mass matrix as
\begin{eqnarray}
M_{Z\gamma}^2\big|_{e} &=&
\left(\begin{array}{cc} g^2 & -g g^\prime \\ -g g^\prime & g^{\prime 2} \end{array}\right)
\mu_{Z\gamma}^2\big|_{e} \,,
\end{eqnarray}
where
\begin{eqnarray}
\label{gbew:mu2:gZ:e}
\mu_{Z\gamma}^2\big|_{e} &=&  -\frac{1}{2}\I  \int\!\frac{\d^d p}{(2\pi)^d}\,\Tr
\bigg\{
\Big[
\big(\Sigma_e^{\vphantom{\dag}}\,\Sigma_e^\dag\big) - \frac{2}{d} p^2
\big(\Sigma_e^{\vphantom{\dag}}\,\Sigma_e^\dag\big)^\prime
\Big] D_{eL}^{2\vphantom{\dag}}
\bigg\} \,.
\end{eqnarray}
Of course, this is (up to the missing factor of $N_c$) the same as the quark expression \eqref{gbew:mu2:gZ:f}.

\subsubsection{Contribution of neutrinos}


Let us continue with the neutrinos. Since this time the relevant generators $T_{\Psi_\nu,3}$, $T_{\Psi_\nu,4}$ do not commute with the self-energy $\boldsymbol{\Sigma}_{\Psi_\nu}$, we cannot make use of the results from Sec.~\ref{gba:ssec:U1N}. In particular, we have to first of all check to condition \eqref{gbm:Aab=0}. The quantity $A_{ab}$ now reads
\begin{eqnarray}
\label{gbew:Aab:nu}
A_{ab} &=& \Tr \bigg\{\phantom{+\,}
T_{\Psi_\nu,a}\,\boldsymbol{\Sigma}_{\Psi_\nu}^\dag\,\boldsymbol{D}_{\Psi_\nu}^\prime\,\bar T_{\Psi_\nu,b}\,\boldsymbol{\Sigma}_{\Psi_\nu}\,\boldsymbol{D}_{\Psi_\nu}^\C
-
T_{\Psi_\nu,a}\,\boldsymbol{\Sigma}_{\Psi_\nu}^\dag\,\boldsymbol{D}_{\Psi_\nu}\,\bar T_{\Psi_\nu,b}\,\boldsymbol{\Sigma}_{\Psi_\nu}\,\boldsymbol{D}_{\Psi_\nu}^{\C\prime}
\nonumber \\ &&
\phantom{\Tr \bigg\{}
\hspace{-0.45em}
+\,
\bar T_{\Psi_\nu,a}\,\boldsymbol{D}_{\Psi_\nu}\,\boldsymbol{\Sigma}_{\Psi_\nu}\, T_{\Psi_\nu,b}\,\boldsymbol{D}_{\Psi_\nu}^{\C\prime}\,\boldsymbol{\Sigma}_{\Psi_\nu}^\dag
-
\bar T_{\Psi_\nu,a}\,\boldsymbol{D}_{\Psi_\nu}^\prime\,\boldsymbol{\Sigma}_{\Psi_\nu}\, T_{\Psi_\nu,b}\,\boldsymbol{D}_{\Psi_\nu}^\C\,\boldsymbol{\Sigma}_{\Psi_\nu}^\dag
\bigg\} \,.
\qquad
\end{eqnarray}
Notice that the generators $T_{\Psi_\nu,3}$, $T_{\Psi_\nu,4}$ have the form (cf.~\eqref{ew1:TPsif34})
\begin{eqnarray}
T_{\Psi_\nu,a} &=& \tau_{\Psi_\nu,a}\,\gamma_5\,P_{+\nu} \,,
\end{eqnarray}
where
\begin{subequations}
\begin{eqnarray}
\tau_{\Psi_\nu,3} &\equiv& -g\phantom{^\prime} \frac{1}{2} \,, \\
\tau_{\Psi_\nu,4} &\equiv& \phantom{-}g^\prime \frac{1}{2} \,,
\end{eqnarray}
\end{subequations}
and $P_{+\nu}$ is given by \eqref{ew1:P+nu}. The quantity $A_{ab}$, \eqref{gbew:Aab:nu}, can be therefore rewritten as
\begin{eqnarray}
\label{gbew:Aab:nu2}
A_{ab} &=&
-\tau_{\Psi_\nu,a} \tau_{\Psi_\nu,b} \Tr \bigg\{\phantom{+\,}
P_{+\nu}\,\boldsymbol{\Sigma}_{\Psi_\nu}^\dag\,\boldsymbol{D}_{\Psi_\nu}^\prime\, P_{+\nu}\,\boldsymbol{\Sigma}_{\Psi_\nu}\,\boldsymbol{D}_{\Psi_\nu}^\C
-
P_{+\nu}\,\boldsymbol{\Sigma}_{\Psi_\nu}^\dag\,\boldsymbol{D}_{\Psi_\nu}\, P_{+\nu}\,\boldsymbol{\Sigma}_{\Psi_\nu}\,\boldsymbol{D}_{\Psi_\nu}^{\C\prime}
\nonumber \\ &&
\phantom{-\tau_{\Psi_\nu,a} \tau_{\Psi_\nu,b}}
\phantom{\Tr \bigg\{}
\hspace{-0.45em}
+\,
P_{+\nu}\,\boldsymbol{D}_{\Psi_\nu}\,\boldsymbol{\Sigma}_{\Psi_\nu}\, P_{+\nu}\,\boldsymbol{D}_{\Psi_\nu}^{\C\prime}\,\boldsymbol{\Sigma}_{\Psi_\nu}^\dag
-
P_{+\nu}\,\boldsymbol{D}_{\Psi_\nu}^\prime\,\boldsymbol{\Sigma}_{\Psi_\nu}\, P_{+\nu}\,\boldsymbol{D}_{\Psi_\nu}^\C\,\boldsymbol{\Sigma}_{\Psi_\nu}^\dag
\bigg\} \,.
\nonumber \\ &&
\end{eqnarray}
Now it turns that the following identities hold:
\begin{subequations}
\begin{eqnarray}
\Tr \bigg\{
P_{+\nu}\,\boldsymbol{\Sigma}_{\Psi_\nu}^\dag\,\boldsymbol{D}_{\Psi_\nu}^\prime\, P_{+\nu}\,\boldsymbol{\Sigma}_{\Psi_\nu}\,\boldsymbol{D}_{\Psi_\nu}^\C
\bigg\}
&=&
\Tr \bigg\{
P_{+\nu}\,\boldsymbol{\Sigma}_{\Psi_\nu}^\dag\,\boldsymbol{D}_{\Psi_\nu}\, P_{+\nu}\,\boldsymbol{\Sigma}_{\Psi_\nu}\,\boldsymbol{D}_{\Psi_\nu}^{\C\prime}
\bigg\} \,,
\\
\Tr \bigg\{
P_{+\nu}\,\boldsymbol{D}_{\Psi_\nu}\,\boldsymbol{\Sigma}_{\Psi_\nu}\, P_{+\nu}\,\boldsymbol{D}_{\Psi_\nu}^{\C\prime}\,\boldsymbol{\Sigma}_{\Psi_\nu}^\dag
\bigg\}
&=&
\Tr \bigg\{
P_{+\nu}\,\boldsymbol{D}_{\Psi_\nu}^\prime\,\boldsymbol{\Sigma}_{\Psi_\nu}\, P_{+\nu}\,\boldsymbol{D}_{\Psi_\nu}^\C\,\boldsymbol{\Sigma}_{\Psi_\nu}^\dag
\bigg\} \,.
\end{eqnarray}
\end{subequations}
Each of these identities can be proven, apart from using the cyclicity of the trace, by inserting $1 = \gamma_0\,\gamma_0$ and using the relations like $\boldsymbol{\Sigma}_{\Psi_\nu}^\dag = \gamma_0\,\boldsymbol{\Sigma}_{\Psi_\nu}\,\gamma_0$ and $\boldsymbol{D}_{\Psi_\nu} = \gamma_0\,\boldsymbol{D}_{\Psi_\nu}^\C\,\gamma_0$, stemming from the Hermiticity condition \eqref{ewa:SgmPsifhc} for the self-energy $\boldsymbol{\Sigma}_{\Psi_\nu}$. Thus, the first and third term in \eqref{gbew:Aab:nu2} cancel with the second and fourth term, respectively, so that $A_{ab}$ vanishes and the condition \eqref{gbm:Aab=0} is indeed satisfied.

Hence we can use the formula \eqref{gbm:M2ab} for the gauge boson mass matrix. The explicit calculation reveals the neutrino contribution to the $M_{Z\gamma}^2$ mass matrix to be of the form
\begin{eqnarray}
M_{Z\gamma}^2\big|_{\nu} &=&
\left(\begin{array}{cc} g^2 & -g g^\prime \\ -g g^\prime & g^{\prime 2} \end{array}\right)
\mu_{Z\gamma}^2\big|_{\nu} \,,
\end{eqnarray}
where $\mu_{Z\gamma}^2|_{\nu}$ can be, likewise in the case of $\mu_{W^\pm}^2|_{\ell}$, \eqref{gbew:mu2:Wpm:ell}, expressed in two ways. The more compact form of $\mu_{Z\gamma}^2|_{\nu}$ reads
\begin{subequations}
\label{gbew:mu2:gZ:nu}
\begin{eqnarray}
\mu_{Z\gamma}^2\big|_{\nu} &=&
-\I\frac{1}{2} \int\!\frac{\d^d p}{(2\pi)^d}\,\Tr\bigg\{
\Big[
\big(\Sigma_{\Psi_\nu,M}^{\vphantom{\dag}}\,\Sigma_{\Psi_\nu,M}^\dag\big) - \frac{2}{d}p^2
\big(\Sigma_{\Psi_\nu,M}^{\vphantom{\dag}}\,\Sigma_{\Psi_\nu,M}^\dag\big)^\prime
\Big]
\big(P_{+\nu}\,D_{\Psi_\nu}\big)\big(P_{+\nu}\,D_{\Psi_\nu}\big)
\nonumber \\ &&
\hspace{3em}
{}- \frac{2}{d}p^2
\big(\Sigma_{\Psi_\nu,M}^{\vphantom{\dag}}\,\Sigma_{\Psi_\nu,M}^\dag\big)
\Big[
\big(P_{+\nu}\,D_{\Psi_\nu}\big)       \big(P_{+\nu}\,D_{\Psi_\nu}\big)^\prime -
\big(P_{+\nu}\,D_{\Psi_\nu}\big)^\prime\big(P_{+\nu}\,D_{\Psi_\nu}\big)
\Big]
\nonumber \\ && 
\hspace{6.55em}
{}+
\Big[
\big(\Sigma_{\Psi_\nu,M}^{\vphantom{\dag}}\,P_{+\nu}^{\vphantom{\dag}}\,\Sigma_{\Psi_\nu,M}^\dag\big) - \frac{2}{d}p^2
\big(\Sigma_{\Psi_\nu,M}^{\vphantom{\dag}}\,P_{+\nu}^{\vphantom{\dag}}\,\Sigma_{\Psi_\nu,M}^\dag\big)^\prime
\Big]
D_{\Psi_\nu}\,P_{+\nu}\,D_{\Psi_\nu}
\bigg\} \,,
\nonumber \\ &&
\end{eqnarray}
while the less elegant form is
\begin{eqnarray}
\mu_{Z\gamma}^2\big|_{\nu} &=&
-\I\frac{1}{2} \int\!\frac{\d^d p}{(2\pi)^d}\,\Tr\bigg\{
\Big[
\big(\Sigma_{\nu D}^{\vphantom{\dag}}\,\Sigma_{\nu D}^\dag+\Sigma_{\nu L}^{\vphantom{\dag}}\,\Sigma_{\nu L}^\dag\big) - \frac{2}{d}p^2
\big(\Sigma_{\nu D}^{\vphantom{\dag}}\,\Sigma_{\nu D}^\dag+\Sigma_{\nu L}^{\vphantom{\dag}}\,\Sigma_{\nu L}^\dag\big)^\prime
\Big]
D_{\nu L}^{2\vphantom{\dag}}
\nonumber \\ &&
\hspace{6.55em}
{}- \frac{2}{d}p^2
\big(\Sigma_{\nu D}^{\vphantom{\dag}}\,\Sigma_{\nu D}^\dag+\Sigma_{\nu L}^{\vphantom{\dag}}\,\Sigma_{\nu L}^\dag\big)
\Big[
D_{\nu L}^{\vphantom{\prime}}\,D_{\nu L}^{\prime} - D_{\nu L}^{\prime}\,D_{\nu L}^{\vphantom{\prime}}
\Big]
\nonumber \\ &&
\hspace{6.55em}
{}+
\Big[
\big( \Sigma_{\nu D}^{\T\vphantom{\dag}}\,\Sigma_{\nu L}^\dag+M_{\nu R}^{\vphantom{\dag}}\,\Sigma_{\nu D}^\dag \big) - \frac{2}{d}p^2
\big( \Sigma_{\nu D}^{\T\vphantom{\dag}}\,\Sigma_{\nu L}^\dag+M_{\nu R}^{\vphantom{\dag}}\,\Sigma_{\nu D}^\dag \big)^\prime
\Big]
D_{\nu L}^{\vphantom{\dag}}\,D_{\nu M}^{\vphantom{\dag}}
\nonumber \\ &&
\hspace{6.55em}
{}- \frac{2}{d}p^2
\big( \Sigma_{\nu D}^{\T\vphantom{\dag}}\,\Sigma_{\nu L}^\dag+M_{\nu R}^{\vphantom{\dag}}\,\Sigma_{\nu D}^\dag \big)
\Big[
D_{\nu L}^{\vphantom{\prime}}\,D_{\nu M}^{\prime} - D_{\nu L}^{\prime}\,D_{\nu M}^{\vphantom{\prime}}
\Big]
\nonumber \\ &&
\hspace{6.55em}
{}+
\Big[
\big( \Sigma_{\nu L}^{\vphantom{\dag}}\,\Sigma_{\nu L}^\dag \big) - \frac{2}{d}p^2
\big( \Sigma_{\nu L}^{\vphantom{\dag}}\,\Sigma_{\nu L}^\dag \big)^\prime
\Big]
D_{\nu L}^{2\vphantom{\dag}}
\nonumber \\ &&
\hspace{6.55em}
{}+
\Big[
\big( \Sigma_{\nu D}^{\T\vphantom{\dag}}\,\Sigma_{\nu D}^{*\vphantom{\dag}} \big) - \frac{2}{d}p^2
\big( \Sigma_{\nu D}^{\T\vphantom{\dag}}\,\Sigma_{\nu D}^{*\vphantom{\dag}} \big)^\prime
\Big]
D_{\nu M}^{\dag}\,D_{\nu M}^{\vphantom{\dag}}
\nonumber \\ &&
\hspace{6.55em}
{}+\Big(
\Big[
\big( \Sigma_{\nu D}^{\T\vphantom{\dag}}\,\Sigma_{\nu L}^{\dag} \big) - \frac{2}{d}p^2
\big( \Sigma_{\nu D}^{\T\vphantom{\dag}}\,\Sigma_{\nu L}^{\dag} \big)^\prime
\Big]
D_{\nu L}^{\vphantom{\dag}}\,D_{\nu M}^{\vphantom{\dag}} + \hc \Big)
\bigg\} \,.
\end{eqnarray}
\end{subequations}
Again, the latter form allows for a crosscheck as one sees more clearly that in the case of Dirac neutrinos ($\Sigma_{\nu L} = M_{\nu R} = 0$ and $D_{\nu M} = 0$) this expression for $\mu_{Z\gamma}^2|_{\nu}$ would be the same as the analogous expressions $\mu_{Z\gamma}^2|_{f}$ for $f=u,d,e$, Eqs.~\eqref{gbew:mu2:gZ:f} and \eqref{gbew:mu2:gZ:e}.

\section{Summary}

We have calculated the gauge boson mass matrix $M^2$ in the basis $A_a^\mu$, $a=1,2,3,4$, using the results from chapter~\ref{chp:gbm} and partially also from chapter~\ref{chp:ablgg}. We found it to be exactly of the expected form
\begin{eqnarray}
M^2 &=& \left(\begin{array}{cc} M_{W^\pm}^2 & 0 \\ 0 & M_{Z\gamma}^2 \end{array}\right) \,,
\end{eqnarray}
with
\begin{eqnarray}
M_{Z\gamma}^2 &=& \left(\begin{array}{cc} g^2 & -g g^\prime \\ -g g^\prime & g^{\prime 2} \end{array}\right) \mu_{Z\gamma}^2 \,,
\\
M_{W^\pm}^2 &=& \left(\begin{array}{cc} g^2 & 0 \\ 0 & g^2 \end{array}\right) \mu_{W^\pm}^2 \,,
\end{eqnarray}
cf.~\eqref{gbew:M2gen} and \eqref{gbew:M2zg:gen34}, \eqref{gbew:M2Wpm:gen12}.

The factor $\mu_{Z\gamma}^2$ was found to be a sum of separate contributions of the up-type and down-type quarks, neutrinos and charged leptons
\begin{eqnarray}
\mu_{Z\gamma}^2 &=& \mu_{Z\gamma}^2\big|_{u}  + \mu_{Z\gamma}^2\big|_{d} + \mu_{Z\gamma}^2\big|_{\nu} + \mu_{Z\gamma}^2\big|_{e} \,,
\end{eqnarray}
where the particular contributions $\mu_{Z\gamma}^2|_{u}$, $\mu_{Z\gamma}^2|_{d}$ and $\mu_{Z\gamma}^2|_{\nu}$, $\mu_{Z\gamma}^2|_{e}$ are given by \eqref{gbew:mu2:gZ:f} and \eqref{gbew:mu2:gZ:nu}, \eqref{gbew:mu2:gZ:e}, respectively, while $\mu_{W^\pm}^2$ is a sum of separate contributions from the quark and lepton doublets:
\begin{eqnarray}
\mu_{W^\pm}^2   &=& \mu_{W^\pm}^2\big|_{q} + \mu_{W^\pm}^2\big|_{\ell} \,,
\end{eqnarray}
where $\mu_{W^\pm}^2|_{q}$, $\mu_{W^\pm}^2|_{\ell}$ are given by \eqref{gbew:mu2:Wpm:q}, \eqref{gbew:mu2:Wpm:ell}, respectively.

The mass spectrum is now given by
\begin{subequations}
\label{gbew:spectrum}
\begin{eqnarray}
M^2_\gamma &=& 0 \,, \\
M^2_Z      &=& (g^2+g^{\prime 2})\mu_{Z\gamma}^2 \,, \\
M^2_{W}    &=& g^2 \mu_{W^\pm}^2 \,,
\end{eqnarray}
\end{subequations}
i.e., it contains the massless photon, the massive $Z$ boson with mass squared $M^2_Z$ and the two massive $W^+$ and $W^-$ bosons with the same mass squared $M^2_{W}$.

Let us finally comment on the $\rho$-parameter, defined in \eqref{ewd:rho} and rewritable using \eqref{gbew:spectrum} as
\begin{eqnarray}
\label{gbew:rho}
\rho &\equiv& \frac{\mu_{W^\pm}^2}{\mu_{Z\gamma}^2} \,.
\end{eqnarray}
Recall that experimentally $\rho$ is close to $1$, which corresponds to approximate custodial symmetry. Since in our case the gauge boson masses $M_W^2$, $M_Z^2$ depend on unknown\footnote{Recall that the present analysis of the gauge boson masses pretends to be as model-independent as possible, i.e., it does not rely on the very mechanism of how the fermion self-energies are actually generated.} details of the momentum-dependencies of the fermion self-energies, we cannot address directly the issue of the value of the $\rho$-parameter. Nevertheless, we can at least crosscheck our results by verifying whether in the case of exact custodial symmetry they yield $\rho = 1$. In the quark sector the custodial symmetry corresponds simply to $\Sigma_u = \Sigma_d$, while in the lepton sector it corresponds to $\Sigma_{\nu D} = \Sigma_e$ and $\Sigma_{\nu L} = M_{\nu R} = 0$ (provided $n = m$, i.e., the number of fermion generations and the number of right-handed neutrinos are the same).\footnote{This situation can be accommodated within the model of strong Yukawa dynamics, discussed in part~\ref{part:ew}, by assuming (on top of $n = m$ and $M_{\nu R} = 0$) that $y_u = y_d$, $y_\nu = y_e$ and $M_S = M_N$, which corresponds to custodially symmetric Lagrangian. One can easily verify that under these assumptions the SD equations indeed allow for the custodially symmetric solution $\Sigma_u = \Sigma_d$, $\Sigma_{\nu D} = \Sigma_e$ and $\Sigma_{\nu L} = 0$ (and $\Pi_S = \Pi_N = \Pi_{SN}$).} Under these assumptions we have $\mu_{Z\gamma}^2|_{u} = \mu_{Z\gamma}^2|_{d} = 2\mu_{W^\pm}^2|_{q}$ and $\mu_{Z\gamma}^2|_{\nu} = \mu_{Z\gamma}^2|_{e} = 2\mu_{W^\pm}^2|_{\ell}$, so that $\mu_{Z\gamma}^2 = \mu_{W^\pm}^2$ and the $\rho$-parameter \eqref{gbew:rho} is consequently indeed equal to $1$.


\chapter{Conclusions}
\label{chp:conclusions}

Nominally the main topic of the thesis was to explore the possibility of breaking a symmetry by strong Yukawa dynamic. First in part~\ref{part:abel} we introduced the raw idea on an example of the Abelian toy model and eventually in part~\ref{part:ew} we applied it on a realistic model of electroweak interactions. Let us now recapitulate shortly our results and make a brief comparison with the main competing models: the SM (and generally all MCS) and the ETC models.

The main dynamical assumption of the presented model (or mechanism) is that the presumably strong Yukawa interactions, linking together the fermions and scalars, give rise to the fermion and scalar propagators that break spontaneously the symmetry in question. One should in particular note that the fermion masses are generated directly in the course of SSB, not as a mere consequence. The fact that the fermion mass generation is in this way intimately connected with SSB can be thus viewed as an appealing feature of the presented model. This is to be contrasted with the situation in MCS, where the SSB is a matter of only the scalar sector alone and the fermion masses may or may not be subsequently generated, depending on whether they have the Yukawa couplings with the condensing scalars. In this respect the presented model is closer to ETC models, where the presence of fermions is also vital for the SSB to happen, although the very mechanism is different.

One of the drawbacks of MCS is certainly the number of free parameters: There are (depending on the number of scalars) at least as many theoretically arbitrary Yukawa coupling constants as fermions, intended to obtain mass. The presented model obviously suffers from this problem too. Even though the Yukawa coupling constants are comfortable in the sense that they explicitly break the chiral and flavor symmetries, their large number suggests that there should exist a more fundamental underlying theory capable of predicting them. In this respect the SM as well as the presented model should be viewed as effective theories valid only up to some scale.

On top of just formulating the model in terms of its particle content, introducing suitable formalism and writing down the equations of motion we also performed its numerical analysis. However, we did not pretend to make any phenomenological predictions but rather aimed merely to demonstrate viability of the proposed mechanism of SSB and eventually to find some of its generic features. This allowed us to make substantial simplifications of the relevant equations: We considered only a subset of the whole possible particle spectrum and looked only for the symmetry-breaking parts of the propagators, while neglecting the symmetry-preserving ones. Moreover, the very fact that the SSB is assumed to be triggered primarily by the scalar and fermion two-point functions, as opposed to the scalar one-point functions in MCS, allowed us to make for the sake of simplicity further approximations: We neglected the scalar self-couplings, needed in MCS but dispensable in the presented model, and also we directly assumed that the scalars do not develop VEVs. These assumptions were operationally manifested by the absence of tadpole (i.e., constant) terms in the SD equations for the propagators of both the scalars and fermions. However, it should be also said honestly that without neglecting the tadpole terms we actually did not manage to find solutions of the SD equations.

The numerical analysis, done under the above assumptions, revealed that the SSB indeed does happen, provided that the Yukawa coupling constants are large enough (larger than certain critical values) so that the dynamics is strong and hence in non-perturbative r\'{e}gime, as anticipated.

One of the particular numerical findings was that the large fermion hierarchy can be accommodated while keeping the Yukawa coupling constants to be of the same order of magnitude (although for the price of certain fine-tuning of their precise values). This rather appealing feature is to be compared with the situation in the MCS (and in particular in the SM): As the fermions masses differ by as much as six orders of magnitude (leaving aside the neutrinos), so do inevitably also the Yukawa coupling constants, which is for dimensionless numbers considered unnatural. Clearly, the point is that while in the MCS the fermions masses depend linearly on the Yukawa coupling constants, in the presented model this dependence is due to its non-perturbative character non-linear.

Another finding was that the scalars generally tend to be heavier than the fermions by orders of magnitude. Namely, we found that they are at least hundred times heavier, but it is conceivable that upon carrying out the numerical analysis with less over-simplifying assumptions they could be even heavier. This is actually reassuring for several reasons. First, the scalars in fact need to be heavy already from phenomenological reasons: in order to be compatible with the suppression of FCNC and in order to be (possibly) able to render the $\rho$-parameter to be close to one. Second reason is rather theoretical: Since in the presented model the scalar masses are not proportional to the scalar self-couplings (and their VEVs), there are no upper limits on their sizes and therefore we do not have to deal with the usual hierarchy problem notorious in the MCS (without SUSY). The large scalar masses can be thus interpreted as indeed being proportional to the theory's cut-off and accordingly the presented model is to be understood as an effective theory valid only up to the scale of the scalar masses.

Obviously, many questions remain unsolved. They are mostly connected with the unknown particle spectrum of the theory. Since the theory is strongly interacting, appearance of bound states must be expected. They are, however, difficult to predict, with the only exception of the \qm{would-be} NG bosons (or, equivalently, the longitudinal polarization states of the $W^\pm$ and $Z$ bosons), whose presence is guaranteed by the existence NG theorem. On top of these, it would be in particular worth knowing whether there exists also an excitation, mimicking the SM Higgs boson and unitarizing the scattering amplitudes. Due to strong and accordingly non-perturbative nature of the dynamics these questions are difficult to answer and the only way to tackle them would be probably to resort to some kind of lattice simulations.

Despite the thesis name, the specific model with strong Yukawa dynamics, developed in parts~\ref{part:abel} and \ref{part:ew}, is by no means its only subject. Equally, if not more important achievements are two model-independent issues, discussed in the following two parts.

In part~\ref{part:mx} we considered the fermion flavor mixing in the situation when instead of constant, momentum-independent fermion mass matrices, occurring in particular in MCS, one has at disposal rather their momentum-dependent generalization, the self-energies. The main motivation for dealing with this issue was of course the above discussed model of strong Yukawa dynamics, where the fermion self-energies serve as agents of the SSB. However, such situation is typical also for other models with dynamical symmetry breaking, like, e.g., the currently developed model \cite{Hosek:2009ys,Benes:2011gi,Smetana:2011tj} of strong gauge flavor dynamics, or, at least in principle, the \qm{more mainstream} ETC models.

Specifically, we considered the case of quarks and investigated how the quark self-energies affect the flavor mixing in interactions of the charged, neutral and electromagnetic currents. Our approach was to calculate first in the leading order in the gauge coupling constants the corresponding amplitudes (i.e., the decays of gauge bosons into the fermion--antifermion pairs) by means of the LSZ reduction formula. In order to make a link with the usual notions like the \qm{CKM matrix} or the \qm{mass-eigenstate basis}, we constructed an effective Lagrangian, corresponding to the calculated amplitudes. We found that the effective CKM matrix defined this way is in general not unitary and that the FCNC, as well as the flavor-changing electromagnetic transitions, can be present already in the leading order in the gauge coupling constants. All these findings are related to the fact that the notion of a mass-eigenstate basis is in this situation merely an effective one and that it cannot be obtained from the weak interaction eigenstate basis by a unitary transformation. It should be, however, stressed that all these peculiarities depend crucially on the details of the quark self-energies momentum dependencies, which have been considered to be virtually arbitrary. In particular, in the special case of constant self-energies (or, equivalently, the mass matrices) the results of our analysis naturally reduce to those familiar ones from the SM (and generally MCS).

In part~\ref{part:gauge} we occupied ourselves with the other model-independent issue, which was the precise mechanism of generation of the gauge boson masses in models like the presented one of the strong Yukawa dynamics. That is to say, we considered a rather general situation when a gauge symmetry is broken down to some of its subgroups by formation of self-energies of the fermions, which are coupled to it. What is important is that the very mechanism of the fermion self-energies generation is not essential, so that the analysis and the outcomes of part~\ref{part:gauge} are applicable on a wider class of theories, including also the mentioned model of gauge flavor dynamics and the ETC models.

The general strategy was to calculate the gauge boson polarization tensor in one loop, with one insertion of the bare vertex and the other of the full vertex, satisfying the WT identity. This is actually nothing else than what was already done in the \qm{classical} references \cite{Jackiw:1973tr,Cornwall:1973ts}, just this time more systematically and under more general assumptions. What was new was the construction of the full vertex, especially of its part that cannot be uniquely determined from the WT identity. Namely, we introduced the new term proportional to the transversal quantity $q^\mu [\slashed{q},\slashed{q}^\prime] - q^2 [\gamma^\mu,\slashed{q}^\prime]$ (where $q$ is the momentum carried by the gauge boson), which has not been considered in the literature yet. We showed that this term is necessary in order to arrive at a \emph{symmetric} gauge boson mass matrix. Taking it into account we also found some minor correction to the Pagels--Stokar formula.

Having said that the new term in the vertex is \emph{necessary} for the gauge boson mass matrix to be symmetric, it must be also added that it is not \emph{sufficient}. Depending on the details of the theory in question (namely on its gauge group and its fermion representations) it may still happen that the gauge boson mass matrix comes out non-symmetric. The point is that within our approach there is actually no known way how to cure this situation. Although in the theories of interest (i.e., in Abelian theories and in the electroweak theory) the gauge boson mass matrices still \qm{miraculously} come out symmetric, in general they do not, which obviously questions our approach (and correspondingly also the approach of Refs.~\cite{Jackiw:1973tr,Cornwall:1973ts}). Investigation of these shortcomings and seeking for their resolution is subject to further research.



\appendix

\chapter{Fermion charge conjugation}
\label{app:charge}

\intro{As the main text deals with Majorana fermions, it relies heavily on the notion of charge conjugation. Thus, in order to make the text reasonably self-contained, we review some facts about it. They will be used mostly in appendix~\ref{app:major} when quantizing Majorana field and in appendix~\ref{app:fermi propag} when introducing the Nambu--Gorkov formalism for fermions.}

\section{Properties of the charge conjugation}

Let $\psi$ be a solution of the \emph{classical} Dirac equation
\begin{eqnarray}
\left( \I\slashed{\partial}-m \right)\psi &=& 0 \,.
\end{eqnarray}
If we demand the \emph{charge conjugated} field $\psi^\C$,
\begin{eqnarray}
\psi^\C &\equiv& C \bar\psi^{\T} \,,
\end{eqnarray}
to be also a solution, then the matrix $C$ must satisfy the relation
\begin{eqnarray}
C^{-1} \gamma_\mu C &=& -\gamma_\mu^{\T} \,.
\label{app:charge_conjugation:main}
\end{eqnarray}
Taking into account this equation together with the properties of gamma matrices under Hermitian conjugation (provided $g_{\mu\nu}$ is given by \eqref{symbols:gmunu})
\begin{eqnarray}
\gamma_\mu^\dag &=& \gamma_0\gamma_\mu\gamma_0 \,,
\label{app:charge_conjugation:hermiticity}
\end{eqnarray}
we arrive at two independent relations for $C$:
\begin{subequations}
\begin{eqnarray}
\gamma_\mu &=& (CC^\dag)\gamma_\mu(CC^\dag)^{-1} \,,\\
\gamma_\mu &=& (CC^*)\gamma_\mu(CC^*)^{-1} \,.
\end{eqnarray}
\end{subequations}
They are evidently satisfied if
\begin{subequations}
\begin{eqnarray}
CC^\dag &=& a \unitmatrix \,, \\
CC^*    &=& b \unitmatrix
\end{eqnarray}
\end{subequations}
for some (non-zero) complex constants $a$, $b$. These are arbitrary at this moment, yet they can be fixed by imposing another conditions that the operation of charge conjugation should fulfil.

The first condition we can impose is the natural requirement that double charge conjugation is an identity:
\begin{eqnarray}
(\psi^\C)^\C &=& \psi \,.
\end{eqnarray}
By requiring this we can easily find that
\begin{eqnarray}
b &=& -1 \,.
\end{eqnarray}

For figuring out the second condition the following observation is crucial: if $\psi$ is a solution of the Dirac equation with positive (negative) energy, then $\psi^\C$ is a solution with negative (positive) energy:
\begin{subequations}
\begin{eqnarray}
(\slashed{p} - m)\psi \>=\> 0 & \iff &  (\slashed{p} + m)\psi^\C \>=\> 0 \,, \\
(\slashed{p} + m)\psi \>=\> 0 & \iff &  (\slashed{p} - m)\psi^\C \>=\> 0 \,.
\end{eqnarray}
\end{subequations}
This suggests that we could identify $u^\C(p)$ with $v(p)$ (and vice versa). Another observation, although without an impact on the determination of $a$, is that charge conjugation does not change the spin of the particle, i.e.,
\begin{subequations}
\label{app:charge_conjugation:spin}
\begin{eqnarray}
(\gamma\slashed{s})\psi \>=\> +\psi & \iff &  (\gamma\slashed{s})\psi^\C \>=\> +\psi^\C \,, \\
(\gamma\slashed{s})\psi \>=\> -\psi & \iff &  (\gamma\slashed{s})\psi^\C \>=\> -\psi^\C \,,
\end{eqnarray}
\end{subequations}
where $s$ is the space-like \emph{spin four-vector}, orthogonal to $p$. To conclude, we see that the charge conjugation interchanges the particle with its antiparticle, but it protect its spin state. Hence we see that $u^\C(p,s)$ is proportional to $v(p,s)$ (and vice versa) and we are free to normalize the operation of charge conjugation in such a way that
\begin{subequations}
\begin{eqnarray}
u^\C(p,s) &=& v(p,s) \,, \\
v^\C(p,s) &=& u(p,s) \,,
\end{eqnarray}
\end{subequations}
assuming that $u(p,s)$ and $v(p,s)$ are properly normalized according to
\begin{subequations}
\begin{eqnarray}
\bar u(p,s)\,u(p,s) &=& \phantom{-}2m \,, \\
\bar v(p,s)\,v(p,s) &=&          - 2m \,.
\end{eqnarray}
\end{subequations}
Then the constant $a$ can be fixed as
\begin{eqnarray}
a &=& 1 \,.
\end{eqnarray}
Summarizing our results, we see that matrix $C$ is unitary,
\begin{eqnarray}
C^\dag &=& C^{-1} \,,
\end{eqnarray}
and antisymmetric \,,
\begin{eqnarray}
\label{app:charge:Casym}
C^\T &=& -C \,.
\end{eqnarray}
We conclude that now the matrix $C$ is defined uniquely up to an arbitrary phase factor.

These results are valid in any representation of gamma matrices (provided that the Hermiticity properties of gamma matrices \eqref{app:charge_conjugation:hermiticity} hold). Sometimes it may happen, however, that by a suitable phase transformation the matrix $C$ can be made real. Then for its inverse there is a nice relation
\begin{eqnarray}
C^{-1} &=& -C \,.
\end{eqnarray}
Incidentally, this happens in the most widely used representations, i.e., in those of Dirac, Weyl and Majorana.

In derivation of \eqref{app:charge_conjugation:spin} we have used the identity
\begin{eqnarray}
\label{app:charge:gmm5}
C^{-1} \gamma_5 C &=& \gamma_5^{\T} \,,
\end{eqnarray}
which follows simply from the basic relation \eqref{app:charge_conjugation:main}. From this we see that the commutation properties of $C$ with $\gamma_5$ are obviously representation-dependent. For example, in Dirac and Weyl representations we have $[C,\gamma_5]=0$, while in Majorana representation there is $\{C,\gamma_5\}=0$.

In order to justify the above statements about reality of $C$ and about the commutation properties of $C$ with $\gamma_5$ in particular representations of gamma matrices we present here the explicit forms of $C$ and $\gamma_5$ in the mentioned representations:
\begin{subequations}
\begin{align}
C &\>=\> \I\gamma^2\gamma^0 =  \left(
\begin{array}{cc}
0 & -\I\sigma_2 \\
-\I\sigma_2 & 0
\end{array}\right)\,,
&&\gamma_5 \>=\> \left(
\begin{array}{cc}
0 & 1 \\
1 & 0
\end{array}\right)\,,
&& \hbox{(Dirac)}
\\
C &\>=\> \I\gamma^2\gamma^0 =  \left(
\begin{array}{cc}
-\I\sigma_2 & 0 \\
0 & \I\sigma_2
\end{array}\right)\,,
&&\gamma_5 \>=\> \left(
\begin{array}{cc}
1 & 0 \\
0 & -1
\end{array}\right)\,,
&& \hbox{(Weyl)}
\\
C &\>=\> \I\gamma^0 =  \left(
\begin{array}{cc}
0 & \I\sigma_2 \\
\I\sigma_2 & 0
\end{array}\right)\,,
&&\gamma_5 \>=\> \left(
\begin{array}{cc}
\sigma_2 & 0 \\
0 & -\sigma_2
\end{array}\right)\,.
&& \hbox{(Majorana)}
\end{align}
\end{subequations}
We see that in all cases $C$ is real and as such it is defined uniquely up to a sign.

Let us consider the eigenstates of the operation of charge conjugation. Since the charge conjugation applied twice is an identity, its eigenvalues should be $\pm1$. Thus, it should be possible to write an arbitrary bispinor $\psi$ as a linear combination of the two charge conjugation eigenstates corresponding to the eigenvalues $\pm1$. Indeed, it is the case: Upon defining\footnote{Convenience of the factors of $\frac{1}{\sqrt{2}}$ in the definitions \eqref{app:charge:psi_12} will be justified later when discussing quantization of Majorana field.}
\begin{subequations}
\label{app:charge:psi_12}
\begin{eqnarray}
\psi_{1} &\equiv&  \frac{1}{\sqrt{2}} \big( \psi^\C + \psi \big) \,, \\
\psi_{2} &\equiv& \frac{\I}{\sqrt{2}} \big( \psi^\C - \psi \big) \,,
\end{eqnarray}
\end{subequations}
we can decompose an arbitrary bispinor $\psi$ and its charge conjugate counterpart $\psi^\C$ as
\begin{subequations}
\label{app:charge:decomp}
\begin{eqnarray}
\psi    &=& \frac{1}{\sqrt{2}} \big( \psi_1+\I\psi_2 \big) \,, \\
\psi^\C &=& \frac{1}{\sqrt{2}} \big( \psi_1-\I\psi_2 \big) \,.
\end{eqnarray}
\end{subequations}
Clearly the fields $\psi_1$ and $\I\psi_2$ are the desired charge conjugation eigenstates corresponding to the eigenvalues $+1$ and $-1$, respectively. Moreover, the fields $\psi_{1,2}$ have the important property of being Majorana fields, since they satisfy the Majorana condition
\begin{eqnarray}
\psi_{1,2}^\C &=& \psi_{1,2} \,.
\label{app:charge:major}
\end{eqnarray}
More issues about the Majorana fields are discussed in appendices~\ref{app:major} and \ref{app:fermi propag}. Now let us only remark that since we can consider the Majorana fields $\psi_{1,2}$ as \qm{real} fermion fields, the decomposition \eqref{app:charge:decomp} is a direct analogue of the decomposition of a complex scalar field $\phi$ into its real and imaginary parts, i.e., $\phi=(\phi_1+\I\phi_2)/\sqrt{2}$, where both $\phi_{1,2}$ are real fields.

Finally, it is also useful to introduce the following definition: Let $A$ be a matrix in the Dirac space; then we define its \emph{charge transpose} $A^\C$ as
\begin{eqnarray}
\label{app:charge:AC}
A^\C &\equiv& C A^\T C^{-1} \,,
\end{eqnarray}
where $C$ is the matrix of charge conjugation. It satisfies
\begin{eqnarray}
(AB)^\C &=& B^\C A^\C \,.
\end{eqnarray}
In this formalism the relations \eqref{app:charge_conjugation:main} and \eqref{app:charge:gmm5} can be compactly rewritten as
\begin{subequations}
\begin{eqnarray}
\gamma_\mu^\C &=& -\gamma_\mu \,,
\\
\gamma_5^\C &=& \gamma_5 \,.
\end{eqnarray}
\end{subequations}

\section{Plane wave solution}

The solution of the classical free Dirac equation can be expressed in terms of plane waves as
\begin{eqnarray}
\psi(x) &=& \sum_{\pm s} \int \! \d^3\threevector{p} \, N_p
\Big[ b(p,s)\,u(p,s)\,\e^{-\I p \cdot x} +  d^*\!(p,s)\,v(p,s)\,\e^{\I p \cdot x}  \Big] \,,
\end{eqnarray}
where the normalization factor $N_p$ is defined as
\begin{eqnarray}
\label{app:charge:Np}
N_p &\equiv& \frac{1}{(2\pi)^{3/2}(2p_0)^{1/2}}
\end{eqnarray}
and the zeroth component $p_0$ of the on-shell four-momentum $p$ is $p_0=\sqrt{\threevector{p}^2+m^2}>0$. The quantities $b(s,p)$, $d(s,p)$ are some undetermined complex numbers with dimension
\begin{eqnarray}
\left[  b(s,p)  \right] \>=\> \left[  d(s,p)  \right] &=& M^{-3/2} \,,
\end{eqnarray}
where $M$ is an arbitrary mass scale. Using the results above, the charge conjugate solution is then
\begin{subequations}
\begin{eqnarray}
\psi^\C(x) &=&
\sum_{\pm s} \int \! \d^3\threevector{p} \, N_p
\Big[ b^*\!(p,s)\,u^\C(p,s)\,\e^{\I p \cdot x} + d(p,s)\,v^\C(p,s)\,\e^{-\I p \cdot x}  \Big]
\\ &=&
\sum_{\pm s} \int \! \d^3\threevector{p} \, N_p
\Big[ d(p,s)\,u(p,s)\,\e^{-\I p \cdot x} +  b^*\!(p,s)\,v(p,s)\,\e^{\I p \cdot x}  \Big] \,.
\end{eqnarray}
\end{subequations}
Hence the charge conjugation at classical level consists effectively only of interchanging
\begin{eqnarray}
b(p,s) &\longleftrightarrow& d(p,s) \,.
\label{app:charge:classical}
\end{eqnarray}

Let us now see how the operation of charge conjugation is implemented at quantum level. The process of quantization consists of promoting the numerical coefficients $b(p,s)$, $d(p,s)$ to operators, acting on the Fock space\footnote{We follow the convention and spell the family name of the Russian  physicist {\cyrrm{Vladimir Aleksandrovich Fok}} as \qm{Fock}, although it would be more appropriate to spell it as \qm{Fok}.} and satisfying certain anticommutation relations. The charge conjugation can now be implemented in terms of the unitary operator $U_C$ as
\begin{eqnarray}
\psi^\C &=& U_C \, \psi \, U_C^\dag \,.
\end{eqnarray}
As a consequence of the requirement $(\psi^\C)^\C=\psi$ the operator $U_C$ is also an involution, i.e.,
\begin{eqnarray}
U_C^{-1} &=& U_C \,.
\end{eqnarray}
This property together with the property of being unitary implies that $U_C$ is Hermitian. The operator $U_C$ commutes with the c-number part\footnote{However, for a numerical multiple $k\in\mathbb{C}$ of $\psi$ it still holds $(k\psi)^\C=k^*\psi^\C$.} of $\psi$  and acts non-trivially only on the creation and annihilation operators in analogy with \eqref{app:charge:classical} as
\begin{subequations}
\label{app:charge:Uc}
\begin{eqnarray}
U_C\,b(p,s)\,U_C^\dag &=& d(p,s) \,, \\
U_C\,d^\dag\!(p,s)\,U_C^\dag &=& b^\dag(p,s) \,.
\end{eqnarray}
\end{subequations}
To be specific, if the solution of the quantized Dirac equation is
\begin{eqnarray}
\label{app:charge:q_plane_waves}
\psi(x) &=& \sum_{\pm s} \int \! \d^3\threevector{p} \, N_p
\Big[ b(p,s)\,u(p,s)\,\e^{-\I p \cdot x} + d^\dag\!(p,s)\,v(p,s)\,\e^{\I p \cdot x}  \Big] \,,
\end{eqnarray}
then for it charge conjugate we have
\begin{eqnarray}
\psi^\C(x) &=& \sum_{\pm s} \int \! \d^3\threevector{p} \, N_p
\Big[ d(p,s)\,u(p,s)\,\e^{-\I p \cdot x} +  b^\dag\!(p,s)\,v(p,s)\,\e^{\I p \cdot x}  \Big] \,.
\end{eqnarray}


\chapter{Quantization of Dirac field}
\label{app:quant}


\intro{The aim of this appendix is to review various approaches to the canonical quantization of the Dirac field and to set up the formalism to be used in the subsequent appendix~\ref{app:major} when quantizing the Majorana field.}

\section{Na\"{\i}ve unconstrained Hamiltonian procedure}

The classical textbook approach to quantize the Dirac field uses the language and methods of the (unconstrained) Hamiltonian mechanics. One begins with the classical\footnote{Throughout this appendix we will consider only \emph{commuting} classical variables. The introduction of anticommuting (Grassmann) classical variables will become indispensable only in appendix~\ref{app:major} when discussing the canonical quantization of Majorana field.} Dirac field, defined by the Lagrangian
\begin{eqnarray}
\eL &=& \bar\psi\I\slashed{\partial}\psi-m\bar\psi\psi \,,
\label{app:quant:eL_nonherm}
\end{eqnarray}
with the dynamical variable $\psi$ being a complex four-component quantity. In accordance with the Hamiltonian mechanics, its conjugate momentum $\pi^\dag$ is defined as\footnote{Notice that we define the canonical momentum conjugate to the dynamical variable $\psi$ as $\pi^\dag$, i.e., \emph{with the dagger}. By this we follow the conventions that the quantity with (without) the dagger is a horizontal (vertical) vector (cf.~the case of $\psi$ and $\psi^\dag$). A similar convention will be adopted also in definitions \eqref{app:quant:pi}, \eqref{app:quant:constr}, \eqref{app:quant:htot} below.}
\begin{eqnarray}
\pi^\dag &\equiv& \frac{\partial\eL}{\partial\dot\psi} \,,
\label{app:quant:pi_gen}
\end{eqnarray}
which leads to
\begin{eqnarray}
\pi^\dag &=& \I\psi^\dag \,.
\label{app:quant:pi}
\end{eqnarray}
The complex conjugate bispinor $\psi^\dag$ is the other independent dynamical variable, with the associated conjugate momentum $\pi$ defined analogously to $\pi^\dag$ (and being actually the Hermitian conjugate of $\pi^\dag$). The space of all $\psi$, $\psi^\dag$, $\pi^\dag$, $\pi$ (as functions of the spatial coordinate $\threevector{x}$ with fixed time coordinate) constitutes the \emph{phase space}. There is an important notion of the Poisson bracket, which is a bilinear antisymmetric map on the space of all smooth functions (or functionals, since we are dealing with a field, i.e., with a system of infinite number of degrees of freedom) on the phase space. For two such functions $f$, $g$, it is defined as (the subscript $\mathrm{P.}$ stands for \qm{Poisson})
\begin{eqnarray}
\big\{ f, g \big\}_{\mathrm{P.}} &\equiv& \Bigg(
\frac{\delta f}{\delta \psi}\frac{\delta g}{\delta \pi^\dag} -
\frac{\delta g}{\delta \psi}\frac{\delta f}{\delta \pi^\dag} \Bigg)
+
\Bigg( \begin{array}{c}
\psi  \rightarrow  \psi^\dag  \\
\pi^\dag  \rightarrow  \pi
       \end{array}
 \Bigg) \,.
\end{eqnarray}
The \emph{fundamental} Poisson brackets are those of the phase space coordinates themselves (understood as the Dirac delta distributions on the phase space) and read
\begin{subequations}
\label{app:quant:poisson}
\begin{eqnarray}
\big\{ \psi_a(t,\threevector{x}), \pi_b^\dag(t,\threevector{y}) \big\}_{\mathrm{P.}} &=& \delta_{ab}\delta^3(\threevector{x}-\threevector{y}) \,,
\\
\big\{ \psi_a(t,\threevector{x}), \psi_b(t,\threevector{y}) \big\}_{\mathrm{P.}} &=& 0 \,,
\\
\big\{ \pi_a^\dag(t,\threevector{x}), \pi_b^\dag(t,\threevector{y}) \big\}_{\mathrm{P.}} &=& 0 \,,
\end{eqnarray}
\end{subequations}
and similarly for $\psi^\dag$, $\pi$. Now the process of canonical quantization consists of two main steps: First, the dynamical variables  $\psi$, $\psi^\dag$ (and consequently also their conjugate momenta $\pi^\dag$, $\pi$) are promoted to be operators on the Hilbert space of states (i.e., on the Fock space). Second, one postulates that they satisfy the equal-time anticommutation relations of the same form as the fundamental Poisson brackets \eqref{app:quant:poisson}, up to an additional factor of $\I$ on the right-hand sides. Using the explicit definition of the conjugate momenta in terms of $\psi$, $\psi^\dag$, these equal-time anticommutation relations read
\begin{subequations}
\label{app:quant:equaltime}
\begin{eqnarray}
\big\{ \psi_a(x), \psi_b^\dag(y) \big\}_{\mathrm{e.t.}} &=& \delta_{ab}\delta^3(\threevector{x}-\threevector{y}) \,, \\
\big\{ \psi_a(x), \psi_b(y) \big\}_{\mathrm{e.t.}} &=& 0 \,, \\
\big\{ \psi_a^\dag(x), \psi_b^\dag(y) \big\}_{\mathrm{e.t.}} &=& 0 \,.
\end{eqnarray}
\end{subequations}

However, the procedure described above is merely a mnemonic approach, leading incidentally to the correct result via an incorrect way. One way of seeing that there is a problem is the following: Instead of the non-Hermitian Lagrangian \eqref{app:quant:eL_nonherm} we could have equally well considered the Hermitian Lagrangian
\begin{eqnarray}
\eL &=& \frac{1}{2}\bar\psi\I\vecleftright{\slashed{\partial}}\psi-m\bar\psi\psi \,,
\label{app:quant:eL_herm}
\end{eqnarray}
which is completely equivalent to \eqref{app:quant:eL_nonherm}, because it differs from \eqref{app:quant:eL_nonherm} only by a total divergence and thus gives the same action and consequently the same equations of motion. However, now the conjugate momentum for the dynamical variable $\psi$ is
\begin{eqnarray}
\pi^\dag &=& \frac{1}{2}\I\psi^\dag \,,
\end{eqnarray}
which differs from the result \eqref{app:quant:pi} by a factor of $1/2$! This factor enters (via the procedure described in the previous paragraph) also the equal-time anticommutation relations of the quantized field. Hence it looks like that there is an ambiguity in the process of the canonical quantization, because it is possible to arrive at two different sets of anticommutation relations (differing by a factor of $1/2$) and it is not \emph{a priori} clear which of them is the correct one.\footnote{One could argue that while differentiating with respect to $\psi$ we considered $\psi^\dag$ to be a constant (and vice versa) and this might be the source of the problem. However, it turns out that this is not really the case. Indeed, doing everything carefully, taking as dynamical variables the real and imaginary parts of $\psi$, i.e., having in total eight real scalar dynamical variables rather than two complex four-component variables $\psi$, $\psi^\dag$ , the result would be the same.}


Let us now localize the source of the problem. When passing from the Lagrangian formalism to the Hamiltonian formalism, one has to determine the operator of Hamiltonian via the dual Legendre transformation. In order to do so, the equation \eqref{app:quant:pi_gen} has to be inverted, i.e., the velocity $\dot\psi$ has to be expressed as a function of the conjugate momentum $\pi^\dag$. This is possible if and only if the Hessian matrix
\begin{eqnarray}
W &\equiv& \frac{\partial^2 \eL}{\partial \dot\psi^2}
\end{eqnarray}
is invertible. (Analogously for $\psi^\dag$ and $\pi$.) In our case of the Dirac Lagrangian (regardless of whether in the Hermitian or non-Hermitian form) the Hessian is not only singular, it is actually identically vanishing, $W\equiv 0$, and the equation \eqref{app:quant:pi_gen} (neither of its component) cannot be inverted.\footnote{However, even though the Hessian is singular, still there is generally no problem in expressing the Hamiltonian only in terms of the dynamical variables and the conjugate momenta and not the velocities. E.g., in our case of identically vanishing Hessian, it will be shown that the Hamiltonian is a function only of the dynamical variables themselves, there is no dependence neither on the conjugate momenta, nor on the velocities.\label{app:quant:fn}} This means that the right-hand side of the equation \eqref{app:quant:pi_gen} has no $\dot\psi$-dependence (otherwise it could have been inverted) and thus it is a constraint of the type $\Phi(\psi,\pi)=0$ on the phase space. In fact, this could have been seen already before when we found that $\pi\sim\psi$: Since $\psi$ and $\pi$ are linearly related to each other, knowing the $\psi$ in a \emph{single} spacetime point allows one to determine $\pi$ in that point.

\section{Dirac constrained Hamiltonian procedure}

We see that the ordinary Hamiltonian approach does not work here because of the presence of constraints on the phase space. There is, however, a method how to deal with such a constrained Hamiltonian system, developed by Dirac \cite{Dirac:1950pj,Dirac:1958jc,Dirac:lectures}. We will not describe here his method in the full generality (interested reader can find details in the original works mentioned above, as well as in \cite{Batlle:1985ss}) but merely apply it for the purposes of the present case of the Dirac field.

In order to show that both the Hermitian and non-Hermitian Lagrangians \eqref{app:quant:eL_herm}, \eqref{app:quant:eL_nonherm} give unambiguously the same quantization, i.e., the same equal-time anticommutation relations \eqref{app:quant:equaltime}, we will consider the Lagrangian of the generic form
\begin{subequations}
\label{app:quant:lag_alpha}
\begin{eqnarray}
\eL &=& \frac{1}{2}\bar\psi\I\vecleftright{\slashed{\partial}}\psi - m\bar\psi\psi + \alpha\frac{\I}{2}\partial_\mu(\bar\psi\gamma^\mu\psi)
\\
&=&
\frac{\I}{2}\Big[ (1+\alpha)\bar\psi\gamma^\mu(\partial_\mu\psi)-(1-\alpha)(\partial_\mu\bar\psi)\gamma^\mu\psi \Big] - m\bar\psi\psi \,.
\end{eqnarray}
\end{subequations}
Here we parameterize by the arbitrary complex parameter $\alpha$ the whole class of equivalence of all Lagrangians, which differ only by a total divergence term. They all give consequently the same Euler--Lagrange equations of motion:
\begin{eqnarray}
\label{app:quant:eom}
(\I\slashed{\partial}-m)\psi &=& 0 \,.
\end{eqnarray}
The choice of $\alpha=1$ corresponds to the non-Hermitian Lagrangian \eqref{app:quant:eL_nonherm}, while $\alpha=0$ yields the Hermitian Lagrangian \eqref{app:quant:eL_herm}, respectively. Since $\alpha$ is completely unphysical parameter, we expect that the resulting equal-time anticommutation relations should have no $\alpha$-dependence. Note that using the na\"{\i}ve approach without constraints there would be the factor of $2/(1+\alpha)$ on the right-hand sides of \eqref{app:quant:equaltime}, which is precisely something we would like to get rid off.

Consider first the conjugate momenta for our dynamical variables $\psi$ and $\psi^\dag$:
\begin{subequations}
\label{app:quant:pi}
\begin{eqnarray}
\psi &\longrightarrow& \pi_1^\dag \>=\> \frac{\partial\eL}{\partial\dot\psi} \>=\> \frac{\I}{2}(1+\alpha)\psi^\dag \,,
\\
\psi^\dag &\longrightarrow& \pi_2 \>=\> \frac{\partial\eL}{\partial\dot\psi^\dag} \>=\> -\frac{\I}{2}(1-\alpha)\psi
\,.
\end{eqnarray}
\end{subequations}
Now we can calculate the canonical (or na\"{\i}ve) Hamiltonian\footnote{We will use freely the same term \qm{Hamiltonian} for both the Hamiltonian density $\mathcal{H}$ and the Hamiltonian $H$ itself (defined as $H=\int\!\d^3\threevector{x}\,\mathcal{H}$).} $\mathcal{H}_{\mathrm{C}}$ as usual via the dual Legendre transformation:
\begin{subequations}
\label{app:quant:ham_can}
\begin{eqnarray}
\mathcal{H}_{\mathrm{C}} &=& \pi_1^\dag\dot\psi+\dot\psi_2\pi_2-\eL
\\
&=& \frac{\I}{2}\Big[ (1+\alpha)\bar\psi\threevector{\gamma}\cdot(\threevector{\nabla}\psi) - (1-\alpha)(\threevector{\nabla}\bar\psi)\cdot\threevector{\gamma}\psi \Big] + m\bar\psi\psi \,.
\end{eqnarray}
\end{subequations}
We see that the canonical Hamiltonian does not depend on the conjugate momenta (as remarked in footnote~\ref{app:quant:fn} on page~\pageref{app:quant:fn}). This means that the corresponding Hamilton equations are not consistent -- they give different dynamics than the correct Euler--Lagrange equations \eqref{app:quant:eom}.

In order to solve the problem we first note that the definitions of the conjugate momenta \eqref{app:quant:pi} are independent on the velocities $\dot\psi$, $\dot\psi^\dag$ and hence they constitute the constraints on the phase space:
\begin{subequations}
\label{app:quant:constr}
\begin{eqnarray}
\phi_1^\dag(\pi_1^\dag,\psi^\dag) &\equiv& \pi_1^\dag-\frac{\I}{2}(1+\alpha)\psi^\dag \>=\> 0 \,,
\\
\phi_2(\pi_2,\psi) &\equiv& \pi_2+\frac{\I}{2}(1-\alpha)\psi \>=\> 0 \,.
\end{eqnarray}
\end{subequations}
Now we introduce the total Hamiltonian, which is the canonical one plus a linear combination of the constraints $\phi_1^\dag$, $\phi_2$:
\begin{eqnarray}
\label{app:quant:htot}
\mathcal{H}_{\mathrm{T}} &\equiv& \mathcal{H}_{\mathrm{C}}+\phi_1^\dag\lambda_1+\lambda_2^\dag\phi_2
\end{eqnarray}
The total Hamiltonian is equivalent to the canonical one on the subspace of the phase space where the solutions of equations of motion lie, since in such a case the constraints are supposed to vanish.

Let us now turn our attention to the \qm{Lagrange multipliers} $\lambda_1$, $\lambda_2^\dag$. We need some condition to determine them. The natural requirement is that the constraints \eqref{app:quant:constr} hold constantly during the time evolution of the system, governed by the total Hamiltonian \eqref{app:quant:htot}. I.e., we demand
\begin{subequations}
\label{app:quant:cond}
\begin{eqnarray}
\dot \phi_1^\dag &=& \big\{ \phi_1^\dag, H_{\mathrm{T}} \big\}_{\mathrm{P.}} \>\exclamation{=}\> 0 \,,
\\
\dot \phi_2 &=& \big\{ \phi_2, H_{\mathrm{T}}  \big\}_{\mathrm{P.}} \>\exclamation{=}\> 0 \,.
\end{eqnarray}
\end{subequations}
Now the question is whether these two conditions are enough to determine both the multipliers $\lambda_1$, $\lambda_2^\dag$. If not, it would mean that in addition to the \emph{primary constraints} \eqref{app:quant:constr} there are some other constraints in the theory. These so called \emph{secondary constraints} can be found via the iterative \emph{Dirac procedure} of consecutive adding new constraints and corresponding Lagrange multipliers to the total Hamiltonian until the requirement of time-independence of all such constraints leads to determination of all Lagrange multipliers.

In the present case of the Dirac field it turns out that the conditions \eqref{app:quant:cond} do really fix the Lagrange multipliers uniquely as
\begin{subequations}
\label{app:quant:lamdas}
\begin{eqnarray}
\lambda_1 &=& +\I\big\{ \phi_2, H_{\mathrm{C}} \big\}_{\mathrm{P.}} \>=\>
\gamma_0\threevector{\gamma}\cdot(\threevector{\nabla}\psi)-\I m \gamma_0\psi \,,
\\
\lambda_2^\dag &=& -\I\big\{ \phi_1^\dag, H_{\mathrm{C}} \big\}_{\mathrm{P.}} \>=\>
-(\threevector{\nabla}\bar\psi)\cdot\threevector{\gamma}+\I m \bar\psi
\end{eqnarray}
\end{subequations}
and consequently, the primary constraints \eqref{app:quant:constr} are the only ones in the Hamiltonian formulation of theory of Dirac field. Using the explicit form of the Lagrange multipliers we can write the total Hamiltonian \eqref{app:quant:htot} as\footnote{We use here for conjugate momenta the same notation of Dirac conjugation as for the bispinors, i.e., $\bar\pi=\pi^\dag\gamma_0$.}
\begin{eqnarray}
\label{app:quant:hamtot}
\mathcal{H}_{\mathrm{T}} &=& \bar\pi_1\threevector{\gamma}\cdot(\threevector{\nabla}\psi) + (\threevector{\nabla}\bar\psi)\cdot\threevector{\gamma}\pi_2 -\I m(\bar\pi_1\psi-\bar\psi\pi_2) \,.
\end{eqnarray}
It is easy to convince oneself that the Hamilton equations of motion of the total Hamiltonian \eqref{app:quant:hamtot} are equivalent to those \eqref{app:quant:eom} of Euler--Lagrange, provided one uses the definitions of the conjugate momenta \eqref{app:quant:pi}.

So far we have only shown how to treat correctly the dynamics at the classical level. Concerning the quantization, we have already seen that postulating the equal-time anticommutation relations as analogues of the fundamental Poisson brackets (which of course hold in the same form \eqref{app:quant:poisson} for all $\alpha$ by the definition) yields inconsistent (because $\alpha$-dependent) equal-time anticommutation relations. Dirac suggested that instead of the Poisson bracket $\{\cdot,\cdot\}_{\mathrm{P.}}$ we should take its generalization -- the \emph{Dirac bracket} $\{\cdot,\cdot\}_{\mathrm{D.}}$. The advantage of the Dirac bracket is that it incorporates the structure of the constraints. For example, using the Dirac bracket the time evolution is generated not by the total Hamiltonian, but merely by the canonical one:
\begin{eqnarray}
\label{app:quant:dirbr1}
\dot f &=& \big\{ f, H_{\mathrm{T}} \big\}_{\mathrm{P.}} \>=\> \big\{ f, H_{\mathrm{C}} \big\}_{\mathrm{D.}} \,.
\end{eqnarray}
Moreover, the Dirac bracket of any function $f$ on the phase space with any second class constraint $\Phi_i$ is vanishing:
\begin{eqnarray}
\label{app:quant:dirbr2}
\big\{ f, \Phi_i \big\}_{\mathrm{D.}} &=& 0 \,.
\end{eqnarray}

Before we proceed to the definition of the Dirac bracket, certain classification of the constraints must be done. Those constraints whose mutual Poisson bracket are all vanishing are called the \emph{first class} constraints. It can be shown that the first class constraints are associated with some non-physical degrees of freedom, they in fact generate gauge symmetries. This type of constraints arises for instance in the Yang--Mills theories. The other constraints (which have at least one non-vanishing Poisson bracket with the others) are called the \emph{second class} constraints.

Now let us introduce the vector $\Phi$ of all second class constraints and calculate the antisymmetric matrix $C$ of the Poisson brackets of all its entries: $C_{ij} \equiv \{\Phi_i,\Phi_j\}_{\mathrm{P.}}$. This matrix can be shown to be regular.\footnote{If it were not, there would exist such a basis in the vector space of the second class constraints that the matrix $C$ would have a diagonal form, with at least one zero on the diagonal (provided we have a special case of diagonalizable matrix $C$). Accordingly there would be some constraint(s) with vanishing Poisson bracket with all other constraints. However, this contradicts our assumption that we are dealing with the second class constraints only.} Now we can define the Dirac bracket of any two functions $f$, $g$ on the phase space as\footnote{We are using here the summation convention also for the space indices. More precisely, yet less clearly, since the proper definition of the matrix $C$ is $C_{ij}(\threevector{x},\threevector{y}) \equiv \{\Phi_i(\threevector{x}),\Phi_j(\threevector{y})\}_{\mathrm{P.}}$ (omitting the time dependence), the second term on the right-hand side of \eqref{app:quant:dirbr_def} should be $\int\!\d^3\threevector{x}\,\d^3\threevector{y}\,\big\{ f, \Phi_a(\threevector{x}) \big\}_{\mathrm{P.}}\,C_{ab}^{-1}(\threevector{x},\threevector{y})\,\big\{ \Phi_b(\threevector{y}), g \big\}_{\mathrm{P.}}$.}

\begin{eqnarray}
\label{app:quant:dirbr_def}
\big\{ f, g \big\}_{\mathrm{D.}} &\equiv& \big\{ f, g \big\}_{\mathrm{P.}}-
\big\{ f, \Phi_a \big\}_{\mathrm{P.}}C_{ab}^{-1}\big\{ \Phi_b, g \big\}_{\mathrm{P.}} \,.
\end{eqnarray}
It is easy to convince oneself that the Dirac bracket not only shares some properties with the Poisson bracket -- it is bilinear, antisymmetric, satisfies the Jacobi identity (and, of course, it reduces to the Poisson bracket in absence of any second class constraints), but it also satisfies the \qm{constraints compatible} conditions advertised above (cf.~\eqref{app:quant:dirbr1} and \eqref{app:quant:dirbr2}).

Let us now turn back to our case of the Dirac field. Explicit calculation reveals that $\{ \phi_1^\dag,\phi_2 \}_{\mathrm{P.}} = -\{ \phi_2,\phi_1^\dag \}_{\mathrm{P.}}=-\I$ and hence the constraints $\phi_1^\dag$, $\phi_2$ are of the second class.\footnote{This is connected to the fact that we were able to determine uniquely the Lagrange multipliers $\lambda_1$, $\lambda_2^\dag$ from the equations \eqref{app:quant:cond}. If the constraints $\phi_1^\dag$, $\phi_2$ were of the first class, the multipliers would remain undetermined (at least at the first stage of the iterative Dirac procedure).} The corresponding matrix $C$ reads
\begin{eqnarray}
C \>=\> C^{-1} &=&
\left(\begin{array}{rr}
0 & -\I \\
\I & 0 \\
\end{array}\right)
\delta^3(\threevector{x}-\threevector{y}) \,.
\end{eqnarray}
Plugging this into the definition of the Dirac bracket, we arrive at the fundamental Dirac brackets (omitting the trivial ones):
\begin{subequations}
\begin{eqnarray}
\big\{ \psi_a(t,\threevector{x}), \pi_{1,b}^\dag(t,\threevector{y}) \big\}_{\mathrm{D.}} &=& \frac{1+\alpha}{2}\delta_{ab}\delta^3(\threevector{x}-\threevector{y}) \,,
\\
\big\{ \psi_a^\dag(t,\threevector{x}), \pi_{2,b}(t,\threevector{y}) \big\}_{\mathrm{D.}} &=& \frac{1-\alpha}{2}\delta_{ab}\delta^3(\threevector{x}-\threevector{y}) \,.
\end{eqnarray}
\end{subequations}
If we now express the conjugate momenta $\pi_1^\dag$, $\pi_2$ in terms of $\psi$, $\psi^\dag$ (Eq.~\eqref{app:quant:pi}), the factors of $(1\pm\alpha)/2$ cancel each other and we arrive at the desired $\alpha$-independent result
\begin{subequations}
\begin{eqnarray}
\big\{ \psi_a(t,\threevector{x}), \I\psi_b^\dag(t,\threevector{y}) \big\}_{\mathrm{D.}} &=& \delta_{ab}\delta^3(\threevector{x}-\threevector{y}) \,,
\end{eqnarray}
\end{subequations}
which leads to the correct equal-time anticommutation relations \eqref{app:quant:equaltime}.

\section{Faddeev and Jackiw method}
\label{app:quant:sec:FJ}

The above described Dirac method is a sort of \qm{classical} method of quantizing constrained Hamiltonian systems. However, there exist an alternative, easier method developed by Faddeev and Jackiw \cite{Faddeev:1988qp,Jackiw:1993in}, which gives the same answers using much less effort. Their method is well suited for systems, whose Lagrangian is linear in the first time derivatives (velocities) and consequently considered singular from the traditional Hamiltonian point of view.

In the Faddeev--Jackiw approach the system of the Dirac field actually turns out to be \emph{unconstrained}. The key observation is that if one understands the phase space as the set of all possible states of the system, or, equivalently, as the set of all initial conditions of the corresponding equations of motion, then for the case of the Dirac field (with first-order Euler--Lagrange equations) the configuration space and the phase space actually coincide. This is the very reason why introducing the conjugate momenta in fact means introducing artificial constraints.

In order to apply the Faddeev--Jackiw method, it is useful first to introduce some new simplifying formalism. Instead of dealing with two independent four-component dynamical variables $\psi$ and $\psi^\dag$ separately, it is useful to combine them to make a new single eight-component variable $\chi$
\begin{eqnarray}
\label{app:quant:chi}
\chi &\equiv& \left(\begin{array}{c} \psi \\ \psi^* \end{array}\right) \,,
\end{eqnarray}
which now constitutes the configuration space (i.e., the phase space).\footnote{Alternatively, we could define $\chi$ as eight-component \emph{real} vector $\chi \equiv \sqrt{2} \bigl(\begin{smallmatrix} \Re\psi \\ \Im\psi \end{smallmatrix} \bigr)$.
However, this is equivalent to our choice \eqref{app:quant:chi}, since both bases are related to each other through a unitary transformation.} (Notice that $\chi$ is actually a variant of the Nambu--Gorkov field, to be discussed in appendix~\ref{app:fermi propag}.) The Lagrangian \eqref{app:quant:lag_alpha} can now be written in terms of $\chi$ as
\begin{eqnarray}
\eL &=& \frac{\I}{2}\chi^\T \Big[ (A^{0}-A^{0\T})+\alpha(A^{0}+A^{0\T}) \Big] \dot \chi - \mathcal{H} \,,
\end{eqnarray}
where the canonical Hamiltonian $\mathcal{H}$, \eqref{app:quant:ham_can}, (we omit the subscript $C$) is now rewritten as
\begin{eqnarray}
\label{app:quant:ham_can_FD}
\mathcal{H} &=& \frac{\I}{2}\chi^\T \Big[ (\threevector{A}-\threevector{A}^{\T})+\alpha(\threevector{A}+\threevector{A}^{\T}) \Big] \cdot (\threevector{\nabla}\chi) + \frac{1}{2}m \chi^\T (B+B^\T) \chi \,.
\end{eqnarray}
The matrices $A^\mu$, $B$ are defined as
\begin{equation}
A^\mu \>\equiv\> \left(\begin{array}{cc} 0 & 0 \\ \gamma_0\gamma^\mu & 0 \end{array}\right) \,,
\qquad
B \>\equiv\> \left(\begin{array}{cc} 0 & 0 \\ \gamma_0 & 0 \end{array}\right) \,,
\end{equation}
or more suggestively, using the familiar notation $\gamma_0=\beta$, $\gamma_0\threevector{\gamma}=\threevector{\alpha}$,
\begin{equation}
A^0 \>=\> \left(\begin{array}{rr} 0 & 0 \\ 1 & 0 \end{array} \right) \,,
\qquad
\threevector{A} \>=\>\left(\begin{array}{rr} 0 & 0 \\ \threevector{\alpha} & 0 \end{array}\right) \,,
\qquad
B \>=\> \left(\begin{array}{rr} 0 & 0 \\ \beta & 0 \end{array}\right) \,.
\end{equation}
The corresponding Euler--Lagrange equations read
\begin{subequations}
\label{app:quant:eom_FD}
\begin{eqnarray}
\I(A^{0}-A^{0\T})\dot \chi &=& \frac{\delta H}{\delta \chi}
\\
&=& \I(\threevector{A}-\threevector{A}^{\T})\cdot (\threevector{\nabla}\chi) + m (B+B^\T) \chi \,.
\end{eqnarray}
\end{subequations}
It is straightforward to check that these equations are equivalent to those in the usual form \eqref{app:quant:eom}.

In analogy with \eqref{app:quant:dirbr1} Faddeev and Jackiw postulated that the time-evolution of a function $f$ on the phase space is governed by the Hamiltonian \eqref{app:quant:ham_can_FD} as
\begin{eqnarray}
\dot f &=& \big\{ f, H \big\}_{\mathrm{F.J.}} \,,
\end{eqnarray}
with the \emph{Faddeev--Jackiw bracket} $\{\cdot,\cdot\}_{\mathrm{F.J.}}$ defined as
\begin{eqnarray}
\big\{ f, g \big\}_{\mathrm{F.J.}} &\equiv&
\frac{\delta f}{\delta \chi_i}\frac{\delta g}{\delta \chi_j}\Omega_{ij} \,.
\end{eqnarray}
For determining the unknown $8 \times 8$ matrix $\Omega$ we notice that the time-evolution of the phase space coordinates themselves is
\begin{eqnarray}
\dot \chi &=& \big\{ \chi, H \big\}_{\mathrm{F.J.}}
\>=\>
\Omega \frac{\delta H}{\delta \chi} \,.
\end{eqnarray}
Now comparing this equation with the equation of motion \eqref{app:quant:eom_FD} we readily see that the matrix $\Omega$ is given by
\begin{eqnarray}
\Omega &=& -\I(A^{0}-A^{0\T})^{-1} \,.
\end{eqnarray}
or explicitly, in terms of $4\times4$ blocks,
\begin{eqnarray}
\label{app:quant:omega}
\Omega &=& \left(\begin{array}{rr} 0 & -\I \\ \I & 0 \end{array}\right) \,.
\end{eqnarray}
It is interesting to note that due to the form of the matrix $\Omega$ the Faddeev--Jackiw bracket is antisymmetric, which implies that it also satisfies the Jacobi identity $\{f, \{g,h\}_{\mathrm{F.J.}}\}_{\mathrm{F.J.}}+\hbox{cycl.}=0$. This, together with the Leibnitz rule $\{f,gh\}_{\mathrm{F.J.}}=g\{f,h\}_{\mathrm{F.J.}}+\{f,g\}_{\mathrm{F.J.}}h$, means that the Faddeev--Jackiw bracket defines a Poisson structure on the phase space in the same way as the Poisson and Dirac brackets do.

As a basis for the quantization we will use, as usual, the \qm{fundamental} Faddeev--Jackiw brackets
\begin{eqnarray}
\big\{ \chi_i(t,\threevector{x}), \chi_j(t,\threevector{y}) \big\}_{\mathrm{F.J.}}
&=& \Omega_{ij}\,\delta^3(\threevector{x}-\threevector{y}) \,,
\end{eqnarray}
which in terms of $\psi$ and $\psi^\dag$, using the definitions of $\Omega$, \eqref{app:quant:omega}, and $\chi$, \eqref{app:quant:chi}, read
\begin{subequations}
\begin{eqnarray}
\big\{ \psi_a(t,\threevector{x}), \psi_b^\dag(t,\threevector{y}) \big\}_{\mathrm{F.J.}} &=& -\I \delta_{ab}\,\delta^3(\threevector{x}-\threevector{y}) \,,
\\
\big\{ \psi_a(t,\threevector{x}), \psi_b(t,\threevector{y}) \big\}_{\mathrm{F.J.}} &=& 0 \,,
\\
\big\{ \psi_a^\dag(t,\threevector{x}), \psi_b^\dag(t,\threevector{y}) \big\}_{\mathrm{F.J.}} &=& 0 \,.
\end{eqnarray}
\end{subequations}
Clearly, these Faddeev--Jackiw brackets give upon quantization rise to the correct equal-time anticommutation relations \eqref{app:quant:equaltime} without the unwanted $\alpha$-dependency (and with less effort than the Dirac procedure).

\chapter{Quantization of Majorana field}
\label{app:major}

\intro{In this appendix we quantize the Majorana field. Although we can use (and will use) for that purpose the technique introduced in the previous appendix on the example of quantizing the Dirac field, there are also certain substantial differences, due to which it is worth dedicating a special appendix to it.}


\section{Necessity of Grassmann variables}

A fermion field $\psi$, satisfying the \emph{Majorana condition} \cite{Majorana:1937vz}
\begin{eqnarray}
\psi &=& \psi^\C \,,
\label{app:major:majorana_cond}
\end{eqnarray}
is called the \emph{Majorana field}. It can be quantized in a similar way as the unconstrained Dirac field, which was done in appendix~\ref{app:quant}. There is one important conceptual difference, however. When quantizing the Dirac field, we started with a Lagrangian of a classical Dirac field. This field was consider to be commuting, the property of being anticommuting was introduced only when promoting the classical field to operator field and introducing the equal-time anticommutation relation. At the classical level there was no problem with the commutation of the field variables, at least not when analyzing the dynamics and introducing various types of brackets. The only problem was that the Hamiltonian was unbounded from below, but this did not concern us (at the classical level; at the quantum level this was cured by the anticommutation of the field operators).

On the contrary, for Majorana field one has to introduce the property of being anticommuting from the very beginning, already at the level of classical Lagrangian. The reason for that is that when imposing the Majorana condition on commuting fields, the Lagrangian itself turns out to be identically vanishing. Let us see it in detail. For the mass term we have (we omit here the unnecessary factor of $m$)
\begin{subequations}
\label{app:maj:commmass}
\begin{eqnarray}
\eL_{\mathrm{mass}} &=& \bar\psi\psi
\label{app:maj:commmass1}
\\ &=& \bar\psi^\C\psi^\C
\label{app:maj:commmass2}
\\ &=& - \psi^\T C^{-1} C \bar\psi^\T
\label{app:maj:commmass3}
\\ &=& - (\bar\psi\psi)^\T
\label{app:maj:commmass4}
\\ &=& - \eL_{\mathrm{mass}} \,.
\label{app:maj:commmass5}
\end{eqnarray}
\end{subequations}
Thus, it must be $\eL_{\mathrm{mass}} = 0$. In the steps in \eqref{app:maj:commmass} we mostly used the results from appendix~\ref{app:charge}, concerning the charge conjugation. The key step, however, was in line \eqref{app:maj:commmass4}, when the assumption of commutativity of fermion fields came into play. If we had assumed rather anticommuting fields, there would be opposite sign in \eqref{app:maj:commmass4}. Let us continue with the kinetic term (again, omitting the factor of $\I$):
\begin{subequations}
\label{app:maj:commkinpoprve}
\begin{eqnarray}
\eL_{\mathrm{kinetic}} &=& \bar\psi\gamma_\mu(\partial^\mu\psi)
\label{app:maj:commkinpoprve1}
\\ &=& \bar\psi^\C\gamma_\mu(\partial^\mu\psi^\C)
\label{app:maj:commkinpoprve2}
\\ &=& -\psi^\T \underbrace{C^{-1}\gamma_\mu C}_{-\gamma_\mu^\T} (\partial^\mu\bar\psi^\T)
\label{app:maj:commkinpoprve3}
\\ &=& [(\partial^\mu\bar\psi) \gamma_\mu \psi]^\T
\label{app:maj:commkinpoprve4}
\\ &=& (\partial^\mu\bar\psi) \gamma_\mu \psi \,.
\label{app:maj:commkinpoprve5}
\end{eqnarray}
\end{subequations}
Most of the steps here were analogous to those in \eqref{app:maj:commmass}. This time the commutativity of $\psi$ was used in line \eqref{app:maj:commkinpoprve4}. On the other hand, the kinetic term \eqref{app:maj:commkinpoprve1} can be rewritten also using the integration by parts as
\begin{eqnarray}
\eL_{\mathrm{kinetic}} &=& -(\partial^\mu\bar\psi)\gamma_\mu\psi + \partial^\mu(\bar\psi\gamma_\mu\psi) \,.
\label{app:maj:commkinpodruhe}
\end{eqnarray}
Thus, as long as we can neglect the total divergence $\partial^\mu(\bar\psi\gamma_\mu\psi)$, we again see that $\eL_{\mathrm{kinetic}}=  - \eL_{\mathrm{kinetic}}$, so that $\eL_{\mathrm{kinetic}} = 0$.

We will thus suppose that the components of the bispinor $\psi$ are anticommuting (Grassmann) variables: $\{\psi_a,\psi_b\}=0=\{\psi_a,\psi_b^*\}$. The Lagrangian reads
\begin{subequations}
\label{app:major:lag_maj}
\begin{eqnarray}
\eL &=& 
\frac{1}{4}\bar\psi\I\vecleftright{\slashed{\partial}}\psi
-\frac{1}{2}m\bar\psi\psi
\\
&=& \frac{1}{2}\bar\psi\I\slashed{\partial}\psi-\frac{1}{2}m\bar\psi\psi \,.
\end{eqnarray}
\end{subequations}
Unlike the case of Dirac field, both of the forms of Lagrangian \eqref{app:major:lag_maj} for Majorana anticommuting field are exactly equal to each other and are perfectly Hermitian. There is no possibility to add a total divergence term of the type $\partial_\mu(\bar\psi\gamma^\mu\psi)$, to the Lagrangian, since for (anticommuting) Majorana fermion the bilinear $\bar\psi\gamma^\mu\psi$ identically vanishes.\footnote{This is related also to the fact that a Majorana field, as being basically a real fermion field, cannot be charged under any $\group{U}(1)$ symmetry, whose Noether current would be otherwise proportional just to the (actually vanishing) quantity $\bar\psi\gamma^\mu\psi$.}

The extra factors of $1/2$ in the Majorana Lagrangian \eqref{app:major:lag_maj} (as compared to the Dirac Lagrangian) are coming from the decomposition of a Dirac field $\psi$ to two Majorana fields $\psi_{1,2}$ (cf.~\eqref{app:charge:decomp}):
\begin{eqnarray}
\label{app:major:decomp}
\psi &=& \frac{1}{\sqrt{2}} \big( \psi_1+\I\psi_2 \big) \,.
\end{eqnarray}
Plugging this decomposition to the Dirac Lagrangian one can rewrite it as a sum of two Majorana Lagrangians \eqref{app:major:lag_maj}.\footnote{As a matter of fact, this is possible again only due to the fact that the field variables anticommute. I.e., in such case upon plugging the decomposition \eqref{app:major:decomp} to the Dirac Lagrangian the result is diagonal in the Majorana fields $\psi_{1,2}$. On the other hand, in commuting case there would be rather off-diagonal terms of (e.g., $\bar\psi_1\psi_2$) instead.} I.e., the Dirac field $\psi$ can be understood as two independent Majorana fields $\psi_{1,2}$ with equal masses. The factor of $1/\sqrt{2}$ in the decomposition \eqref{app:major:decomp} ensures that the creation and annihilation operators of the eventually quantized Majorana fields $\psi_{1,2}$ are properly normalized, as will be shown below.



\section{Quantization}

For quantizing the Majorana field we will use the method of Faddeev and Jackiw, described in appendix~\ref{app:quant}. When quantizing the Dirac field, our independent dynamical variables were $\psi$ and $\psi^*$. Now these variable are no more independent, in fact they are proportional to each other as a consequence of the Majorana condition \eqref{app:major:majorana_cond}. Hence we will take $\psi$ as our only dynamical variable. The Lagrangian \eqref{app:major:lag_maj} can be rewritten only in terms of $\psi$ (using the Majorana condition and the assumption of anticommutation) as
\begin{eqnarray}
\eL &=& -\frac{\I}{2} \psi^\T C^{-1}\gamma_0\dot\psi-\mathcal{H} \,,
\end{eqnarray}
with the Hamiltonian
\begin{eqnarray}
\mathcal{H} &=&
-\frac{\I}{2} \psi^\T C^{-1}\threevector{\gamma}\cdot(\threevector{\nabla}\psi) -\frac{m}{2}\psi^\T C^{-1}\psi \,,
\end{eqnarray}
where $C$ is the matrix of charge conjugation, introduced in appendix~\ref{app:charge}. The corresponding Euler--Lagrange equations are
\begin{subequations}
\label{app:major:eom}
\begin{eqnarray}
-\I C^{-1}\gamma_0\dot\psi &=& \frac{\delta H}{\delta \psi}
\\
&=& -\I C^{-1}\threevector{\gamma}\cdot(\threevector{\nabla}\psi) -m C^{-1}\psi \,,
\end{eqnarray}
\end{subequations}
which can be rewritten in the usual covariant form as the Dirac equation
\begin{eqnarray}
\label{app:major:eom_dirac}
(\I\slashed{\partial}-m)\psi &=& 0 \,.
\end{eqnarray}

Now we will determine the fundamental Faddeev--Jackiw brackets (see Sec.~\ref{app:quant:sec:FJ} of the previous appendix), which will later serve as a basis for the quantization of the system:
\begin{eqnarray}
\big\{ \psi_i(\threevector{x}), \psi_j(\threevector{y}) \big\}_{\mathrm{F.J.}} &=&\Omega_{ij}\,\delta^3(\threevector{x}-\threevector{y}) \,.
\end{eqnarray}
The unknown $4 \times 4$ matrix $\Omega$ will be determined from the requirement that the time evolution is given by
\begin{eqnarray}
\dot\psi &=& \big\{ \psi, H \big\}_{\mathrm{F.J.}} \>=\> \Omega \frac{\delta H}{\delta \psi} \,.
\end{eqnarray}
Comparing this with the equations of motion \eqref{app:major:eom} we find the matrix $\Omega$ as
\begin{eqnarray}
\Omega &=& (-\I C^{-1} \gamma_0)^{-1} \>=\> \I \gamma_0 C
\end{eqnarray}
and arrive at the fundamental Faddeev--Jackiw brackets
\begin{eqnarray}
\label{app:major:fj1}
\big\{ \psi_a(\threevector{x}), \psi_b(\threevector{y}) \big\}_{\mathrm{F.J.}} &=&  \I(\gamma_0 C)_{ab}\,\delta^3(\threevector{x}-\threevector{y}) \,.
\end{eqnarray}
(Note that in contrast to the case of the Dirac field, discussed in appendix~\ref{app:quant}, now as a consequence of anticommutation of $\psi$ the Faddeev--Jackiw brackets are symmetric, since $(\gamma_0 C)^\T=\gamma_0 C$.) Moreover, we can use just derived brackets of the type $\{\psi,\psi\}_{\mathrm{F.J.}}$, \eqref{app:major:fj1}, to derive also those of the type $\{\psi^*,\psi^*\}_{\mathrm{F.J.}}$ and $\{\psi,\psi^*\}_{\mathrm{F.J.}}$, using only the Majorana condition and the properties of the charge conjugation:
\begin{subequations}
\label{app:major:fj2}
\begin{eqnarray}
\big\{ \psi_a^*(\threevector{x}), \psi_b^*(\threevector{y}) \big\}_{\mathrm{F.J.}} &=&  \I(C^{-1}\gamma_0)_{ab}\,\delta^3(\threevector{x}-\threevector{y}) \,,
\\
\big\{ \psi_a(\threevector{x}), \psi_b^*(\threevector{y}) \big\}_{\mathrm{F.J.}} &=&  -\I\delta_{ab}\,\delta^3(\threevector{x}-\threevector{y}) \,.
\end{eqnarray}
\end{subequations}

Applying the prescription of canonical quantization on the above Faddeev--Jackiw brackets \eqref{app:major:fj1} and \eqref{app:major:fj2} we readily arrive at the equal-time anticommutation relations\footnote{Again, only one of the following three anticommutators is independent, the other two can be derived from it using the Majorana condition.}
\begin{subequations}
\label{app:major:et}
\begin{eqnarray}
\big\{ \psi_a(x), \psi_b^\dag(y) \big\}_{\mathrm{e.t.}} &=&
\delta_{ab}\,\delta^3(\threevector{x}-\threevector{y}) \,,
\label{app:major:et1}
\\
\big\{ \psi_a(x), \psi_b(y) \big\}_{\mathrm{e.t.}} &=&
-(\gamma_0 C)_{ab}\,\delta^3(\threevector{x}-\threevector{y}) \,,
\label{app:major:et2}
\\
\big\{ \psi_a^\dag(x), \psi_b^\dag(y) \big\}_{\mathrm{e.t.}} &=&
-(C^{-1}\gamma_0)_{ab}\,\delta^3(\threevector{x}-\threevector{y}) \,.
\label{app:major:et3}
\end{eqnarray}
\end{subequations}
The first anticommutator \eqref{app:major:et1} of the type $\{\psi,\psi^\dag\}_{\mathrm{e.t.}}$ is the same as in the Dirac case. However, the other two of the type $\{\psi,\psi\}_{\mathrm{e.t.}}$, \eqref{app:major:et2}, and $\{\psi^\dag,\psi^\dag\}_{\mathrm{e.t.}}$, \eqref{app:major:et3}, while trivial in the Dirac case, are now non-trivial, which is novel and important feature of the Majorana field.


\section{Creation and annihilation operators}

Recall that general solution of the quantized Dirac equation in the plane wave expansion (cf.~\eqref{app:charge:q_plane_waves}) reads
\begin{eqnarray}
\label{app:major:solution_dirac}
\psi(x) &=& \sum_{\pm s} \int\!\d^3\threevector{p} \, N_p
\Big[ b(p,s)\,u(p,s)\,\e^{-\I p \cdot x} + d^\dag\!(p,s)\,v(p,s)\,\e^{\I p \cdot x} \Big] \,,
\end{eqnarray}
where the annihilation and creation operators $b(p,s)$, $b^\dag\!(p,s)$ and $d(p,s)$, $d^\dag\!(p,s)$ satisfy the well known anticommutation relations\footnote{We list here (as well as below in \eqref{app:major:anti_rel_maj}) only the independent non-trivial anticommutators.}
\begin{subequations}
\label{app:major:anti_rel_dir}
\begin{eqnarray}
\big\{ b(p,s), b^\dag\!(p^{\prime},s^{\prime}) \big\} &=&
\delta_{ss^{\prime}} \, \delta^{3}(\threevector{p}-\threevector{p}^{\prime}) \,,
\\
\big\{ d(p,s), d^\dag\!(p^{\prime},s^{\prime}) \big\} &=&
\delta_{ss^{\prime}} \, \delta^{3}(\threevector{p}-\threevector{p}^{\prime}) \,.
\end{eqnarray}
\end{subequations}
These anticommutation relations are implied by the equal-time anticommutation relations of the Dirac field \eqref{app:quant:equaltime}.

We have seen that the Majorana field is a solution of the ordinary Dirac equation \eqref{app:major:eom_dirac} constrained by the Majorana condition \eqref{app:major:majorana_cond}. Thus, applying the Majorana condition on \eqref{app:major:solution_dirac} we readily arrive at the general Majorana solution of the Dirac equation \cite{Bilenky:1987ty}:
\begin{eqnarray}
\label{app:major:solution_majorana}
\psi(x) &=& \sum_{\pm s} \int \! \d^3\threevector{p} \, N_p
\Big[ a(p,s)\,u(p,s)\,\e^{-\I p \cdot x} + a^\dag\!(p,s)\,v(p,s)\,\e^{\I p \cdot x} \Big] \,.
\end{eqnarray}
Now the Majorana equal-time anticommutation relations \eqref{app:major:et} imply that the annihilation and creation operators $a(p,s)$, $a^\dag\!(p,s)$ satisfy
\begin{eqnarray}
\label{app:major:anti_rel_maj}
\big\{ a(p,s), a^\dag\!(p^{\prime},s^{\prime}) \big\} &=&
\delta_{ss^{\prime}} \, \delta^{3}(\threevector{p}-\threevector{p}^{\prime}) \,.
\end{eqnarray}
For completeness let us also note that the unitary operator of the charge conjugation $U_C$ (introduced in appendix~\ref{app:charge}) now acts trivially on Majorana creation and annihilation operators (cf.~Eq.~\eqref{app:charge:Uc}):
\begin{eqnarray}
U_C \, a(p,s) \, U_C^\dag &=& a(p,s) \,.
\end{eqnarray}

Return now to the decomposition \eqref{app:major:decomp} of a Dirac field $\psi$ to a sum of two Majorana fields $\psi_{1,2}$. Denoting the annihilation operators of the Majorana fields $\psi_{1,2}$ as $a_{1,2}(p,s)$ and plugging the plane wave expansions \eqref{app:major:solution_dirac}, \eqref{app:major:solution_majorana} into the decomposition \eqref{app:major:decomp}, we see that the annihilation operators $a_{1,2}(p,s)$ are expressed in terms of $b(p,s)$, $d(p,s)$ as
\begin{subequations}
\begin{eqnarray}
a_{1}(p,s) &=& \frac{1}{\sqrt{2}}  \big( d(p,s) + b(p,s) \big)  \,, \\
a_{2}(p,s) &=& \frac{\I}{\sqrt{2}} \big( d(p,s) - b(p,s) \big)  \,.
\end{eqnarray}
\end{subequations}
Now it is straightforward to calculate the anticommutation relations $\{a_{i}(p,s),a_{i}^\dag\!(p,s)\}$, $i=1,2$, using the Dirac anticommutation relation \eqref{app:major:anti_rel_dir} and to check that they do correspond to the Majorana anticommutation relations \eqref{app:major:anti_rel_maj}, including the correct factor of $1$ on the right-hand side of \eqref{app:major:anti_rel_maj}. This is the very reason why we have included the factor of $1/\sqrt{2}$ in the definition of the Majorana fields $\psi_{1,2}$ in the decomposition \eqref{app:major:decomp}.

\section{Propagators}

Another novel property of the Majorana field, important when doing perturbation expansions and using the Wick's theorem, is that apart from the contractions of the type $\contraction[0.5ex]{}{\psi}{}{\bar\psi}\psi\bar\psi$ and $\contraction[0.5ex]{}{\bar\psi}{}{\psi}\bar\psi\psi$ there are also contractions of the type $\contraction[0.5ex]{}{\psi}{}{\psi}\psi\psi$ and $\contraction[0.5ex]{}{\bar\psi}{}{\bar\psi}\bar\psi\bar\psi$. The contractions of the type $\contraction[0.5ex]{}{\psi}{}{\bar\psi}\psi\bar\psi$ and $\contraction[0.5ex]{}{\bar\psi}{}{\psi}\bar\psi\psi$ are the same as for the Dirac field:\footnote{We are using compact matrix notation, more precisely we should write $\bra{0}T \big\{ \psi_a(x)\bar\psi_b(y) \big\} \ket{0} = \I G_{ab}(x-y)$.}
\begin{eqnarray}
\label{app:major:contractions_old}
\bra{0}T \big\{ \psi(x) \, \bar\psi(0) \big\} \ket{0} &\equiv& \I G(x) \,.
\end{eqnarray}
The contractions of the type $\contraction[0.5ex]{}{\psi}{}{\psi}\psi\psi$ and $\contraction[0.5ex]{}{\bar\psi}{}{\bar\psi}\bar\psi\bar\psi$ are now easily calculated by straightforward application of the Majorana condition to \eqref{app:major:contractions_old}:
\begin{subequations}
\label{app:major:contractions_new}
\begin{eqnarray}
\bra{0}T \big\{ \psi(x) \, \psi^\T\!(0) \big\} \ket{0} &=& -\I G(x)C \,,
\\
\bra{0}T \big\{ \bar\psi^\T\!(x) \, \bar\psi(0) \big\} \ket{0} &=& \I C^{-1}G(x) \,.
\end{eqnarray}
\end{subequations}
This result can be calculated also directly by inserting the plane wave expansion of the Majorana field \eqref{app:major:solution_majorana} to the left-hand sides of \eqref{app:major:contractions_new}, using the integral representation of the Heaviside step function and taking into account the relations
\begin{subequations}
\begin{eqnarray}
\sum_{\pm s} u(p,s)\,v^\T\!(p,s) &=& -(\slashed{p}+m)C \,,
\\
\sum_{\pm s} v(p,s)\,u^\T\!(p,s) &=& -(\slashed{p}-m)C \,,
\\
\sum_{\pm s} \bar u^\T\!(p,s)\,\bar v(p,s) &=& C^{-1}(\slashed{p}-m) \,,
\\
\sum_{\pm s} \bar v^\T\!(p,s)\,\bar u(p,s) &=& C^{-1}(\slashed{p}+m) \,.
\end{eqnarray}
\end{subequations}

Let us now investigate in more detail the propagator $G(x)$, \eqref{app:major:contractions_old}, under the assumption of Majorana condition \eqref{app:major:majorana_cond}:
\begin{subequations}
\begin{eqnarray}
\I G(x)  &=& \bra{0}T \big\{ \psi(x) \, \bar\psi(0) \big\} \ket{0}
\\ &=& \bra{0}T \big\{ \psi^\C(x) \, \bar\psi^\C(0) \big\} \ket{0}
\\ &=& -C \bra{0}T \big\{ \bar\psi^\T(x) \, \psi^\T(0) \big\} \ket{0} C^{-1}
\\ &=& C \bra{0}T \big\{ \psi(0) \, \bar\psi(x) \big\} \ket{0}^\T C^{-1}
\\ &=& \I G^\C(-x) \,,
\end{eqnarray}
\end{subequations}
where $G^\C$ is defined in \eqref{app:charge:AC}. We have thus arrived at the important property of the propagator of a Majorana field:
\begin{subequations}
\label{app:maj:Gmaj}
\begin{eqnarray}
G(x)  &=& G^\C(-x) \,.
\end{eqnarray}
The same also holds in the momentum representation (we use the same symbol $G$ for both the position and momentum representations):
\begin{eqnarray}
G(p)  &=& G^\C(-p) \,.
\end{eqnarray}
\end{subequations}

\chapter{Fermion propagator}
\label{app:fermi propag}

\intro{In this comprehensive appendix we discuss various issues connected with propagators of fermion fields. We first assume the most general case of arbitrary number of left-handed and right-handed fermions and catalogue all possible ways how to comprise them into a single field, allowing for more compact treatment. Out of these we pick two, the standard Dirac field and the Majorana field in the Nambu--Gorkov formalism, and discuss them in more detail. We also eventually show how to switch, under certain conditions, between the two descriptions. Finally, we also discuss in some detail some special issues for propagator of the Dirac field, namely its diagonalization and asymptotic behavior.}

\section{General case}

Let $\psi_L$ be an $m$-plet of left-handed and $\psi_R$ an $n$-plet of right-handed fermions. We can now construct the following $16$ bilinears:
\begin{subequations}
\label{frm:bilinears}
\begin{equation}
\psi_L \bar\psi_L \,, \quad
\psi_R \bar\psi_R \,, \quad
\psi_L \bar\psi_R \,, \quad
\psi_R \bar\psi_L \,, \quad
\end{equation}

\begin{equation}
(\psi_L)^\C (\bar\psi_L)^\C \,, \quad
(\psi_R)^\C (\bar\psi_R)^\C \,, \quad
(\psi_L)^\C (\bar\psi_R)^\C \,, \quad
(\psi_R)^\C (\bar\psi_L)^\C \,, \quad
\end{equation}

\begin{equation}
(\psi_L)^\C \bar\psi_L \,, \quad
(\psi_R)^\C \bar\psi_R \,, \quad
(\psi_L)^\C \bar\psi_R \,, \quad
(\psi_R)^\C \bar\psi_L \,, \quad
\end{equation}

\begin{equation}
\psi_L (\bar\psi_L)^\C \,, \quad
\psi_R (\bar\psi_R)^\C \,, \quad
\psi_L (\bar\psi_R)^\C \,, \quad
\psi_R (\bar\psi_L)^\C \,. \quad
\end{equation}
\end{subequations}

%
%
%

\noindent Out of each of these $16$ bilinears $\psi_1 \bar\psi_2$ we can make the full ($\langle \psi_1 \bar\psi_2 \rangle$) of 1PI ($\langle \psi_1 \bar\psi_2 \rangle_{\mathrm{1PI}}$) propagator (we use the shorthand notation $\langle \psi_1 \bar\psi_2 \rangle \equiv \int\! \d^4 x \, \bra{0}T[\psi_1(x)\, \bar\psi_2(0)] \ket{0} \e^{-\I p \cdot x}$). The intention of this appendix is to systemize somehow these propagators and to find out some compact notation for them.

First we investigate the Lorentz structure of the propagators. Since the propagators depend only on momentum $p$, they can only be linear combination of four independent Lorentz matrices $\slashed{p}$, $\slashed{p}\gamma_5$, $1$, $\gamma_5$,  or, in more convenient basis, of $\slashed{p}P_L$, $\slashed{p}P_R$, $P_L$, $P_R$. The coefficients of the linear combination are Lorentz scalars, i.e., they can depend only on $p^2$. At the same time, the coefficients are matrices in the flavor space.

For the full propagators we can employ the chiral projectors $P_L$, $P_R$ and use the fact that multiplication of fermion fields with the chiral projectors \qm{commutes} with the $T$-product. I.e., for instance, we have $\langle \psi_R \bar\psi_R \rangle = \langle P_R \psi_R \bar\psi_R P_L \rangle = P_R \langle \psi_R \bar\psi_R \rangle P_L$. Therefore the quantity $\langle \psi_R \bar\psi_R \rangle$ must be of the form $\slashed{p} P_L$ (times something containing no gamma matrices), since all other possibilities ($\slashed{p} P_R$, $P_L$ and $P_R$) vanish after enclosing with $P_R$ and $P_L$.

On the other hand, for 1PI propagators this approach is no longer possible. A better approach is to view the 1PI propagators as two-point interaction vertices, stemming from the Lagrangian. E.g., the Lagrangian $\eL = \bar\psi_R \I \slashed{\partial} \psi_R = \bar\psi_R P_L \I \slashed{\partial} P_R \psi_R = \bar\psi_R \I \slashed{\partial} P_R \psi_R$, viewed as an interaction Lagrangian, gives rise to the \qm{two-point interaction vertex} $\langle \psi_R \bar\psi_R \rangle_{\mathrm{1PI}} = \I \slashed{p} P_R$.

%
%

Applying these (rather mnemonic) rules to all possible propagators (both full and 1PI), constructible out of the bilinears \eqref{frm:bilinears}, we obtain:


\noindent
\begin{tabular*}{\textwidth}{@{}p{0.5\textwidth}@{} @{}p{0.5\textwidth}@{}}
{
\begin{subequations}
\label{frm:brckt:A}
\begin{align}
\langle \psi_R \bar\psi_R \rangle &\ =\  \I \slashed{p}P_L \, A_{RR} \,, & n &\times n \,, \\
\langle \psi_L \bar\psi_L \rangle &\ =\  \I \slashed{p}P_R \, A_{LL} \,, & m &\times m \,, \\
\langle \psi_L \bar\psi_R \rangle &\ =\  \I            P_L \, A_{LR} \,, & m &\times n \,, \\
\langle \psi_R \bar\psi_L \rangle &\ =\  \I            P_R \, A_{RL} \,, & n &\times m \,,
\end{align}
\end{subequations}
}
&
{
\begin{subequations}
\label{frm:brckt:a}
\begin{align}
\quad
\langle \psi_R \bar\psi_R \rangle_{\mathrm{1PI}} &\ =\  \I \slashed{p}P_R \, a_{RR} \,, & n &\times n \,, \\
\langle \psi_L \bar\psi_L \rangle_{\mathrm{1PI}} &\ =\  \I \slashed{p}P_L \, a_{LL} \,, & m &\times m \,, \\
\langle \psi_L \bar\psi_R \rangle_{\mathrm{1PI}} &\ =\  \I            P_R \, a_{LR} \,, & m &\times n \,, \\
\langle \psi_R \bar\psi_L \rangle_{\mathrm{1PI}} &\ =\  \I            P_L \, a_{RL} \,, & n &\times m \,,
\end{align}
\end{subequations}
}
\end{tabular*}

\vspace{-5mm}

\noindent
\begin{tabular*}{\textwidth}{@{}p{0.5\textwidth}@{} @{}p{0.5\textwidth}@{}}
{
\begin{subequations}
\label{frm:brckt:B}
\begin{align}
\langle (\psi_R)^\C (\bar\psi_R)^\C \rangle &\ =\  \I \slashed{p}P_R \, B_{RR} \,, & n &\times n \,, \\
\langle (\psi_L)^\C (\bar\psi_L)^\C \rangle &\ =\  \I \slashed{p}P_L \, B_{LL} \,, & m &\times m \,, \\
\langle (\psi_L)^\C (\bar\psi_R)^\C \rangle &\ =\  \I            P_R \, B_{LR} \,, & m &\times n \,, \\
\langle (\psi_R)^\C (\bar\psi_L)^\C \rangle &\ =\  \I            P_L \, B_{RL} \,, & n &\times m \,,
\end{align}
\end{subequations}
}
&
{
\begin{subequations}
\label{frm:brckt:b}
\begin{align}
\quad
\langle (\psi_R)^\C (\bar\psi_R)^\C \rangle_{\mathrm{1PI}} &\ =\  \I \slashed{p}P_L \, b_{RR} \,, & n &\times n \,, \\
\langle (\psi_L)^\C (\bar\psi_L)^\C \rangle_{\mathrm{1PI}} &\ =\  \I \slashed{p}P_R \, b_{LL} \,, & m &\times m \,, \\
\langle (\psi_L)^\C (\bar\psi_R)^\C \rangle_{\mathrm{1PI}} &\ =\  \I            P_L \, b_{LR} \,, & m &\times n \,, \\
\langle (\psi_R)^\C (\bar\psi_L)^\C \rangle_{\mathrm{1PI}} &\ =\  \I            P_R \, b_{RL} \,, & n &\times m \,,
\end{align}
\end{subequations}
}
\end{tabular*}

\vspace{-5mm}

\noindent
\begin{tabular*}{\textwidth}{@{}p{0.5\textwidth}@{} @{}p{0.5\textwidth}@{}}
{
\begin{subequations}
\label{frm:brckt:C}
\begin{align}
\quad
\langle (\psi_R)^\C \bar\psi_R \rangle &\ =\  \I            P_L \, C_{RR} \,, & n &\times n \,, \\
\langle (\psi_L)^\C \bar\psi_L \rangle &\ =\  \I            P_R \, C_{LL} \,, & m &\times m \,, \\
\langle (\psi_L)^\C \bar\psi_R \rangle &\ =\  \I \slashed{p}P_L \, C_{LR} \,, & m &\times n \,, \\
\langle (\psi_R)^\C \bar\psi_L \rangle &\ =\  \I \slashed{p}P_R \, C_{RL} \,, & n &\times m \,,
\end{align}
\end{subequations}
}
&
{
\begin{subequations}
\label{frm:brckt:c}
\begin{align}
\langle (\psi_R)^\C \bar\psi_R \rangle_{\mathrm{1PI}} &\ =\  \I            P_R \, c_{RR} \,, & n &\times n \,, \\
\langle (\psi_L)^\C \bar\psi_L \rangle_{\mathrm{1PI}} &\ =\  \I            P_L \, c_{LL} \,, & m &\times m \,, \\
\langle (\psi_L)^\C \bar\psi_R \rangle_{\mathrm{1PI}} &\ =\  \I \slashed{p}P_R \, c_{LR} \,, & m &\times n \,, \\
\langle (\psi_R)^\C \bar\psi_L \rangle_{\mathrm{1PI}} &\ =\  \I \slashed{p}P_L \, c_{RL} \,, & n &\times m \,,
\end{align}
\end{subequations}
}
\end{tabular*}

\vspace{-5mm}

\noindent
\begin{tabular*}{\textwidth}{@{}p{0.5\textwidth}@{} @{}p{0.5\textwidth}@{}}
{
\begin{subequations}
\label{frm:brckt:D}
\begin{align}
\quad
\langle \psi_R (\bar\psi_R)^\C \rangle &\ =\  \I            P_R \, D_{RR} \,, & n &\times n \,, \\
\langle \psi_L (\bar\psi_L)^\C \rangle &\ =\  \I            P_L \, D_{LL} \,, & m &\times m \,, \\
\langle \psi_L (\bar\psi_R)^\C \rangle &\ =\  \I \slashed{p}P_R \, D_{LR} \,, & m &\times n \,, \\
\langle \psi_R (\bar\psi_L)^\C \rangle &\ =\  \I \slashed{p}P_L \, D_{RL} \,, & n &\times m \,,
\end{align}
\end{subequations}
}
&
{
\begin{subequations}
\label{frm:brckt:d}
\begin{align}
\langle \psi_R (\bar\psi_R)^\C \rangle_{\mathrm{1PI}} &\ =\  \I            P_L \, d_{RR} \,, & n &\times n \,, \\
\langle \psi_L (\bar\psi_L)^\C \rangle_{\mathrm{1PI}} &\ =\  \I            P_R \, d_{LL} \,, & m &\times m \,, \\
\langle \psi_L (\bar\psi_R)^\C \rangle_{\mathrm{1PI}} &\ =\  \I \slashed{p}P_L \, d_{LR} \,, & m &\times n \,, \\
\langle \psi_R (\bar\psi_L)^\C \rangle_{\mathrm{1PI}} &\ =\  \I \slashed{p}P_R \, d_{RL} \,, & n &\times m \,.
\end{align}
\end{subequations}
}
\end{tabular*}
The factors of $\I$ are just conventional. The form factors $A$, $B$, $C$, $D$, $a$, $b$, $c$, $d$ are matrices in the flavor space (with indicated dimensions) and may depend only on $p^2$. This dependence is not explicitly indicated. For the special case of momentum-independent form factors $a$, $b$, $c$, $d$ the 1PI propagators are equivalent to the Lagrangian
\begin{eqnarray}
\eL &=& \phantom{+\,}
\bar\psi_R \I \slashed{\partial} a_{RR} \psi_R +
\bar\psi_L \I \slashed{\partial} a_{LL} \psi_L +
\bar\psi_R                       a_{RL} \psi_L +
\bar\psi_L                       a_{LR} \psi_R
\nonumber \\ && +\,
(\bar\psi_R)^\C \I \slashed{\partial} b_{RR} (\psi_R)^\C +
(\bar\psi_L)^\C \I \slashed{\partial} b_{LL} (\psi_L)^\C +
(\bar\psi_R)^\C                       b_{RL} (\psi_L)^\C +
(\bar\psi_L)^\C                       b_{LR} (\psi_R)^\C
\nonumber \\ && +\,
(\bar\psi_R)^\C                       c_{RR} \psi_R +
(\bar\psi_L)^\C                       c_{LL} \psi_L +
(\bar\psi_R)^\C \I \slashed{\partial} c_{RL} \psi_L +
(\bar\psi_L)^\C \I \slashed{\partial} c_{LR} \psi_R
\nonumber \\ && +\,
\bar\psi_R                       d_{RR} (\psi_R)^\C +
\bar\psi_L                       d_{LL} (\psi_L)^\C +
\bar\psi_R \I \slashed{\partial} d_{RL} (\psi_L)^\C +
\bar\psi_L \I \slashed{\partial} d_{LR} (\psi_R)^\C \,.
\end{eqnarray}

However, not all of these propagators are independent or unconstrained. Using the properties of the charge conjugation (see appendix~\ref{app:charge}) we can write the dependencies between the form factors as

\noindent
\begin{tabular*}{\textwidth}{@{}p{0.5\textwidth}@{} @{}p{0.5\textwidth}@{}}
{
\begin{subequations}
\label{frm:cond:BAT}
\begin{eqnarray}
B_{RR} &=&  A_{RR}^\T \,, \\
B_{LL} &=&  A_{LL}^\T \,, \\
B_{LR} &=&  A_{RL}^\T \,, \\
B_{RL} &=&  A_{LR}^\T \,,
\end{eqnarray}
\end{subequations}
}
&
{
\begin{subequations}
\label{frm:cond:baT}
\begin{eqnarray}
b_{RR} &=&  a_{RR}^\T \,, \\
b_{LL} &=&  a_{LL}^\T \,, \\
b_{LR} &=&  a_{RL}^\T \,, \\
b_{RL} &=&  a_{LR}^\T \,,
\end{eqnarray}
\end{subequations}
}
\end{tabular*}

\vspace{-5mm}

\noindent
\begin{tabular*}{\textwidth}{@{}p{0.5\textwidth}@{} @{}p{0.5\textwidth}@{}}
{
\begin{subequations}
\label{frm:cond:CCT}
\begin{eqnarray}
C_{RR} &=& C_{RR}^\T \,, \\
C_{LL} &=& C_{LL}^\T \,, \\
C_{LR} &=& C_{RL}^\T \,,
\end{eqnarray}
\end{subequations}
}
&
{
\begin{subequations}
\label{frm:cond:ccT}
\begin{eqnarray}
c_{RR} &=& c_{RR}^\T \,, \\
c_{LL} &=& c_{LL}^\T \,, \\
c_{LR} &=& c_{RL}^\T \,,
\end{eqnarray}
\end{subequations}
}
\end{tabular*}

\vspace{-5mm}

\noindent
\begin{tabular*}{\textwidth}{@{}p{0.5\textwidth}@{} @{}p{0.5\textwidth}@{}}
{
\begin{subequations}
\label{frm:cond:DDT}
\begin{eqnarray}
D_{RR} &=&  D_{RR}^\T \,, \\
D_{LL} &=&  D_{LL}^\T \,, \\
D_{LR} &=&  D_{RL}^\T \,,
\end{eqnarray}
\end{subequations}
}
&
{
\begin{subequations}
\label{frm:cond:ddT}
\begin{eqnarray}
d_{RR} &=&  d_{RR}^\T \,, \\
d_{LL} &=&  d_{LL}^\T \,, \\
d_{LR} &=&  d_{RL}^\T \,.
\end{eqnarray}
\end{subequations}
}
\end{tabular*}

%

Even though one takes into account the fact that not all of the propagators are independent of one another, there are still quite a lot of independent propagators. Nevertheless, it turns out that it is not necessary to treat them all separately. It is possible to construct a new field $\Psi$ out of the original fields $\psi_L$, $\psi_R$ in such a way that its propagator $\langle\Psi\bar\Psi\rangle$, $\langle\Psi\bar\Psi\rangle_{\mathrm{1PI}}$ contains all the propagators listed above.

\begin{table}[t]
\begin{center}
\begin{tabular}{|c|c|c|}
\hline
& $m=n$ & $m \neq n$
\\ \hline && \\
Dirac case & $\Psi_1 \equiv \psi_L+\psi_R$ & $\Psi_2 \equiv \left(\begin{array}{c} \psi_L \\ \psi_R \end{array}\right)$
\\ && \\ \hline && \\
Majorana case & $\Psi_3 \equiv \left(\begin{array}{c}
\Psi_1 \\
\Psi_1^\C \\
\end{array}\right) =
\left(\begin{array}{c}
\psi_L + \psi_R \\
(\psi_L)^\C + (\psi_R)^\C \\
\end{array}\right)$ & $\Psi_4 \equiv \Psi_2+\Psi_2^\C =
\left(\begin{array}{c}
\psi_L + (\psi_L)^\C \\
\psi_R + (\psi_R)^\C \\
\end{array}\right)$
\\ && \\ \hline
\end{tabular}
\caption[Ways of organizing $\psi_L$, $\psi_R$ into a single field.]{Four possibilities how to organize the fields $\psi_L$, $\psi_R$ into a single field, based on two independent criteria. We discuss in more detail only the fields $\Psi_1$ and $\Psi_4$, denoted in the text as $\psi$ and $\Psi$, respectively, together with the relations between them.}
\label{app:frm:tab:Psi1234}
\end{center}
\end{table}

The are basically four ways (denoted in Tab.~\ref{app:frm:tab:Psi1234} as $\Psi_1$, $\Psi_2$, $\Psi_3$, $\Psi_4$) how to construct the field $\Psi$, based on two independent criterions: First criterion is whether $m = n$ or $m \neq n$. Second and more important criterion is whether we demand invariance of the propagator under the phase (i.e., $\group{U}(1)$) transformation
\begin{subequations}
\label{frm:phase:psiLpsiR}
\begin{eqnarray}
\group{U}(1)\,:\qquad \psi_L &\TransformsTo& [\psi_L]^\prime \ =\ \e^{\I \theta} \, \psi_L \,, \\
\group{U}(1)\,:\qquad \psi_R &\TransformsTo& [\psi_R]^\prime \ =\ \e^{\I \theta} \, \psi_R \,.
\end{eqnarray}
\end{subequations}
The point is that this invariance forbids the propagators of the type $\langle\psi_1^\C\bar\psi_2^{\vphantom{\C}}\rangle$, $\langle\psi_1^{\vphantom{\C}}\bar\psi_2^\C\rangle$. Thus, if this invariance holds, there are less propagators to be included in $\langle\Psi\bar\Psi\rangle$.

We will not discuss here all three possibilities listed in Tab.~\ref{app:frm:tab:Psi1234}. Considering the applications in the main text, we will analyze here in more detail only the Dirac case with $m = n$ and the Majorana case with $m \neq n$, i.e., the fields denoted in Tab.~\ref{app:frm:tab:Psi1234} as $\Psi_1$ and $\Psi_4$, which we rename for our purposes as $\psi$ and $\Psi$, respectively. We will investigate the \qm{anatomy} of the propagators $\langle\psi\bar\psi\rangle$ and $\langle\Psi\bar\Psi\rangle$ and show that they really incorporate all the particular propagators that they should. Finally, we will also see how the \emph{most constrained} field $\psi$ can be implemented as a special case of the \emph{most general} field $\Psi$.

\section{Dirac field}
\label{frm:sec:Dirac}

We will investigate first the most special, or most constrained case, requiring satisfaction of both conditions mentioned above: The same number of the left-handed and the right-handed fermions and at the same time invariance of their propagators under the phase transformation \eqref{frm:phase:psiLpsiR}. On the other hand, however, it is also the most familiar case, as it applies to all charged fermions.


\subsection{General treatment}

\subsubsection{Propagator in general}

Since $n = m$, we can define new field $\psi$,
\begin{eqnarray}
\label{app:frm:defpsi}
\psi &\equiv& \psi_L + \psi_R \,,
\end{eqnarray}
and its full, free and 1PI propagator:
\begin{eqnarray}
\I\,G_\psi &=& \langle \psi \bar\psi \rangle \,, \\
\I\,S_\psi &=& \langle \psi \bar\psi \rangle_0 \,, \\
-\I\,\boldsymbol{\Sigma}_\psi &=& \langle \psi \bar\psi \rangle_{\mathrm{1PI}} \,.
\end{eqnarray}
These are related to one another as
\begin{eqnarray}
\boldsymbol{\Sigma}_\psi &=& S_\psi^{-1} - G_\psi^{-1} \,.
\end{eqnarray}

The full and 1PI propagators have the form
\begin{subequations}
\label{app:frm:psiG}
\begin{eqnarray}
\I\,G_\psi &=& \langle \psi \bar\psi \rangle \\
&=&
\langle \psi_R \bar\psi_R \rangle +
\langle \psi_L \bar\psi_L \rangle +
\langle \psi_L \bar\psi_R \rangle +
\langle \psi_R \bar\psi_L \rangle
\label{app:frm:psiGbrkt}
\\
& = &
\I \Big( \slashed{p}P_L \, A_{RR} + \slashed{p}P_R \, A_{LL} + P_L \, A_{LR} + P_R \, A_{RL} \Big)
\end{eqnarray}
\end{subequations}
and
\begin{subequations}
\label{app:frm:psiSgm}
\begin{eqnarray}
-\I\,\boldsymbol{\Sigma}_\psi &=& \langle \psi \bar\psi \rangle_{\mathrm{1PI}} \\
&=&
\langle \psi_R \bar\psi_R \rangle_{\mathrm{1PI}} +
\langle \psi_L \bar\psi_L \rangle_{\mathrm{1PI}} +
\langle \psi_L \bar\psi_R \rangle_{\mathrm{1PI}} +
\langle \psi_R \bar\psi_L \rangle_{\mathrm{1PI}}
\label{app:frm:psiSgmbrkt}
\\
& = &
\I \Big( \slashed{p}P_R \, a_{RR} + \slashed{p}P_L \, a_{LL} + P_R \, a_{LR} + P_L \, a_{RL} \Big) \,,
\end{eqnarray}
\end{subequations}
respectively. One can easily see that all the four particular propagators in \eqref{app:frm:psiGbrkt} and in \eqref{app:frm:psiSgmbrkt} are invariant under the phase transformation \eqref{frm:phase:psiLpsiR}, which now in terms of the field $\psi$ read
\begin{eqnarray}
\group{U}(1)\,:\qquad \psi \ \TransformsTo \ [\psi]^\prime &=& \e^{\I \theta} \, \psi \,.
\end{eqnarray}

\subsubsection{Free Lagrangian and propagator}

Let us now consider the most general free Lagrangian, made of the fields $\psi_L$, $\psi_R$ and invariant under the phase transformation \eqref{frm:phase:psiLpsiR}:
\begin{eqnarray}
\eL &=&
\bar\psi_L \I \slashed{\partial}\psi_L + \bar\psi_R \I \slashed{\partial}\psi_R
- \bar\psi_L m_{D} \psi_R - \bar\psi_R m_{D}^\dag \psi_L \,,
\end{eqnarray}
where $m_D$ is in principle arbitrary complex $m \times m = n \times n$ matrix. The subscript $D$ stands for \qm{Dirac}, since these are the \qm{Dirac mass terms}. In terms of $\psi$ the free Lagrangian can be easily rewritten as
\begin{eqnarray}
\eL &=& \bar\psi \I \slashed{\partial}\psi - \bar\psi (m_{D}^\dag P_L+m_{D} P_R) \psi \,.
\end{eqnarray}
If we denote
\begin{eqnarray}
\label{frm:def:bldm}
\boldsymbol{m} &\equiv& m_{D}^\dag P_L+m_{D} P_R \,,
\end{eqnarray}
we can write it even more compactly as
\begin{eqnarray}
\eL &=& \bar\psi \I \slashed{\partial}\psi - \bar\psi \boldsymbol{m} \psi \,.
\end{eqnarray}

The free propagator $\I S_\psi = \langle \psi\bar\psi \rangle_0$ can be easily achieved by inverting the free Lagrangian:
\begin{subequations}
\label{frm:Spsi}
\begin{eqnarray}
S_\psi & = & \big[ \slashed{p} - (m_{D}^\dag P_L+m_{D}P_R) \big]^{-1}
\\ & = &
(\slashed{p}+m_{D})(p^2-m_{D}^\dag m_{D})^{-1}P_L +
(\slashed{p}+m_{D}^\dag)(p^2-m_{D}m_{D}^\dag)^{-1}P_R \,.
\end{eqnarray}
\end{subequations}
In terms of $\boldsymbol{m}$, \eqref{frm:def:bldm}, we can write also
\begin{subequations}
\label{frm:Spsibdl}
\begin{eqnarray}
S_\psi & = & \big[ \slashed{p} - \boldsymbol{m} \big]^{-1}
\\ &=& (\slashed{p}+\boldsymbol{m}^\dag)(p^2-\boldsymbol{m} \, \boldsymbol{m}^\dag)^{-1}
\\ &=& (p^2-\boldsymbol{m}^\dag \boldsymbol{m})^{-1}(\slashed{p}+\boldsymbol{m}^\dag) \,.
\end{eqnarray}
\end{subequations}

\subsection{Simplifying assumptions}
\label{frm:ssec:Dirac:simp}

With respect to the applications in the main text, we are now going to make some simplifying assumptions concerning the free and 1PI propagators and to arrive at expressions the corresponding full propagator.

\subsubsection{Hermiticity}

We first make the assumption about Hermiticity of the self-energy $\boldsymbol{\Sigma}_\psi$:
\begin{eqnarray}
\label{frm:bldSgmpsiherm}
\boldsymbol{\Sigma}_\psi &=& \boldsymbol{\bar \Sigma}_\psi
\end{eqnarray}
(where $\boldsymbol{\bar \Sigma}_\psi \equiv \gamma_0 \, \boldsymbol{\Sigma}_\psi^\dag \, \gamma_0$). Notice that the free propagator already satisfies analogous condition:
\begin{eqnarray}
S_\psi &=& \bar S_\psi \,.
\end{eqnarray}
This is in fact just the condition for the free Lagrangian to be Hermitian. Consequently, since the full propagator can be expressed in terms of those free and 1PI as $G_\psi = (S_\psi^{-1}-\boldsymbol{\Sigma}_\psi)^{-1}$, the condition \eqref{frm:bldSgmpsiherm} for the self-energy induces an analogous condition for the full propagator:
\begin{eqnarray}
G_\psi &=& \bar G_\psi \,.
\end{eqnarray}
These conditions for the 1PI and full propagators imply the following relations among their, until now independent, components:
\begin{subequations}
\begin{eqnarray}
A_{RR} &=& A_{RR}^\dag \,, \\
A_{LL} &=& A_{LL}^\dag \,, \\
A_{RL} &=& A_{LR}^\dag
\end{eqnarray}
\end{subequations}
for the full propagators and
\begin{subequations}
\begin{eqnarray}
a_{RR} &=& a_{RR}^\dag \,, \\
a_{LL} &=& a_{LL}^\dag \,, \\
a_{RL} &=& a_{LR}^\dag
\end{eqnarray}
\end{subequations}
for the 1PI propagator.

\subsubsection{No wave-function renormalization}

We may set
\begin{subequations}
\begin{eqnarray}
a_{RR} &=& 0 \,, \\
a_{LL} &=& 0 \,,
\end{eqnarray}
\end{subequations}
and rename the remaining coefficients $a_{RL} = a_{LR}^\dag$ as
\begin{subequations}
\begin{eqnarray}
a_{LR} &=& -\Sigma_D      \,, \\
a_{RL} &=& -\Sigma_D^\dag \,.
\end{eqnarray}
\end{subequations}
The subscript $D$ stands for \emph{Dirac}. The self-energy $\boldsymbol{\Sigma}_\psi$, \eqref{app:frm:psiSgm}, then recasts as
\begin{eqnarray}
\label{frm:Sgmpsi}
\boldsymbol{\Sigma}_\psi &=& \Sigma_D^\dag\,P_L+\Sigma_D\,P_R \,.
\end{eqnarray}
Now if we further assume that the bare propagator is just $S_\psi^{-1} = \slashed{p}$, the components of the full propagator $G_\psi$, \eqref{app:frm:psiG}, are
\begin{subequations}
\begin{eqnarray}
A_{RR} &=& D_R \,, \\
A_{LL} &=& D_L \,, \\
A_{LR} &=& \Sigma_D      \, D_R \ =\ D_L \, \Sigma_D      \,,
\label{frm:ALR} \\
A_{RL} &=& \Sigma_D^\dag \, D_L \ =\ D_R \, \Sigma_D^\dag \,,
\label{frm:ARL}
\end{eqnarray}
\end{subequations}
where we denoted
\begin{subequations}
\label{frm:defDRDL}
\begin{eqnarray}
D_R &\equiv& \big(p^2-\Sigma_D^\dag\,\Sigma_D^{\phantom{\dag}}\big)^{-1} \,, \\
D_L &\equiv& \big(p^2-\Sigma_D^{\phantom{\dag}}\,\Sigma_D^\dag\big)^{-1} \,.
\end{eqnarray}
\end{subequations}
Let us explicitly state the commutation relation
\begin{eqnarray}
\label{frm:SgmDDR=DLSgmD}
\Sigma_D \, D_R &=& D_L \, \Sigma_D \,,
\end{eqnarray}
used in \eqref{frm:ALR} and \eqref{frm:ARL}. Also note that $D_L$, $D_R$ trivially commute with $\gamma^\mu$, since $D_L$, $D_R$ do not contain any $\gamma_5$.

It is convenient to introduce also another notation. Let us define
\begin{subequations}
\label{frm:def:bldDRDL}
\begin{eqnarray}
\boldsymbol{D}_R &\equiv&
\big(p^2-\boldsymbol{\Sigma}_\psi^\dag\,\boldsymbol{\Sigma}_\psi^{\phantom{\dag}}\big)^{-1}
\ = \  D_L\,P_R + D_L\,P_L \,, \\
\boldsymbol{D}_L &\equiv&
\big(p^2-\boldsymbol{\Sigma}_\psi^{\phantom{\dag}}\,\boldsymbol{\Sigma}_\psi^\dag\big)^{-1}
\ = \  D_L\,P_L + D_R\,P_R \,.
\end{eqnarray}
\end{subequations}
The commutation relation of $\boldsymbol{D}_L$, $\boldsymbol{D}_R$ with $\boldsymbol{\Sigma}_\psi$ reads
\begin{eqnarray}
\boldsymbol{D}_L\,\boldsymbol{\Sigma}_\psi &=& \boldsymbol{\Sigma}_\psi\,\boldsymbol{D}_R \,,
\end{eqnarray}
which is much the same as the commutation relation \eqref{frm:SgmDDR=DLSgmD}. On the other hand, the commutation relations with $\gamma^\mu$ are now non-trivial, due to presence of $\gamma_5$ in $\boldsymbol{D}_L$, $\boldsymbol{D}_R$:
\begin{subequations}
\begin{eqnarray}
\gamma^\mu\,\boldsymbol{D}_L &=& \boldsymbol{D}_R\,\gamma^\mu \,, \\
\gamma^\mu\,\boldsymbol{D}_R &=& \boldsymbol{D}_L\,\gamma^\mu \,.
\end{eqnarray}
\end{subequations}

\subsubsection{Expressions of the full propagator}

Consider now the full propagator $G_\psi$, given in terms of $\boldsymbol{\Sigma}_\psi$ as
\begin{eqnarray}
G_\psi &=& \big( \slashed{p} - \boldsymbol{\Sigma}_\psi \big)^{-1} \,.
\end{eqnarray}
The inversion can be done in terms of $\Sigma_D$, with the chiral projectors shown explicitly, as
\begin{eqnarray}
\label{frm:Gpsipodruhe}
G_\psi &=&
\big(\slashed{p}+\Sigma_{D}\big)D_R\,P_L + \big(\slashed{p}+\Sigma_{D}^\dag\big)D_L\,P_R \,,
\end{eqnarray}
which is analogous to the expression \eqref{frm:Spsi} of the free propagator $S_\psi$. Using the definitions \eqref{frm:def:bldDRDL} it is also possible to express $G_\psi$ in terms of $\boldsymbol{\Sigma}_\psi$ in more compact way, with the chiral projectors \qm{hidden}:
\begin{subequations}
\label{frm:Gpsi}
\begin{eqnarray}
G_\psi
&=& \big(\slashed{p}+\boldsymbol{\Sigma}_\psi^\dag\big)\boldsymbol{D}_L \\
&=& \boldsymbol{D}_R\big(\slashed{p}+\boldsymbol{\Sigma}_\psi^\dag\big) \,,
\end{eqnarray}
\end{subequations}
in analogy with the expression \eqref{frm:Spsibdl} for the free propagator $S_\psi$.


\subsubsection{Diagrammatics}

Let us finally state here the Feynman rules for the self-energy \eqref{frm:Sgmpsi} and the full propagator \eqref{frm:Gpsi}. The self-energy line is
\begin{subequations}
\begin{eqnarray}
\langle \psi \bar \psi \rangle_{\mathrm{1PI}}
\ = \
\begin{array}{c}
\scalebox{0.85}{\includegraphics[trim = 10bp 12bp 19bp 11bp,clip]{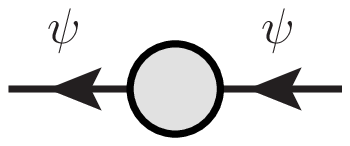}}
\end{array}
&=& -\I\big(\Sigma_D^\dag\,P_L+\Sigma_D\,P_R\big)
\\ &=& - \I \boldsymbol{\Sigma}_\psi
\end{eqnarray}
\end{subequations}
and its chiral components read
\begin{subequations}
\begin{eqnarray}
\langle \psi_L \bar \psi_R \rangle_{\mathrm{1PI}}
\ = \
\begin{array}{c}
\scalebox{0.85}{\includegraphics[trim = 10bp 12bp 19bp 11bp,clip]{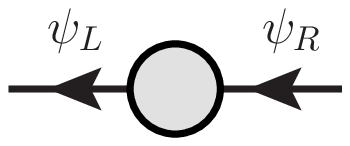}}
\end{array}
&=& - \I\,\Sigma_D \,P_R \,,
\\
\langle \psi_R \bar \psi_L \rangle_{\mathrm{1PI}}
\ = \
\begin{array}{c}
\scalebox{0.85}{\includegraphics[trim = 10bp 12bp 19bp 11bp,clip]{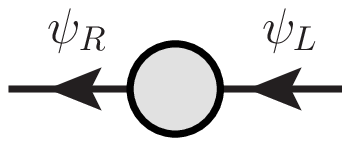}}
\end{array}
&=& - \I\,\Sigma_D^\dag \,P_L \,.
\end{eqnarray}
\end{subequations}

For the full propagator we have
\begin{subequations}
\begin{eqnarray}
\langle \psi \bar \psi \rangle
\ = \
\begin{array}{c}
\scalebox{0.85}{\includegraphics[trim = 10bp 12bp 19bp 11bp,clip]{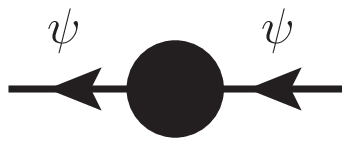}}
\end{array}
&=& \I\big( \slashed{p}+\Sigma_D \big)D_{R}\,P_L+\I\big( \slashed{p}+\Sigma_D^\dag \big)D_{L}\,P_R
\\ &=& \I\big(\slashed{p}+\boldsymbol{\Sigma}_\psi^\dag\big)\boldsymbol{D}_L
\ = \
\I\boldsymbol{D}_R\big(\slashed{p}+\boldsymbol{\Sigma}_\psi^\dag\big) \,,
\end{eqnarray}
\end{subequations}
with the chiral component not including $\slashed{p}$:
\begin{subequations}
\begin{eqnarray}
\langle \psi_L \bar \psi_R \rangle
\ = \
\begin{array}{c}
\scalebox{0.85}{\includegraphics[trim = 10bp 12bp 19bp 11bp,clip]{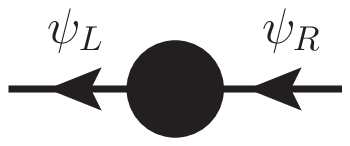}}
\end{array}
&=& \I\,\Sigma_D\,D_{R}\,P_L \ =\  \I\,D_{L}\,\Sigma_D\,P_L \,,
\\
\langle \psi_R \bar \psi_L \rangle
\ = \
\begin{array}{c}
\scalebox{0.85}{\includegraphics[trim = 10bp 12bp 19bp 11bp,clip]{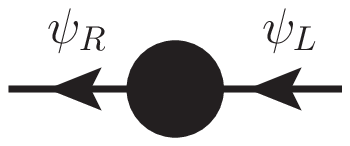}}
\end{array}
&=& \I\,\Sigma_D^\dag\,D_{L}\,P_R \ =\  \I\,D_{R}\,\Sigma_D^\dag\,P_R \,,
\end{eqnarray}
\end{subequations}
and the chiral component proportional to $\slashed{p}$:
\begin{subequations}
\begin{eqnarray}
\langle \psi_L \bar \psi_L \rangle
\ = \
\begin{array}{c}
\scalebox{0.85}{\includegraphics[trim = 10bp 12bp 19bp 11bp,clip]{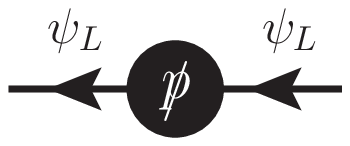}}
\end{array}
&=& \I\,\slashed{p}\,D_{L}\,P_R \,,
\\
\langle \psi_R \bar \psi_R \rangle
\ = \
\begin{array}{c}
\scalebox{0.85}{\includegraphics[trim = 10bp 12bp 19bp 11bp,clip]{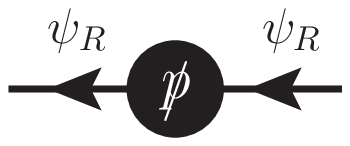}}
\end{array}
&=& \I\,\slashed{p}\,D_{R}\,P_L \,.
\end{eqnarray}
\end{subequations}

\section{Majorana field in the Nambu--Gorkov formalism}
\label{app:frm:Maj}

Now we relax both assumptions made in the previous section, i.e., we do not anymore require invariance of the propagators of $\psi_L$, $\psi_R$ under the phase transformation \eqref{frm:phase:psiLpsiR} and we assume that the numbers of $\psi_L$ and $\psi_R$ are in general different: $m \neq n$.

\subsection{General treatment}

\subsubsection{Propagator in general}

We define the new field $\Psi$, called the \emph{Namby--Gorkov field}, as
\begin{eqnarray}
\label{frm:Psidef}
\Psi &\equiv& \left(\begin{array}{c} \psi_L + (\psi_L)^\C \\ \psi_R + (\psi_R)^\C \end{array}\right)
\end{eqnarray}
and denote its propagators as
\begin{eqnarray}
\I\,G_\Psi &=& \langle \Psi \bar\Psi \rangle \,, \\
\I\,S_\Psi &=& \langle \Psi \bar\Psi \rangle_0 \,, \\
-\I\,\boldsymbol{\Sigma}_\Psi &=& \langle \Psi \bar\Psi \rangle_{\mathrm{1PI}} \,.
\end{eqnarray}
Again, the self-energy is related to the full and free propagators as
\begin{eqnarray}
\label{frm:Psipropsrel}
\boldsymbol{\Sigma}_\Psi &=& S_\Psi^{-1} - G_\Psi^{-1} \,.
\end{eqnarray}

Clearly, the definition \eqref{frm:Psidef} is consistent with the assumption $m \neq n$. Moreover, the propagators really do incorporate the particular propagators like, e.g., $\langle \psi_L^\C \bar\psi_R \rangle$, breaking the invariance under \eqref{frm:phase:psiLpsiR}. Let us see it explicitly. The full propagator reads
\begin{subequations}
\label{frm:GPsi}
\begin{eqnarray}
\I\,G_\Psi &=& \langle \Psi \bar\Psi \rangle \\
&=& \hphantom{+\,}
\left(\begin{array}{cc}
\langle (\psi_L)^\C (\bar\psi_L)^\C \rangle & \langle (\psi_L)^\C \bar\psi_R \rangle \\
\langle \psi_R (\bar\psi_L)^\C \rangle & \langle \psi_R \bar\psi_R \rangle \\
\end{array}\right)
+
\left(\begin{array}{cc}
\langle \psi_L \bar\psi_L \rangle & \langle \psi_L (\bar\psi_R)^\C \rangle \\
\langle (\psi_R)^\C \bar\psi_L \rangle & \langle (\psi_R)^\C (\bar\psi_R)^\C \rangle \\
\end{array}\right)
\qquad
\nonumber \\
&&
+\,
\left(\begin{array}{cc}
\langle \psi_L (\bar\psi_L)^\C \rangle & \langle \psi_L \bar\psi_R \rangle \\
\langle (\psi_R)^\C (\bar\psi_L)^\C \rangle & \langle (\psi_R)^\C \bar\psi_R \rangle \\
\end{array}\right)
+
\left(\begin{array}{cc}
\langle (\psi_L)^\C \bar\psi_L \rangle & \langle (\psi_L)^\C (\bar\psi_R)^\C \rangle \\
\langle \psi_R \bar\psi_L \rangle & \langle \psi_R (\bar\psi_R)^\C \rangle \\
\end{array}\right)
\\
& = &
\I \Big( \slashed{p}P_L \, \mathcal{A} + \slashed{p}P_R \, \mathcal{B} + P_L \, \mathcal{C} + P_R \, \mathcal{D} \Big) \,,
\end{eqnarray}
\end{subequations}
while the 1PI is
\begin{subequations}
\label{frm:SgmPsi}
\begin{eqnarray}
-\I\,\boldsymbol{\Sigma}_\Psi &=& \langle \Psi \bar\Psi \rangle_{\mathrm{1PI}} \\
&=&
\hphantom{+\,}
\left(\begin{array}{cc}
\langle \psi_L \bar\psi_L \rangle_{\mathrm{1PI}} & \langle \psi_L (\bar\psi_R)^\C \rangle_{\mathrm{1PI}} \\
\langle (\psi_R)^\C \bar\psi_L \rangle_{\mathrm{1PI}} & \langle (\psi_R)^\C (\bar\psi_R)^\C \rangle_{\mathrm{1PI}} \\
\end{array}\right)
+
\left(\begin{array}{cc}
\langle (\psi_L)^\C (\bar\psi_L)^\C \rangle_{\mathrm{1PI}} & \langle (\psi_L)^\C \bar\psi_R \rangle_{\mathrm{1PI}} \\
\langle \psi_R (\bar\psi_L)^\C \rangle_{\mathrm{1PI}} & \langle \psi_R \bar\psi_R \rangle_{\mathrm{1PI}} \\
\end{array}\right)
\nonumber \\
&&
+\,
\left(\begin{array}{cc}
\langle (\psi_L)^\C \bar\psi_L \rangle_{\mathrm{1PI}} & \langle (\psi_L)^\C (\bar\psi_R)^\C \rangle_{\mathrm{1PI}} \\
\langle \psi_R \bar\psi_L \rangle_{\mathrm{1PI}} & \langle \psi_R (\bar\psi_R)^\C \rangle_{\mathrm{1PI}} \\
\end{array}\right)
+
\left(\begin{array}{cc}
\langle \psi_L (\bar\psi_L)^\C \rangle_{\mathrm{1PI}} & \langle \psi_L \bar\psi_R \rangle_{\mathrm{1PI}} \\
\langle (\psi_R)^\C (\bar\psi_L)^\C \rangle_{\mathrm{1PI}} & \langle (\psi_R)^\C \bar\psi_R \rangle_{\mathrm{1PI}} \\
\end{array}\right)
\nonumber \\ &&
\\
& = &
\I \Big( \slashed{p}P_L \, \mathcal{A}_{\mathrm{1PI}} + \slashed{p}P_R \, \mathcal{B}_{\mathrm{1PI}} + P_L \, \mathcal{C}_{\mathrm{1PI}} + P_R \, \mathcal{D}_{\mathrm{1PI}} \Big) \,,
\end{eqnarray}
\end{subequations}
where we denoted

\noindent
\begin{tabular*}{\textwidth}{@{}p{0.5\textwidth}@{} @{}p{0.5\textwidth}@{}}
{
\begin{subequations}\label{frm:defmathcalABCD}\begin{eqnarray}
\mathcal{A} &\equiv&
\left(\begin{array}{cc}
B_{LL} & C_{LR} \\
D_{RL} & A_{RR} \\
\end{array}\right) \,,
\\
\mathcal{B} &\equiv&
\left(\begin{array}{cc}
A_{LL} & D_{LR} \\
C_{RL} & B_{RR}  \\
\end{array}\right) \,,
\\
\mathcal{C} &\equiv&
\left(\begin{array}{cc}
D_{LL} & A_{LR}  \\
B_{RL} & C_{RR}  \\
\end{array}\right) \,,
\\
\mathcal{D} &\equiv&
\left(\begin{array}{cc}
C_{LL} & B_{LR} \\
A_{RL} & D_{RR} \\
\end{array}\right) \,,
\end{eqnarray}
\end{subequations}
}
&
{
\begin{subequations}\label{frm:defmathcalABCD1PI}\begin{eqnarray}
\mathcal{A}_{\mathrm{1PI}} &\equiv&
\left(\begin{array}{cc}
a_{LL} & d_{LR} \\
c_{RL} & b_{RR}  \\
\end{array}\right) \,,
\\
\mathcal{B}_{\mathrm{1PI}} &\equiv&
\left(\begin{array}{cc}
b_{LL} & c_{LR} \\
d_{RL} & a_{RR} \\
\end{array}\right) \,,
\\
\mathcal{C}_{\mathrm{1PI}} &\equiv&
\left(\begin{array}{cc}
c_{LL} & b_{LR} \\
a_{RL} & d_{RR} \\
\end{array}\right) \,,
\\
\mathcal{D}_{\mathrm{1PI}} &\equiv&
\left(\begin{array}{cc}
d_{LL} & a_{LR}  \\
b_{RL} & c_{RR}  \\
\end{array}\right) \,.
\end{eqnarray}
\end{subequations}
}
\end{tabular*}

Indeed, we can see the propagators, non-invariant under \eqref{frm:phase:psiLpsiR}, are really included. In fact, \emph{all} of the possible propagators \eqref{frm:brckt:A}, \eqref{frm:brckt:B}, \eqref{frm:brckt:C}, \eqref{frm:brckt:D} and \eqref{frm:brckt:a}, \eqref{frm:brckt:b}, \eqref{frm:brckt:c}, \eqref{frm:brckt:d}, that can be made out of the fields $\psi_L$, $\psi_R$, are included in the propagators \eqref{frm:GPsi} and \eqref{frm:SgmPsi}, respectively. In this sense the formalism $\Psi$ is the most general one.

For completeness, let us also derive how the $\group{U}(1)$ transformation \eqref{frm:phase:psiLpsiR} looks in terms of the field $\Psi$:
\begin{subequations}
\label{app:frm:phasePsi}
\begin{eqnarray}
\group{U}(1)\,:\qquad
\Psi \>=\> \left(\begin{array}{c} \psi_L + (\psi_L)^\C \\ \psi_R + (\psi_R)^\C \end{array}\right)
\quad\TransformsTo\quad
\hspace{-4em}
&& \nonumber \\
{[\Psi]}^\prime
&=&
\left(\begin{array}{c} \e^{\I \theta}\,\psi_L + \e^{-\I \theta}\,(\psi_L)^\C \\ \e^{\I \theta}\,\psi_R + \e^{-\I \theta}\,(\psi_R)^\C \end{array}\right)
\\ &=&
\left(\begin{array}{cc} \e^{\I \theta} P_L + \e^{-\I \theta} P_R & 0 \\ 0 & \e^{\I \theta} P_R + \e^{-\I \theta} P_L \end{array}\right)
\left(\begin{array}{c} \psi_L + (\psi_L)^\C \\ \psi_R + (\psi_R)^\C \end{array}\right)
\nonumber \\ &&
\\ &=&
\left(\begin{array}{cc} \e^{- \I \gamma_5 \theta} & 0 \\ 0 & \e^{\I \gamma_5 \theta} \end{array}\right) \Psi
\\ &=&
\e^{-\I \gamma_5\sigma_3\,\theta}\,\Psi \,.
\end{eqnarray}
\end{subequations}
The matrix $\sigma_3$ acts of course in the Nambu--Gorkov doublet space.

The field $\Psi$ is a Majorana field, since it satisfies the Majorana condition \eqref{app:charge:major}:
\begin{eqnarray}
\Psi^\C & = & \Psi \,,
\end{eqnarray}
as can be readily seen from its definition \eqref{frm:Psidef}. As shown in appendix~\ref{app:major}, the full propagator $G_\Psi$ must therefore satisfy the condition
\begin{eqnarray}
\label{frm:majcondGPsi}
G_\Psi(p) &=& G_\Psi^\C(-p) \,.
\end{eqnarray}
The same condition must be satisfied also by the free propagator $S_\Psi$, which is after all merely a special case of $G_\Psi$ in the case of no interactions. Thus, self-energy $\boldsymbol{\Sigma}_\Psi$ must satisfy it too:
\begin{eqnarray}
\label{frm:majcondbldsgmPsi}
\boldsymbol{\Sigma}_\Psi(p) &=& \boldsymbol{\Sigma}_\Psi^\C(-p) \,.
\end{eqnarray}
The conditions \eqref{frm:majcondGPsi} and \eqref{frm:majcondbldsgmPsi} for $G_\Psi$ and $\boldsymbol{\Sigma}_\Psi$ are in fact equivalent to the conditions \eqref{frm:cond:BAT}, \eqref{frm:cond:CCT}, \eqref{frm:cond:DDT} and \eqref{frm:cond:baT}, \eqref{frm:cond:ccT}, \eqref{frm:cond:ddT}, respectively, discussed already above. In terms of the matrix formalism \eqref{frm:defmathcalABCD} and \eqref{frm:defmathcalABCD1PI} they can be more compactly rewritten as

\noindent
\begin{tabular*}{\textwidth}{@{}p{0.5\textwidth}@{} @{}p{0.5\textwidth}@{}}
{
\begin{subequations}\label{frm:mathcalABCDcond}
\begin{eqnarray}
\mathcal{A} & = & \mathcal{B}^\T \,, \\
\mathcal{C} & = & \mathcal{C}^\T \,, \\
\mathcal{D} & = & \mathcal{D}^\T \,,
\end{eqnarray}
\end{subequations}
}
&
{
\begin{subequations}\label{frm:mathcalABCD1PIcond}
\begin{eqnarray}
\mathcal{A}_{\mathrm{1PI}} & = & \mathcal{B}_{\mathrm{1PI}}^\T \,, \\
\mathcal{C}_{\mathrm{1PI}} & = & \mathcal{C}_{\mathrm{1PI}}^\T \,, \\
\mathcal{D}_{\mathrm{1PI}} & = & \mathcal{D}_{\mathrm{1PI}}^\T \,.
\end{eqnarray}
\end{subequations}
}
\end{tabular*}

\subsubsection{Free Lagrangian and propagator}

The most general free Lagrangian of the fields $\psi_L$, $\psi_R$, non-invariant under \eqref{frm:phase:psiLpsiR}, reads
\begin{eqnarray}
\label{frm:eLpsiLpsiRnoninv}
\eL &=&
\bar\psi_L \I \slashed{\partial}\psi_L + \bar\psi_R \I \slashed{\partial}\psi_R
- \Big( \bar\psi_L m_{D} \psi_R + \frac{1}{2}\bar\psi_L m_{L} (\psi_L)^\C + \frac{1}{2}(\bar\psi_R)^\C m_{R} \psi_R + \hc \Big) \,.
\qquad
\end{eqnarray}
Here $m_D$ is a rectangular $m \times n$ matrix, while the $m_L$, $m_R$ are square matrices with dimensions $n \times n$, $m \times m$, respectively. Moreover, the matrices $m_L$, $m_R$ can be taken without loss of generality symmetric:
\begin{subequations}
\begin{eqnarray}
m_L &=& m_L^\T \,, \\
m_R &=& m_R^\T \,,
\end{eqnarray}
\end{subequations}
since their antisymmetric parts do not contribute to the Lagrangian. Let us see it on an example of, say, $m_R$:
\begin{subequations}
\label{appfrm:mR}
\begin{eqnarray}
(\bar\psi_R)^\C m_{R} \psi_R &=& -\psi_R^\T C^{-1} m_{R} \psi_R
\label{appfrm:mR1} \\
&=& -\big[\psi_R^\T C^{-1} m_{R} \psi_R\big]^\T
\label{appfrm:mR2} \\
&=& -\psi_R^\T C^{-1} m_{R}^\T \psi_R
\label{appfrm:mR3} \\
&=& (\bar\psi_R)^\C m_{R}^\T \psi_R \,.
\label{appfrm:mR4}
\end{eqnarray}
\end{subequations}
(In the second line, \eqref{appfrm:mR2}, we used the antisymmetricity \eqref{app:charge:Casym} of the matrix $C$ of charge conjugation, which compensated for the minus sign due to anti-commuting character of the fermion field, occurring when taking the transpose.) Therefore the antisymmetric part of $m_R$ must vanish in the bilinear $\bar\psi_R m_{R} (\psi_R)^\C$. For $m_L$ the argument would be the same.

In terms of the field $\Psi$ the free Lagrangian \eqref{frm:eLpsiLpsiRnoninv} can be rewritten as
\begin{eqnarray}
\label{frm:eLPsinoninv}
\eL &=& \frac{1}{2}\bar\Psi \I \slashed{\partial}\Psi - \frac{1}{2} \bar\Psi \boldsymbol{m} \Psi
+ \frac{1}{2}\partial_\mu \big(\bar\psi_L\gamma^\mu\psi_L\big) + \frac{1}{2}\partial_\mu \big(\bar\psi_R\gamma^\mu\psi_R\big) \,,
\end{eqnarray}
where we defined the matrix $\boldsymbol{m}$,
\begin{eqnarray}
\label{frm:Psi:bldm}
\boldsymbol{m} &\equiv& m^\dag P_L + m P_R \,,
\end{eqnarray}
in terms of the symmetric matrix $m$:
\begin{eqnarray}
\label{frm:Psi:m}
m &\equiv& \left(\begin{array}{cc} m_{L} & m_{D} \\ m_{D}^\T & m_{R} \end{array}\right) \,.
\end{eqnarray}
The total divergencies in \eqref{frm:eLPsinoninv} do not contribute to the action and we will accordingly dismiss them in the following.

The free propagator $S_\Psi$ is now obtained easily by inverting the free Lagrangian \eqref{frm:eLPsinoninv}. Likewise in the Dirac case, we can express it either in terms of $m$, \eqref{frm:Psi:m}, as
\begin{subequations}
\label{frm:SPsi}
\begin{eqnarray}
S_\Psi & = & \big[ \slashed{p} - (m^\dag P_L+m P_R) \big]^{-1}
\\ & = &
(\slashed{p}+m_)(p^2-m^\dag m)^{-1}P_L +
(\slashed{p}+m^\dag)(p^2-mm^\dag)^{-1}P_R \,,
\end{eqnarray}
\end{subequations}
or more compactly, in terms of $\boldsymbol{m}$, \eqref{frm:Psi:bldm}, as
\begin{subequations}
\label{frm:SPsibdl}
\begin{eqnarray}
S_\Psi & = & \big[ \slashed{p} - \boldsymbol{m} \big]^{-1}
\\ &=& (\slashed{p}+\boldsymbol{m}^\dag)(p^2-\boldsymbol{m} \, \boldsymbol{m}^\dag)^{-1}
\\ &=& (p^2-\boldsymbol{m}^\dag \boldsymbol{m})^{-1}(\slashed{p}+\boldsymbol{m}^\dag) \,.
\end{eqnarray}
\end{subequations}
Notice that both expressions \eqref{frm:SPsi} and \eqref{frm:SPsibdl} for $S_\Psi$ are formally the same as their Dirac counterparts \eqref{frm:Spsi} and \eqref{frm:Spsibdl}, respectively, for $S_\psi$. One can verify that $S_\Psi$ indeed satisfies the condition
\begin{eqnarray}
S_\Psi(p) &=& S_\Psi^\C(-p) \,,
\end{eqnarray}
due to obvious symmetricity of $m$,
\begin{eqnarray}
m & = & m^\T \,,
\end{eqnarray}
or equivalently, due to
\begin{eqnarray}
\boldsymbol{m} & = & \boldsymbol{m}^\C \,.
\end{eqnarray}

\subsection{Simplifying assumptions}

\subsubsection{Hermiticity}

We are again free to demand
\begin{eqnarray}
\label{frm:bldSgmPsiherm}
\boldsymbol{\Sigma}_\Psi & = & \boldsymbol{\bar\Sigma}_\Psi \,.
\end{eqnarray}
Since the free propagator $S_\Psi$ already satisfies $S_\Psi = \bar S_\Psi$, the condition \eqref{frm:bldSgmPsiherm} implies, by means of the relation \eqref{frm:Psipropsrel}, similar condition for $G_\Psi$:
\begin{eqnarray}
\label{frm:GPsiherm}
G_\Psi & = & \bar G_\Psi \,.
\end{eqnarray}
Assuming \eqref{frm:GPsiherm} and \eqref{frm:bldSgmPsiherm} we obtain, on top of the conditions \eqref{frm:mathcalABCDcond} and \eqref{frm:mathcalABCD1PIcond}, also the following conditions for the particular components \eqref{frm:defmathcalABCD}, \eqref{frm:defmathcalABCD1PI} of the propagators:

\noindent
\begin{tabular*}{\textwidth}{@{}p{0.5\textwidth}@{} @{}p{0.5\textwidth}@{}}
{
\begin{subequations}
\begin{eqnarray}
\mathcal{A} & = & \mathcal{A}^\dag \,, \\
\mathcal{B} & = & \mathcal{B}^\dag \,, \\
\mathcal{C} & = & \mathcal{D}^\dag \,,
\end{eqnarray}
\end{subequations}
}
&
{
\begin{subequations}
\begin{eqnarray}
\mathcal{A}_{\mathrm{1PI}} & = & \mathcal{A}_{\mathrm{1PI}}^\dag \,, \\
\mathcal{B}_{\mathrm{1PI}} & = & \mathcal{B}_{\mathrm{1PI}}^\dag \,, \\
\mathcal{C}_{\mathrm{1PI}} & = & \mathcal{D}_{\mathrm{1PI}}^\dag \,.
\end{eqnarray}
\end{subequations}
}
\end{tabular*}

\subsubsection{No wave-function renormalization}

We may set
\begin{subequations}
\begin{eqnarray}
\mathcal{A}_{\mathrm{1PI}} & = & 0 \,, \\
\mathcal{B}_{\mathrm{1PI}} & = & 0
\end{eqnarray}
\end{subequations}
and rename the remaining self-energy $\mathcal{C}_{\mathrm{1PI}} = \mathcal{D}_{\mathrm{1PI}}^\dag$ as
\begin{subequations}
\begin{eqnarray}
\mathcal{D}_{\mathrm{1PI}} &=& -\Sigma_\Psi      \,, \\
\mathcal{C}_{\mathrm{1PI}} &=& -\Sigma_\Psi^\dag \,.
\end{eqnarray}
\end{subequations}
The self-energy $\boldsymbol{\Sigma}_\Psi$ hence takes the form
\begin{eqnarray}
\label{frm:bldSgmPsipodruhe}
\boldsymbol{\Sigma}_\Psi &=& \Sigma_\Psi^\dag\,P_L+\Sigma_\Psi\,P_R \,.
\end{eqnarray}
As an aside, notice that
\begin{eqnarray}
\boldsymbol{\Sigma}_\Psi^{\C} &=& \Sigma_\Psi^*\,P_L+\Sigma_\Psi^\T\,P_R \,,
\end{eqnarray}
so that the condition \eqref{frm:majcondbldsgmPsi} is in terms of $\Sigma_\Psi$ equivalent to
\begin{eqnarray}
\Sigma_\Psi &=& \Sigma_\Psi^\T \,,
\end{eqnarray}
where we ignore the momentum argument, since for $\Sigma_\Psi$, as being a function of $p^2$, the change of sign in \eqref{frm:majcondbldsgmPsi} does not matter.

The chiral components $\mathcal{A}$, $\mathcal{B}$, $\mathcal{C}$, $\mathcal{D}$ of the full propagator $G_\Psi$ can be now expressed as
\begin{subequations}
\begin{eqnarray}
\mathcal{A} &=& D_\Psi^\T \,, \\
\mathcal{B} &=& D_\Psi    \,, \\
\mathcal{C} &=& \Sigma_\Psi      \, D_\Psi^\T \ =\ D_\Psi \, \Sigma_\Psi      \,, \\
\mathcal{D} &=& \Sigma_\Psi^\dag \, D_\Psi \ =\ D_\Psi^\T \, \Sigma_\Psi^\dag \,,
\end{eqnarray}
\end{subequations}
where we denoted
\begin{subequations}
\label{app:frm:DPsiDpsiT}
\begin{eqnarray}
D_\Psi &\equiv& \big(p^2-\Sigma_\Psi^{\phantom{\dag}}\,\Sigma_\Psi^\dag\big)^{-1} \,.
\end{eqnarray}
Notice that due to the symmetricity of $\Sigma_\Psi$ we have
\begin{eqnarray}
D_\Psi^\T &=& \big(p^2-\Sigma_\Psi^\dag\,\Sigma_\Psi^{\phantom{\dag}}\big)^{-1} \,,
\end{eqnarray}
\end{subequations}
so that there is no need to introduce independent denotations (e.g., $D_{\Psi L}$, $D_{\Psi R}$, in analogy with $D_{L}$, $D_{R}$, \eqref{frm:defDRDL}) for the two quantities \eqref{app:frm:DPsiDpsiT}. This time the commutation relation of $D_\Psi$ with $\Sigma_\Psi$ reads
\begin{eqnarray}
\label{frm:commSgmPsi}
\Sigma_\Psi  \, D_\Psi^\T &=& D_\Psi \, \Sigma_\Psi
\end{eqnarray}
and the commutation relation of $D_\Psi$ with $\gamma^\mu$ is again of course trivial.

Like in the previous section, it is again useful to define
\begin{subequations}
\label{app:frm:DPsiDpsiC}
\begin{eqnarray}
\boldsymbol{D}_\Psi &\equiv& \big(p^2-\boldsymbol{\Sigma}_\Psi^{\phantom{\dag}}\,\boldsymbol{\Sigma}_\Psi^\dag\big)^{-1}
\ = \
D_\Psi\,P_L + D_\Psi^\T\,P_R \,.
\end{eqnarray}
Note that since
\begin{eqnarray}
\boldsymbol{D}_\Psi^\C &\equiv& \big(p^2-\boldsymbol{\Sigma}_\Psi^\dag\,\boldsymbol{\Sigma}_\Psi^{\phantom{\dag}}\big)^{-1}
\ = \
D_\Psi^\T\,P_L + D_\Psi\,P_R \,,
\end{eqnarray}
\end{subequations}
we again do not need to introduce two independent denotations for the two quantities \eqref{app:frm:DPsiDpsiC}, in contrast to the Dirac case \eqref{frm:def:bldDRDL}. The commutation relation \eqref{frm:commSgmPsi} translates in terms of $\boldsymbol{D}_\Psi$ as
\begin{eqnarray}
\boldsymbol{\Sigma}_\Psi  \, \boldsymbol{D}_\Psi^\C &=& \boldsymbol{D}_\Psi \, \boldsymbol{\Sigma}_\Psi \,.
\end{eqnarray}
Commutation relation with $\gamma^\mu$ is this time non-trivial:
\begin{subequations}
\begin{eqnarray}
\gamma^\mu \, \boldsymbol{D}_\Psi &=& \boldsymbol{D}_\Psi^\C \, \gamma^\mu \,,  \\
\gamma^\mu \, \boldsymbol{D}_\Psi^\C &=& \boldsymbol{D}_\Psi \, \gamma^\mu \,.
\end{eqnarray}
\end{subequations}

\subsubsection{Expressions of the full propagator}

The full propagator of the field $\Psi$,
\begin{eqnarray}
G_\Psi &=& \big( \slashed{p} - \boldsymbol{\Sigma}_\Psi \big)^{-1} \,,
\end{eqnarray}
can be again expressed in terms of $\Sigma_\Psi$, with the chiral projectors shown explicitly, as
\begin{eqnarray}
\label{frm:GPsipodruhe}
G_\Psi &=&
\big(\slashed{p}+\Sigma_{\Psi}\big)D_\Psi^\T\,P_L + \big(\slashed{p}+\Sigma_{\Psi}^\dag\big)D_\Psi\,P_R \,,
\end{eqnarray}
which is analogous to the expression \eqref{frm:SPsi} of the free propagator $S_\Psi$. Using the definition \eqref{app:frm:DPsiDpsiC} it also possible to express $G_\Psi$ in terms of $\boldsymbol{\Sigma}_\Psi$ in more compact way, with the chiral projectors \qm{hidden}:
\begin{subequations}
\begin{eqnarray}
G_\Psi
&=& \big(\slashed{p}+\boldsymbol{\Sigma}_\Psi^\dag\big)\boldsymbol{D}_\Psi \\
&=& \boldsymbol{D}_\Psi^\C\big(\slashed{p}+\boldsymbol{\Sigma}_\Psi^\dag\big) \,,
\end{eqnarray}
\end{subequations}
in analogy with the expression \eqref{frm:SPsibdl} for $S_\Psi$.

\subsubsection{Propagators in the Nambu--Gorkov doublet space}

Let us introduce some denotation for the components of the self-energy $\Sigma_\Psi$ in the Nambu--Gorkov doublet space $\Psi$:
\begin{eqnarray}
\label{frm:SgmPsimatrix}
\Sigma_\Psi &=&
\left(\begin{array}{cc} \Sigma_{L} & \Sigma_{D} \\ \Sigma_{D}^\T & \Sigma_{R} \end{array}\right) \,,
\end{eqnarray}
where the components $\Sigma_{L}$, $\Sigma_{R}$ are symmetric matrices, so that the $\Sigma_\Psi$ is symmetric too. Let us also introduce some denotation for the corresponding blocks of $D_\Psi$:
\begin{subequations}
\label{frm:DPsiblocks}
\begin{eqnarray}
D_\Psi &=& (p^2-\Sigma_\Psi^{\phantom{\dag}}\,\Sigma_\Psi^{\dag})^{-1}
\label{frm:DPsiblocks1} \\
&=&
\left(\begin{array}{cc}
p^2-(\Sigma_{L}    \,\Sigma_{L}^* + \Sigma_{D}    \,\Sigma_{D}^\dag) &
-   (\Sigma_{L}    \,\Sigma_{D}^* + \Sigma_{D}    \,\Sigma_{R}^*)    \\
-   (\Sigma_{D}^\T \,\Sigma_{L}^* + \Sigma_{R}    \,\Sigma_{D}^\dag) &
p^2-(\Sigma_{R}    \,\Sigma_{R}^* + \Sigma_{D}^\T \,\Sigma_{D}^*)
\end{array}\right)^{-1}
\label{frm:DPsiblocks2} \\
&\equiv& \left(\begin{array}{cc} D_{L} & D_{M} \\ D_{M}^\dag & D_{R}^\T \end{array}\right) \,.
\label{frm:DPsiblocks3}
\end{eqnarray}
\end{subequations}
It is possible to invert $D_\Psi$, \eqref{frm:DPsiblocks2}, explicitly, i.e., to express the blocks $D_{L}$, $D_{R}$, $D_{M}$ in terms of $\Sigma_{L}$, $\Sigma_{R}$, $\Sigma_{D}$. One can use for this purpose the formula for the block-wise inversion
\begin{subequations}
\begin{eqnarray}
\left(\begin{array}{cc}
A & B \\
C & D \\
\end{array}\right)^{-1}
&=&
\left(\begin{array}{cc}
\big(A-BD^{-1}C\big)^{-1} &
-\big(A-BD^{-1}C\big)^{-1}BD^{-1} \\
-D^{-1}C\big(A-BD^{-1}C\big)^{-1} &
D^{-1}+D^{-1}C\big(A-BD^{-1}C\big)^{-1}BD^{-1} \\
\end{array}\right)
\nonumber \\ &&
\\
&=&
\left(\begin{array}{cc}
A^{-1}+A^{-1}B\big(D-CA^{-1}B\big)^{-1}CA^{-1} &
-A^{-1}B\big(D-CA^{-1}B\big)^{-1} \\
-\big(D-CA^{-1}B\big)^{-1}CA^{-1} &
\big(D-CA^{-1}B\big)^{-1} \\
\end{array}\right) \,,
\nonumber \\ &&
\end{eqnarray}
\end{subequations}
holding provided $A$ and $D$ are square matrices; one can choose the appropriate form of the inversion according to which of the inversions $\big(A-BD^{-1}C\big)^{-1}$, $D^{-1}$ or $\big(D-CA^{-1}B\big)^{-1}$, $A^{-1}$ do exist and which do not. However, for general $\Sigma_{L}$, $\Sigma_{R}$, $\Sigma_{D}$ the explicit forms of $D_{L}$, $D_{R}$, $D_{M}$ would not be neither very elegant nor illuminating. Nevertheless, in order to get some feeling about it, we are going to do it for two special cases: We consider vanishing the Dirac self-energy $\Sigma_D$ and non-vanishing Majorana self-energies $\Sigma_L$, $\Sigma_R$, and vice versa:
\begin{itemize}
  \item Let both $\Sigma_L = 0$, $\Sigma_R = 0$. Then
\begin{subequations}
\begin{eqnarray}
D_{L} &=& (p^2-\Sigma_D^{\phantom{\dag}}\,\Sigma_D^{\dag})^{-1} \,, \\
D_{R} &=& (p^2-\Sigma_D^{\dag}\,\Sigma_D^{\phantom{\dag}})^{-1} \,, \\
D_{M} &=& 0 \,.
\end{eqnarray}
\end{subequations}
Note that in this case the form of $D_{L}$, $D_{R}$ in terms of $\Sigma_D$ coincides with the definition \eqref{frm:defDRDL} of $D_{L}$, $D_{R}$ in the context of a Dirac field.
  \item Let $\Sigma_D = 0$. Then
\begin{subequations}
\begin{eqnarray}
D_{L} &=& (p^2-\Sigma_L^{\phantom{\dag}}\,\Sigma_L^{\dag})^{-1} \,, \\
D_{R} &=& (p^2-\Sigma_R^{\dag}\,\Sigma_R^{\phantom{\dag}})^{-1} \,, \\
D_{M} &=& 0 \,.
\end{eqnarray}
\end{subequations}
\end{itemize}

We have seen that in both cases $D_M=0$. This is not a coincidence. One can see from \eqref{frm:DPsiblocks} clearly that $D_M$ is proportional to the off-diagonal blocks of $p^2-\Sigma_\Psi^{\phantom{\dag}}\,\Sigma_\Psi^{\dag}$ (both of which are related to each other only by the Hermitian conjugation):
\begin{eqnarray}
\label{app:frm:DMpropto}
D_{M} &\propto& \Sigma_{L}\,\Sigma_{D}^* + \Sigma_{D}\,\Sigma_{R}^* \,.
\end{eqnarray}
This should be understood as
\begin{eqnarray}
\Sigma_{L}\,\Sigma_{D}^* + \Sigma_{D}\,\Sigma_{R}^* = 0  &\Rightarrow& D_{M} = 0 \,.
\end{eqnarray}

Let us finally show how the relation \eqref{frm:commSgmPsi} is translated in terms of the Nambu--Gorkov blocks \eqref{frm:DPsiblocks}:
\begin{subequations}
\begin{eqnarray}
D_L    \, \Sigma_L + D_M      \, \Sigma_D^\T &=& \Sigma_L \, D_L^\T + \Sigma_D    \, D_M^\T \,, \\
D_R^\T \, \Sigma_R + D_M^\dag \, \Sigma_D    &=& \Sigma_R \, D_R    + \Sigma_D^\T \, D_M^*  \,, \\
D_M    \, \Sigma_R + D_L      \, \Sigma_D    &=& \Sigma_L \, D_M^*  + \Sigma_D    \, D_R \,.
\end{eqnarray}
\end{subequations}

\subsubsection{Diagrammatics}

The line corresponding to the 1PI propagator $\boldsymbol{\Sigma}_\Psi$ is
\begin{subequations}
\begin{eqnarray}
\langle \Psi \bar \Psi \rangle_{\mathrm{1PI}}
\ = \
\begin{array}{c}
\scalebox{0.85}{\includegraphics[trim = 10bp 12bp 19bp 11bp,clip]{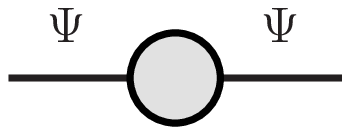}}
\end{array}
&=& -\I\big(\Sigma_\Psi^\dag\,P_L+\Sigma_\Psi\,P_R\big)
\\ &=& -\I\boldsymbol{\Sigma}_\Psi \,.
\end{eqnarray}
\end{subequations}
Notice that it has no arrows, as the field $\Psi$ is real. The lines corresponding to the chiral components of $\boldsymbol{\Sigma}_\Psi$ (i.e., corresponding to the fields $\psi_L$, $\psi_R$) are
\begin{subequations}
\begin{eqnarray}
\langle \psi_L \bar \psi_R \rangle_{\mathrm{1PI}}
\ = \
\begin{array}{c}
\scalebox{0.85}{\includegraphics[trim = 10bp 12bp 19bp 11bp,clip]{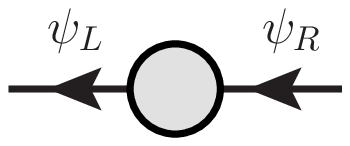}}
\end{array}
&=& - \I\,\Sigma_{D} \,P_R \,,
\\
\langle \psi_R \bar \psi_L \rangle_{\mathrm{1PI}}
\ = \
\begin{array}{c}
\scalebox{0.85}{\includegraphics[trim = 10bp 12bp 19bp 11bp,clip]{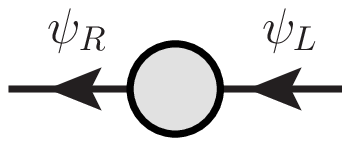}}
\end{array}
&=& - \I\,\Sigma_{D}^\dag \,P_L \,,
\\
\langle \psi_R (\bar \psi_R)^\C \rangle_{\mathrm{1PI}}
\ = \
\begin{array}{c}
\scalebox{0.85}{\includegraphics[trim = 10bp 12bp 19bp 11bp,clip]{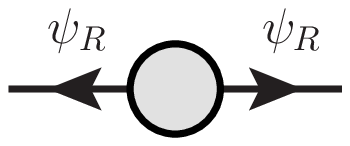}}
\end{array}
&=& - \I\,\Sigma_{R}^\dag \,P_L \,,
\\
\langle (\psi_R)^\C \bar \psi_R \rangle_{\mathrm{1PI}}
\ = \
\begin{array}{c}
\scalebox{0.85}{\includegraphics[trim = 10bp 12bp 19bp 11bp,clip]{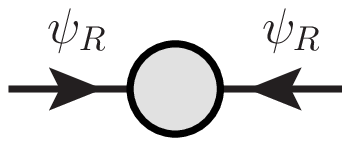}}
\end{array}
&=& - \I\,\Sigma_{R} \,P_R \,,
\\
\langle \psi_L (\bar \psi_L)^\C \rangle_{\mathrm{1PI}}
\ = \
\begin{array}{c}
\scalebox{0.85}{\includegraphics[trim = 10bp 12bp 19bp 11bp,clip]{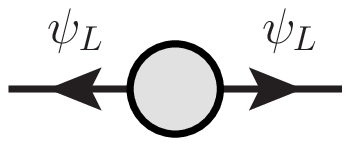}}
\end{array}
&=& - \I\,\Sigma_{L} \,P_R \,,
\\
\langle (\psi_L)^\C \bar \psi_L \rangle_{\mathrm{1PI}}
\ = \
\begin{array}{c}
\scalebox{0.85}{\includegraphics[trim = 10bp 12bp 19bp 11bp,clip]{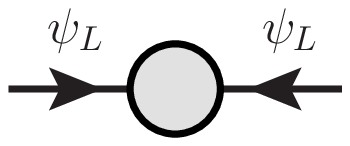}}
\end{array}
&=& - \I\,\Sigma_{L}^\dag \,P_L \,.
\end{eqnarray}
\end{subequations}

The full propagator $G_\Psi$ is
\begin{subequations}
\begin{eqnarray}
\langle \Psi \bar \Psi \rangle
\ = \
\begin{array}{c}
\scalebox{0.85}{\includegraphics[trim = 10bp 12bp 19bp 11bp,clip]{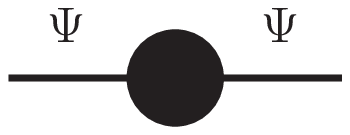}}
\end{array}
&=& \I\big(\slashed{p}+\Sigma_\Psi \big)D_{\Psi}^\T\,P_L+\I\big( \slashed{p}+\Sigma_\Psi^\dag\big)D_{\Psi}\,P_R
\\
&=& \I\big(\slashed{p}+\boldsymbol{\Sigma}_\Psi^\dag\big)\boldsymbol{D}_\Psi \ = \  \I\boldsymbol{D}_\Psi^\C\big(\slashed{p}+\boldsymbol{\Sigma}_\Psi^\dag\big) \,,
\end{eqnarray}
\end{subequations}
again without the arrows. The chiral components without $\slashed{p}$ are
\begin{subequations}
\begin{eqnarray}
\langle \psi_L \bar \psi_R \rangle
\ = \
\begin{array}{c}
\scalebox{0.85}{\includegraphics[trim = 10bp 12bp 19bp 11bp,clip]{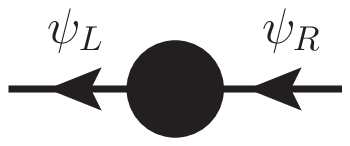}}
\end{array}
&=&
\I \big( \Sigma_{D}\,D_{R} + \Sigma_{L}\,D_{M}^* \big) P_L \ =\
\I \big( D_{L}\,\Sigma_{D} + D_{M}\,\Sigma_{R}   \big) P_L \,,
\nonumber \\ &&
\\
\langle \psi_R \bar \psi_L \rangle
\ = \
\begin{array}{c}
\scalebox{0.85}{\includegraphics[trim = 10bp 12bp 19bp 11bp,clip]{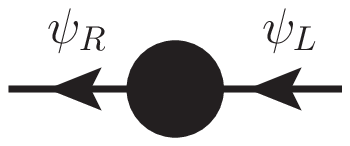}}
\end{array}
&=&
\I \big( D_{R}\,\Sigma_{D}^\dag + D_{M}^\T\,\Sigma_{L}^\dag   \big) P_R \ =\
\I \big( \Sigma_{D}^\dag\,D_{L} + \Sigma_{R}^\dag\,D_{M}^\dag \big) P_R \,,
\nonumber \\ &&
\\
\langle \psi_L (\bar \psi_L)^\C \rangle
\ = \
\begin{array}{c}
\scalebox{0.85}{\includegraphics[trim = 10bp 12bp 19bp 11bp,clip]{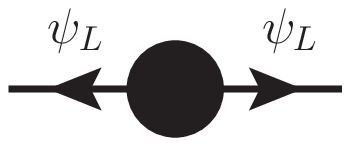}}
\end{array}
&=&
\I \big( \Sigma_{L}\,D_{L}^\T + \Sigma_{D}\,D_{M}^\T \big) P_L \ =\
\I \big( D_{L}\,\Sigma_{L}    + D_{M}\,\Sigma_{D}^\T \big) P_L \,,
\nonumber \\ &&
\\
\langle (\psi_L)^\C \bar \psi_L \rangle
\ = \
\begin{array}{c}
\scalebox{0.85}{\includegraphics[trim = 10bp 12bp 19bp 11bp,clip]{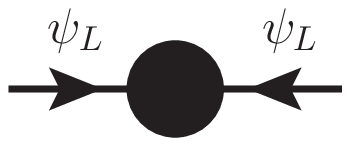}}
\end{array}
&=&
\I \big( D_{L}^\T\,\Sigma_{L}^\dag + D_{M}^*\,\Sigma_{D}^\dag \big) P_R \ =\
\I \big( \Sigma_{L}^\dag\,D_{L}    + \Sigma_{D}^*\,D_{M}^\dag \big) P_R \,,
\nonumber \\ &&
\\
\langle \psi_R (\bar \psi_R)^\C \rangle
\ = \
\begin{array}{c}
\scalebox{0.85}{\includegraphics[trim = 10bp 12bp 19bp 11bp,clip]{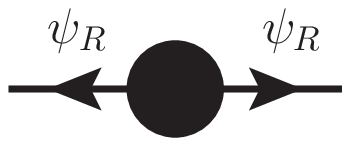}}
\end{array}
&=&
\I \big( \Sigma_{R}^\dag\,D_{R}^\T + \Sigma_{D}^\dag\,D_{M} \big) P_R \ =\
\I \big( D_{R}\,\Sigma_{R}^\dag    + D_{M}^\T\,\Sigma_{D}^* \big) P_R \,,
\nonumber \\ &&
\\
\langle (\psi_R)^\C \bar \psi_R \rangle
\ = \
\begin{array}{c}
\scalebox{0.85}{\includegraphics[trim = 10bp 12bp 19bp 11bp,clip]{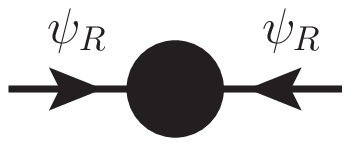}}
\end{array}
&=&
\I \big( D_{R}^\T\,\Sigma_{R} + D_{M}^\dag\,\Sigma_{D} \big) P_L \ =\
\I \big( \Sigma_{R}\,D_{R}    + \Sigma_{D}^\T\,D_{M}^* \big) P_L
\nonumber \\ &&
\end{eqnarray}
\end{subequations}
and proportional to $\slashed{p}$ are
\begin{subequations}
\label{app:frm:slashedp}
\begin{eqnarray}
\langle \psi_L \bar \psi_L \rangle
\ = \
\begin{array}{c}
\scalebox{0.85}{\includegraphics[trim = 10bp 12bp 19bp 11bp,clip]{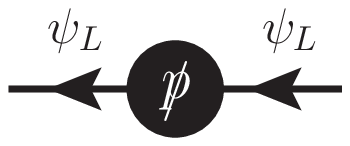}}
\end{array}
&=&
\I\,\slashed{p}\, D_L \,P_R \,,
\\
\langle \psi_R \bar \psi_R \rangle
\ = \
\begin{array}{c}
\scalebox{0.85}{\includegraphics[trim = 10bp 12bp 19bp 11bp,clip]{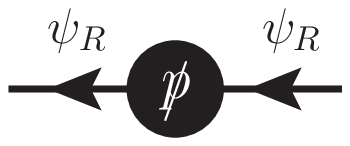}}
\end{array}
&=&
\I\,\slashed{p}\, D_R \,P_L \,,
\\
\langle (\psi_L)^\C \bar \psi_R \rangle
\ = \
\begin{array}{c}
\scalebox{0.85}{\includegraphics[trim = 10bp 12bp 19bp 11bp,clip]{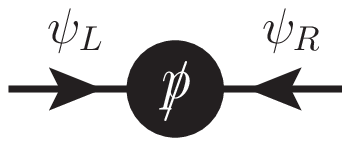}}
\end{array}
&=&
\I\,\slashed{p}\, D_M^* \,P_L \,,
\\
\langle \psi_R (\bar \psi_L)^\C \rangle
\ = \
\begin{array}{c}
\scalebox{0.85}{\includegraphics[trim = 10bp 12bp 19bp 11bp,clip]{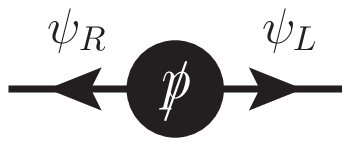}}
\end{array}
&=&
\I\,\slashed{p}\, D_M^\T \,P_L \,,
\\
\langle (\psi_R)^\C \bar \psi_L \rangle
\ = \
\begin{array}{c}
\scalebox{0.85}{\includegraphics[trim = 10bp 12bp 19bp 11bp,clip]{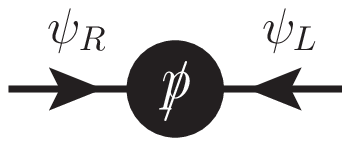}}
\end{array}
&=&
\I\,\slashed{p}\, D_M^\dag \,P_R \,,
\\
\langle \psi_L (\bar \psi_R)^\C \rangle
\ = \
\begin{array}{c}
\scalebox{0.85}{\includegraphics[trim = 10bp 12bp 19bp 11bp,clip]{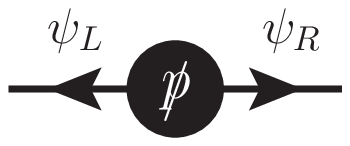}}
\end{array}
&=&
\I\,\slashed{p}\, D_M \,P_R \,.
\end{eqnarray}
\end{subequations}

\section[Relations between the Dirac and Nambu--Gorkov fields]{Relations between the Dirac and \\Nambu--Gorkov fields}
\label{app:frm:Rel}

The Nambu--Gorkov formalism $\Psi$ is more general than the more usual $\psi$ one, as it releases all the special assumptions made when working with $\psi$. Therefore the quantities written in terms of $\psi$ must be expressible in terms of $\Psi$. And vice versa, the quantities written in the $\Psi$ basis should be expressible in the $\psi$ basis in the special case of $n = m$ and with the invariance under \eqref{frm:phase:psiLpsiR}. This section is devoted to the problem of translating quantities from one basis to the other.

\subsection{Basic relations}

We start by stating the basic relations between the fields $\psi$ and $\Psi$:
\begin{subequations}
\label{frm:psiPsirels}
\begin{eqnarray}
\Psi &=& P \, \psi + \bar P^\dag \, \psi^\C \,,
\label{frm:Psiinpsi} \\
\psi &=& P^\dag \, \Psi \,,
\label{frm:psiinPsi}
\end{eqnarray}
\end{subequations}
where we introduced the quantity $P$ as
\begin{subequations}
\label{app:fermi:Pdef}
\begin{eqnarray}
P &\equiv& \left(\begin{array}{c} P_L \\ P_R \end{array}\right) \,,
\end{eqnarray}
so that
\begin{eqnarray}
\bar P   &=& \big(P_R,P_L\big) \,, \\
P^\dag     &=& \big(P_L,P_R\big) \,, \\
\bar P^\dag &=& \left(\begin{array}{c} P_R \\ P_L \end{array}\right) \,.
\end{eqnarray}
\end{subequations}
Taking into account the definition of $P$, the relations \eqref{frm:psiPsirels} can be seen rather directly from the explicit expression of $\Psi$ and $\psi$ in terms of $\psi_L$, $\psi_R$, Eqs.~\eqref{frm:Psidef} and \eqref{app:frm:defpsi}, respectively.

The quantities \eqref{app:fermi:Pdef} satisfy
\begin{subequations}
\label{app:fermi:PP01}
\begin{eqnarray}
P^\dag P \ = &1& = \ \bar P \bar P^\dag \,, \\
\bar P P \ = &0& = \ P^\dag \bar P^\dag \,,
\end{eqnarray}
\end{subequations}
as well as the relation
\begin{eqnarray}
\label{frm:PdagPC}
P^\dag &=& P^\C \,.
\end{eqnarray}
In practical calculations one may find useful the expressions
\begin{subequations}
\begin{eqnarray}
P  \bar P &=&
\left(\begin{array}{cc} 0 & P_L \\ P_R & 0 \end{array}\right) \>=\>
\frac{1}{2} (\sigma_1 - \I \gamma_5 \sigma_2) \,, \\
\bar P^\dag  P^\dag &=&
\left(\begin{array}{cc} 0 & P_R \\ P_L & 0 \end{array}\right) \>=\>
\frac{1}{2} (\sigma_1 + \I \gamma_5 \sigma_2) \,, \\
P  P^\dag &=&
\left(\begin{array}{cc} P_L & 0 \\ 0 & P_R \end{array}\right) \>=\>
\frac{1}{2} (1 - \gamma_5 \sigma_3) \,, \\
\bar P^\dag  \bar P &=&
\left(\begin{array}{cc} P_R & 0 \\ 0 & P_L \end{array}\right) \>=\>
\frac{1}{2} (1 + \gamma_5 \sigma_3) \,.
\end{eqnarray}
\end{subequations}
We will occasionally call the quantities \eqref{app:fermi:Pdef} the \emph{projectors}, although, strictly speaking, only their combinations $P P^\dag$ and $\bar P^\dag \bar P$ are true projectors.

\subsection{Propagators}

From the above considerations one can infer the relations between the full propagators $G_\Psi$ and $G_\psi$:
\begin{subequations}
\begin{eqnarray}
G_\Psi(p) &=& P \, G_\psi(p) \, \bar P + \bar P^\dag \, G_\psi^\C(-p) \, P^\dag \,,
\label{frm:GPsiinGpsi} \\
G_\psi(p) &=& P^\dag \, G_\Psi(p) \, \bar P^\dag \,,
\label{frm:GpsiinGPsi}
\end{eqnarray}
\end{subequations}
as well as the relations between the self-energies $\boldsymbol{\Sigma}_\Psi$ and $\boldsymbol{\Sigma}_\psi$:
\begin{subequations}
\begin{eqnarray}
\boldsymbol{\Sigma}_\Psi(p) &=& \bar P^\dag \, \boldsymbol{\Sigma}_\psi(p) \, P^\dag + P \, \boldsymbol{\Sigma}_\psi^\C(-p) \, \bar P \,,
\\
\boldsymbol{\Sigma}_\psi(p) &=& \bar P \, \boldsymbol{\Sigma}_\Psi(p) \, P \,.
\end{eqnarray}
\end{subequations}
Notice that the \qm{Majorana} symmetries \eqref{frm:majcondGPsi} and \eqref{frm:majcondbldsgmPsi} of these expressions for $G_\Psi(p)$ and $\boldsymbol{\Sigma}_\Psi(p)$, respectively, are evident upon taking into account the relation \eqref{frm:PdagPC}. Explicitly we can write $G_\Psi(p)$ and $\boldsymbol{\Sigma}_\Psi(p)$ in terms of the chiral components of the most general forms \eqref{app:frm:psiG} and \eqref{app:frm:psiSgm} of $G_\psi(p)$ and $\boldsymbol{\Sigma}_\psi(p)$, respectively, as
\begin{subequations}
\begin{eqnarray}
G_\Psi(p) &=& \phantom{-} \left(\begin{array}{ll}
\slashed{p}(A_{LL}^\T\,P_L+A_{LL}\,P_R) & \phantom{\slashed{p}(}A_{RL}^\T\,P_R+A_{LR}\,P_L \\
\phantom{\slashed{p}(}A_{LR}^\T\,P_L+A_{RL}\,P_R & \slashed{p}(A_{RR}^\T\,P_R+A_{RR}\,P_L)
\end{array}\right) \,,
\\
\boldsymbol{\Sigma}_\Psi(p) &=& - \left(\begin{array}{ll}
\slashed{p}(a_{LL}\,P_L+a_{LL}^\T\,P_R) & \phantom{\slashed{p}(}a_{LR}\,P_R+a_{RL}^\T\,P_L \\
\phantom{\slashed{p}(}a_{RL}\,P_L+a_{LR}^\T\,P_R & \slashed{p}(a_{RR}\,P_R+a_{RR}^\T\,P_L)
\end{array}\right) \,.
\label{frm:bldSgmPsi}
\end{eqnarray}
\end{subequations}

Assume now that the self-energy $\boldsymbol{\Sigma}_\psi(p)$ has the special form \eqref{frm:Sgmpsi}:
\begin{eqnarray}
\label{frm:bldSgmpsi}
\boldsymbol{\Sigma}_\psi &=& \Sigma_D^\dag \, P_L + \Sigma_D \, P_R \,.
\end{eqnarray}
The matrix expression \eqref{frm:bldSgmPsi} for $\boldsymbol{\Sigma}_\Psi$ then acquires the form
\begin{eqnarray}
\boldsymbol{\Sigma}_\Psi &=&
\left(\begin{array}{cc}
0 & \boldsymbol{\Sigma}_\psi \, P_R + \boldsymbol{\Sigma}_\psi^\C \, P_L \\
\boldsymbol{\Sigma}_\psi \, P_L + \boldsymbol{\Sigma}_\psi^\C \, P_R & 0 \\
\end{array}\right) \,,
\end{eqnarray}
which is expressible in the form \eqref{frm:bldSgmPsipodruhe},
\begin{eqnarray}
\boldsymbol{\Sigma}_\Psi &=& \Sigma_\Psi^\dag \, P_L + \Sigma_\Psi \, P_R \,,
\end{eqnarray}
with $\Sigma_\Psi$ given in terms of $\Sigma_D$ as
\begin{eqnarray}
\Sigma_\Psi &=& \left(\begin{array}{cc} 0 & \Sigma_D \\ \Sigma_D^\T & 0 \end{array}\right) \,.
\end{eqnarray}
Notice that this corresponds to the general form \eqref{frm:SgmPsimatrix} of $\Sigma_\Psi$ up to the missing Majorana components $\Sigma_L$, $\Sigma_R$.

Similarly can be treated the full propagators. Assuming the $G_\psi(p)$ to be of the form \eqref{frm:Gpsipodruhe}
\begin{eqnarray}
G_\psi &=& \big(\slashed{p}+\Sigma_D\big)D_R\,P_L+\big(\slashed{p}+\Sigma_D^\dag\big)D_L\,P_R \,,
\end{eqnarray}
corresponding to $\boldsymbol{\Sigma}_\psi$ of the form \eqref{frm:bldSgmpsi}, we find $G_\Psi(p)$ to be
\begin{eqnarray}
G_\Psi &=& \left(\begin{array}{cc}
\slashed{p}(D_L^\T\,P_L+D_L\,P_R) & (\Sigma_D\,D_R) P_L+(\Sigma_D^\dag\,D_L)^\T P_R \\
(\Sigma_D\,D_R)^\T P_L + (\Sigma_D^\dag\,D_L) P_R & \slashed{p}(D_R\,P_L+D_R^\T\,P_R)
\end{array}\right) \,.
\end{eqnarray}
Comparing this with the expression \eqref{frm:GPsipodruhe} for $G_\Psi$ we find
\begin{eqnarray}
D_\Psi &=& \left(\begin{array}{cc} D_L & 0 \\ 0 & D_R^\T \end{array}\right) \,,
\end{eqnarray}
which is just a special case of \eqref{frm:DPsiblocks3} with vanishing Majorana component $D_M$.

\subsection{Vertex}

\subsubsection{Full vertex}

Although this appendix is predominantly devoted to the fermion propagators, it is worth spending few words here also about the fermion-fermion-gauge boson vertex, due to applications in the main text. The full three-point function in question has in both bases $\psi$, $\Psi$ the same general structure
\begin{subequations}
\begin{eqnarray}
\langle A^\mu \psi \bar\psi \rangle &\sim& G_\psi(p^\prime) \, \Gamma_\psi^\mu(p^\prime,p) \, G_\psi(p) \,, \\
\langle A^\mu \Psi \bar\Psi \rangle &\sim& G_\Psi(p^\prime) \, \Gamma_\Psi^\mu(p^\prime,p) \, G_\Psi(p) \,.
\end{eqnarray}
\end{subequations}
We omit here the gauge boson propagators as well as the gauge indices at the vertex functions, as they are irrelevant for the present discussion.

We will now derive the relation between the proper vertices $\Gamma_\psi^\mu(p^\prime,p)$, $\Gamma_\Psi^\mu(p^\prime,p)$, like we did before for the self-energies $\boldsymbol{\Sigma}_\psi(p)$, $\boldsymbol{\Sigma}_\Psi(p)$. Taking into account the expression \eqref{frm:Psiinpsi} of $\Psi$ in terms of $\psi$ and the properties of the charge conjugation, we arrive at
\begin{eqnarray}
\label{frm:3GPsi}
G_\Psi(p^\prime) \, \Gamma_\Psi^\mu(p^\prime,p) \, G_\Psi(p) &=&
P \, G_\psi(p^\prime) \, \Gamma_\psi^\mu(p^\prime,p) \, G_\psi(p) \, \bar P +
\bar P^\dag \, G_\psi^\C(-p^\prime) \, \Gamma_\psi^{\mu\C}(-p,-p^\prime) \, G_\psi^\C(-p) \, P^\dag \,.
\nonumber \\ &&
\end{eqnarray}
In deriving it one must also take into account properly the definition of the Fourier transform and be consistent with assignments of the momenta in both terms on the right-hand side of \eqref{frm:3GPsi}, as well as the fact that $\langle A^\mu \psi \bar\psi^\C \rangle = \langle A^\mu \psi^\C \bar\psi \rangle = 0$ due to assumed Dirac character of the field $\psi$.\footnote{It is unnatural to assume that the propagator is invariant under phase transformation, while at the same time the vertex is not.} On the other hand, we may also use the expression \eqref{frm:GPsiinGpsi} of $G_\Psi(p)$ in terms of $G_\psi(p)$ to arrive at
\begin{eqnarray}
\label{frm:3GPsipodruhe}
G_\Psi(p^\prime) \, \Gamma_\Psi^\mu(p^\prime,p) \, G_\Psi(p) &=&
\nonumber \\ && \hspace{-4em}
P \, G_\psi(p^\prime) \, \bar P \, \Gamma_\Psi^\mu(p^\prime,p) \, P \, G_\psi(p) \, \bar P +
\bar P^\dag \,G_\psi^\C(-p^\prime) \, P^\dag \, \Gamma_\Psi^{\mu}(p^\prime,p) \, \bar P^\dag \, G_\psi^\C(-p) \, P^\dag \,.
\nonumber \\ &&
\end{eqnarray}
(Again, the \qm{cross terms}, proportional to $G_\psi^\C(p^\prime) \ldots G_\psi(p)$, $G_\psi(p^\prime) \ldots G_\psi^\C(p)$, are not considered for similar arguments.) We can now compare the two expressions \eqref{frm:3GPsi}, \eqref{frm:3GPsipodruhe} and use again the projectors \eqref{app:fermi:Pdef}: First we multiply the them from left by $P^\dag$ and from right by $\bar P^\dag$ and then from left by $\bar P$ and from right by $P$. The projectors project out two separate equations:
\begin{eqnarray}
\bar P \, \Gamma_\Psi^\mu(p^\prime,p) \, P &=& \Gamma_\psi^\mu(p^\prime,p) \,, \\
P^\dag \, \Gamma_\Psi^{\mu}(p^\prime,p) \, \bar P^\dag &=& \Gamma_\psi^{\mu\C}(-p,-p^\prime) \,.
\end{eqnarray}
This implies that the $\Gamma_\Psi^\mu(p^\prime,p)$ in terms of $\Gamma_\psi^\mu(p^\prime,p)$ is given by
\begin{subequations}
\label{frm:relGmmpsiGmmPsi}
\begin{align}
\Gamma_\Psi^\mu(p^\prime,p)
&\ =\
\bar P^\dag \, \Gamma_\psi^\mu(p^\prime,p) \, P^\dag + P \, \Gamma_\psi^{\mu\C}(-p,-p^\prime) \, \bar P \,.
\label{app:fermi:GPsi_in_Gpsi}
\\
\intertext{Conversely, one can now easily obtain expression for $\Gamma_\psi^\mu(p^\prime,p)$ in terms of}
\Gamma_\psi^\mu(p^\prime,p) &\ =\  \bar P \, \Gamma_\Psi^\mu(p^\prime,p) \, P
\end{align}
\end{subequations}
by applying the projectors \eqref{app:fermi:Pdef} on \eqref{app:fermi:GPsi_in_Gpsi} and taking into account the relations \eqref{app:fermi:PP01}.

\subsubsection{Bare vertex and generators}

The formul{\ae} \eqref{frm:relGmmpsiGmmPsi} can be now exploited by considering the bare vertices
\begin{subequations}
\label{frm:Gmmbare}
\begin{eqnarray}
\Gamma_\psi^\mu(p^\prime,p)\big|_{\mathrm{bare}} &=& g\gamma^\mu t_\psi \,, \\
\Gamma_\Psi^\mu(p^\prime,p)\big|_{\mathrm{bare}} &=& g\gamma^\mu t_\Psi \,,
\end{eqnarray}
\end{subequations}
where $t_\psi$, $t_\Psi$ are the symmetry generators of the symmetry $\group{G}$ in question in the respective bases (recall that we suppress the gauge indices) and $g$ is a gauge coupling constant. This corresponds to the interaction Lagrangian
\begin{subequations}
\label{frm:eLint}
\begin{eqnarray}
\eL &=& g \bar\psi \gamma^\mu t_\psi \psi A_\mu \\ &=& g\frac{1}{2}\bar\Psi \gamma^\mu t_\Psi \Psi A_\mu
\end{eqnarray}
\end{subequations}
and to the symmetry transformation
\begin{subequations}
\begin{eqnarray}
\group{G}\,:\qquad \psi  & \TransformsTo & {[\psi]}^\prime \>=\> \e^{ \I \theta t_{\psi}}\, \psi \,,
\\
\group{G}\,:\qquad \Psi  & \TransformsTo & {[\Psi]}^\prime \>=\> \e^{ \I \theta t_{\Psi}}\, \Psi \,.
\end{eqnarray}
\end{subequations}

Upon plugging the bare vertices \eqref{frm:Gmmbare} into \eqref{frm:relGmmpsiGmmPsi} we arrive at the generator $t_\Psi$ expressed in terms of $t_\psi$:
\begin{eqnarray}
\label{app:frm:tPsi-tpsinonmtr}
t_\Psi  &=& P \, t_\psi \, P^\dag - \bar P^\dag \, \bar t_\psi^\C \, \bar P \,,
\end{eqnarray}
and vice versa:
\begin{eqnarray}
\label{app:frm:tpsi-tPsi}
t_\psi &=& P^\dag \, t_\Psi \, P \,.
\end{eqnarray}
Since $t_\psi$ is in general a linear combination of only $\unitmatrix$ and $\gamma_5$ (or, equivalently, $P_L$ and $P_R$), we can write the matrix form of $t_\Psi$, \eqref{app:frm:tPsi-tpsinonmtr}, as
\begin{eqnarray}
\label{app:frm:tPsi-tpsi}
t_\Psi &=&
\left(\begin{array}{cc}
t_\psi\,P_L - \bar t_\psi^\C\,P_R & 0 \\
0 & t_\psi\,P_R - \bar t_\psi^\C\,P_L
\end{array}\right) \,.
\end{eqnarray}

The Lagrangian \eqref{frm:eLint} in the basis $\psi$ can be also expressed in the chiral bases $\psi_L$, $\psi_R$ as
\begin{eqnarray}
\eL &=& g \bar\psi_L \gamma^\mu t_{\psi_L} \psi_L A_\mu + g \bar\psi_R \gamma^\mu t_{\psi_R} \psi_R A_\mu \,,
\end{eqnarray}
where $t_{\psi_L}$, $t_{\psi_R}$ do not contain any $\gamma_5$. It corresponds to the symmetry transformations
\begin{subequations}
\begin{eqnarray}
\group{G}\,:\qquad \psi_L  & \TransformsTo & {[\psi_L]}^\prime \>=\> \e^{ \I \theta t_{\psi_L}} \, \psi_L \,, \\
\group{G}\,:\qquad \psi_R  & \TransformsTo & {[\psi_R]}^\prime \>=\> \e^{ \I \theta t_{\psi_R}} \, \psi_R \,.
\end{eqnarray}
\end{subequations}
The generators $t_\psi$ and $t_{\psi_L}$, $t_{\psi_R}$ are related to each other by
\begin{eqnarray}
t_\psi &=& t_{\psi_L}\,P_L + t_{\psi_R}\,P_R \,.
\end{eqnarray}
Now we can express the generator $t_\Psi$ in terms of $t_{\psi_L}$, $t_{\psi_R}$ as
\begin{eqnarray}
\label{app:frm:tPsi-tpsiLtpsiR}
t_\Psi &=&
\left(\begin{array}{cc}
t_{\psi_L}\,P_L - t_{\psi_L}^\T\,P_R & 0 \\ 0 & t_{\psi_R}\,P_R - t_{\psi_R}^\T\,P_L
\end{array}\right) \,.
\end{eqnarray}

\section{Diagonalization of the Dirac propagator}
\label{frm:sec:diag}

Regarding the application in the main text (namely in chapter~\ref{chp:mx}) we will now consider diagonalization of the propagator $G_\psi$ of the Dirac field $\psi$, introduced in section~\ref{frm:sec:Dirac}. We will not do it in full generality but rather restrict ourselves to the special case discussed in section~\ref{frm:ssec:Dirac:simp}.

\subsection{Diagonalization}

Consider the Dirac self-energy $\boldsymbol{\Sigma}_\psi$ of the form \eqref{frm:Sgmpsi}. Using the \emph{bi-unitary transformation} (which is a special case of the more general \emph{singular value decomposition}) we can diagonalize its part $\Sigma_D$ as\footnote{We now indicate explicitly the momentum dependencies, as they are going to be important for the present discussion.}
\begin{eqnarray}
\label{eq:definition_UV}
\Sigma_D(p^2) &=& V^\dag(p^2) \, M(p^2) \, U(p^2) \,,
\end{eqnarray}
where $U$, $V$ are some unitary matrices and $M$ is a diagonal, real, non-negative matrix:
\begin{eqnarray}
\label{def:M_diag_nonconst}
M(p^2) &=& \diag\Big(M_1(p^2),M_2(p^2), \ldots, M_{n}(p^2)\Big) \,.
\end{eqnarray}

It is convenient to introduce unitary matrix
\begin{eqnarray}
\label{frm:dg:Xdef}
X(p^2) &\equiv& V^\dag(p^2)\, P_L + U^\dag(p^2)\, P_R \,,
\end{eqnarray}
as it will allow us to write more compact formul{\ae}, without the necessity to use explicitly the chiral projectors $P_{L}$, $P_{R}$. It can be used to diagonalize $\boldsymbol{\Sigma}_\psi$ as
\begin{eqnarray}
\label{frm:SgmdiagX}
\boldsymbol{\Sigma}_\psi(p^2) &=& \bar X^\dag(p^2) \, M(p^2) \, X^\dag(p^2) \,,
\end{eqnarray}
where $\bar X \equiv \gamma_0 X^\dag \gamma_0$. Then the propagator $G_\psi$ can be diagonalized as
\begin{eqnarray}
\label{eq:S_diagonalized}
G_\psi(p) &=&  X(p^2) \frac{\slashed{p}+M(p^2)}{p^2-M^2(p^2)} \bar X(p^2) \,.
\end{eqnarray}
(This expression is correct, since the matrices in the nominator and denominator commute with each other, as they are both diagonal.)

The spectrum is easily revealed by looking for the poles of the propagator $G_\psi(p)$. Thus, taking into account its diagonalized form \eqref{eq:S_diagonalized}, we have to solve the pole equation
\begin{eqnarray}
\label{eq:pole_eq_general}
\det \! \Big( p^2 - M^2(p^2) \Big) &=& 0 \,.
\end{eqnarray}
Due to the diagonality of $M(p^2)$ we have $\det \! \big( p^2 - M^2(p^2) \big) = \prod_{i=1}^n \big( p^2 - M_i^2(p^2) \big)$ and consequently the equation \eqref{eq:pole_eq_general} decouples into $n$ partial pole equations
\begin{eqnarray}
\label{eq:pole_eq_partial}
p^2 - M_i^2(p^2) &=& 0 \quad\quad (i=1,\dots,n) \,.
\end{eqnarray}
We will for the sake of simplicity assume that each partial pole equation \eqref{eq:pole_eq_partial} has exactly one solution $p^2=m_i^2$,
\begin{eqnarray}
m_i^2 - M_i^2(m_i^2) &=& 0 \,,
\end{eqnarray}
which is necessarily non-negative due to reality of $M(p^2)$.

We can now expand the function $M_i^2(p^2)$ about $p^2=m_i^2$ as
\begin{eqnarray}
M_i^2(p^2) &=& M_i^2(m_i^2) + (p^2-m_i^2)M_i^{2\prime}(m_i^2) + \mathcal{O}\big((p^2-m_i^2)^2\big)
\end{eqnarray}
(where $M_i^{2\prime}(m_i^2)$ is the derivative of $M_i^2(p^2)$ with respect to $p^2$ at $m_i^2$), so that the left-hand side of the pole equation \eqref{eq:pole_eq_partial} reads
\begin{eqnarray}
p^2 - M_i^2(p^2) &=& (p^2-m_i^2)\big(1-M_i^{2\prime}(m_i^2)\big) + \mathcal{O}\big((p^2-m_i^2)^2\big) \,.
\end{eqnarray}
Thus the asymptotic behavior of $(p^2 - M_i^2(p^2))^{-1}$ around $p^2=m_i^2$ reads
\begin{eqnarray}
\frac{1}{p^2 - M_i^2(p^2)} &\xrightarrow[p^2 \rightarrow m_i^2]{}&
\frac{1}{1-M_i^{2\prime}(m_i^2)}\frac{1}{p^2-m_i^2}+ \mbox{regular terms} \,.
\end{eqnarray}
We now make the simplifying assumption, consistent with our systematic neglecting of the wave-function renormalization throughout the text, that the derivative $M_i^{2\prime}(m_i^2)$ vanishes:
\begin{eqnarray}
M_i^{2\prime}(m_i^2) &=& 0 \,.
\end{eqnarray}
Under this assumption we can write the asymptotic behavior of the full propagator $G_{\psi}(p)$, \eqref{eq:S_diagonalized}, for the momentum going on-shell as\footnote{There is no summation over the fermion index $i$. Any summations over the fermion indices will be always denoted explicitly.}
\begin{subequations}
\label{eq:S_asymptotics}
\begin{eqnarray}
G_\psi(p) & \xrightarrow[p^2 \rightarrow m_i^2]{} & \phantom{-}\frac{\mathcal{U}_i(p)\,\mathcal{\bar U}_i(p)}{p^2-m_i^2} + \mbox{regular terms} \,, \quad
\\
G_\psi(-p) & \xrightarrow[p^2 \rightarrow m_i^2]{} & -\frac{\mathcal{V}_i(p)\,\mathcal{\bar V}_i(p)}{p^2-m_i^2} + \mbox{regular terms} \,, \quad\quad
\end{eqnarray}
\end{subequations}
where we denoted
\begin{subequations}
\label{eq:definition_mathcal_UV}
\begin{eqnarray}
\mathcal{U}_i(p) &\equiv&  X(m_i^2) \, e_i \, u_i(p) \,,
\\
\mathcal{V}_i(p) &\equiv&  X(m_i^2) \, e_i \, v_i(p)
\end{eqnarray}
\end{subequations}
and their Dirac conjugate defined in the usual way as $\mathcal{\bar U}=\mathcal{U}^\dag\gamma_0$, $\mathcal{\bar V}=\mathcal{V}^\dag\gamma_0$ (interpretation of these symbols is discussed more below in section~\ref{subsec:interpretation_U_V}). The symbol $e_i$ is the $i$'th canonical basis vector of $n$-dimensional flavor vector space, i.e., with the $j$'th component given by $(e_i)_j=\delta_{ij}$. Symbols $u_i(p)$, $v_i(p)$ are the standard bispinor solutions of the momentum-space Dirac equation\footnote{We suppress the polarizations indices in Eqs.~\eqref{eq:def:u_v} as well as sums over them in Eqs.~\eqref{eq:S_asymptotics}.}
\begin{subequations}
\label{eq:def:u_v}
\begin{eqnarray}
(\slashed{p}-m_i)\,u_i(p) &=& 0 \,,
\\
(\slashed{p}+m_i)\,v_i(p) &=& 0 \,.
\end{eqnarray}
\end{subequations}

Having defined the momentum-dependent matrices $V(p^2)$, $U(p^2)$ (Eq.~\eqref{eq:definition_UV}), it is now useful to define their momentum-independent counterparts $\tilde V$, $\tilde U$ in such a way that their elements on position $i,j$ are given by
\begin{subequations}
\label{eq:def:U_V_tilde}
\begin{eqnarray}
(\tilde V)_{ij} &=&  (V(m_i^2))_{ij} \,,
\\
(\tilde U)_{ij} &=&  (U(m_i^2))_{ij} \,,
\end{eqnarray}
\end{subequations}
i.e., explicitly
\begin{subequations}
\begin{eqnarray}
\tilde V &=&
\left(\begin{array}{cccc}
    V_{11}(m_1^2) & V_{12}(m_1^2) & \cdots & V_{1n}(m_1^2) \\
    V_{21}(m_2^2) & V_{22}(m_2^2) &        & V_{2n}(m_2^2) \\
    \vdots        &               & \ddots & \vdots        \\
    V_{n1}(m_n^2) & V_{n2}(m_n^2) & \cdots & V_{nn}(m_n^2) \\
\end{array}\right)
\,,
\\ \nonumber && \phantom{1} \\
\tilde U &=&
\left(\begin{array}{cccc}
    U_{11}(m_1^2) & U_{12}(m_1^2) & \cdots & U_{1n}(m_1^2) \\
    U_{21}(m_2^2) & U_{22}(m_2^2) &        & U_{2n}(m_2^2) \\
    \vdots        &               & \ddots & \vdots        \\
    U_{n1}(m_n^2) & U_{n2}(m_n^2) & \cdots & U_{nn}(m_n^2) \\
\end{array}\right)
\,.
\end{eqnarray}
\end{subequations}
We can also for convenience define the constant matrix $\tilde X$ as
\begin{eqnarray}
\label{eq:def:X_tilde}
\tilde X &\equiv& \tilde V^\dag P_L + \tilde U^\dag P_R \,.
\end{eqnarray}
Obviously, for constant (momentum-independent) $U$, $V$ we have $\tilde V=V$, $\tilde U=U$ and consequently $\tilde X = X$. In this case the matrices $\tilde V$, $\tilde U$ and $\tilde X$ are also unitary, which need not to be true in general.

\subsection{Interpretation of the $\mathcal{U}$, $\mathcal{V}$ symbols}
\label{subsec:interpretation_U_V}

Let us add a brief comment on how to interpret the symbols $\mathcal{U}_i$, $\mathcal{V}_i$. Assume for that purpose that the self-energy $\Sigma_D$ is a \emph{constant} (i.e., momentum-independent) matrix, i.e., effectively a mass matrix in the Lagrangian. Then the plane-wave solutions to the Dirac equation
\begin{eqnarray}
\big(\I\slashed{\partial}-\boldsymbol{\Sigma}_\psi\big)\psi &=& 0
\end{eqnarray}
with positive and negative energy (we assume $p_0>0$) read
\begin{subequations}
\begin{eqnarray}
\psi_{+}(x) &=& \mathcal{U}(p) \, \e^{-\I p \cdot x} \,,
\\
\psi_{-}(x) &=& \mathcal{V}(p) \, \e^{+\I p \cdot x} \,,
\end{eqnarray}
\end{subequations}
where the quantities $\mathcal{U},\mathcal{V}$ satisfy
\begin{subequations}
\begin{eqnarray}
\big(\slashed{p}-\boldsymbol{\Sigma}_\psi\big)\,\mathcal{U}(p) &=& 0 \,,
\\
\big(\slashed{p}+\boldsymbol{\Sigma}_\psi\big)\,\mathcal{V}(p) &=& 0 \,.
\end{eqnarray}
\end{subequations}
Now using $\boldsymbol{\Sigma}_\psi=\bar X^\dag M X^\dag$ with $M=\diag(m_1,\ldots,m_{n})$ (i.e., momentum-independent version of Eq.~\eqref{frm:SgmdiagX}) we arrive at
\begin{subequations}
\begin{eqnarray}
\mathcal{U}(p) &=& \sum_i X \, e_i \,u_i(p) \>\equiv\> \sum_i \mathcal{U}_i(p) \,,
\\
\mathcal{V}(p) &=& \sum_i X \, e_i \,v_i(p) \>\equiv\> \sum_i \mathcal{V}_i(p) \,,
\end{eqnarray}
\end{subequations}
which (for momentum-independent $X$) coincides with definitions \eqref{eq:definition_mathcal_UV}. Thus, we can understand the symbol $\mathcal{U}_i(p)$ ($\mathcal{V}_i(p)$) as the polarization vector of the fermion (antifermion) of $i$'th flavor with mass $m_i$, or as a generalization of the usual polarization vector $u_i(p)$ ($v_i(p)$) in the case of multicomponent fermion field $\psi$.

\chapter{Nambu--Gorkov formalism for scalars}
\label{app:sclr}

\intro{In this appendix we redo for scalars the analysis done in the previous appendix for fermions, although this time in much more modest way. That is to say, we consider an unspecified number of complex scalar fields and look for a notation (scalar version of the Nambu--Gorkov formalism), allowing for a compact treatment of their propagators and other quantities.}


\section{Nambu--Gorkov doublet}

Consider $n$ complex scalar fields $\phi_i$, $i=1,\ldots,n$, organized into the $n$-plet $\phi$:
\begin{eqnarray}
\phi &\equiv& \left(\begin{array}{c} \phi_1 \\ \vdots \\ \phi_n \end{array}\right) \,.
\end{eqnarray}
Assume that the theory containing this multi-component field $\phi$ is non-invariant under the phase transformation
\begin{eqnarray}
\label{ngsc:phasephi}
\group{U}(1)\,:\qquad \phi &\TransformsTo& [\phi]^\prime \ =\ \e^{\I \theta} \, \phi \,,
\end{eqnarray}
at this moment regardless whether due to explicit or spontaneous symmetry breaking. In any case, non-invariance under \eqref{ngsc:phasephi} means that apart from the propagators of the type $\langle \phi\phi^\dag \rangle$, invariant under \eqref{ngsc:phasephi}, there will be also non-vanishing propagators of the type $\langle \phi\phi^\T \rangle$, non-invariant under \eqref{ngsc:phasephi}.

In order to treat this situation, we introduce, similarly to the case of fermions (section~\ref{app:frm:Maj} of previous appendix), the Nambu--Gorkov field $\Phi$ for scalars, defined in terms of the field $\phi$ as
\begin{eqnarray}
\label{app:sclr:NG}
\Phi &\equiv& \left(\begin{array}{c} \phi \\ \phi^\C \end{array}\right) \,.
\end{eqnarray}
Here $\phi^\C$ is the charge conjugate of $\phi$, defined as\footnote{This time, in contrast to fermions, we do not dedicate a separate appendix to the charge conjugation of scalars.}
\begin{eqnarray}
\label{app:sclr:phiC}
\phi^\C &\equiv& \phi^{\dag\T} \>=\> \left(\begin{array}{c} \phi_1^\dag \\ \vdots \\ \phi_n^\dag \end{array}\right) \,.
\end{eqnarray}
Thus, $\Phi$ is explicitly given as
\begin{eqnarray}
\Phi &=& \left(\begin{array}{c} \phi \\ \phi^{\dag\T} \end{array}\right)
\>=\> \left(\begin{array}{c} \phi_1 \\ \vdots \\ \phi_n \\ \phi_1^\dag \\ \vdots \\ \phi_n^\dag \end{array}\right) \,.
\end{eqnarray}
Notice that charge conjugation of $\Phi$ is given as
\begin{eqnarray}
\Phi^\C &=& \left(\begin{array}{c} \phi^\C \\ \phi \end{array}\right) \,,
\end{eqnarray}
or in other words, it is just a linear combination of $\Phi$ itself:
\begin{eqnarray}
\label{app:sclr:NGcond}
\Phi^\C &=& \sigma_1 \Phi \,.
\end{eqnarray}
where $\sigma_1$ operates in the two-dimension Nambu--Gorkov space. Compare this relation with analogous Majorana condition \eqref{app:major:majorana_cond} for fermions.

\section{Free Lagrangian}

Assume that the free Lagrangian of the field $\phi$ is
\begin{eqnarray}
\label{ngsc:eLfreephi}
\eL_{\mathrm{free}} &=& (\partial_\mu\phi)^\dag(\partial^\mu\phi) - \phi^\dag M_\phi^2 \phi \,,
\end{eqnarray}
where $M^2$ is a Hermitian $n \times n$ matrix. Notice that we assume for simplicity, regarding the applications in the main text, that the free Lagrangian \eqref{ngsc:eLfreephi} is actually invariant under the phase transformation \eqref{ngsc:phasephi}. In terms of the new field $\Phi$ it acquires the form
\begin{equation}
\label{ngsc:eLfreePhi}
\eL_{\mathrm{free}} = \frac{1}{2}(\partial_\mu\Phi)^\dag(\partial^\mu\Phi)-\frac{1}{2}\Phi^\dag M_\Phi^2 \Phi \,,
\end{equation}
where
\begin{eqnarray}
M_\Phi^2 &\equiv& \left(\begin{array}{cc} M_\phi^2 & 0 \\ 0 & M_\phi^{2\T} \end{array}\right) \,.
\end{eqnarray}

The free propagator of the field $\phi$,
\begin{eqnarray}
\I \, D_\phi &=& \langle \phi \phi^\dag\rangle_0 \,,
\end{eqnarray}
corresponding to the Lagrangian \eqref{ngsc:eLfreephi}, reads of course
\begin{eqnarray}
D_\phi &=& \big(p^2-M_\phi^2\big)^{-1} \,.
\end{eqnarray}
The corresponding free propagator of the field $\Phi$,
\begin{eqnarray}
\I \, D_\Phi &=& \langle \Phi \Phi^\dag\rangle_0 \,,
\end{eqnarray}
is easily expressed in terms of $D_\phi$ as
\begin{eqnarray}
D_\Phi &=& \left(\begin{array}{cc} D_\phi & 0 \\ 0 & D_\phi^\T \end{array}\right) \,.
\end{eqnarray}

\section{Propagators}

Consider now the full and 1PI propagators of the Nambu--Gorkov field $\Phi$:
\begin{subequations}
\label{ngsc:GPhi}
\begin{eqnarray}
\I \, G_\Phi &=& \langle \Phi\Phi^{\dag} \rangle
\label{ngsc:GPhi1} \\
&=&
\left(\begin{array}{cc}
\langle\phi\phi^\dag\rangle & \langle\phi\phi^{\C\dag}\rangle \\
\langle\phi^\C\phi^\dag\rangle & \langle\phi^\C\phi^{\C\dag}\rangle
\end{array}\right)
\label{ngsc:GPhi2} \\
&\equiv& \I \left(\begin{array}{cc} A & B \\ C & D \end{array}\right)
\label{ngsc:GPhi3}
\end{eqnarray}
\end{subequations}
and
\begin{subequations}
\label{ngsc:PiPhi}
\begin{eqnarray}
-\I\,\boldsymbol{\Pi}_\Phi &=& \langle \Phi\,\Phi^\dag \rangle_{\mathrm{1PI}}
\label{ngsc:PiPhi1} \\
&=&
\left(\begin{array}{cc}
\langle\phi\phi^\dag\rangle_{\mathrm{1PI}} & \langle\phi\phi^{\C\dag}\rangle_{\mathrm{1PI}}
\label{ngsc:PiPhi2} \\
\langle\phi^\C\phi^\dag\rangle_{\mathrm{1PI}} & \langle\phi^\C\phi^{\C\dag}\rangle_{\mathrm{1PI}}
\end{array}\right)
\\ &\equiv& -\I \left(\begin{array}{cc} a & b \\ c & d \end{array}\right) \,,
\label{ngsc:PiPhi3}
\end{eqnarray}
\end{subequations}
respectively. Notice that both propagators indeed include the components invariant under \eqref{ngsc:phasephi} (i.e., the diagonal entries in the matrix forms \eqref{ngsc:GPhi2}, \eqref{ngsc:PiPhi2}), as well as the components non-invariant under \eqref{ngsc:phasephi} (the off-diagonal entries). Notice also that the expressions \eqref{ngsc:GPhi}, \eqref{ngsc:PiPhi} diagrammatically correspond to
\begin{eqnarray}
\begin{array}{c}
\scalebox{0.85}{\includegraphics[trim = 10bp 12bp 19bp 11bp,clip]{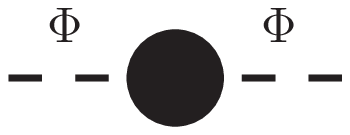}}
\end{array}
&=&
\left(\begin{array}{cc}
\begin{array}{c}
\scalebox{0.85}{\includegraphics[trim = 10bp 12bp 19bp 11bp,clip]{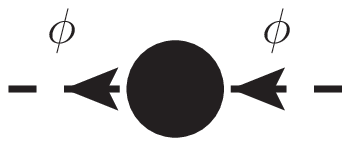}}
\end{array}
&
\begin{array}{c}
\scalebox{0.85}{\includegraphics[trim = 10bp 12bp 19bp 11bp,clip]{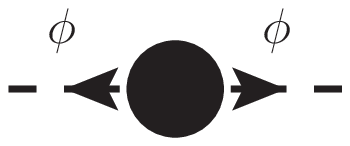}}
\end{array}
\\
\begin{array}{c}
\scalebox{0.85}{\includegraphics[trim = 10bp 12bp 19bp 11bp,clip]{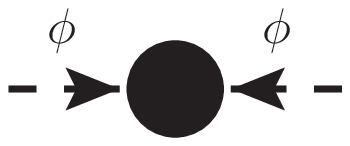}}
\end{array}
&
\begin{array}{c}
\scalebox{0.85}{\includegraphics[trim = 10bp 12bp 19bp 11bp,clip]{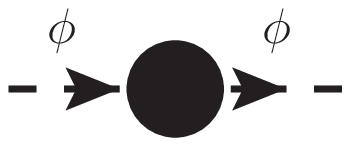}}
\end{array}
\end{array}\right) \,,
\\ \nonumber && \\
\begin{array}{c}
\scalebox{0.85}{\includegraphics[trim = 10bp 12bp 19bp 11bp,clip]{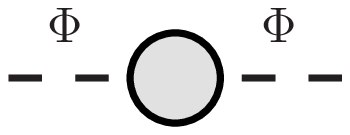}}
\end{array}
&=&
\left(\begin{array}{cc}
\begin{array}{c}
\scalebox{0.85}{\includegraphics[trim = 10bp 12bp 19bp 11bp,clip]{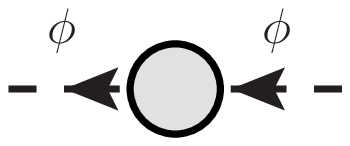}}
\end{array}
&
\begin{array}{c}
\scalebox{0.85}{\includegraphics[trim = 10bp 12bp 19bp 11bp,clip]{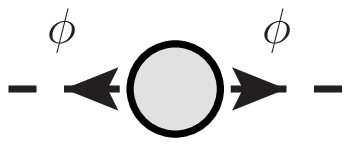}}
\end{array}
\\
\begin{array}{c}
\scalebox{0.85}{\includegraphics[trim = 10bp 12bp 19bp 11bp,clip]{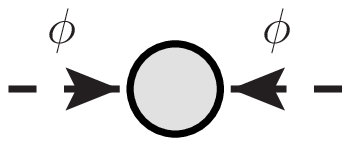}}
\end{array}
&
\begin{array}{c}
\scalebox{0.85}{\includegraphics[trim = 10bp 12bp 19bp 11bp,clip]{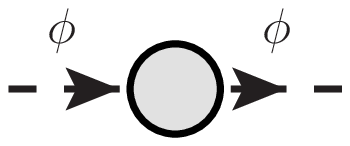}}
\end{array}
\end{array}\right) \,.
\end{eqnarray}


The condition \eqref{app:sclr:NGcond} for $\Phi$ induces the following conditions for the propagators:
\begin{subequations}
\label{ngsc:prpNG}
\begin{eqnarray}
G_{\Phi} &=& \sigma_1 \, G_{\Phi}^\T \, \sigma_1 \,, \\
\boldsymbol{\Pi}_\Phi &=& \sigma_1 \, \boldsymbol{\Pi}_\Phi^\T \, \sigma_1 \,.
\end{eqnarray}
\end{subequations}
Imposing these conditions on the matrix forms \eqref{ngsc:GPhi3} and \eqref{ngsc:PiPhi3} we find
\begin{subequations}
\begin{eqnarray}
B &=& B^\T \,, \\
C &=& C^\T \,, \\
D &=& A^\T
\end{eqnarray}
\end{subequations}
and
\begin{subequations}
\begin{eqnarray}
b &=& b^\T \,, \\
c &=& c^\T \,, \\
d &=& d^\T \,,
\end{eqnarray}
\end{subequations}
respectively.

Moreover, one can assume that the propagators are Hermitian:
\begin{subequations}
\label{ngsc:prpherm}
\begin{eqnarray}
G_{\Phi} &=& G_{\Phi}^\dag \,, \\
\boldsymbol{\Pi}_\Phi &=& \boldsymbol{\Pi}_\Phi^\dag \,.
\end{eqnarray}
\end{subequations}
This yields
\begin{subequations}
\begin{eqnarray}
A &=& A^\dag \,, \\
C &=& B^\dag \,, \\
D &=& D^\dag \,,
\end{eqnarray}
\end{subequations}
and
\begin{subequations}
\begin{eqnarray}
a &=& a^\dag \,, \\
c &=& b^\dag \,, \\
d &=& d^\dag \,.
\end{eqnarray}
\end{subequations}

As a result of the two conditions \eqref{ngsc:prpNG} and \eqref{ngsc:prpherm} we obtain
\begin{eqnarray}
G_\Phi &=&  \left(\begin{array}{cc} A & B \\ B^\dag & A^\T \end{array}\right) \,,
\\
\boldsymbol{\Pi}_\Phi &=& \left(\begin{array}{cc} a & b \\ b^\dag & a^\T \end{array}\right) \,,
\end{eqnarray}
where
\begin{subequations}
\begin{eqnarray}
A &=& A^\dag \,, \\
B &=& B^\T \,,
\end{eqnarray}
\end{subequations}
and
\begin{subequations}
\begin{eqnarray}
a &=& a^\dag \,, \\
b &=& b^\T \,.
\end{eqnarray}
\end{subequations}

\section{Another basis}

The relation \eqref{app:sclr:NGcond} resembles the Majorana condition \eqref{app:major:majorana_cond} for fermions. Indeed, while Majorana fermion field is a real field, so is also the scalar field $\Phi$, satisfying condition \eqref{app:sclr:NGcond}. This can be seen more clearly in another basis. The complex field $\phi$ can be decomposed into its real and imaginary part
\begin{eqnarray}
\phi &=& \frac{1}{\sqrt{2}} (\phi_{R} + \I\phi_{I})
\end{eqnarray}
in such a way that
\begin{eqnarray}
\phi^\C &=& \frac{1}{\sqrt{2}} (\phi_{R} - \I\phi_{I}) \,.
\end{eqnarray}
Now we can define a new, strictly real field $\Phi^\prime$ in terms of $\phi_{R}$, $\phi_{I}$:
\begin{eqnarray}
\Phi^\prime &\equiv& \left(\begin{array}{c} \phi_{R} \\ \phi_{I} \end{array}\right) \,.
\end{eqnarray}
It is now straightforward to see that the fields $\Phi$ and $\Phi^\prime$ are actually related by the linear transformation
\begin{eqnarray}
\Phi &=& U \Phi^\prime \,,
\end{eqnarray}
where
\begin{eqnarray}
U &\equiv& \frac{1}{\sqrt{2}} \left(\begin{array}{rr} 1 & \I \\ 1 & -\I \end{array}\right)
\end{eqnarray}
is a unitary matrix. Now in terms of $\Phi^\prime$ the condition \eqref{app:sclr:NGcond} just reads
\begin{eqnarray}
\Phi^{\prime\C} &=& \Phi^\prime \,.
\end{eqnarray}

As we do not use the basis $\Phi^\prime$ extensively in the main text, we do not present here expression for the propagators and other quantities in its terms. Nevertheless, let us, just for curiosity, observe how the free Lagrangian \eqref{ngsc:eLfreePhi} looks in it:
\begin{eqnarray}
\eL_{\mathrm{free}} &=& \frac{1}{2}(\partial_\mu\Phi^\prime)^\dag(\partial^\mu\Phi^\prime) - \frac{1}{2}\Phi^{\prime\dag} M_{\Phi^\prime}^{2} \Phi^\prime \,,
\end{eqnarray}
where
\begin{subequations}
\begin{eqnarray}
M_{\Phi^\prime}^{2} &=& U^\dag M_{\Phi}^{2} U
\\
&=& \frac{1}{2}
\left(\begin{array}{cc}
M_\phi^{2}+M_\phi^{2\T} & \I(M_\phi^{2}-M_\phi^{2\T}) \\
-\I(M_\phi^{2}-M_\phi^{2\T}) & M_\phi^{2}+M_\phi^{2\T}
\end{array}\right) \,.
\end{eqnarray}
\end{subequations}
Notice that mass matrix $M_{\Phi^\prime}^{2}$ in the strictly real basis $\Phi^\prime$ is not only Hermitian, but also real, due to Hermiticity of $M_\phi^{2}$.






\end{document}